\documentclass[12pt]{book}

\def\Re{\mathop{\mathgroup\symoperators Re}\nolimits}
\def\Im{\mathop{\mathgroup\symoperators Im}\nolimits}

\begin{document}

\title{ELECTRONIC STRUCTURE, CORRELATION EFFECTS
AND PHYSICAL PROPERTIES OF d- AND f-TRANSITION
METALS AND THEIR COMPOUNDS}
\author{By V.Yu.Irkhin and Yu.P.Irkhin}
\date{\ }

\maketitle

\chapter*{PREFACE}

There exist several reasons to pick out the physics of transition group
metals as a separate branch of solid state physics. The first one is the
variety of properties of transition metals (TM) and their compounds, which
are not only much more complicated, but have some important peculiarities in
comparison with simple metals. In particular, ferromagnetic ordering takes
place practically in TM and their compounds only. The second reason is the
importance of investigating TM both from theoretical and practical point of
view. TM yield an example of a strongly interacting many-electron system,
which does not enable one to introduce in a simple way an effective
one-electron potential. Thus we deal with the full many-particle quantum
problem which requires application of the all modern theoretical physics
methods.

The term ``transition metals'' has two meanings. In the narrow sense of the
word, TM are elements with partly occupied 3d, 4d and 5d-shells, which form
large periods in the Mendeleev periodic table. Sometimes this term is applied
to all the elements with partly filled inner electron shells
(transition-group elements), including 4f (rare earths, RE) and 5f
(actinides). As a rule, we shall use the notion ``transition metals'' in the
narrow sense, but all the classes of transition-group elements will be
discussed in the book.

We have 24 TM, 13 RE and 8 actinides, so that about one half of elements
belong to TM in the broad sense (at the same time, there exist only 25 simple
metals). Among TM we find most practically important metals which exhibit
maximum strength, melting temperature etc. The most well-known example is Fe:
up to now we live in the iron age. More and more applications find rare
earths. For example, the intermetallic compound SmCo$_5$ yields a basis for
best permanent magnets. Actinidies are widely used in nuclear energetics.

A rather large number of books is available which treat metal physics. They
are devoted mainly to simple metals and contain usually some separate
paragraphs concerning TM. True, detailed monograps [15-17] discuss properties
of rare-earth metals and their compounds. At the same time, similar books on
d-metals seem to be absent. This is probably explained by that
systematization of corresponding large material, which may be found in
original and review papers on particular topics, is a more difficult problem.
However, in our opinion, it is useful to collect most important results on TM
physics and discuss some general regularities.

Let us list some distinctive physical properties of TM and their compounds:

(i) large binding energies (high strength and melting temperature)

(ii) large electronic specific heat (heavy electron masses)

(iii) strong magnetism: large paramagnetic susceptibility and sometimes
ferro- or antiferromagnetic ordering

(iv) superconductuvity, often with high $T_{{\rm c}}$

(v) anomalous transport properties (in particular, extraordinary
halvanomagnetic effects).

The book is devoted to consideration of these non-trivial physical
properties, especial attention being paid to connection with the electronic
structure. (The latter term includes both the properties of partially filled
d- and f-shells and the anomalies of band structure.) We do not pretend to
consider all the variety of TM compounds and alloys, but try to illustrate
some interesting physical phenomena, which are not pronounced for elemental
metals, by some bright examples.

The book contains, where possible, a simple physical discussion of a number
of problems. At the same time, we use widely in last three Chapters such
methods of theoretical physics as the second quantization, atomic
representation, Green's function approach. These methods permit to apply
microscopic many-electron models which describe systems with strong
interelectron correlations. Besides traditional questions of the solid state
physics, we treat some modern topics, e.g. magnetism of highly correlated and
low-dimensional electron systems, anomalous properties of exotic rare earth
and actinide systems (Kondo lattices, heavy-fermion compounds), formation of
unusual quantum states with non-trivial excitation spectrum etc. More
difficult mathematical aspects of these topics are considered in Appendices
A-P. Inclusion of numerous rather long Appendices makes the ``topological''
structure of the book somewhat non-trivial and non-traditional. However, such
a structure reflects many-sided connections which exist between various
branches of TM physics.

At present the TM theory is far from completeness, and a number of important
problems are still not resolved. Therefore the treatment of some TM
properties may seem to be not so clear and logical as that of corresponding
issues of general solid state physics in classical textbooks and monographs
[1-14]. However, we believe that description of the modern complicated
situation in the TM physics is justified since this may excite the interest
in the unsolved questions and stimulate further researches.

The plan of the book is as follows. In the introductory Chapter 1 we treat
atomic aspects of TM physics which, unlike the case of simple metals, are
rather important since d- and especially f-states retain in a large measure
atomic features. Detailed mathematical consideration of some related
questions is given in Appendices A-C. In particular, we review applications
of the Racah's angular momentum formalism and Hubbard's many-electron
operator representation in the solid state theory, which are seldom discussed
in the literature on the metal theory.

Chapter 2 considers the electronic structure of TM from the ``band'' side.
We review briefly methods of band structure calculations, including the
density functional approach, with especial attention to TM peculiarities.
Besides that, we discuss some simple model approaches to the band spectrum
and related experimental (especially spectral) data. We consider also
theoretical and experimental results concerning the Fermi surfaces. In
Chapter 3 we discuss thermodynamical properties of TM: cohesive energy and
related properties, stability of crystal structures, and specific heat,
electronic contributions being treated in details.

Chapter 4 deals with magnetic properties. Here we discuss various theoretical
models describing highly correlated d- and f-electrons. These models permit
all-round consideration of the complicated metallic magnetism problem which
includes the atomic (``localized'') and band (``itinerant'') aspects of
d-electron behaviour. A large number of Appendices, related to this Chapter
(D-K), demonstrate concrete practical applications of the many-electron
models, mainly within the simple method of double-time retarded Green's
functions.

Chapter 5 is devoted to transport phenomena in TM which demonstrate a number
of peculiarities in comparison with simple metals, e.g., occurrence of
spontaneous (anomalous) effects. Quantitative treatment of these effects is
performed with the use of density-matrix approach in the operator form
(Appendix M).

Finally, in Chapter 6 we treat some questions of the anomalous f-compound
theory. In particular, we discuss various mechanisms for occurence of
``heavy'' electron mass and the problem of competition between the Kondo
effect and magnetic interactions. We consider also model descriptions of
electronic structure in two-dimensional highly-correlated systems including
copper-oxide high-$T_{{\rm c}}$ superconductors.

We intended to consider in the book all main properties of TM and
corresponding theoretical concepts. However, the choice and volume of the
material in different Chapters are determined to some extent by scientific
interests of the authors. In particular, we pay a great attention to the
theory of magnetism and transport phenomena, but discuss in less details
lattice properties, and almost do not touch superconductivity (the latter
topic develops now extensively and is widely discussed in modern review and
monograph literature). We list once more some examples of non-traditional
questions which are treated in the book: the influence of density of states
singularities on electron properties; many-electron description of strong
itinerant magnetism; the problem of quenching and unquenching of orbital
magnetic magnetic momenta in solids and their significance for magnetic
anisotropy; microscopic theory of anomalous transport phenomena in
ferromagnets.

The book is partly based on the lection course of the transition metal
physics, which was read for a number of years at the Ural State University.

Except for numerical estimations, we use often in formulas the system of
units with $e=k_{{\rm B}}=\hbar =1$. We hope that the book may be of interest
for researchers which work in the solid state physics and for beginners, both
for theorists and experimentators. Some part of the material may be used in
lection courses for students.

The authors are grateful to A.Zarubin, A.Nasibullin and A.Katanin for the
help in preparing the electronic version of the book.

\tableofcontents

\chapter{INTRODUCTION}

\section{Partly filled atomic shells and electron localization in transition
metals}

The ``hydrogen-like'' scheme of energy levels depending on one-electron
principal and orbital quantum numbers $n$ and $l$ corresponds to consecutive
filling of atomic shells with increasing $n$ and $l$. (The ``accidental''
degeneracy of the levels with different $l$ for a given $n$ in the purely
Coulomb potential $1/r$, which is connected with a dynamical symmetry, is
lifted for many-electron atoms where the potential $r$-dependence is
modified.) However, in the periodic table this sequence is violated several
times. This results in formation of large periods of transition elements with
partly filled d(f)-shells and peculiar physical properties.

To first time, such a situation occurs in the fourth period where filling of
4s-states starts in potassium, the 3d-shell remaining empty. This tendency
holds also in calcium (configuration of valence electrons is 4s$^2$), and
filling of 3d-shell starts only for the next element, i.e.scandium
(configuration 3d 4s ). This element opens the 3d transition group (iron
group). The filling of the 3d-shell goes in a not quite regular way (Table
1.1). So, chromium has the atomic confuiguration 3d$^1$4s$^2$ (instead of
3d$^4$s$^2$), and copper 3d$^{10}$4s (instead of 3d$^9$4s$^2$). One can see
that a tendency to formation of configurations d$^0$, d$^5$ and d$^{10}$
exists. The stability of these configurations is apparently connected with
that they correspond to zero summary orbital moment of all the d-electrons,
i.e. to a spherically symmetric electron density. It should be noted that in
some cases copper demonstrates a considerable contribution of d$^9$s
configuration, which is connected with an appreciable s-d hybridization.
Therefore we shall discuss Cu side by side with transition metals.

A situation, similar to 3d metals, takes place for 4d, 5d, 4f and 5f
transition metals groups (Tables 1.1, 1.2). Thus the filling of d- and
f-shells is delayed, and, after it begins, the electrons from the 4s, 4p
shells with higher energies do not pass into ``free'' d(f)-states. These
phenomena are connected with inapplicability of the simple one-electron
picture, which is based on the hydrogen atom theory, for many-electron atoms.

Consider the radial atomic potential $V_l(r)$, ``felt'' by electrons. This is
an effective potential which is obtained after averaging two- particle
interaction among electrons and including the rotational motion energy
(centrifugal potential):
$$
V_l(r)=-\frac{Z(r)}r+\frac{l(l+1)}{r^2},
\eqno{(1.1)}
$$
where $Z(r)$ is an effective nuclear charge, which depends on the electron
coordinate $r$ (measured in Bohr radia). For the hydrogen atom we have
$Z(r)=1$ and $V_l(r)$ has the usual form with a minimum at $r_0=l(l+1)$
(Fig.1.1). This picture holds in hydrogen-like atoms where the condition
$Z(r)=$ const is satisfied to high accuracy. However, in more complicated
situations the dependence $Z(r)$ becomes important because of non-uniform
screening of nuclear potential. It is this dependence that may lead to an
anomalous form of the function $V_l(r)$ for a large centrifugal term (i.e.
sufficiently large values of $l$). In particular, for
$$
Z(r)=\left\{
\begin{array}{ll}
A/r & ,\qquad r_1<r<r_2 \\
\mathrm{const} & ,\qquad r<r_1,\quad r>r_2
\end{array}
\right. ,
\eqno{(1.2)}
$$
$V_l(r)$ may have two minima separated by a potential barrier. Concrete
calculations demonstrate that $V(r)$ may even become positive in an interval
$[r_1,r_2]$. The results for Ba and La are shown in Fig.1.2. In fact, the
interval $[r_1,r_2]$ corresponds to the position of orbits of 5s- and
5p-electrons which screen strongly the nucleus. In other words, strong
repulsion between 4f and s,p-electrons occurs in the region of localization
of the latter.

In the presence of two minima, the electron density may be concentrated in
any one, depending on the form of $Z(r)$. The energy of the corresponding
states is, generally speaking, considerably different, as one can see from
Fig.1.2. At passing from Ba to La, the lanthanoide collapse of 4f-states,
i.e. a sharp decrease of 4f-shell radius takes place. Further, for cerium
4f-states become more energetically favourable than 5s, 5p and 6s ones, the
maximum of 4f-electron density passes to the first minimum and filling of
4f-levels starts.

Thus the main reason of irregular filling of atomic levels are interelectron
interaction and appreciable value of the orbital energy at $l\neq 0$. These
effect become important from $l=2$ and are strongly pronounced for rare
earths ($l=3$). In Fig.1.3 the data on energy levels of external electrons
and their density distribution are shown for the gadolinium atom.

Now we treat the question about true capacity of d(f)-states which remain in
transition elements, from the point of view of the simple one-electron
theory, partially unfilled (despite filling of higher shells). According to
the latter theory, each shell with the orbital quantum number $l$ may contain
$2(2l+1)$ electrons, their energies being equal because of spherical
symmetry. However, this is is not true when interelectron correlations are
taken into account. This may be performed in the many-configuration
approximation [20] where a Hartree-Fock-type potential, which depends on
electron configuration, is introduced.

We present a simple illustration of this effect. Write down the energy of $n$
electrons in a $l$-shell (the principal quantum number is dropped for
brevity)
$$
E_l^n=n\varepsilon _l+\frac 12n(n-1)Q_l ,
\eqno{(1.3)}
$$
where $\varepsilon _l$ is the sum of the kinetic energy and additive part of
the potential energy, $Q_l$ is the interelectron repulsion for the shell.
The energies of the configurations $l^{n+1}$ and $l^nl^{\prime }$ (e.g.,
d(f)$^{n+1}$ and d(f)$^n$s) read
$$
E_l^{n+1} =(n+1)\varepsilon _l+\frac 12n(n+1)Q_l ,
\eqno{(1.4)}
$$
$$
E_l^n =n\varepsilon _l+\varepsilon _{l^{\prime }}+\frac
12n(n-1)Q_l+nQ_{ll^{\prime }}
$$
with $Q_{ll^{\prime }}$ being the repulsion between $l$ and $l^{\prime
}$-electrons. One can see that the configuration $l$ has the higher energy
provided that
$$
Q_l-Q_{ll^{\prime }}>(\varepsilon _{l^{\prime }}-\varepsilon _l)/n .
\eqno{(1.5)}
$$
In transition elements, the intrashell repulsion is much stronger than the
intershell one, $Q_{sd(f)}\ll Q_{d(f)}$, and the difference $\varepsilon
_s-\varepsilon _{d(f)}$ is not too large. Therefore the maximum possible
filling of the d(f)-shell is not energetically favourable.

The atomic picture of 4f-electrons which are well localized holds in metals.
A possible exception is given by cerium, which opens the 4f-series, so that
the f-electron lies apparently near the centrifugal potential barrier. The
$\gamma -\alpha $ transition in the metallic Ce is assumed now to be
connected with the tunneling of the f-electron through the barier and its
appreciable delocalization (see 6.5). A similar situation occurs in Sm, Nd
and Pr under high pressure of order of 1Mbar [21]. Besides that, in some rare
earth compounds f-electrons become partly delocalized because of
hybridization mixing with conduction electrons.

The description of d-states in solids is rather difficult. Unlike
4f-electrons, they demonstrate both ``localized'' and itinerant features.
Such a behaviour is determined by the corresponding atomic potential, which
differs essentially from that for s,p-electrons. The presence of a potential
barrier results in an appreciable localization of d-states and decrease of
overlap between d-functions at different lattice sites. However, the density
of d-electrons still lies partly outside the potential barrier. As a result,
their kinetic energy turns out to be considerable, and the corresponding
bandwidth values are comparable with those for s,p-electrons.

The second factor, which determines the d-electron spectrum, is the Coulomb
repulsion between them. This depends strongly on the number of d-electrons
inside the barrier. Therefore the effective one-electron energies and
wavefunctions should be rather sensitive to the many-electron configuration
d$^{n\pm 1}$s$^{1\mp 1}$ , and the degeneracy is lifted:
$$
\varepsilon _d=\varepsilon _d(d^n),\qquad \varepsilon _d(d^n)<E_F<\varepsilon
_d(d^{n+1})\simeq \varepsilon _d(d^n)+Q
\eqno{(1.6)}
$$
(the Fermi energy $E_F$ determines the band filling). In Appendix C we
discuss these effects in more detail by using many-electron Hamiltonian and
angular momentum theory.

Owing to increase of the d-shell radius, the localization degree of
d-electrons decreases when we move in the periodic Table both from left to
right and from up to below. Thus 3d-electrons are considerably more localized
than 4d- and especially 5d-electrons. This explains the fact that magnetic
ordering, which is connected with existence of pronounced local moments,
occurs for 3d-metals only. A similar difference in the localization degree
takes place between 4f-electrons in rare earths and 5f-electrons in
actinides. In light actinides (from Th to Pu) 5f-electrons form wide energy
bands and may be considered as itinerant, and in actinides with higher atomic
numbers they are localized. However, the delocalization in Am, Cm, Bc and Cf
may occur under pressure [22,23].

\section{Atomic and band approaches in the transition element theory}

Most important peculiarity of transition metals from the theoretical point of
view is an important role of electron correlations. There exist a number of
approaches to the problem of treating many-electron (ME) systems with strong
correlations. The first one was the self-consistent field (Hartree-Fock)
approximation at solving the Schroedinger equation for ME atoms, which
yielded satisfactory quantitative results. The Hartree-Fock method allows to
take into account ME atomic terms, but its full version requires the solution
of a complicated system of integro-differential non-linear equations [20], so
that its direct generalization on solids (systems of large number of atoms)
is hardly possible. Main successes of the solid state theory were connected
with first-principle one-electron band structure calculations. Modern
versions of this approach, which are based on the spin-density functional
method (2.3), enable one to obtain a precise description of ground state
characteristics [24].

At the same time, the band theory is wittingly insufficient for strongly
localized f-states, and also at treating some physical phenomena, e.g.
magnetism (especially at finite temperatures) and metal-insulator
(Mott-Hubbard) transition [25]. Such oversimplified versions of the
Hartree-Fock method, as the Stoner mean-field theory of itinerant magnetism,
turned out to be not too successful. Later, some shortcomings of the Stoner
theory were improved by semiphenomenological spin-fluctuation theories [26],
which, however, do not take into account in most cases correlation effects in
the ground state. To treat electron correlations, various perturbation
approaches in the electron-electron interaction (e.g., diagram techniques)
were used. Besides that, the Fermi-liquid theory was proposed to describe
systems with strong electron interactions [27]. However, this theory is
violated in the cases where the correlations result in a reconstruction of
the ground state, e.g. for systems with a Mott-Hubbard gap [25].

Another approach to the problem of electron correlations was proposed by
Hubbard [28-31] who considered the simplest microscopic model including the
strong on-site Coulomb repulsion. Starting from the atomic limit, Hubbard
carried out a decoupling of Green's functions and obtained an interpolation
solution, describing both atomic and band limits for s-states [28] and the
simplest model of degenerate band [29]. In the subsequent paper [30] this
solution was improved to describe the metal-insulator transition. In the
paper [31] Hubbard proposed a general formalism of ME X-operators (atomic
representation) which enables one to take into account intraatomic
interactions in the zeroth-order approximation. This formalism for real
atomic configurations is considered in Appendix A.

It is clear from the physical point of view that interelectron correlations
are most important for the electrons of the same atomic shell (the equivalent
electrons). The modern form of the theory of atomic spectra is based on the
Racah formalism for angular momenta (see, e.g., [20]). In some cases,
calculations may be considerably simplified by using the irreducible tensor
operator technique in the second quantization representation [32] (related
questions are considered in Appendix B). This powerful mathematical technique
introduces the representation of many-electron quantum numbers $\Gamma
=\{{SL\mu M\}}$ instead of one-electrons ones, $\gamma =\{{lm\sigma \}}$,
$$
\mathbf{S}=\sum_i\mathbf{s}_i,\qquad \mathbf{L}=\sum_i\mathbf{l}_i
\eqno{(1.7)}
$$
being the total angular spin and orbital momenta and $\mu $ and $M$ their
projections. Then numerous possible combinations of $\gamma $-sets for a
partly occupied shell are replaced by the sets of $\Gamma $. The total number
of ME state is the same, but the energy degeneracy is lifted, so that in most
physical problems one can retain only the lowest ME term. According to Hund's
rules, this term corresponds to maximum $L$ and $S.$ Within this approach,
the problem of electrostatic interaction in the system is reduced to
calculating a few Slater integrals $F^{(p)}$ (Appendix C). These integrals
may be calculated with the use of atomic wavefunctions [33] or determined
from experimental data on energies of atomic terms (see [34]).

Information on the energy of the Hund atomic terms may be obtained from the
values of atomic ionization potentials. Table 1.1 presents the values of
third ionizaton potentials (of M$^{2+}$ ions) in d-rows, which characterize
the binding of d-electrons (first and second ionization potentials correspond
as a rule to moving away of s-electrons from the d$^n$s$^2$ configuration).
One can see that with increasing atomic number the binding in the d-rows
increases. Such a dependence explains the corresponding increase of
d-electron localization in solids, which is discussed in previous Section.

In solids the values of the Slater integrals become modified. In the recent
paper [35] the lowering $\Delta F^{(p)}/F^{(0)}$, which is due to the change
in correlation effects (in particular, in interconfiguration interaction) in
a crystal, was calculated. At $p=2$, the values of this quantity of 0.22 and
0.16 were obtained for Ti$^{2+}$(d$^2$) and Ni$^{2+}$ (d$^8$) ions
respectively.

Besides the band approach, the ME approach with inclusion of the atomic
spectroscopy mathematical technique should be very useful for the development
of quantitative theory of solids. Such an approach is applicable in the case
of strong electron localization and especially effective for equivalent
electron shells of d- and f-ions with a complicated term structure. It is
important for the solid state theory that the second quantization
representation (unlike the standard atomic theory) enables one to consider
processes with changing electron number in a shell or an atom (Appendix C).
An account of band energy dependence on many-electron quantum numbers
corresponds to the degenerate Hubbard model (Appendices C,H).

The localization degree of 4f-electrons is sufficient to justify using the
atomic picture. On the other hand, for d- and 5f-electrons the crystal field
effects result in at least partial quenching of orbital momenta and
destruction of atomic terms. However, atomic structure is important for local
effects (e.g., in the case of d-metal impurities). Atomic description may be
useful also at considering some physical phenomena in periodic crystals
(e.g., strong magnetism with pronounced local magnetic moments, formation of
an insulating state in d-compounds).

An interesting many-electron term effect for a pure metal is the 6eV
sattelite in the XPS spectrum of nickel which may be attributed to the
multiplet strucutre of the d$^8$-configuration (see also 2.6). Another
example is provided by Auger spectra for d-states which exhibit clearly the
term structure. This is due to large localization of d-electrons in the
presence of highly correlated two-hole atomic states which occur after the
core hole decay. The Coulomb interaction of d-holes $U$ in Cu is so large
that the two-hole final state is split off from the one-hole band states.
Therefore one can distinguish in the Cu spectrum (Fig.1.4) all the terms of
the d$^8$ configuration.

A somewhat less pronounced multiplet structure of the d-configuration is
observed in the Auger spectrum of metallic nickel. The value of $U$ in Ni is
of order of the one-hole bandwidth and the two-hole states mix strongly with
band states (the initial state is a mixture of d and d configurations).
Fig.1.5 shows an interpretation of the L$_3$VV spectrum in terms of the
multiplet structure corresponding to the main peak $^1$G with
$$
U(^1G)=F^{(0)}+\frac 4{49}F^{(2)}+\frac 1{441}F^{(4)}.
\eqno{(1.8)}
$$
The comparison of XPS and L$_3$VV spectra for nickel and its compounds [37]
permitted to estimate the value of $U$ and determine its dependence on
crystal surrounding. Experimental energies of the XPS satellite correlate
with the energy of the L$_3$VV peak in all the cases. A fit of the Slater
integral yields
$$
F^{(0)}=1.7,\quad F^{(2)}=9.6,\quad F^{(4)}=6.4
$$
(in eV) which leads to
$$
U(^1S)=6.3,\quad U(^1G)=2.5,\quad U(^3P)=7.85,
\quad U(^1D)=1.6,\quad U(^3F)\simeq 0.
$$
The considerable decrease in the value of F in comparison with atomic one
demonstrates strong screening of the Coulomb interaction by conduction
electrons. The overlap of the atomic $^3F$ state with bandlike states may
influence appreciably electronic properties of nickel.

Besides the energy characteristics, the transition propabilities in spectral
measurements are also of interest for investigating the ME term structure
features in the solid state. The probabilities of the Auger processes
L$_{2,3}$M$_{45}$M$_{45}$ for the d$^8$ multiplet of Cu were calculated in
[38] with the use of modified wavefunctions of continuum Auger electrons
(using the plane waves yielded a drastic disagreement with experimental
data). The results are presented in Table 1.3. One can see that the
probabilities are not proportional to term multiplicities, which indicates a
considerable change of atomic states in the metal.

The values of Slater integrals with account of Coulomb interaction screening
by extra-atomic electrons in two-hole states, which were obtained in [38],
are
$$
F^{(2)}=10.0,\qquad F^{(4)}=5.6 .
$$
They are appreciably suppressed in comparison with the atomic values
$$
F^{(2)}=11.65,\qquad F^{(4)}=7.18 .
$$

\section{Crystal field and orbital momenta in solids}

As discussed above, electrostatic interactions in free transition-element
atoms result in occurrence of partly filled d(f)-shells, their states being
characterized by many-electron quantum numbers. Now we consider the role of
extra interactions which are present in crystals containing such atoms. In
particular, we discuss the problem of role of orbital magnetic momenta in
solids.

The orbital momentum $L$ is one of principal atomic characteristics. This
quantum number determines the ground state of the atom according to the
second Hund's rule and the general picture of excited terms. Existence of
orbital magnetic momenta (OMM) in ME atomic configurations should result in a
number of important effects in solids. In particular, one might expect the
occurence of the orbital magnetism with large magnetic momenta and strong
anisotropy. Such a situation does take place in rare-earth systems [16,39]
where magnetism is determined by the total angular momentum $\mathbf{J=L+S}$.
Orbital momenta play also an important role in actinides because of strong
spin-orbital coupling.

In most d-metals and their compounds, the contribution of OMM to physical
properties is considerably smaller (e.g., for Ni the investigation of L
absorption edge [40] yielded $l=0.05m$). This is explained by that
d-electrons are influenced by crystal field (CF) considerably stronger than
well-localized 4f-electrons. The ME levels of d-ions are split by CF, so that
OMM become quenched: $\langle \mathbf{L}\rangle =0$ in not too high magnetic
fields. However, other interactions in the crystal unquench them partly, and
(even small) OMM that arise are important for the anisotropy of physical
properties due to coupling to the lattice. In particular, strong anisotropy
of paramagnetic susceptibility owing to the orbital contribution takes place
in a number of hexagonal transition metals (see a detailed discussion in
Sect.4.2).

In the problem of quenching of orbital momenta, three basic interactions
should be taken into account: the Coulomb interaction $V_Q$, the spin-orbit
coupling $V_{so}$ and the CF potential $V_{cf}$. There occur three cases.
For a weak crystal field where
$$
V_{cf}\ll V_{so}\ll V_Q
\eqno{(1.9)}
$$
we have the coupling scheme
$$
\sum_i\mathbf{s}_i=\mathbf{S},\qquad \sum_i\mathbf{l}_i\mathbf{=}\mathbf{L},
\qquad \mathbf{L+S=J} .
\eqno{(1.10)}
$$
Then OMM are unquenched (the whole quantum number set is $\{SLJM{\}}$). The
coupling between $\mathbf{L}$ and $\mathbf{S}$ is rigid, and CF orients the
total momentum $\mathbf{J}$ in some crystallographic directions, which
results in a large magnetic anisotropy. The weak field approximation is valid
for rare-earth ions where 4f-shells are strongly screened from $V_{cf}$ by
s,p,d shells, and the value of $V_{so}$ is appreciable due to large nuclear
charge $Z$.

For the intermediate crystal field
$$
V_{so}\ll V_{cf}\ll V_Q
\eqno{(1.11)}
$$
$\mathbf{L}$ and $\mathbf{S}$ still persist, but the momentum $\mathbf{J}$ is
not formed because of CF splitting. Then OMM are oriented in easy crystal
directions and spin momenta are coupled to them by a weak spin-orbit
coupling. At inclusion of an external magnetic field $H$ in a hard direction,
the deviation of spin momentum from the easy direction will be proportional
to the ratio $\mu _BH/V_{cf}$. The orbital contribution to magnetization will
be negligible because of smallness of the ratio $\mu _BLH/V_{cf}$ in any
realistic fields. The intermediate-field situation apparently takes place for
d-impurities in some salts [41] and for some non-metallic d-compounds. This
case is compatible with orbital ordering, which exists in a number of
d-systems [42].

Finally, for the strong crystal field
$$
V_{so}\ll V_Q\ll V_{cf}
\eqno{(1.12)}
$$
the Russel-Saunders coupling between d-electrons is totally destroyed. Then
their energy levels are to be considered as one-electron ones which are split
by CF. The Coulomb interaction should be now treated as a perturbation for
the states with a new symmetry, and orbital momenta are transformed into
quasimomenta which are determined by the corresponding irreducible
representations of the point group. The strong field situation is usually
assumed to take place for transition metals and their alloys. However, the
estimation $V_{cf}\sim V_Q$ seems to be more realistic in such systems. This
leads to difficulties at developing a consistent quantitative theory.

The CF theory is well developed for paramagnetic ions in non-metallic
crystals where a general qualitative analysis of the d- shell structure can
be made on the basis of the local point group, corresponding to a given site.
General expansion of the CF potential in spherical functions has the form
$$
V_{cf}=\sum_{\lambda \mu }A_\lambda ^\mu \,\overline{r^\lambda }\,Y_{\lambda
\mu }(\theta ,\phi ).
\eqno{(1.13)}
$$
For a $l-$shell, it is sufficient to retain in (1.13) the terms with $\lambda
\leq 2l$ only; $\lambda $ is even in the presence of inversion symmetry. The
crystal potential can be also expanded in cubic harmonics $V_\lambda ^\mu $.
For $\lambda =2$
$$
V_2^0=3z^2-r^2,\quad V_2^2=x^2-y^2,\quad V_2^1=xz,
\quad V_2^{-1}=yz,\quad V_2^{-2}=xy.
\eqno{(1.14)}
$$
In many cases, it is convenient to use the method of Stevens equivalent
momentum operators which have the same matrix elements as spherical
functions. For example,
$$
V_2^0 =\alpha _J\,\overline{r^2}[3(J^z)^2-J(J+1)],
\qquad V_2^2=\frac 12\alpha _J\,\overline{r^2}[(J^{+})^2+(J^{-})^2],
$$
$$
V_2^1 =V_2^2=\frac 12\alpha _J\,\overline{r^2}(J^xJ^z+J^zJ^x),
$$
where $\alpha _J$ (and similar factors $\beta _J$, $\gamma _J$ for $\lambda
=4,6$) are the Stevens proportionality coefficients which depend on the
configuration $l^n$ [16,41,43]. The Stevens operator representation holds
beyond the multiplet with a given $J$ (or $L$) only. Of course, the operator
(1.5), unlike (1.7), has also matrix elements between different terms. The
method of equivalent operators in problems with changing $J$ or $L$ is more
complicated (see [41]).

Consider concrete examples of d$^n$ configurations in the crystal field.
Since various situations may take place for d-electrons in CF, constructing
interpolation schemes is useful [43].

The picture of the two-hole spectrum for the Ni$^{2+}$ ion (configuration
d$^8$) is shown in Fig.1.6. We use Bethe's notations $^{2S+1}\Gamma _i$ for
irreducible representations of the cubic group: $\Gamma _1$ ($A_{1g})$ and
$\Gamma _2$ ($A_{2g}$) are one-dimensional, $\Gamma _3$ ($E_g$) is
two-dimensional, $\Gamma _4$ ($T_{1g}$) and $\Gamma _5$ ($T_{2g}$) are
three-dimensional. An arbitrary position of one-electron levels is chosen; an
analogue of the Hund rule takes place for the effective orbital momentum in
CF [41]. One can see that the initial orbital momentum $L=3$ is quenched in
intermediate crystal field provided that the lowest state is the orbital
singlet $^3\Gamma _2$. For three-dimensional ``Hund'' representations
$^3\Gamma _4$ and $^3\Gamma _5$, quenching is absent. In the strong CF,
electron orbital momenta are quenched provided that the lowest one-electron
state is e (which corresponds to metallic nickel). We see that, a regrouping
of $e_g$ and t states takes place at increasing Coulomb interaction, which
may result in a new ground state with unquenched OMM.

The ion Fe$^{2+}$ (configuration d$^6$) posseses the lowest term $^5D$ in the
intermediate CF and the configuration $(e_g)^4(t_{2g})^2$ in the strong field
(the case of the configuration $(t_{2g})^6$ is trivial). The corresponding
interpolation scheme is shown in Fig.1.7. We see that in the intermediate CF
quenching takes place only provided that the level $^3\Gamma _3$ is lower. In
the strong CF, the state $^3\Gamma _5$, corresponding to three-dimensional
$t_{2g}$ representation, has an unquenched orbital quasimomentum.

The examples considered demonstrate that, generally speaking, in the case of
a high lattice symmetry a local CF results in only partial degeneracy lift
which cannot provide total quenching of OMM. Even in the case where such a
quenching does take place, OMM may be partly unquenched by the Coulomb
interaction which has off-diagonal matrix elements between different point
group irreducible representations, both for intermediate and strong CF cases.
E.g., the matrix elements of the type $\langle e_g|Q|t_{2g}\rangle $ unquench
OMM in the state $e_g$. On the other hand, in the case of intermediate CF,
off-diagonal matrix elements of crystal field itself may mix different terms
and unquench OMM of the ground state term.

Thus the local CF potential does not explain almost total quenching of OMM in
transition metals and their alloys. The quenching mechanism owing to the
periodic lattice potential will be considered in Sect.4.8.1. This potential
results in an appreciable kinetic energy of d-electrons. The formation of
rather wide energy d-bands leads also to destruction of localized spin
momenta and suppression of magnetism in most d-metals.

The role of OMM effects in highly-correlated d-compounds is debatable. Often
the description of the electronic structure of Mott insulators (e.g,
transition metal oxides and sulphides) is performed within the density
functional approach, which does not take into account ME term structure
[44,45]. At the same time, there exist attempts of interpretation of their
optical properties within a picture including spectroscopic terms with
account of CF splitting [46,13]. Detailed investigations of optical and X-ray
spectra (Sect.2.5), which contain information on selection rules in $L$, seem
to be useful for solving these problems. It should be noted that the orbital
contributions may be different for the configurations s(p)d$^{n-1}$ and d$^n$
which are mixed by crystal field.

The resonance photoemission spectrum of CuO is shown in Fig.1.8. The
resonance intensity allows the identification of the various atomic
multiplets of the d$^8$ final state. Figs.1.9 and 1.10 [45] show a comparison
of the XPS valence band spectra of MnO and NiO with the results of cluster
calculations. The parameters of cluster calculations are shown in diagrams.
One can see from Fig.1.10 the strong hybridization between the d$^7 $ and
d$^8$ L$^{-1}$ final state.

\chapter{BAND THEORY}

Band theory is now a large and rather independent branch of solid state
physics, which uses the whole variety of modern computational methods. The
corresponding problems are discussed in detail in a number of monographs and
reviews [12,13,52-57]. Nevertheless, we include this material in the book
with two main purposes (i) to consider in a simple form main band
calculation methods bearing in mind their physical basis (ii) to concretize
the advantages and drawbacks of various methods as applied to transition
metals. Further we discuss the results of band calculations for TM and
compare them with experimental data.

Band theory deals with description of electron spectrum in a regular lattice
with a periodic crystal potential,
$$
V(\mathbf{r})=V(\mathbf{r+R)=}\sum_{\mathbf{R}}v\mathbf{(r-R)}
$$
where $v(\mathbf{r})$ is the potential of an ion, \textbf{R} are the lattice
sites. According to the Bloch theorem, one-electron states in a periodic
potential may be chosen in the form
$$
\psi _{n\mathbf{k}}(\mathbf{r)=}e^{i\mathbf{kR}}u_{n\mathbf{k}}(\mathbf{r)}
\eqno{(2.1)}
$$
They are classified by the quasimomentum \textbf{k} and form a set of energy
bands with the index $n$.

At considering the band spectrum, two opposite approaches are possible. The
first one starts from the picture of isolated atoms and uses the crystal
potential of neighbour atoms as a perturbation, the intraatomic potential
being taken into account in the zeroth-order approximation. The simplest
method of band calculations is the tight-binding method which uses
one-electron atomic wave functions $\phi _\gamma (\mathbf{r})$. The
corresponding wavefunctions of a crystal are written as
$$
\phi _{\gamma \mathbf{k}}(\mathbf{r})=\sum_{\mathbf{R}}e^{i\mathbf{kR}}\phi
_\gamma (\mathbf{r-R})
\eqno{(2.2)}
$$
The band spectrum in this approximation contains a set of bands originating
from atomic levels. In the case of the s-band for the simple cubic lattice
in the nearest-neighbour approximation we have
$$
E_{\mathbf{k}}=\varepsilon _s+2\beta (\cos k_x+\cos k_y+\cos k_z)
\eqno{(2.3)}
$$
where
$$
\beta =\int d\mathbf{r}\phi ^{*}(\mathbf{r)}v(\mathbf{r)}\phi (\mathbf{r+a)}
\eqno{(2.4)}
$$
is the transfer integral. The width of the energy bands is determined by the
overlap of atomic wavefunctions at neighbour lattice sites and decreases
rapidly for inner shells. As a rule, the bands, which originate from
different levels, overlap considerably. Taking into account off-diagonal
matrix elements and combining the atomic functions with $\gamma \neq \gamma
^{\prime }$ we come to the method of linear combination of atomic orbitals
(LCAO). This method was widely used in early band calculations. A version of
this method was developed by Slater and Coster [58]. An important drawback
of the LCAO method is non-orthogonality of atomic functions at different
lattice sites (see Apendix C) and absence of delocalized states with
positive energies in the basis. Thus it is difficult to describe weakly
localized electron states in metals within this method.

The opposite limiting case corresponds to the picture of nearly free
electrons with a large kinetic energy, so that the whole periodic crystal
potential may be considered as a perturbation which results in formation of
gaps (forbidden bands) in the energy spectrum. Contrary to the tight-binding
approximation, the strength of crystal potential determines the widths of
gaps rather than of electron bands.

Free electrons have the wavefunctions and the energy spectrum
$$
\psi _{\mathbf{k}}(\mathbf{r)=}e^{i\mathbf{kr}}\equiv |\mathbf{k}\rangle ,
\qquad E_{\mathbf{k}}=\frac{\hbar ^2k^2}{2m}
\eqno{(2.5)}
$$
The matrix elements of the periodic lattice potential are given by
$$
\langle \mathbf{k|}V|\mathbf{k}^{\prime }\rangle =\sum_{\mathbf{g}}V_{
\mathbf{g}}\delta _{\mathbf{k-k}^{\prime }\mathbf{,g}}
\eqno{(2.6)}
$$
where $\mathbf{g}$ are reciprocal lattice vectors which are defined by
$$
\mathbf{gR=}2\pi n,\qquad n=0,\pm 1,...
\eqno{(2.7)}
$$
For most values of \textbf{k}, corrections to the spectrum may be calculated
to second order of perturbation theory
$$
\Delta E_{\mathbf{k}}=\sum_{\mathbf{g}}\frac{|V_{\mathbf{g}}|^2}{E_{\mathbf{k}}
-E_{\mathbf{k-g}}}
\eqno{(2.8)}
$$
However, the denominators of the terms in (2.8) with
$E_{\mathbf{k}}=E_{\mathbf{k-g}}$, i.e.
$$
2(\mathbf{kg})=g^2
\eqno{(2.9)}
$$
vanish and we have to calculate the spectrum with the use of the
Brillouin-Wigner perturbation theory which yields
$$
\left|
\begin{array}{ll}
E-E_{\mathbf{k}} & V_{\mathbf{g}}^{*} \\
V_{\mathbf{g}} & E-E_{\mathbf{k-g}}
\end{array}
\right| =0
\eqno{(2.10)}
$$
or
$$
E=\frac 12(E_{\mathbf{k}}+E_{\mathbf{k-g}})\pm [\frac 14(E_{\mathbf{k}}
-E_{\mathbf{k-g}})^2+|V_{\mathbf{g}}|^2]^{1/2}
\eqno{(2.11)}
$$
Thus we obtain the splitting of the spectrum with formation of the energy
gap
$$
\Delta E_{\mathbf{g}}=2|V_{\mathbf{g}}|
\eqno{(2.12)}
$$
The equation (2.9) just determines the boundaries of the Brillouin zone
where the gaps occur.

The nearly-free electron approximation is applicable provided that the
crystal potential is small in comparison with the bandwidth $W$. Since the
matrix elements of crystal potential are of order of a few eV, this
condition may be satisfied for external s and p-electrons with $W\sim 10$
eV. On the other hand, d-electrons form more narrow bands and disturb
considerably the conduction band because of s-d hybridization. Thus this
approximation describes satisfactorily the conduction electron states in
simple (in particular, alcaline) metals only.

In most cases both above pictures are insufficient to obtain a quantitative
description of the electron spectrum. Especially difficult is this problem
for d-electrons in transition metals, which are characterized by an
intermediate degree of localization.

It should be noted that the shortcomings of the simplest band calculation
methods occur because of the approximate solution of the problem in any real
calculations. Using an infinite full set either of localized or extended
wavefunctions would in principle provide an exact result (of course, in the
one-electron approximation only). However, the use of a physically
reasonable basis enables one to obtain much better results for a finite set.
Evidently, the true wavefunctions should have an atomic character in the
region near the lattice site and are close to plane waves in the outer
space. Therefore modern methods of band calculations (e.g., the methods of
augmented and orthogonalized plane waves), which are discussed briefly in
the next sections, use combinations of atomic functions and plane waves.

The second difficulty of the band theory is connected with the problem of
electron correlations, which are large for transition metals. The Coulomb
interaction among electrons, which has a two- particle nature and prevents
reducing the many-electron problem to the one-electron one. In principle,
the possibility of such a reducing is provided by the Hohenberg-Kohn theorem
[59]. This theorem guarantees the existence of an unique density functional
which yields the exact ground state energy. However, explicit constructing
of this functional is a very complicated problem. In the situation, where
correlations are not too strong, the Coulomb interaction may be taken into
account by introducing a self-consistent potential which depends on electron
density. These questions are treated in Sect.2.3.

The correlation effects are especially important for narrow d- and f-bands.
The strong intrasite Coulomb repulsion may lead to splitting of one-electron
bands into many-electron subbands, and the usual band description (even in
the tight-binding approach) is inapplicable. In particular, in the case of
one conduction electron per atom the Mott-Hubbard transition takes place
with increasing interatomic distances, so that electrons become localized at
lattice sites. In such a situation, we have to construct the correlation
Hubbard subbands with the use of the atomic statistics and many-electron
quantum numbers (Appendices C, H). Development of first-principle band
calculation methods in the many-electron representation is an exciting
problem. Such calculations might be based on a modification of the
eigenfunctions of a spherically symmetic potential, which are used, e.g., in
APW and OPW methods, by expessing them in terms of many-electron functions
of the atomic problem. In particular, the dependence of one-electron radial
functions on many-electron quantum numbers $S,L$ occurs in the full
Hartree-Fock approximation (see Sect.2.3). Picking out the functions, which
correspond to lowest (Hund) atomic terms should modify considerably the
results of band calculations.

\section{Orthogonalized plane wave method and pseudopotential}

Main shortcoming of the nearly-free electron method is a poor description of
strong oscillations of electron wavefunctions near atomic cores in terms of
not too large set of plane waves. This shortcoming is removed in the
orthogonalized plane wave method (OPW) where the wavefunctions are chosen to
be orthogonal with respect to core states:
$$
\chi _{\mathbf{k}}(\mathbf{r})=e^{i\mathbf{kr}}+\sum_cb_c\phi
_{\mathbf{k}}^c(\mathbf{r})
\eqno{(2.13)}
$$
where $\phi _{\mathbf{k}}^c(\mathbf{r})$ are the tight-binding Bloch
functions originating from the core levels. The condition 
$$
\int d\mathbf{r}\chi _{\mathbf{k}}(\mathbf{r})\phi _{\mathbf{k}}^c(\mathbf{r})
\equiv \langle \chi |c\rangle =0
\eqno{(2.14)} 
$$
yields 
$$
b_c=-\int d\mathbf{r}e^{-i\mathbf{kr}}\phi _{\mathbf{k}}^c(\mathbf{r})\equiv
-\langle \mathbf{k}|c\rangle 
\eqno{(2.15)} 
$$
The orthogonalized plane wave (2.13) oscillates rapidly near atomic nucleus
and is close to the plane wave between the atoms. The trial wavefunction of
the crystal is chosen as a superposition of OPW: 
$$
\psi _{\mathbf{k}}=\sum_{\mathbf{g}}\alpha _{\mathbf{k-g}}\chi _{\mathbf{k-g}}
\eqno{(2.16)} 
$$
(\textbf{g} are reciprocal lattice vectors). The coefficients $a$ are
determined from the Schroedinger equation. The matrix elements of the
crystal potential for OPW (unlike those for plane waves) turn out to be
small, so that it is sufficient to take in the expansion (2.16) not too many
terms. It should be noted that the OPW basis is overfilled so that the
increasing the number of functions in the set does not necessarily increase
the accuracy of calculations.

Further development of the OPW method led to the idea of introducing a weak
pseudopotential which permits (unlike the real crystal potential) the use of
perturbation theory. To this end one introduces the pseudowavefunction 
$$
\widetilde{\psi }_{\mathbf{k}}=\sum_{\mathbf{g}}\alpha _{\mathbf{k-g}}e^{i(%
\mathbf{k-g)r}}=\psi _{\mathbf{k}}+\sum_c\langle c|\widetilde{\psi }\rangle
\phi _{\mathbf{k}}^c
\eqno{(2.17)} 
$$
which coincides with the true wavefunction outside the core region but does
not exhibit oscillations, which are due to strong core level potential,
within it. Substituting (2.17) into the usual Schroedinger equation $H\psi
=E\psi $ we get 
$$
H\widetilde{\psi }_{\mathbf{k}}+\sum_c(E-E_c)\langle c|\widetilde{\psi }
\rangle \phi _{\mathbf{k}}^c=E\widetilde{\psi }_{\mathbf{k}}
\eqno{(2.18)} 
$$
which may be represented in the form of a new Schroedinger equation 
$$
(-\frac{\hbar ^2}{2m}\Delta +W)\widetilde{\psi }_{\mathbf{k}}=
E\widetilde{\psi }_{\mathbf{k}}
\eqno{(2.19)} 
$$
where the non-local energy-dependent pseudopotential operator $W$ is defined
by 
$$
W=V(\mathbf{r})+V^R
\eqno{(2.20)} 
$$
$$
V^R\widetilde{\psi }_{\mathbf{k}}(\mathbf{r)}=\sum_c(E-E_c)\langle c|
\widetilde{\psi }\rangle \phi _{\mathbf{k}}^c
\eqno{(2.21)} 
$$
For $E>E_c$, the matrix elements of the potential $V^R$
$$
\int d\mathbf{r}\widetilde{\psi }_{\mathbf{k}}^{*}(\mathbf{r})V^R\widetilde{%
\psi }_{\mathbf{k}}(\mathbf{r)}=\sum_c(E-E_c)|\langle c|\widetilde{\psi }%
\rangle |^2
\eqno{(2.22)} 
$$
are positive. At the same time, the matrix elements of the true periodic
potential $V$ are negative and large in the absolute value in the core
region. Thus the diagonal matrix elements of $W$ turn out to be positive and
small because of partial cancellation of two terms in (2.20). In particular,
the pseudopotential does not result, unlike the real crystal potential, in
formation of bound states, corresponding to core levels. Therefore the
pseudopotential method describes conduction electron states only.

Generally speaking, the pseudopotential approach is not simpler than the
original OPW method since it requires introducing a non-local potential
function which is not defined in a unique way. Therefore one uses frequently
empirical or model local pseudopotentials which are fitted to experimental
data, rather than first-principle ones. Perturbation theory with the use of
such potentials enables one to describe the whole variety of properties of
simple metals [55].

The standard form of the pseudopotential theory is inapplicable both in
noble and transition metals because d-electrons are much stronger localized
than s,p-electrons. To describe d-electron states one introduces into (2.17)
the contribution from d-functions [9]
$$
|\psi \rangle =|\widetilde{\psi }\rangle -\sum_c\langle c|\widetilde{\psi }%
\rangle |c\rangle +\sum_da_d|d\rangle 
\eqno{(2.23)} 
$$
where, as well as for simple metals, the pseudowavefunction $\widetilde{\psi 
}$ is a linear combination of plane waves. One has to take into account the
difference between potentials of free ion and of ion in a crystal,
$$
\delta v(\mathbf{r})=v^{ion}(r)-v(\mathbf{r}) 
$$
Define the operator $V^h$ by 
$$
V^h|d\rangle =\delta v|d\rangle -|d\rangle \langle d|\delta v|d\rangle 
\eqno{(2.24)} 
$$
Then we obtain 
$$
a_d=-\langle d|\widetilde{\psi }\rangle +\frac{\langle d|V^h|\widetilde{\psi 
}\rangle }{E_d-E} 
$$
and the Schroedinger equation for the pseudowavefunction takes the form 
$$
(-\frac{\hbar ^2}{2m}\Delta +W)|\widetilde{\psi }\rangle +\sum_d\frac{%
V^h|d\rangle \langle d|V^h|\widetilde{\psi }\rangle }{E-E_d}=E|\widetilde{%
\psi }\rangle 
\eqno{(2.25)} 
$$
where the pseudopotential operator of the transition metal is given by
$$
W|\widetilde{\psi }\rangle =V|\widetilde{\psi }\rangle
+\sum_{i=c,d}(E-E_i)|i\rangle \langle i|\widetilde{\psi }\rangle
+\sum_d(|d\rangle \langle d|V^h|\widetilde{\psi }\rangle +V^h|d\rangle
\langle d|\widetilde{\psi }\rangle )
\eqno{(2.26)} 
$$
The most important peculiarity of eq.(2.25) is the presence of the terms,
which have a resonance denominator $E-E_d$. The physical origin of this is
as follows. As discussed in Sect.1.1, the atomic potential for d-electrons
may have a centrifugal barrier where $v(r)$ is positive. Therefore there
exist bound d-levels merged in the continuous spectrum. The strong
scattering of continuous spectrum electron by localized levels takes place
with the phase shift of $\pi /2$. Such a situation may be also described by
the s-d hybridization model. The corresponding energy spectrum is strongly
distorted and should be calculated by exact diagonalization which yields
$$
E_{\mathbf{k}}^{1,2}=\frac 12(\frac{\hbar ^2k^2}{2m}+E_d)\pm [\frac 14(\frac{%
\hbar ^2k^2}{2m}-E_d)^2+|V_{\mathbf{k}}^h|^2]^{1/2}
\eqno{(2.27)} 
$$
Far above the resonance the hybridization potential may be considered, as
well as the usual pseudopotential, within the framework of perturbation
theory. This permits a satisfactory description withiin the modified
pseudopotential approach [60] of noble metals where the d-resonance,
although crossing conduction band, lies well below $E_F$.

In transition metals the resonance lies near the Fermi energy and
perturbation theory is inapplicable. Thus one has to pick out the
singularity, corresponding to the resonance. A method of completely
orthoghonalized plane waves [61], application of which to transition metals
is discussed in [57,62], gives in principle a possibility to eliminate this
difficulty. However, this method is rather complicated and goes far beyond
the original concept of the OPW method. Generally, such approaches require
in a sense exact diagonalization of a matrix and lose in fact the main
advantage of the pseudopotential idea - applicability of perturbation
theory. More recent applications of the pseudopotential method for band
calculations of transition metals are considered in [63,64].

There exist a number of model pseudopotential forms, in particular including
explicitly the d-resonance. However, they do not provide as a rule
sufficiently satisfactory description of TM. Recently a model
pseudopotential was successfully applied to explain properties of iridium
[65]. The reasons of the good fit to experimental data for this particular
d-metal are not quite clear.

\section{Augmented plane wave (APW) and Korringa-Kohn-Rostoker (KKR)
methods}

The idea of other modern band calculation methods is to describe electron
states in different space regions in a proper way using different bases. To
this end, one builds around each site an atomic sphere with the volume which
is somewhat smaller than that of the lattice cell, so that spheres centered
at different sites do not overlap. The crystal potential at each site is
taken in the so called muffin-tin (MT) form
$$
v_{MT}(\mathbf{r})=\left\{ 
\begin{array}{ll}
v(r) & ,\qquad r<r_{MT} \\ 
v_c & ,\qquad r>r_{MT}
\end{array}
\right. 
\eqno{(2.28)} 
$$
Inside the sphere the potential is supposed to be spherically symmetric, and
the corresponding wavefunctions for a given energy may be expanded in
spherical harmonics
$$
\phi (\mathbf{r})=\sum_{lm}C_{lm}R_l(r,E)Y_{lm}(\widehat{\mathbf{r}}),\qquad
r<r_{MT}
\eqno{(2.29)} 
$$
($R_l$ are the solutions of the corresponding radial Schroedinger equation)
which permits to simplify greatly the calculations due to using the angular
momentum technique.

Since beyond the spheres the MT potential is constant, the basis
wavefunctions $\phi (\mathbf{r})$ may be chosen in the form of plane waves.
Using the expansion (C.28) we may join continuosly the solutions at the
sphere boundaries and determine the coefficients $C_{lm}=C_{lm}(\widehat{%
\mathbf{k}})$
$$
C_{lm}=i^l(2l+1)\frac{j_l(kr_{MT})}{R_l(r_{MT},E)}Y_{lm}^{*}(\widehat{%
\mathbf{k}})
\eqno{(2.30)} 
$$
Then the function (2.29) is called the augmented plane wave (APW). The
wavefunction of the crystal are searched as a linear combination of APW's
with the same energy:
$$
\psi _{\mathbf{k}}(\mathbf{r})=\sum_{\mathbf{g}}\alpha _{\mathbf{k-g}}
\phi _{\mathbf{k-g}}(\mathbf{r})
\eqno{(2.31)} 
$$
Substituting (2.31) into the Schroedinger equation one obtains the system of
linear equations. Poles of its determinant yield the dispersion law $E%
\mathbf{(k})$ which contains the crystal potential in a complicated way. To
obtain accurate results, it is sufficient to restrict oneself by not too
large number of terms in (2.31). Practically, one uses up to 100 APW's, and
for a larger set the solution is as a rule stabilized.

The application of the APW method to d-bands requires an account of a larger
number of APW's in comparison with s,p-bands. However, such band
calculations were performed for all the transition d-metals [53,54]. For 4d-
and 5d-metals, relativistic effects (in particular, spin-orbital
interaction) are important. The corresponding version of the APW method (the
RAPW method) was developed by Loucks [66].

The Green's function method, developed by Korringa, Kohn and Rostoker (KKR),
uses, as well as APW method, the expansion in spherical functions inside the
MT spheres (2.28). At the same time, the wavefunctions outside them are
constructed not from plane waves, but also from spherical waves which are
scattered from other sites. Formally, this is achieved by using the integral
form of the Schroedinger equation ($\kappa =|E|^{1/2}$)
$$
\psi (\mathbf{r})=-\frac 1{4\pi }\int d\mathbf{r}^{\prime }\frac{\exp
(i\kappa |\mathbf{r-r}^{\prime }|)}{|\mathbf{r-r}^{\prime }|}V(\mathbf{r}%
^{\prime })\psi (\mathbf{r}^{\prime })
\eqno{(2.32)} 
$$
Using the Bloch theorem condition 
$$
\psi _{\mathbf{k}}(\mathbf{r+R})=\psi _{\mathbf{k}}(\mathbf{r})e^{i\mathbf{kR%
}} 
$$
one reduces (2.32) to the form
$$
\psi _{\mathbf{k}}(\mathbf{r})=-\frac 1{4\pi }\int d\mathbf{r}^{\prime }G_{%
\mathbf{k}}(\mathbf{r-r}^{\prime },E)v(\mathbf{r}^{\prime })\psi _{\mathbf{k}%
}(\mathbf{r}^{\prime })
\eqno{(2.33)} 
$$
where the integration goes over one lattice cell and
$$
G_{\mathbf{k}}(\mathbf{r},E)=\sum_{\mathbf{R}}\frac{\exp (i\kappa |\mathbf{%
r-R}|)}{|\mathbf{r-R}|}e^{i\mathbf{kR}}
\eqno{(2.34)} 
$$
is the lattice Green's function.

The functional describing the system and corresponding to (2.33) has the form
$$
\Lambda =\int d\mathbf{r}|\psi _{\mathbf{k}}(\mathbf{r})|^2+\frac 1{4\pi
}\int \int d\mathbf{r}d\mathbf{r}^{\prime }\psi _{\mathbf{k}}^{*}(\mathbf{r}%
)v(\mathbf{r})G_{\mathbf{k}}(\mathbf{r-r}^{\prime },E)v(\mathbf{r}^{\prime
})\psi _{\mathbf{k}}(\mathbf{r}^{\prime })
\eqno{(2.35)} 
$$
Substituting (2.29) into (2.35) yields [52]
$$
\Lambda =\sum_{lm,l^{\prime }m^{\prime }}(A_{lm,l^{\prime }m^{\prime
}}+\kappa \delta _{ll^{\prime }}\delta _{mm^{\prime }}\cot \eta
_l)C_{lm}C_{l^{\prime }m^{\prime }}
\eqno{(2.36)} 
$$
Here $\eta _l$ is the $l$-dependent phase shift owing to the MT-potential,
$$
\cot \eta _l=\frac{\kappa r_{MT}n_l^{\prime }(\kappa
r_{MT})-D_l(E)n_l(\kappa r_{MT})}{\kappa r_{MT}j_l^{\prime }(\kappa
r_{MT})-D_l(E)j_l(\kappa r_{MT})}
\eqno{(2.37)} 
$$
where $j_l$ and $n_l$ are the spherical Bessel and Neumann functions,
$$
D_l(E)=r_{MT}R_l^{\prime }(r_{MT},E)/R_l(r_{MT},E)
\eqno{(2.38)} 
$$
are the logarithmic derivatives of the radial wave function. The energy
spectrum is determined by zeros of the determinant of the matrix $\Lambda $.
The structural constants $A$ are given by
$$
A_{lm,l^{\prime }m^{\prime }}=4\pi \kappa \sum_{\mathbf{R\neq 0}}e^{i\mathbf{%
kR}}\sum_{l^{\prime \prime }m^{\prime \prime }}[n_{l^{\prime \prime
}}(\kappa R)-ij_{l^{\prime \prime }}(\kappa R)]Y_{l^{\prime \prime
}m^{\prime \prime }}^{*}(\widehat{\mathbf{R}})\widetilde{C}_{lm,l^{\prime
\prime },m^{\prime \prime }}^{l^{\prime }m^{\prime }}
\eqno{(2.39)} 
$$
(the quantities $\widetilde{C}$ are defined in (C.9)). These constants do
not depend on crystal potential and may be calculated once for all for a
given lattice.

The advantage of the KKR method in comparison with the APW one is the
decoupling of structural and atomic factors. For the same lattice
potentials, the KKR and APW methods yield usually close results. The
intimate relation of these methods is discussed in [52].

The main difficulty of the KKR method is the energy dependence of the
structural constants. However, this dependence may be eliminated if we use
so called atomic sphere approximation (ASA) where the volume of the sphere
in the MT potential (2.28) is put to be equal to the lattice cell volume, $%
r_{MT}=s$, so that the volume of the intersphere region vanishes. Then one
puts for simplicity the kinetic energy beyond the atomic spheres, $E-v_c$ to
be zero, and the ASA-KKR equations take a very simple form [56] 
$$
\det |S_{lm,l^{\prime }m^{\prime }}(\mathbf{k})-P_l(E)\delta _{ll^{\prime
}}\delta _{mm^{\prime }}|=0
\eqno{(2.40)} 
$$
where the potential function is given by
$$
P_l(E)=2(2l+1)\frac{D_l(E)+l+1}{D_l(E)-l}
\eqno{(2.41)} 
$$
and the structural constants are energy independent,
$$
S_{lm,l^{\prime }m^{\prime }}(\mathbf{k}) =(4\pi )^{1/2}(-1)^{m^{\prime
}+1}(-i)^\lambda \frac{(2l^{\prime }+1)(2l+1)}{(2\lambda +1)} 
$$
$$
\times \frac{(\lambda +\mu )!(\lambda -\mu )!}{(l^{\prime }+m^{\prime
})!(l^{\prime }-m^{\prime })!(l+m)!(l-m)!}\sum_{\mathbf{R\neq 0}}
e^{i\mathbf{kR}}Y_{\lambda \mu }^{*}(\widehat{\mathbf{R}})(s/R)^{\lambda +1} 
\eqno{(2.42)} 
$$
where 
$$
\lambda =l+l^{\prime },\qquad \mu =m-m^{\prime } 
$$
The wavefunctions corresponding to the MT-potential with zero kinetic energy 
$E-v_c$ may be chosen in the form
$$
\phi _{lm}(\mathbf{r,}E) =i^lY_{lm}(\widehat{\mathbf{r}})\phi _l(r\mathbf{,%
}E)
\eqno{(2.43)}
$$
$$
\phi _l(r\mathbf{,}E) =R_l(r,E)\times \left\{ 
\begin{array}{ll}
1 & ,\qquad r<s \\ 
\frac{D_l+l+1}{2l+1}\left( \frac rs\right) ^l+\frac{l-D_l}{2l+1}\left( \frac
rs\right) ^{-l-1} & ,\qquad r>s
\end{array}
\right.  
$$
where the coefficients are determined from the condition of joining the tail
continuosly and differentiably at the boundary of the MT-sphere. It is
convenient to substract the term which describes the diverging spherical
wave. Then we obtain the so-called MT-orbitals
$$
\chi _l(r\mathbf{,}E)=\phi _l(r\mathbf{,}E)\times \left\{ 
\begin{array}{ll}
\frac{\phi _l(r\mathbf{,}E)}{\phi _l(s\mathbf{,}E)}-\frac{D_l+l+1}{2l+1}%
\left( \frac rs\right) ^l & ,\qquad r<s \\ 
\frac{l-D_l}{2l+1}\left( \frac rs\right) ^{-l-1} & ,\qquad r>s
\end{array}
\right. 
\eqno{(2.44)} 
$$
Condition of cancellation of tails of the MT orbitals, which originate from
other sites, leads again to the KKR equations.

A modification of LCAO method which uses instead of atomic functions the
MT-orbitals is called the LCMTO method, the linear combinations of
MT-orbitals being constructed similar to (2.1). Substituting these trial
functions into the Schroedinger equation we obtain the LCMTO equations [67].

In the general APW, KKR and LCMTO methods the matrix elements are functions
of energy. Therefore, at calculating eigenvalues one has to compute the
determinants in each point of $k$-space for a large number values of $E$ (of
order of 100) which costs much time. To simplify the calculation procedure
Andersen [67] proposed to expand the radial wavefunctions at some energy
value to linear terms in $E$. Then the Hamiltonian and overlap matrices do
not depend on energy. The error owing to the linearization does not as a
rule exceed that owing to the inaccuracy in the crystal potential (in
particular, owing to the MT- approximation). The accuracy may be increased
by account of higher-order terms in the expansion. The linear methods (LMTO
and LAPW) permitted to perform band calculations of a large number of
transition metal compounds with complicated crystal structures [57].

Using the MT-orbital basis gives a possibility to carry out band
calculations beyond MT-approximation for the potential itself. Such an
approach (the full-potential LMTO method) enables one to improve
considerably the accuracy of calculations. Very good results are achieved
also within the framework of the full-potential LAPW method [68].

\section{The Hartree-Fock-Slater and density functional approaches to
the problem of electron correlations}

The above consideration of the electron spectrum was performed within the
one-electron approximation. Main purpose of the calculation methods
discussed was joining of atomic-like solutions near the lattice sites with
wavefunctions in the outer space where the crystal potential $V(r)$ differs
strongly from atomic one. The Coulomb interaction among conduction electrons
was assumed to be included in $V(r)$. Such an inclusion should be performed
in a self-consistent way. In the simplest approximation, this may be
achieved by introducing into the Schroedinger equation the averaged Coulomb
potential,
$$
\varepsilon _i\psi _i(\mathbf{r})=[-\frac{\hbar ^2}{2m}\Delta +V(\mathbf{r}%
)+V_C(\mathbf{r})]\psi _i(\mathbf{r})
\eqno{(2.45)} 
$$
$$
V_C(\mathbf{r})=-e\int d\mathbf{r}^{\prime }
\frac{\rho (\mathbf{r}^{\prime })}{|\mathbf{r-r}^{\prime }|},\qquad 
\rho (\mathbf{r})=\sum_i{}^{\prime }|\psi _i(\mathbf{r})|^2
\eqno{(2.46)} 
$$
where the sum goes over occupied electron states and integration includes
summaton over spin coordinate.

The Hartree equations (2.45) do not include the exchange interaction which
results from antisymmetry of electron wavefunctions. Substituting the
antisymmetrized product of one-electron wavefunctions (Slater determinant)
(A.2) in the Schroedinger equation for the many- electron system with the
Hamiltonian (C.1) and minimizing the average energy
$$
\mathcal{E}=\langle \Psi |H|\Psi \rangle /\langle \Psi |\Psi \rangle 
$$
we obtain the Hartree-Fock equations
$$
\varepsilon _i\psi _i(\mathbf{r}) =[-\frac{\hbar ^2}{2m}\Delta +V(\mathbf{r%
})+V_C(\mathbf{r})]\psi _i(\mathbf{r}) 
$$
$$
-e^2\sum_j{}^{\prime }\psi _j(\mathbf{r})\int d\mathbf{r}^{\prime }\frac{%
\psi _j^{*}(\mathbf{r}^{\prime })\psi _i(\mathbf{r}^{\prime })}{|\mathbf{r-r}%
^{\prime }|}\delta _{\sigma _i\sigma _j}
\eqno{(2.47)}
$$
where $\varepsilon _i$ are the Lagrange multipliers. The exchange term (last
term in the left-hand side of (2.47)) has a non-local form, i.e. is not
reduced to multiplication of the wavefunction $\psi _i$ by a potential
function. It should be noted that the terms, describing the electron
self-interaction ($i=j$) are mutually cancelled in the Coulomb and exchange
contributions to (2.47). Due to orthogonality of spin wave functions, the
exchange interaction occurs between electron with parallel spins only.

Parameters of the Hartree-Fock equation depend, generally speaking, not only
on one-electron quantum numbers, but on the whole set of many-electron
quantum numbers in a given state. For example, in the many-electron atom,
the coeficients at the Slater parameters (except for $F^0$) in the multipole
expansion of the Coulomb interaction (C.19) depend explicitly on the
many-electron $SL$-term. This results in the dependence of the radial
wavefuctions on the quantum numbers $S,L$. The corresponding expression for
the parameters $\varepsilon _i$ may be represented as [20]
$$
\varepsilon _l=E(l^nSL)-\sum_{S^{\prime }L^{\prime }}(G_{S^{\prime
}L^{\prime }}^{SL})^2\widetilde{E}(l^{n-1}S^{\prime }L^{\prime })%
\eqno{(2.48)} 
$$
where $\widetilde{E}$ is the energy of the ``frozen'' ion with the same
radial functions as the $n$-electron atom, $G$ are the fractional parentage
coefficients. Evidently, $\widetilde{E}$ exceeds the energy of the true ion,
which is calculated from the Hartree-Fock equation. In systems with large
number of electrons $N$ the differences $E-\widetilde{E}$ may be neglected
since they give corrections of order of $1/N$ only [9]. The quantity (2.48)
differs from the energy, measured in spectral experiments
$$
\varepsilon _l(SL,S^{\prime }L^{\prime })=E(l^nSL)-\sum_{S^{\prime
}L^{\prime }}E(l^{n-1}S^{\prime }L^{\prime })
\eqno{(2.49)} 
$$
which corresponds to a concrete transition $SL\rightarrow S^{\prime
}L^{\prime }$ (experiments with not too high resolution yield the Lorentz
broadening which differs from (2.48)).

In the band theory the dependence on ME quantum numbers is usually
neglected. Then the parameters $\varepsilon_i$ have the meaning of
one-electron energies, $- \varepsilon_i$ being equal to the ionization
energy of the corresponding state in the crystal (the Koopmans theorem [9]).

Obviously, the Koopmans theorem is inapplicable for partially filled shells
in transition metals where the energy of a level depends essentially on
filling of other states (see Sect.1.1). The self- consistent field method,
which considers one-electron levels only, does not take into account such
correlation effects. Nevertheless, simple averaged exchange-correlation
potentials, expressed in terms of electron density, are widely used to
calculate electron structure of partially filled d-bands in transition
metals. When applied to free atoms, this approach yields atomic levels with
a non-integer filling. Altough not quite physically correct, such a picture
permits a satisfactory description of atomic and molecular spectra [69]. To
calculate the energy of optical transitions, one takes the self- consistent
potential calculated in the so-called transition state which corresponds to
the level filling of 1/2.

The solution of the Schroedinger equation with a non-local potential
resulting from the exchange interaction is a very complicated problem. The
case of free electron may be considered analytically. Performing the
integration for plane waves we obtain from (2.46)
$$
\varepsilon _{\mathbf{k}}\psi _{\mathbf{k}}(\mathbf{r})=[\frac{\hbar ^2k^2}{%
2m}+V_x(k)]\psi _{\mathbf{k}}(\mathbf{r})
\eqno{(2.50)} 
$$
with the local exchange potential
$$
V_x(k)=-8\left( \frac 3{8\pi }\rho \right) ^{1/3}F(k/k_F)
\eqno{(2.51)} 
$$
where $\rho =N/V$ is the electron density, 
$$
F(z)=\frac 12+\frac{1-z^2}{4z}\ln \left| \frac{1+z}{1-z}\right| 
\eqno{(2.52)}
$$
The logarithmic singularity in $F(z)$, and, consequently, in the electron
spectrum results in a non-physical behaviour of the density of states near
the Fermi level, $N(E\rightarrow E_F)\rightarrow 0$. This singularity is due
to the long-range character of the Coulomb interaction and should be in fact
removed by correlation effects which result in screening. Thus the
Hartree-Fock approximaton turns out to be insufficient to describe the free
electron gas because of the strong $k$-dependence. To avoid this difficulty
Slater [69] proposed to use the potential which describes the motion of an
electron in an averaged exchange field.

The exchange potential for a given state may be represented in the form
$$
V_{xi}(\mathbf{r})=-e^2\sum_j{}^{\prime }\delta _{\sigma _i\sigma _j}\int d%
\mathbf{r}^{\prime }\frac{\psi _i^{*}(\mathbf{r})\psi _j^{*}(\mathbf{r}%
^{\prime })\psi _j(\mathbf{r})\psi _i(\mathbf{r}^{\prime })}{|\mathbf{r-r}%
^{\prime }|}/[\psi _i^{*}(\mathbf{r})\psi _i(\mathbf{r})]
\eqno{(2.53)} 
$$
The expression (2.53) may be interpreted as the potential of a charge with a
value of $e$ which is removed from the hole surrounding the $i$-th electron.
We may write for the averaged potential 
$$
V_x(\mathbf{r})=-e^2\sum_{ij}{}^{\prime }\delta _{\sigma _i\sigma _j}\int d%
\mathbf{r}^{\prime }\frac{\psi _i^{*}(\mathbf{r})\psi _j^{*}(\mathbf{r}%
^{\prime })\psi _j(\mathbf{r})\psi _i(\mathbf{r}^{\prime })}{|\mathbf{r-r}%
^{\prime }|}/\sum_i{}^{\prime }\psi _i^{*}(\mathbf{r})\psi _i(\mathbf{r})%
\eqno{(2.54)} 
$$
Averaging the free-electron gas potential (2.51) over $k$ we obtain 
$$
\overline{F}=\int_0^1z^2dzF(z)/\int_0^1z^2dz=\frac 34
\eqno{(2.55)} 
$$
$$
V_x=-6\left( \frac 3{8\pi }\rho \right) ^{1/3}
\eqno{(2.56)} 
$$
Starting from the expression (2.56), Slater introduced the local
exchange-correlation potential with the same dependence on electron density
$$
V_{xS}(\mathbf{r})=-6\left( \frac 3{8\pi }\rho (\mathbf{r})\right) ^{1/3}%
\eqno{(2.57)} 
$$
An alternative approach to the problem of constructing exchange- correlation
potential [70] uses the calculation of the total energy. Carrying out the
integration of one-electron energy for the electron gas one obtains
$$
\mathcal{E} =2\sum_{k<k_F}[\frac{\hbar ^2k^2}{2m}+\frac 12V_x(k)] 
$$
$$
=\frac{e^2}{2a_B}[\frac 35(k_Fa_B)^2-\frac 3{2\pi }k_Fa_B]
$$
$$
=\left[ \frac{2.21}{(r_s/a_B)^2}-\frac{0.916}{r_s/a_B}\right] \mathrm{Ry}
\eqno{(2.58)}
$$
where $\mathrm{Ry}=e^2/2a_B=$13.6 eV, 
$$
r_s=\left( \frac 3{4\pi \rho }\right) ^{1/3}
\eqno{(2.59)} 
$$
is the radius of the sphere with the volume which corresponds to one
electron for a given uniform density $\rho $. When measured in the units of
Bohr radius $a_B$, $r_s$ is of order of the ratio of potential energy of an
electron to its mean kinetic energy. The second summand in (2.58), which
comes from the exchange interaction, is comparable with the first one for
realistic metallic values $r_s=2\div 6$. In the high-density limit the
problem of electron gas may be treated more rigorously by expanding $%
\mathcal{E}$ in $r_s$ [12]:
$$
\mathcal{E}=\left[ \frac{2.21}{(r_s/a_B)^2}-\frac{0.916}{r_s/a_B}+0.622\ln 
\frac{r_s}{a_B}-0.096+O\left( \frac{r_s}{a_B}\right) \right] \mathrm{Ry}
\eqno{(2.60)} 
$$
The corrections in (2.60) describe correlation effects beyond the
Hartree-Fock approximation.

Taking into account the exchange energy of the uniform gas per particle and
neglecting correlations we obtain the Gaspar-Kohn-Sham exchange potential
[70]
$$
V_{GKS}(\mathbf{r})=-4\left( \frac 3{8\pi }\rho (\mathbf{r})\right) ^{1/3}%
\eqno{(2.61)} 
$$
This potential differs from the Slater approximation (2.57) by a factor of
2/3. This fact is due to that the operations of varying the total energy and
of statistical averaging do not commute. The expression (2.60) may be also
obtained from the potential (2.51) by putting $k=k_F$ ($F=1/2$).

A more general exchange-correlation potential ($x\alpha $-potential) may be
written as 
$$
V_{x\alpha }(\mathbf{r})=V_{xS}(\mathbf{r})
\eqno{(2.62)} 
$$
The optimization of the parameter $\alpha <1$ enables one to take into
account effects of electron correlations. The potential (2.61) was
succesfully applied for calculations of electron structure of large number
of solids [69,57]. However, the reasons of this success were not clear until
the density functional method was developed. The results of $x\alpha $%
-calculations for pure metals turn out to be not too sensitive to the choose
of $\alpha $. Earlier calculations with $\alpha =1$ yielded results, which
agreed satisfactory with experimental data and are close to those for
optimal $\alpha $. This is apparently due to that the error in the total
energy for $\alpha =1$ is partially compensated by increasing the depth of
the potential well. However, the correct choose of $\alpha $ is important
for metallic and insulating compounds. For transition metals the atomic
densities are usually taken for d$^{n-1}$s configuration in the 3d- and
4d-series and for the d$^{n-2}$s$^2$configuration in the 5d-series, the
optimal values of $\alpha $ lying in the interval 0.65$\div $0.8.

An important step in the theory of electron correlations in solids was made
by Hohenberg and Kohn [59] who proved the general theorem, according to
which the ground state energy is an unique functional of electron density.
Kohn and Sham [70] represented this density functional (DF) in the form
$$
\mathcal{E}[\rho (\mathbf{r})] =\int d\mathbf{r}\rho (\mathbf{r})V_{ext}(%
\mathbf{r})+\frac 12\int \int d\mathbf{r}d\mathbf{r}^{\prime }\frac{\rho (%
\mathbf{r})\rho (\mathbf{r}^{\prime })}{|\mathbf{r-r}^{\prime }|}  
$$
$$
+T[\rho (\mathbf{r})]+\mathcal{E}_{xc}[\rho (\mathbf{r})]
\eqno{(2.63)}
$$
where $V_{ext}$ is the external potential of lattice ions, $T$ is the
kinetic energy of non-interacting electrons with the density $\rho $, the
functional $\mathcal{E}_{xc}$corresponds to exchange and correlation
contributions. The DF appoach may be generalized to consider spin-polarized
systems.

Representing the electron density in the form (2.46) and varying (2.63) with
respect to $\psi _i$ we obtain
$$
\varepsilon _i\psi _i(\mathbf{r})=[-\frac{\hbar ^2}{2m}\Delta +V_{ext}(%
\mathbf{r})+V_C(\mathbf{r})+V_{xc}(\mathbf{r})]\psi _i(\mathbf{r})%
\eqno{(2.64)} 
$$
where the Lagrange multipliers $\varepsilon _i$ correspond to the
quasiparticle energies, the exchange-correlation potential $V_{xc}$ is
defined by varying $\mathcal{E}_{xc}$with respect to density,
$$
V_{xc}(\mathbf{r})=\delta \mathcal{E}_{xc}[\rho (\mathbf{r})]/\delta \rho (%
\mathbf{r})
\eqno{(2.65)} 
$$
Thus the many-electron problem is reduced to the one-particle one with the
local exchange-correlation potential. However, it should be stressed that
the DF approach guarantees the correct value of the ground state energy only
and describes exactly the distribution of charge and spin densities and
related physical quantitities. Here belong elastic characteristics and
exchange parameters, which are expressed in terms of corresponding energy
differences, saturation magnetic moment etc. On the other hand, the
Kohn-Sham quasiparticles do not coincide with true electrons, so that the
wavefunctions $\psi _i$ do not have a direct physical meaning. Therefore the
possibility of calculation of the whole excitation spectrum is, generally
speaking, not justified. Especially doubtful is the description of the
states far from $E_F$ and of finite temperature behaviour. Nevertheless, the
DF method is widely used to consider various physical properties.

The true functional (2.63) is of course unknown, and one has to use some
approximations. The formally exact expession for $\mathcal{E}_{xc}$ may be
represented in the form of the Coulomb interaction of an electron with the
exchange-correlation hole which surrounds it:
$$
\mathcal{E}_{xc}[\rho (\mathbf{r})]=\frac 12\int d\mathbf{r}\rho (\mathbf{r}%
)\int d\mathbf{r}^{\prime }\frac{\rho _{xc}(\mathbf{r,r}^{\prime }-\mathbf{r}%
)}{|\mathbf{r-r}^{\prime }|}
\eqno{(2.66)} 
$$
The corresponding electron density may be obtained by integration over the
coupling constant $\lambda $ [71,72]
$$
\rho _{xc}(\mathbf{r,r}^{\prime }-\mathbf{r})=\rho (\mathbf{r}^{\prime
})\int_0^1d\lambda [g(\mathbf{r,r}^{\prime },\lambda )-1]
\eqno{(2.67)} 
$$
where $g$ is the pair correlation function. Since the hole contains one
electron, the charge density should satisfy the sum rule
$$
\int d\mathbf{r}^{\prime }\rho _{xc}(\mathbf{r,r}^{\prime }-\mathbf{r})=-e%
\eqno{(2.68)} 
$$
The functional $\mathcal{E}_{xc}$ depends in fact on the spherically
averaged charge density, which follows from the isotropic nature of the
Coulomb interaction [72]. Indeed, the variable substitution in (2.66) allows
to perform integration over angles of the vector $r-r^{\prime }$:
$$
\mathcal{E}_{xc} =\frac 12\int d\mathbf{r}\rho (\mathbf{r})\int R^2dR\frac{%
\overline{\rho }_{xc}(r\mathbf{,}R)}R 
$$
$$
\overline{\rho }_{xc}(r\mathbf{,}R) =\int d\Omega \rho _{xc}(\mathbf{r,}R)%
\eqno{(2.69)}
$$
Thus approximate expessions for $\mathcal{E}_{xc}$ can give exact results
even if the detailed description of $\rho _{xc}$ (in particular, of its
non-spherical part) is inaccurate.

Further we have to concretize the form of the correlation function. In the
local density approximation (LDA), which is used as a rule in band
calculations, $\rho _{xc}$ is taken in the same form, as for uniform
electron density with the replacement $\rho \rightarrow \rho (r)$ and the
corresponding correlation function $g_0$:
$$
\rho _{xc}(\mathbf{r,r}^{\prime }-\mathbf{r})=\rho (\mathbf{r}%
)\int_0^1d\lambda [g_0(|\mathbf{r-r}^{\prime }|,\lambda ,\rho (r))-1]%
\eqno{(2.70)} 
$$
Substituting (2.70) into (2.69) we obtain the LDA functional, which has the
local form
$$
\mathcal{E}_{xc}[\rho ]=\int d\mathbf{r}\rho (\mathbf{r})\varepsilon
_{xc}(\rho )
\eqno{(2.71)} 
$$
where $\mathcal{\varepsilon }_{xc}$ is the exchange and correlation energy
per electron for the interacting uniform electron gas. This quantity was
calculated numerically to a high accuracy and various interpolation
analytical formulae are available. Therefore the LDF approach enables one to
construct various local exchange-correlation potentials. The simplest
exchange potential is given by the $x\alpha $-expression (2.61). This
expression may be generalized as 
$$
V_{xc}(\mathbf{r})=\beta (r_s/a_B)V_{GKS}(\mathbf{r})
\eqno{(2.72)} 
$$
Thus the $x\alpha $-method becomes justified in the density functional
theory, the parameter $\alpha $ being dependent of electron density. The
value of $\alpha $ changes from 2/3 to 0.85 as $r_s$ increases from zero
(high-density limit) to values which are typical for metals in interatomic
region ($r_s/a_B=4$). An example of an interpolation approximation is the
Hedin-Lundqvist potential [73] which corresponds to
$$
\beta _{HL}(z)=1+0.0316z\ln (1+24.3/z)
\eqno{(2.73)} 
$$
The functional of local spin density (LSD) approximation which is applied to
consider magnetic transition metals reads 
$$
\mathcal{E}_{xc}[\rho _{\uparrow },\rho _{\downarrow }]=\int d\mathbf{r}\rho
(\mathbf{r})\varepsilon _{xc}(\rho _{\uparrow }(\mathbf{r}),\rho
_{\downarrow }(\mathbf{r}))
\eqno{(2.74)} 
$$
The simplest LSDA von Barth-Hedin potential [74] has in the first order in
spin polarization $m(\mathbf{r})=n_{\uparrow }(\mathbf{r})-n_{\downarrow }(%
\mathbf{r})$ the form
$$
V_{xc}^\sigma (\mathbf{r})=\frac 13\sigma \delta (n)\frac{m(\mathbf{r})}{n(%
\mathbf{r})}V_{GKS}(\mathbf{r})
\eqno{(2.75)} 
$$
where $n(\mathbf{r})=n_{\uparrow }(\mathbf{r})+n_{\downarrow }(\mathbf{r})$
and the parameter $\delta $ takes into account correlation effects. In the
high-density limit $\delta =1$, and $\delta =0.55$ for $r_s=4$.

The LD approximation oversimplifies greatly the general DF approach which is
based on the Hohenberg-Kohn theorem. In particular, LDA does not take into
account correctly the dependence of $\rho _{xc}(\mathbf{r,r}^{\prime })$ on $%
\rho (\mathbf{r})$ in the whole space (compare (2.67) with (2.70)). However,
it provides good results in the case of a slow varying density $\rho (%
\mathbf{r})$. Besides that, this approximation works well in the
high-density limit since it includes correctly the kinetic energy of the
non-interacting system.

In the original papers [70], attempts were made to take into account
non-uniform corrections to LDA by expanding $\rho _{xc}$with respect to grad 
$\rho (\mathbf{r})$. However, such an expansion yields small contributions
only and does not hold in the case of strongly non-uniform electron systems
(e.g., for 4f-electrons in rare earths and narrow d-bands), and is
therefore, strictly speaking, not justified for TM. Other modifications of
LDA are discussed in the review [72]. The problem of account of multiplet
structure in the density functional approach is treated in [75,76].

One of drawbacks of LDA is the inexact (unlike the full Hartree- Fock
approximation) cancellation of the unphysical interaction of an electron
with itself. In this connection, the so-called self-interaction correction
(SIC) approach [72] was developed. The SIC functional within the LSDA reads
$$
\mathcal{E}_{SIC}=\mathcal{E}_{LSD}[\rho _{\uparrow }(\mathbf{r}),\rho
_{\downarrow }(\mathbf{r})]-\sum_{i\sigma }\delta \mathcal{E}_{i\sigma }%
\eqno{(2.76)} 
$$
The substracted SI correction for the orital $i\sigma $ has the form
$$
\delta \mathcal{E}_{i\sigma }=\frac 12\int \int d\mathbf{r}d\mathbf{r}%
^{\prime }\frac{\rho _{i\sigma }(\mathbf{r})\rho _{i\sigma }(\mathbf{r}%
^{\prime })}{|\mathbf{r-r}^{\prime }|}+\mathcal{E}_{xc}[\rho _{i\sigma },0]%
\eqno{(2.77)} 
$$
where the first term is the self-interaction energy, and the second term is
the LSD approximation to the exchange-correlation energy of a fully
spin-polarized system with the density $\rho _{i\sigma }(\mathbf{r})$. The
functional (2.76) is exact for a system with one electron. The SI corections
are negligible for plane-wave-like states due to smallness of $\rho
_{i\sigma }(\mathbf{r})$. However, they may be important for description of
strongly localized states. The SIC approach was succsesfully applied to
model systems with Hubbard correlations and to transition-metal oxides
[72,77].

\section{Discussion of band calculation results}

Band structure calculations yield the whole picture of electron spectrum of
a given substance. The corresponding dependences $E(\mathbf{k})$ and
densities of states may be found in [24,78]. The calculations of [24] are
performed for cubic symmetry only. Besides that, the 5d-metals are omitted
since they require relativistic methods. In the handbook [78], real crystal
structure for all the transition metals was taken into account. The
Slater-Coster parametrization of the band spectrum was applied, the basis
containing $t_{2g}$ and $e_g$ states being used even for metals with the
hexagonal structure. Bibliography on earlier band structure calculations is
given in Sect.2.7 (see Table 2.6). Some examples of densities of states for
transition metals are shown in Figs.2.1-2.7.

Real accuracy of the detailed and complicated information on the electron
spectrum turns out often to be in fact insufficient. Therefore it is
instructive to consider some general characteristics of band spectra
(classification and widths of energy bands, position of the Fermi level, the
value of the density of states $N(E_F)$) within simple model concepts.

The KKR method gives a possibility to separate the structural and dynamical
aspects of the band structure. Using the atomic sphere approximation, which
is discussed in Sect.2.2, enables one to obtain so-called canonical bands
which depend on the crystal structure only. These bands are obtained by
neglecting off-diagonal structural constants with $l\neq l^{\prime }$ in the
KKR equations (2.40), considering the potential functions $P_l(E)$ as
independent variables and diagonalizing the matrix $S_{lm,l^{\prime
}m^{\prime }}(\mathbf{k})$ for each value of $l$. Then for the set $n,l$ we
get $2l+1$ unhybridized energy bands $E_{ni}(\mathbf{k})$ defined by
$$
S_{li}(\mathbf{k})=P_l(E),\qquad i=1,2,...2l+1
\eqno{(2.78)}
$$
The canonical bands $S_l(\mathbf{k})$ for the bcc, fcc and hcp lattices are
shown in Figs.2.8-2.10, and the corresponding densities of states in
Fig.2.11. Because of the infinite range of s-type MT-orbitals, a pure
canonical s-band diverges at the centre of the Brillouin zone,
\[
S_s(\mathbf{k})\rightarrow -6(ks)^{-2}+\mathrm{const}
\]

Therefore one has to use in Figs.2.8-2.10 the free-electron-like scale $%
[1-(2/\pi )^2S_s]^{-1}$. The width of a canonical band is estimated from the
second moment:
$$
\widetilde{W}_l =(12S_l^2)^{1/2},\qquad
S_l^2=\frac 1{2l+1}\sum_{i=1}^{2l+1}\sum_{\mathbf{k}}S_{li}^2(\mathbf{k})
$$
$$
=2^{l+2}(2l+1)\frac{(4l-1)!!}{l!(2l)!}\sum_{\mathbf{R\neq }%
0}(s/R)^{2(2l+1)}
\eqno{(2.79)}
$$
and depends only on the number of atoms in coordination spheres. For the
bcc, fcc and ideal hcp lattices one has respectively
\begin{eqnarray*}
\widetilde{W}_l &=& 18.8,\ 18.7,\ 18.6 \\
\widetilde{W}_l &=& 23.8,\ 23.5,\ 23.5
\end{eqnarray*}
The canonical bands may provide a basis for further analysis of the band
structure. They reflect an important experimental fact: resemblance of band
structures of crystals with the same lattice. Real energy bands for various
elements are obtained from the canonical bands by specifying the functions $%
P_l(E)$. The hybridization of bands with different $l$ does not influence
strongly the spectrum provided that the bands do not cross each other.
However, for crossing s,p,d-bands in transition metals the hybridization may
lead to qualitative changes.

Consider the problem of the shape of energy bands, in particular the origin
of peaks in the density of states $N(E)$. As follows from the formula
$$
N(E)=\int d\mathbf{k}\delta (E-E_{\mathbf{k}})=\int \frac{dS}{|\mathrm{grad}_{%
\mathbf{k}}E_{\mathbf{k}}|}
\eqno{(2.80)}
$$
a peak of $N(E)$ corresponds to a nearly flat region (e.g., extremum) of the
surface $E(\mathbf{{k})}$ with a constant energy in the $\mathbf{k}$-space.
Such peculiarities of the spectrum occur in the one-dimensional case where $%
dE/dk=0$ at the boundaries of the Brillouin zones. As a result, DOS has
divergences of the form
$$
N(E)=|E-E_b|^{-1/2}
\eqno{(2.81)}
$$
According to the Van Hove theorem, in the general situation, DOS
singularities (divergence of $N(E)$ or $N^{\prime }(E)$) occur at some
points $E_c$ by topological reasons. In the two-dimensional case, these
singularities are logarithmic. The $N(E)$ pictures demonstrating the Van
Hove singularities were obtained for simple cubic, fcc and bcc lattices for
an s-band in the nearest-neighbour approximation [79] (Figs.2.12, 2.13). The
divergence of DOS at the band centre in the bcc lattice and the logarithmic
singularity in the fcc lattice at the band bottom are removed when one
includes next-nearest neighbours (Fig.2.14). Thus in the three-dimensional
case the usual Van Hove singularities have as a rule the one-sided
square-root form
$$
\delta N(E)=|E-E_c|^{1/2}\theta (\pm (E-E_c))
\eqno{(2.82)}
$$
and are too weak for explaining sharp DOS peaks. However, the peaks may form
at merging of Van Hove singularities along some lines in the $k$-space. As
demonstrated in [81], this leads to two-sided logarithmic singularities
$$
\delta N(E)=-\ln |E-E_c|
\eqno{(2.83)}
$$
As follows from analysis of calculations [24,78], the merging along the $P-N$
line in the bcc lattice results in formation of the ``giant'' Van Hove
singularities (2.83) in Li, V, Cr, Fe, Ba [81]. A similar situation takes
place for fcc Ca and Sr.

Some peaks in the canonical DOS's may be identified with the Van Hove
singularities. Despite a general similarity, the DOS, obtained in realistic
band calculations, contain a more number of peaks than canonical DOS's. The
additional peaks in calculated DOS pictures, which take into account
off-diagonal matrix elements (hybridization) between s,p,d-bands, may be
connected with splitting of a peak at the band intersection.

A comparison of canonical bands with real band calculations which take into
account hybridization is performed in the review [56]. The effect of
hybridization turns out to depend strongly on the mutual position of the
bands. As an example, Fig.2.15 shows the density of d-states for Nb. The
effect of p-d hybridization is weak since p- and d-band are well separated.
The s-d hybridization effect is more appreciable since s- and d-bands
overlap (as well as in other transition metals). However, the s-band is
influenced stronger than the degenerate d-band with a large electron
capacity. Another typical example is provided by molybdenum where the
splitting of some peaks occurs due to s-d hybridization (Fig.2.16) [82].

Now we discuss the position and width of energy bands. These characteristics
of band structure may be investigated either by direct using results of
numerical calculations or by parametrization within simple models which
neglect unimportant details. The latter approach permits to obtain a more
clear physical picture and to treat the problem analytically.

The simplest model of transition metals is the Friedel model [83] which
considers their electron system as containing free s-electrons and the
narrow d-band with the constant density of states of $10/W_d$. The lowering
of electron energy at formation of the d-band from the level is given by the
expression

$$
\delta E_{band}=5W_d\left[ -\frac{n_d}{10}+\left( \frac{n_d}{10}\right)
^2\right]
\eqno{(2.84)}
$$
A model description of band spectrum was proposed by Harrison [13]. Using
the set of one s-function and five d-functions in the form of MT orbitals
(2.44) he reduced the potential matix elements to two parameters - the width
of d-band $W_d$ and its position with respect to s-band bottom $E_d$.

As follows from (2.44), the $R$-dependence of the intersite matrix elements
of the potential has the form
$$
V_{ll^{\prime }m}=C_{ll^{\prime }m}=(r_{MT}/R)^{l+l^{\prime }+1}
\eqno{(2.85)}
$$
where $l,l^{\prime }$ are the orbital quantum numbers and $m=\sigma ,\pi
,\delta $ are determined by the angle between the direction of the orbital
and the vector which connects the atoms. In particular,
$$
V_{ss\sigma }\sim R^{-1},\quad
V_{sp\sigma }\sim R^{-2},\quad
V_{pp\sigma ,\pi }\sim R^{-3},\quad
V_{ddm}\sim R^{-5}
\eqno{(2.86)}
$$
(Note that in the free-electron approximation all the matrix elements are
proportional to $R^{-2}$ which results in a divergence after volume
integration). In an explicit form one may write down
$$
V_{ddm}=\eta _{ddm}\frac{\hbar ^2r_d^3}{mR^5}
\eqno{(2.87)}
$$
where the d-state radius $r_d$ is a characteristics of the element, $\eta $
are dimensionless constants which are tabulated in [13] (see also
consideration of f-elements in paper [84]). The values of $r_d$ are obtained
from comparison with first-principle band calculations. The quantities $W_d$
are obtained in terms of the matrix elements with the use of standard
formulas. For example, for the bcc lattice with account of nearest and
next-nearest interactions,
\[
W_d=-\frac 83V_{dd\sigma }^{(1)}+\frac{32}9V_{dd\pi }^{(1)}-3V_{dd\sigma
}^{(2)}+4V_{dd\pi }^{(2)}
\]
The values of $E_d$ and $W_d$ in the d-rows, together with the results of
band calculations [78] and experimental data [85], are presented in the
Table 2.1. One can see that $E_d$ and $W_d$ increase from Sc to V and
decrease from V to Cu.

In the pseudopotential theory, one can derive also the relation between the
effective masses of s- and d-electrons [13]
$$
\frac{m_s}{m_d}=\left( 1+2.91\frac m{m_d}\right) ^{-1}
\eqno{(2.88)}
$$
A number of theoretical models were proposed to describe the electronic
structure of ferromagnetic 3d-metals Fe, Co and Ni [85-88]. These models are
based on the idea of separating of localized and itinerant d-states with
different symmetry (e.g., of $t_{2g}$ and $e_g$ states).

Now we consider some general regularities in the results of band
calculations. These results agree qualitatively with above-discussed simple
model notions. In agreement with the idea of canonical bands, the DOS
pictures for elements of one column in the periodic table demonstrate a
similarity. This is illustrated by comparison of DOS for hcp metals Ti, Zr
and Hf (Figs.2.1-2.3). A similar situation takes place for Sc and Y (hcp),
V, Nb and Ta (bcc), Cr, Mo and W (bcc). In Tc and Re (hcp) the heights of
the DOS peaks (especially of the lower ones) differ appreciably. The same
difference occurs between Ru and Os (hcp), Rh and Ir (fcc), Pd and Pt (fcc).
The canonical band idea works also for elements of different columns with
the same lattice, the position of the Fermi level being changed roughly
according the rigid band model. This may be illustrated by comparison DOS of
V and Cr (bcc), Rh and Pt (fcc).

The values of partial DOS of s,p,d-type, which are given in Table 2.2,
demonstrate that d-states in transition metals are dominating ones. They
make up about 70-90\% of the total DOS at the Fermi level. One can see also
from Table 2.2 that large and small values of $N_d(E_F)$ alternate when
passing to a neighbour element. This rule holds even irrespective of the
crystal structure and demonstrates an important role of atomic
configurations. The regularity is to some extent violated in the end of
d-periods. This may be explained by ferromagnetism of Fe, Co, and Ni and
violation of regularity in d-shell filling. It should be also noted that the
numbers of d-electrons per ion in metals exceed by about unity the atomic
values [78].

To end this section we discuss the problem of accuracy of band calculations.
This problem is connected with a number of approximations: choose of initial
atomic configuration, the density functional approach for states which are
far from $E_F$ , the local approximation, the form of crystal potential
(including exchange- correlation contribution), the calculation method used,
an account of relativistic effects etc. The role of these approximations may
be clarified by comparing results of various calculations. Such a comparison
is performed in numerous review papers and monographs (see e.g. [57]). For
example, influence of initial atomic configuration is illustrated by band
calculations of vanadium (Fig.2.17). The width of s-band differs by two
times for the configurations 3d$^3$4s$^2$ and 3d$^4$4s$^1$. The Table 2.3
shows results of band energy calculation at some Brillouin zone points in Pd
by different methods [91-94] and corresponding experimental results [92].

We may conclude that the accuracy of energy determination makes up about 0.1$%
\div $0.3eV. The disagreemet with experimental data may be about 0.5eV. At
comparing the calculated band spectrum with angle-resolved emission and de
Haas - van Alphen effect data, one has often to shift it by such values. In
temperature units, the corresponding uncertainty is of order 10$^3$ K which
makes difficult precise calculations of thermodynamic properties within the
band approach.

\section{Experimental investigations of band structure: spectral data}

A number of quantities, which are obtained in band calculations (e.g., the
positin of energy bands and the shape of the Fermi surface, the electronic
structure near the Fermi level) may be analyzed and compared with results of
experimental investigations.

There exist a large number of spectral methods for investigating electron
structure. They may be divided in emission and absorption spectroscopy
methods. The first ones enable one to obtain data on the filled part of the
band (below $E_F$ ), and the second on the empty part (above $E_F$ ). Both
the optical and X-ray regions may be investigated. Let us list main spectral
methods and corresponding notations [40]:

PES Photoelectron (Photoemission) Spectroscopy

XPS X-ray Photoelectron Spectroscopy

UPS Ultraviolet Photoelectron Spectroscopy

IPES Inverse Photoemission Spectroscopy

AES Auger Electron Spectroscopy

BIS Bremsstrahlung Isochromat Spectroscopy

EELS Electron Energy Loss Spectroscopy

MXD Magnetic X-ray Dichroism

MXS Magnetic X-ray Scattering

SEC Secondary Electron Spectroscopy

XAS X-ray Absorption Spectroscopy

XANES X-ray Absorption Near Edge Spectroscopy

EXAFS Extended X-ray Absorption Fine Structure

When the spectroscopies are performed angle-resolved and spin-polarized,
acronyms AR and SP are added.

An advantage of the X-ray spectroscopy is the simple separation electron
states with different angular moments. By virtue of the selection rule $%
\Delta l=\pm 1$, $K$-spectra (transitions from 1s-states of inner core
levels of an atom) contain information on p-states of higher bands, and $L$%
-spectra (transitions from 2p-states) on s and d-states. Due to large
difference in the energies of $K$- and $L$-shells, these spectra are well
separated. This simplifies the comparison with band calculations which also
yield separately s,p,d-contributions. The optical part of the spectrum
corresponds to transitions between external overlapping energy bands or
inside them. Therefore the qualitative analysis of the optical spectra is
much more complicated. We do not discuss in detail these spectra of
transition metals where the interpretation is especially difficult (see
[95]).

On the other hand, the resolution of X-ray spectra in the energy measurement
is smaller than for optical spectra because of large intrinsic width of
internal levels (of order of 1eV, see the handbook [96]). The situation is
somewhat better only for soft X-ray spectra (e.g., for the $N$-spectra the
width of the internal 4f-level makes up about 0.15eV). Besides that, there
exist the contribution to level width which is due to finite lifetime of
electrons and holes. This contribution increases with increasing distance
from the Fermi energy:
$$
\Gamma _{\mathbf{k}}\sim (E_{\mathbf{k}}-E_F)^2  
\eqno{(2.89)}
$$
Provided that the density of states in a band contains a symmetric maximum
which does not overlap with contributions of other bands, its position may
be determined with a satisfactory accuracy (about 0.1 eV). However, such a
determination is prevented in the presence of overlapping bands. Usually
comparison of X-ray data with calculated spectra is performed by smoothing
out the band structure with account of the level width.

Occupied states are investigated in PES and ultravilolet PS methods. The
former yields, in particular, information on $K,L,M,N$ and $O$-spectra of
3d, 4d and 5d-metals. Very promising is the angle resolved photoemission
method which yields information on not only the density of state, but also
the spectrum $E(\mathbf{k})$. Methods of X-ray spectroscopy for empty states
include XAS, XANES, BIS, IPS. The relation between the IPES and BIS methods
is rougly the same as between the ultraviolet PS and X-ray PS ones.

Large widths of internal levels in $K,L$ and $M$ PES-spectra prevent as a
rule do resolve the density of state structure in the conduction band. The
comparison on$K,L$ and $M$-spectra of vanadium with results of theoretical
calculations are shown in Fig.2.18. One can see that the complicated DOS
structure [78] becomes smeared. At the same time, total width of the
spectrum coincides roughly with the calculated bandwidth after account of
the level broadening. So, the width is 8 eV for the $K$-spectrum and 6 eV
for the $L$ spectrum, the calculated bandwidth being equal to 4 eV.

The second typical example is the spectrum of zirconium (Fig.2.19). Here the 
$L_{III}$ spectrum (4d-2p transitions) has the width of 3.4 eV, and the $M_V$
spectrum (3d-4d transitions) the width of 2.9eV. High-energy maximum
corresponds to the $M_{IV}$ spectrum, and low-energy maximum at 6-7eV has
probably a satellite origin. Altough the width of the internal 4p-level is
considerably smaller than that of the 2p-level, the experimental broadening
is approximately the same for the $L_{III}$ and $N_{III}$ spectra. Possibly,
this is due to larger contributions from the s-4p transitions to the $%
N_{III} $ spectrum.

For next elements of 3d- and 4d-series, the widths of internal levels for $%
K,L,M$ spectra increase so that their quality becomes worse. The situation
is more favourable for the 5d series where the spectra $N_{VI,VII}$ (5d-4f)
and $O_{II,III}$ (5d-5p, 5s-5p) with a small internal level width occur.
Because of the closeness of 4f and 5d levels, four lines lie in the interval
of the order of 10eV. Nevertheless, because of small widths of f and 5p
levels, these lines are clearly resolved in the beginning of the 5d series,
emission bands being rather contrast and intensive. In the end of the
series, the intensity of lines decreases and their structure becomes
unclear. For the $N_{VI,VII}$ bands an opposite tendency takes place
(Fig.2.20).

An important role in the spectra of 5d-metals belongs to the spin-orbital
interaction. According to the selection rules, the transitions into internal
doublets f$_{5/2}$ , f$_{7/2}$ go from d$_{3/2}$ and d$_{5/2}$-levels of the
conduction band. Analysis of experimental data demonstrates that the
distributions of d$_{3/2}$ and d$_{5/2}$ -states in the conduction band are
different. Fig.2.21 shows the $N_{VI,VII}$ -spectra of iridium and platinum
with resolution of $N_{VI}$ (i.e. d$_{3/2}$) and $N_{VII}$ (d$_{5/2}$)
states. The main maximum $b_2$ is connected with states of both types. The
high-energy maximum $b_3$ is due to d$_{3/2}$ -states, and the low-energy
one $b_1$ to the d$_{5/2}$ -states. Thus the d$_{5/2}$ -states dominate at
the d-band bottom and the d$_{3/2}$ -states near the Fermi level. It is
interesting that the spectra of Ir and Pt are satisfactory explained by
atomic calculations which indicates a considerable localization of
5d-electrons in these metals.

A more complicated situation occurs for the $N_{VI,VII}$ -spectra in the
middle of the 5d-series (Ta, W, Re, Os), see [57]. The experimental width of
the $O_{III}$ -spectrum turns out to be larger than the calculated one. This
is connected with extra interactions and many-electron effects (e.g., the
influence of the vacancy in the $O_{III}$ -shell).

The ARP method takes into account not only energy, but also quasimomentum of
an electron in the conduction band. Consequently, it enables one to
determine experimentally the dispersion law $E(\mathbf{{k})}$ for various
branches, which gives a possibility to calculate the density of states to
high accuracy. This method was applied to investigate Cu [97], Ag [98], Au,
Pt [99], Pd [90], Ir [100].

The results for the spectrum of Cu in the $<211>$
direction [97] agree well with the free electron model. Himpsel and Eastmann
[90] measured $E(\mathbf{{k})}$ for Pd in the $<111>$ direction to accuracy
of 0.1$\div $0.2eV, the accuracy in the determining $\mathbf{k}$ being about
5\% of $k_{\max }$ (Table 2.3). One can see that agreement is not too good.
A similar situation takes place as for Pt and Au [99].

Although investigating the occupied part of the conduction band is more
important for electron properties of a metal, data for the empty part of the
band from absorption spectra are also useful from the point of view of
comparison with band calculations.

The BIS data for 3d and 4d-metals were obtained and compared with band
calculations in the paper [102] (see Table 2.4). The total resolution was
about 0.7eV, and the lifetime broadening at $E_F$ about 0.25eV. On the
whole, the band shapes obtained are close to those for canonical bands. The
agreement with calculated $N(E)$ turns out to be better than for XPS spectra
which are strongly distorted in the beginning of periods [103]. Besides
that, BIS specta do not exhibit satellites which correspond to the d$^{n+1}$
configuration. The d-bandwidth increases up to the middle of periods by
about 25\%. The s-d hybridization is important near the band bottom and is
most strong for hcp metals. Peaks near $E_F$ , which correspond to the empty
d-band, are followed by a structureless plateau and at 7-10eV by a step,
which is interpreted as the van Hove singularity. The peaks (d-band tails)
at $E_F$ are observed even in Cu and Ag. Two peaks are observed for Sc, Ti,V
and Zr, and one peak in other metals, which is in agreement with band
calculations. Strong asymmetry of the peak is observed in molybdenum in
agreement with theoretical predictions. The position of 1.6eV peak in Fe
corresponds to DOS in ferromagnetic rather than paramagnetic state
(Figs.2.22, 2.23). The relative d-band BIS intensity in the 3d-series was
derived and demonstrate to have a maximum in the middle of the series.

Calculated spin-resolved densities of states of ferromagnetic transition
metals Fe and Ni [24] are shown in Figs.2.24,2.25. Of great interest for the
theory of both the electron structure and magnetism is the investigation of
spin splitting $\Delta $. In principle, such investigations may be carried
out within the standard PES and IPES methods. However, these methods provide
only a rough estimations of averaged values of $\Delta $. According to band
calculations, $\Delta $ depends considerably on $E$ and $\mathbf{k}$, so
that angle resolved methods are required for obtaining detailed information.
Data on spin and angle resolved photoemission were obtained for Fe
[103-107], Ni [103,107-109] and Co [103].

According to the review [40], there exist in Fe both collapsing and
non-collapsing spin subbands which demonstrate different types of behaviour
at crossing the Curie point. Fig.2.26 shows spin-polarized angle-resolved
spectra of Fe in the (100) direction near the $\Gamma $-point at $h\nu =60$
eV. These data demonstrate the energy and temperature dependence of $\Delta $.

The ultraviolet photoemission spectrum of Fe was investigated by Pessa et al
[110]. The observed peaks at 0.58 and 2.4 eV below the Fermi energy may be
interpreted within the calculated density of states of ferromagnetic iron
provided that one shifts the latter by 0.5eV to higher energies. The value
of $N(E_F)$ for Fe obtained in [110] coincides approximately with the
results of band calculations [194], but is considerably smaller than that
determined from electronic specific heat [266] and paramagnetic
susceptibility (see the detailed discussion in the corresponding Chapters).
It should be noted that an error of order 0.1 eV in the energy measurement
may influence strongly the $N(E_F)$ value. Comparison of results [110] with
the spin-polarized spectra [111] confirms that the 0.58eV peak belongs to
the majority spin subband. The energy dependence of spin polarization is not
explained by the theoretical density of states. The experimental estimation
of spin splitting yields the value not more than 1.9eV which is considerably
smaller than theoretical one (2.35 eV). No shift of the spectrum at the
Curie point is observed which may indicate retaining of strong short-range
order above $T_C$ .

The d-bandwidth of Ni from the ARP data [108] makes up 3.4 eV at the point $%
L $, the theoretical value being about 4.5eV. The measured value of the
exchange splitting is 0.31 eV at 293K (the theoretical value is 0.7eV) and
about 0.2eV above the Curie point (Fig.2.27). Evidence of the existence of
spontaneous spin splitting above $T_C$ in Ni was obtained by the positron
annihilation technique [112].

Spin-resolved photoemission data on the short-range magnetic order above $%
T_C $ in Co were obtained in [113]. The problem of spontaneous spin
splitting in iron group metals may be also investigated in optical
experiments [114].

Recently, the Magnetic X-Ray Dichroism (MXD) method was applied to determine
the conduction electron spin polarization [40]. The L$_{23}$ absorption edge
(2p$\rightarrow $3d transitions) was investigated for Ni, and the asymmetry
in photon absorption with different polarization permitted to find both spin
and orbital momenta which turned out to be $s=0.52$ $\mu _B$ and $l=0.05\mu
_B$ . These values are close to theoretical ones, which are obtained with
account of many-electron effects [115]. According to the latter paper, the
ground state of the Ni ion is a superposition of the 3d$^{10}$ , 3d$^9$ and
3d$^8$ states with the weights of 15-20\%, 60$\div $70\% and 15$\div $20\%
respectively.

Consider the question about the origin of low-energy satellites in spectra
of some d-metals. Such satellites are observed in X-ray emission spectra of
Ni (at 6eV below $E_F$ ) (see, e.g., [116]) and Zr [57]. There exist the
mechanisms of satellite formation which are connected with vacancies at
excited atoms, Auger processes, shake up potential changes etc.(see [57]).
We discuss more interesting many-electron mechanisms which are specific for
transition metals.

Liebsh [117] treated the Coulomb interaction between two d-holes in Ni.
Introducing atomic Slater integrals $F^{(p)}$ ($p=0,2,4$), he obtained in
the low density approximation additional peaks in the spectrum and compared
them with experimental data. Besides the 6eV satellite, the theory yields
the second satellite which lies lower by 2eV. The occurence of satellites
resulted in narrowing of the d-band since the total number of states is
constant. The exchange splitting obtained turned out to be different for $%
e_g $ and $t_{2g}$ states (0.37 and 0.21 eV), the average value 0.3eV being
in satisfactory agreement with experiment. Penn [118] performed calculations
within the degenerate Hubbard model. The fitted value of the Hubbard
parameter was $U=2$ eV.

The authors of the paper [119] treated this problem exactly at restricting
to four point in the Brillouin zone, which is equivalent to consideration of
a four-atom cluster. An agreement with experimental data was obtained for $%
U=4.3$ eV and the ratio of the Hund and Coulomb parameters $J/U=1/7$.

A more consistent consideration of the satellite formation problem should
include many-electron terms (the satellite corresponds to the multiplet
structure of the configuration d${^8}$). The corresponding data of the Auger
spectroscopy which contain more information on many- electron effects are
discussed in Sect.1.2.

The ARP investigations of $\gamma$-cerium [120] yielded two dispersionless
density of states peaks near $E_F$ which did not obey the one-electron
selection rule. They were attributed to atomic-like 4f-states.

To describe high-energy spectroscopy data for cerium and its compounds,
Gunnarsson and Schoenhammer [121] proposed a model, based on the
one-impurity Anderson Hamiltonian and including effects of d-f
hybridization. Further this model was applied to consider spectra of
actinides and Ti (see [40]).

\section{Band calculations of rare earths and actinides}

One-electron band calculations of rare earths are performed by the same
methods as for 5d-metals, the role of relativistic effects becoming here
still more important because of strong spin-orbital interaction. Since the
4f-shell lies well below the conduction band, one may assume that the
picture of this band is similar for all the 4f-elements (especially for
heavy rare earths). This is confirmed by data of photoemision spectroscopy
from the 6s,5d band [122]. However, attempts of description in the
one-electron approach of highly-correlated f-electron themselves meet with
serious problems.

Results of band structure calculations of various RE metals by
non-relativistic and relativistic augmented plane wave method (APW and RAPW)
are presented in the review [19] (see also the discussion of electronic
structure for heavy rare earths [15]). For most RE metals (except for Ce,
Sm, Eu, Tm and Yb) the conclusions are as follows. The 4f-band lies by about
51$\div 1$0 eV below the conduction (5s,6sp) band and has the width about
0.05eV. The 5d-band moves up relative to the conduction band with increase
of the atomic number from La to Lu, the d-electron density of states at $E_F$
increasing from 1.9 to 2.5 eV$^{-1}$atom$^{-1}$ .

For elements with unstable valence Ce, Eu and Yb, especial attention was
paid to investigation of 4f-level position $\varepsilon _f$ . A strong
sensitivity to $x$ was found in APW calculations of Ce in 4f$^{2-x}$5d$^x$6s$%
^2$ ($x=0.5\div 2.0$) configurations [123]. The 4f-level is considerably
lower than 5d-6s band for $x=0$, but rapidly becomes higher and broader with
increasing $x$. The calculated in [124] distance $E_F-\varepsilon _f$ equals
to 0.36 and 0.24 Ry, and 4f-bandwidth to 0.01 and 0.02 Ry for $\gamma $- and 
$\alpha $-cerium respectively.

The full calculation of band structure of Ce in the linear RAPW method with
account of correction to MT-approximation was performed by Pickett et al
[125]. The width of 4f-band is about 1 eV and increases by about 60\%  at
the $\gamma -\alpha $ transition. The occupation numbers of f-electron
(about 1.1) changes weakly under presure. This leads to the conclusion that
the $\gamma -\alpha $ transition is accompanied by delocalization of
4f-electrons rather than by promotion into sd-band.

Relativistic APW calculation for Yb in 4f$^{14}$6s$^2$ configuration [126]
yielded unexpectedly the 4f-level position which is by 0.1Ry higher than the
6s conduction band (but still lower than $E_F$ by 0.2$\div $0.3Ry). A small
gap near $E_F$ was obtained in contradiction with experimental situation. In
the self-consistent calculation by Koelling (see [127]), $E_F\simeq
\varepsilon _f$ and the gap in the spectrum vanishes.

APW calculation of 4f-band in thulium [128] with account of 5s6d-4f
hybridization yielded $\varepsilon _{5d}-\varepsilon _{4f}=0.69$Ry, $%
\varepsilon _{6s}-\varepsilon _{4f}=1.27$Ry and the f-bandwidth of order 
10$^{-2}$ eV.

According to the first calculation of ferromagnetic gadolinium with account
of 4f-states within the self-consistent RAPW method by Harmon [127], the
f-level for spin down states (with the width about 0.03Ry) lies slightly
above $E_F$ because of large spin splitting. This should lead to giant
values of $N(E_F)$, and, consequently, of electronic specific heat, which
drastically contradicts to experimental data. Similar conclusions were
obtained in [129-131] for Tb, Dy and Gd. These results demonstrate
inadequacy of usual band calculations within local-spin-density
approximation for 4f-states. This problem is discussed in the recent papers
[132] with application with gadolinium.

In a number of works, attempts to take into account correlation effects were
made. Authors of papers [133-134] obtained a picture of 4f- and
conduction-electron states with account of correlation, screening and
relaxation. The f-level energies for the ground and excited configurations 
f$^n$ and f$^{n+1}$ were determined by
$$
\Delta _{-}=E(f^n)-E(f^{n-1}),\qquad \Delta _{+}=E(f^{n+1})-E(f^n)
\eqno{(2.90)}
$$
Calculations were performed within non-relativistic [133] and relativistic
[134] approximations.

The crystal potential in [133] was constructed from renormalized
Hartree-Fock functions for the configurations 4f$^n$5d$^{m-1}$6s$^1$ and 4f$%
^{n\pm 1}$5d $^{m-1\mp 1}$6s$^1$ ($m=2,3$). The renormalization (transfer of
atomic wavefunction tails inside the Wigner-Seitz cell) resulted in a strong
change of density for 5d, 6s electrons and, because of screening effects, in
a considerable (by about 0.5Ry) increase of one-electron f-energies. The
values of $\Delta _{+}$ were corrected by introducing correlation effects
from comparison with spectral data (the correlation energy was assumed to be
the same in the cases of free atom and crystal). Although the correlation
energy in each state may be large, their difference
$$
\xi ^{+}=E_{corr}(f^nd^2s)-E_{corr}(f^{n+1}ds)  
\eqno{(2.91)}
$$
makes up about 0.1Ry only. The results of calculation of the quantity 
$\xi^{+}$ are shown in Fig.2.28.

The edges of 5d- and 6s-bands were determined from the zero of the
wavefunction and its derivative at the Wigner-Seitz cell boundary. The
results turned out to be somewhat different from those of standard
calculations. The 4f-bandwidth made up 0.03Ry for $\gamma$-Ce, 0.01Ry for Pr
and Nd and smaller than 0.005Ry for other rare earths.

One of important drawbacks of the calculation [133] was using of f-level
energies $\varepsilon_f$ which are averaged over many- electron terms. At
the same time, spectral data, which were used, corresponded to transitions
between ground states with definite $S$ and $L$. Since the difference in
energies of terms reaches 5eV, the error in the correlation energy may be
large. A detailed calculation of the term structure for the f$^{n-1}$
configuration was performed in [135]. The corresponding line spectrum should
be observed in experiments with high resolution. As follows from equation
(2.48), the averaged energy $\varepsilon_{4f}$ may differ considerably from
the ionization energy with the final state corresponding to the Hund term.
The difference may be presented in the form
$$
(-2\frac bn+\Delta b)F^{(2)},\qquad 
\Delta b=b(f^nSL)-b(f^nS^{\prime }L^{\prime })
\eqno{(2.92)}
$$
where the coefficient $b($f$^n$ $SL)$ determines the contribution of the
Slater integral $F^{(2)}$ to the $SL$-term energy, $2bF^{(2)}/n$ is the
Coulomb energy per electron in the Hartree- Fock approximation, which is
averaged over terms. In the rare earth series the value of $\Delta b$
changes from 0 (f$^0$-configuration) to 113 (f$^7$ -configuration), and $%
\Delta b(f^{n+7})=\Delta b(f^n)$. Since the values of $F^{(2)}\simeq 0.05$eV
are not too small, this contribution to the f-electron energy should be
important.

In the relativistic calculations [134], an account of term structure was
also performed. Final results for the effective one- electron energies
differ weakly from those obtained in [133] (Table 2.5), but the changes in
correlation corrections are appreciable. In particular, the correlation
energy difference $\xi ^{+}$, which is negative for most rare earths
according to [133], becomes positive. One can interpret these data as
increase of f-f correlation with increasing number of f-electrons.

Although the absolute calculated values of $\varepsilon _{4f}$ and $E_F$ are
not quite reliable because of numerous approximations, general regularities
in the rare-earth series seem to be qualitatively correct. This is confirmed
by data on photoelectron spectra for the f-shell [136,137].

A correct calculation should take into account also the change of
wavefunctions in the final configuration f$^{n-1}$. The total relaxation in
the final state (f$^n$d$^m$ $\rightarrow $ f$^{n-1}$d$^{m+1}$ transition,
the hole in the 4f-shell is completely screened by additional electron in
5d-state) was assumed in [133-134]. Such an assumption explains the decrease
of 4f-electron binding energy in comparison with the Hartree-Fock
approximation. An attempt of the calculation in absence of relaxation (f$^n$d%
$^m$ $\rightarrow $ f$^{n-1}$d$^m$ transitions) [134] agrees worse with
experimental data.

Now we discuss briefly the results of band calculations for 5f- elements
[138,139,57]. Band structure of actinides is to some extent similar to that
of rare earths, but the strong spin-orbital coupling is accompanied with
considerable delocalization of f-states. Thus the spectrum is appreciably
disturbed owing to overlap and hybridization between s,p,5d-states with
5f-bands. Concrete forms of bandstructure and f-level positions in the
self-consistent calculations depend strongly on the choose of the
exchange-correlation potential. In the RAPW calculation of thorium by Keeton
and Loucks [140] with the Slater potential ($\alpha =1$) the position of
5f-band (in the middle of the valence band) contradicted to de Haas - van
Alphen data. However, the calculation [141] with the Gaspar-Kohn-Sham
potential ($\alpha =2/3$) yielded the correct 5f-band position (well above
the Fermi level). Thus the band structure of Th is similar to that of
d-metals. Despite the absence of f-electrons, the shape of d-band is
appreciably influenced by d-f hybridization.

According to [138], the 5f-bandwidth decreases with increasing atomic number
from uranium ( 0.4Ry) to plutonium ( 0.3Ry). The electron spectrum of U near 
$E_F$ contains s-d conduction band which is strongly hybridized with a
rather wide 5f-band. The 5f-bands in Pu are considerably more flat for 
$\alpha =1$ than for $\alpha =2/3.$ This indicates that Pu lies on the
boundary of the f-electron localization.

In the self-consistent calculation [142], the f-band turns out to be
narrower than in [138]. The density of states of Pu contains sharp peaks
with the height about 150Ry$^{-1}$. The total value of $N(E_F)$ is 123.6
states/Ry (the f-contribution is about 50 states/Ry), which corresponds to
linear term in electronic specific heat with $\gamma $ = 21mJ/mol K$^2$ .
The experimental value makes up about 50 mJ/mol K$^2$, so that the
calculation seems to be qualitatively satisfactory, but account of
correlation enhancement is needed.

Relativistic calculations of band structure in heavy actinides (Am, Cm, Bc)
was also performed by Freeman and Koelling [138]. As compared to light
actinides, the width of 7s-band increases, 5f-bands become rather narrow and
flat, the sd-5f hybridization being suppressed. Thus a strong localization
of 5f-electrons takes place. As well as for rare earths, usual band
description becomes in such a situation inapplicable. An important role of
correlations is confirmed by high sensitivity of the spectrum to the number
of f-electrons. The calculation for Am in the configuration 5f$^7$7s$^2$
(instead of the correct one, 5f$^6$6d$^1$7s$^2$ ) yields a shift of
5f-states by 0.5Ry. The delocalization of 5f-electrons in heavy actinides
under pressure was investigated in the band calculations [143].

\section{Fermi surface}

One of most powerful tools to verify the results of band calculations is
investigation of the Fermi surface (FS). The structure of FS determines a
number of electronic properties of metals, in particular their anisotropy.
The shape of FS depends on the geometry of Brillouin zones, crystal
potential and interelectron correlations. The volume under FS is fixed by
the Luttinger theorem: it has the same value as for the non-interacting
electron system. It should be noted that this theorem is valid only provided
that the Fermi-liquid picture (Landau theory) holds, and may be violated in
system where strong Hubbard correlations result in splitting of one-electron
bands (e.g., in systems which exhibit correlation-driven metal-insulator
transitions [25]), so that the statisitics of one-particle excitations
changes.

In band calculations, FS is obtained as the constant energy surface which
separates occupied and empty states at $T=0$. If FS lies far from the
boundaries of the Brillouin zones (e.g., at small band filling), it differs
weakly from that of free electrons and has nearly spherical shape. At
approaching the zone boundaries, the influence of crystal potential becomes
stronger and anisotropy of FS increases. Especial strong singularities of FS
occur at crossing a boundary. Since a Brillouin zone contains two electron
states per atom (with account of the spin quantum number), in most cases one
has to use several zones to place all the conduction electrons. Due to the
anisotropy of the lattice and, consequently, of the electrons spectrum,
zones with higher numbers start to be filled up when the lower zones are
only partially filled, so that FS crosses several zones.

In transition d-metals, the valence electrons include d-electrons. An
important feature of d-states is that they are highly anisotropic and poorly
described by the free electron approximation even far from the Brillouin
zone boundaries. Therefore the shape of the Fermi suface is very complicated
and one has to use a large number of zones to describe it. Usually one
considers Fermi surfaces of transition metals in the reduced zone picture.

\subsection{Methods of Fermi surface investigation and de Haas-van Alphen 
effect}

Experimental information on FS may be obtained from investigation of
anisotropic electron characteristics of a metal. Most widely used methods of
constructing FS which are based on magnetic field effects in electron
spectrum. To first place, here belong so-called oscillation effects - de
Haas - van Alphen (dHvA) effect (oscillations of magnetic susceptibility)
and Shubnikov - de Haas effect (oscillations of conductivity), and also a
number of magnetoacoustical and resonance effects [10].

Microscopical basis for the oscillations effects is the Landau quantization
in magnetic field. It is well known that at inclusion of the field $H_z$ the
orbital motion of free electron is quantizised in the $xy$-plane. The
quantization condition is written in terms of the area of the circumference
with the radius $(k_x^2+k_y^2)^{1/2}$: 
$$
S_\nu =\pi (k_x^2+k_y^2)=\frac{2\pi eH}{\hbar c}(v+\frac 12)  
\eqno{(2.93)}
$$
where $\nu =0,1,...$ are the numbers of Landau levels. The corresponding
condition for energy reads
$$
E_\nu \equiv E(k_x,k_y,k_z)-E(k_z)=\hbar \omega _c(v+\frac 12)  
\eqno{(2.94)}
$$
with
$$
E(k_z)=\frac{\hbar ^2k_z^2}{2m},\qquad \omega _c=\frac{eH}{mc}  
\eqno{(2.95)}
$$
being the kinetic energy of an electron moving in $z$-direction and the
classical cyclotron frequency which describes rotation of an electron in the 
$xy$-plane. Therefore the quantization in magnetic field results in
occurrence of an oscillatory contribution to the electron energy. Comparing
(2.93) and (2.94) we may write down
$$
S_\nu =\frac{2\pi m_c}{\hbar ^2}E_\nu  
\eqno{(2.96)}
$$
where
$$
m_c=\frac{\hbar ^2}{2\pi }\frac{\partial S_\nu }{\partial E_\nu }  
\eqno{(2.97)}
$$
the expression (2.97) being valid for an arbitrary dispersion law of
conduction electrons [10-12]. Thus at considering the electron motion in
magnetic field the free-electron mass m is replaced by the cyclotron mass $%
m_c$. Unlike the band effective mass, which is defined at a point of the $%
\mathbf{k}$-space, $m_c$ is defined for a closed curve coresponding to an
electron trajectory on the Fermi surface in the magnetic field.

The difference of energies for two neighbour Landau levels reads
$$
E_{\nu +1}-E_\nu =\hbar \omega _c=\frac{2\pi eH}{\hbar c}\left( \frac{%
\partial S}{\partial E}\right) ^{-1}  
\eqno{(2.98)}
$$
Since the numbers $\nu $ are large and the energy differences are small in
comparison with energies themselves, we may put
$$
\frac \partial {\partial E}S(E_\nu )=\frac{S(E_{\nu +1})-S(E_\nu )}{E_{\nu
+1}-E_\nu }  
\eqno{(2.99)}
$$
As follows from (2.98), (2.99), the areas of orbits in the $k$-space differ
by the constant value
$$
\Delta S=\hbar \omega _c\frac{2\pi eH}{\hbar c}  
\eqno{(2.100)}
$$
or 
$$
S(E_\nu ,k_z)=(\nu +\lambda )\Delta S  
\eqno{(2.101)}
$$
where $\lambda $ does not depend on $\nu $ (typically, $\lambda =1/2$).

Quantization of electron motion in $xy$-plane leads to that the
quasicontinuous distribution of electron states is influenced by a discrete
dependence on Landau level numbers. Therefore the resulting electron density
of states $N(E)$ in the external magnetic field will exhibit an oscillating
behaviour. To demonstrate the occurrence of oscillations we suppose that the 
$\nu$-th Landau level crosses $E_F$ at some $H$ for a given $k_z$ , i.e.
$$
E(k_z)+\hbar \omega _c(\nu +\lambda )=E_F  
\eqno{(2.102)}
$$
At slight decreasing $H$, $\omega _c$ increases and $N(E)$ decreases.
However, with further decreasing field the Fermi level is crossed by the ($%
\nu +1$)-th Landau level and $N(E)$ takes again the maximum value at $%
H=H^{\prime }$. Subtracting the equalities (2.93) for two adjacent Landau
levels one can see that the oscillating behaviour of $N(E)$ (and of related
physical properties) as a function of the inverse magnetic field has the
period
$$
\Delta \left( \frac 1H\right) =\frac 1H-\frac 1{H^{\prime }}=\frac{2\pi e}{%
\hbar c}[S_v(E_F)]^{-1}  
\eqno{(2.103)}
$$
To complete the consideration we have to sum over $k_z$ which was up to now
fixed. Since the energy $E$ depends on $k_z$, for most $k_z$ the condition
(2.102) will be satisfied at different values of $H$ and the summary
oscillating contribution will be absent. However, this condition will
approximately hold for a finite segment of $k_z$ values provided that the
energy varies very weakly in the $k_z$ direction, i.e.
$$
\frac{\partial E_{\mathbf{k}}}{\partial k_z}=0  
\eqno{(2.104)}
$$
or, which is equivalent,
$$
\frac{\partial S_\nu (k_z)}{\partial k_z}=0  
\eqno{(2.105)}
$$
Thus the Landau quantization results in an oscillating behaviour of the
electron density of states $N(E)$, this quantity having a singularity if $E$
equals to the energy of a stationary (extremal) orbit determined by (2.105).
The value of $N(E_F)$ which determines most electron properties of a metal
should be singular at
$$
S_{extr}(E_F)=(\nu +\lambda )\frac{2\pi eH}{\hbar c}  
\eqno{(2.106)}
$$
This permits to determine the extremal cross section areas by measuring the
oscillation period.

At $T=0$, near each energy $E_0$ which satisfies the quantization condition, 
$N(E)\sim |E-E_0|^{-1/2}$ (as well as in the case of one-dimensional Van
Hove singularity). In real situations, the oscillation effects are smeared
due to thermal excitations at the Fermi surface over the energy interval of
order $T$ or due to impurity scattering. The oscillations are observable at 
$$
k_BT<\hbar \omega _c\sim \frac{e\hbar }{mc}H=2\mu _BH  
\eqno{(2.107)}
$$
Usually one has to use the magnetic fields of order 10 kG at temperatures of
a few K.

To obtain the oscillating contributions to physical quantities one has to
perform the integration over $k_z$. In the case of free electrons the result
for the magnetization at $T = 0$ reads [144]
$$
M=\frac{eE_F}{4\pi ^2\hbar c}(m\mu _BH)^{1/2}\sum_p\frac{(-1)^p}{p^{3/2}}%
\sin \left( \frac{\pi pE_F}{\mu _BH}-\frac \pi 4\right)  
\eqno{(2.108)}
$$
The oscillation amplitude may be repesented in the form
$$
\overline{M}\sim \frac 1{2\pi ^3}\left( \frac{\mu _BH}{E_F}\right) ^{1/2}\mu
_B  
\eqno{(2.109)}
$$
and, for small $H$, is large in comparison with the smooth part of the
magnetization (of order of $H$). The diamagnetic susceptibility $\chi =dM/dH$
may be of order of unity, so that its oscillations are very strong.

The corresponding results for an arbitrary dispersion law are analyzed in
detail in the monograph [10]. The oscillation period is obtained by the
replacement
\[
\pi E_F/\mu _B\rightarrow \frac{\hbar c}eS_{extr}(E_F) 
\]
in (2.108). Thus investigations of magnetic susceptibitlity oscillations (de
Haas - van Alphen effect) allow one to determine the areas of extremal cross
sections of the Fermi surface. The temperature dependence of magnetization
oscillation amplitude enables one to determine the cyclotron mass:
$$
\overline{M}\sim T\exp (-2\pi ^2m_ck_BT/e\hbar H)  
\eqno{(2.110)}
$$
In ferromagnetic metals, oscillation effects take place on the background of
large spontaneous magnetization $M_s$. In particular, cross sections of the
Fermi surface, which correspond to both spin projections, should be
observed. For iron, the oscillation amplitude is by approximately nine
orders smaller than $M_s$ [145], but is still several order larger than
which can be measured by most sensitive modern technique.

\subsection{Experimental and theoretical results on the Fermi surfaces}

Investigation of the Fermi surface in TM is a considerably more difficult
problem in comparison with simple metals. The reasons for this are as
follows. High orbital degeneracy of d-electrons (unlike s,p-electrons)
results in a large number of sheets of FS. Some of these sheets correspond
to large cyclotron masses, which requires using strong magnetic fields in de
Haas - van Alphen measurements. For magnetically ordered metals (Fe, Co, Ni,
Mn, Cr) the picture is complicated by the presence of spin splitting of
conduction band states. Besides that, there exist technical problems at
preparing the samples which possess sufficient purity to observe
oscillations. As a result, the situation in the fermiology of TM is far from
total clearness - there are uncertainties in experimental results for a
number of elements and ambiguities in theoretical models. Several d-metals
(Sc, Hf, Mn, Tc, Ru) and many rare-earths and actinidies are practically not
investigated. Data on cross sections with large $m_c$ are often absent.

From the theoretical point of view, there exist problems connected with the
applicability of one-electron picture in systems where electron correlations
play an important role. Even within one- electron band calculations the
results are rather sensitive to the choose of the crystal potential, so that
the shape of FS may exhibit essential (in particular, topological) changes.
This leads to difficulties in the interpretation of experimental data (e.g.,
in identification of different FS sheets and their details). The accuracy of
band calculations does not as a rule exceed 0.5eV. In a number of case, one
has to fit theoretical results to experimental data by modifying some
parameters, e.g., by shifting the Fermi level position or changing values of
the gaps between different branches of electron spectrum.

However, some important results are now well established. First of all,
existence of FS sheets with heavy electron effective masses, which are
characteristic for TM and their compounds, should be mentioned. (Especially
large values were found for so-called heavy-fermion rare-earth and actinide
systems which are discussed in detail in Chapter 6.) Thus FS investigations
yield direct proof of delocalization of d-electrons. In particular, these
data demonstrate itinerant character of strong magnetism in iron group
metals.

General information about the Fermi surfaces of TM (and, for comparison, of
Li, Na and Ca) may be found in the Table 2.6 which contains data on
classifications of sheets, stationary cross sections $S$ and corresponding
effective masses $m_c$. Since there exist a large number of different data
for $S$ and $m_c/m$, the results presented should not be considered as
unambiguous. As a rule, we write down later data. In many cases, we restrict
ourselves to writing down the interval of corresponding values. Most data on 
$S$ are taken from the handbook [146], which contains the information up to
1981 and the detailed bibliography. More recent data are included provided
that they seem to be important. Unlike [146], we give all the values of $S$
in units of \AA$^{-2}$. The transition from oscillation
frequencies in Tesla (or periods in T$^{-1}$) may be made according to the
formula
\[
S(\mbox{\AA}^{-2})=9.55\cdot 10^{-5} F(\mathrm{T})
\]
The values of $m_c/m$, which are also included in the Table 2.6, are of
especial interest for transition metals.

Before discussing the FS characteristics of metals of each group (and,
separately, of ferromagnetic TM Fe, Co, and Ni) we make some preliminary
notes. As well as in the case of simple metals, correspondence between Fermi
surfaces of TM with the same lattice structure may be established. In the
rigid band approximation, the modification of FS reduces to a shift of the
Fermi energy. This circumstance was used often (especially in the early
works) for constructing FS at lack of theoretical calculations [11]. For
example, the FS of Rh was obtained by Coleridge from that of Ni. The FS of
bcc Fe was used to determine FS of a number of bcc TM (e.g., Cr, Mo, W). The
type of FS is fairly good retained within a given column of the periodic
system. However, with increasing the atomic number $Z$, the role of
spin-orbital interaction increases. The latter may change qualitatively the
form and topology of FS: because of degeneracy lift for some energy bands,
additional gaps in the spectrum occur and some open orbits may be
transformed into closed ones. Deviations of FS in TM from that for free
electrons become more appreciable with increasing $Z$ in a given period,
which agrees with the general statement about increasing the d-electron
localization.

The role of d-electrons in the FS formation may be characterized by the
value of $m_c/m$. One can see that this quantity increases in each d-period
from left to right. In the beginning of d-periods (up to the Fe column)
usually $m_c^{th}/m<1$ (except for Ti where $m_c^{th}/m=$ 1.95), but maximum
values of $m_c^{th}/m$ are considerably larger. The latter fact is
apparently connected with effective mass enhancement (electron-phonon, spin
fluctuation etc.) which is not taken into account in band calculations. At
the same time, a satisfactory agreement between theory and experiment takes
place for cross section areas. Starting from the Co column, large values $%
m_c^{th}/m=3.02\div 10.53$ occur, but the experimental values turn out to be
smaller (e.g, for Ni, $m_c^{th}/m\simeq 8$, $m_c^{exp}/m\simeq 1.9$; the
largest values belong to palladium). Probably this is due the heaviest
electron masses are still not found. Unfortunately, incompletness of
experimental data does not permit to trace a regularity of $m_c/m$ values in
the periodic table.

\subsubsection{Li, Na, Ca}

To pick out peculiar features of TM, it is instructive to discuss first
alcaline metals which are satisfactorily described by the free electron
model (FEM). In particular, FEM yields correctly all the FS cross sections
and effective masses of sodium (both experimental and theoretical). However,
for lithium there exist some deviations, especially in $m_c$ values. This is
explained by a strong localization of the valence electron in Li near the
nucleus, which is due to weakness of nuclear charge screening by the 1s$^2$
shell (a similar effect takes place for d-electrons in TM since their mutual
screening is also small). With increasing $Z$, disagreement with FEM
increases and is appreciable for caesium.

The band spectrum and Fermi surface of calcium, which stands before d-metals
in the periodic system, already differ qualitatively from predictions of
FEM. Not all predicted stationary cross sections are experimentally
observed. This may be explained by the presence of the above-discussed DOS
peak near the Fermi level. Very strong are the deviations from FEM near the
band top. Probably, they are connected with the hybridization with d-states
which lie somewhat higher.

\subsubsection{Sc, Y, La}

Unfortunately, experimental data on d-elements of first column are poor
because of absence of pure samples. Calculated FS of Y and Sc are similar
and have multiconnected form with open orbits. They are stongly disturbed in
comparison with FEM. Results of various calculations differ considerably,
and their verification is difficult because of lack of experimental
information.

\subsubsection{Ti, Zr, Hf}

The results of various band calculations demonstrate considerably different
results. Thus unambigous theoretical models of FS are absent for both Ti and
Zr. In the work on dHvA effect in Zr [164], an attempt of interpretation in
the nearly-free electron model was performed. In the papers [159-161],
experimental and theoretical cross section and $m_c/m$ values for Ti are
presented. The calculated FS is very sensitive to small variarions of $E_F$
position. The maximum value of $m_c^{exp}/m$ is 2.8. The FS of Hf is not
investigated, but theory predicts its similarity to FS of Zr.

\subsubsection{V, Nb, Ta}

First attempt to construct the Fermi surfaces of these bcc metals was made
in the rigid band model by using the calculation [192] for the bcc Fe. DHvA
measurements in Nb and Ta [170,171] yielded two oscillation periods which
were related to holes at the point $N$. Other periods were attributed to
``jungle gym'' features. A detailed comparison with theory for Nb and Ta was
carried out in [172]. A satisfactory agreement (about 5\%) was obtained for
the cross sections areas of the types 2 and 3. The holes in the second zone
at the point $\Gamma $ were observed in dHvA experiments [171]. Maximum
values of $m_c/m$ is 1.8 which agrees with the data on specific heat.

\subsubsection{Cr, Mo, W}

The Fermi surface of chromium is strongly influenced by antiferromagnetism.
The spin-density waves are connected with definite pieces of FS [11]. One of
recent calculation was performed by Kulikov and Kulatov [181]. Unlike
previous papers, appreciable deviations of from the sphere were obtained. It
was found that the electron masses of the ball orbits exceeds considerably
those of the ellipsoid orbits.

To understand main features of Mo and W one has to remember that the Fermi
level lies at a minimum of density of states. Thus the role of d-states is
not so strong as in other TM. Probably, this is the reason for that orbits
with large $m_c/m$ are absent in both experiment and theory. For all the
four FS sheets the values $m_c/m<1$ are characteristic. First calculations
of FS in Cr, Mo, W used the band calculation for Fe [192]. More recent
calculations [176,177] supported in general the rigid band model. However,
because of the strong spin-orbital interaction in W, the hole surface $h_H$
in the third zone and electron surface $e_{\Gamma H}$ in the fifth zone
vanish or strongly diminish. The LAPW calculation [178] confirms these old
results.

\subsubsection{Mn, Tc, Re}

$\alpha $-manganese, which has a complicated cubic structure and is
antiferromagnetic at low temperatures, is poorly investigated, as well as
the radioactive technetium. Band calculations of bcc $\gamma $-Mn were made
in [188,189]. At the same time, there exist detailed results on rhenium, the
theory [190] being in good agreement with experiment. Some uncertainty takes
place for open orbits $e(8)$ (cylinder or tor?). Besides $m_c/m\simeq 2.8$,
there are very light masses $m_c/m\simeq 0.07$.

\subsubsection{Fe, Co, Ni}

Extensive investigations of FS in Fe and Ni were performed by using dHvA and
halvanomagnetic effects. It turned out that the dHvA data may be obtained in
an usual way, although the oscillations are periodic in $1/B = 1/(H + 4\pi
M) $ rather than in $1/H$. However, full picture of the Fermi surface is up
to now absent.

Experimental data for Fe and Ni are discussed the review [145]. The main
conclusion is the corroboration of itinerant electron ferromagnetism
picture: the sum of volumes, corresponding to spin up and down surfaces,
agrees with the total number of conduction electrons, and the difference
with the magnetization. It should be, however, noted that this conclusion is
based not only on experimental results, but uses also some model notions.
Besides that, the picture of spin-split bands is complicated by the
spin-orbital interaction. The latter results in the degeneracy lift at
points of accidental crossing of different spin and orbital subbands.

A detailed comparison of the dHvA results for Ni with the theoretical
calculations is performed in [145]. The theoretical and experimental
pictures demonstrate similarities. The s-d hybridization turns out to be
important for understanding the experimental situation.The quasiellipsoidal
pocket $X_5$ has an unusual form which is connected with the anisotropic
interaction with the neighbour $X_2$ band. The largest experimental mass $%
m_c=1.8m$ seem to correspond to the $X_{2\downarrow }$ pocket. The band
calculation yields for the large belly surface in the [001] direction $%
m^{th}/m\simeq 8$ (from the electronic specific heat, $m^{*}/m\simeq 10$).
Unlike Pd and Pt, such large values were not observed in dHvA experiments.

The situation with FS of iron is considerably less clear. In particular, no
theoretical calculations explain its shape in detail, and there exist
several theoretical model to describe this [145].

Calculation of FS of cobalt was performed in [197]. In subsequent papers
[198-200] detailed theoretical information on stationary cross sections and
effective masses was obtained. Experimentally, a few of dHvA oscillation
frequencies were found [201-203]. Two of them, $\alpha$ and $\beta$,
correspond to cyclotron masses of $0.08m$ and $0.2m$.

\subsubsection{Ru, Os}

DHvA data for ruthenium [212] was interpreted in terms of the rigid band
model by using the band structure of rhenium (see also [213]). For osmium,
two groups of dHvA frequencies were found which difffer by two orders of
magnitude [214]. The largest ratio $m_c/m$ is 1.5. The dependence 
$m_c^{exp}/m$ on the area $S$ turns out to be linear which corresponds to
the quadratic dispersion law $E(\mathbf{{k})}$.

\subsubsection{Rh, Ir, Pd, Pt}

Despite the difference in the d-electron number, it is convenient to combine
the d$^7$ and d$^8$ metals since they have the same fcc lattice and possess
similar properties. Early experiments on dHvA effect for Rh and Ir [216,217]
and Pd and Pt [223,224] were interpreted in the rigid band model with the
use of results for Cu and Ni. The model of the Fermi surface obtained is
shown in Fig.2.29. The calculation in the relativistic APW method was
performed in [215]. Comparison of theoretical and dHvA experimental values
demonstrates an appreciable enhancement of $m_c$ in Pt and Pd for light
electron and holes. However, dHvA frequencies corresponding to heavy holes
at open sheets (which yield about 80\% of $N(E_F)$) with $m_c^{th}/m=9.1$
and 6.23 were not observed in early experimental works. Only in the paper
[225], the values $m_c/m=5.7\div 8.1$ were found for these holes in Pd.
These values are still somewhat lower in comparison with the theoretical
ones [215] and are by about 15\% smaller than those required to explain data
on electronic specific heat. At the same time, data on cross section areas
[225] agree well with the theoretical predictions.

\subsubsection{Rare earths and actinides}

Fermi surfaces of 4f- and 5f-metals were calculated in a number of works
(see [11]). On the other hand, direct experimental data are not numerous.
DHvA measurements in ytterbium [229] demonstrated the presence of
hyperboloid pieces of FS. DHvA investigations of gadolinium [230 ] yielded
the estimation of the spin splitting $\Delta = 0.57$ eV.

Provided that 4f-states (that lie as a rule well below the conduction band)
are not taken into account, FS of all the rare earth metals should be
similar for the same crystal lattice. A FS model for the hcp structure which
describes Gd and Tb is presented in Fig.2.30. The Fermi surface lies within
third and fourth zones and is multiconnected. The shape of Fermi surface is
in general similar to that of yttrium. At passing to heavy rare earths,
changes occur which are connected with the degeneracy lift at some Brillouin
zone points owing to the relativistic interactions. According to [231], the
latter result in that the ``arms'' at the point $M$ vanish, and the ``arms''
at the point $L$ become touching, so that the similarity with FS of ittrium
increases. Thus one may expect that, at increasing atomic number, the Fermi
surfaces of heavy hcp rare earths change their shape between those of Gd and
Y. A number of attempts were made to use information on FS in rare earths
for considering concrete physical effects. To first place, here belong
explanation of complicated non-collinear magnetic structures (Sect.4.7).

Actinides (5f-elements) are investigated still less than rare earths. Unlike
4f-electrons, 5f-electrons, at least in light actinides, are delocalized and
take part in the electron spectrum near the Fermi level. First FS
calculations within the relativistic APW method for Th and Ac were performed
in [232]. DHvA data for thorium, obtained in [233], were used to construct
an empirical FS model. The latter contains a nearly spherical sheet at the
centre of the zone and six closed sheets at the points $X$ along the (100)
directions. This model turned out to contradict the band calculation [140].
Some contradictions were eliminated in the subsequent calculation [234].

\chapter{THERMODYNAMIC PROPERTIES}

\section{Cohesive energy and related properties}

Condensation of atoms from the gaseous states into liquid and solid ones
occurs due to various interatomic forces. Main contribution to the binding
energy comes from valence electrons which have large orbit radii, so that
the wavefunctions at different lattice sites overlap appreciably. For
metals, the valence electrons become itinerant and propagate freely in a
crystal, the energy gain being determined by their interaction with
positively charged ions.

As discussed in Chapter 1, d-electrons (and sometimes f-electrons) in
transition group metals exhibit the behaviour which is intermediate between
the localized and itinerant one. In particular, the collapse of d- and
f-orbits in the beginning of periods takes place, which is accompanied by
formation of high-energy states in the potential well (Fig.1.2). For
d-elements the energy is close to that of valence sp-electrons (although the
radius of d-shell is much smaller than the radii of sp-orbits). For
4f-electrons the well is much deeper, so that they may contribute
appreciably to the binding energy only at the beginning of period, e.g., for
cerium.

From the thermodynamical point of view, the total cohesive energy may be
experimentally determined from the Born-Haber cycle. Thus main physical
characteristics of the binding energy are the melting and boiling
temperatures, $T_m$ and $T_b$ , and heats of fusion and sublimation, $\Delta
H_f$ and $\Delta H_s$. Besides that, the stability of the crystal is
characterized by the density (or atomic volume), ionization potentials, work
function, Debye temperature, elastic moduli, activation energy for
diffusion, specific heat etc. The corresponding values may be found in
reviews and standard handbooks, e.g., [235-239]. Some of these quantities
for d-, 4f- and 5f- metals are presented in Tables 3.1-3.3 and shown in
Figs.3.1-3.8. As follows from comparison of the values of $T_m$ for d-metals
with those for preceeding and following sp-elements, d-electrons yield an
appreciable contribution to the binding energy. At the same time, the
entropy change at melting $\Delta S_m$ in d-metals differs weakly from that
in simple metals. In the periodic table, values of $\Delta S_m$ vary in a
wide region (from 0.47 kcal/mol K for white phosphorus to 7.13 kcal/mol K
for Si). However, these regions become narrow for element groups with a
given lattice: $\Delta S_m$ (in kcal/mol K) equals to 1.76$\pm $0.34 for bcc
crystals, 2.28$\pm $0.23 for fcc crystals, 2.33$\pm $0.23 for hcp crystals
and 6.50$\pm $0.23 for the diamond structure [235].

The change of the thermodynamic and lattice parameters in the d-series
enables one to investigate the dependence of the contribution of d-electrons
on their number $n$. Most strong influence of d-electrons may be seen in the
beginning of d-periods. The atomic volume decreases sharply (almost by two
times) at passing from Ca to Sc, which is connected with the collapse of
d-orbits. With further increasing $n$, $V_a$ continues to decrease, but
increases in Cu and Zn where the d-shell is completely filled. The values of
the thermodynamic characteristics $T_m$ , $T_b$ , $\Delta H_f$ and $\Delta
H_s$ , and elastic moduli of d-metals also exceed corresponding values for
simple metals and demonstrate a smooth $n$-dependence. However, unlike $V_a$,
these quantities have maxima in the middle of 3d-period (between V and Mn
for different characteristics; a similar situation takes place for 4d and
5d-periods). We shall see in following Chapters that purely electronic
characteristics (electronic specific heat coefficient $\gamma $ and
paramagnetic susceptibility $\chi $) have a quite different (oscillating)
$n$-dependence.

Most simple explanation of such a behaviour of thermodynamic properties is
existence of several different factors which influence the total energy. In
particular, with increasing the number of d-electrons, d-levels become
lower, so that the participation of d- states in the cohesion decreases.
Besides that, the Coulomb repulsion among d-electrons also increases with $n$.

In the simple Friedel model the electron energy (2.84) has a minimum at $n =
5$, i.e. for half-filled d-shell. This result is explained by that the
energy of d-electrons decreases for the lower half of d-band and is
increases for the upper one. Such a picture is in agreement with the
situation in 3d, 4d and 5d periods except for an appreciable deviation for
manganese.

Since the d-band width $W_d$ is inversely proportional to some power of the
interatomic distance $r_0$, a dependence with a maximum for $n = 5$ should
take place for the compressibility which is determined in terms of
$dW_d/dr_0$.

An attempt to explain the anomaly of $E_{coh}$ in Mn with account of
correlation effect were performed by Oles [240]. He started from the earlier
paper [241] where the cohesive energy was estimated as the energy difference
beween the atomic limit and condensed state. We illustrate the idea of this
explanation in the simplest case of s-band where the estimation gives
$$
E_{coh}=\frac n2\left( 1-\frac n2\right) W-\frac{n^2}4U
\eqno{(3.1)}
$$
where $U$ is the Hubbard parameter, $n<1$. The expression (3.1) has a
maximum at band filling
\[
n=\frac W{W+U}<1
\]
Thus a maximum is shifted from the middle of the band because of the Coulomb
interaction. Since the the picture should be symmetric for the second half
of the band (electron-hole symmetry), a minimum at $n=1$ occurs.

Unlike d-metals, most thermal and elastic characterisitics of rare earths
(Table 3.2, Fig.3.6) have only a weak (nearly linear) dependence on the
number of f-electrons $n_f$ . This agrees with the statement about a weak
participation of 4f-electrons in the chemical bond. At the same time, $T_b$
and $\Delta H_0^0$ (Fig.3.4) demonstrate a rather pronounced $n_f$%
-dependence. As discussed in [235], this fact may be connected with the
difference in electron configurations of the solid and gaseous phases for
some 4f-elements. It is interesting that this dependence is reminiscent of
the curves for the 4f- electron energy which are obtained from spectral
characteristics and band calculations [133-135] (see Sect.2.6). One can see
that the most strong change in $T_b$ and $\Delta H_0^0$ takes place at
passing from Eu and Gd (unfortunately, there is only a rough estimation of $%
T_b$ for Gd). In this sense, the situation is d and 4f-periods is similar
and may be related to high stability of the half-filled shells with the
configurations d$^5$ and f$^7$.

The situation in actinides (where 5f-electrons have an intermediate degree
of localization as compared to d and 4f electrons) is not clear in detail.
According to Table 3.3, the atomic volume decreases coniderably from Ac to
Pu (there is an appreciable anomaly for Am). However, other thermodynamic
and lattice characteristics do not exhibit a regular dependence on $n_f$. An
accurate consideration should take into account the role of 6d-electrons.
The number of these for free atoms changes non-monotonously from 1 in Ac to
zero in Pu with a maximum value of 2 in Th. However, the situation in
metallic phase changes considerably since 5f-states in light actinides where
5f-electrons are strongly delocalized and hybridized with d-electrons.
Experimental data for heavy actinides where 5f-electrons are well localized
are insufficient to discuss regularities in thermodynamic properties.

Tables 3.1-3.3 demonstrate existence of some correlations between different
thermodynamic and lattice characteristics. A number of theoretical and
empirical relations, which describe these correlations, were established.

A correlation between the Debye temperature and $T_m$ was obtained by
Lindemann (1910). It is based on the idea that melting takes place when the
atomic thermal displacement amplitude reaches some critical value $u = x_m d$
with $d$ being the lattice constant. The corresponding calculation may be
performed within the Debye model for lattice vibrations [6]. The
displacement $u$ of an ion $\nu$ may be written in Fourier representation
$$
u_{\nu \alpha }(t)=\sum_{\mathbf{q}}u_{\mathbf{q}\alpha }
\exp (i\mathbf{qR} _\nu -i\omega _{\mathbf{q}\alpha }t)
\eqno{(3.2)}
$$
(\textbf{q} is the wavevector, $\alpha $ is the polarization). Using the
virial theorem we obtain the expression for the vibration energy
$$
\mathcal{E}=\sum_{\nu \alpha }M|\dot u _{\nu \alpha }|^2
\eqno{(3.3)}
$$
where $M$ is the ion mass. On substituting (3.2) into (3.3) we obtain
$$
\mathcal{E}=\sum_{\mathbf{q}\alpha }M|\dot u _{\mathbf{q}\alpha }
|^2\omega _{\mathbf{q}\alpha }^2\equiv \sum_{\mathbf{q}\alpha }
E_{\mathbf{q}\alpha }
\eqno{(3.4)}
$$
Application of the Bose statisitics yields for the averaged square of
vibration amplitude
$$
\overline{|u_{\mathbf{q}\alpha }|^2}=\overline{E_{\mathbf{q}\alpha }}
/(M\omega _{\mathbf{q}\alpha })=\hbar (N_{\mathbf{q}\alpha }+\frac
12)/(M\omega _{\mathbf{q}\alpha })
\eqno{(3.5)}
$$
Performing the integraton at high temperatures where $N_{\mathbf{q}%
}=T/\omega _{\mathbf{q}}$ in the Debye approximation ($\omega _{\mathbf{q}%
}=sq,q<q_D$) we get for the average square of atomic displacement
$$
\overline{u^2} =3\overline{|u_{\nu \alpha }|^2}=3\sum_{\mathbf{q}}
\overline{|u_{\mathbf{q}\alpha }|^2}
$$
$$
=\frac{9k_BT}{M\omega _D^2}=\frac{9\hbar ^2T}{k_BM\theta _D^2}
\eqno{(3.6)}
$$
Then the melting temperature is estimated as
$$
T_m=x_mMd^2k_B\theta _D^2/(9\hbar ^2)
\eqno{(3.7)}
$$
For most solids the parameter $x_m$ equals to 0.2$\div $0.25. This
universality supports strongly the adequacy of the Lindemann formula.

An empirical relation between the melting temperature $T_m$ and cohesive
energy reads [235]
$$
T_m=\frac{0.08E_{coh}}{3R\ln 2.045}
\eqno{(3.8)}
$$
where $R$ is the gas constant. The relation between the thermal expansion
coefficient $\alpha $ and $T_m$ has the form [235]
$$
\alpha T_m=K
\eqno{(3.9)}
$$
For fcc, bcc and hcp crystals $K=0.0197\div 0.0051$, and for crystals with
the diamond structure $K=0.0039\div 0.0018$. This correlation may be
justified by that both a and $T_m$ are connected with an anharmonic
contribution to lattice vibrations.

From this point of view, it is instructive to reconsider the Lindemann
correlation. Write down the expansion of the potential energy of a lattice
with account of anharmonic terms
$$
V(u)=V_0=\frac 12au^2-bu^3-cu^4  
\eqno{(3.10)}
$$
(we confine ourselves for simplicity to the one-dimensional case). Then the
average displacement reads
$$
\overline{u}=\frac{3b}{a^2}k_BT  
\eqno{(3.11)}
$$
Assuming that melting takes place at $\overline{u}=u_m=y_md$ we obtain the
relation
$$
k_BT_m=\frac{a^2}{3b}u_m  
\eqno{(3.12)}
$$
Using the relation
$$
\alpha =\frac 1d\frac{d\overline{u}}{dT}=\frac{3b}{da^2}k_B  
\eqno{(3.13)}
$$
we derive the result (3.9) with $K=\overline{u}/d=y_m$, so that $K$ is just
the relative average displacement. An experimental verification of this
statement is of interest.

In a similar way, the heat of fusion should be connected with the volume
change at melting $\Delta V_m$ . However, systematic data for $\Delta S_m$
in transition metals are absent.

There exists also the relation between the melting temperature, the shear
modulus $\mu = C_{11} - C_{12}$ and the atomic volume
$$
rT_m/(\mu V_a)=\mathcal{L}  
\eqno{(3.14)}
$$
where the constant $\mathcal{L}$ is the Leibfried number, and a similar
relation for the heat of fusion
$$
\Delta H_f/(\mu V_a)=\mathcal{B}  
\eqno{(3.15)}
$$
where $\mathcal{B}$ is the Bragg number. Since the entropy of fusion $\Delta
H_f/T_m$ is approximately equal to $R$ (the Richards rule [235]), we have $%
\mathcal{L}\simeq \mathcal{B}$. However, the empirical values differ
appreciably: $\mathcal{L}=0.042,\mathcal{B}=0.034$. One uses also the
modified Leibfried number $\mathcal{L}^{\prime }=K\mathcal{L}$ which takes
into account the lattice structure ($K=1.76$ for the bcc lattice and 2.29
for fcc and hcp lattices). The value of $\mathcal{L}^{\prime }$ changes from
0.019 (Cr) to 0.041 (V) in the 3d-period and from 0.02 (Mo) to 0.05 (Nb) in
the 4d-period. One can see from the Table 3.1 that the deviations of $%
\mathcal{L},\mathcal{L}^{\prime }$ and $\mathcal{B}$ from their average
values have a rather regular character, so that correlation with electronic
characteristics may be established (e.g., with specific heat, see
Sect.3.3.2).

Although thermodynamic and lattice properties of metals are often considered
in a phenomenological way, there exist also a large number of microscopical
calculations within the band theory. The results on the cohesive energies,
atomic volumes and bulk moduli for 3d and 4d row metals are presented in
[24], and the atomic volumes and bulk moduli for 4d and 5d metals are given
in [56].

Fig.3.8 shows, besides the measured cohesive energy $E_{coh}$ of d-metals,
the calculated valence bond energy
\[
E_{vb}=E_{coh}+\Delta E_{at} 
\]
where $\Delta E_{at}$ is the preparation energy required to take the atom to
a state corresponding to the non-magnetic ground-state configuration of the
metal. This quantity is calculated as
\[
\Delta E_{at}=E_p-E_{sp}^{LDA} 
\]
Here $E_p$ is the preparation energy required to take the atom to the
magnetic ground state configuration, which may be obtained from experimental
data [34]; the spin-polarization energy is the LSDA equivalent of the first
Hund's rule energy. It is $\Delta E_{at}$ which behaves irregularly in the
d-rows, and $E_{vb}$ turns out to vary smoothly (approximately as a
parabolic function).

Analysis of cohesive energy of all the 3d- and 4d transition metals, and
calculations of equilibrium lattice constant and bulk modulus of two
representative elements, Cu and Ti, were performed in [242] within a
renormalized-atom method (where the atomic valence wavefunction are
truncated at the Wigner-Seitz radius and multiplied by a renormalization
constant). Cohesive energy of a transition metal was presented as a sum of
five terms: (i) the atomic preparation energy required to excite a free atom
from its ground state to the d$^{n-1}$s configuration appropriate to the
crystal, $\Delta E_{at}$(ds) - $E_{cr}$(d$^{n-1}$s) (ii) the difference in
total Hartree-Fock energy between the free d$^{n-1}$s atom and the
renormalized atom (iii) the difference between the average energy of a
free-electron band containing one electron and that of the renormalized atom
s-level (iv) the change in one-electron energies which results from the
broadening of the renormalized atom d-level into the d-band (v) the change
in one-electron energy due to the hybridization between the conduction and
d-bands. These contributions are shown in Fig.3.9. The problem of large
cohesive energy in Cu, which exceeds by more than two times that of K, is
discussed. This fact may be explained by the d-electron contribution owing
to the s-d hybridization.

In the paper [243], results of calculation of the Debye temperature,
Grueneisen constant and elastic moduli are presented for 14 cubic metals
including TM. These results are obtained in the APW LDF approach with the
use of Morse potential parametrization. The theoretical expression
$$
k_B\theta _D=(48\pi ^5)^{1/6}\hbar s(r_0B\rho /M)^{1/2},\qquad 
s=(B/\rho )^{1/2} 
\eqno{(3.16)}
$$
($r_0$ is is the rigid-lattice equilibrium separation between ions, $\rho $
is the density, $B$ is the bulk modulus, $s$ is the sound velocity) strongly
overestimates the value of $\theta _D$ . Therefore the expressions (3.16)
should be modified by introducing a scaling factor:
\[
s=0.617(B/\rho )^{1/2},\qquad \theta _D=41.63(r_0B/M)^{1/2} 
\]
with $r_0$ in atomic units, $B$ in kbars. The equilibrium value of $r_0$ is
obtained by minimization of the free energy (including temperature-dependent
vibrating-lattice contributions to entropy). The results for the temperature
coefficient of linear expansion
$$
\alpha (T)=d\ln r_0/dT  
\eqno{(3.18)}
$$
are shown in Fig.3.10. The theory describes well the difference between the
``soft'' alcaline metals and ``hard'' transition metals. The calculated
Debye temperatures determine the position of the ``knee'', and the
Grueneisen constants $\gamma $ do the high-temperature amplitude of the $%
\alpha (T)$ curves. The theoretical values of $\gamma $ turn out to be to
high for Li and Sr and too low for Al, V, Cu and Nb. Authors of [243]
conclude that the LDF approach may be used for description not only of
ground states characteristics, but also of thermal properties.

Ab initio calculation of work function for 37 metals (including 3d, 4d and
5d series) was recently performed in the LMTO-ASA method [244]. The results
agree within 15\% with experimental data.

\section{Crystal structure}

Crystal structures of elements in the periodic system at $T = 0$ under
normal pressure are shown in Fig.3.11.

A number of elements exhibit phase transformations with a change of crystal
structure (Table 3.4). In particular, for d-metals the closely packed fcc
and hcp structures, which are stable in the ground state, transform often at
high temperatures into a less dense bcc structure. This phenomenon may be
explained by more soft phonon spectrum of the bcc lattice. Thus with
increasing temperature the free energy for the bcc structure decreases more
rapidly owing to a large negative entropy contribution and this structure
becomes more favourable than the closely packed ones.

Most rare earth metals have at low temperatures the hcp structure and
transform with increasing $T$ into the bcc one [16,246]. A tendency to
decreasing the temperature interval for the bcc phase existence takes place
with increasing the atomic number. Cerium has a double hcp structure at $T=0$
and a complicated phase diagram, praseodymium a double hcp structure,
samarium a rhomboedric structure, and europium possesses the bcc lattice
already at $T=0$. There exist also data on existence of fcc phases for Pr,
Nd and Yb [246,247].

The light actinide metals with delocalized 5f-electrons have the structures
which are similar to those of d-metals, but the phase diagrams are more
complicated. For example, plutonium has the largest variety of crystals
modifications among the periodic table elements. The structures of heavy
actinides (starting from americium) with localized 5f-electrons are
reminiscent of those of lanthanides.

Consider some approaches to the problem of determining crystal structures in
metals from the point of view of electronic structure.

The simplest empirical correlation was established by Hume-Rothery [248] who
found that crystal structures of stable phases for a wide class of metallic
alloys were determined by the mean number of valence electrons per atom, $n$. 
A typical example is given by the Cu-Zn system [6]. From $n=1$ (pure
copper) to $n=1.38$ the alloy has the fcc lattice ($\alpha $-phase) At
larger $n$ there occurs the region of coexistence of $\alpha $ and $\beta $
(bcc) phases, and starting from $n=1.48$ the pure $\beta $-phase is stable.
The $\gamma $-phase with a complicated cubic structure exists at 
$1.58<n<1.66 $. Finally, the hcp $\varepsilon $-phase becomes stable at 
$n=1.75.$

A geometrical interpretation of this regularity may be obtained by
considering the position of the Fermi surface in the recirpocal space with
respect to the Brillouin zone boundaries [249]. Appearance of an energy gap
in the electron spectrum near Fermi level, which takes place near these
boundaries, is energetically favourable. Indeed, it results in lowering of
the electron system total energy since the energy of occupied states becomes
lower. In the one-dimensional case, this effect results in distortion of a
lattice with the formation of an insulating state (Peierls instability).
Calculating the volume of the sphere, which is inscribed into the first
Brillouin zone, we obtain the electron numbers $n = 1.36$ for the fcc
lattice, 1.48 for the bcc lattice and 1.69 for the ideal hcp lattice in a
satisfactory agreement with above experimental data.

Thus the Hume-Rothery rule relates the stability of a crystal structure to
the touching of the Fermi surface with the Brillouin zone boundary where
electron spectrum is maximally disturbed. Unfortunately, justification of
this rule for simple metals within the qualitative band theory in the
nearly-free electron approximation (pseudopotential theory) meets with
difficulties: the singularity in the total energy turns out to be very weak
in the three-dimensional situation (see discussion in [55]). However, one
may expect that the influence of Brillouin zone boundary becomes appreciably
stronger if density of states peaks are present near the Fermi surface.

For transition d-metals, the Hume-Rothery rule usually does not hold (at the
same time, it may be applicable for rare earth alloys [16]). A different
concept of correlation between crystal structure and $n$ was put forward by
Engel and Brewer (see [250,251]). According to their theory, the type of
lattice is determined by the number of s,p electrons. So, Na, Mg and Al,
which possess one, two and three sp-electrons respectively, have bcc, hcp
and fcc lattices. An attempt was made to generalize this concept to
d-metals. In particular, the configuration d$^5$s (the same as for isolated
atoms) is attributed to metals of the middle of d-periods (Cr, Mo, W) with
the bcc structure. Formation of the same structure in V and its analogues is
explained by high statistical weight of the d$^4$s-configuration (unlike
isolated atoms). The hcp structure in the metals of III and IV groups
corresponds to dominating role of configurations dsp and and d$^2$sp
respectively. The difference between bcc structure of iron (configuration d$%
^7$s) and hcp metals Ru and Os is related to strong s-d hybridization in the
former.

At the same time, the applicability of the Brewer-Engel picture is rather
restricted, so that it was criticized [251,252]. Especially unsatisfactory
is this theory for fcc metals. In particular, it yields incorrect
predictions for alcaline earths Ca and Sr (s$^2$-configuration) and noble
metals Cu, Ag, Au which have one s-electron. Generally, the supposition that
d-electrons take part in chemical bonding but do not influence the lattice
symmetry does not seem to be physically reasonable.

Of course, the direct calculation of the total energy from the real band
electron spectrum is more justified for predicting stable structures,
although this requires picking out small energy differences for various
structures (of order of 10$^{-2}$-10$^{-1}$ eV) on the background of large
binding energy (of order of 10 eV). As a rule, such differences are smaller
than real accuracy of band calculations. In particular, many-electron
effects should give appreciable contributions and be taken into account in a
quantitative theory of crystal structures. The problem of structural
transitions at finite temperatures is still more complicated since the
density functional approach becomes, generally speaking, inapplicable, and
collective excitations of the lattice and electron subsystems should
contribute to the entropy term. However, standard band calculations were
rather successfully applied in the theory of crystal structures.

Stabiltity of the bcc phase in the middle of the d-periods is connected with
double peak structure of the density of states for this lattice. Maximal
energy gain is achieved at filling of the lower peak. As follows from band
calculations [24,78], this peak is filled in vanadium at $n_d\simeq 4$
(which is by unity larger than in free V atom). For fcc and hcp lattices,
the density of states has a more complicated many-peak structure, so that
they should be stable at the beginning or end of d-periods.

From the atomic level point of view, one may consider in cubic lattices the
states which have $e_g$ and $t_{2g}$ origin. The $e_g$-states lie lower than 
$t_{2g}$ ones in the fcc lattice and higher in the bcc lattice. Since the $%
e_g$-band contains four electrons and the $t_{2g}$-band six electrons, the
fcc structure should be favourable up to $n_d=4,$ and the bcc structure at $%
n_d$ between 4 and 6. This agrees to some extent with the situation in the
3d-period (the bcc structure between Ti and Fe). However, Sc and Ti ($n_d<4$%
) have the hcp (rather than fcc) structure, the stability of which is not
described in terms of $t_{2g}$- and $e_g$-states.

Pettifor [253] calculated the energies of three basic structures (fcc, bcc
and hcp) the 3d-metal series within the model pseudopotential approach with
two d-resonance parameters. The stability of the crystal structures as a
function of the number of d-electrons $n_d=n-2$ is described by the
following sequence 
$$
\begin{tabular}{lllllll}
$n_d$     & 1-1.5 & 1.5-2 & 2-4.5 & 4.5-6.5 & 6.5-7.9 \\
Structure & fcc   & hcp   & bcc   & hcp     & fcc    
\end{tabular}
$$
These results just correspond to the experimental situation in the 3d-row.

Skriver [245] evaluated the energy differences for various lattices in the
3d, 4d and 5d series and for rare-earths (Figs.3.12-3.16), and also for
actinides, by using the force theorem approach in the LMTO method. The
results by Pettifor were confirmed, and general sequence hcp-bcc-hcp-fcc in
other d-periods was also reproduced. Investigations of stability of crystal
structures in the lanthanide series yielded the sequence
$$
\begin{tabular}{l}
hcp - Sm-type - double hcp - fcc - fcc$^{\prime }$
\end{tabular}
$$
The type of the lattice is determined by the number of 5d-electrons which is
changed from 2 (La) to 1.5 (Lu) (Fig.3.14).

The one-electron band calculations [245] turn out to agree qualitatively
with experimental data and give correctly the crystal structure of 35 metals
from 42 investigated (in particular, the theory fails for Au, Mn, Fe and
Co). As to quantitative agreeement (values of the phase diagram parameters,
e.g. transition temperatures and pressures), the problem is considerably
more difficult, so that satisfactory results are obtained for simple metals
only. The calculated energy differences for d-metals [245] turn out to be by
3-5 times greater than enthalpy differences obtained from study of binary
phase diagrams (Fig.3.16).

Structural and cohesive energies of 5d-elements were calculated in [254]
within the linear augmented-Slater-type-orbital method. Recent calculation
of total energies for 3d, 4d and 5d metals [255] was performed by scalar
(semirelativistic) RAPW method (unlike the calcullations by Skriver [245],
the frozen potential approximation was not used). The conclusions agree on
the whole with the results for lattice parameters and bulk moduli [24] and
crystal structure stability [245], relativistic corrections being not too
important. However, the experimental structure for Au was obtained in this
calculation.

Total energy calculations of crystal structure stability, equilibrium volume
and bulk moduli for light actinides Th, Pa and U within the full-potential
approach [256] yielded good agreement with experimental data.

\section{Specific heat}

In the previous sections of this Chapter we have treated energy
characteristics of a crystal which yield a general description of the
electron and ion systems. Here we consider the specific heat, which enables
one to obtain a more detailed information about electron and lattice
spectrum.

\subsection{Lattice specific heat}

Investigation of specific heat played an important role in the development
of quantum solid state theory. We remind briefly of principal results.
Specific heat at constant volume is defined in terms of the average energy
$$
c_V=\left( \frac{\partial \overline{\mathcal{E}}}{\partial T}\right) _V 
\eqno{(3.19)}
$$
In the classical picture, each degree of freedom of a particle contributes $%
k_BT/2$ to the thermal energy. Thus the energy of a three- dimensional
crystal which contains $N$ ions and $n$ electrons is given by
$$
\overline{\mathcal{E}}=3Nk_BT+\frac 32nk_BT  
\eqno{(3.20)}
$$
(the average energy of ions, which are considered as oscillators, is doubled
owing to potential energy). Thus the lattice specific heat of a monoatomic
crystal should be equal to 3$R$ ($R=k_BN_A=8.3$ J/mol K is the gas constant, 
$N_A$ is the Avogadro number). This result corresponds to the Dulong-Petit
law which was established for a wide number of solids. However,
experimentally this law holds as a rule in a not too wide region near the
room temperature. With decreasing temperature, $c_V$ decreases and tends, in
agreement with the general Nernst theorem, to zero at $T\rightarrow 0$. On
the other hand, $c_V$ increases considerably at high $T$.

The quantum theory of lattice contribution to specific heat was developed by
Einstein (1906) and Debye (1910). Einstein considered the crystal as a
system of independent oscillators with the constant frequency $\omega_E$ .
Then application of the Bose statistics yields
$$
\overline{\mathcal{E}}=3N\hbar \omega _E[N_B(\omega _E)+\frac 12]  
\eqno{(3.21)}
$$
with 
\[
N_B(\omega )=(\exp \frac{\hbar \omega }{k_BT}-1)^{-1} 
\]
Then
$$
c_V=3R\left( \frac{\hbar \omega _E}{k_BT}\right) ^2\exp \left( -\frac{\hbar
\omega _E}{k_BT}\right) \left/ \left[ \exp \left( -\frac{\hbar \omega _E}{%
k_BT}\right) -1\right] ^2\right.  
\eqno{(3.22)}
$$
Thus specific heat equals to $3R$ at high temperatures and is exponentially
small at low $T$. Experimental data demonstrate a more slow decreasing of
specific heat. This is due to that the Einstein model is well applicable
only for optical branches of lattice oscillations (phonons), frequencies of
which depend weakly on the wavevector. The calculation of acoustic branch
contribution with the model dispersion law
$$
\omega _{\mathbf{q}}=sq\quad (q<q_D),\qquad 
\omega _D=sq_D=k_B\theta _D/\hbar 
\eqno{(3.23)}
$$
was performed by Debye. The result reads
$$
c_V=3R\left( \frac T{\theta _D}\right) ^33\int_0^{\theta _D/T}\frac{e^zz^4dz%
}{(e^z-1)^2}  
\eqno{(3.24)}
$$
At high $T$ we come again to the Dulong-Petit law. At low $T\ll \theta _D$
specific heat obeys the Debye $T^3$-law
$$
c_V=12\frac{\pi ^4}5R\left( \frac T{\theta _D}\right) ^3  
\eqno{(3.25)}
$$
A more accurate calculation of the phonon contribution to specific heat in a
wide temperature region requires using a realistic dispersion law of phonons
which may be very complicated. As an example, experimental data for acoustic
branches in Nb and Mo are shown in Fig.3.17. One can see that the dependence 
$\omega (\mathbf{q})$ in Mo turns out to be non-monotonous. Anomalies in $%
\omega (\mathbf{q})$ curves are connected with strong electron-phonon
interaction in transition metals.

Data on specific heat of simple and transition metals in a wide temperature
region are presented, and regularities in $c(T)$ behaviour for different
columns of the periodic table are discussed in the handbook [239]. The
dependences $c(T)$ within a column turn out to be similar and differ mainly
by singularities owing to structural and magnetic transitions. As an
example, Fig.3.18 shows specific heat of vanadium where such phase
transitions are absent, so that the dependence $c(T)$ has a simple form. One
can see that the Dulong-Petit law holds in a not too wide temperature region
and is violated at both low and high $T$. More complicated behaviour takes
place in Zr (Fig.3.19) where the structural $\alpha -\beta $ transformation
is present, and in heavy rare earths (Fig.3.20) which have a complicated
magnetic phase diagram.

The jump of specific heat at the melting point, $\Delta c_m=c(T>T_m)-c(T<T_m)
$, is as a rule positive for transition metals which exhibit phase
transition (including light rare earths, Fe, Co, Ni) and negative for metals
which do not (e.g., for Nb, Ta, although a small positive $\Delta c_m$ is
observed in vanadium). It should be noted that the situation is different
for simple metals: $\Delta c_m>0$ for Zn, Cd, Al, Ga, In, Pb despite the
absence of structural phase transitions.

In the liquid phase, specific heat of d-metals may reach the values of 4-9$R$%
, which are considerably larger than those for simple metals (see, e.g., the
discussion in [258]).

The increase of $c$ in comparison with 3$R$, observed above the room
temperature in most substances, is usually attributed to anharmonicity
effects which result in occurence of $T$-linear terms in the
high-temperature lattice specific heat. We illustrate this phenomenon in the
simplest one-dimensional model with the potential (3.10) (a more detailed
discussion is given in [4]). Write down the partition function of an ion as
$$
Z(T)=\int dpdu\exp \left( [-\frac{p^2}{2M}+V(u)]/k_BT\right)  
\eqno{(3.26)}
$$
Integrating over momentum $p$ and expanding in the anharmonicity parameters
up to $b^2$ and $c$ we derive
$$
Z(T) =(2\pi k_BTM)^{1/2}\int du\exp \left( -\frac 12au^2/k_BT\right)  
$$
$$
\times \left[ 1+\frac{bu^3}{k_BT}+\frac 12\left( \frac{bu^3}{k_BT}\right)
^2+\frac{cu^4}{k_BT}\right]   
\eqno{(3.27)}
$$
The integration over $u$ yields
$$
Z(T)=2\pi \left( \frac Ma\right) ^{1/2}k_BT\left[ 1-\left( \frac{3c}{a^2}-%
\frac{15b^2}{a^3}\right) k_BT\right]  
\eqno{(3.28)}
$$
Then the specific heat is given by
$$
c(T) =-3N_A\frac \partial {\partial T}\left( T^2\frac{\partial \ln Z(T)}{%
\partial T}\right)  
$$
$$
=3R\left[ 1+\frac 6a\left( \frac{5b^2}{2a^2}-\frac ca\right) k_BT\right] 
\eqno{(3.29)}
$$
Since $a/b\sim b/c\sim d$ and the average displacement is given by (3.11),
the $T$-linear corrections owing to triple and quartic anharmonisms are of
the same order of magnitude and may be estimated as $(\overline{u}/d)3R$.
Therefore the anharmonicity may hardly explain the increase of specific heat
at high temperatures in d-metals, which may be of order 100\%, so that this
problem needs further investigations.

\subsection{Electronic specific heat}

Besides the violation the Dulong-Petit law at low temperatures, the second
difficulty of the classical theory of specific heat concerned the electronic
contribution to specific heat. According to (3.20), this contribution should
be equal to $3R/2$ for a metal with one electron per atom. However, such
large contributions were never found in experiment. Only a small $T$-linear
electron term in specific heat is observed practically at any temperatures.

This contradiction was resolved only in twentieth years of our century after
formulation of the Fermi-Dirac statistics for electrons. The corresponding
distribution function reads
$$
f(E)=(\exp \frac{E-\zeta }{k_BT}+1)^{-1}  
\eqno{(3.30)}
$$
where $\zeta =\zeta (T)$ is the chemical potential, $\zeta (0)$ being equal
to the maximum energy of occupied states - the Fermi energy $E_F$. The
latter quantity is determined, according to the Pauli principle, by the
number of electrons:
$$
n=\int_{-\infty }^\infty dEf(E)N(E)  
\eqno{(3.31)}
$$
(in this Section the density of states $N(E)$ is determined for both spin
projections). In metals, $E_F/k_B$ is large and makes about 10$^4$-10$^5$K.
This explains qualitatively small value of electronic specific heat. Indeed,
only electrons in a narrow layer with a width about $\pi k_BT$ near the
Fermi energy may change their energy and take part in thermal excitations,
and most electrons are ``frozen''. The number of the ``thermal'' electrons
is estimated as
$$
n^{*}/n\sim k_BTN(E_F)\sim k_BT/E_F  
\eqno{(3.32)}
$$
so that
$$
\delta \overline{\mathcal{E}}_e(T)\sim \frac 32n^{*}k_BT\sim k_B^2T^2N(E_F) 
\eqno{(3.33)}
$$
and 
\[
c_e\sim k_B^2N(E_F)T 
\]
A more accurate calculation (see below (3.40)) yields
$$
c_e=\gamma T,\qquad \gamma =\frac{\pi ^2}3k_B^2N(E_F)  
\eqno{(3.34)}
$$
Thus electronic specific heat should be linear in a wide temperature
interval $0<T<E/k_B$ (practically, up to temperatures which exceed the
melting points). The electronic contribution may be picked out at low $T$
where the lattice contribution rapidly decreases. To this end one writes down
$$
c_V/T=\gamma +\alpha T^2  
\eqno{(3.35)}
$$
and extrapolates the experimental dependence $c_V/T$ vs. $T^2$ to zero
temperature.

According to (3.34), the value of $\gamma $ is determined by the most
important characteristic of electron system in a metal - the density of
states at the Fermi level. Since transition metals are characterized by
large values of $N(E_F)$, their $\gamma ^{\prime }$s (Table 3.5) are
considerably greater than for simple metals. Especially large values are
observed in d-compounds with N(E) peaks at the Fermi level. Here belong the
superconductors with the A-15 structure (for example, $\gamma $ = 33 mJ/mol K%
$^2$ for V$_3$Ga, see also Fig.6.1).

The experimental values of $\gamma $ may be compared with those obtained
from band calculation results, 
\[
\gamma \ (\mathrm{mJ/mol K}^2)=0.1734\cdot N(E_F)\ (\mathrm{states/Ry}) 
\]
As one can see from Table 3.5, a distinct correlation between these values
takes place. As well as $N(E_F)$, $\gamma _{exp}$ demonstrates a ``toothed''
behaviour as a function of the d-electron number - large values for the
configurations d$^n$ with odd $n$ and small values for even $n$ (the ratio $%
\gamma _{exp}/\gamma _{theor}$ does not reveal such a behaviour). In the
middle of d-series, $\gamma _{exp}/\gamma _{theor}$ is somewhat smaller than
in the beginning or in the end.

The theoretical values of $\gamma $ turn out to be as a rule smaller than
experimental ones. This difference is larger for transition elements. So,
for Sc, Ti, V, Y, Zr, Nb, Hf, Ta one has $\gamma _{exp}/\gamma _{theor}\sim
2,$ and for neighbour simple metals Ca, Ba, Cu, Ag, Au this ratio is close
to unity. Thus the deviation may be partially attributed to correlation
effects which are more important in transition metals. In this connection we
mention the recent first-principle band calculation beyond the density
functional method [259] where a considerable change in $N(E_F)$ values is
obtained, so that a better agreement with experimental data on $\gamma $
takes place.

Besides that, one has to bear in mind the inaccuracy in band calculations.
Since the density of states in TM has a sharp energy dependence, small
errors in the Fermi level position may influence strongly the value of $%
N(E_F)$. The relation $\gamma _{exp}<\gamma _{theor}$ for chromium is
probably connected with the influence of antiferromagnetic ordering which is
not taken into account in band calculations, but may disturb considerably
electron structure near the Fermi level (formation of the antiferromagnetic
gap).

The enhancement of $\gamma _{exp}$ in comparison with $\gamma _{theor}$ may
be explained by effects of interaction of conduction electrons with various
elementary excitations in a crystal. Most frequently one discusses the
electron-phonon interaction which results in occurence of a factor $%
1+\lambda $ in the electron effective mass, and, consequently, in $\gamma $.
The theoretical calculation of $\lambda $ is a very difficult problem since
this requires a detailed information on characteristics of electron and
phonon subsystems. In transition metals the electron-phonon interaction is
strong due to a strong dependence of total energy on the lattice parameter.
The large value of $\lambda $ leads also to superconductivity with rather
high $T_c$ [257]. It should be noted that $\lambda $ should exhibit an
appreciable temperature dependence. Calculations of this dependence were
performed both in the Einstein and Debye models by Grimvall [260] and Osuzu
[261] respectively. Results of [261] are shown in Fig.3.21. One can see that 
$\lambda (T)$ decreases rapidly at $T>0.2\theta _D$.

Another important mechanism of specific heat enhancement is the interaction
with spin fluctuations. This mechanism, which is especially important for
weakly magnetic and nearly magnetic transition metals and their compounds
(e.g., $\gamma $ = 56 mJ/mol K$^2$ for the nearly antiferromagnetic system
TiBe$_2$ [507]), is discussed in Sect.4.4 and Appendix G.

An interesting example is metallic praseodymium where the interaction with
crystalline electric field excitations is supposed to provide the
enhancement $m^{*}/m\simeq 4$ [263]. Such effects of spin and charge
fluctuations in anomalous rare-earth and actinide systems are considered in
Sect.6.3.

Temperature dependence of electronic specific heat in transition metals
turns out to be strong. Corresponding experimental results are shown in
Fig.3.22. It should be noted that picking out the electronic contribution on
the background of the lattice one was performed in [262] within the Debye
approximation. The latter is not quite satisafactory at high $T$ where $c_e$
makes up about 10\% of the total specific heat.

The most simple reason for the dependence $\gamma (T)$ is smearing of the
Fermi function with temperature. To consider this effect we consider the
expressions for the number of particles (3.31) and for average energy
$$
\overline{\mathcal{E}}_e=\int_{-\infty }^\infty EdEf(E)N(E)  
\eqno{(3.36)}
$$
Further we use the expansion of the Sommerfeld integrals
$$
\int_{-\infty }^\infty dEf(E)F(E)  
$$
$$
=\int_{-\infty }^\zeta dEF(E)+\frac{\pi ^2}6(k_BT)^2F^{\prime }(\zeta )+
\frac{7\pi ^4}{360}(k_BT)^4F^{\prime \prime }(\zeta )+...  
\eqno{(3.37)}
$$
which is obtained with the use of integrating by part. Then we derive
$$
\zeta (T)=E_F-\frac{\pi ^2}6(k_BT)^2N^{\prime }(E_F)/N(E_F)+...  
\eqno{(3.38)}
$$
$$
\overline{\mathcal{E}}_e=\int_{-\infty }^{N(E_F)}dEN(E)+\gamma (T)T 
\eqno{(3.39)}
$$
where
$$
\gamma (T) =\frac{\pi ^2}3k_BN(E_F)\left\{ 1+\frac 12(\pi k_BT)^2\right. 
$$
$$
\times \left. \left[ \frac 75\frac{N^{\prime \prime }(E_F)}{N(E_F)}
-\left( \frac{N^{\prime }(E_F)}{N(E_F)}\right) ^2\right] \right\}  
\eqno{(3.40)}
$$
The temperature corrections, although being formally small, are especially
important in the case of a strong dependence $N(E)$, which is typical for
transition metals. The first term in the square brackets means that finite-$%
T $ specific heat is determined by the density of states, averaged over the
energy interval of order of $\pi k_BT$ near $E_F$. In particular, at $%
N^{\prime }(E_F)=0,N^{\prime \prime }(E_F)>0$ (minimum of the function $N(E)$%
) $\gamma $ increases with $T$, and at $N^{\prime \prime }(E)<0$ decreases
with $T$. This result agrees qualitatively with the band calculations
results and experimental data on the sign of the $T$-dependence (Fig.3.22):
for odd d$^n$ configurations (V, Ta, Nb, Pt, Pd) the great values of $\gamma 
$, which correspond to a $N(E)$ maximum, decrease with $T$, and for even
configurations (Zr, Ti, Cr, Mo, W) the small values of $\gamma $ correspond
to a $N(E)$ minimum and increase with $T$. The second term in square
brackets of (3.40) occurs because of the temperature dependence of the
chemical potential (3.38).

Shimizu [262] carried out numerical calculations to compare the experimental
dependence $\gamma (T)$ with those following from the first-principle
(Fig.3.23) and empirical (Fig.3.24) densities of states. Although a
qualitative agreement takes place, the theoretical dependences turn out to
be weaker and shifted towards high temperatures.

Provided that the function $N(E)$ possess a singular behaviour near $E_F$
(narrow peaks), the expansion (3.37) does not hold and the analytical
calculation is more complicated. The calculation of the dependence $\gamma
(T)$ in the presence of a DOS peak was performed in [264]. The simplest
model of $N(E)$ with triangle peaks at $E = E_i$ was used,
$$
\delta N(E) =\sum_i\delta N_i(E)  
\eqno{(3.41)} 
$$
$$
\delta N(E) =\left\{ 
\begin{array}{ll}
A_i(E-E_i) & ,\qquad E<E_i \\ 
B_i(E_i-E) & ,\qquad E>E_i
\end{array}
\right.
$$
Unlike the expansion (3.40), the relative temperature variation of
$$
\gamma (T)=\frac 1{T^2}(I_2-I_1^2/I_0)  
\eqno{(3.42)}
$$
with
$$
I_n=\int_{-\infty }^\infty dE\left( -\frac{\partial f(E)}{\partial E}\right)
(E-E_F)^nN(E)  
\eqno{(3.43)}
$$
may be large. Using the position of the peaks $E_i$ and the jumps of $%
N^{\prime }(E)$ at these points as parameters, one can restore from the
experimental dependence of $\gamma $ the density of states and compare this
with the calculated one. Such an analysis was performed in [264] for the
A-15 structure superconductors Nb$_3$Sn and V$_3$Si.

Besides the one-electron mechanism, the dependence $\gamma (T)$ is
determined by the above-discussed temperature dependences of electron-phonon
enhancement and correlation mechanisms which are expected to be rather
strong for d-bands. The relative role of these effects in transition metals
is still not investigated in detail.

\subsection{Specific heat of magnetic metals}

In magnetically ordered crystals, contributions to entropy and specific heat
occur which are due to destruction of magnetic ordering with increasing
temperature. Experimental separation of the magnetic entropy may be
performed with the use of magnetocaloric effect (change of temperature at
adiabatic magnetizing by the external magnetic field) [265].

In the model of localized spins with the value $S$ the total magnetic
entropy change is given by
$$
\mathcal{S}_{mag}(T\rightarrow \infty )=\int_0^\infty \frac{dT}%
Tc_{mag}(T)=R\ln (2S+1)  
\eqno{(3.44)}
$$
The magnetic specific heat $c_{mag}$ is especially important at the magnetic
phase transition points $T_M=T_C(T_N)$ where it has a singularity. In the
simplest mean-field approximation $c_{mag}$ is exponentially small at low
temperatures, exhibits a jump at crossing $T_M$ and vanishes in the
paramagnetic region, so that expression (3.44) holds at arbitary $T>T_M$. In
fact, the presence of short-range order, which is neglected in the
mean-field theory, results in that $c_{mag}$ is finite above $T_M$ due to
interaction with spin fluctuations. Besides that, fluctuations result in a
weak power-law divergence of $c_{mag}$ at $T_M$. The experimental
temperature dependences of specific heat in ferromagnetic transition metals
are shown in Figs.3.25, 3.26.

At low temperatures (well below $T_M$) there exist contributions owing to
spin-wave (magnon) excitations (Appendix E). The corresponding average
energy is given by
$$
\overline{\mathcal{E}}_{sw}(T)=\sum_{\mathbf{q}}\hbar \omega _{\mathbf{q}}N_{%
\mathbf{q}}  
\eqno{(3.45)}
$$
with $\omega _{\mathbf{q}}$ being the magnon frequency. Since in
ferromagnets $\omega _{\mathbf{q}}\sim q^2$ at $q\rightarrow 0$, we have 
$$
\overline{\mathcal{E}}_{sw}(T)\sim T^{5/2}/T_C^{3/2},\qquad 
c_{sw}\sim (T/T_C)^{3/2}  
\eqno{(3.46)}
$$
At the same time, for antiferromagnets the dispersion law is linear and we
have
$$
\overline{\mathcal{E}}_{sw}(T)\sim T^4/T_N^3,\qquad 
c_{sw}\sim (T/T_N)^3 
\eqno{(3.47)}
$$
The spin-wave contributions are present both in magnetic insulators amd
metals. In metals the linear electronic specific heat dominates at low
temperatures and the spin-wave corrections are less important. However, the
electronic contribution is appreciably influenced by magnetic ordering.
Since an enhancement of the linear $\gamma T$-term takes place due to
interaction with spin fluctuations (Sect.4.4), the strict separation of
magnetic and electronic contribu- tions is, generally speaking, impossible.
Some peculiar terms in specific heat of conducting ferromagnets which are
due to incoherent (non-quasiparticle) contributions are discussed in
Appendix G.

The electron density of states in the ordered phase of ferromagnetic metals
(below the Curie temperature $T_C$) differs considerably from that in the
paramagnetic phase because of spin splitting. As a rule, total $%
N(E_F)=N_{\uparrow }(E_F)+N_{\downarrow }(E_F)$ is smaller in the
ferromagnetic state since the ordering results in a shift of the peak, which
is responsible for the Stoner instability, from the Fermi level. At the same
time, in nickel the value of $\gamma (T<T_C)=$ 7.0 mJ/mol K$^2$ [266] is
larger than $\gamma (T>T_C)=$ 5.8 mJ/mol K$^2$ [267]. This leads to the
conclusion about an important role of correlation effects.

Consider the simplest theory of magnetic specific heat. In the mean-field
approximation the magnetic (exchange) energy is expressed in terms of
magnetization
$$
\mathcal{E}_{mag}=-IM^2  
\eqno{(3.48)}
$$
where $I$ is an exchange parameter. Then the magnetic specific heat reads
$$
c_{mag}=-IM\frac{dM}{dT}  
\eqno{(3.49)}
$$
In metals one has to take into account, besides the exchange energy, the
kinetic energy of conduction electrons which depends on magnetization
through the exchange splitting. The total energy in the Stoner model (see
Sect.4.3) may be transformed as
$$
\overline{\mathcal{E}} =\sum_{\mathbf{k}\sigma }E_{\mathbf{k}}
n_{\mathbf{k}\sigma }+In_{\downarrow }n_{\downarrow } 
$$
$$
=\sum_{\mathbf{k}\sigma }E_{\mathbf{k}\sigma }
n_{\mathbf{k}\sigma }-In_{\downarrow }n_{\downarrow }  
$$
$$
=\sum_\sigma \int dEEf(E)N_\sigma (E)-I\left( \frac{n^2}4-M^2\right) 
\eqno{(3.50)} 
$$
where
$$
n_\sigma =\frac n2+\sigma M,\qquad 
E_{\mathbf{k}\sigma }=E_{\mathbf{k}}-\sigma IM  
$$
$$
N_\sigma (E) =\frac 12N(E+\sigma IM)  
\eqno{(3.51)}
$$
Differentiating with respect to $T$ one obtains
$$
c(T)=c_{\downarrow }(T)+c_{\downarrow }(T)+\frac{\pi ^2}6\left[ \frac{%
N_{\uparrow }^{\prime }(E_F)}{N_{\uparrow }(E_F)}-\frac{N_{\downarrow
}^{\prime }(E_F)}{N_{\downarrow }(E_F)}\right] T^2I\frac{dM}{dT}  
\eqno{(3.52)}
$$
where $c_\sigma (T)$ is the usual electronic specific heat, corresponding to
the density of states $N_\sigma (E)$. One can see that the term of the type
(3.49) is cancelled by the corresponding contribution from the kinetic
energy. The magnetic specific heat is determined by the second term in
(3.52). Only electrons in the energy layer of order of T near the Fermi
level contribute to $c_{mag}$ so that it contains a small factor of order of 
$(k_BT/E_F)^2$ and depends strongly on the DOS shape. The jump of specific
heat at the Curie point is given by [26]
$$
\Delta c=\frac 23(\pi T_C)^2\{[N^{\prime }(E_F)\frac{\partial \zeta }
{\partial M}+I^2MN^{\prime \prime }(E_F)]\}_{T=T_C}  
\eqno{(3.53)}
$$
The situation may change somewhat in the presence of DOS peaks.
Consideration of specific heat of Ni with account of the realistic DOS
structure is performed in [268]. As well as in the Heisenberg magnets, spin
fluctuations result in a considerable mofifications of $c_{mag}(T)$
behaviour at high temperatures [26,268].

\chapter{MAGNETIC PROPERTIES}

Strong magnetism is one of the most important peculiarities of transition d-
and f-elements. Iron group metals (Fe, Co and Ni) exhibit ferromagnetic
ordering and large spontaneous magnetization, chromium and manganese are
antiferromagnetic --- magnetic moments are ordered, but the summary
magnetization is zero. Ferro- and antiferromagnetism are characteristic for
most rare earth metals. Many alloys and compounds of transition metals also
possess strong magnetic properties and have wide technical applications.

Basic magnetic characteristics of strongly magnetic substances are as follows

\noindent (i) the value of magnetic moment in the ground state

\noindent (ii) behaviour of spontaneous magnetization with temperature and
magnetic field; values of the ordering temperature $T_C(T_N)$ and magnetic
anisotropy constants; the saturation field $H_a$

\noindent (iii) the type of temperature dependence of magnetic
susceptibility (Pauli of Curie-Weiss) and its anisotropy; presence of local
moments above $T_C$ ; values of the effective moment (Curie constant) and
paramagnetic Curie temperature.

All these characterictistics are described by some microscopical parameters
(value of atomic magnetic moment in the localized model of magnetism or the
exchange splitting in the itinerant electron model, exchange parameters,
spin-orbital interaction constant). The main purpose of the theory is
comparison of these parameters with observable physical properties. At
present, alternative and in some respects contradicting explanations of
transition metal magnetic properties exist, and the problem of their strong
magnetism is far from the final solution.

\section{Exchange interactions and the Heisenberg model for localized spins}

The history of theoretical investigation of ferromagnetism was rather
dramatic. First attempts to explain this phenomenon gave rise to development
of the theory of exchange interactions within the Dirac-Heisenberg model for
localized spins. Although further theoretical developments demonstrated
inapplicability of this model to metallic magnets, we discuss this model to
first place.

Usual relativistic magnetic interactions between localized moments at
different lattice sites corresponds to the energy
\[
\mu _B^2/a^3\sim 10^{-17}\ \mathrm{erg/cm}^3\sim 10^{-1}\ \mathrm{K}
\]
so that magnetic ordering should be destroyed by thermal motion at very low
temperatures. However, transition metals of iron group and a number of their
compounds possess high values of ordering temperature (Curie point $T_C$) of
order of 10$^2$ -10$^3$K. Such large interactions between magnetic moments
are not explained by classical electrodynamics and require a quantum
mechanical treatment.

In the absence of spin-orbital interaction the wave function of a
two-electron system has the form of the product of spin and coordinate
functions
$$
\Psi (\mathbf{r}_1s_{1,}\mathbf{r}_2s_2)=\Psi (\mathbf{r}_1\mathbf{r}_2)\chi
(s_1s_2)
\eqno{(4.1)}
$$
and, according to the Pauli principle, should be antisymmetric with respect
to permutation of electrons. For symmetric (antisymmetric) spin function,
which corresponds to the total spin $\mathbf{S=s}_1\mathbf{+s}_2$ of unity
(zero), the trial coordinate wavefunctions may be written in the form
$$
\Psi (\mathbf{r}_1\mathbf{r}_2)=\frac 1{\sqrt{2}}[\psi _1(\mathbf{r}_1)\psi
_2(\mathbf{r}_2)\mp \psi _1(\mathbf{r}_2)\psi _2(\mathbf{r}_1)]
\eqno{(4.2)}
$$
respectively. Calculating the first-order perturbation theory correction to
the total energy from the electrostatic interaction we obtain
$$
E=\int d\mathbf{r}_1d\mathbf{r}_2\Psi ^{*}(\mathbf{r}_1\mathbf{r}_2)\frac{e^2%
}{|\mathbf{r}_1-\mathbf{r}_2|}\Psi (\mathbf{r}_1\mathbf{r}_2)=Q\mp J
\eqno{(4.3)}
$$
where we have introduced Coulomb integral
$$
Q=\int d\mathbf{r}_1d\mathbf{r}_2\rho _1(\mathbf{r}_1)
\frac 1{|\mathbf{r}_1-\mathbf{r}_2|}\rho _2(\mathbf{r}_2),\qquad
\rho (\mathbf{r})=-e\psi (\mathbf{r})\psi \mathbf{(r)}
\eqno{(4.4)}
$$
and the exchange integral, which differs from (4.4) by permutation
(exchange) of electrons
$$
J=\int d\mathbf{r}_1d\mathbf{r}_2\psi _1^{*}(\mathbf{r}_1)\psi _2^{*}(%
\mathbf{r}_2)\frac{e^2}{|\mathbf{r}_1-\mathbf{r}_2|}\psi _{_1}(\mathbf{r}%
_2)\psi _2(\mathbf{r}_1)
\eqno{(4.5)}
$$
Unlike the relativistic magnetic interaction, the exchange interaction has
the electrostatic nature and is considerably larger. At the same time,
exchange interaction between electrons at different lattice sites is
strongly reduced by the squared factor of overlap of the corresponding
wavefunctions. We may estimate
\[
J\sim Q(10^{-2}-10^{-3})\sim 10^{-3}\ \mathrm{eV}\sim 10\ \mathrm{K}
\]
The expression (4.3) may be rewritten in the form of the spin Hamiltonian
$$
\mathcal{H}=Q-J(\frac 12+2\mathbf{S}_1\mathbf{S}_2)=\left\{
\begin{array}{ll}
Q-J&,\qquad S_{tot}=1 \\
Q+J&,\qquad S_{tot}=0
\end{array}
\right.
\eqno{(4.6)}
$$
where
$$
2\mathbf{S}_1\mathbf{S}_2=S_{tot}^2-2s^2=S_{tot}(S_{tot}+1)-\frac 32
\eqno{(4.7)}
$$
Besides the ``potential'' exchange interaction (4.5), there exists another
exchange mechanism which is due to kinetic energy of electrons. In the
simplest situation of s-band, we obtain (Appendix D)
$$
\mathcal{H}^{^{\prime }}=-\frac{2t^2}U(2\mathbf{S}_1\mathbf{S}_2-\frac 12)
\eqno{(4.8)}
$$
with $t$ being the transfer integral and $U$ the intraatomic Coulomb
repulsion. This mechanism results in an antiferromagnetic interaction since
the gain in the kinetic energy is achieved at antiparallel orientation of
electron spins. As demonstrate numerical calculations (see, e.g., [265]), at
realistic interatomic distances this contribution as a rule prevails over
the ferromagnetic potential exchange. Thus the localized model does not
explain ferromagnetism of iron group metals.

The derivation of Heisenberg model for orbitally degenerate atomic shells
and arbitrary values of spin $S$ is presented in Appendix D. It turns out
that in this case the exchange Hamiltonian contains the interaction of not
only spin, but also orbital moments. Exchange mechanisms in rare earths
owing to indirect interaction via conduction electrons is discussed in
Appendix K. In most cases, real exchange mechanisms are complicated, so that
we have to consider the Heisenberg Hamiltonian as an effective one.

Consider the system of interacting magnetic moments which is described by
the Heisenberg Hamiltonian in the magnetic field $H$
$$
\mathcal{H}=-g\mu _BH\sum_iS_i^z-\sum_{ij}J_{ij}\mathbf{S}_i\mathbf{S}_j
\eqno{(4.9)}
$$
with $J_{ij}>0$ for nearest neighbours. Direct calculation of the
magnetization $M=gm_B\langle S^z\rangle $ (hereafter we put $g\mu _B=1$) in
the model (4.9) is prevented by the non-linearity of the interspin exchange
interaction in spin operators. To simplify the problem, we can linearize the
Hamiltonian by introducing the effective mean field $H^{*}$ which is
expressed in terms of magnetization
$$
\mathcal{H}=-\widetilde{H}S^z=-(H+H^{*})S^z,\qquad
S^z=\sum_iS_i^z
\eqno{(4.10)}
$$
where
$$
H^{*}=\mathcal{\lambda }\left\langle S^z\right\rangle,\qquad
\lambda =2\sum_iJ_{ij}=2J_0,\qquad
J_0=J_{\mathbf{q}=0}
\eqno{(4.11)}
$$
(in the nearest-neighbour approximation $J_0=zJ$ with $z$ being the number
of the neighbour spins). In the case of spin $S=1/2$ we have
$$
\left\langle S^z\right\rangle =\frac 12(n_{\uparrow }-n_{\downarrow}),\qquad
n_{\uparrow }+n_{\downarrow }=1
\eqno{(4.12)}
$$
Using the Gibbs distribution
$$
n_\sigma =\exp \frac{\sigma \widetilde{H}}T\left/ [\exp \frac{\widetilde{H}}%
T+\exp (-\frac{\widetilde{H}}T)]\right.
\eqno{(4.13)}
$$
we obtain the self-consistent mean-field equation for the magnetization
$$
\langle S^z\rangle =\frac 12\tanh \frac{H+\lambda \langle S^z\rangle }{2T}
\eqno{(4.14)}
$$
The result (4.14) is easily generalized to the case of arbtitrary $S$
$$
\langle S^z\rangle =SB_S\left( \frac{S\widetilde{H}}T\right)
\eqno{(4.15)}
$$
where
$$
B_S(x)=\frac d{dx}\ln \sum_{m=-S}^Se^{-\frac{mx}S}
$$
$$
=(1+\frac 1{2S})\coth (1+\frac 1{2S})x-\frac 1{2S}\coth \frac x{2S}
\eqno{(4.16)}
$$
is the Brillouin function. In particular cases
$$
B_{\frac 12}(x)=\tanh x,\qquad
B_\infty (x)=\coth x-\frac 1x
\eqno{(4.17)}
$$
The equation (4.15) has non-trivial solutions in the zero magnetic field
provided that $T<T_C$, the Curie temperature being given by
$$
T_C=S(S+1)\lambda =\frac 23S(S+1)J_0
\eqno{(4.18)}
$$
Using the expansion
\[
\coth x=\frac 1x+\frac x3-\frac{x^3}{45}+...
\]
we derive near the Curie point
$$
\langle S^z\rangle =\sqrt{\frac 53}\frac{S(S+1)}{(S^2+S+\frac 12)}\sqrt{%
1-\frac T{T_C}}
\eqno{(4.19)}
$$
At low $T\ll T_C$ equation (4.15) may be solved by iterations and yields an
exponential behaviour of the magnetization
$$
\langle S^z\rangle =S[1-S^{-1}\exp (-2T_C/T)]
\eqno{(4.20)}
$$
A more accurate description of the low-temperature behaviour is obtained in
the spin-wave theory (Appendix E).

Consider the case of high temperatures $\mu _BH\ll T$. Performing expansion
of (4.15) in $H$ we obtain
\[
\langle S^z\rangle =\frac{S(S+1)H}{3T[1-\frac 13\lambda S(S+1)/T]}
\]
Thus we obtain the Curie-Weiss law for the paramagnetic susceptibilty $\chi
=M/H$
$$
\chi (T)=\frac C{T-\theta }
\eqno{(4.21)}
$$
The Curie constant $C$ is expressed in terms of the effective atomic moment,
$$
C=\frac 13\mu _{eff}^2,\qquad mu _{eff}=g\mu _B\sqrt{S(S+1)}
\eqno{(4.22)}
$$
The paramagnetic Curie temperature turn out to coincide in the simple model
under consideration with $T_C$:
$$
\theta =\lambda C=\frac 23\lambda S(S+1)J_0
\eqno{(4.23)}
$$
The case of antiferromagnetic exchange interaction $J<0$ is considered in a
similar way. In such a case, neighbour spins are ordered in an antiparallel
way and the equation for the sublattice magnetization has the same form
(4.15) with $J\rightarrow -J$. The magnetic susceptibility obeys the
Curie-Weiss law with a negative value of $\theta $. In the nearest-neighbour
approximation, the Neel temperature $T_N$, where magnetic ordering
disappears, equals to $|\theta |$.

\section{Magnetic susceptibility of paramagnetic transition metals}

Now we return to discussion of situation in transition metals. Most
d-elements (3d-metals Sc, Ti, V and all the 4d and 5d-metals) do not possess
magnetic ordering. However, their magnetic properties differ considerably
from those of simple (e.g. alcaline) metals which demonstrate a weak Pauli
paramagnetism with $\chi \sim 10$ emu/mol. The values of paramagnetic
susceptibility of d-metals, which are presented in the Tables 4.1, 4.2 are
by an order of magnitude larger.

There exist some regularities in the behaviour of $\chi$ 
in the transition metal
rows. For metals of one group in the periodic system, the susceptibility
decreases with increasing atomic number. The change of $\chi $ in a given
period turns out to be non-monotonous: as a rule, metals with even number of
d-electrons have lower susceptibility than those with odd number (Fig.4.1,
Tables 4.1,4.2). Thus a correlation with electronic specific heat takes
place. The temperature dependences of $\chi $ (Figs.4.2-4.6) are appreciably
stronger than in simple metals. The sign of $d\chi /dT$ changes as a rule at
passing to neighbour element. So, for metals of IV group (Ti, Zr and Hf),
which have a small susceptibility, $\chi $ increases with $T$, and for
metals of the V group (V, Nb, Ta) $d\chi /dT<0$. The presented data for $%
d\chi /dT$ correspond to not too low temperatures, since at low $T$ the
behaviour of $\chi $ may be masked by the Curie-Weiss contribution from
magnetic impurities. However, for Pd the intrinsic susceptibility has a
maximum at 80K, and for Y at 300K.

Now we discuss various contributions to magnetic susceptibility of
transition metals. To calculate the paramagnetic susceptibility we write
down the magnetization of band electrons in a magnetic field $H$
$$
M=\mu _B\int dEN(E)\left[ f(E-\mu _BH)-f(E+\mu _BH)\right]
\eqno{(4.24)}
$$
(the density of states $N(E)$ is put for one spin projection). Expanding
(4.24) in $H$ we derive
$$
\chi =\frac MH=-2\mu _B^2\int dEN(E)\frac{\partial f(E)}{\partial E}
$$
$$
=2\mu _B^2\left[ N(\zeta )+\frac{\pi ^2}6N^{\prime \prime }(\zeta
)(k_BT)^2\right]
\eqno{(4.25)}
$$
Performing expansion in $\zeta -E_F$ and taking into account the temperature
dependence of the chemical potential (3.38) we obtain
$$
\chi (T)=2\mu _B^2N(E_F)\left\{ 1+\frac 16(\pi k_BT)^2\left[ \frac{N^{\prime
\prime }(E_F)}{N(E_F)}-\left( \frac{N^{\prime }(E_F)}{N(E_F)}\right)
^2\right] \right\}
\eqno{(4.26)}
$$
Thus spin susceptibility, as well as specific heat, of conduction electrons
is determined by the density of states at the Fermi level and is to leading
order approximation temperature independent (Pauli paramagnetism). Such a
behaviour, which is drastically different from the Curie-Weiss law for
localized electrons, is explained by the Pauli principle: only a small part
(of order of $k_BT/E_F$) electrons contribute to $\chi $. The dependence $%
\chi (T)$, which is obtained after account of next-order terms in $T/E_F$ ,
turns out to be weak.

One can see that the sign of $d\chi /dT$ should be determined by the shape
of density of states. Thus large values and strong $T$-dependence of $\chi $
in d-metals may be related to the presence of narrow d-bands near the Fermi
level.

Due to presence of partially filled degenerate bands in d-metals, orbital
contributions to susceptibility turn out to be important, which are
analogous to the Van Vleck contribution for localized magnetic ions. The
total paramagnetic susceptibility for a degenerate band may be represented
in the form [271,270]
$$
\chi =\chi _S+\chi _L+\chi _{SL}
\eqno{(4.27)}
$$
with
$$
\chi _S=2\mu _B^2\sum_{m\ m^{\prime }\mathbf{k}}\frac{f(E_{m\mathbf{k}%
})-f(E_{m^{\prime }\mathbf{k}})}{E_{m^{\prime }\mathbf{k}}-E_{m\mathbf{k}}}%
|\langle m\mathbf{k}|2\mathbf{S}|m^{\prime }\mathbf{k}\rangle |^2
\eqno{(4.28)}
$$
$$
\chi _L=2\mu _B^2\sum_{m\ m^{\prime }\mathbf{k}}\frac{f(E_{m\mathbf{k}%
})-f(E_{m^{\prime }\mathbf{k}})}{E_{m^{\prime }\mathbf{k}}-E_{m\mathbf{k}}}%
|\langle m\mathbf{k}|\mathbf{L}|m^{\prime }\mathbf{k}\rangle |^2
\eqno{(4.29)}
$$
$$
\chi _{SL}=2\mu _B^2\sum_{m\ m^{\prime }\mathbf{k}}\frac{f(E_{m\mathbf{k}%
})-f(E_{m^{\prime }\mathbf{k}})}{E_{m^{\prime }\mathbf{k}}-E_{m\mathbf{k}}}%
\langle m\mathbf{k}|L|m^{\prime }\mathbf{k}|\rangle \langle m\mathbf{k}%
|2S|m^{\prime }\mathbf{k}\rangle
\eqno{(4.30)}
$$
where $m$ is the band index. A simple estimation of the orbital contribution
in terms of the number of d-electrons $n$ has the form
$$
\chi _L\sim 2\mu _B^2\frac{n(10-n)}{5W}
\eqno{(4.31)}
$$
where $W$ is a characteristic energy difference between occupied and empty
states, which equals approximately to the d-bandwidth. Separation of spin
and orbital contributions may be performed by combined studying the
dependences of $\chi (T)$ (with account of Stoner enhancement) and
electronic specific heat $c_e(T)$, the orbital contribution being proposed
to be weakly temperature-dependent [270]. Spin susceptibility may be also
picked out by measuring the Knight shift.

According to the review [270], calculated orbital and spin susceptibilities
of d-metals are of the same order of magnitude, the orbital susceptibility
of d-electrons exceeding the spin susceptibility of s-electrons. The
spin-orbit contribution is negligible, except for 5d-metals. Band
calculations of the orbital and spin-orbital susceptibility in bcc
transition metals are presented in [272].

Data on orbital contributions are also obtained by measuring the
gyromagnetic (magnetomechanical) ratio with the use of the Einstein - de
Haas effect [273]
$$
g^{\prime }=\frac{M_L+M_S}{M_L+\frac 12M_S}
\eqno{(4.32)}
$$
This quantity differs somewhat from the $g$-factor
$$
g=\frac{M_L+M_S}{\frac 12M_S}
\eqno{(4.33)}
$$
which is measured in magnetic resonance experiments (it is taken into
account in (4.33) that the orbital mechanical momentum is not conserved in
the lattice). Thus the factors $g$ and $g^{\prime }$ are connected by
$$
\frac 1g+\frac 1{g^{\prime }}=1,\qquad
g^{\prime }=g/(g-1)
\eqno{(4.34)}
$$
In the case of small orbital magnetization,
$$
g=2(1+\frac{M_L}{M_S}),\qquad
g^{\prime }=2(1-\frac{M_L}{M_S})
\eqno{(4.35)}
$$
Investigations of gyromagnetic ratio in paramagnetic metals are much more
difficult than for ferromagnets since one has to measure very small rotation
angles. The results of the experiments [274] are as follows
$$
\begin{tabular}{llllll}
              & V    & Nb   & Ta   & Pd   & Pt   \\
$g^{\prime }$ & 1.18 & 1.05 & 1.02 & 1.77 & 1.62
\end{tabular}
$$
Strong deviations of values from 2 show a presence of appreciable orbital
contributions. The measurements for iron group metals and ferromagnetic
alloys yield $g-2\sim 2-g\sim 5-10\%$ [273].

Finally, the diamagnetic contribution to $\chi$ in the one-band
approximation is given by [275]
$$
\chi _{dia}=\frac{e^2}{6c}\sum_{\mathbf{k}}\frac{\partial f(E_{\mathbf{k}})}{%
\partial E_{\mathbf{k}}}\left[ \frac{\partial ^2E_{\mathbf{k}}}{\partial
k_x^2}\frac{\partial ^2E_{\mathbf{k}}}{\partial k_y^2}-(\frac{\partial ^2E_k%
}{\partial k_x\partial k_y})^2\right]
\eqno{(4.36)}
$$
(the magnetic field is directed along the $z$-axis). In the effective mass
approximation where $E_{\mathbf{k}}=k^2/2m^{*}$ we obtain
$$
\chi _{dia}=-\frac 23\mu _B^2N\left( E_F\right) (m/m^{*})^2
\eqno{(4.37)}
$$
where $c$ is the light velocity. Unlike the paramagnetic contribution, the
diamagnetic susceptibility is inversely proportional to electron effective
mass m since the influence of magnetic field on the orbital motion is
directly proportional to electron velocity; for free electrons $\chi
_{dia}=-\chi _{para}/3.$ For the effective mass of d-electrons is large, the
diamagnetic contribution may be as a rule neglected.

Investigations of single crystals of paramagnetic d-metals with the hcp
structure demonstrated a considerable anisotropy of $\chi $ (Table 4.2). In
Y, Re, Ru, Os, $\Delta \chi =\chi _{\Vert }-\chi _{\bot }<0$, a typical
difference being tens of percents (however, for osmium, $\Delta \chi /\chi
\simeq 3$). For Ti, Zr and Hf, $\Delta \chi >0$ and the anisotropy is also
large. As a rule (with the exception of Os), the sign of $\Delta \chi $ is
the same for the elements of one row, $\Delta \chi $ being positive for the
atomic configuration d$^n$ with even $n$ and negative for odd $n$. This may
indicate a partial adequacy of the atomic picture.

Reliable theory of susceptibility anisotropy is now absent. Anisotropic
contributions may have orbital, spin-orbital and diamagnetic origin. The
estimations of anisotropy of the orbital contibution to $\chi $ in the band
theory for hexagonal metals ($\Delta \chi \sim \sqrt{8/3}-c/a=1.633-c/a)$
[276] yielded the values $\Delta \chi /\chi \sim 0.02$ which are too small
to explain experimental data. Recently [277] an attempt to calculate Dc for
Sc and Y within relativistic calculations has been performed. Due to
relativistic interactions, anisotropic part of spin susceptibility was
obtained with $\chi _{\Vert }>\chi _{\bot }$, which disagrees with
experimental data. According to [277], the agreement with experimental data
should be restored after allowance for Van Vleck-type contributions. These
conclusions agree on the whole with results of earlier calculations
[278,279].

\section{Itinerant electron ferromagnetism and the Stoner theory}

Magnetic characteristics (values of paramagnetic and ferromagnetic Curie
temperatures and of corresponding magnetic moments) for some ferromagnetic
d- and f-metals and their compounds are presented in the Table 4.3. The
correlation between the values of Curie temperature and the ratio of
paramagnetic and ferromagnetic moments $p_C/p_s$ is demonstrated by the
Rhodes-Wolfarth curve (Fig.4.7). In the Heisenberg model with spin $S$, $%
p_C=p_s=2S$, so that $p_C/p_s$ should be equal to unity. This condition is
well satisfied for insulating localized-spin ferromagnets EuO and CrBr$_3$
and f-metals (some deviations in rare-earth metals may be explained by
polarization of conduction electrons). One can see that for Fe and Ni $%
p_C/p_s$ somewhat exceeds 1, and the values of $p_C$ and $p_s$ correspond to
no values of $S$, so that magnetic moment turns out to be non-integer.

In metallic d-magnets, $p_C/p_s$ is rougly inversely proportional to $T_C$.
For intermetallic compounds with very small saturation moment $M_s$ (ZrZn$_2$%
, Sc$_3$In, Ni$_3$Al), which are called weak itinerant ferromagnets, $%
p_C/p_s $ is very large, so that the situation is drastically different from
the localized magnetism picture. On the contrary, for Heusler alloys T$_2$%
MnZ, TMnZ we have $p_C/p_s<1$. The latter case is discussed in Sect.4.4.

One can also see from Table 4.3 that the values of $T_C$ and q in some
systems are quite different. (Especially strange is the situation for the
compound CeRh$_2$B$_2$; this belongs to anomalous systems where the Kondo
effect plays apparently an important role, see Chapter 6.) Within the
Heisenberg model, a not great difference only may be explained by
next-nearest neighbour exchange interactions or fluctuation effects
(Appendix E).

Thus the localized spin model is adequate for some insulator d- and
f-compounds and for f-subsystem in rare earths, but does not describe
magnetic properties of transition d-metals. In particular, the Heisenberg
model does not explain non-integer values of magnetic moments and, in the
case of magnetic ordering, the difference between values of the moments in
the ground state and at high temperatures (as determined from the Curie
constant).

To solve these problems, Stoner [285] put forward the model of itinerant
magnetism based on the exchange interaction among band electrons, which was
treated in the mean field approximation. The Fermi statistics of band
electrons (contrary to atomic statistics of localized electrons in the
Heisenberg model) leads to a radical change of magnetic properties. The
magnetic ordering results in an increase of kinetic energy, but decreases
the potential (exchange) energy. Thus the conditions for ferromagnetism to
occur are rather restricted.

First calculations by Stoner [285] used free-electron model and did not take
into account correlation effects. The system of equations for the
magnetization and number of electrons of an itinerant electron system in the
molecular field $H = H + 2IM$ ($I$ is the exchange parameter for itinerant
electrons) has the form
$$
M=\frac 12(n_{\uparrow }-n_{\downarrow })\,\,,\,\,\,\,\,\,\,n=n_{\uparrow
}+n_{\downarrow },\,\,n_\sigma =\int dEN(E)f(E-\frac 12\sigma \widetilde{H})
\eqno{(4.38)}
$$
This system is an analogue of the equations (4.13). For free-electron
density of states
\[
N(E)=\frac 34nE^{1/2}/E_F^{3/2}
\]
we have
$$
n_{\uparrow },_{\downarrow }=\frac 34\left( \frac T{E_F}\right)
^{3/2}F\left[ \zeta \pm \frac 12(H+2IM)/T)\right]
\eqno{(4.39)}
$$
with
\[
F(y)=\int_0^\infty \frac{dzz^{1/2}}{e^{z-y}+1}
\]
At high temperatures we may expand the Fermi integrals to obtain
$$
M=\frac{H+2IM}T\frac{F^{\prime }(\zeta )}{F(\zeta )}n
\eqno{(4.40)}
$$
Then the susceptibility reads
$$
\chi =\frac{\chi _p}{1-2I\chi _p}
\eqno{(4.41)}
$$
with
\[
\chi _p=\frac nT\frac{F^{\prime }(\zeta )}{F(\zeta )}
\]
The condition of ferromagnetism (the Stoner criterion)
$$
2I\chi _p>1
\eqno{(4.42)}
$$
corresponds to divergence of the ground state susceptibility, and the Curie
temperature is estimated as
$$
2I\chi _p(T_C)=1,\,\,\,\,\,\,\,\,\,\,\,\,T_C\sim (2I\chi _p(0)-1)^{1/2}E_F
\eqno{(4.43)}
$$
Now we have to consider the quantity $F^{\prime }/F$. If this would be
weakly temperature dependent, the expression (4.41) would yield above $T_C$
the Curie-Weiss law
$$
\chi (T)=\frac C{T-\theta },\,\,\,\,\,\,\,\,\,\,\,\,\,\,C=n\frac{F^{\prime }}%
F\,\,,\,\,\,\,\,\,\,\,\,\,\,\theta =2IC
\eqno{(4.44)}
$$
\thinspace However, such a situation takes place only at very high
temperatures $T>E_F$ which exceed in fact melting points. On the contrary,
provided that the ratio $T_C/E_F$ is small, i.e. $2I\chi (0)$ is close to
unity, the susceptibilty above $T_C$ has the usual Pauli (square)
temperature dependence. Thus the Stoner theory does not explain in fact the
Curie-Weiss law.

Solving the Stoner equations for the parabolic band (4.39) at $T = 0$
demonstrates that at
$$
I>2^{2/3}E_F/n
\eqno{(4.45)}
$$
there exist solutions which describe the saturated ferromagnetic state with $%
M_0=n/2$. At
$$
2^{2/3}E_F/n>I>\frac 4{3n}E_F
\eqno{(4.46)}
$$
we have non-saturated ferromagnetism with partially filled spin down
subband, and at smaller $I$ magnetic ordering is absent. In the limit of
small ground state magnetization we have the relation
$$
\frac{2\sqrt{2}}{3\pi }\frac{M_0}n\sim \frac{T_C}{E_F}
\eqno{(4.47)}
$$
Thus $M_0$ is rougly proportional to the Curie point, in agreement with the
Rhodes-Wohlfarth correaltion (Fig.4.7).

The temperature dependence of magnetization at low $T$ in the saturated
state turns out to be exponential
$$
M=\frac n2\left[ 1-\frac{3\sqrt{\pi }}4\frac T{E_F}\exp \left( -\frac{E_F}%
T\left( \frac{In}{E_F}-2^{2/3}\right) \right) \right]
\eqno{(4.48)}
$$
At the same time, corrections, which are proportional to $T^2$, occur in the
non-saturated state owing to electron-hole spin-flip processes (Stoner
excitations):
$$
M\simeq M_0\left[ 1-\frac 12\left( \frac T{T_C}\right) ^2\right]
\eqno{(4.49)}
$$
Near the Curie point the magnetization has the square-root behaviour, as
well as in the mean-field approximation in the Heisenberg model:
$$
\left( M/M_0\right) ^2=3(1-T/T_C)\left[ 1+\frac 23\left( 2\pi \right)
^{-1/2}\left( E_F/T_C\right) ^{3/2}\right]
\eqno{(4.50)}
$$
Generally speaking, the results of the solution of the Stoner equations
(4.38) turn out to be rather sensitive to the density of states shape. In
particular, for the ``rectangle'' band with constant density of states,
considered in [286], only saturated ferromagnetic solutions are possible.

Consider the results of the Stoner theory for an arbitrary bandstructure.
The criterion of ferromagnetism reads
$$
IN(E_F)>1
\eqno{(4.51)}
$$
Provided that $IN(E_F)-1$ is small, the saturation magnetization equals [26]
$$
M_0=N(E_F)\left\{ \frac{2\left[ IN(E_F)-1)\right] }{\left[ N^{\prime
}(E_F)/N(E_F)\right] ^2-\frac 13N^{\prime \prime }(E_F)/N(E_F)}\right\}
^{1/2}
\eqno{(4.52)}
$$
According to (4.41), the value of $T_C$ is
$$
T_C^2=\frac 6{\pi ^2}\frac{IN(E_F)-1}{IN(E_F)}\left\{ \left[ \left[ \frac{%
N^{\prime }(E_F)}{N(E_F)}\right] ^2-\frac{N^{\prime \prime }(E_F)}{N(E_F)}%
\right] \right\} ^{-1}
\eqno{(4.53)}
$$
The susceptibility above $T_C$ is presented as
$$
\frac 1{\chi (T)}=\frac 1{\chi _p(T)}-\frac 1{\chi _p(T_C)}=\frac{2I}{\left[
IN(E_F)-1\right] ^2}\left( \frac{T^2}{T_C^2}-1\right)
\eqno{(4.54)}
$$
Of course, the expansion of the Fermi integrals (3.37), used at derivation
of (4.52)-(4.54), holds only at $T_C\ll E_F$ and for a smooth (near the
Fermi energy) density of states.

Modern numerical calculations with account of correlation effects (see,
e.g., [287]) demonstrate that ferromagnetic ordering in the free electron
system is possible at unrealistic large values of $r_s$ (small densities).
Thus the occurence of ferromagnetism is intimately related to the crystal
lattice potential which modifies considerably the shape of electron density
of states. In particular, small widths of d-bands result in a considerable
localizaton of d-electrons and decrease of the kinetic energy, and
consequently, favour magnetic ordering. First-principle band calculations
[24] demonstrate that the values of the Stoner parameter I vary smoothly and
not too strongly in the periodic table. At the same time, the values of $%
N(E_F)$ vary by orders of magnitude because of presence of density of states
peaks for d-electrons. As a rule, such peaks are connected with the Van Hove
electron spectrum singularities (Sect.2.4). In particular, for non-magnetic
iron (Fig.4.8) the peak occurs due to merging of the singularities for
$e_g$-states along the $P-N$ line [288]. The situation for Ni is discussed in
[289]. For the weak itinerant ferromagnet ZrZn$_2$ the fulfilment of the
Stoner criterion is connected with the presence of DOS peaks owing to $C15$
crystal structure [290].

Calculated values of the Stoner parameter turn out to describe
satisfactorily ferromagnetism of transition metals in the ground state. On
the other hand, the experimental data, which are discussed in Sect.4.2,
demonstrate that the conclusions of the Stoner theory for finite
temperatures turn out to be poor. First, this theory may yield only very
weak dependences $\chi (T)$. At the same time, experimental data on
transition metals demonstrate as a rule more strong dependences. One can see
from Figs.4.2-4.6 that in the temperature interval under consideration $\chi
(T)$ changes by tens of percents, whereas expression (4.41) yields the
change of order of $(T/E_F)^2\ll $ 1. In a number of cases (Fe, Co, Ni, Pt,
Pd), the Curie-Weiss law, i.e. the linear dependence $\chi (T)$
approximately holds. Second, experimental values of $T_C$ for iron group
metals (of order of 10$^3$ K) are considerably smaller than those predicted
by (4.53).

To treat the latter problem in more detail, we consider the ratio of the
ground state spin splitting $\Delta _0=2IM_0$ to the Curie temperature,
$\delta =\Delta _0/T_C$ . In the mean field Stoner theory this quantity turns
out to be rather stable. This situation is reminiscent of the BCS theory of
superconductivity where the ratio of the energy gap to the critical
temperature is an universal constant, $2\Delta /T_c$ = 3.52. Deviations from
this value, which occur, e.g., in high-$T_c$ superconductors, demonstrate
inapplicability of the simple BCS theory.

At calculating $\delta$ for a ferromagnet with a smooth density of states we
may use the expressions (4.52), (4.53) to obtain [291]
$$
\delta =\frac{2\pi }{\sqrt{3}}\left\{ \frac{\left[ N^{\prime
}(E_F)/N(E_F)\right] ^2-\frac 13N^{\prime \prime }(E_F)/N(E_F)}{\left[
N^{\prime }(E_F)/N(E_F)\right] ^2-N^{\prime \prime }(E_F)/N(E_F)}\right\}
^{1/2}
\eqno{(4.55)}
$$
Restricting ourself to the case $N^{\prime \prime }<0$ we get
$$
\frac{2\pi }{\sqrt{3}}\leq \delta \leq 2\pi
\eqno{(4.56)}
$$
In the case $N^{\prime \prime }>0$ (where ferromagnetism is a rule unstable
since the Fermi level is near the density of states minimum) $\delta $ may
become still smaller. Deviations of $\delta $ from the result (4.55) provide
a measure of correlation (fluctuation) effects. The corresponding data
according to various authors are presented in Table 4.4. The largest value
of $\Delta $ are obtained by Gunnarsson [292]. The result of [293] is
appreciably smaller because of an account of the magnetization dependence
for the exchange-correlation energy.

One can see that the experimental values of $\delta $ exceed considerably
theoretical ones and lie in the interval (4.56) for nickel only. However,
the values of $\Delta _0$ and $T_C$ themselves do not agree with experiment
in the latter case too. Besides that, angle-resolved photoemission and other
spectral data (Sect.2.5) demonstrate an appreciable short-range order above $%
T_C$ , which is not described by the mean-field theory.

The difficulties of the Stoner theory require new physical ideas. One of
simple ways to obtain a more strong temperature dependence of paramagnetic
susceptibility and lower the Curie point is changing the dependence $\chi
_p(T)$. Evidently, presence of a $T$-linear term in the latter would give
(after allowance for the Stoner enhancement) the Curie-Weiss law for $\chi
(T)$. The dependence $\chi _p(T).$ for complicated electron densities of
states was analyzed in [294]. The expansion (4.26) is quite correct for
smooth functions $N(E)$. However, the situation may be changed drastically
in the case of a sharp energy dependence, e.g., in the presence of a break
owing to the van Hove singularity.

Consider the simplest model of $N(E)$ with a break
$$
N(E) =\widetilde{N}(E)+N^{\prime }(E)(E-E_0)
\eqno{(4.57)}
$$
$$
N^{\prime }(E) =\left\{
\begin{array}{c}
A_1,\,\,\,\,\,\,E<E_0 \\
A_2,\,\,\,\,\,\,E>E_0
\end{array}
\right. \,\,\,\,\,
$$
where $\widetilde{N}(E)\,\,$is a smooth function. Then exact integration in
(4.25) gives
$$
\chi _p(T)=\frac 12\left[ N(\zeta )+(A_2-A_1)T\ln (1+\exp (-\frac{\mid
E_0-\zeta \mid }T)+O\left( \frac{T^2}{\zeta ^2}\right) \right]
\eqno{(4.58)}
$$
The effect of singularity vanishes at $|E_0-\zeta |\gg T$.

According to the band calculations for palladium, there exists the van Hove
singularity with $|E_0-E_F|=300$K. To describe correctly the experimental
behaviour of $\chi (T)$ in Pd (in particular, the position of the maximum),
it is necessary to shift the singularity to the Fermi level up to $%
|E_0-E_F|=100$ K.

Presence of DOS peaks at the Fermi level, which is, as discussed above,
typical for both strong and weak itinerant ferromagnets, may influence
appreciably the $\chi (T)$ behaviour and the $T_C$ value even in the Stoner
theory of ferromagnetism. In particular, the triangle peak model was used to
describe properties of weak itinerant ferromagnets [280]. Under condition of
the linear dependence $\chi ^{-1}(T)$ owing to a DOS singularity the Curie
temperature of a ferromagnetic metal is estimated as
$$
T_C\sim (2I\chi _p(0)-1)E_F
\eqno{(4.59)}
$$
and may be considerably smaller than (4.53).

\section{Spin-fluctuation theories}

A more radical way to improve the results of the Stoner theory for
temperature dependences of magnetic properties is going out beyond the
one-electron mean-field theory. Formally, the Stoner model corresponds to
the interaction Hamiltonian of an infinite range,
$$
H_{int}=In_{\uparrow }n_{\downarrow }=
I(\frac{n^2}4-\langle S^z\rangle ^2),\qquad
n_\sigma =\sum_{\mathbf{k}}n_{\mathbf{k}\sigma }
\eqno{(4.60)}
$$
so that the mean-field approximation is exact. A more rich physics is
described by models which include electron correlations. The simplest among
them is the Hubbard model (G.1) with the on-site Coulomb repulsion $U$. The
Stoner approximation for the electron spectrum in this model corresponds to
the simplest Hartree-Fock decoupling and may be justified only in the small-$%
U$ limit which is inconsistent with the ferromagnetism criterion. Therefore
correlation (spin-fluctuation) effects, generally speaking, turn out to be
important.

As well as for Heisenberg magnets, low-temperature thermodynamics of
itinerant magnets is determined by spin-wave contributions (see, e.g.,
[295]). Spin-wave theory of metallic ferro- and antiferromagnets is
considered in Appendix G. The simple spin-wave treatment enables one also to
reproduce some of the results of spin-fluctuation theories.

Spin-fluctuation theories, which describe thermodynamics of itinerant
magnets in a wide temperature region, were developed during last twenty
years [26, 296-302]. They turned out to be especially successful for weak
itinerant ferromagnets which are characterized by extremely small values of
saturation magnetic moment $M_0$ and large ratios $p_C /p_s$ (Table 4.3,
Fig.4.7). These substances correspond to a near vicinity of the Stoner
instability and possess strong spin fluctuations. However, unlike systems
with local magnetic moments, the fluctuations are localized not in real, but
in reciprocal space (near the wavevector $q = 0$).

Presence of the small parameter $2I\chi -1$ yields a possibility of an
analytical treatment of weak itinerant ferromagnetism. The simple RPA
expression for magnetic susceptibilty (G.12), which follows from the
spin-wave theory, turns out to be thermodynamically inconsistent. In the
theory of Moriya and Kawabata [296] the expression for paramagnetic
susceptibility was taken in the form
$$
\chi (T)=\frac{\chi _p(T)}{1-2I\chi _p(T)+\lambda (T)}
\eqno{(4.61)}
$$
where $\lambda (T)$ is proportional to the squared amplitude of spin
fluctuations. Calculation of $\lambda $ in a self-consistent way by using
the expression for the free energy yields in the immediate vicinity of $T_C$
\[
\lambda (T)\sim (T/E_F)^{4/3}
\]
Thus $\lambda $ dominates over $2I\chi $ in the temperature dependence of
the inverse susceptibility $\chi ^{-1}(T)$. With increasing $T$ the
dependence $\chi ^{-1}(T)$ becomes practically linear. It should be noted
that the Curie-Weiss behaviour in the spin-fluctuation approach has a quite
different origin in comparison with the Heisenberg model. The Curie-Weiss
law holds approximately not only for weakly ferromagnetic, but also for
almost ferromagnetic metals where
\[
0<1-2I\chi _p(0)\ll 1
\]
e.g., for Pd, HfZr$_2$. The Curie constant is determined by the electron
structure near the Fermi level only and, unlike the Heisenberg magnets, does
not depend on the saturation magnetization.

At $0<2I\chi _p(0)-1\ll 1$ fluctuations result in a considerable decrease of
the Curie temperature in comparison with the Stoner theory:
$$
T_C\sim \left[ 2I\chi _p(0)-1\right] ^{3/4}E_F
\eqno{(4.62)}
$$
Near $T_C$ one has
$$
M(T)\sim \left( T_C^{4/3}-T^{4/3}\right) ^{1/2}
\eqno{(4.63)}
$$
Corresponding results for a weak itinerant antiferromagnet have the form
$$
T_N\sim \left[ 2I\chi _{\mathbf{Q}}-1\right] ^{2/3},\,\,\,\,M_{\mathbf{Q}%
}(T)\sim (T_N^{3/2}-T^{3/2})^{1/2}
\eqno{(4.64)}
$$
Here $\chi _{\mathbf{Q}}$ and $M_{\mathbf{Q}}$ are the non-uniform
(staggered) susceptibility of the non-interacting system and the sublattice
magnetization, which correspond to the wavevector of the antiferromagnetic
structure $\mathbf{Q}$, $2I\chi _{\mathbf{Q}}-1\ll 1.$ The electronic specific
heat of weakly and almost ferromagnetic metals is substantially enhanced by
spin fluctuations:
$$
C(T)\sim T\ln (\min [|1-IN(E_F)|,\frac{E_F}T])
\eqno{(4.65)}
$$
(4.65) (see also Appendix G). In the case of a weak antiferromagnet, the
logarithmic factor is absent, and one has for the spin-fluctuation
contribtion [26]
$$
\delta C(T)\sim |1-2I\chi _{_{\mathbf{Q}}}|^{1/2}T
\eqno{(4.66)}
$$

Thus main difficulties of the Stoner theory are eliminated for weak
itinerant magnets. A number of attempts to generalize the spin-fluctuation
approach on strong itinerant ferromagnets (e.g., transition metals) were
performed [26,297,300-302]. In particular, an interpolation theory within
the path integral method is considered in [26]. An important shortcoming of
such theories is that spin-fluctuation effects are taken into account for
finite-temperature effects only, but the ground state is described by the
Stoner theory. This approach is inapplicable in the case of strong
correlations which is discussed in Sect.4.6.

Generally speaking, equations of spin-fluctuation theories are rather
complicated and require numerical calculations. By these reasons, simple
approaches, which provide analytical expressions, are of interest. A version
of such approach may be based on introducing the effective temperature $%
T_{eff}$ which is related to the effect of smearing the electron densty of
states by spin fluctuations [303]. This quantity may be substituted into
Fermi distribution functions at calculating thermodynamical averages.
Consider the local approximation for spin fluctuations [300-303]. For a
fivefold degenerate d-band the average square of fluctuation amplitude in
the paramagnetic region is estimated as
$$
v^2=\frac{3UT}{10(1-2U\chi _0)}
\eqno{(4.67)}
$$
where $U$ is the on-site Hubbard repulsion,
$$
\chi _0=-\frac 1{2\pi }\int dEf(E)\Im \left[ G(E)\right] ^2
\eqno{(4.68)}
$$
is the local susceptibility, the quantity
$$
G(E)=\int dE^{\prime }\frac{N(E^{\prime })}{E-E^{\prime }-\Sigma (E)}
\eqno{(4.69)}
$$
being the site-digonal Green's function, $\Sigma (E)$ the self-energy. Note
that, unlike the Stoner criterion $2U\chi _p>1$, the condition $2U\chi _0>1$
which corresponds to the formation of local magnetic moments (or to
occurrence of the Hubbard splitting, which means a reconstruction of the
ground state) is usually not satisfied in such approaches.

The effective smeared density of states is given by
$$
\widetilde{N}(E)=-\frac 1\pi \int \frac {dE^{\prime }N(E^{\prime })
\Im \Sigma (E)}{\left[ E-E^{\prime }-\Re (E)\right] ^2
+\left[\Im \Sigma (E)\right] ^2}
\eqno{(4.70)}
$$
To lowest order in fluctuations
$$
\Im \Sigma (E)=-\pi N(E)v^2
\eqno{(4.71)}
$$
so that the fluctuations result in the Lorentz smearing. At calculating the
integrals which include the Fermi function, this smearing may be taking into
account by the effective smearing the Fermi distribution
$$
\frac{\partial f(E)}{\partial E}=-\frac 1\pi \int dE^{\prime }\frac{\partial
f(E^{\prime })}{\partial E^{\prime }}\frac{\Im \Sigma (E^{\prime })}{%
\left[ E^{\prime }-E-\Re \Sigma (E^{\prime })\right] ^2+\left[ \Im
\Sigma (E^{\prime })\right] ^2}
\eqno{(4.72)}
$$
Using near the Fermi energy the linear approximation
$$
\Re \Sigma (E)=\Sigma ^{\prime }E,\qquad
f(E)\simeq \frac 12-\frac E{4T}
\eqno{(4.73)}
$$
\thinspace \thinspace \thinspace \thinspace we obtain the simple analytical
result
$$
\frac{\partial \widetilde{f}(E)}{\partial E} =-\frac{\kappa ^{-1}}{4T}
=-\frac 1{4T_{eff}},
\eqno{(4.74)}
$$
$$
\kappa  =\left[ \frac 2\pi (1-\Sigma ^{\prime })^{-1}\tan ^{-1}\frac{
2T(1-\Sigma ^{\prime })}{\pi v^2N(E_F)}\right] ^{-1}
$$
The effective temperature $T=\kappa T$ defined by (4.74) may be considerably
larger than $T$. Then the Curie temperature $T=T_S/k_B$ is appreciably
renormalized in comparison with the Stoner value $T_S$ . The effect of
lowering the Curie temperature becomes more appreciable in the presence of
DOS singularities.

Another approach based on introducing a spin-fluctuation temperature was
developed by Mohn and Wohlfarth [304]. Following to [305], they used the
Ginzburg-Landau expansion for the inverse susceptibility
$$
\frac HM=A+BM^2+B(3\langle m_{\parallel }^2\rangle +2\langle m_{\perp
}^2\rangle )
\eqno{(4.75)}
$$
where, as well as in the Stoner theory,
$$
A=-\frac 1{2\chi (0)}(1-\frac{T^2}{T_s^2})\,\,\,,\,\,\,B=\frac 1{2\chi
(0)M_0^2}
\eqno{(4.76)}
$$
and $\chi (0)$ is the enhanced susceptibility given by
$$
\chi (0)^{-1}=N_{\uparrow }^{-1}(E_F)+N_{\downarrow }^{-1}(E_F)-\frac I2
\eqno{(4.77)}
$$
The equilibrium magnetic moment is determined as
\[
M_0^2=|A/B|_{T=0}
\]
and are the mean suare values of the longitudinal and transverse local
fluctuating moments (the coefficients in (4.75) are determined by the
condition of vanishing of the scalar products $\mathbf{(Mm)}$ in the free
energy). A renormalization of the coefficients $A$ and $B$ owing to the last
term in (4.75) was considered by expanding in powers of magnetization
$$
B\left( 3\langle m_{\parallel }^2\rangle +2\langle m_{\perp }^2\rangle
\right) =a_1-a_2M^2+a_3M^4
\eqno{(4.78)}
$$
The expression for $a_1$ may be expessed in terms of the spin-fluctuation
susceptibility
$$
a_1(T)=(1-t_C^2)\chi _{sf}^{-1}(T)\,,\,\,\,\,\,\,\,\,\,\,t_C\equiv \frac{T_C%
}{T_S}
\eqno{(4.79)}
$$
where $T_C$ is the true value of the Curie temperature, renomalized by spin
fluctuations. At $T=T_C$ we have
$$
\chi _{sf}(T_C)=\frac{\langle m^2\rangle }{3T_C},
\eqno{(4.80)}
$$
$$
\langle m^2\rangle  =\frac 13\left( 3\langle m_{\parallel }^2\rangle
+2\langle m_{\perp }^2\rangle \right)
$$
The amplitude of the fluctuating moment at $T_C$ is given by
$$
\langle m^2\rangle _{T_C}=\frac 35M_0^2(1-t_C^2)
\eqno{(4.81)}
$$
so that
$$
a_1(T_C)=5T_C/M_0^2
\eqno{(4.82)}
$$
The equation for $T_C,\,A^{\prime }=A+a_1=0$, yields
$$
\frac{T_C^2}{T_S^2}+\frac{T_C}{T_{sf}}-1=0
\eqno{(4.83)}
$$
where
$$
T_{sf}=M_0^2/10\chi (0)
\eqno{(4.84)}
$$
is a characteristic temperature describing the influence of fluctuations.

The expression (4.83) differs somewhat from the corresponding result of the
more rigorous theory for weak itinerant ferromagnets by Lonzarich and
Taillefer [305]
$$
\left( \frac{T_C}{T_S}\right) ^2+\left( \frac{T_C}{T_{sf}}\right) ^{4/3}-1=0
\eqno{(4.85)}
$$
where the $T^{4/3}$-dependence is due to influence of a finite cutoff
wavevector.

The paramagnetic susceptibility within the approach [304] is given by
$$
2\chi (0)/\chi (T)=t_C^2(T/T_C)^2+(1-t_C^2)(T/T_C)-1
\eqno{(4.86)}
$$
The curves $\chi (T)$ interpolate between a Curie-Weiss law (spin
fluctuations predominate, $t_C=0$) and the Stoner square law
(single-particle excitations play the main role). The Curie constant is
determined entirely by the ferromagnetic susceptibility $\chi (0)$.

Determining $T_S$ and $\chi (0)$ from band calculations, Mohn and Wohlfarth
calculated the values of Curie temperature for iron group metals (Table 4.5)
in a good agreement with experiment (however, a considerable uncertainty of
the parameters used should be noted). Calculations for some Y-Co and Y-Fe
compounds and borides were also performed. The results turn out to be close
to the universal curve which shows the dependence of $T_C/T_{sf}$ as a
function of $T_S/T_{sf}$ (Fig.4.9).

The results enable one to estimate the role of fluctuation effects in
various metals. The value of $t_C$ may be a quantitative measure of these
effects: according to [304], we have for $t_C>0.5$ the Stoner systems and
for $t_C<0.5$ the fluctuation systems. Large values of $M_0$ suppress
fluctuations owing to strong molecular field, and the high-field
susceptiblity $\chi (0)$ works in the opposite direction by lowering $T$ and
favouring magnetic moment fluctuations at low temperatures. Although the
value of M in iron exceeds by 15 times that in nickel, the value of $\chi
(0) $ in Ni is smaller since it is a ``stronger'' ferromagnet which is close
to the saturation limit. Therefore fluctuation effects in both metals turn
out to be comparable. At the same time, the large value of $M_0$ in cobalt
is sufficient to suppress fluctuation effects, so that T is even larger than
$T_S$ , and $T_C$ is determined mainly by single-particle excitations. It
should be noted that these conclusions differ from the resuts of [291],
according to which fluctuations play an important role in Fe and Co, whereas
Ni is satisfactorily described by the Stoner theory with account of DOS
singlularities.

\section{Electronic structure and properties of half-metallic ferromagnets}

Opposite to weak itinerant ferromagnets is the limiting case of ``strong''
ferromagnets with a large spin splitting. In the old Stoner theory
(Sect.4.3), a strongly ferromagnetic solution, where the spin splitting
exceeds the Fermi energy and one spin subband is empty or completely filled,
was considered. It was believed that such a situation (for the hole states)
corresponds to ferromagnetic nickel. However, modern band calculations [24]
within the LSDF approach disproved this assumption (the spin up density of
states at the Fermi level turned out to be small but finite, Fig.2.25).

At the same time, band calculations did lead to discovery of real magnets,
which are similar to strong Stoner ferromagnets. Calculation by de Groot et
al of the band structure for the Heusler alloy NiMnSb [306], PtMnSb
[306-308] with the $C_{1b}$ (MgAgAs) structure demonstrated that the Fermi
level for one of spin projections lies in the energy gap. Since these
systems behave for one of spin projections as insulators, they were called
``half-metallic ferromagnets'' (HMF). Later a similar picture was obtained
for CoMnSb [309], ferrimagnet FeMnSb [310], antiferromagnet CrMnSb [308].
Band structure calculations for a large group of ferro- and
antiferromagnetic Heusler alloys from another series T$_2$MnZ (T = Co, Ni,
Cu, Pd) with the $L2_1$ structure demonstrated that a state, which is close
to HMF ($N(E)$ is practically zero), takes place in systems Co$_2$MnZ with Z
= Al, Sn [311] and Z = Ga, Si, Ge [312]. (In the $L2_1$ structure, all the
four sublattices in the fcc lattice are filled by the atoms T, Mn and Z,
whereas in the the $C_{1b}$ structure some of the positions are empty, so
that the symmetry lowers to tetraedric one). Besides that, a half-metallic
state was found in band calculations of CrO$_2$ (rutile structure)
[313,314], UNiSn ($C_{1b}$ structure) [315,316], Fe$_3$O$_4$ [317].

The situation of strongly different spin up and spin down states, which is
realized in HMF, is of interest for the general theory of itinerant
magnetism [318]. The scheme of the ``half-metallic'' state formation in
Heusler alloys may be described as follows [306,308,311,316]. At neglecting
hybridization of T- and Z-atom states, d-band of manganese for above
structures is characterized by a wide energy gap between bonding and
antibonding states. Due to strong intraatomic (Hund) exchange for manganese
ions, in the ferromagnetic state spin up and down subbands are moved apart
considerably. One of spin subbands comes closely to ligand p-band, and
therefore the corresponding gap is partially or completely smeared by p-d
hybridization. Other subband retains the gap, and the Fermi level may find
itself at the latter, The energy gap in other subband retains and may
coincide under certain conditions with the Fermi level, which just yields
the HMF state. For the $C_{1b}$ structure we have the true gap, and for the $%
L2_1$ structure a deep pseudogap. This is connected with a considerable
change in the character of the p-d hybridization (especially between p and $%
t_{_2g}$ states) in the absence of inversion centre, as it takes place for
the $C_{1b}$ structure. Thus the latter structure is more favorable for the
HMF state.

According to [324], similar factors are responsible for the gap in the
partial density of states for one of manganese position (Mn(I)) in the
compound Mn$_4$N, the structure of which is obtained from the T$_2$MnZ
structure by removing some atoms. A qualitatively similar mechanism, which
is based on the strong Hund exchange and hybridization between d-states of
chromium with p-states of oxygen, is considered in [313] for CrO$_2$. As
discussed in [311], stability of the ferromagnetic state itself is a
consequence of difference in the p-d hybridization for states with opposite
spin projections. To describe such a situation, Kuebler et al [311] have
introduced the term ``covalent magnetism'' and stressed the distinction from
the picture of the spectrum in the Stoner model where the densities of
states with oposite spin projections differ by the constant spin splitting.
The results of band structure calculations are shown in Figs.4.10-4.13.

The interest in half-metallic ferromagnets (HMF) was connected to the first
place with their unique magnetooptical properties [307,308], which are
intimately related to the electronic structure near the Fermi level (absence
of states with one spin projection), which results in strong asymmetry of
optical transitions (see the discussion in Sect. 5.7.3). In particular,
large magnetooptical Kerr effect was observed in PtMnSb (for other compounds
of this series the effect is smaller because of smallness of relativistic
interactions for light atoms).

Besides that, the HMF are important in connection with the problem of
obtaining the large saturation magnetic momentum, since, evidently, their
electron spectrum is favourable for maximum spin polarization (further
increase of the spin splitting in the half-metallic state does not result in
an increase of magnetic moment). In fact, the electronic structure, which is
reminiscent of the half-metallic one (a deep minimum of DOS at $E_F$ for $%
\sigma =$ $\downarrow $) was found in the system of alloys Fe-Co [319] and
the systems R$_2$Fe$_{17}$, R$_2$Fe$_{14}$B with the record value of $M_0$
[320,321]. Such a minimum is typical for systems with well-defined local
magnetic moments and takes place also in pure iron (Fig.2.24).
Well-pronounced minima for the states with $\sigma =\downarrow $ occur in
the compounds RCo$_5$ (R = Y, Sm, Gd) [322]. A comparison of magnetic
properties of a large series of alloys Y$_n$Co$_m$, Y$_n$Fe$_m$ with their
electronic structure, calculated by the recursion method, was performed in
[323]; the state, close to HMF, was obtained for YCo$_5$, YCo$_7$, Y$_2$Co$%
_{17}$. The small partial values of $N_{\uparrow }(E_F)$ were also found in
Mn$_4$N, Fe$_4$N for the crystallographic Mn(I) position [324-326]. 

Recently the interest in the half-metallic ferromagnetism has been greatly
increased in connection with the discovery of giant magnetoresistivity
in ferromagnetic manganites [688].

Magnetic properties of some half-metallic Heusler alloys are presented in
the Table 4.3. These compounds possess large ground state moment and high
Curie temperature. Strong ferromagnetism of Heusler alloys is mainly due to
local moments of well-separated Mn atoms. An interesting feature is that the
Rhodes-Wolfarth ratio $p_C/p_s$ exceeds considerably unity. Moreover, the
effective moment in the paramagnetic state, determined from the paramagnetic
susceptibility, decreases appreciably with temperature.

This behaviour may be explained by that the change of electronic structure
in half-metallic magnets at destruction of the magnetic ordering is
especially strong - the gap in the electron spectrum vanishes. The
temperature dependence of magnetic moment in the paramagnetic state may be
due to short-range magnetic order (local densities of states are similar to
those in the ferromagnetic state). From the many-electron point of view, the
decrease of the local moment with increasing temperature is connected with
the absence of corrections to ground state magnetization of the type (G.49).
However, such corrections do occur at high temperatures.

From the theoretical point of view, HMF are also characterized, due to
quasimomentum conservation law, by absence of decay of a spin wave into a
pair electron-hole with opposite spins (Stoner excitations). Therefore spin
waves are well defined in the whole Brillouin zone, similar to Heisenberg
ferromagnets and degenerate ferromagnetic semiconductors. Thus effects of
electron-magnon interactions (spin-polaron effects) are not masked, unlike
usual itinerant ferromagnets, by the Stoner excitations and may be studied
in a pure form. At present, experimental data on neutron scattering in the
Heusler alloys Pd$_2$MnSn, Ni$_2$MnSn [327] and Cu$_2$MnAl [328] are
available. Spin waves turn out to be well defined over the whole Brillouin
zone, which makes up, according to [26], a criterion of validity of
localized moment model.

From the point of view of band theory, the smallness of magnon damping may
be explained by that the partial Mn-atom spin-up density of states is small
since the corresponding subband is almost empty [311]. One may expect that
the damping will be still smaller provided that the Fermi level for one of
spin projections lies in the energy gap. Thus a purposeful investigation of
spin-wave damping at comparing results for various Heusler alloys from the T$%
_2$MnZ and TMnZ series is of great interest for verification of the theory.

As discussed above, the p-d hybridization plays an important role in the
Heusler alloys, so that they should be described within a generalized
Anderson-lattice model (Sect.6.7). However, this model is reduced in
limiting cases to the Hubbard or s-d exchange models. Within the latter
models, HMF are described as saturated ferromagnets where the spin splitting
$\Delta$ exceeds the Fermi energy.

Consider the picture of density of states of a HMF within the s-d exchange
model with account of correlation effects [329]. Writing down the expansion
of the Dyson equation (G.30) we obtain
$$
N_\sigma (E) =-\frac 1\pi \Im \sum_{\mathbf{k}}G_{\mathbf{k}\sigma }(E)
=\sum_{\mathbf{k}}\delta (E-t_{\mathbf{k}\sigma })
$$
$$
-\sum_{\mathbf{k}}\delta ^{\prime }(E-t_{\mathbf{k}\sigma })\Re
\Sigma _{\mathbf{k}\sigma }(E)-\frac 1\pi \sum_{\mathbf{k}}\frac{\Im
\Sigma _{\mathbf{k}\sigma }(E)}{(E-t_{\mathbf{k}\sigma })^2}
\eqno{(4.87)}
$$
The second term in the right-hand side of (4.87) describes renormalization
of quasiparticle energies. The third term, which arises from the branch cut
of the self-energy $\Sigma _{\mathbf{k}\sigma }(E)$, describes the
incoherent (non-quasiparticle) contribution owing to scattering by magnons.
One can see that this does not vanish in the energy region, corresponding to
the ``alien'' spin subband with the opposite projection $-\sigma $. The
picture of density of states for the empty conduction band is shown in
Fig.4.14. The $T^{3/2}$-dependence of the magnon contribution to the residue
(G.53), which follows from (G.34), i.e. of the effective electron mass in
the lower spin subband, and an increase with temperature of the incoherent
tail from the upper spin subband result in a strong temperature dependence
of partial densities of states $N(E)$, the corrections being of opposite
sign. The corresponding behaviour of conduction electron spin polarization $%
P(T)\simeq \langle S^z\rangle $ (see (G.39)) is confirmed by experimental
data on field emission from ferromagnetic semiconductors [330] and transport
properties of half-metallic Heusler alloys [331].

The picture of $N(E)$ near the Fermi level in HMF (or degenerate
semiconductor) turns out to be also non-trivial. If we neglected magnon
frequencies in the denominators of (G.34), the partial density of incoherent
states should occur by a jump above or below the Fermi energy for $I>0$ and $%
I<0$ respectively owing to Fermi distribution functions. An account of
finite magnon frequencies $\omega _{\mathbf{q}}=Dq^2$ ($D$ is the spin
stiffness constant) leads to smearing of these singularities on the energy
interval $\omega _{\max }\ll E_F$ (Figs.4.15, 4.16), $N(E_F)$ being equal to
zero. For $|E-E_F|\ll \omega _{\max }$ we obtain [329,332]
$$
\frac{N_{-\alpha }(E)}{N_\alpha (E)}=\frac 1{2S}\left| \frac{E-E_F}{\omega
_{\max }}\right| ^{3/2}\theta (\alpha (E-E_F)),\qquad
\alpha =\mathrm{sign}\,I
\eqno{(4.88)}
$$
With increasing $|E-E_F|$, $N_{-\alpha }/N_\alpha $ tends to a constant value
which is of order of $I^2$ within perturbation theory. In the strong
coupling limit where $|I|\rightarrow \infty $ we have
$$
\frac{N_{-\alpha }(E)}{N_\alpha (E)}=\frac 1{2S}\theta (\alpha
(E-E_F)),|E-E_F|\gg \omega _{\max }  \eqno{(4.89)}
$$
Similar calculations for the Hubbard ferromagnet with strong correlations
and electron concentration $n>1$ yield [333]
$$
N_{\uparrow }(E) =\sum_{\mathbf{k}\sigma }f(t_{\mathbf{k+q}})\delta (E-t_{%
\mathbf{k+q}}+\omega _{\mathbf{q}})
$$
$$
=\left\{
\begin{array}{ll}
N_{\downarrow }(E) & ,\qquad E_F-E\gg \omega _{\max } \\
0 & ,\qquad E>E_F
\end{array}
\right.  
\eqno{(4.90)}
$$
(cf. (J.22)). This result has a simple physical meaning. Since the current
carriers are spinless doubles (doubly occupied sites), the electrons with
spins up and down may be picked up with an equal probability from the states
below the Fermi level of doubles. On the other hand, according to the Pauli
principle, only the spin down electrons may be added in the singly occupied
states in the saturated ferromagnet.

Thus the spin polarization $P(E)$ in strong itinerant ferromagnets changes
sharply near $E_F$. In the case of an antiferromagnetic s-d exchange (or in
a Hubbard ferromagnet) there exist occupied non-quasiparticle states which
may result in electron depolarization in photoemission experiments. At the
same time, empty non-quasiparticle states may be observed by inverse
photoemission. Since emission experiments have not too high energy
resolution, appreciable deviatons from 100\% spin polarization should be
observed. Data on photemission in the HMF NiMnSb [334] have yielded the
polarization about 50\%.

The incoherent contribution to the density of states should also result in
peculiar contributions to thermodynamic and transport properties. The
non-quasiparticle terms in electronic specific heat are discussed in
Appendix G (see (G.65)-(G.67)). Since the impurity resistivity is determined
by the dependence $N(E)$ near the Fermi level, it contains also
non-quasiparticle contributions (Sect.5.3). Asymmetry of $N(E)$ should lead
to an appreciable contribution to thermoelectric power.

Peculiarities of electronic structure of HMF should be clearly observed at
investigation of nuclear magnetic relaxation. The $T$-linear Korringa
contribution to the longitudinal nuclear relaxation rate, which dominates as
a rule for ferromagnetic metals, is determined by the transverse spin
susceptibity and given by the Moriya formula [335,336]
$$
1/T_1 =-\frac{A^2T}{2\pi \omega _n}\Im \sum_{\mathbf{q}}\langle
\langle S_{\mathbf{q}}^{+}|S_{-\mathbf{q}}^{-}\rangle \rangle _\omega
$$
$$
=\pi A^2TFN_{\uparrow }(E_F)N_{\downarrow }(E_F)  
\eqno{(4.91)} 
$$
where $\omega _n$ is the MNR frequency, $\gamma _n$ is the nuclear
hyromagnetic ratio, $A$ is the hyperfine interaction constant, $F$ is the
exchange enhancement factor. (A detailed discussion and comparison of
different contributions to $1/T_1,$ including orbital ones, is given in
paper [335].) At the same time, the transverse relaxation rate contains a
contribution from the longitudinal susceptibility:
$$
1/T_2=1/2T_1+\frac \pi 2A^{\prime 2}TF^{\prime }[N_{\uparrow
}^2(E_F)+N_{\downarrow }^2(E_F)]  \eqno{(4.92)}
$$
(generally speaking, $A^{\prime }$ and $F^{\prime }$ differ from $A$ and $F$%
). Neglecting for simplicity the exchange splitting and spin dependence of
hyperfine interaction matrix elements, we derive
$$
\frac{1/T_2}{1/T_1}=\frac{[N_{\uparrow }^2(E_F)+N_{\downarrow }^2(E_F)]}{%
4N_{\uparrow }(E_F)N_{\downarrow }(E_F)}>1  \eqno{(4.93)}
$$
One can see that $1/T_1=1/T_2$ in paramagnetic metals, but the ratio (4.93)
should considerably differ from unity in itinerant ferromagnets with an
appreciable dependence $N(E)$. In fact, $1/T_2$ exceeds by several times $%
1/T_1$ in iron and nickel [336].

In HMF the contribution (4.91) vanishes, which is connected with absence of
processes of magnon decay into Stoner excitations, whereas $1/T_2$, which is
determined by electron transitions without spin flips, should have usual
behaviour. To obtain the temperature dependence of $1/T_1$ in HMF one has to
consider the contribution of two-magnon processes. Using (G.24) we obtain
[337,338]
$$
1/T_1^{(2)} =\frac{A^2TS}{\pi \omega _n}\sum_{\mathbf{q}}\frac{\gamma _{%
\mathbf{q}}^{(2)}(\omega _n)}{\omega _{\mathbf{q}}^2}  
$$
$$
=\frac{12\pi ^{1/2}}S\left( \frac{v_0}{16\pi ^2}\right) ^2\zeta (\frac 32)%
\frac{T^{5/2}}{D^{7/2}} (k_{F \uparrow}^2+k_{F \downarrow}^2) 
\eqno{(4.94)}
$$
Qualitatively, the dependence $T^{5/2}$ may be interpreted as the Korringa
law with the value $N_{\downarrow }(E_F)$ being replaced with the
``thermal'' value of density of non-quasiparticle states, which is
proportional to $T^{3/2}$. Experimentally, strong deviations from the linear
Korringa law were observed at measuring $1/T_1$ in the HMF NiMnSb [339]. At
not too low temperatures $T>250$K ($T_C=750$K) the dependence of the form
\[
1/T_1(T)=aT+bT^{3.8}
\]
was obtained.

In the above-discussed ferrimagnet Mn$_4$N with $T_C=750$K the nuclear
magnetic relaxation was investigated for the Mn(I) position (a narrow NMR
line was obtained only for this position) [340]. At low temperatures $T<77$K
the behaviour $1/T_1(T)$ is linear, and at higher temperatures a square law
holds.

\section{Magnetism of highly-correlated d-systems}

The most difficult case for standard approaches in the itinerant electron
magnetism theory (band calculations, spin-fluctuation theories) is that of
systems where strong interelectron correlations lead to a radical
reconstruction of the electron spectrum - formation of the Hubbard's
subbands. Examples of such systems are oxides and sulphides of transition
metals with a large energy gap, e.g. MeO (Me = Ni, Co, Mn), NiS$_2$ [25],
basis systems for copper-oxide high-$T_c$ superconductors La$_2$CuO$_4$ and
YBa$_2$Cu$_3$O$_6$. At low temperatures these compounds are
antiferromagnetic, so that the gap may be described as the Slater gap owing
to AFM ordering. However, the gap retains in the paramagnetic region and has
therefore a Mott-Hubbard nature.

Standard band calculations within the density functional method [341] did
not yield satisfactory explanation of insulating properties of the MeO
compounds. The reasonable value of gap in MeO and CuO$_2$-planes may be
obtained by using special calculation methods which take into account in a
sense the Hubbard correlations (e.g., the self-interaction correction (SIC)
approach [77,342] which considers the attraction of an electron to the field
of the hole formed, Sect.2.3). First-principle calculations [343] yield the
reasonable values of the Hubbard parameter in TM oxides, $U=6\div 10$eV.

The Hubbard splitting is apparently present also in some metallic
ferromagnets, e.g., the solid solution Fe$_{1-x}$Co$_x$S$_2$ which has the
pyrite structure, CrO$_2$ [26]. For the latter compound, corresponding
direct optical data are present [344]. The spontaneous spin splitting above
the Curie point, which is observed in iron group metals, may be also
interpreted as related to the Hubbard subbands.

To treat the problem of ferromagnetic ordering in narrow bands from
experimental point of view, we discuss the system Fe$_{1-x}$Co$_x$S$_2$.
According to [345], its electronic structure is rather simple: all the
electrons, responsible both for conductivity and magnetism, belong to the
same narrow $e_g$-band. CoS$_2$ is a ferromagnetic metal with strong
correlations, and FeS$_2$ is a diamagnetic insulator. Recent band
calculations of these compounds [346] demonstrate that CoS$_2$ has an almost
half-metallic ferromagnet structure.

Experimental investigations of magnetic properties of Fe$_{1-x}$Co$_x$S$_2$
have been performed in [347]. The most salient feature is the onset of
ferromagnetism at surprisingly small electron concentrations $n=x<0.05$. The
magnetic moment equals to 1$\mu _B$ in the wide concentration region $%
0.15<n<0.95$, and magnetic state is non-saturated for $n<0.15$. However,
unlike usual weak itinerant ferromagnets, there is no indications of an
exchange enhancement of the Pauli susceptibility above $T_c$, and the
Curie-Weiss law works well at arbitrary electron concentrations, the Curie
constant being proportional to $n$. Such a behaviour cannot be explained
within one-electron Stoner-like approaches, e.g. the $T$-matrix
approximation [348], and demonstrates an important role of local magnetic
moments (LMM). The inconsistency of one-electron approach for
highly-correlated systems is demonstrated in Appendix H.

The problem of the description of LMM is a crucial point of the strong
itinerant ferromagnetism theory. In the spin-fluctuation theories [26], LMM
are introduced essentially {\itshape{ad\ hoc}} (e.g., the static approximation 
in the
path integral, which corresponds to replacement of the translation-invariant
system by the disordered one with random magnetic fields). On the other
hand, the many-electron (atomic) picture describes LMM naturally. The role
of strong electron correlations in the formation of LMM can be qualitatively
demonstrated in the following way. If the Hubbard intrasite repulsion $U$ is
large enough, the electron spectrum contains the Hubbard subbands of singly
and doubly occupied states. At $n<1$, the number of doubles is small and
tends to zero at $U\rightarrow \infty $. Then the singly occupied states
make up LMM, and the empty site (holes) are the current carriers.

The simplest way to describe the formation of Hubbard subbands is
calculation of electron Green's function with the use of Hubbard atomic
representation (Appendix H). The formation of the Hubbard subbands
contradicts the Fermi-liquid picture, in particular the Luttinger theorem
about the conservation of the volume under the Fermi surface: each from two
Hubbard subbands, originating from a free-electron band, contains one
electron state per spin. Thus the energy spectrum of itinerant electron
systems with LMM, contrary to weak itinerant magnets, differs essentially
from that of a normal Fermi liquid. In the simplest Hubbard-I approximation,
magnetic ordering results in narrowing of spin subbands rather than in a
constant spin splitting (see (H.10)). Therefore the condition of
ferromagnetism in such systems should not coincide with the Stoner criterion
which corresponds to instability of the non-magnetic Fermi liquid with
respect to a small spin polarization. Below we discuss the problem of
ferromagnetism in strongly correlated Hubbard systems.

Rigorous investigation of ferromagnetism in the Hubbard model with $%
U\rightarrow \infty $ was performed by Nagaoka [349]. He proved that the
ground state for simple cubic and bcc lattices in the nearest-neighbour
approximation with electron number $N_e=N+1$ ($N$ is the number of lattice
sites) possesses maximum total spin, i.e. is saturated ferromagnetic. The
same statement holds for the fcc and hcp lattices with the transfer integral
$t<0$, $N_e=N+1$, or $t>0,N_e=N-1$. (For other sign combinations, the ground
state is more complicated because of divergence of the density of states at
the band edge.) The physical sense of the Nagaoka theorem is rather simple.
For $N_e=N,U=\infty $ each sites is singly occupied and the motion of
electrons is impossible, so that the energy of the system does not depend on
spin configuration. At introducing of an excess electron or hole, its
kinetic energy turns out to be minimum for the uniform ferromagnetic spin
alignment since this does not prevent their motion. It should be, however,
noted, that in fact the proof of the Nagaoka theorem uses non-trivial
topological consideration. In particular, it does not work in the
one-dimensional case where the dependence of the kinetic energy on spin
configurations is absent because of absence of closed trajectories [349].
Evidently, the picture of saturated ferromagnetism is preserved at small
finite concentrations of current carriers.

In the case of half-filled band ($N_e=N$), $|t|\ll U$ the ground state is
antiferromagnetic because of the Anderson's kinetic exchange interaction
(Appendix D). This interactions is due to the gain in the kinetic energy at
virtual transitions of an electron to a neighbour site, which are possible
provided that the electron at this site has an opposite spin directions. In
systems with finite $U$ and $N_e\neq N$, a competition beween ferro- and
antiferromagnetic ordering occurs (it should be noted that, as discussed in
Appendix D, the kinetic antiferromagnetic interaction emerges even in the
formal limit $U=\infty $ due to non-orthogonality corrections, 
see also Ref.[727]). As follows from the calculation of the spin-wave 
energy [349], ferromagnetism preserves provided that
$$
|t|/U<\alpha n  
\eqno{(4.95)}
$$
where the constant $\alpha $ $\sim 1$ depends on the lattice structure. At
the same time, antiferromagnetism is stable at $N_e=N$ only. It was supposed
in early papers that a canted magnetic structure is formed in the
intermediate region [350]. However, numerical calculations [351]
demonstrated that in fact a phase separation takes place into insulating
antiferromagnetic and metallic ferromagnetic regions, all the current
carriers being localized in the latter ones. Such a phenomenon seems to be
observed in some highly doped magnetic semiconductors [352].

The problem of electron and magnon spectrum of a Hubbard ferromagnet with
nearly half-filled band may be considered rigorously in the spin-wave
temperature region (Appendix J). To obtain a simple interpolation scheme for
arbitrary current carrier concentrations and temperatures, we use the
simplest Hamiltonian of the Hubbard model with $U\rightarrow \infty $, $n<1$
in the many-electron representation with inclusion of the external magnetic
field $H$:
$$
\mathcal{H}=\sum_{\mathbf{k}\sigma }\varepsilon _{\mathbf{k}}X_{-\mathbf{k}%
}(0\sigma )X_{\mathbf{k}}(\sigma 0)-\frac H2\sum_i[X_i(++)-X_i(--)]
\eqno{(4.96)}
$$
with $\varepsilon _{\mathbf{k}}=-t_{\mathbf{k}}$. This Hamiltonian describes
the motion of current carriers (holes) on the background of local magnetic
moments - singly occupied sites. Following to [353], we consider the dynamic
magnetic susceptibility (H.14). Its calculation (Appendix H) yields
$$
G_{\mathbf{q}}(\omega ) =\left( 2\langle S^z\rangle +\sum_{\mathbf{k}}%
\frac{(\varepsilon _{\mathbf{k-q}}-\varepsilon _{\mathbf{k}})(n_{\mathbf{%
k\uparrow }}-n_{\mathbf{k-q\downarrow }})}{\omega -H-E_{\mathbf{k\uparrow }%
}+E_{\mathbf{k-q\downarrow }}}\right)  
$$
$$
\times \left( \omega -H-\sum_{\mathbf{k}}\frac{(\varepsilon _{\mathbf{k-q}%
}-\varepsilon _{\mathbf{k}})(\varepsilon _{\mathbf{k}}n_{\mathbf{k\uparrow }%
}-\varepsilon _{\mathbf{k-q}}n_{\mathbf{k-q\downarrow }})}{\omega -H-E_{%
\mathbf{k\uparrow }}+E_{\mathbf{k-q\downarrow }}}\right) ^{-1} 
\eqno{(4.97)} 
$$
where the Hubbard-I approximation energies $E_{\mathbf{k}\sigma }$ and the
occupation numbers $n_{\mathbf{k}\sigma }$ are given by (H.12).

At calculating the static magnetic susceptibility $\chi $, one has to treat
carefully the limits $H\rightarrow 0$, $\omega \rightarrow 0$, $q\rightarrow 0$
because of non-ergodicity of the ferromagnetic ground state. Indeed, simple
putting $H=\omega =q=0$, which was made in some papers [354-356], yielded
only the Pauli susceptibility rather than the Curie-Weiss behavior, which is
physically unreasonable. To avoid the loss of the Curie-Weiss contribution
from local moments, we apply the approach, employed by Tyablikov [357] for
the Heisenberg model (see Appendix E). Using the sum rules
\[
n_\sigma =\frac{1-c}2+\sigma \langle S^z\rangle =\sum_{\mathbf{q}}\langle
X_{-\mathbf{q}}(\sigma ,-\sigma )X_{\mathbf{q}}(-\sigma ,\sigma )\rangle
\]
and the spectral representation (E.18) we derive the equation for
magnetization
$$
\langle S^z\rangle =\frac{1-c}2+\frac 1\pi \int_{-\infty }^\infty d\omega
N_B(\omega )\sum_{\mathbf{q}}G_{\mathbf{q}}(\omega )  \eqno{(4.98)}
$$
For small concentration of holes $c\ll 1$ one obtains, in agreement with the
consideration by Nagaoka [349], the saturated ferromagnetism with
$$
\langle S^z\rangle =\frac{1-c}2-\sum_{\mathbf{p}}N_B(\omega _{\mathbf{p}})
\eqno{(4.99)} 
$$
$$
\omega _{\mathbf{p}} =\sum_{\mathbf{k}}(\varepsilon _{\mathbf{k-p}%
}-\varepsilon _{\mathbf{k}})f(\varepsilon _{\mathbf{k}})  
$$
The magnon spectrum in (4.99) coincides with the exact result to lowest
order in the inverse nearest-neighbour number $1/z$ (Appendix J).

The equation (4.98) can be simplified under the condition $\langle
S^z\rangle \ll 1$, which holds both in the paramagnetic region ($\langle
S^z\rangle =\chi H$, $H\rightarrow 0$) and for $n\ll 1$ at arbitrary
temperatures. Expansion of denominator and numerator of (4.9) in $\langle
S^z\rangle $ and $H$ has the form
$$
G_{\mathbf{q}}(\omega )=\frac{\omega A_{\mathbf{q}\omega }+\langle
S^z\rangle B_{\mathbf{q}\omega }+HC_{\mathbf{q}\omega }}{\omega -\langle
S^z\rangle D_{\mathbf{q}\omega }-HP_{\mathbf{q}\omega }}  \eqno{(4.100)}
$$
Here the quantity
$$
A_{\mathbf{q}\omega }=\sum_{\mathbf{k}}\frac{f_{\mathbf{k-q}}-f_{\mathbf{k}}%
}{\omega +E_{\mathbf{k-q}}-E_{\mathbf{k}}}\left( 1+\frac 2{1+c}\sum_{\mathbf{%
k}}\frac{E_{\mathbf{k-q}}f_{\mathbf{k-q}}-E_{\mathbf{k}}f_{\mathbf{k}}}{%
\omega +E_{\mathbf{k-q}}-E_{\mathbf{k}}}\right) ^{-1}  \eqno{(4.101)}
$$
determines the value of the effective magnetic moment, and
$$
D_{\mathbf{q}0} =-\sum_{\mathbf{k}}\frac{8E_{\mathbf{k}}}{(1+c)^2}\left(
E_{\mathbf{k}}\frac{\partial f_{\mathbf{k}}}{\partial E_{\mathbf{k}}}-E_{%
\mathbf{k+q}}\frac{f_{\mathbf{k-q}}-f_{\mathbf{k}}}{\omega +E_{\mathbf{k-q}%
}-E_{\mathbf{k}}}\right) 
$$
$$
\times \left( 1+\frac 2{1+c}\sum_{\mathbf{k}}\frac{E_{\mathbf{k-q}}f_{%
\mathbf{k-q}}-E_{\mathbf{k}}f_{\mathbf{k}}}{\omega +E_{\mathbf{k-q}}-E_{%
\mathbf{k}}}\right) ^{-1}  
$$
$$
\equiv J_0^{eff}-J_{\mathbf{q}}^{eff}  
\eqno{(4.102)}
$$
describes the effective exchange interaction owing to motion of current
carriers. Roughly speaking, this differs from the RKKY interaction by the
replacement of the s-d(f) exchange parameter to the transfer integral. Such
a replacement is characteristic for the narrow-band limit.

The ground state magnetization for small electron concentration $n\ll 1$ is
given by
$$
S_0=\frac n2=\frac 1\pi \int_0^\infty d\omega \Im \sum_{\mathbf{q}}A_{%
\mathbf{q}\omega }  
\eqno{(4.103)}
$$
The non-zero value of $S_0$ occurs because of the second term in brackets in
(4.11) and is formally small in the parameter $1/z$ (in the simplest ``Debye
model'' for the electron spectrum, $\varepsilon _{\mathbf{k}}=a+bk^2$, $k<k_D$, 
one obtains $S_0=n/8$ [353]). Thus we have for small n the non-saturated
ferromagnetic state, in agreement with the experimental data [347].
Calculation of the critical electron concentration for onset of
ferromagnetism apparently requires more advanced approximations.

At $T>T_C$, putting in (4.100) $\langle S^z\rangle =\chi H$ one obtains the
equation for the paramagnetic susceptility
$$
\frac{1-c}2+\frac 1\pi \int_0^\infty d\omega \cot \frac \omega {2T}\Im 
\sum_{\mathbf{q}}A_{\mathbf{q}\omega }  
$$
$$
=T\sum_{\mathbf{q}}\left( \frac{\chi B_{\mathbf{q}0}+C_{\mathbf{q}0}}{\chi
D_{\mathbf{q}0}+P_{\mathbf{q}0}}+A_{\mathbf{q}0}\right)  
\eqno{(4.104)}
$$
The Curie temperature is determined by the condition $\chi (T_C)=\infty $.
At small $n$ we have (cf.(E.25))
$$
T_C=\frac{S_0}2\left( D_{\mathbf{q}0}^{-1}\right) ^{-1}  \eqno{(4.105)}
$$
Expanding the right-hand side of (4.104) in $\chi $ at $T_C\ll T\ll E_F$ one
gets
$$
\chi =-\frac 14\sum_{\mathbf{k}}\frac{\partial f(\varepsilon _{\mathbf{k}})}{%
\partial \varepsilon _{\mathbf{k}}}+\frac C{T-\theta }  \eqno{(4.106)}
$$
where the first term corresponds to the Pauli contribution and the second
one to the Curie-Weiss contribution of LMM,
$$
C=\frac 12S_0,\qquad \theta =\sum_{\mathbf{q}}D_{\mathbf{q}0}>T_C  
\eqno{(4.107)}
$$
being the Curie constant and the paramagnetic Curie temperature. To lowest
order approximation in $n$,
$$
\theta \simeq T_C\simeq C\frac{v_0}{2\pi ^2}m^{*}k_F\varepsilon _{\max }^2
\eqno{(4.108)}
$$
with
\[
k_F=(3\pi ^2n)^{1/3},\qquad 
frac 1{m^{*}}=\left. \frac{\partial ^2\varepsilon _{\mathbf{k}}}
{\partial k^2}\right| _{\varepsilon _{\mathbf{k}}
=\varepsilon _{\max }}
\]
Thus the many-electron approach provides a simple derivation of the
Curie-Weiss law. This approach may be generalized to the narrow-band s-d
model [353].

The above concepts may be useful for general theory of metallic magnetism
too. Obviously, the assumption about strong (in comparison with the total
bandwidth) interelectron repulsion is not valid for transition d-metals.
However, it may hold for some electron groups near the Fermi level. This
idea was used in [353] to treat ferromagnetism of iron group metals. The
narrow-band Hubbard model was applied to describe the group of ``magnetic''
states which form the narrow density of states peak owing to the ``giant''
Van Hove singularities (Sect.2.4). Correlations for these states are to be
strong because of small peak width $\Gamma \simeq 0.1$eV. The rest of s,p,d
electrons form broad bands and are weakly hybridized with ``magnetic''
electrons of the peak. The peak states were assumed to be responsible for
LMM formation and other magnetic properties in Fe and Ni. Such a model
enables one to explain simply the low (as compared to the Fermi energy)
values of the Curie temperature which is, as follows from the above
consideration, of order of $\Gamma $.

In the ferromagnetic phase the splitting of spin up and down peaks is $%
\Delta \simeq 1\div 2$eV$\ll \Gamma $ and structures of both peaks are
similar. Since the lower peak is completely filled, the situation for
``magnetic'' electrons turns out to be close to the saturated
ferromagnetism, i.e. to the half-metallic state in the usual Hubbard model
with large $U$. In such a situation one may expect strong (even in
comparison with the Heusler alloys) non-quasiparticle effects. These may be
important for the explanation of experimental data on spin polarization,
obtained by photo- [358] and thermionic [359] emission, which contradict
drastically to band calculation results for Fe and Ni.

\section{Magnetism of rare earths and actinides}

Since well-localized 4f-electrons retain their magnetic moment in a crystal,
rare earth metals exhibit strong magnetism. All the ``light'' rare earths
(RE) from Ce to Eu have at low temperatures complicated antiferromag netic
structures. All the ``heavy'' RE from Gd to Tm (except for Yb which is a
Pauli paramagnet) are ferromagnetic at low temperatures. With increasing
temperature, the ferromagnetic ordering is as a rule changed by
antiferromagnetic spiral structures, except for Gd which passes immediately
into paramagnetic state. The period of spiral structures turns out to be
appreciable temperature dependent, the spiral angle decreasing with lowering
$T$.

Unlike d-ions, 4f-ions are as a rule rather weakly influenced by the crystal
field (Sect.1.3). Therefore magnetic moments of rare earth metals should be
close to those of the free ion R . However, there are a number of important
exceptions, especially among light rare earth elements. Consider magnetic
structures of RE metals [16,17,246].

Cerium exhibits AFM ordering in the $\beta$ (double hcp) phase with $T_N =
12.5$K. The ordered moment 0.62 $\mu_B$ is small in comparison with that for
Ce$^{3+}$ ion (2.51$\mu_B$). Probably this is connected with strong crystal
field influence on the not too stable f-shell.

Magnetism of praseodymium is determined to a large measure by existence of
low-lying excited states of Pr$^{3+}$ ion. Praseodymium exhibits AFM
ordering below 25K with ordered moment about 1$\mu_B$ .

Neodymium has a complicated AFM structure. The moments at hexagonal sites of
the double hcp lattice order at $T_N = 19.2$K, their value being modulated
along the [1010] axis in the basal plane with the amplitude of 2.3$\mu_B$ .
The moments at cubic sites (in neighbour planes) order at 7.8K, the
modulation amplitude being 1.8$\mu_B$ . These moment values are considerably
smaller than that of free ion (3.2$\mu_B$ ) owing to crystal field.

Preliminary neutron investigations of promethium [360] demonstrate below 98K
a possible existence of ferromagnetic domains with ordered moment of 0.24$%
\mu_B$ . These data require a verification.

Crystal structure of samarium contains sites with cubic and hexagonal
symmetry. Magnetic moments in the hexagonal layers order in a collinear AFM
structure at 106K, and in cubic layers at 13.8K. The ordered moments are
extremely small (about 0.1$\mu_B$).

Europium, which has the bcc structure, possesses below 90K a helcoidal AFM
strucure with the [100] axis and ordered moment of 5.9$\mu_B$ .

Magnetic structures of heavy rare earths are shown in Fig.4.17. Gadolinium
is ferromagnetic with $T_C = 293$K (see also Table 4.3). Below 232 K, there
occurs a finite angle of the ferromagnetic spiral, which reaches the
maximium value of 75$^o$ at 180 K and decreases up to 32$^o$ at low
temperatures.

Terbium and dysprosium are feromagnetic below 221K and 85K respectively. At
higher $T$, they have a simple helicoidal structure, the spiral angle being
also appreciably temperature dependent.

Holmium has below 20K a ferromagnetic spiral structure with
$$
\mu _i^x=\mu _{\perp }\cos \mathbf{QR}_i,\qquad
\mu _i^y=\mu _{\perp }\sin \mathbf{QR}_i,\qquad
\mu _i^z=\mu _{\parallel }
\eqno{(4.109)}
$$
and at higher $T$ the helicoidal structure.

In erbium, the ferromagnetic spiral state ($T<20$K) changes into a
complicated spiral stucture
$$
\mu _i^x=\mu _{\perp }\cos \mathbf{QR}_i,\qquad
\mu _i^y=\mu _{\perp }\sin \mathbf{QR}_i,\qquad
\mu _i^z=\mu _{\parallel }\cos \mathbf{QR}_i
\eqno{(4.110)}
$$
In the interval 52K$<T<84$K, a static longitudinal spin wave state occurs
with
$$
\mu _i^x=\mu _i^y=0,\qquad
\mu _i^z=\mu \cos \mathbf{QR}_i,
\eqno{(4.111)}
$$
Thulium has at low $T$ an antiphase domain structure, spins of four layers
being in up direction and of three layers in down direction. As the result,
average magnetic moment makes up 1$\mu _B$ , altough the moment at each site
is 7$\mu _B$ . Above 32K the static longitudinal spin wave state is realized.

Ytterbium is diamagnetic below 270K and paramagetic above 270K.
Paramagnetism is connected with presence of an ``intrinsic'' fracture of
trivalent ions Yb$^{3+}$ .

Lutetium trivalent ions have a completely filled 4f-shell and do not
posssess magnetic moment. Therefore Lu is a Pauli paramagnet.

Theory of magnetic properties of rare earths was extensively developed
during 60s-70s [15,16,265]. Here we discuss briefly some points with
especial attention to connections beween magnetism and electronic structure.

Although, unlike d-metals, magnetic moments in 4f-metals have an atomic
nature, electronic structure of conduction band influences also magnetic
ordering since the exchange interaction between well-separated f-spins at
different lattice sites is mediated by current carriers (the
Ruderman-Kittel-Kasuya-Yosida, RKKY, indirect interaction). In the real
space this interaction has oscillating and long-range behaviour (Appendix K).

The wavevector of magnetic structure $\mathbf{Q}$ is determined by the
maximum of the function $J(\mathbf{q})$. Phenomenologically, formation of
spiral magnetic structures may be described in the Heisenberg model with
interactions between nearest and next-nearest layers $J_1$ and $J_2$ where
$$
\cos Qc=-J_1/4J_2
\eqno{(4.112)}
$$
However, in the free-electron approximation, $J(\mathbf{q})$, as well as the
susceptibility of the electron gas,
$$
\chi (\mathbf{q}) =\frac 12\sum_{\mathbf{k}}\frac{f(E_{\mathbf{k}})-f(E_{%
\mathbf{k+q}})}{E_{\mathbf{k+q}}-E_{\mathbf{k}}}
$$
$$
=\frac{m^{*}k_F}{8\pi ^2}\left( 1+\frac{4k_F^2-q^2}{4k_Fq}\ln \left| \frac{%
q+2k_F}{q-2k_F}\right| \right)
\eqno{(4.113)}
$$
has a maximum at $q=0$, which corresponds to the collinear ferromagnetic
ordering. Yosida and Watabe [361] obtained a shift of the minimum to finite
$q$ by an account of reciprocal lattice vectors $\mathbf{g}$ (i.e. of the
crystal structure), other factors being taken in the free-electron theory.
Calculation for the hcp lattice with $\mathbf{Q}$ being parallel to the
$c$-axis yielded
\[
Q/2k_F\simeq 0.11,\qquad
Qc=48^o
\]
which agrees with experimental data for Tm, Er and Ho ($Qc=51^o$). However,
this theory does not explain the magnetic structures of Tb and Dy which have
the same lattice.

Further development of the theory was connected with an account of the real
shape of the Fermi surface. It was realized that $J(\mathbf{q})$ may have
minimum in the cases where the vector $\mathbf{q}$ connects large congruent
pieces of the Fermi surface (the ``nesting'' condition). Such a situation
takes place in the itinerant antiferromagnet Cr. It was supposed in [362]
that in 4f-metals $\mathbf{Q}$ corresponds to the vector of the ``arms'' near
the point $L$ in the Brillouin zone of rare earths (Sect.2.7). Change of
this vector in the rare earth series explains satisfactorily the
experimental tendency in the spiral angle. In particular, absence of the
``arms'' in gadolinium corresponds to ferromagnetic ordering.

Besides the topology and shape of the Fermi surface, an important role
belongs to energy gaps near the superzone boundaries, which are determined
by the magnetic structure. The most strong energy gain takes place provided
that the gaps occur near the Fermi surface (the situation which is
reminiscent of the Hume-Rothery rule in the theory of crystal structures,
sect.3.3). The influence of superzone boundary on the spiral period was
studied by Elliott and Wedgwood [363]. They demonstrated that the value of $%
Q $, obtained in [361], decreases with increasing the energy gap. This fact
explains readily the observed $T$-dependence of the spiral angle with
temperature, since the sublattice magnetization decreases with $T$.

An alternative mechanism which influences the spiral angle was proposed by
de Gennes [364]. He took into account the finite value of the mean free path
l for conduction electrons which mediate the exchange interaction among
4f-moments. The damping of electron states, which is due to scattering,
weakens the longe-range antiferromagnetic components of the oscillating RKKY
interaction and favours the ferromagnetic ordering. Therefore $Q$ should
decrease with decreasing $l$. Since the value of localized spin decreases
from Gd to Tb, the magnetic resistivity changes from 10$\mu \Omega \cdot $cm
in Tm to 120$\mu \Omega \cdot $cm in Gd, which corresponds to $l$ of 40 and 5
\AA. Thus the strong magnetic scattering may explain
the absence of antiferromagnetism in gadolinium. The increase of $l$ with
lowering $T$ should give the tendency which is opposite to the experimental
data and to influence of superzone boundaries. Combined effect of both
mechanisms on the spiral angle was considered by Miwa [365].

Fedro and Arai [366] investigated the influence of higher order Kondo-like
logarithmic corections (see (G.73)) to the electron spectrum on the spiral
angle. They obtained the strong effect for the negative sign of the s-f
exchange parameter $I$, as well as in the Kondo problem (Chapter 6).
However, this assumption is unrealistic for elemental rare earths where $I >
0$. Besides that, an account of spin dynamics results in a considerable
smearing of the logarithmic Kondo divergences [367].

Phenomenological treatment of formation of commensurate and incommensurate
magnetic structures in rare earth and of their temperature evolution was
performed by Dzyaloshinsky [368].

Last time, inhomogeneities of spiral structures were investigated. In
particular, so called ``bunching'' and ``spin sleep'' phenomena were
discovered (see, e.g., [369]). Methods of Moessbauer and $\mu $SR
spectroscopy are extensively applied to this problem [370].

Magnetic properties of actinide series metals are determined by localization
of 5f-electrons. As discussed in Sect 1.1, its degreee increases with
increasing atomic number. Light actinides (Th, Pa, U) exhibit usual Pauli
paramagnetism. These metals are satisfactorily described by the model where
itinerant 5f-electrons form wide bands so that the Stoner criterium is not
satisfied. Neptunium and plutonium have apparently, similar to Pt and Pd,
``nearly ferromagnetic'' properties. Their magnetic susceptibility has a
non-monotonous temperature dependence at low $T$, and appreciable
spin-fluctuation $T^2$-contributions to resistivity and enhancement of
specific heat are observed.

Magnetic susceptibility of americium, although being somewhat larger,
depends weakly on temperature. However, its magnetic properties may be
already described within the model of localized 5f-electrons since the Am$%
^{3+}$ ion (configuration 5f$^6$) possesses, as well as Sm$^{3+}$ ion, the
ground term with zero value of $J$, the first excited state with $J=1$ lying
higher by about 500K. Thus americium should exhibit van Vleck paramagnetism.

In Cm, Bc and Cf (other heavy actinides are not investigated) the
Curie-Weiss law holds at high temperatures, the effective magnetic moments
being in agreement for those of corresponding three-valent ions [371]. Thus
the Russel-Saunders coupling scheme turns out to be a good approximation.
Neutron scattering data demonstrate that the double-hcp $\alpha $-curium is
ordered antiferromagnetically below 50K. At the same time, the metastable
fcc $\beta $-curium possesses below 205K a finite saturation moment about 0.4%
$\mu _B$ and probably has a ferrimagnetic order. $\alpha $-Bc is
antiferromagnetic with $T_N=22\div 34$K according to data of various
authors. The data on magnetism of $\alpha $-Cf are contradictory. In the
work [372], AFM ordering below about 60K was found which was changed by
ferro- or ferrimagnetic ordering below 32K. Only one transition into
ferromagnetic (or ferrimagnetic) state with $\mu _0=6.1\mu _B$ was observed
at 51K in [373]. No magnetic ordering was found in the cubic $\beta $-Bc and 
$\beta $-Cf.

Actinide intermetallic and insulating compounds demonstrate a rich variety
of magnetic properties. According the so-called Hill criterion (see [371]),
the f-f overlap ceases and formation of magnetic moment takes place at
sufficiently large distances between the actinide ions exceeding 
$3.25\div 3.50$ \AA (Fig.4.18).

Quantitative theory of magnetic structure of actinides is now developed
insufficiently. Large spin-orbit coupling results in strong orbital
contributions to magnetic properties. For example, the relativistic
spin-polarized self-consistent band calculation of $\delta $-Pu [374]
yielded the following values of magnetic momenta (in $\mu _B$): 
$M_S=5.5$, $M_L=-2.4$, $M_J=M_S+M_L=2.1$ whereas the experimental value of 
$M_J$
makes up about 1$\mu _B$ . There is an open question about the accuracy of
the atomic sphere approximation (ASA). The effect of the non-sphericity of
the potential should be appreciable for orbital contributions. Band
calculation of hcp Bk [375] yielded the value $M_L=2.36$, which is
comparable with the spin magnetic momentum (but is still smaler than the
atomic value for the f$^8$ configuration). This demonstrates considerable
narrowing of 5f-band in heavy actinides.

\section{Magnetic anisotropy}

In zero magnetic field a ferromagnetic sample becomes divided in regions
(domains) which are magnetized along the easy magnetization axes of the
crystal, so that the total magnetic moment is zero. Existence of the easy
and hard directions is just determined by crystalline magnetic anisotropy
(MA) which is characterized by the dependence of the total energy for
magnetically saturated states on the magnetization direction.

Investigation of MA is of interest by two reasons. First, MA determines main
technical properties of ferromagnetic materials - size and shape of magnetic
domains, magnetization process in the external field, coercitive force etc.
Second, MA reflects weak, but important relativistic interactions among
electrons in a crystal and study of corresponding effects gives a valuable
information for the microscopic theory. As well as exchange integrals, the
values of MA parameters are not explained by dipole forces. It is believed
now that the main source of MA is the spin-orbital interaction. The orders
of magnitude of MA parameters turn out to be different for iron group metals
and rare earths, which is directly related to the different role of various
interactions for corresponding ions.

From a phenomenological point of view, the anisotropy energy may be expanded
in spherical functions
$$
\mathcal{E}_a=\sum_{\lambda \mu }C_\lambda ^\mu Y_\lambda ^\mu  
\eqno{(4.114)}
$$
One uses also often (especially for hexagonal and cubic crystals) the
expansion
$$
\mathcal{E}_a^{hex} =K_0+K_1\sin ^2\theta +K_2\sin ^4\theta +K_3\sin
^6\theta   
$$
$$
+K_4\sin ^6\theta \cos 6\phi  
$$
$$
\mathcal{E}_a^{cub} =K_0+K_2(\alpha _x^2\alpha _y^2+\alpha _y^2\alpha
_z^2+\alpha _z^2\alpha _x^2)+K_2\alpha _x^2\alpha _y^2\alpha _z^2  
\eqno{(4.115)} 
$$
with $\alpha _i=M_i/M,$ the constant $K_1$ being zero for cubic crystals.
The constants $C$ (or $K$) depend on external parameters like temperature,
pressure etc.

Values of MA constants for iron group metals and gadolinium are given in
Table 4.6. Gd possesses, unlike other rare earths, zero atomic orbital
momentum and is similar in this respect to cobalt. Therefore MA of Gd is
rather small, the corresponding experimental data being contradictory and
the constant $K_1$ demonstrating a non-monotonous temperature dependence
with the sign change [381]. Other rare earths have MA constants which are by
one-two order of magnitude larger (Table 4.7).

Theoretical calculation of the signs and values of the MA constants is a
complicated problem which is far from complete solution. First-principle
calculations of MA in iron group metals turn out to be highly sensitive to
their electronic structure, so that reliable results are now absent.

Especially difficult is the explanation of the MA temperature dependences.
In the phenomenological approach, the MA constants are proportional to the
corresponding powers of magnetization. In particular, performing the
simplest averaging procedure we obtain the Akulov-Zener law [265]
$$
K_n(T)=K_n(0)\left[ \frac{M(T)}{M(0)}\right] ^{n(2n+1)}  
\eqno{(4.116)}
$$
so that
$$
K_1\sim M^3,\qquad 
K_2\sim M^{10}  
\eqno{(4.117)}
$$
A more accurate treatment of the dependences $K_i(T)$ in the spin-wave
theory of a Heisenberg ferromagnet was performed in papers [376-378]. These
results describe well the behaviour of MA constants in insulators (e.g.,
yttrium garnets). However, experimental $T$-dependences in ferromagnetic
d-metals turn out to be more complicated. The behaviour $K_2(T)$ in iron and
nickel corresponds to higher powers of magnetization than given by (4.117).
The dependence $K_3(T)$ in iron is non-monotonous. A change in $K_1(T)$ sign
occurs in cobalt near 500 K [265] (note that the hcp-fcc transition takes
place at 700K).

\subsection{Quenching of orbital momenta by periodic lattice
potential and magnetic anisotropy of d-metals}

As discussed above, mechanisms of magnetic anisotropy are intimately related
to spin-orbital interactions and orbital magnetic contributions: in the
absence of SOI electron spins do not ''feel'' their orientation. In contrast
with rare earths, a very strong crystal field in the 3d-magnets destroys the
atomic structure of the $SL$-terms of the d$^n$ configurations and
particularly the multiplet structure of the total angular momentum $J$.
Therefore we have to consider the orbital momenta of single electrons.
Experimental results demonstrate that they are almost completely quenched ($%
\langle l\rangle $ is close to zero for the ground state in not too strong
magnetic fields). In the case of the atomic levels this occurs when the
ground state is a singlet (i.e. spherically symmetric). For d-electrons in a
crystal, this can take place, e.g., in the presence of a crystal field of
sufficiently low point symmetry, which splits off a sublevel $m=0$ from
other states (cf.Sect.1.3). However, real ferromagnetic metals are
characterized by high symmetry and the quenching of $l$ is clearly due to a
different mechanism associated with the periodic crystal potential.

The idea of this quenching mechanism is as follows [379,39]. The degeneracy
of atomic levels, which is retained by the local crystal potential, becomes
lifted at spreading them into the energy bands in the $\mathbf{k}$-space.
Indeed, the dispersion law of two bands $E_1(\mathbf{k})$ and $E_2(\mathbf{k}%
)$, corresponding to atomic levels $E_1$ and $E_2$ is, generally speaking,
different. The accidental degeneracy takes place only for those $\mathbf{k}$
where the bands are crossing. The electron orbital momentum operator 
$\mathbf{l=[rk]}$ is diagonal in quasimomentum $\mathbf{k}$. Then all the
diagonal matrix elements $l_{ii}$ between crystal functions $\psi _{\mathbf{k%
}}^i$ vanish because of their symmetry (or antisymmetry) in the sign of
projection $m$, whereas the off-diagonal matrix elements $l_{ij}$ , although
non-zero, correspond to different energies $E_1(\mathbf{k})$ and $E_2(%
\mathbf{k})$ for a fixed $\mathbf{k}$. This means that the contribution of the
latter matrix elements to $\langle l\rangle $ in the presence of a
perturbation with matrix elements $h_{ij}$ is of order of $l/\Delta _{ij}(%
\mathbf{k})$ (with $\Delta _{ij}(\mathbf{k})=E_i(\mathbf{k})-E_j(\mathbf{k})$%
) and, generally speaking, small.

Consider the simple example of the triple representation $t_{2g}$ for a
cubic crystal. It contains three functions $\psi ^i$ which are linear
combinations of the functions with specific values of $m$: 
$$
\psi ^{1,2}=\frac 1{\sqrt{2}}(\phi _1\pm \phi _{-1}),\psi ^3=\frac 1{\sqrt{2}%
}(\phi _2-\phi _{-2})  
\eqno{(4.118)}
$$
As is clear from the symmetry properties, we have $l=0$ for all $i$. On the
other hand, e.g., $\langle \psi ^1|l^z|\psi ^2\rangle =1$. With account of
the point crystal potential only we have the degenerate local level
situation and the existence of $l_{ij}^z\neq 0$ means that in fact the
orbital momenta are not quenched. Indeed, in a magnetic field $H_z$ a
regrouping of the functions takes place and the degeneracy is lifted by the
values of the Zeeman energy $g\mu _BmH_z$. Thus effective orbital momenta
with the value $l=1$ occur. However, for the crystal functions of the band
type the degeneracy of $\psi _{\mathbf{k}}^i$ is lifted for almost all
values of $\mathbf{k}$. This means that, as long as $g\mu _BmH_z\ll $ $\Delta (%
\mathbf{k})$, strong regrouping of the functions in a field $H_z$ is
energetically unfavourable and the value of orbital momentum unquenched by
the field is proportional to a small quantity $g\mu _BmH_z/\Delta (\mathbf{k}%
)$.

Intrinsic $l$-dependent mechanisms also mix the states $\psi _{\mathbf{k}}^i$
and thus unquench $l$. Especially important from this poit of view is the
spin-orbit coupling (Appendix L) since this connects the vectors $\mathbf{s}$
and $\mathbf{l}$. This yields a small unquenched momentum $\widetilde{l}\sim
\lambda /\Delta $, which becomes oriented in the crystal. Again, the spin
momentum is oriented due to spin orbit coupling, so that the magnetic
anisotropy energy $\mathcal{E}_a$ is proportional to $\lambda ^2$ for
uniaxial crystals (and to $\lambda ^4$ for cubic crystals). In fact, $%
\mathcal{E}_a$ corresponds to the spin-orbital coupling energy for the
unquenched momentum $\widetilde{l}$: 
$$
\mathcal{E}_a\simeq \widetilde{\mathcal{E}}_{so}=\lambda (\widetilde{\mathbf{%
l}}\mathbf{s})\sim \lambda ^2/\Delta   
\eqno{(4.119)}
$$
For $\lambda \sim 10^{-14}$erg and $\Delta \sim E_F\sim $ $10^{-11}$erg we
have $\mathcal{E}_a\sim $ $10^{-11}$erg $\sim 10^{-1}$K.

The magnetization process is then as follows (Fig.4.19). In zero field the
vectors $\mathbf{s}$ and $\widetilde{\mathbf{l}}$ are oriented along the easy
axis $z$. The application of field $H_x$ in a hard direction tilts the spin
momentum from the $z$-axis as the ratio $H_x/\widetilde{\mathcal{E}}_{so}$
increases. (For simplicity, we do not take into account the contribution of
the l -component, which is possible provided that corresponding splitting is
large.) The direction of l remains practically unchanged, since this is
fixed by strong CF ($\mathcal{E}_{cr}\gg \widetilde{\mathcal{E}}_{so}$ ). At 
$H=H_a\sim \widetilde{\mathcal{E}}_{so}/\mu _B$ the coupling between the
vectors $\widetilde{\mathbf{l}}$  and $\mathbf{s}$ is broken and the
magnetization is saturated, its value being smaller by the quantity 
$\widetilde{l}$ than along the easy direction. The quantity $H_a$ is called
the anisotropy field and is an important characteristic of the magnetic
hardness of a material.

Concrete calculations of magnetic anisotropy of transition metals with
account of realistic band structure are discussed in papers [380,381]. The
degeneracy points in the $\mathbf{k}$-space, where $\Delta (\mathbf{k})=0$,
turn out to play an important role, quantitative results being strongly
influenced by band structure details. The unusual orientation dependence of
the magnetic anisotropy energy in nickel was attributed in [382] to
existence of very small pockets of the Fermi surface, which undergo dramatic
changes in size as a function of magnetization [145]. However, such pockets
were not found from dHvA or halvanomagnetic effects data. Recent LMTO-ASA
calculations of anisotropy energy in iron group metals [383] have yielded
the values, which are small in comparison with experimental data.

The orbital contributions increase noticeably for surface and impurity
states, and in layered systems, since here the quenching mechanism owing to
periodic lattice potential becomes less effective. So, an appreciable
enhancement of orbital magnetic momenta (OMM) at surfaces of Fe, Co, Ni was
obtained in the band calculation [384]. Further investigations of these
problems within many-electron models are required.

The unquenching of OMM due to Coulomb interaction, discussed above in
Sect.1.3, should be once more mentioned. The latter tends to satisfy the
Hund rules. Up to now, this mechanism was practically misregarded (see,
however, the paper [385] where such a possibility was considered within a
spin-density functional approach). On the other hand, intersite orbital
exchange interaction (Appendix D) can also unquench OMM and induce an
orbital ordering. Corresponding quantitative calculations and comparison of
the electrostatic mechanisms with the spin-orbital one would be very
interesting.

To conclude this Section, we discuss some examples of OMM effects in
d-systems. Orbital momenta seem to play an important role in electron
structure and physical properties of high-$T_c$ superconductors. A strong
anisotropy of c is observed above T [386]. Its sign and absolute value may
be used to investigate the electron structure near $E_F$. The calculation
[387] shows that the sign of Dc corresponds to the contribution of
off-diagonal matrix elements between the states $x^2-y^2$ at $E_F$ and the $%
xy$-states, which lie below by about 1eV. Recent investigations of polarized
X-ray spectra of copper ions [388] confirm that the bands of the type $%
x^2-y^2$ with a small admixture of $z$-type bands are present at the Fermi
surface. Thus unquenching of OMM is due to transitions between the states $%
2^{-1/1}(|2\rangle \pm |-2\rangle ).$

A number of d-compounds (e.g., with perovskite, spinel and rutile structure)
contain Jahn-Teller ions which possess an orbitally degenerate ground state.
Strong anisotropy of magnetization and g-factors takes place in such systems
(e.g., for the ions Cu$^{2+}$, $g_{\parallel }~=2.4$, $g_{\perp }=2.08$
[274]). A gain in the total energy is achieved by the degeneracy lift at
symmetry lowering which is due to lattice distortion. A ``ferro''- or
``antiferromagnetic'' orbital ordering may accompany the cooperative
Jahn-Teller effect. Description of such phenomena and their interpretation
is usually given within an one-electron approach [42]. Apparently, in a
number of cases the interaction of ion L-momenta with crystal field should
be considered in the many-electron scheme with account of spin-orbital
coupling. However, such a consideration seems to be up to now not performed.

\subsection{Magnetic anisotropy of rare earths}

Now we consider the situation in rare earths [39]. For 4f-electrons the
energy of crystal field E is small in comparison with not only the
electrostatic interaction, but also with the spin-orbital energy, so that
the the total atomic quantum numbers $S,L,J$ are retained also in a crystal
(Sect.1.3). Consequently, the magnetization process in RE magnets occurs in
a different way as compared to the case of d-magnets.

In the case of easy-axis anisotropy, at $H=0$ the vector $\mathbf{J=L+S}$
lies along the $z$-axis governed by the minimum of $\mathcal{E}_{cr}$ for
the angular momentum $\mathbf{L}$. In a field $H_x$ the coupling between 
$\mathbf{L}$ and $\mathbf{S}$ is retained (because 
$\mathcal{E}_{cr}\ll \mathcal{E}_{so}$ ) and they rotate as a whole, 
approaching the $x$-direction. The
angle of rotation is determined by the ratio $g\mu _BJH/\mathcal{E}_{cr}$.
Although the energy $\mathcal{E}_{cr}\sim 10^{-2}\div 10^{-3}$eV is the
smallest among the energies of other interactions for 4f-electrons, it is
considerably higher (by two or three orders of magnitude) than the effective
energy $\widetilde{\mathcal{E}}_{so}$  responsible for the magnetic
anisotropy of 3d-metals with quenched orbital momenta. This accounts for the
giant anisotropy exhibited by rare-earth magnets.

The existence of unquenched OMM results in another important consequence in
the theory of magnetic properties of rare earths. As demonstrated in
Appendix K, the s-f exchange interaction leads, in the second order of
perturbation theory, not only to an indirect f-f exchange of the Heisenberg
(or de Gennes) type, but also results to the exchange interactions, which
are determined by orbital momenta (see (K.11)). Such interactions (in
contrast to spin exchange) readily become anisotropic after allowance for
the anisotropy of the crystal. The contributions of the anisotropic exchange
to the magnetic anisotropy energy were calculated in [389]. The total result
for the hexagonal lattice with the parameters $c$ and $a$ reads
$$
\mathcal{E}_{cr} =(K_1^{cr}+K_1^{exch})\cos ^2\theta +...  
$$
$$
K_1^{cr} =\alpha _JJ(J-\frac 12)\frac{Z^{eff}e^2}a\frac{\langle
r_f^2\rangle }{a^2}1.2(c/a-\sqrt{8/3})  
\eqno{(4.120)} 
$$
$$
K_1^{exch} \sim (g-1)D_1J^2I_{sf}^2N(E_F)  
$$
Here $\alpha _J$ is the Stevens parameter, $Z^{eff}$ is the effective ion
charge, $\langle r_f^2\rangle $ is the averaged square of the f-shell
radius, $D_1$ is defined in (K.10). The expressions (4.120) yield estimates
of the order of magnitude of $K_1^{cr}$ and $K_1^{exch}$. Since for heavy
rare earths $\alpha _J\sim $ 10$^{-2}$ $\div $10$^{-3}$, we get $%
K_1^{cr}\sim $ 10$^7\div $ 10$^8$erg/cm$^3$. Then, $D_1\sim 10^{-2}$, so
that $K_1^{exch}\sim $ 10$^6\div $ 10$^7$erg/cm$^3$ .

Thus the magnetic anisotropy of rare earths is one or two order of magnitude
higher than that of the most strongly anisotropic hexagonal d-magnets. This
difference is a consequence of the fact that, for RE, magnetic anisotropy is
determined by the electrostatic interactions of the crystal field or an
anisotropic exchange type and not by the weak spin-orbital interaction (as
for d-magnets).

The MA constant $K_1$ and the anisotropy of the paramagnetic Curie
temperature $\Delta \theta =\theta _{\parallel }-\theta _{\perp }$ of heavy
of rare earths are given in Table 4.7. The latter quantity is found by
extrapolating the Curie-Weiss law $\chi ^{-1}=C^{-1}(T-\theta )$ for
corresponding crystal directions up to the temperatures where $\chi ^{-1}=0$%
. For clarity, a comparison with the experimental resuts is made separately
for the crystal field and exchange mechanisms. The largest difference
between two theoretical values is found in the case of Tm, which is
therefore of the greatest interest from the point of view of the magnetic
anisotropy.

Although a complete comparison with experiment is difficult because of
absence of precise data, the crystal field contribution is probably dominant
and the anisotropic exchange contributes only 10-20\%. A reliable
experimental determination of the latter would be of fundamental interest in
the theory of exchange interactions. In contrast with the single-ion
crystal-field mechanism, the anisotropic exchange gives rise to a two-ion
anisotropy so that it can be separated on the basis of the alloy composition
dependence. Therefore experimental investigations of magnetic anisotropy of
RE alloys may be useful. It is worthwhile also to mention the methods for
detecting anisotropic exchange, which are based on hyperfine interactions
[390].

Now we discuss the sign of magnetic anisotropy. Here the theory provides
precise predictions. The signs of both $\alpha _J$ and $D_1$ are reversed on
transition from f$^3$(f$^{10}$) to f$^4$(f$^{11}$) configurations in the
first (second) halves of the RE series and also on transition from the first
to the second half. If we bear in mind that Pm has not yet been
investigated, and Eu and Yb have cubic lattices, the agreement with the
experimental results is quite satisfactory.

This mathematical result has a clear physical meaning. The magnetic
anisotropy is related to the magnitude and orientation of the orbital
components of the total orbital momenta in an electric crystal field. One
can see from the Table 4.8 that, besides the trivial electron-hole symmetry
in the values of $L$ between the first and second halves of the series,
there is also a symmetry within each half associated with the occupancy of
orbital quantum states. For example, f$^1$- and f$^6$-states have the same
value $L=3$ and it might seem that the anisotropy should also be the same.
However, we should take into account that $L($f$^1$) is the angular momentum
of one electron, whereas $L$(f$^6$) is the orbital momentum of a hole in the
spherical f$^7$ configuration characterized by $L=0$. Thus the anisotropy of
the electric charge will be opposite for the f$^1$ and f$^6$-configurations
(Fig.4.20).

The same reversals of the sign of the first anisotropy constant in the RE
series are as a rule observed not only for pure RE metals, but also for
their alloys and compounds. Consider the situation in the practically
important intermetallic systems RCo$_5$ . The theory predicts the
orientational plane-axis transition for the compounds with R = Ce, Pr, Nd,
Tb, Dy, Ho. Indeed, in the case of these elements, in the ground state the
angular momentum $\mathbf{J}$ is oriented in the basal plane of the
hexagonal lattice. The Co ions have an easy $c$-axis, but because of the
exchange interaction with the R ions they are oriented so that their
magnetic moments are aligned in a plane at low temperatures. As a sample is
heated, the magnetization of the R sublattice decreases faster than that of
the Co sublattice. Thus the gain in the anisotropy energy of the Co
sublattice, which is characterized by a larger magnetization, becomes
dominating, the plane-axis transition takes place at some temperature [391].
Such transitions are observed experimentally [392].

The possibility of occurrence of strong local anisotropy effects in RE
alloys because of lowering of the local symmetry of the environment for a
given RE ion is also of interest. An effect of this kind was considered in
[393] with application to the system RCo$_x$Ni$_{5-x}$. It was found that,
besides the usual anisotropy constants corresponding to the macroscopic
(average) symmetry of the alloy, there are also constants of a local type.
After averaging over various possible configurations of the nearest
neighbours, these local constants contribute to the observed macroscopic
anisotropy of a crystal giving rise to a specific dependence on the
composition.

It is worthwhile to mention another factor which strongly affects both the
magnitude and sign of the anisotropy. This is the geomentric factor of the
hcp lattice $\sqrt{8/3}-c/a=1.633-c/a$. For all the RE metals, $%
c/a=1.58-1.61<1.63$. However, in principle we can change the sign of the
anisotropy by changing this factor.

\chapter{TRANSPORT PROPERTIES}

Besides practical significance, electronic transport phenomena in metals are
rather important from the theoretical point of view. They have played a
great role in formulation of principal statements of the modern quantum
solid state physics. In particular, the classical theory does not explain
the zero value of electrical resistivity at $T = 0$ which should be
determined by cross sections of electron scattering by ions. In the quantum
theory, electrons in a periodical crystal are described by Bloch states with
a definite wavevector $\mathbf{k}$ and carry current without a loss of
energy. The scattering of electrons is determined by various mechanisms in
electron-lattice system which disturb the periodicity (impurities, thermal
vibrations etc.). It turns out also that, as well as for electronic specific
heat and magnetic properties, the degenerate statistics of current carriers
and the existence of the Fermi surface are crucial for transport phenomena.

The theoretical description of transport phenomena includes a number of
parameters: concentration $n$ and charge sign of current carriers, effective
mass $m^{*}$, electron velocity at the Fermi level $v_F$, parameters of
various interactions: electron-lattice, electron-impurity, electron-spin,
electron-electron. These interactions influence not only energy spectrum,
but also the mean free path $l$ and the relaxation time $\tau $. The
simplest expressions for the conductivity and the Hall coefficient
$$
\sigma =\frac{e^2n\tau }{m^{*}},\qquad
R=-\frac 1{nec}
\eqno{(5.1)}
$$
enable one to determine $n$ and $\tau $ (the effective mass may be
determined from de Haas-van Alphen effect data).

In real substances the situation is more complicated than that described by
(5.1). This is connected with the presence of several groups of current
carriers, combined influence of different scattering mechanisms etc. Main
problem of the microscopic theory in such a stuation is separating these
factors and determining their relative contributions. This problem is rather
important for transition metals which demonstrate a great variety of
transport phenomena. In this Chapter we pay especial attention to
peculiarities which occur in comparison with simple metals. Let us list some
of these peculiarities.

\noindent 1) Existence of current carriers with strongly different
characteristics (``s'' and ``d''-electrons). Although 4f-electrons usually
do not take part in conductivity, they may influence the current carriers
due to s-f hybridization, especially in some rare-earth compounds. Besides
that, current carriers with opposite spins have different characteristics in
ferromagnetic metals.

\noindent
2) Presence of internal partially filled d-shells (narrow bands) may be an
subsidiary scattering source (due to the s-d transitions).

\noindent
3) Occurence of additional scattering mechanisms by d- and f-shell
magnetic moments.

\noindent
4) Anomalous (spontaneous) transport phenomena (e.g., the Hall,
Faraday, Kerr and $\Delta \rho /\rho $ effects), which are connected with
the magnetization of d(f)-ions rather than with external magnetic field.

\noindent
5) Anomalous behaviour of transport coefficients near magnetic
ordering points.

\noindent
6) Large correlation effects and possibility of electron localization in
narrow d-bands.

Thus main peculiarities of transition d-metals are connected with
intermediate degree of d-electron localization. This fact leads to that
d-electrons play simultaneously two roles: of current carriers and of
scattering centres. Below we analyze a number of scattering mechanisms which
are specific for transition metals.

\section{General classification of transport phenomena}

The densities of charge and heat current $\mathbf{j}$ and $\mathbf{W}$, produced
by the external electric field $\mathbf{E}=$ grad$\phi $ and temperature
gradient, are determined by the equations
$$
j_\alpha =\sigma _{\alpha \beta }E_\beta +\lambda _{\alpha \beta }
\mathrm{grad}_\beta T
\eqno{(5.2)}
$$
$$
W_\alpha =q_\alpha -\phi j_\alpha =\nu _{\alpha \beta }E_\beta -\kappa
_{\alpha \beta }\mathrm{grad}_\beta T
\eqno{(5.3)}
$$
where the electric and thermal conductivities $\sigma $ and $\kappa $, and
the coefficients $\lambda $ and $\nu $ are tensor quantities, which
determine transport properties of the crystal (summation over repeated
indices is assumed). The term $\phi \mathbf{j}$, which yields transfer of
electron potential energy in the electric field, is subtracted in (5.3) from
the total energy current $\mathbf{q}$.

At considering transport phenomena, it is important to take into account the
Onsager symmetry principle. Within the framework of linear theory (small
deviations of the system from the equilibrium state) the coefficients in the
system
$$
\frac{\partial X_i}{\partial T}=\sum_j\gamma _{ij}\frac{\partial \mathcal{S}%
}{\partial X_j}
\eqno{(5.4)}
$$
($X_i$ are quantities determining the state of the system, $\mathcal{S}$ is
the density of entropy) are symmetric in indices $i,j$:
$$
\gamma _{ij}=\gamma _{ji}
\eqno{(5.5)}
$$
Equations (5.2),(5.3) are a particular case of the system (5.4) with
$$
\frac{\partial X_1}{\partial T} =\mathbf{j},
\frac{\partial X_2}{\partial T}=\mathbf{W},
\eqno{(5.6)}
$$
$$
\frac{\partial \mathcal{S}}{\partial X_1} =\frac{\mathbf{E}}T,
\frac{\partial \mathcal{S}}{\partial X_2}=-\frac{\mathrm{grad}T}{T^2}
$$
The relations (5.6) are readily obtained from the expression
$$
\frac{d\mathcal{S}}{dT}=\int d\mathbf{r}\frac{\mathrm{div}\mathbf{q}}T
=\int d\mathbf{r}\frac{\mathbf{Ej}}T\mathbf{-}\int d\mathbf{r}
\frac 1{T^2}\mathbf{W}\mathrm{grad}T
\eqno{(5.7)}
$$
In the presence of the magnetic field $H$ the transport coefficients are
functions of $H$. Since the equations of motion in non-magnetic crystals are
invariant under simultaneous replacement $H\rightarrow -H$, $t\rightarrow -t$,
the Onsager relations read
$$
\sigma _{\alpha \beta }(H)=\sigma _{\beta \alpha }(-H),\qquad
\kappa _{\alpha \beta }(H)=\kappa _{\beta \alpha }(-H),\qquad
\nu _{\alpha \beta }(H)=T\lambda _{\beta \alpha }(-H)
\eqno{(5.8)}
$$
The transport coefficients may be expanded in $H$, the transport phenomena
being subdivided into even and odd ones with respect to magnetic field.
Practically, it is sufficient to retain linear and quadratic terms only to
obtain from (5.1), (5.2)
$$
E_\alpha =(\rho _{\alpha \beta }+R_{\alpha \beta \gamma }H_\gamma
+r_{\alpha \beta \gamma \delta }H_\gamma H_\delta )j_\beta
$$
$$
+(\alpha _{\alpha \beta }+Q_{\alpha \beta \gamma }H_\gamma +\Delta \alpha
_{\alpha \beta \gamma \delta }H_\gamma H_\delta )\mathrm{grad}_\beta T
\eqno{(5.9)}
$$
$$
W_\alpha =(\alpha _{\alpha \beta }+Q_{\alpha \beta \gamma }H_\gamma
+\Delta \alpha _{\alpha \beta \gamma \delta }H_\gamma H_\delta )j_\beta
$$
$$
+(-\kappa _{\alpha \beta }+L_{\alpha \beta \gamma }H_\gamma +B_{\alpha
\beta \gamma \delta }H_\gamma H_\delta )\mathrm{grad}_\beta T
\eqno{(5.10)}
$$
Here $\rho $ is the electical resistivity tensor, $R$ is the Hall
coefficient, $r$ determines the magnetoresistivity
($\Delta \rho /\rho $-effect), $\alpha $ is the thermoelectric power coefficient, the
Nernst-Ettingshausen coefficient $Q$ determines the change of thermoelectric
power in magnetic field, and the coefficients $\Delta \alpha $ describe the
longitudinal and transverse magnetothermoelectric effect. In the equation
for the thermal current (5.10), $\kappa $ is the thermal conductivity
tensor, $L$ determines the occurence of transverse temperature gradient in
magnetic field (the Righi-Leduc effect), $B$ describes thermal
magnetoresistivity; other coefficients coincide with those in (5.9) due to
the Onsager relations.

Now we discuss magnetically ordered crystals. In ferromagnets we have to
take into account, besides the external magnetic field, the magnetization $%
\mathbf{M}$. In the cases of antiferro- or ferrimagnetic ordering we have to
consider the magnetization of each magnetic sublattice. In the simplest
two-sublattice case it is convenient to use the variables $\mathbf{M=M}_1%
\mathbf{+M}_2$ and $\mathbf{L=M}_1\mathbf{-M}_2$. Expanding
transport coefficients in $\mathbf{H,M,L}$ we write down by analogy with
(5.9), (5.10)
$$
E_\alpha (\mathbf{H,M,L}) =E_\alpha (0)+(\sum_iR_{\alpha \beta \gamma
}^iY_\gamma ^i+\sum_{ij}r_{\alpha \beta \gamma \delta }^{ij}Y_\gamma
^iY_\delta ^j)j_\beta
$$
$$
+(\sum_iN_{\alpha \beta \gamma }^iY_\gamma ^i+\sum_{ij}\Delta \alpha
_{\alpha \beta \gamma \delta }^{ij}Y_\gamma ^iY_\delta ^j)\mathrm{grad}_\beta T
\eqno{(5.11)}
$$
$$
W_\alpha (\mathbf{H,M,L}) =W_\alpha (0)+(\sum_iN_{\alpha \beta \gamma
}^iY_\gamma ^i+\sum_{ij}\Delta \alpha _{\alpha \beta \gamma \delta
}^{ij}Y_\gamma ^iY_\delta ^j)Tj_\beta
$$
$$
+(\sum_iL_{\alpha \beta \gamma }^iY_\gamma ^i+\sum_{ij}B_{\alpha \beta
\gamma \delta }^{ij}Y_\gamma ^iY_\delta ^j)\mathrm{grad}_\beta T
$$
where the indices $i,j=H,M,L$ and $\mathbf{Y}_i\mathbf{=H,M,L}$. A concrete
form of the tensor coefficients in (5.11) may be found for each crystal
structure [394].

One can see that, besides usual transport coefficients, there exist the Hall
and $\Delta \rho /\rho $ effects owing to the spontaneous magnetization.
Such effects are called spontaneous or extraordinary (anomalous). From the
phenomenological point of view, the spontaneous effects are similar to the
corresponding effect in non-magnetic crystals. However, in fact they possess
essential peculiarities. As a rule, the corresponding coefficients are large
in comparison with those for ``normal'' effects. So, the spontaneous Hall
coefficient $R^M$ in metals depends strongly on temperature, its absolute
value at high temperatures exceeding by several orders that of $R^H$. Thus
the expressions (5.11) may not in fact be treated as just a formal
expansion, and the spontaneous effects may not be reduced to normal ones,
e.g., by introducing an effective field.

Thus we meet with the problem of constructing a microscopic theory of
spontaneous transport effects on the basis of new mechanisms of interaction
of current carriers with the lattice ions in magnetic crystals. Evidently,
one of such mechanims is the exchange interaction. Surprisingly,
spin-orbital interaction (Appendix L) plays also an important role in a
number of effects.

Of interest are also terms in (5.11) which contain the vector $\mathbf{L}$.
In particular, the $R^LL$-term results in the even Hall effect and the $%
r^{HL}HL$-term in the odd $\Delta \rho /\rho $-effect. Consider the
corresponding experimental data. At investigating the Hall effect in the
ferrimagnet Mn$_5$Ge$_2$ [395] it was found that the Hall coefficient $R$
changes its sign at the compensation temperature where the magnetizations of
sublattices are opposite. This phenomenon was explained in paper [396].
According to (5.11),
$$
R=\frac{E_y}{j_xM_z}=R^M+R^L\frac{L_z}{M_z}
\eqno{(5.12)}
$$
One can see that the second term in (5.12) diverges at the compensation
point ($M_z=0$), $L_z$ changing its sign, which yields an effective sign
change in $R_s$. Thus existence of the compensation point enables one to
separate the ferromagnetic and antiferromagnetic Hall effects. The odd $%
\Delta \rho /\rho $-effect in Mn$_5$Ge$_2$ was picked out in the paper [397].

The situation in true antiferromagnets where magnetic sublattices are
equivalent is different. The terms, which are linear in $L$, should vanish
in crystals where the sublattices are connected with crystal symmetry
transformations. However, these terms may occur in crystals where the period
of magnetic structure coincides with the crystallographic period and
magnetic sublattices transform one into another under the antisymmetry
operations (i.e. usual symmetry operations combined with time reversion).
Since these operations change the sign of magnetic moment, the $L$-linear
terms turn out to be not forbidden by symmetry requirements. According to
[398], such a situation takes place in the hematite Fe$_2$O$_3$ where the
even Hall effect was observed [399]. In this case one may also expect the
odd magnetoresistivity
$$
\left( \frac{\Delta \rho }\rho \right) _{odd}=\frac{\rho (H)-\rho (-H)}{%
2\rho (0)}\propto H[L(H)-L(-H)]
\eqno{(5.13)}
$$
which was also found in hematite [400]. It should be noted that the
microscopic theory of $L$-linear effects is up to now absent.

Recently, the linear term in the magnetoresistivity was observed in the
normal phase of the high-$T_c$ system YBa$_2$Cu$_3$O$_y$ [401], the value of
this term strongly increasing in the narrow region of oxygen concentration $%
6.88<y<6.95$. The authors relate the effect with existence of dynamically
correlated antiferromagnetic regions in these sample.

\section{Calculation of transport coefficients}

To evaluate transport coefficients (e.g., resistivity) we need to know the
scattering probability which is determined by the values and $\mathbf{k}$%
-dependences of interaction matrix elements. The simplest method to solve
this problem is consideration of the transport equation for the electron
distribution function in a crystal in the presence of external fields. In
the stationary regime, the evolution of this function is described by the
equation
$$
\frac{df}{dt}=\left( \frac{\partial f}{\partial t}\right) _{field}+\left(
\frac{\partial f}{\partial t}\right) _{collis}=0
\eqno{(5.14)}
$$
so that the effect of acceleration by external electric field $E_x$ is
balanced by collisions for some non-zero electron velocity $v_x$ along the
field direction. To linear approximation in the field we may write down
$$
f=f_0+f_1
\eqno{(5.15)}
$$
\[
\left( \frac{\partial f}{\partial t}\right) _{field}=\left( \frac{\partial
f_0}{\partial t}\right) _{field}=\frac{\partial f_0}{\partial E_{\mathbf{k}}}%
=-eE_x\frac{\partial f_0v_x}{\partial E_{\mathbf{k}}}\frac{\partial E_{%
\mathbf{k}}}{\partial \mathbf{k}}\frac{d\mathbf{k}}{dt}
\]
where $f_0$ is the equilibrium Fermi distribution and $f_1$ is the linear
correction. Introducing the scattering probability $W_{\mathbf{kk}^{\prime
}} $ we represent the equation (5.14) in the form
$$
\sum_{\mathbf{k}^{\prime }}(W_{\mathbf{kk}^{\prime }}f_{1\mathbf{k}}-W_{%
\mathbf{kk}^{\prime }}f_{1\mathbf{k}^{\prime }})=-eE_xv_x\frac{\partial f_{0%
\mathbf{k}}}{\partial E_{\mathbf{k}}}
\eqno{(5.16)}
$$
Thus we have obtained the integral equation for the function $f_{1\mathbf{k}}$.
After its solution, the electric current and conductivity are
calculated as
$$
j_x=-e\int d\mathbf{k}v_xf_1,\qquad
\sigma _{xx}=j_x/E_x
\eqno{(5.17)}
$$
The quantity $W$ may be calculated for each scattering mechanism
(impurities, phonons, spin inhomogeneities etc.). In the case of the
independent mechanisms
$$
\rho (T)=\rho _i+\rho _{ph}+\rho _{mag}+...
\eqno{(5.18)}
$$
The additivity of various mechanism contributions is called the Matthiessen
rule. Generally speaking, deviations from this rule occur which are
connected with interference of different scattering processes.

An alternative method for calculating the transport relaxation time is the
use of the Kubo formula for conductivity [402]
$$
\sigma _{xx}=\frac 1{2T}\int_{-\infty }^\infty dt\langle j_x(t)j_x\rangle
\eqno{(5.19)}
$$
where
$$
\mathbf{j}=-e\sum_{\mathbf{k}\sigma }\mathbf{v}_{\mathbf{k}\sigma }c_{%
\mathbf{k}\sigma }^{\dagger }c_{\mathbf{k}\sigma },\mathbf{v}_{\mathbf{k}%
\sigma }=\frac{\partial E_{\mathbf{k}\sigma }}{\partial \mathbf{k}}
\eqno{(5.20)}
$$
is the current operator. Representing the total Hamiltonian in the form $%
\mathcal{H}=\mathcal{H}_0+\mathcal{H}^{\prime }$, the correlator in (5.19)
may be expanded in the perturbation $\mathcal{H}^{\prime }$ [403,404]. In
the second order we obtain for the electrical resistivity [403]
$$
\rho _{xx}=\sigma _{xx}^{-1}=\frac{k_BT}{\langle j_x^2\rangle }\int_0^\infty
dt\langle [j_x,\mathcal{H}^{^{\prime }}(t)][\mathcal{H},j_x]\rangle
\eqno{(5.21)}
$$
where $\mathcal{H}^{\prime }(t)$ is calculated with the Hamiltonian $%
\mathcal{H}_0$. Provided that the perturbation Hamiltonian has the form
$$
\mathcal{H}^{\prime }=\sum_{\mathbf{kk}^{\prime }\sigma \sigma ^{\prime }}%
\widehat{W}_{\mathbf{kk}^{\prime }}^{\sigma \sigma ^{\prime }}c_{\mathbf{k}%
\sigma }^{\dagger }c_{\mathbf{k}^{\prime }\sigma ^{\prime }}
\eqno{(5.22)}
$$
(in particular, for the phonon and magnon scattering) we obtain
$$
\rho =\frac{k_BT}{\langle j_x^2\rangle }e^2\sum_{\mathbf{kk}^{\prime }\sigma
\sigma ^{\prime }}(\mathbf{v}_{\mathbf{k}\sigma }-\mathbf{v}_{\mathbf{k}%
^{\prime }\sigma ^{\prime }})^2\int_0^\infty dt\langle \widehat{W}_{\mathbf{%
kk}^{\prime }}^{\sigma \sigma ^{\prime }}(t)\widehat{W}_{\mathbf{k}^{\prime }%
\mathbf{k}}^{\sigma ^{\prime }\sigma }\rangle \exp [i(E_{\mathbf{k}\sigma
}-E_{\mathbf{k}^{\prime }\sigma ^{\prime }})t]
\eqno{(5.23)}
$$
with
$$
\langle j_x^2\rangle =e^2\sum_{\mathbf{k}\sigma }(v_{\mathbf{k}\sigma
}^x)^2n_{\mathbf{k}\sigma }(1-n_{\mathbf{k}\sigma })
\eqno{(5.24)}
$$
This approach is equivalent to solution of the transport equation by the
variational method [7,8].

The transport equation for elastic impurity scattering is considered in
Appendix M. The result for the impurity resistivity in the lowest-order Born
approximation has the form
$$
\rho _i=\sigma _i^{-1}=\frac{e^2n\tau _i}{m^{*}},\qquad
\tau _i=\frac{2\pi \hbar ^4}{n_i\overline{\phi }^2(2m^{*})^{3/2}E_F^{1/2}}
\eqno{(5.25)}
$$
here $\overline{\phi }$ is the average impurity potential, $n_i$ is the
impurity concentration. Thus the resistivity is temperature-independent.

The case of the scattering by acoustical phonons may be considered with the
use of either the transport equation (5.15) [1] or the formula (5.23) [7,8].
The situation is most simple in the case of high $T>\theta _D$. Then the
phonon frequency is small in comparison with temperature, so that the
scattering is elastic and and we may put for the phonon occupation numbers
\[
N_{\mathbf{q}}=N_B(\omega _{\mathbf{q}})\simeq k_BT/\omega _{\mathbf{q}}
\]
Then the resistivity is proportional to the number of scattering particles,
as well as in the case of impurity scattering, and is linear in temperature:
$$
\sigma _{ph}=\rho _{ph}^{-1}=\frac{e^2n}{\pi ^3\hbar ^3}\frac M{k_F}\left(
\frac kC\frac{dE_{\mathbf{k}}}{d\mathbf{k}}\right) _{k_F}k_B\theta _D\frac{%
\theta _D}T
\eqno{(5.26)}
$$
where $M$ is the ion mass and
$$
C=\frac{\hbar ^2}{2m}\int d\mathbf{r}|\mathrm{grad}u_{\mathbf{k}}|^2
\eqno{(5.27)}
$$
is the Bloch constant, the function $u_{\mathbf{k}}$ being determined by
(2.1).

At $T\ll \theta _D$ the number of phonons decreases rapidly with lowering $T$. 
Since the phonons with small $q$ yield main contribution, replacement $%
\omega _{\mathbf{q}}/T=x$ in the corresponding integral yields $q^2N_{%
\mathbf{q}}dq\sim T^3$. Besides that, only a small part of the electron
quasimomentum $k$, of order of $q^2/k^2$, is lost at scattering by long-wave
phonons. This results in an extra factor of $\omega _{\mathbf{q}}^2/k^2\sim
T^2$. Thus the total scattering probability is proportional to $T^5$ (the
Bloch law). An interpolation expression for the phonon contribution may be
presented in the form [1]
$$
\sigma _{ph}=\frac{e^2n}{4\pi ^3\hbar ^3}\frac M{k_FI(T)}\left( \frac kC%
\frac{dE_{\mathbf{k}}}{d\mathbf{k}}\right) _{k_F}^2k_B\theta _D\left( \frac{%
\theta _D}T\right) ^5
\eqno{(5.28)}
$$
where
$$
I(T)=\int_0^{\theta _D/T}\frac{e^xx^5dx}{(e^x-1)^2}
\eqno{(5.29)}
$$
At low temperatures
\[
I(T)\simeq I(0)=5\int_0^\infty \frac{x^4dx}{e^x-1}=124.4
\]
The Bloch theory of resistivity describes well experimental data for simple
metals in a wide temperature region. So, for Ag and Cu the agreement is
within 5\%.

\section{Resistivity}

Examples of $\rho (T)$ dependences for transition metals are shown in
Figs.5.1-5.8 [239]. Some of the Figures show also different contributions to
resistivity (impurity, $\rho _i$, phonon, $\rho _{ph}$, magnetic, $\rho
_{mag}$, Mott s-d scattering, $\rho _{sd}$, electron-electron, $\rho _{ee}$,
etc.) which are discussed below.

The values of both $\rho $ and $d\rho /dT$ are considerably larger than for
simple metals. There exist two types of $\rho (T)$ behaviour - convex and
concave one. The type is determined to some extent by the column number in
the periodic table: the convex dependence is observed for atomic d$^n$%
-configurations with odd $n$ (such a behaviour takes place also in most
rare-earth metals). However, this regularity is not universal. In
particular, the concave dependence is observed in Cr, Mo, W (d$^4$) and Ru,
Os (d$^6$), but not in Ti, Zr, Hf (d$^2$). The convex behaviour reflects a
saturation tendency which becomes more pronounced with increasing absolute
values of $\rho $.

The connection of conductivity with the energy dependence of density of
states may be obtained in the simplest approximation of the relaxation time $%
\tau $ [8] where
$$
f_1=-\tau \mathbf{Ev}\frac{\partial f_0}{\partial E_{\mathbf{k}}}
\eqno{(5.30)}
$$
so that
$$
\mathbf{j} =-\frac 13e^2\mathbf{E}\sum_{\mathbf{k}}\tau v^2\frac{\partial
f_0}{\partial E_{\mathbf{k}}}
$$
$$
=-\frac{e^2E}{24\pi ^3}\int d\varepsilon \frac{\partial f_0(\varepsilon )}{%
\partial \varepsilon }\int \frac{dS}{|\mathrm{grad}_{\mathbf{k}}\varepsilon |}%
\tau v^2
$$
$$
\equiv \int d\varepsilon \sigma (\varepsilon )\left( -\frac{\partial
f_0(\varepsilon )}{\partial \varepsilon }\right) \mathbf{E}
\eqno{(5.31)}
$$
For the spherical Fermi surface
$$
\sigma (\varepsilon )=-\frac{4e^2}{3m^{*}}\varepsilon N(\varepsilon )\tau
(\varepsilon )
\eqno{(5.32)}
$$
To lowest-order approximation we have
$$
\sigma =\frac{j_x}{E_x}=\sigma (E_F)=\frac{e^2n\tau (E_F)}{m^{*}}
\eqno{(5.33)}
$$
Using the expansion (3.37) we obtain
$$
\sigma =\sigma (E_F)+\frac{\pi ^2}6(k_BT)^2\left( \frac{\partial ^2\sigma
(\varepsilon )}{\partial \varepsilon ^2}\right) _{\varepsilon =E_F}
\eqno{(5.34)}
$$
The second term in (5.34) should be important at high temperatures.

The role of the $N(E)$ dependence in transport properties for an arbitrary
electron system (including one with strong correlations) may be illustrated
by simple consideration of impurity scattering [405]. To this end we expand
the one-electron Green's function to second order in the impurity potential
$V$
$$
\langle \!\langle
c_{\mathbf{k}}|c_{\mathbf{k}^{\prime }}^{\dagger }
\rangle \!\rangle _E=\delta _{\mathbf{kk}^{\prime }}G_{\mathbf{k}}(E)
+G_{\mathbf{k}}(E)T_{\mathbf{kk}^{\prime}}(E)
G_{\mathbf{k}^{\prime }}(E),
$$
$$
T_{\mathbf{kk}^{\prime }}(E) =V+V^2\sum_{\mathbf{p}}G_{\mathbf{p}}(E)+...
\eqno{(5.35)}
$$
where $G$ are exact Green's functions for the ideal crystal. Then, at
neglecting vertex corrections, the transport relaxation time is determined
from the imaginary part of $T$-matrix
$$
\tau ^{-1}(E)=-2V^2\Im \sum_{\mathbf{p}}G_{\mathbf{p}}(E)=2\pi V^2N(E)
\eqno{(5.36)}
$$
which yields the required connection with the density of states.

\subsection{Electron-electron scattering}

The low-temperature resistivity of most transition metals is satisfactorily
described by the formula
$$
\rho =\rho _0+AT^2+BT^5
\eqno{(5.37)}
$$
According to [406], the fitting yields negligible values of the coefficients
at $T,T^3$ and $T^4$ terms (however, the $T^4$-term was observed in V and Ta
[407]). The $T^5$-term is to be attributed to electron-phonon scattering,
and the $T^2$-term, which dominates at $T<10$K, may be connected with
various mechanisms. The simplest among them is the electron-electron
scattering. The considerations of this mechanism were performed in a number
of papers [408-411]. It was demonstrated that in the case of a single group
of current carriers the scattering is possible provided that umklapp
processes are taken into account. These processes lead to an momentum
transfer from electrons to lattice so that the conservation law
\[
\mathbf{k}_1+\mathbf{k}_2-\mathbf{k}_1^{\prime }\mathbf{-k}_2^{\prime }=%
\mathbf{g}
\]
holds at non-zero reciprocal lattice vectors $\mathbf{g}$. The calculations
in the case of the screened Coulomb interaction, which use solving the
transport equation by the variational approach [7], yield the estimate for
the electron-electron resistivity
$$
\rho _{ee}\simeq \frac{\pi ^2z^{\prime }}{32}\frac{e^2}{v_FE_F}G^2\frac{gk_F%
}{\kappa ^2}\left( \frac{k_BT}{E_F}\right) ^2
\eqno{(5.38)}
$$
where $\kappa $ is the inverse screening radius, $z^{\prime }$ is the
nearest-neighbour number in the reciprocal lattice, $G$ is an analogue of
the atomic form factor (e.g., in X-ray scattering theory):
$$
G\simeq \frac 1{v_0}\int d\mathbf{r}e^{\mathbf{gr}}|u(\mathbf{r})|^2
\eqno{(5.39)}
$$
with $u(\mathbf{r})$ being the Bloch modulation function, $v_0$ the lattice
cell volume. The occurrence of the factor $(T/E_F)^2$ is connected with that
the scattering is possible in a narrow layer near the Fermi level with the
width of order of $T$ only. The formula (5.38) may be presented in the form
$$
\rho _{ee}=\left( \frac{e^2n\tau _e}{m^{*}}\right) ^{-1},\tau _e=\frac{\hbar
E_{kin}}{E_{coul}^2}\left[ \frac{\pi ^2z^{\prime }}{32}G^2\frac{gk_F}\kappa
\left( \frac{k_BT}{E_F}\right) ^2\right] ^{-1}
\eqno{(5.40)}
$$
with
\[
E_{kin}=\frac{\hbar ^2}{m^{*}a_0^2},\qquad
E_{coul}=\frac{e^2}{a_0}
\]
being effective kinetic and potential energies of the electron gas, $a_0$
the lattice constant. Estimating numerical coefficients in (5.38) for a
simple metal we obtain [7]
$$
\rho _{ee}\sim 5\cdot 10^{-3}G^2\left( \frac{k_BT}{E_F}\right) ^2\Omega
\cdot \mathrm{cm}
\eqno{(5.41)}
$$
This quantity would yield an appreciable part of the room-temperature
resistivity provided that $G\sim 1$. However, apparently $G\ll 1$ which is
confirmed by absence of the $T^2$-term in simple metals even at low
temperatures.

The contribution of electron-electron scattering with account of the $k$%
-dependence of the relaxation time was considered by Schroeder [411]. Then
the effect is determined by the quantity
$$
\mathbf{k}_1\tau (\mathbf{k}_1)+\mathbf{k}_2\tau (\mathbf{k}_2)-\mathbf{k}%
_1^{\prime }\tau (\mathbf{k}_1^{\prime })\mathbf{-k}_2^{\prime }\tau (%
\mathbf{k}_2^{\prime })
\eqno{(5.42)}
$$
which does not vanish at $\mathbf{g}=0$.

The role of electron-electron mechanism may increase in the presence of
several carrier groups where scattering is possible without Umklapp
processes. For spherical Fermi sheets of s- and d-electrons with the
effective masses $m_s,m_d$ an analogue of the expression (5.38) reads
$$
\rho _{ee}\simeq \frac{\pi ^4}{16}\frac{e^2}{v_sE_F^s}\frac{k_F}\kappa
\left( \frac{v_s-v_d}{v_d}\right) ^2\left( \frac{k_BT}{E_F^s}\right) ^2
\eqno{(5.43)}
$$
so that the small factor $G^2$ is absent, and there occurs a large factor 
$(v_s/v_d)^2$.

The expression (5.43) may explain the value of the observed $T^2$
-contribution in transition metals. The corresponding experimental situation
by the beginning of 70s is described in the review [406]. Measurement of
low-temperature resistivity of pure d-metals permitted to determine the
coefficients $A$ at the $T^2$-term and perform comparison with various
theoretical models. An important question is the correlation between $A$ and
electron density of states. This question was discussed by Rice [410]. As
follows from (5.43), $\rho _{ee}$ should be proportional to squared
d-electron mass, i.e to squared coefficient at linear specific heat $\gamma $.
Fig.5.9 displays the coefficient $A$ as a function of $\gamma ^2$ for some
d-metals. An approximate relation between these quantities is
$$
A(\mu \Omega \ \mathrm{ cm/K}^2)\simeq
0.4\cdot 10^{-6}[\gamma \ (\mathrm{mJ/mol} \cdot \mathrm{K}^2)]^2
\eqno{(5.44)}
$$
The correlation beween $A$ and $\gamma $ is also evident from the Table 5.1.
However, the value of the coefficient in (5.44) may vary considerably. For
Zr, Ta, Hf, W this coefficient makes up (2$\div $10)$\cdot $10$^{-6}$.
Possibly, this is connected with insufficient purity of the samples. For
very pure samples of W and Re ($\rho $(273K)/$\rho $(4.2K)$\sim $10$^4$-10$%
^5 $) one has $A\sim 10^{-6}\gamma ^2$.

More later investigations demonstrated that the scattering at the sample
surface may be important in pure metals provided that the mean free path is
large in comparison with sample size (e.g, with diameter of a wire). The
interference of electron-phonon and surface scattering results in occurrence
of a $T^2$-like contribution [414,8]. It was demonstrated in papers by
Volkenshtein et al [412] that the value of A in pure W, Re, Os is determined
by the dimensional effect, and the contribution of electron-electron
scattering is smaller than 0.05$\cdot $10$^{-12}\Omega \cdot $cm/K$^2$ In
impure samples (e.g., in Ta and V) a considerable contribution may originate
from scattering by thermal vibrations of impurity ions, according to the
theory [415].

At the same time, in a number of other pure transition metals the
coefficient $A$ remains large even after excluding the dimensional effect.
Especially surprising is the situation for Mo where $A=1.2\cdot
10^{-12}\Omega \cdot $cm/K$^2$ exceeds by several tens the value for an
analogous metal W. This strong difference may be hardly explained by the
electron-electron scattering since the values of $\gamma $ differ by 2.5
times only. Thus the question about origin of the $T^2$-contribution in
paramagnetic transition metals remains open. Further investigation of this
problem is prevented by absence of reliable experimental data on some
d-metals, in particular for two first columns of periodic table. Data for Sc
and Ti are absent, and the purity of investigated samples of Zr and Hf is
insufficient. Data for Y [412] yielded a large value of $A\sim
10^{-10}\Omega \cdot $cm/K$^2$. Since the value of $\gamma $ in Y is rather
high, this result supports the idea of $A-\gamma ^2$ correlation.

From the theoretical point of view, various interference mechanisms for
different scattering processes may be considered. However, more simple
interpretations of observed correlations between $\rho $ and other
characteristics seem to be not excluded.

As will be discussed in Sect.5.3.3, an alternative mechanism for $T^2$%
-dependence of resistivity in ferromagnetic metals is scattering by spin
waves. Separation of electron-electron and electron-magnon scattering is a
rather difficult problem. Since the $T^2$-term is comparable in
ferromagnetic and paramagnetic metals, main role is often attributed to the
electron-electron mechanism (see, e.g.,[406]). However, spin-density
fluctuations, which occur at finite temperatures even in paramagnets, should
result in scattering of current carriers due to exchange interaction. This
scattering may be treated as enhancement of the exchange part of
electron-electron scattering by spin fluctuations. Such an enhancement
should be appreciable in metals like Pd. Thus it is difficult to
discriminate electron-electron and spin-fluctuation contributions even in
principle. Quantitative theoretical description of spin-fluctuation effects
in elemental paramagnetic d-metals is now absent. Spin-fluctuation
resistivity enhancement in weakly and almost magnetic metals (Sect.4.4) is
considered in [416,26]. In weakly and nearly ferromagnetic metals we have
$$
\rho _{mag}\sim |1-IN(E_F)|^{-1/2}T^2
\eqno{(5.45)}
$$
In the antiferromagnetic case
$$
\rho _{mag}\sim |1-2I\chi _{\mathbf{Q}}|^{-1/2}T^2
\eqno{(5.46)}
$$
(cf.(4.65),(4.66)). At the boundary of the magnetic instability, where the
prefactors in (5.45), (5.46) diverge, the type of the temperature dependence
changes:
$$
\rho _{mag}\sim
\left\{
\begin{array}{ll}
T^{5/3} & ,\qquad FM \\
T^{3/2} & ,\qquad AFM
\end{array}
\right.
\eqno{(5.47)}
$$
These results are confirmed by data on resistivity of some d-compounds (see
[26]). A Stoner-type renormalization of electron-electron scattering
amplitude by spin-spin interactions and its role in resistivity of
transition metals is considered in [417].

\subsection{Mott s-d scattering mechanism}

An independent mechanism for resistivity of transition metals is the s-d
scattering considered by Mott [418]. This mechanism is based on the
assumption that an appreciable part of scattering corresponds to transitions
of main current carriers (s-electrons) to the unoccupied part of d-band, the
states of which possess much smaller mobility. Such interband transitions
may occur due to any scattering mechanism (impurities, phonons, spin
excitations etc.), their probability being large because of high density of
d-states near the Fermi level. At the same time, the inverse d-s transitions
may be neglected at calculating conductivity since $N_s(E_F)\ll N_d(E_F)$
(electron numbers in the subbands are restored mainly due to thermal
relaxation). Thus we may write down
\[
\sigma =\sigma _s+\sigma _d=\rho _s^{-1}+\rho _d^{-1}
\]
with
$$
\rho _s=\rho _{ss}+\rho _{sd},\qquad
\rho _d=\rho _{dd}+\rho _{ds}\simeq \rho _{dd}
\eqno{(5.48)}
$$
The Mott's model is widely used to explain concentration dependences of
resistivity of transition metal alloys. In particular, the resistivity
decreases at filling of d-shells of transition metal ions by electrons of
another alloy component, which results in prohibition of s-d transitions.

The s-d scattering owing to phonons was also considered by Wilson [479,2].
In the case $T\gg \theta _D$ he obtained the result
$$
\rho _{sd}^{ph}=\left( \frac 3{4\pi }\right) ^{1/3}\frac{m_s^{1/2}m_d}{%
e^2nMa_0}\frac{(E_F^s)^{3/2}}{k_B\theta _D^2}T\left( 1-\frac{\hbar s\Delta
k_{sd}}{k_B\theta _D}\right)
\eqno{(5.49)}
$$
where $s$ is the sound velocity, $\Delta k_{sd}=k_F^s-k_F^d$ is the minimum
scattering quasimomentum of a phonon, as determined by the momentum
conservation law. Temperature corrections to (5.49) are obtained by the
expansion of the Fermi distribution functions which yields
$$
\rho _{sd}^{ph}=aT[1-b\left( \frac{k_BT}{E_F}\right) ^2]
\eqno{(5.50)}
$$

The formulas (5.49), (5.50) may explain values and temperature dependence of
transition metal resistivity at high $T$. On the other hand, at low $T$
phonons with small quasimomenta play the dominant role, so that we should
have
$$
\rho _{sd}^{ph}\sim \exp \left( -\frac{\hbar s\Delta k_{sd}}{k_BT}\right)
\eqno{(5.51)}
$$
(if one takes into account the overlap of s- and d-sheets of the Fermi
surface, $\rho _{sd}^{ph}\sim T^5$). However, such a strong decrease is not
observed experimentally, which leads to difficulties of the theory. In
particular, Wilson [2] claimed that s-d transitions do not play an important
role in the resisitvity of transition metals.

Generally, s-d transitions may take place for all scattering mechanisms and
are especially important for elastic processes. At low temperatures the
resistivity is determined mainly by impurity scattering. Due to strong
energy dependence of density of states in d-band, this contribution may
exhibit a considerable temperature dependence according to (5.34). For the
parbolic s- and d-bands one obtains [8]
$$
\rho _{sd}^i(T)=\rho _{sd}^i(0)\left[ 1-\frac{\pi ^2}6\left( \frac{k_BT}{%
E_d-E_F}\right) ^2\right]
\eqno{(5.52)}
$$
Unlike the electron-electron contribution $\rho _{ee}$, $\rho _{sd}$ is
proportional to the density of states of d-electrons at $E_F$ and
consequently to first power of $\gamma $. This difference may be used to
separate the s-d scattering contribution. One can see from Table 5.1 that
the ratio of the $T$-linear resistivity term to g in the beginning of
periods (of order of 10) exceeds considerably that in the end of periods.
Possibly this tendency is connected with different role of s-d transitions.
This seems be important for Sc and Ti columns, decreases appreciably in the
V column and further on (as well as for other properties, the Mn column
makes up an exception). Such a behaviour agrees with that of the coefficient
$A$ at the $T^2$-term and reflects the tendency towards lowering and
narrowing of d-bands to the end of periods. The narrowing may result in
hampering s-d transitions because of energy and quasimomentum conservation
laws.

\subsection{Resistivity of magnetic metals}

Existence of magnetic moments in transition elements results in additional
factors which influence the behaviour of current carriers in external
electric field. Firstly, thermal fluctuations in the system of magnetic
moments provide a new scattering mechanism owing to s-d exchange
interaction. Secondly, electron spectrum of magnetic crystals depends
appreciably on spontaneous magnetization (or sublattice magnetization in
antiferromagnets) and, consequently, on temperature.

The first effect may be described by introducing an additional contribution
to resistivity,
$$
\rho _{tot}=\rho _{}+\rho _{mag}
\eqno{(5.53)}
$$
The second effect cannot be described by the simple expression (5.53). In
the simplest case where the influence of magnetic ordering is small we may
perform the expansion in magnetization to obtain
$$
\rho _{tot}(M)=\rho _{tot}(0)+aM^2
\eqno{(5.54)}
$$
Unlike (5.53), the sign of the second term in (5.54) needs not to be
positive. The expansion (5.54) does not hold in the cases where the gap in
the spectrum modifies strongly the states near the Fermi level. In such
situations the occurence of magnetic splitting may result in considerable
anomalies of resistivity and other transport properties at the magnetic
ordering point.

For the antiferromagnetic structure with the wavevector $\mathbf{Q}$, the
disturbation of electron spectrum by magnetic ordering is especially strong
at $\mathbf{2k=Q}$. Then we obtain from (G.70)
$$
E_{\mathbf{k}}^{1,2}=t_{\mathbf{k}}\pm |I\overline{S}|
\eqno{(5.55)}
$$
This disturbation may influence strongly transport properties provided that
the Fermi level coincides with the antiferromagnetic gap. The corresponding
resistivity anomaly at the Neel point was discussed in [420]. The result in
the mean-field approximation has the form
$$
\rho (T)=aT+b[1-\overline{m}^2(T)]+c\overline{m}^2(T)T
\eqno{(5.56)}
$$
where
\[
\overline{m}(T)=\overline{S}/S\sim (1-T_N)^{1/2}
\]
is the relative sublattice magnetization. With lowering temperature, this
contribution increases rapidly at passing $T_N$ and may lead to a $\rho (T)$
maximum. Such a behaviour is observed in $\alpha $-Mn (Fig.5.10), Cr
(Fig.5.3) and rare-earth metals [265,406]. An alternative explanation of the
maximum is based on the critical scattering near the second-order magnetic
transition point. Investigations of Dy and Ho in strong magnetic fields
[421] demonstrated that the resistivity decreases sharply at field-induced
transition into ferromagnetic state with disappearance of the spiral
magnetic superstructure.

Consider exchange scattering of conduction electrons by spin disorder within
the s-d exchange model. The result for the magnetic resistivity at high
temperatures in the mean-field approximation is given by (M.84). For spin $S
= 1/2$ it takes the form
$$
\rho _{mag}=\frac{9\pi }2\frac{m^{*}}{ne^2}\frac{I^2}{E_F}(\frac 14-%
\overline{S}^2)
\eqno{(5.57)}
$$
In far paramagnetic region we have for arbitrary $S$
$$
\rho _{mag}=\frac{3\pi }2\frac{m^{*}}{ne^2}\frac{I^2}{E_F}S(S+1)
\eqno{(5.58)}
$$
The result of the type (M.84) was first obtained by Kasuya [422]. It
explains rather well experimental data on the temperature dependence of
resistivity of ferromagnetic metals near the Curie point. For rare earth
metals, the expression (5.58) with the replacement
\[
S(S+1)\rightarrow (g-1)^2J(J+1)
\]
describes satisfactory the change of high-temperature spin-disorder
resisitivity in the 4f-series [16].

Consider magnetic scattering at low temperatures. Passing to magnon
operators with the use of the Holstein-Primakoff representation (E.1) we
obtain from (5.23)
$$
\rho =\frac{\pi k_BT}{\langle j_x^2\rangle }2I^2Se^2\sum_{\mathbf{kq}}(v_{%
\mathbf{k\uparrow }}^x-v_{\mathbf{k+q\downarrow }}^x)^2N_{\mathbf{q}}n_{%
\mathbf{k\uparrow }}(1-n_{\mathbf{k+q\downarrow }})\delta (E_{\mathbf{%
k\uparrow }}-E_{\mathbf{k+q\downarrow }}+\omega _{\mathbf{q}})
\eqno{(5.59)}
$$
Integrating over $\mathbf{k}$ we derive the result for resistivity
$$
\rho =C_1T^2\int_{T_0/T}^\infty \frac{xdx}{\sinh x}+C_2T_0T\ln \coth \frac{%
T_0}{2T}
\eqno{(5.60)}
$$
where the constants $C_i$ are determined by the electron spectrum, $C_2$
being non-zero only for a non-parabolic electron spectrum, the quantity
$$
T_0\sim T_Cq_0^2\sim (I/E_F)^2T_C
\eqno{(5.61)}
$$
coincides with the boundary of the Stoner continuum $\omega _{-}$ (Appendix
G.1), $q_0=2|IS|/v_F$ is the threshold vector for the one-magnon scattering
processes. At very low temperatures $T<T_0$ the one-magnon resistivity
(5.60) is exponentially small since the quasimomentum and energy
conservation law cannot be satisfied at characteristic thermal magnon
quasimomenta. At $T\gg T_0$ we have
$$
\rho _0(T)\sim T^2N_{\mathbf{\uparrow }}(E_F)N_{\mathbf{\downarrow }}(E_F)
\eqno{(5.62)}
$$
(the second term in (5.60) yields small corrections of order $T_0T\ln T$ and
is usually neglected).The same results follow from the solution of the
transport equation (Appendix M.3). Thus spin-wave scattering in the wide
temperature interval $T_0<T<T_C$ results in the square temperature
dependence of resistivity. The difference from the electron-phonon
scattering (see (5.28)) is explained by the square disperson law of magnons,
so that their number is proportional to $T^{3/2}$ rather than $T^3$. 

The $T^2$ -dependence was obtained by Turov [425] and Kasuya [426], and
further confirmed by many authors. However, at very low temperatures in
ferromagnetic transition metals there exist contributions to resistivity,
which are proportional to $T^{3/2}$ or $T$ [265,406]. Linear temperature
corrections owing to relativistic interactions were found in [425]. However,
such corrections are too small to explain the experimental data. An attempt
was made in [288] to explain the $T^{3/2}$-term by the non-quasiparticle
contributions to the impurity resistivity which occurs due to strong energy
dependence of incoherent states near the Fermi level. Indeed, taking into
account (5.36) we obtain the correction to conductivity
\[
\delta \sigma (E)\sim -V^2\int dE\left( -\frac{\partial f(E)}{\partial E}%
\right) \delta N(E)\sim -T^{3/2}
\]
It should be noted that the $T^2$-term (5.62) is absent in the case of a
half-metallic ferromagnets (HMF, Sect.4.5) where the states with one spin
projection only exist at the Fermi level and one-magnon scattering processes
are forbidden in the whole spin-wave region. This seems to be confirmed by
comparing experimental data on resistivity of Heusler alloys TMnSb (T =
Ni,Co,Pt,Cu,Au) and PtMnSn [331]. The $T^2$-contribution from one-magnon
processes to resistivity for half-metalic systems (T = Ni, Co, Pt) was
really not picked out, whereas the dependences $\rho (T)$ for ``usual''
ferromagnets were considerably steeper (Fig.5.11). In the HMF situation, as
well as for usual ferromagnets at $T<T_0$, the resistivity is determined by
two-magnon scattering processes. These result in the weak $T^{7/2}$%
-dependence of resistivity [427] (see also [428]) which is due to vanishing
of the electron-magnon scattering amplitude at zero magnon wavevector
(Appendix G.1).

Consider the situation in rare-earth metals which are ferromagnetic at low
temperatures. Because of strong anisotropy, the dispersion law of spin waves
differs from that in d-metals. The magnon spectrum in rare earths contains a
gap of order $T^{*}\sim 10$K; in the absence of anisotropy in the basal
plane we have the linear law $\omega _{\mathbf{q}}\sim q$. The gap results
in occurence of an exponential factor $\exp (-T^{*}/T)$ in the magnetic
resistivity. For the linear dispersion we obtain dependence $\rho \sim T^4$
[429] instead of $T^2$ since each power of $q$ yields at integration a
factor of $T/T_C$ (instead of $(T/T_C)^{1/2}$ at $\omega _{\mathbf{q}}\sim
q^2$). The latter result was confirmed by the experimental dependence $\rho
\sim T^{3.7}$ for gadolinium in the region 4-20K [430].

Using the formula (5.23) with the Hamiltonian of the s-d model (G.2) in the
spin-wave region, we obtain for low-temperature magnetic resistivity of
antiferromagnetic metals
$$
\rho =\frac{\pi k_BT}{\langle j_x^2\rangle }2I^2Se^2\sum_{\mathbf{kq}}(v_{%
\mathbf{k}}^x-v_{\mathbf{k+q}}^x)^2n_{\mathbf{k}}(1-n_{\mathbf{k+q}})[N_{%
\mathbf{q}}(u_{\mathbf{q}}+v_{\mathbf{q}})^2)
$$
$$
\times \delta (E_{\mathbf{k}}-E_{\mathbf{k+q}}+\omega _{\mathbf{q}})+N_{%
\mathbf{q+Q}}(u_{\mathbf{q+Q}}-v_{\mathbf{q+Q}})^2)\delta (E_{\mathbf{k}}-E_{%
\mathbf{k+q}}+\omega _{\mathbf{q+Q}})]
\eqno{(5.63)}
$$
The resistivity at very low temperatures is determined by contributions of
small $q$ in (5.63), i.e. by transitions inside antiferromagnetic subbands
(Appendix G.2). Due to the linear dispersion law of magnons, such
transitions result, as well as electron-phonon scattering, in a $T^5$%
-dependence of resistivity [428]. (The earlier result $\rho \sim T^4$ [426]
was erroneous since the coefficients of the Bogoliubov transformation
(E.10), which have an essential $q$-dependence, were not taken into
account.) Because of singularity in $uv$-transformation coefficients, the
contribution from the region of small $|\mathbf{q-Q}|$ (i.e. of the
intersubband contibutions) is, generally speaking, larger. However, as well
as for ferromagnets, it is impossible to satisfy the quasimomentum
conservation law at $\mathbf{q\rightarrow Q}$ because of antiferromagnetic
splitting, so that this contribution is in fact exponentially small at
$$
T<T_0=\omega (q_0)\sim (|IS|/E_F)T_N
\eqno{(5.64)}
$$
with $q_0=2|IS|/v_F$ being the threshold value of $|\mathbf{q-Q}|$. (Note
that the boundary temperature is not so small as for a ferromagnet, (5.61).)
At higher temperatures $T>T_0$ the intersubband contributions yield the $T^2$%
-behaviour of resistivity [433]. In the two-dimensional situation these
contributions become $T$-linear which may explain the characteristic
dependence $\rho (T)$ in high-$T_c$ superconductors.

\subsection{Resistivity of transition metal alloys}

Investigation of transport phenomena in alloys as a function of
concentration of various components gives a possibilty to obtain information
on their electron structure. In particular, data on concentration dependence
of residual resistivity $\rho (c)$ in transition metal alloys yields an
important information on change of d-states.

In disordered alloys of simple metals with the same valence (metals with
different valence do not form usually the continuous solid solution series),
the Nordheim rule usually holds
$$
\rho (c) =\rho _0c(1-c)
\eqno{(5.65)}
$$
$$
\frac 1{\rho _0}\left( \frac{d\rho (c)}{dc}\right) _{c=0,1} =\pm 1,\qquad
\frac 1{\rho _0}\left( \frac{d^2\rho (c)}{dc^2}\right) _{c=0,1}=\pm 2
$$
In TM alloys the symmetry of $\rho (c)$ curve is violated (see Fig.5.12). An
explanation of this violation can be obtained within the Mott s-d transition
model. To demonstrate this we write down the conductivity of the alloy A$%
_{1-c}$B$_c$ , with A being a transition metal and B a simple metal, in the
form
$$
\sigma =\sigma _s+\sigma _d\simeq \sigma _s=\rho _s^{-1},\qquad
\rho _s=\rho _{ss}+\rho _{sd}
\eqno{(5.66)}
$$
Taking into account the relations
$$
\rho _{ss}=\rho _s^0c_B(1-c_B),\qquad
\rho _{sd}=ac_B(1-c_B)N_d(E_F)
\eqno{(5.67)}
$$
and the concentration dependence
$$
N_d(E_F)\simeq c_B(1-c_B)N_A(E_F)
\eqno{(5.68)}
$$
we obtain [434]
$$
\rho \simeq \rho _s=\rho _s^0c_B(1-c_B)+\widetilde{a}c_B(1-c_B)^2
\eqno{(5.69)}
$$
which yields a deviation from the Nordheim rule.

Other factors which lead to violation of this rule may exist. So, for Cu-Ni
alloys, d-electrons are at localized levels for small nickel concentrations
and form band states for $c > 40\%$ only [435]. At the critical
concentration, an abrupt change of the alloy properties takes place.

\subsection{Two-current model of ferromagnetic metals}

The two-current models considers two types of current carriers in an
itinerant ferromagnet which have different spin projections. Unlike the Mott
model, where the current of d-states is neglected, contributions of both
types of states are comparable. Effects of strong spin polarization are
especially important for d-bands, but may be appreciable for s-type current
carriers too.

The phenomenological treatment of the model is rather simple [436]. The
total current and, consequently, conductivity, are represented as a sum of
contributions from majority and minority current carriers:
$$
j=\sigma E,\qquad
\sigma =\sigma _{\uparrow }+\sigma _{\downarrow }
\eqno{(5.70)}
$$
Then the resistivity takes the form
$$
\rho =\sigma ^{-1}=\left( \frac 1{\rho _{\uparrow }}+\frac 1{\rho
_{\downarrow }}\right) ^{-1}=\frac{\rho _{\uparrow }\rho _{\downarrow }}{%
\rho _{\uparrow }+\rho _{\downarrow }}
\eqno{(5.71)}
$$
which corresponds to a parallel junction. Further we have to take into
account transitions between both the types of carriers. These are due to
spin-flip scattering processes by spin inhomogeneities, spin waves, magnetic
impurities etc. Then we have
$$
\rho _{\uparrow }=\rho _{\uparrow }^0+\rho _{\uparrow \downarrow },\qquad
\rho _{\downarrow }=\rho _{\downarrow }^0+\rho _{\downarrow \uparrow }
\eqno{(5.72)}
$$
(consecutive junction). On substituting (5.72) into (5.71) we obtain
$$
\rho =\frac{\rho _{\uparrow }^0\rho _{\downarrow }^0+\rho _{\downarrow
}^0\rho _{\uparrow \downarrow }+\rho _{\uparrow }^0\rho _{\downarrow
\uparrow }+\rho _{\uparrow \downarrow }\rho _{\downarrow \uparrow }}{\rho
_{\uparrow }^0+\rho _{\uparrow }^0+\rho _{\uparrow \downarrow }+\rho
_{\downarrow \uparrow }}
\eqno{(5.73)}
$$
Such an approach may be applied to describe transport effects in
ferromagnetic transition metals and their alloys. Unfortunately, reliable
microscopic calculations of the quantities which enter (5.73) are hardly
possible, and it is more convenient to determine them from experimental
data. One uses often the data on the deviation from the Matthiessen rule.

To illustrate this approach we consider a ternary alloy M$_{1-x-y}$ A$_x$ B$%
_y$. We have for the impurity resistivity ($\rho _{\uparrow \downarrow
}=\rho _{\downarrow \uparrow }=0$)
$$
\rho _{AB}=\rho _A+\rho _B+\Delta \rho
\eqno{(5.74)}
$$
where $\rho _{A,B}$ are the resistivities of the corresponding binary alloys,
$$
\Delta \rho =(\alpha _A-\alpha _B)^2\rho _A\rho _B[(1+\alpha _A)^2\alpha
_B\rho _A+(1+\alpha _B)^2\alpha _A\rho _B]^{-1}
\eqno{(5.75)}
$$
with
\[
\alpha _{A,B}=\rho _{A,B\downarrow }/\rho _{A,B\uparrow }
\]
Varying the concentrations $x$ and $y$ one can experimentally determine the
parameters $\alpha _{A,B}$.

At finite temperatures, additivity of the impurity residual resistivity $%
\rho _i$ and $T$-dependent contributions for a binary alloy is also violated:
$$
\rho _{\uparrow ,\downarrow }(T) =\rho _{i\uparrow ,\downarrow }
+\rho _{l\uparrow ,\downarrow }(T)
\eqno{(5.76)}
$$
$$
\rho (T) =\rho _i\left[ 1+\left( \frac{\alpha -\mu }{1+\alpha }\right)
^2\right] +\rho _l(T)+\left( \frac{\alpha -1}{\alpha +1}\right) ^2\rho
_{\uparrow \downarrow }(T)
$$
with
$$
\alpha  =\rho _{i\downarrow }/\rho _{i\uparrow },\qquad
\mu _l=\rho _{l\downarrow }/\rho _{l\uparrow },
\eqno{(5.77)}
$$
$$
\rho _i =\frac{\rho _{i\uparrow }\rho _{i\downarrow }}{\rho _{i\uparrow }
+\rho _{i\downarrow }},\qquad
\rho _l=\frac{\rho _{l\uparrow }\rho _{l\downarrow }}{\rho _{l\uparrow }
+\rho _{l\downarrow }}
$$
where $l$ stands for the index of a concrete temperature-dependent
scattering mechanism.

Using experimental data on ternary and binary alloys, parameters $\alpha _A$,
$\alpha _B$, $\mu _l$, $\rho _i$, $\rho _l(T)$ and $\rho _{\uparrow
\downarrow }$ were determined for the ferromagnetic transition metals [436].
In a number of cases, the parameters $\alpha $ deviate strongly from unity.
As an example, data on the alloy {\itshape{Ni}\/}Co$_{1-x}$Rh$_x$ are shown in
Fig.5.13. Due to strong non-linearity of the $\rho (x)$ dependence, the
parameters $\rho _{\uparrow }$ and $\rho _{\downarrow }$ turn out to be
considerably different. Note that the deviations from the linear dependence
at small $x$ cannot be explained by the Nordheim rule.

In a metal with transition metal impurities the scattering of conduction
electrons into d-resonance impurity states near the Fermi level plays the
dominant role. The behavior of the quantity $\alpha $ in the 3d-impurity
series for Fe and Ni hosts is presented in the Table:
$$
\begin{tabular}{lllllllll}
T & Ti & V & Cr & Mn & Fe & Co & Ni & Cu \\ 
$\alpha $ ({\itshape{Ni}\/}T) & 1.9 & 0.52 & 0.3 & 10 & 15 & 20 & - & 3.4 \\ 
$\alpha $ ({\itshape{Fe}\/}T) & 0.38 & 0.12 & 0.22 & 0.13 & - & 2.1 & 5 & -
\end{tabular}
$$
As discussed in [436], qualitatively this behavior may be
explained on the basis of the Friedel concept of the virtual bound d-state.
The spin-up nickel d-band is practically filled, and only spin down states
may screen the impurity charge perturbation. For early 3d-impurities (Cr, V,
Ti) in Ni the spin-up impurity d-state is repelled above the spin-up d-band,
so that the impurity density of states $g_{d\uparrow }(E_F)$ is rather
large, the magnetic moment being opposite to the host one. At the same time,
for strongly magnetic Co, Fe and Mn impurities only s-states are present at $%
E_F$ for spin up. The value of $g_{d\downarrow }(E_F)$ is rather large for
all the 3d-impurities in the Ni host. Thus $\alpha $ strongly increases at
passing from the first half of 3d-series to the second one.

Magnetic moments of 4d-impurities are created mainly by host magnetization,
the magnetization perturbation being strongly delocalized. The behavior of $%
\alpha $ for the first half 4d-series impurities in Ni is similar to that in
3d-series.

In pure iron the Fermi level is just below the spin-up d-band. Therefore in
the Fe host the repulsive impurity potential pushes up the spin up d-level
for Mn, Cr, V and Ti impurity series through the Fermi energy, the magnetic
moment decreasing. At the same time, for Co and Ni impurities $g_{d\uparrow
}(E_F)$ is small. For all the 3d-impurities in iron $g_{d\downarrow }(E_F)$
is rather small. The qualitative consideration of the problem may be
performed also within the Anderson impurity model (see [717]).
Quantitatively, the local density of states and other characteristics of the
impurity electronic structure are obtained from band calculations. Such
calculations were performed for a large number of impurities (including 3d-
and 4d-impurities) in the Ni and in the Fe host [718-722]. One can see from
the results that even for 3d-impurities in Ni the simple Anderson-model
picture is in fact poor: the local impurity DOS $g_{d\sigma }(E)$ is
strongly influenced by the host. Besides that, the Fe host, which has
(unlike the Ni host) a large magnetization, determines in a great measure
formation of magnetic moment even for later 3d-impurities. The 4d-impurity
states are strongly hybridized with the valence host states so that the
picture of narrow virtual bound impurity d-state is not applicable. This is
especially obvious for Y, Zr and Nb impurities in Ni where $g_\sigma (E_F)$
for both $\sigma $ are quite small and differ by a small exchange spin
splitting only [719]. However, for Tc to Pd impurities rather sharp
hybridization peaks are present. As compared to Mn, Fe and Co, the spin up
peaks for Tc, Ru and Rh are broader and extent somewhat above $E_F$ . The
most consistent calculation of resistivity may be performed on the basis of
the scattering phase shift analysis on the base of band structure
calculations. Such calculations were carried out for d-impurities in Cu
[723], the role of magnetic moments being neglected, and in Ni [724]. In the
latter case the agreement with experiment was not quite satisfactory. A
simplified estimations in the case of a ferromagnetic host with appreciable
magnetization perturbations was performed in [717] by using the results of
band structure calculations and the Friedel sum rule.

A general (with neglecting crystal-field effects) expression for resistivity
per impurity in terms of the phase shifts $\eta _{l\sigma }$ has the form
$$
\rho _\sigma =\rho _{u\sigma }\sum_l(l+1)\sin ^2(\eta _{l\sigma }-\eta
_{l+1,\sigma })  
\eqno{(5.78)}
$$
where $\rho _{u\sigma }$ is the resistivity unitarity limit for a given spin
projection per scattering channel (see Sect.6.1). In the approximation of
free conduction electrons we have 
\[
\rho _{u\sigma }=2m^{*}/\pi z_\sigma eg_\sigma 
\]
where $m^{*},z_\sigma $ and $g_\sigma $ are the effective mass,
concentration and density of states of conduction electrons at the Fermi
level for a given spin projection, $e$ is the electronic charge. To estimate
the phase shifts we use for each spin projection (spin-flip processes are
neglected) the Friedel sum rule 
$$
\Delta n_\sigma =\frac 1\pi \sum_l(2l+1)\eta _{l\sigma }  
\eqno{(5.79)}
$$
The changes of electron numbers are determined by 
\[
\Delta n_{\uparrow ,\downarrow }=\frac 12(\Delta Z\pm \Delta M)
\]
where $\Delta Z$ is excess charge introduced by the impurity ion, i.e. the
difference between impurity and host atomic numbers, $\Delta M$ is the total
magnetization change induced by the impurity (in $\mu _B$). Similar to the
standard Friedel approach for non-magnetic hosts, only d-scattering ($l=2$)
contribution can be taken into account. This is generally speaking,
sufficient for rough estimations. However, in a number of cases
s,p-contributions play an important role. In particular, for the impurities
in the Ni host, which do not destroy strong ferromagnetism ($\Delta
Z=-\Delta M$), the density of spin-up d-states near the Fermi level is
rather small, and their disturbation by impurity is practically absent
[720]. A similar situation takes place for d-impurities from the beginning
of d-series where d-states are almost empty and $\Delta Z+\Delta M=-10$. The
sp-contributions are also important for non-magnetic sp-impurities which
introduce a strong charge perturbation.

The values of $\Delta M$ and some data on the partial contributions $\Delta
n_{l\sigma }=(2l+1)\eta _{l\sigma }/\pi $ may be obtained from the band
calculations results [720] for the Ni host and in [721] for the Fe host.
First we discuss the case of the Ni host. To perform calculations we have to
specify the values of $\rho _{u\sigma }$. For the Cu host one puts usually $%
\rho _u=\rho _{u\sigma }/2=3.8\mu \Omega $ cm/at\% [725]. As demonstrate
band structure calculations [24], the value of total density of states with
spin up for metallic nickel is two times larger, but the s-contribution is
considerably smaller than for Cu: although the spin up d-subband is
practically filled, its tail dominates at $E_F$, so that $g_{d\uparrow
}/g_{s\uparrow }\sim 10$. Still larger d-contributions occur for spin down
where a large density of states peak is present at $E_F$ and $g_{d\downarrow
}/g_{s\downarrow }\sim 100$. Thus, although d-electrons are usually assumed
to possess small mobility, the problem of their contribution to conductivity
should be investigated.

For non-magnetic impurities Cu, Zn, Ga, Ge in Ni the partial values of $%
\Delta n_\sigma $ may be estimated from results of Ref.720. The calculations
according to (5.78) with $\rho _{u\uparrow }=15\mu \Omega $ cm/at\% yield
satisfactory values of $\rho $. However, the estimation of $\rho
_{\downarrow }$ with the same value of $\rho _{u\downarrow }$ yields very
large values of $\alpha $. To reduce $\alpha $ up to reasonable values we
have to put $\rho _{u\uparrow }/\rho _{u\downarrow }\simeq 5$.

Thus the estimations of Ref.[717] lead to the conclusion that d-electrons
make an important contribution to conductivity, especially to the spin-down
current. Indeed, the s-electron contribution cannot provide such a large
ratio $\rho _{u\uparrow }/\rho _{u\downarrow }$; besides that, this is
expected to yield an opposite tendency since $g_{s\uparrow }>g_{s\downarrow }
$ [24]. Note that in the situation of several conduction electron groups the
quantities $\rho _{u\sigma }$ should be considered as phenomenological
fitting parameters. Similar statement about the important role of d-states
in electronic transport was made by Kondorsky [502,503] on the basis of data
on the anomalous Hall effect.

Now we consider briefly the case of the Fe host. Unlike nickel, iron is not
a strong ferromagnet since d-states with both spin projections are present
at $E_F$. Moreover, the Fermi level lies in a pseudogap for the spin down
d-states. Therefore one may expect that spin up d-states should make a
larger contribution to the transport properties (besides that, band
calculations [24] give a strong spin polarization of sp-electrons at $E_F$).
Indeed, for {\itshape{Fe}\/}Ni and {\itshape{Fe}\/}Co systems we have 
$\Delta M>0$, but $\alpha >1$, 
so that we may estimate $\rho _{u\uparrow }/\rho _{u\downarrow }=0.2$.

At present, the two-current model was also applied to consider anisotropy of
electrical resistivity in magnetic field, temperature dependence of the
normal Hall effect, thermoelectric power, magnetoresistivity, transport
effects in bulk samples [726] and multilayers (especially, giant $\Delta
\rho /\rho $-effect [437]).

\section{Thermoelectric power}

In the presence of a temperature gradient, the electric and heat currents
are given by the linear relations (5.1), (5.2). For $j=0$ we obtain
$$
E=-\frac \lambda \sigma \,\mathrm{grad}T\equiv \alpha \,\mathrm{grad}T
\eqno{(5.80)}
$$
where $\alpha $ is the absolute differential thermoelectric power.

The coefficients in (5.1), (5.2) are determined by the disturbance of the
equilibrium distribution function by external fields. Provided that the
$\mathbf{k}$-dependent relaxation time $\tau $ may be introduced,
$$
f_{1\mathbf{k}}=-\tau _{\mathbf{k}}v_{\mathbf{k}}\left[ eE+(\varepsilon _{%
\mathbf{k}}-E_F)\frac 1T\mathrm{grad}T\right]
\eqno{(5.81)}
$$
Using the expressions for electric and heat current, we obtain
\[
\sigma =e^2K_0,\qquad
\lambda =\nu /T=eK_1/T
\]
where
$$
K_n=-\frac 13\sum_{\mathbf{k}}v_{\mathbf{k}}^2\tau _{\mathbf{k}}(\epsilon _{%
\mathbf{k}}-E_F)^n\frac{\partial n_{\mathbf{k}}}{\partial \varepsilon _{%
\mathbf{k}}}
\eqno{(5.82)}
$$
Then we have
$$
\alpha =-\frac{K_1}{eTK_0}
\eqno{(5.83)}
$$
After expanding (5.83) in $T/E_F$ up to the second order we obtain
$$
\alpha =-\frac \pi 3\frac{k_B^2T}e\left( \frac{\partial \ln \sigma (E)}{%
\partial E}\right) _{E=E_F}
\eqno{(5.84)}
$$
where $\sigma (E)$ is the conductivity as a function of the Fermi level
position (see (5.31),(5.32)). Thus the thermopower is expressed in terms of
conductivity and its energy derivative.

The sign of $\alpha$ is determined by the sign of electric charge (or
effective mass). In particular, it should be reversed when the Fermi level
approaches the Brillouin zone boundaries (a becomes positive).

In the simplest approximation we may write down
$$
\sigma (E)=e^2n(E)\tau (E)/m^{*}
\eqno{(5.85)}
$$
where $n(E)$ is the number of electrons in the $\mathbf{k}$-space under the
surface with a given $E=E_F$, so that $dn(E)/dE$ equals to the density of
states $N(E)$. Then we have
$$
\alpha =-\frac{\pi ^2}3\frac{k_B}ek_BT\left[ \frac{N(E)}{n(E)}+\frac 1{\tau
(E)}\frac{\partial \tau (E)}{\partial E}\right] _{E=E_F}
\eqno{(5.86)}
$$
The expression (5.86) contains the concentration contribution which is
determined by the number of electrons, and the relaxation contribution which
depends on the function $\tau (E)$. The value of the first contribution is
estimated as
$$
\alpha \sim -\frac{\pi ^2}3\frac{k_B}{ne}\frac{k_BT}{E_F}\approx -0.9\cdot
10^2\frac{k_BT}{E_F}\frac{\mu \mathrm{V}}{\mathrm{K}}
\eqno{(5.87)}
$$
which agrees roughly with experimental data ($\alpha \sim $ 1$\mu $V/K at $%
k_BT/E_F\sim 10$). At the same time, in semiconductors 
$\alpha$ does not contain
the small factor $k_BT/E_F$ and is considerably larger.

The dependences $\tau (E)$ are different for various scattering mechanisms.
At high temperatures,
\[
\tau (E)\sim E^{3/2},\qquad
n(E)\sim E^{3/2},\qquad
\sigma(E)\sim E^3
\]
At low temperatures, where impurity scattering dominates and the electron
mean free path is constant,
\[
\tau (E)\sim E^{-1/2},\qquad
\sigma (E)\sim E
\]
Then (5.84) yields
$$
\alpha =-\frac{\pi ^2k_B^2T}{eE_F}\cdot \left\{
\begin{array}{ll}
 1   &,\qquad T>\theta _D \\
 1/3 &,\qquad T\ll \theta _D
\end{array}
\right.
\eqno{(5.88)}
$$
The relation (5.88) gives reasonable results for Na and K which are
described by the free electron model, so that $\alpha $ and the Hall
coefficient are negative. However, generally speaking, such simple
dependences do not reproduce experimental data even for simple metals (see
Fig.5.14). One can see that $\alpha $ may become positive and in most cases
the dependences $\alpha (T)$ are non-monotonous. The high-temperature
behaviour is usually attributed to the effect of phonon drag [8]. The
corresponding contribution in the free electron model may be presented in
the form
$$
\alpha _{ph}=-\frac{c_{ph}}{3ne}=\frac 13c_{ph}R
\eqno{(5.89)}
$$
where $c_{ph}$ is the lattice specific heat, $R$ is the Hall coefficient.

The dependences $\alpha (T)$ in transition metals are still more
complicated. Especially large values of $\alpha$ are observed in Pd and Pt.
An important fact is a similarity in the $\alpha (T)$ behaviour within each
column of the periodic table. The generalized dependences $\alpha (T)$ are
shown in Fig.5.15. A correlation between signs of $\alpha$ and the Hall
coefficient is as a rule absent, which demonstrates inapplicability of
simple theories to these quantities in transition metals. In some cases
(e.g., for La near the a-b transformation and for Ti near 500K), a
simultaneous sign change takes place in $R(T)$ and $\alpha (T)$. In other
cases (e.g., for Sc and Hf) the sign inversion temperatures are considerably
different. Sometimes the change of $\alpha (T)$ sign is not accompanied with
that for $R(T)$ (however, one has to bear in mind that $R(T)$ is measured as
a rule in a more narrow temperature interval). As well as other transport
properties, thermoelectric power exhibits also anomalies at the magnetic
ordering points (Fig.5.16).

To explain complicated $\alpha (T)$ behavior one has apparently to take into
account presence of several current carrier groups and scattering
mechanisms. One of the possible approaches is using the Mott model of s-d
transitions. In this model, main contribution to scattering is connected
with transitions of s-electrons into the d-band. The corresponding
relaxation time is proportional to the inverse density of d-states. Assuming
that the relaxation contribution in (5.86) dominates we derive
$$
\alpha (T)=-\frac{\pi ^2}{3e}k_B^2T\left( \frac{\partial \ln N_d(E)}
{\partial E}\right) _{E=E_F}
\eqno{(5.90)}
$$
The expression (5.90) is used frequently to describe the data on the
thermopower for metals in the end of periods (e.g., Pd and Pt) and for
alloys like Cu-Ni, Pd-Ag. It reproduces satisfactorily the temperature and
concentration dependences of 
$\alpha$ in the cases where a filling of transition
metal d-band by electrons of second alloy component takes place near its top
where the value of $dN_d(E)/dE$ is very large.

We may conclude that the investigation of the $\alpha (T)$ behavior permits
to study the d-electron density of states. As discussed above, a sharp
dependence $N(E)$ influences considerably a number of physical properties of
transition metals. However, the anomalies in thermopower are expected to be
especially sensitive to electron structure details due to presence of the
factor $d\tau (E)/dE$ in (5.86). However, one has to exclude other factors
which may influence $\alpha (T)$.

In the presence of density of states singularities (e.g., Van Hove ones),
the standard scheme of calculating $\alpha (T)$ should be modified. This is
due to that the expansion of the integrals which determine thermopower (see
(5.83)) in $T/E_F$ becomes impossible for peaks with the width of order $%
k_BT $. Direct integration was performed in paper [439] with application to
palladium. A triangle model with a jump of $dN(E)/dE$ near $E_F$ was used.
Although an agreement with experimental data was obtained at high
temperatures, the maximum at low temperatures remained not explained. This
maximum may be due to the phonon drag effect [7] (however, the signs of $R_0$
and $\alpha$ are opposite in this temperature region). Freezing out of the
s-d transitions and occurence of a many-electron mechanism at low
temperatures may also play a role in the $\alpha (T)$ dependence.

Non-monotonous dependences $\alpha (T)$ are observed also in a number of
transition metal compounds, e.g., copper-oxide systems which provide a basis
for high-$T_c$ superconductors (see experimental data [440,441] and
theoretical considerations [405,442]), Kondo lattices and heavy-fermion
compounds (see discussion in Sect.6.4). In such systems, very narrow density
of states peaks near the Fermi level have apparently a correlation
(many-electron) origin and are not obtained in band calculations.

\section{The Hall effect}

Investigation of electric properties of metals in an external magnetic field
lead to discovery of a number of interesting physical effects which provided
a powerful tool for analyzing their electron structure. This was realized
already in the beginning of modern solid state physics development. So, the
physical basis of galvanomagnetic effects (the Hall effect and
magnetoresistivity) is described correctly in the classical monograph [1].
However, concrete methods of the Fermi surface reconstruction with the use
of such effects were developed only after the extensive theoretical work in
50-60s (see [10]).

Modern theory explains well most effects in normal metals. At the same time,
the situation in transition metals is less satisfactory, which is connected
with their complicated electronic structure and the presence of spontaneous
magnetization (in ferromagnets).

The Hall effect in transition metals includes in fact two effects of
different microscopic nature

\noindent (i) the normal effect connected with the Lorentz force

\noindent (ii) the anomalous (spontaneous) Hall effect owing to spin-orbit
coupling.

\noindent The corresponding Hall constants are defined by
$$
E_y=R_0j_xH_z+R_1j_xM_z,\qquad
R_1=R^M\equiv 4\pi R_s
\eqno{(5.91)}
$$
Since $R_s\gg R_0$, the anomalous Hall effect dominates in ferromagnetic
crystals. The Hall effect in magnetically ordered metals will be considered
in Sect.5.7.1. Here we note that the spontaneous effect takes place in
paramagnets too. Putting in (5.91) $M=\chi H$ we obtain
$$
E_y/j_x=R_0^{*}H_z,\qquad
R_0^{*}=R_0+4\pi \chi R_s\,  
\eqno{(5.92)}
$$
Thus the spontaneous Hall effect may be picked out in the case of a strong
dependence $\chi (T)$ and is appreciable for large $\chi $, especially near
the Curie point.

Consider the simple quasiclassical theory of the normal Hall effect. Writing
down the phenomenological equations 
$$
j_x=\sigma _{xx}E_x+\sigma _{xy}E_y,\qquad
j_y=\sigma _{yx}E_x+\sigma _{yy}E_y  
\eqno{(5.93)}
$$
where off-diagonal components are determined by the magnetic field we obtain
for cubic crystals ($\sigma _{xx}=\sigma _{yy}=\rho ^{-1}$)
$$
j_x=\left( \sigma _{xy}-\frac{\sigma _{xx}\sigma _{yy}}{\sigma _{yx}}\right)
E_y,\qquad 
E_y\simeq -\rho ^2\sigma _{yx}j_z,\qquad
\sigma _{yx}^2\ll \rho ^{-2}  
\eqno{(5.94)}
$$
The Hall coefficient is introduced by
$$
R=\frac{E_y}{j_xH_z}=-\rho ^2\frac{\sigma _{yx}(H_z)}{H_z}  
\eqno{(5.95)}
$$
In the presence of the field $H_x$ and the Lorentz force
\[
F_y=-\frac ec\left[ \mathbf{vH}\right] _y=\frac ecv_xH_z 
\]
the off-diagonal component $\sigma _{yx}$ is connected with the diagonal one
by the dimensionless parameter $\omega _c\tau $ ($\omega _c$ is the
cyclotron frequency, $\tau $ is the relaxation time):
$$
\sigma _{yx}=\omega _c\tau \sigma _{xx}  
\eqno{(5.96)}
$$
so that the conductivity tensor in the field $H_z$ has the form 
$$
\widehat{\sigma }=\frac{e^2n\tau /m^{*}}{1+(\omega _c\tau )^2}\left( 
\begin{array}{ccc}
1 & \omega _c\tau  & 0 \\
\omega _c\tau  & 1 & 0 \\
0 & 0 & 1+\left( \omega _c\tau \right) ^2
\end{array}
\right)   
\eqno{(5.97)}
$$
It is interesting that according to (5.96) $\sigma _{yx}\ll \rho ^{-2}$.
Then we obtain from (5.95)
$$
R=-\rho ^2\frac{e\tau \sigma _{xx}}{m^{*}c}=-\frac 1{enc}  
\eqno{(5.98)}
$$
Thus in the simplest approximation the normal Hall effect coefficient is a
constant which does not depend on the scattering mechanism. We shall see
below that the situation changes drastically for the spontaneous Hall
effect. In particular, in the case of the phonon scattering (Appendix M.2)
the expansion of s in the scattering amplitude may start, in contrast with
(5.96), from zero-order terms, so that $R_s\sim \rho ^2$ .

Now we consider the experimental situation. In simple (in particular,
alcaline) metals the Hall effect is satisfactorily described by the formula
(5.98) which may be used to determine the carrier concentration $n$.
However, in polyvalent metals, where the Fermi surface crosses the
boundaries of the first Brillouin zone and several current carrier group
exist, there occur considerable deviations and a $T$-dependence arises [8].
Thus the Hall effect may be used in principle to determine characterisitics
of electron structure and the number and mobility of current carriers in
various regions of the Brillouin zone. The behaviour of the Hall coefficient
in the d-series is shown in Fig.5.17.

Unlike $\Delta \rho /\rho $-effect (Sect.5.6), the Hall effect in
non-ferromagnetic transition metals has essentially peculiar features. The
normal Hall coefficients of TM exhibit strong temperature dependences which
are in some cases non-monotonous. The forms of the $R_0(T)$ behaviour are as
a rule similar in the periodic table columns. These dependences are shown in
Figs.5.18-5.28. A weak $T$-dependence is observed for Mo and W only. The
local minimum of $R_0(T)$ dependence in Mn (Fig.5.29) is probably connected
with antiferromagnetism of this metal.

In principle, the complicated behaviour $R_0(T)$ may be explained by the
presence of several groups of current carriers. Using (5.97) we obtain in
the case of two groups
$$
R =\left( R^{(1)}\sigma _1^2+R^{(2)}\sigma _2^2\right) /\left( \sigma
_1+\sigma _2\right) ^2  
$$
$$
=\rho ^2\left( R^{(1)}\sigma _1^2+R^{(2)}\sigma _2^2\right)   
\eqno{(5.99)} 
$$
where
\[
R^{(i)}=-\frac 1{en_ic},\qquad
\sigma _i=\frac{e^2n_i\tau _i}{m_i^{*}}
\]
The result (5.99) may give a strong temperature dependence only provided
that the dependences $\sigma _1(T)$ and $\sigma _2(T)$ are appreciably
different. This may take place at low $T$, e.g., where light carriers
(s-electrons) are mainly scattered by phonons, and heavy carriers
(d-electrons with large density of states) due to electron-electron
collisions. For example, this mechanism may be responsible for a maximum of $%
|R_0(T)|$ in Cu at low temperatures (Fig.5.30). Similar (but appreciably
more pronounced maxima) are present in Pd and Pt. It is known that in all
these three metals two groups of carriers with considerably different
effective masses exist which correspond to neck and belly Fermi surfaces.
Ziman [444] proposed that the increase of $|R_0(T)|$ in Cu below 100K is
connected with freezening out Umklapp processes, the lifetime of light-mass
belly states becoming lengthened with respect to one on the neck region,$%
\tau _B/\tau _N\gg 1$ whereas\thinspace \thinspace $\tau _B\sim \tau _N$
\thinspace at room temperatures. These arguments are confirmed by the
calculations [445]. The occurrence of the $|R_0(T)|$ maximum may be
connected with the transition to strong field regime $\omega _c\tau \gg 1$
and with strong anisotropy of $\tau $. This hypothesis seems to be confirmed
by investigations of single crystals (see discussion in [443]). At the same
time, the explanation of the maximum in Pd and Pt at $T\sim 200$K is more
difficult.

At high $T\sim 100-1000$K the electron-phonon mechanism domiinates for all
the carriers in paramagnetic metals. Therefore the quantitative explanation
of the strong $R_0(T)$ change (by several times in Sc, Ti, Zr, Hf, V, Re) is
hardly possible on the basis of (5.99). The non-monotonous dependence $R_0(T)
$ in V and Ta (Fig.5.28) is discussed in [446]. The authors claim that usual
Umklapp processes do not explain the $R_0(T)$ minimum at $T=20-30$K. since
the temperatures of their freezing in these metals make up about 300 and
200K respectively for closed sheets, and for open sheets the Umkplapp
processes do not freeze out up to $T=0$. Only Umklapp processes between
closed hole sheets $h_N(3)$ and the open hole surface in the $\Gamma NH$
plane, which are localized in the region of minimum distance between the
sheets, yield reasonable values of the freezing temperature $T^{*}$. Below $%
T^{*}$ the anisotropy of scattering decreases which results in  $R_0(T)$
increase. Adding impurities suppresses the anisotropy too and leads to
vanishing of $R_0(T)$ minima in agreement with experimental data. This
interpretation agrees with the theory [415] which considers the anisotropy
of inequlibrium distribution function.

Another mechanism of strong dependence $R(T)$ in paramagnetic metals may be
the influence of the anomalous Hall effect. As follows from (5.92), such a
dependence may be connected with the dependences $\chi (T)$ and $R_s(T)$.
This possibility was noted in early papers, in particular, in connection
with the anomalies at the Curie point. Kondorsky [447] used this idea for
explaining the $R(T)$ behaviour in Zr and Re. Of course, reliable
independent methods for determining $R_s$ in non-ferromagnetic metals are
absent. The expression (5.92) may give required values provided that $%
R_s/R_0\sim 10^2-10^4$ , i.e. $R_s$ is nearly the same or somewhat larger in
comparison with ferromagnetic metals. This assumption seems to be reasonable
since the spin-orbital coupling has the same order of magnitude for all the
transition metals.

The expression (5.92) may explain also concrete temperature dependences $R(T)
$. As discussed in Sect.4.2, $\chi $ increases with $T$ in the periodic
table columns with even configurations d$^n$ and decreases for odd
configurations. At the same time, $R_s(T)$ increases in absolute value as $%
\rho ^2(T)$ due to electron-phonon scattering (Sect.5.7.1). Then we obtain
$$
R_0^{*}=R_0+4\pi \chi (T)bT^2  
\eqno{(5.100)}
$$
The sign of the temperature correction in (5.100) is determined by the sign
of $R_s$. In fact, the value of $b$ is not known exactly and should be used
as a fitting parameter. Then we may reproduce various experimental
dependences $R(T)$. For example, one has to put $R_0<0,R_s>0$ for Ti, $%
R_0>0,R_s>0$ for Zr, $R_0>0,R_s<0$ for Re with $R_s\sim \rho ^2(T)\sim T^2$
in all the cases. Unfortunately, it is difficult to separate the normal and
anomalous contributions. Apparently, the value of R may be determined from
low-temperature data for high-purity samples where $R_s\rightarrow 0$.

\section{Magnetoresistivity}

The resistivity change in magnetic field (magnetoresistivity, $\Delta \rho
/\rho $ effect) corresponds to square terms in H, i.e. is an even effect.
One distinguishes the longitudinal (H ~ E) and transverse (H E) effects. As
well as the Hall effect, magnetoresistivity occurs due to distortion of
electron traectory owing to external field or magnetization. Naively, one
might assume that this distortion should diminish the velocity component v
along the electric field and result in a decrease of current, i.e. an
increase of resistivity. In fact, the situation is more complicated since
one has to take into account the Hall field E which compensates the
influence of magnetic field.

The change of resistivity in magnetic field may be described by the Frank law
$$
\Delta \rho /\rho \equiv \frac{\rho (H)-\rho (0)}{\rho (0)}=\frac{BH^2}{%
1+CH^2}  
\eqno{(5.101)}
$$
In weak fields ($CH^2\ll 1$) we have the square-law increase, and in strong
fields the resistivity is saturated, 
\[
\Delta \rho /\rho =B/C=\mathrm{const}
\]
In some situations, the linear dependence $\rho (H)$ (the Kapitza law) is
observed in strong fields (see [1,10]). Besides that, for Mo, Re, Pt, Fe and
Pd the experimental dependences may be fitted as [448] 
$$
\Delta \rho /\rho \sim H^m,\qquad m<2  
\eqno{(5.102)}
$$

According to the Kohler rule,
$$
\Delta \rho /\rho =f\left( \frac H\rho \right)   
\eqno{(5.103)}
$$
with  $f$  being an universal function. The Kohler plots for some simple and
transition metals are shown in Fig.5.31. The Kohler rule holds in wide
intervals of r and H almost in all cases.

Let us try to estimate the values of the coefficients $B$ and $C$ in (5.101)
in the simple theory which considers the motion of an electron in a crystal
in external electric and magnetic fields. We use the expansion in $T/E_F$
and the relaxation time approximation. Then the longitudinal $\Delta \rho
/\rho $ effect turns out to be absent to the second-order terms, and for the
transverse effect one obtains
$$
B=\frac{\pi ^2}3\left( \frac{e\overline{l}}{m^{*^2}}\frac{k_BT}{\overline{v}%
^3}\right) ^2,\qquad
C=\left( \frac{e\overline{l}}{m^{*}\overline{v}}\right) ^2  
\eqno{(5.104)}
$$
where $l=v\tau $ is the electron mean free path, the averages are taken over
the Fermi surface. The result (5.104) may be interpreted in terms of
competition between the circular motion under the influence of the Lorentz
force and the straightforward motion in the electric field on the path $l$.
The former is characterized by the orbit radius
\[
r_c=\overline{v}/\omega _c=m^{*}\overline{v}c/eH 
\]
Thus we have
$$
CH^2=\left( \overline{l}/r_c\right) ^2,\qquad
BH^2=\frac{\pi ^2}3\left( \overline{l}/r_c\right) ^2
\left( \frac{k_BT}{m^{*}\overline{v}^2}\right) ^2  
\eqno{(5.105)}
$$
The occurence of the ratio $\overline{l}/r_c\sim H/\rho $ illustrates the
Kohler rule. In weak fields ($\overline{l}\ll r_c$) the increase of $\rho $
is determined by that the spiral motion of an electron in the field $H_z$
results in decreasing of the path in the x-direction between collisions. At
large fields,
$$
\Delta \rho /\rho =\frac BC=\frac{\pi ^2}3\left( \frac{k_BT}{m^{*}\overline{v%
}^2}\right) ^2  
\eqno{(5.106)}
$$
Although the result (5.106) is in agreement with the law (5.101), the value
of the coefficient $B\sim (T/E_F)^2$ is considerably smaller than the
experimental one, and the strong temperature dependence is in fact not
observed. The difficulties of the above simple theory are eliminated when
one includes a dispersion of the relaxation time $\tau (\mathbf{k})$. Then
the average value
$$
\overline{\tau ^n}=\sum_{\mathbf{k}}\frac{\partial n_{\mathbf{k}}}{\partial
E_{\mathbf{k}}}v_{\mathbf{k}}^2\tau (\mathbf{k})/\sum_{\mathbf{k}}\frac{%
\partial n_{\mathbf{k}}}{\partial E_{\mathbf{k}}}v_{\mathbf{k}}^2 
\eqno{(5.107)}
$$
differs from $\overline{\tau ^n}$, and we obtain the non-zero effect at $T=0$:
$$
\Delta \rho /\rho =Q^2\left( \frac{eH}{m^{*}c}\right) ^2,\qquad
Q=\left[ \overline{\tau ^3}\overline{\tau }-(\overline{\tau ^2})^2\right] 
/\left( \overline{\tau }\right) ^2 
\eqno{(5.108)}
$$
Since 
\[
\overline{\tau ^3}\overline{\tau }\geq \left( \overline{\tau ^2}\right) ^2
\]
$\Delta \rho /\rho $ is positive.

Temperature dependence of$\Delta \rho /\rho $ may be calculated
provided that the factorization
$$
\tau (\mathbf{k})=\varphi (T)\chi (\mathbf{k})  
\eqno{(5.109)}
$$
is possible (this may take place, e.g., at high temperatures). Then we derive
$$
\Delta \rho /\rho \sim \sigma ^2(T)H^2  
\eqno{(5.110)}
$$
with $\sigma (T)$ being the conductivity. Thus $\Delta \rho /\rho $ is
expected to increase with lowering $T$. The relation (5.110) is in agreement
with the Kohler rule. It should be noted that the temperature dependence of
(5.110) (a decrease with $\rho $) contradicts to expressions (5.101),
(5.104) for a single group of current carriers, which yield an increase as $%
\rho ^2$ with increasing $T$. Thus the experimental $T$-dependences of $%
\Delta \rho /\rho $-effect correspond to essentially anisotropic relaxation
time or to existence of several current groups.

Consistent calculation of the parameter $Q$, which is defined in (5.108), is
a very complicated problem. More convenient is the simple estimation from
experimental data with the use of
$$
Q=\frac 1H\left( \Delta \rho /\rho \right) ^{1/2}\frac \rho {R_0} 
\eqno{(5.111)}
$$
where $R$ is the Hall coefficient. For most metals the value of $Q$ makes up
from 1 to 4. An important exception are semimetals As, Sb, Bi where $Q\sim
10^2-10^3$. The giant $\Delta \rho /\rho $-effect in these substances is
connected with their anomalous electronic structure [10]. Some transition
metals also possess large value. E.g., for zirconium $Q=20$, which may be
connected with strong anisotropy for hcp crystals.

Higher-order approximations which take into account the 
$\mathbf{k}$-dependence of electron velocity permit to obtain the longitudinal 
$\Delta \rho /\rho $-effect. The corresponding correction to the distribution
function is given by [1] 
$$
f_{\mathbf{k}}^{(3)} =-\tau ^3\frac{e^3}{\hbar ^2c^2}EH^2\frac{\partial n_{%
\mathbf{k}}}{\partial \varepsilon _{\mathbf{k}}}\left[ v_y\left( \frac{%
\partial v_y}{\partial k_x}\frac{\partial v_z}{\partial k_z}-\frac{\partial
v_z}{\partial k_y}\frac{\partial v_x}{\partial k_z}+\right. \right.  
$$
$$
\left. \left. v_y\frac{\partial ^2v_z}{\partial k_x^2}-v_x\frac{\partial
^2v_z}{\partial k_x\partial k_y}\right) +x\longleftrightarrow y\right]  
\eqno{(5.112)}
$$
According to experiments by Kapitza (see [1,10]), the longitudinal effect
may be comparable with the transverse one.

In 50s-60s, investigations of galvanomagnetic effects were widely applied to
reconstruction of the Fermi surface shape of metals. In strong magnetic
fields where
\[
\omega _c\tau =eH\tau /m^{*}c\gg 1 
\]
three types of the $\Delta \rho /\rho $-effect behaviour were found

\noindent (i) $\Delta \rho /\rho $ is saturated at arbitrary orientation of 
$\mathbf{H}$ in the crystal

\noindent (ii) $\Delta \rho /\rho $ is not saturated at arbitrary
orientation of $\mathbf{H}$

\noindent (iii) $\Delta \rho /\rho $ is saturated for some orientations of 
$\mathbf{H}$ and continues to increase in strong fields for other orientations.

The type of behaviour depends on that whether a given metal is compensated,
i.e. the number of electrons equals to that of holes, $n_e = n_h$.
Evidently, the compensation situation is impossible for odd number of
conduction electrons per atom. Among transition elements, the compensated
metals are Ti, Cr, Mo, W, Re, Fe, Os, Ni, Pd, Pt, and Sc, V, Nb, Ta are
uncompensated (for other d-metals the data in [10,443] are absent).

As follows from (5.92), at $\omega _c^{e,h}\tau \gg $ 1 the off-diagonal
conductivities do not depend on $\tau $ and $m^{*}$, and the Hall fields
are given by
$$
E_y^{e,h}=-\frac 1{n_{e,h}ec}j_xH_z  
\eqno{(5.113)}
$$
so that 
$$
\sigma _{yx}=\sigma _{yx}^e+\sigma _{yx}^h=(n_h-n_e)\frac{ec}H  
\eqno{(5.114)}
$$
At $n_e=n_h$ the quantity $\sigma _{yx}$ vanishes and
$$
\sigma _{xx}\simeq ne^2\left( \frac{\tau _e}{m_e}+\frac{\tau _h}{m_h}\right)
\frac 1{\omega _c^e\omega _c^h\tau _e\tau _h}  
\eqno{(5.115)}
$$
so that$\rho _{xx}\sim H^2$ . Thus in the case of a compensated
metal $\Delta \rho /\rho $ does not exhibit saturation.

In a similar way one may explain existence of peculiar field directions in a
crystal where saturation is absent [10,144]. To this end we have to consider
open orbits in some direction v (which correspond to an open k orbit in the 
$k$-space). Then the field $\mathbf{H}$ will not influence such electron states
and we obtain again $\rho _{xx}\sim H^2$ . The linear Kapitza law in strong
fields may be connected with the anisotropy of $\Delta \rho /\rho $-effect
in single crystals. Such a behaviour is obtained after averaging over
orientations with account of open orbits [144].

The above classification may be changed by magnetic ordering because of
lifting spin degeneracy. In particular, the compensation may become violated
for partial occupation numbers with a given spin projection.

\section{Anomalous transport effects in ferromagnetic metals}

\subsection{The extraordinary Hall effect}

According to (5.91), the Hall resistivity in ferromagnetic crystals is given
by
$$
\rho _H=E_y/j_x=R_0H+R_1M=R_0B+4\pi R_sM
\eqno{(5.116)}
$$
where $B=H+4\pi M$ is the ``true'' macroscopic field in the substance,
$$
R_1=4\pi (R_0+R_s),\qquad
R_s=R_1/4\pi -R_0
\eqno{(5.117)}
$$
The presence of the peculiar Hall coefficient $R_s$ in ferromagnets was
established in first experimental investigations of the Hall effects (see
[265,384]). Already in 1881 (the ``normal'' effect was discovered in 1879)
Hall found the influence of the magnetization on the transverse electric
field, which occured in an external magnetic field. At measuring the
electric field as a function of magnetic induction $B$ in nickel, he noted
that the slope of the linear increase was changed after magnetic saturation.

A typical field dependence $\rho _H(H)$ in a ferromagnet is shown in
Fig.5.32. At $0<H<H_c$ ($H_c$ is the field where magnetic domains become
fully oriented) the magnetization increases rapidly from 0 to $M_s$, so that
$\rho _H(H_c)=4\pi (R_s+R_0)M_s$ and the slope of the curve is determined by
the quantity $4\pi (R_0+R_s)$. At $H>H_c$ the slope is determined by the
usual coefficient $R_0$ and weak high-field susceptibility (paraprocess).
The coefficient $R_s$ determined in such a way turns out to exceed $R_0$ by
several order of magnitude, which may be seen explicitly from the break of
the plot $\rho _H(B)$. This demonstrates wittingly the existence of the
extraordinary Hall effect.

Besides that, attempts to define the Hall coefficient in ferromagnets in the
standard form $\rho_H = RB$ resulted in quite strange temperature
dependences of the Hall coefficient (in particular, in a jump at the Curie
point). Therefore Pugh [449] proposed to express the Hall field in terms of
the magnetization. Kikoin [450] investigated the temperature dependence of
the Hall field of nickel in a wide region below and above $T_C$. The results
yielded convincing evidences for existence of the spontaneous Hall effect.
Its temperature dependence was described by the expression
$$
R_1(T)=a\left[ M^2(0)-M^2(T)\right]
\eqno{(5.118)}
$$
It was also found that the relation with the resistivity $R(\rho )\sim \rho $
takes place in iron group metals for varying temperature and impurity
concentration where $n=1.2\div 2$ (see [394,457]). Experimental data on the
temperature dependence of $R_s$ in iron group metals are shown in Fig.5.33.

The extraordinary contribution to the Hall effect in ferromagnets occurs
even in the paramagnetic phase where the magnetization is given by $M = \chi
H$ and
$$
R=R_0+\chi R_1
\eqno{(5.119)}
$$
Despite the small value of $\chi \sim $ $10^{-3}$, the addition to $R_0$ may
be noticeable. The extraordinary (anomalous) Hall effect is observed also in
antiferromagnets and paramagnets. Its value is especially large in compounds
with high values of $\chi $, including Kondo lattices and heavy fermion
systems (Chapter 6) [451,452].

First theoretical consideration of the anomalous Hall effect (AHE) was
carried out by Rudnitskii [453]. He demonstrated that the simplest
supposition about the deviation of magnetized conduction electrons in the
field induced by electric current may not explain the value of the effect,
leading to quantities which are smaller by three orders of magnitude than
the experimental ones. Further, he put forward the idea to explain the
effect by the spin-orbit interaction. The corresponding energy
$$
\mathcal{E}_{so}=\left\langle \mathcal{H}_{so}\right\rangle ,\qquad
\mathcal{H}_{so}=\lambda \mathbf{ls}=\lambda [\mathbf{rp]s}
\eqno{(5.120)}
$$
is proportional to the magnetization $\langle S^z\rangle =M$ and yields the
force
\[
\mathbf{F}_{so}=\frac \partial {\partial \mathbf{r}}\mathcal{E}_{so}\sim
\left[ \mathbf{pM}\right]
\]
which is similar to the Lorentz force. The estimation
\[
\mathcal{E}_{so}=\mu _BH_{so}\sim 10^{-13}\ \mathrm{erg},\qquad
H_{so}\sim 10^7\ \mathrm{Oe}
\]
just yields the effective magnetic field required.

Already in the paper [453] the basic question about the averaging of the
spin-orbit interaction over a crystal with possible zero result was
considered. Indeed, the periodic SOI itself does not provide the Hall
effect, but one has to introduce inhomogeneities which scatter current
carriers and lead to an asymmetry of $y$ and $-y$ directions. Thus the value
of $R_s$ correlates with the electrical resistivity.

The first attempt of quantum calculation of the Hall coeffucient in
ferromagnets with account of SOI was carried out [454]. However, the authors
of this paper did not take into account symmetry properties of SOI matrix
elements. In fact, no linear in SOI corrections to the electron distribution
function exist, so that the result of the calculation [454] should vanish.

The absence of the AHE within the lowest order approximation in the
transport equation was demonstrated by Karplus and Luttinger [455]. To
obtain a non-zero effect, they considered dynamical corrections to the
electron energy in the electric field (i.e. to the field term) owing to
interband matrix elements of SOI and of velocity. The corresponding
off-diagonal conductivity does not depend on the scattering mechanism. Since
$$
R_s\approx \frac 1{4\pi }R_1\approx -\frac 1{4\pi M}\frac{\sigma _{yx}(M)}{%
\sigma _{xx}^2}
\eqno{(5.121)}
$$
we obtain
$$
R_s=\alpha \rho _{xx}^2
\eqno{(5.122)}
$$
where the constant $\alpha $ does not depend on temperature. The result
(5.122) was widely used to fit experimental data.

Being important for the development of the theory, the calculations [455]
still had incomplete and rather artifical character. Indeed, they did not
take into account corrections to the collision term which have lower order
in the scattering amplitude and may yield larger contributions. Thus the
theory of AHE required a more consistent consideration.

A step in this direction was made in papers by Kohn and Luttinger [458]
where a new method of obtaining transport equations with the use of
equations of motion for the density matrix was proposed. In [459] this
method was applied for calculation of AHE owing to scattering of
magnetically polarized current carriers by impurity centres. Unlike [457],
contributions owing to collision terms (``skew scattering'') occured in both
the lowest and next orders with respect to the scattering amplitude. The
final result has the form
$$
R_s^i=\alpha \rho _i+\beta \rho _i^2
\eqno{(5.123)}
$$
where $\rho _i$ is the impurity resistivity of a sample. The ratio of the
first term in (5.123) to the second one equals $E_F/(3n_i\overline{\phi })$
where $n_i$ is the concentration of impurity atoms, $\overline{\phi }$ is
the impurity potential. At small $n_i$ this ratio is large, so that almost
linear dependence $R_s(\rho _i)$ should be observed.

Detailed derivation of the $\rho $-linear term is considered in Appendix
M.1. The expression for the coefficient a in (5.123) reads
$$
\alpha =-\frac{\mu _B^2k_F^3}{18\pi \Delta ^2}\frac{\rho _{eff}\overline{%
\varphi }}{M(0)}
\eqno{(5.124)}
$$
where the effective charge density $\rho _{eff}$ is defined by (M.52) and
includes spin-orbital parameter, $\Delta $ is the splitting of energy
levels, which is of order of bandwidth. Some peculiarities of this
expression should be discussed. The sign of $R_s$ is determined by signs of
not only current carrier charge and SOI, but also the potential $\overline{%
\phi }$. Therefore this sign may be reversed by changing the sign of
impurity charge. The influence of impurities on the Hall coefficient is
stronger than that on resistivity since $R_s$ is proportional to the third
power of scattering amplitude.

The results (5.123) cannot be easily extended by replacing r to the total
resistivity (such a replacement may be made only in the second term). Thus a
concrete consideration of various scattering mechanisms was needed. As
discussed above, the explanation of AHE is based on the spin-orbital
interaction which results in occurence of transverse contribution to current
even for isotropic scattering. Since SOI is linear in magnetization, we have
to calculate corrections to the distribution function which are linear in
SOI. On the other hand, for non-degenerate wavefunctions (i.e. for quenched
orbital momenta) the operator $\mathcal{H}_{so}$ has only off-diagonal
matrix elements, so that linear corrections to the electron energy are
absent. Therefore the transport equation in the Born approximation, which
depends only on the electron energy and the squared scattering amplitude,
does not yield AHE. Thus we have to consider higher-order transport
equations for various scattering mechanisms (Appendix M).

The phonon mechanism was considered by Irkhin and Shavrov [460]. The
lowest-order transport equation which yields the phonon scattering
contribution to AHE is the equation of the second order in the perturbation
Hamiltonian $\mathcal{H}^{\prime }$. According to (M.72), the corresponding
expression for the spontaneous Hall coefficient reads
$$
R_s^{ph}=-\frac 23g\mu _B^2\frac{e^2n\hbar }{\Delta ^2}\rho
_{eff}t\left\langle \frac 1{m^{*}}\right\rangle \frac{\rho ^2}{M(0)}
\eqno{(5.125)}
$$
where $t$ is the number of subbands. Introducing the effective spin-orbital
field
$$
\mu _BH_{so}=\mathcal{E}_{so}=-\frac \pi 3\mu _B^2\rho _{eff}  
\eqno{(5.126)}
$$
we derive
$$
R_s^{ph}=2t\frac{\mathcal{E}_{so}}{\Delta ^2}\frac{e^2n\hbar }{m^{*}}\frac{%
\rho _{ph}^2}{M(0)}  
\eqno{(5.127)}
$$
Thus the relation $R_s^{ph}\sim \rho _{ph}^2$ takes place. The expression
(5.127) may be rewritten in the form 
$$
R_s^{ph}=\pm 2t\frac \hbar {\tau _{ph}}\frac{\mathcal{E}_{so}}{\Delta ^2}%
\frac{\rho _{ph}}{M(0)}  
\eqno{(5.128)}
$$
so that the sign of $R_s$ is determined by the sign of $m^{*}$: $R_s<0$ for
electrons and $R_s>0$ for holes. Putting $\mathcal{E}_{so}\sim $ $10^{-14}$
erg, $\Delta \sim 10^{-12}$ erg, $\rho \sim $ $10^{-5}$ $\Omega $ cm, $\tau
\sim $ $10^{-13}$ s we estimate $R_s\sim $ $10^{-13}$ $\Omega $ cm/G.
Further theoretical investigations of the phonon mechanisms were performed
in papers [461,462]. The role of two-phonon scattering processes was
investigated in [463].

The square dependence $R_s$ $(\rho )$ was also found in the paper [464] for
the scattering by spin inhomogeneities; the principal linear term was not
found because of too simple decoupling of spin correlators. Kondo [465]
considered the latter problem in the framework of the s-d exchange model
with account of the proper SOI among localized electrons. He did not derive
the transport equation, but used the equations by Kohn and Luttinger [458]
for the impurity scattering. Thus the inelastic part of the scattering was
not taken into account. The result by Kondo for d-metals reads
$$
R_s^{mag}\sim \frac \lambda \Delta \frac{\left( m^{*}\right) ^{5/2}}{%
E_F^{1/2}\hbar ^4e^2}G_2\left( G_0^2-\frac 43G_0G_1\right) \frac{%
\left\langle \left( S^z-\left\langle S^z\right\rangle \right)
^3\right\rangle }{\left\langle S^z\right\rangle }  
\eqno{(5.129)}
$$
where $G$ are the s-d exchange ``Slater'' integrals of the type (K.5), $%
\Delta $ is the energy difference for magnetic d-electrons. Main shortcoming
of the calculation [465] was using d-electron states with unquenched orbital
momenta. At the same time, the unquenching of orbital momenta in d-metals is
connected with the same SOI which is responsible for AHE.

This point was taken into account by Abelskii and Irkhin [466] who
considered within the two-band s-d model two types of SOI: ``proper'' d-d
interaction $\lambda $ and the interaction s-spin -- d-orbit $\lambda
^{\prime }$ (see Appendix L). The magnetized d-electrons were supposed
itinerant and described in the tight-binding approximation with a small
bandwidth, so that their orbital momenta were almost quenched. The result of
this paper for high temperatures for spin $S=1/2$ reads (see (M.99))
$$
R_s^{mag}=\frac{9\pi }{64}\left( \frac 1{E_F}\right) ^2\frac{\rho _\lambda }{%
M(0)}\left( \frac 14-\langle S^z\rangle ^2\right)  
\eqno{(5.130)}
$$
where
\[
\rho _\lambda =\lambda _{eff}\frac{m^{*}}{e^2n\hbar },\qquad
\lambda _{eff}\approx \lambda \frac{I^{(1)}}{\Delta E}\frac 23
\overline{l_z}+\lambda ^{^{\prime }} 
\]
Taking into account the expression for the spin-disorder resitivity (5.57)
this result may be presented in the simple form
$$
R_s^{mag}=\pm \frac 3{16}\frac{\lambda _{eff}}{E_F}\frac{\rho _{mag}}{M(0)} 
\eqno{(5.131)}
$$
where the + and - signs correspond to electron and hole conductivity. Thus
we obtain a simple connection between $\rho _{mag}$ and $R_s$.

The contribution of the electron-electron scattering to the extraordinary
Hall effect was calculated in [467]. The result reads
$$
R_s^{ee}\sim \lambda \left( \frac T{E_F}\right) ^4  
\eqno{(5.132)}
$$

Separation of various contributions to AHE may be carried out by
investigating the dependence $R_s$($T$) at crossing the Curie point, since
in the far paramagnetic region the magnetic scattering mechanism is
saturated. Another possible way is considering the dependence of $R_s$ on
the magnetic field in the ferromagnetic phase [468]. Since the magnetic
field suppresses spin disorder, $d\rho _{mag}/dH<0$ in the paraprocess
region. Therefore the signs of $R_s$ and $dR_s$ $/dH$ should be opposite
provided that magnetic mechanism dominates. According to [468], for nickel
the signs of both $R_s$ and $dR_s/dH$ are negative. This may be explained by
the large phonon contribution. However, this does not quite agree with the
dependence $R_s$ ($T$) above $T_C$ (the rapid increase of $R_s$ is not
observed).

An attempt to formulate a new picture of AHE was made by Berger [469] who
considered the ``side-jump'' scattering of the electron wave packet under
influence of various mechanisms. The universal result $R_s\sim \rho ^2$ is
obtained in such an approach. It was demonstrated later [470] that the
side-jump scattering is just another formulation of the Luttinger's skew
scattering (corrections to the field term).

On the whole, experiments at high temperatures are in agreement with the
above theoretical results. However, the measurements at low $T$ do not yield
a linear dependence between ln$R_s$ and ln$\rho $. Thus a special
consideration of the spin-wave region, where the mean field approximation is
inapplicable, is required. The calculation by Kagan and Maksimov [471]
yielded the result $R_s\sim T^4$ . In the paper [472] another contribution,
which was due to energy dependence of the distribution function, was found.
It turns out that, because of cancellation of lowest-order terms, such
contributions to AHE (unlike resistivity) are the most important ones. The
final result of [472] reads (see Appendix M.3)
$$
R_s=\pm \frac{3\pi }{512}\frac{\hbar I^2}{e^2k_FM(0)}\left( k\frac{dE_{%
\mathbf{k}}}{dk}\right) _{k=k_F}^{-2}\left[ A\left( \frac T{T_C}\right)
^4+B\left( \frac T{T_C}\right) ^3\right]  
\eqno{(5.133)}
$$
where the constants 
$$
A=1.1\overline{l}\frac{I^{(1)}}{\Delta E},\qquad
B=0.8\lambda ^{^{\prime }}  
\eqno{(5.134)}
$$
are determined by ``proper'' and ``improper'' SOI respectively. One can see
that a simple connection between the Hall coefficient and resistivity is
violated at low temperatures and the dependence $R_s(T)$ may be rather
complicated. This is confirmed by experimental data on Fe, Co, Ni (Fig.5.33)
and diluted ferromagnetic alloys [473], which demonstrate non-monotonous
temperature dependences of $R_s$ whereas $\rho $ behaves monotonously.

A number of calculations of the anomalous Hall effect with account of
realistic band structure (in particular, the Fermi surface topology and
anisotropy) were performed [474,475]. The results show the strong influence
of the Fermi surface details on the anomalous Hall effect at low
temperatures.

The Hall effect in 4f-metals merits a special consideration. Due to variety
of magnetic structures and complicated phase diagrams in rare earths, the
role of various factors in the Hall effect may be investigated here. In
particular, the influence of antiferromagnetism and strong magnetic
anisotropy becomes important. On the other hand, the situation is somewhat
simlplified in comparison with d metals because of a distinct separation
between current carriers (s,p,d electrons) and magnetic f-electrons, which
permits to separate proper and improper spin-orbital interaction effects.

The normal and anomalous Hall effects in 4f-metals were extensively
investigated starting from 60s. Detailed investigations were performed for
heavy rare earths (a review is given the monograph [15]). The temperature
dependences of the Hall coefficient for heavy rare-earth metals are shown in
Fig.5.34. Main distinctive features of the anomalous Hall effect in
comparison with 3d-metals are as follows

\noindent (i) The dependence $R_s(T)$ has a more complicated non-monotonous
character. However, the temperature of extrema do not coincide with points
of transitions between magnetic structures. A change in $R_s(T)$ sign takes
place for Tb and Dy at $T\sim $0.8-0.9$T_N$ .

\noindent 
(ii) A good proportionality between $R_s$ and magnetic resistivity takes
place (Fig.5.35).

\noindent (iii) R is highly anisotropic: the values in different
crystallographic directions vary by several times.

\noindent (iv) The dependence $R_s(T)$ is not influenced by ferro-antiferro
transitions.

It would be instructive to compare the $R_s$ behavior in AFM region with
data on other antiferromagnets, e.g., Cr and Mn [443]. However, this
comparison is hampered by that the latter data are insufficient to separate
the normal and anomalous contributions. Theory of AHE in rare-earths is
based on the above-discussed general theory but some extra factors should be
taken into account.

The terms with the vector products $[\mathbf{k},\mathbf{k}^{\prime }]$,
which arise from the matrix elements of the conduction electron orbital
momenta $(\mathbf{l})_{\mathbf{kk}^{\prime }}$, describe the anisotropic
electron scattering. Such terms correspond to the coupling of conduction
electron current in the external electric field to the momentum J and yield
therefore the anomalous Hall effect. The Hall coefficient is proportional to 
$A(g-2)$, which corresponds to the interaction of electron orbital momenta
with the localized orbital momenta $(2-g)\mathbf{J}$ (see (B.20)). This
picture is different from that in d-metals where the anomalous Hall effect
is due to weak spin-orbit coupling. For f-electrons, this coupling is strong
(of order 10 eV), which enables us to consider only one $J$-multiplet, so
that the spin-orbit coupling constant does not enter explicitly the results.
It is worthwhile to mention that one of first papers, devoted to derivation
of the Hamiltonian of the type (K.4), was the paper by Kondo [465] on the
theory of the anomalous Hall effect (which preceeded to the famous papers on
the Kondo effect).

In the antifferromagnetic region one has to determine the saturation
magnetization in terms of high-field susceptibility $\chi _s$ = $dM/dH$.
Then one obtains the following equations for the Hall resistivity $\rho _H$
$$
\rho _H(H_{dm})=4\pi \left( R_0+R_s\right) M_s  
\eqno{(5.135)}
$$
$$
\frac{\partial \rho _H}{\partial H}=R_0+4\pi \left[ R_s+R_0(1-N)\right] \chi
_s  
\eqno{(5.136)}
$$
where $H=4\pi NM_s$ , $N$ is the demagnetization factor. Thus one can
separate the normal and anomalous Hall coefficients by measuring $M_s$, 
$\chi _s$, $N$ and $\rho _H(H).$ It should be noted that in the AFM region 
$\chi $ is not saturated up to $H\sim 40$ kOe.

Although the spin-disorder scattering mechanism [465,466] yields the
behaviour $R_s(T)\sim \rho _{mag}(T)$ which is really observed at low
temperatures, such simple theory does not explain changes in the $R_s(T)$
sign. Attempts to do this were performed by a number of authors [476-478].

The idea by Maranzana [476] was using higher-order terms in the s-f
Hamiltonian derived by Kondo (see also Appendix K). Besides the main
anisotropic-scattering term
$$
i\lambda _1(2-g)\left[ \mathbf{k,k}^{\prime }\right] J_\nu c_{\mathbf{k}%
\alpha }^{+}c_{\mathbf{k}^{^{\prime }}\alpha }=i\lambda _1\left[ \mathbf{k,k}%
^{\prime }\right] L_\nu c_{\mathbf{k}\alpha }^{+}c_{\mathbf{k}^{^{\prime
}}\alpha }  
\eqno{(5.137)}
$$
the terms were considered, which had the structure
$$
i\lambda _2\langle \mathbf{J}_\nu \mathbf{J}_\nu \rangle \langle \left[ 
\mathbf{k,k}^{\prime }\right] \mathbf{\sigma }_{\alpha \beta }\rangle c_{%
\mathbf{k\alpha }}^{+}c_{\mathbf{k}^{^{\prime }}\beta }  
\eqno{(5.138)}
$$
the scalar product of tensors being defined by
$$
\left\langle AB\right\rangle \left\langle CD\right\rangle =\frac 12\left\{
\left( AC\right) \left( BD\right) +\left( AD\right) \left( BC\right) -\frac
13\left( AB\right) \left( CD\right) \right\}  
\eqno{(5.139)}
$$
As a result, higher powers of momentum operators occur, which result after
averaging in new functions of magnetization. In particular, there exist the
contribution to $R_s$ which is proportional to the second derivative of the
Brillouin function,
$$
M_S=\left\langle \left( S^z-\left\langle S^z\right\rangle ^3\right)
\right\rangle =-J^3B_J^{^{^{\prime \prime }}}(y)=-J^3f_1\left( \frac
T{T_N}\right)  
\eqno{(5.140)}
$$
where
$$
y=\frac{3\left\langle J^z\right\rangle }{J+1}\frac{T_N}T+\frac{g_J\mu _BH}{%
k_BT}  
\eqno{(5.141)}
$$
Therefore this contribution is non-monotonous near $T_N$. Giovannini [477]
introduced a still more complicated function
$$
M_4=J^4\left\{ B_J^{^{\prime \prime \prime }}(y)+2B_J^{^{\prime \prime
}}(y)+2\left[ B_J^{^{\prime }}(y)\right] ^2\right\} \equiv
J^4f_2(T/T_N)/B_J(y)  
\eqno{(5.142)}
$$
which contains the higher-order derivatives of $B_J$. This led to the
dependence of the form
$$
\rho _H=C_1f_1(T/T_N)+C_2f_2(T/T_N)  
\eqno{(5.143)}
$$
The constants $C_1\sim \lambda _1$ and $C_2\sim \lambda _2$ were used as
fitting parameters. For Tb and Dy, $C_2/C_1\sim $ 30 and 9 respectively,
whereas $C_2$ = 0 for gadolinium. The value of $C_1$ for Gd corresponds to
the unreasonably large spin-orbital constant $\lambda \simeq $ 0.5eV. The
values of $R_s$in paramagnetic regions, which are obtained from the
estimations of $C_1$ and $C_2$, turn out to be not quite satisfactory. It
should be also noted that the $C_2$-term corresponds to a rather higher
order in the s-f Hamiltonian and hardly may give a dominating contribution.
Therefore, despite a qualitative explanation of the $R_s(T)$ behaviour,
the mechanism discussed is debatable.

Somewhat later Fert [478] considered the influence of the side-jump
scattering which corresponds to corrections to the field term in the
transport equation. This mechanism yields some renormalizations of the
coefficients at the Brillouin function derivatives. However, the value of
AHE in gadolinium remains unexplained.

The anomalous Hall efect in heavy-fermion systems is discussed in the papers
[479].

\subsection{Magnetoresistivity in the presence of spontaneous
magnetization}

The $\Delta \rho /\rho $-effect in ferromagnets has important peculiarities.
Its value may be of order 10$^{-2}$, which is much greater than in usual
metals, and have both positive and negative sign (Fig.5.36). The Kohler rule
is usually not satisfied.

An important circumstance of the situation in a magnetized sample is
possibility of the non-zero effect in the absence of the external magnetic
field. The spontaneous effects are masked in multidomain samples where the
average magnetization is zero. The single magnetic domain forms in the
fields above the field of technical saturation $H_c$. In the low-field
region (below $H_c$ ) the effect is due to a change in relative volumes of
domains with $M\Vert j $ and $M\bot j$:
$$
\rho (H)=\frac 1V\left( V_{\Vert }\rho _{\Vert }+V_{\bot }\rho _{\bot
}\right)  
\eqno{(5.144)}
$$
For $H=0$ we have
$$
\rho =\frac 13\rho _{\Vert }+\frac 23\rho _{\bot }  
\eqno{(5.145)}
$$
so that
$$
\Delta \rho _{\Vert }(H_c)/\Delta \rho _{\bot }(H_c)=-2  
\eqno{(5.146)}
$$
(Akulov's rule for even transport effects [265]).

Above the technical saturation $H>H_c$, the field dependence of resistivity
is considerably weaker and determined by the paraprocess (the field
dependence of magnetization $M=M_s+\chi H$). The dependence $\rho (H)$ is
owing to suppression of spin disorder, which results in a decrease of
exchange scattering of current carriers. This effect should take place also
in the paramagnetic region $T>T_C$, resulting in a negative contribution to $%
\Delta \rho /\rho $. Experimental data are often described by the equation
$$
\Delta \rho /\rho =a(M_s^2-M^2(H))  
\eqno{(5.147)}
$$
so that
$$
\Delta \rho /\rho =-a_1H-a_2H^2  
\eqno{(5.148)}
$$
Thus we obtain the linear $\Delta \rho /\rho $-effect. It should be noted
that similar field dependences of $\Delta \rho /\rho $ may take place also
in antiferromagnetic metals. For example, the behaviour $\Delta \rho /\rho
\sim $ $H^{3/2}$ was found in Fe$_3$Pt [480].

As it is clear from the above consideration, the $\Delta \rho /\rho $-effect
in a single domain is determined by the difference of $\rho _{\bot }$ and $%
\rho _{\Vert }$~, i.e. by dependence of resistivity on the angle between
vectors $\mathbf{j}$ and $\mathbf{M}$. This dependence turns out by one-two
order of magnitude stronger than in the usual $\Delta \rho /\rho $ -effect.
Therefore the Lorentz force does not explain the effect quantitatively. The
most natural relevant microscopical mechanism is, as well as for the
extraordinary Hall effect, the spin- orbital interaction. The extraordinary $%
\Delta \rho /\rho $-effect, which is quadratic in $M$, occurs in the second
order of perturbation theory in this interaction.

Unlike the Hall effect, the microscopic theory of the $\Delta \rho /\rho $%
-effect in ferromagnets is not developed in detail, and the whole physical
picture is still absent. Some calculations with account of different
scattering mechanisms and the spin-orbital interaction were performed
starting from 50s [265]. Smit [481] and Marsocci [482] investigated the Mott
s-d transition mechanism. Kondo [465] treated the scattering by magnetic
inhomogeneities (as well as for the Hall effect, the picture of unquenched
orbital momenta was used).

Vu Dinh Ky [483] considered the transport equation for the impurity
scattering. In the second order in SOI, corrections, which are proportional
to$\mathbf{[kM]}^2$ , occur both in collision and scattering terms. They
yield the resisitivity anisotropy required. The final result is rather
cumbersome and may be represented in the form
$$
\Delta \rho /\rho \sim \lambda ^2M^2  
\eqno{(5.149)}
$$
so that $\Delta \rho /\rho $ decreases with temperature and vanishes above
the Curie point. A numerical estimation was made in [483] by using the
comparison with the anomalous Hall effect. Since the Hall resistivity $\rho
_H=4\pi R_sM$ occurs in the first order in $\lambda $, but contains an
extra power of the impurity potential $\overline{\phi }$, we have
$$
\Delta \rho /\rho _H\sim H_{so}/\overline{\phi }  
\eqno{(5.150)}
$$
Putting for nickel $H_{so}\sim $ 10$^{-13}$ erg and $\overline{\phi }\sim $
10$^{-14}$erg one obtains $\Delta \rho /\rho $ $\sim $10 in a rough
agreement with the experimental data ($\rho _H/\rho \sim 0$.5\%, $\Delta
\rho /\rho \sim $ 3\%).

The phenomenological consideration of magnetoresistivity in ferromagnetic
metals may be performed within the two-current model with strongly different
currents $j_{\uparrow }$ and $j_{\downarrow }$ [436].

\subsection{Magnetooptical effects}

Magnetooptical (MO) effects in ferromagnetic transition metals are closely
related to galvanomagnetic effects. Experimentally, the Faraday and Kerr
polarization plane rotation angles in ferromagnets are by several orders
larger than in paramagnetic metals. They are proportional to magnetization
rather than to magnetic field and strongly decrease above the Curie point.

Microscopic mechanisms of the large MO effects are connected with
spin-orbital interaction. In particular, the Faraday effect is analogous to
the high-frequency extraordinary Hall effect. Although the spontaneous Hall
effect is determined in the static limit by the magnetization, separation of
magnetic and electric characterisitics at high frequencies becomes
impossible, so that the similarity with magnetooptical effects occurs.
Proportionality of the Hall field and the Faraday rotation was established
already in 1893 by Kundt [484].

From the phenomenological point of view, polarization plane rotation is due
to gyrotropy, i.e. presence of antisymmetric contribution, in dielectric and
magnetic permeability tensors $\varepsilon _{\alpha \beta }(\omega )$ and $%
\mu _{\alpha \beta }(\omega )$. In the optical region the magnetic gyrotropy
is connected mainly with anisotropy of gyroelectric properties, the 
$q$-dependence of $\varepsilon (\mathbf{q},\omega )$ playing an important role.

First physical explanation of MO effects in ferromagnets was given by Hulme
[485] and Kittel [486], and the theory of the frequency dependence was
developed by Argyres [487] and Cooper [488]. An account of electron
scattering permitted to extend the theory to the low-frequency region
[489,490] and the case of low temperatures [491,381].

Consider a simple theory of the MO Kerr effect. At reflection from the
magnetic medium with the complex refraction factor $\widetilde{n}=n+ik$ and
off-diagonal conductivity $\sigma _{xy}$, the light with the frequency 
$\omega $ changes its polarization by the Kerr angle
$$
\theta _K=\frac{4\pi }\omega \left( A\Im \sigma _{xy}
+B\Re \sigma _{xy}\right) /\left( A^2+B^2\right)   
\eqno{(5.151)}
$$
where
\[
A=n^3-3nk^2-n,\qquad
B=-k^3+3n^2k-k
\]
At small damping $k\ll n$ the value of $\theta _K$ is determined mainly by 
$\Im \sigma _{xy}$ . In the simplest case of a cubic structure with the
magnetization vector which is parallel to the (001) plane, the Argyres
formula takes the form [316]
$$
\Im \sigma _{xy} =\frac \pi \omega \sum_{\mathbf{k},m\neq m^{\prime
}}[F_{m^{\prime }m\uparrow }^{xy}(\mathbf{k)}n_{\mathbf{k}m^{\prime
}\uparrow }(1-n_{\mathbf{k}m\uparrow })\delta (\omega -\omega _{mm^{\prime
}\uparrow }\left( \mathbf{k}\right) ) 
$$
$$
-F_{m^{\prime }m\downarrow }^{xy}(\mathbf{k)}n_{\mathbf{k}m^{\prime
}\downarrow }(1-n_{\mathbf{k}m\downarrow })\delta (\omega -\omega
_{mm^{\prime }\downarrow }\left( \mathbf{k}\right) )]  
\eqno{(5.152)}
$$
where $m$ is the band index,
\[
\omega _{mm^{\prime }\sigma }(\mathbf{k)=}\varepsilon _{\mathbf{k}\mathsf{%
m\sigma }}-\varepsilon _{\mathbf{k}m^{\prime }\sigma } 
\]
is the interband transition frequency, n is the Fermi distribution function,
$$
F_{m^{\prime }m\sigma }^{xy}(\mathbf{k)=}2i\sum_{m^{\prime \prime }}\left[ 
\frac{(l_{m^{\prime \prime }m}^z)^{*}}{\omega _{m^{\prime }m^{\prime \prime
}\sigma }}p_{m^{\prime \prime }m}^xp_{mm^{\prime }}^y+\frac{l_{m^{\prime
\prime }m}^z}{\omega _{m^{\prime }m^{\prime \prime }\sigma }}p_{m^{\prime
\prime }m^{\prime }}^xp_{m^{\prime }m}^y\right]   
\eqno{(5.153)}
$$
\begin{eqnarray*}
p_{m^{\prime }m}^\alpha  &=&\langle \mathbf{k}m^{\prime }\sigma \mid
-i\partial /\partial x_\alpha \mid \mathbf{k}m\sigma \rangle  \\
l_{m^{\prime }m}^z &=&\langle \mathbf{k}m^{\prime }\sigma 
\mid \xi l_z\mid \mathbf{k}m\sigma \rangle 
\end{eqnarray*}
\thinspace \thinspace $l_z$ is the orbital momentum operator $z$-projection,
\[
\xi =\frac 2{rc^2}\frac{\partial V_{eff}}{\partial r} 
\]
with $V_{eff}$ being the effective potential for conduction electrons.

The formula (5.152) demonstrates a strong dependence of the Kerr angle on
electronic structure and magnetic ordering. Thus MO effects are promising
from the point of view of comparison with band structure calculations.
Comparison of magnetooptical properties with results of band calculations is
performed, e.g., for nickel [381], Fe-Co alloys [492] and gadolinium [493].

An almost total compensation of the first and second term in the square
brackets of (5.152) takes place provided that the spectrum e depends weakly
on $\sigma $. At the same time, the effect is large in strongly
ferromagnetic case. In particular, for half-metallic ferromagnets (Sect.4.4)
at $\omega <\delta _\sigma $ ($\delta _{\sigma }$ is the gap for the spin
projection $\sigma $) the corresponding term in (5.152) vanishes, so that
one may expect large values of Kerr rotation. Indeed, for the system 
NiMnSn$_{1-x}$ Sb$_x$ the intensity of peaks in the frequency dependence of 
$\theta _K$ decreases sharply with increasing $x$, 
i.e. as the Fermi level goes out
of the gap [494]. According to (5.153), the angle $\theta _K$ is
proportional to the spin-orbit coupling, i.e. increases for heavy elements.
Therefore one may expect that the HMF which contain platinum should have
larger values of $\theta _K$. Indeed, giant values $\theta _K\simeq 0.15^0$
(for the red light), which exceed considerably the values for
NiMnSb, were observed the compound PtMnSb [307,308] (the results of the
calculation are given in [316]). Note, however, that according to [495] main
difference between electronic structures of HMF's PtMnSb and NiMnSb, which
results in smaller value of $\theta _K$ in the latter compounds, is
connected not so much with the spin- orbit matrix element values, as with a
shift of some energy levels owing to ``scalar'' relativistic effects
(velocity dependence of mass and the Darvin correction). In this sense, the
simplest assumption about the direct connection between the spin-orbit
coupling strength and the Kerr rotation is not quite adequate.

Record values of $\theta _K$ might be observed in the ferromagnetic phase of
the compound UNiSn, which should also have a half-metallic structure
[315,316]. However, experimentally this turns out to be antiferromagnetic
[496,497] (see discussion in Sect.6.6). Nevertheless, investigation of
isostructural compounds containing actinides (e.g., UCoSn, UPdSn) is of
interest from this point of view.

The polar, meridional and equatorial Faraday and Kerr effects, as well as
the gyrotropic effect (change of reflected light intensity with
magnetization) in ferromagnetic metals were studied systematically in
[498-500]. In the case where the magnetization is perpendicular to the plane
of the ingoing light wave it is posssible to determine simultaneously real
and imaginary part of the tensors $\varepsilon $ and $\mu $ [500]. Both
intraband (indirect) and interband transitions turn out to play a role. The
latter are important near resonance absorption frequencies, so that the
frequency dependence of magnetooptical effects is a $1/\omega $-hyperbola
with peaks owing to interband transitions. A simple expression for the
intraband contribution to off-diagonal magnetic permeability may be
presented in the form [499]
$$
\mu _{xy}(\omega )=-\frac \hbar {2m^{*}c^2}i\overline{\kappa }\frac{\omega
_p^2}\omega   
\eqno{(5.154)}
$$
(5.154) where $\omega _p$ is the plasma frequency, $\overline{\kappa }$ is
an averaged (over the Fermi surface) dimensionless parameter which
determines the correction to electron quasimomentum owing to spin-orbital
interaction. For $\omega \sim 10^{14}$s$^{-1}$ one has $\mu _{xy}\sim
10^{-6}-10^{-4}$ in agreement with experimental data.

It is instructive to establish a correlation between temperature dependences
of magnetooptical and galvanomagnetic effects and compare quantitatively the
corresponding microscopic SOI parameters. In the paper [489], the relation
between the extraordinary Hall coefficient $R_s$ and the MO parameters
defined by
$$
\mathbf{j} =\sigma _1(\omega )\mathbf{E}+\sigma _2(\omega )\left[ \frac{%
\mathbf{M}}{\widetilde{M}}\mathbf{E}\right]  
$$
$$
\sigma _2(\omega )/\sigma _1(\omega ) \equiv iq  
\eqno{(5.155)}
$$
was obtained in the form
$$
4\pi R_sM=-\frac{\sigma _2\left( 0\right) }{\sigma _1^2\left( \omega \right) }  
\eqno{(5.156)}
$$
Using the expression for the one-band conductivity
$$
\sigma _1(\omega )=\sigma _1(0)\frac \gamma {\gamma +i\omega }  
\eqno{(5.157)}
$$
where $\gamma =1/\tau $ is the relaxation rate we obtain
$$
\Re q=-4\pi R_sM\sigma _1(0)\frac \omega \gamma ,\qquad
\Im q=4\pi R_sM\sigma _1(0)  
\eqno{(5.158)}
$$
so that the signs of $\Re q$ and $\Im q$ are opposite and determined by the sign
of $R_s$ , which agrees with the data [501,451]. At the same time, in the
case of the normal MO effect the signs of $\Re q$ and $\Im q$ coincide. The
formulas (5.159) provide the correct order of magnitude for $\Im q$. Since
$$
R_s\symbol{126}\left[ \sigma _1(0)\right] ^{-2},\qquad
\gamma \sim \left[ \sigma _1(0)\right] ^{-1}  
\eqno{(5.159)}
$$
we obtain from (5.158) the temperature dependences
$$
\Re q \sim M(T),\qquad
\Im q \sim \frac{M(T)}{\sigma (T)}  
\eqno{(5.160)}
$$
Their verification is hampered by that the experimental $T$-dependences
[501] correspond to the resonance region. Therefore investigation of the
long-wave region would be of interest.

MO effects in the X-ray region seem also to be promising to investigate the
band structure. In particular, the magnetic X-ray dichroism (MXD) effect is
discussed in Sect.2.5.

\subsection{Thermomagnetic effects}

Besides above-discussed electric, thermoelectric and galvanomagnetic
effects, there exist a number of effects owing to combined action of the
fields $\mathbf{E,H}$ and $\mathrm{grad}T$ [7,8,265]. Although not
investigated now in detail, the thermomagnetic effects (TME) may be of
interest since they provide an additional information about microscopic
transport mechanisms in solids.

Difficulties in studying TME increase for transition metals, especially in
the case of magnetic ordering. First paper in this direction have
demonstrated that TME are described by the same concepts and are determined
by the same microscopic parameters as galvanomagnetic effects. In a number
of cases one can perform a separation of ``normal'' and ``spontaneous'' TME.
Therefore a correlation between TME and galvanomagnetic effects should exist
which may be used to verify values of the microscopic parameters and
separate different scattering mechanisms. Some attempts of this kind were
made (see [475]).

At present, most investigated TME are the Nernst-Ettingshausen and
Righi-Leduc effects (see also Sect.5.1). The transverse Nernst-Ettingshausen
effect is the occurrence of the electric power $E_y$ in the presence of the
magnetic field $H_z$ and temperature gradient in the x-direction. Similar to
the Hall effect, we have
$$
E_y^{NE}=(Q_0B_z+4\pi Q_sM_z)grad_xT  
\eqno{(5.161)}
$$
where $Q_0$ and $Q_s$ are the normal and spontaneous Nernst-Ettingshausen
coefficients. The general expression (5.1) under the conditions
$$
\mathbf{j}=0,\qquad
\partial T/\partial y=\partial T/\partial z=0  
\eqno{(5.162)}
$$
yields
$$
E_y\simeq \frac{\lambda _{xx}\sigma _{yx}-\lambda _{yx}\sigma _{xx}}{\sigma
_{xx}^2}\frac{\partial T}{\partial x}  
\eqno{(5.163)}
$$

The experimental dependence $Q_s(T)$ turn out to be stronger than $Q_0(T)$. 
This dependence is described by the empirical equation [502]
$$
Q_s(T)=-T(\alpha +\beta \rho )  
\eqno{(5.164)}
$$
where $\alpha $ is determined by impurities. A derivation of the formula
(5.164) with the use of the density-matrix appoach was performed by
Kondorsky [503] by analogy with the Hall effect. The case of alloys was
considered in papers [504]. The result (5.164) was confirmed also within the
``side-jump'' mechanism approach by Berger [204].

The spontaneous Righi-Leduc effect (occurrence of grad$_yT$ in the presence
of grad$_xT$ and $M_z$) and the Ettingshausen effect (occurence of grad$_yT$
in the presence of electric current $j_x$ and magnetization $M_z$ ) were
also investigated [475]. The normal Righi-Leduc coefficient is known to be
expressed in terms of the Hall coefficient and conductivity,
$$
A_0=\sigma R_0=\frac{e\tau }{m^{*}c}  
\eqno{(5.165)}
$$
Search of a similar relation for the spontaneous coefficient is of interest.

\chapter{THE KONDO EFFECT AND PROPERTIES OF ANOMALOUS d- AND f-COMPOUNDS}

In this Chapter we consider physics of some types of 4f- and 5f-compounds
which exhibit anomalous electronic properties. Here belong so-called
heavy-fermion, Kondo lattice and intermediate valence systems; in some
aspects close is physics of some d-systems, in particular of copper-oxide
high-$T_{\mathrm{c}}$ superconductors where strong electron correlation
effects in CuO$_2$ planes take place. We shall demonstrate applications of
many-electron models to the description of unusual physical phenomena in
these substances. Of course, we do not claim to cover completely this topic
which is very wide and rapidly develops, but consider some selected
questions which are determined by the authors' scientific interests.

Most exotic properties are characteristic for heavy-fermion compounds. They
possess giant values of effective electron mass, which are manifested most
brightly in the huge linear specific heat. In a somewhat arbitrary
definition of heavy-fermion systems, the boundary value of $\gamma $ = 400
mJ/mol K$^2$ was established. Besides that, large paramagnetic
susceptibility at low temperatures and large coefficient at the $T^2$-term
in resistivity are observed.

An especial interest in the heavy-fermion compounds was stimulated by the
discovery of unconventional superconductivity in CeCu$_2$Si$_2$, UBe$_{13}$,
UPt$_3$. The superconducting state is characterized by an anisotropic
(non-zero angular momentum) pairing and is possibly not caused by
electron-phonon interaction [505,506]; often superconductivity coexists with
antiferromagnetic ordering.

The properties of the ``classical'' heavy-fermion systems CeAl$_3$, CeCu$_6$,
CeCu$_2$Si$_2$, UBe$_{13}$, UPt$_3$, U$_2$Zn$_{17}$, UCd$_{11}$,
NpBe$_{13} $ are considered in detail in the review [507]. For recent years, a
number of ternary Ce-based compounds with huge (of order 1 J/mol$\cdot $K$^2$
and larger) values of $\gamma $, e.g., 2.5 J/mol K$^2$ for CeInPt$_4$, 1.2
J/mol$\cdot $K$^2$ for CeInCu$_2$, were synthesized. Besides that many Ce,
Yb and U-based systems possess ``moderate'' value of g (of order 100 mJ/mol
K$^2$). Data on electronic specific heat and magnetic properties of some
anomalous rare earth and actinide compounds, and corresponding bibliography
are given in Table 6.1 (see also the reviews [512,520,545-547]).

As well for transition metals, the ratio of the coefficient at the
$T^2$-term in resisitivity to $\gamma ^2$ is universal, but has a magnitude
about
25 times larger: $A/\gamma ^2\sim 10^{-5}\mu \Omega $ cm (mol K/mJ)$^2$ .
This correlation is seen in Fig.6.1 [548]. For comparison, the data on
d-systems with large $\gamma $ (A15 structure compounds which exhibit
superconductivity with moderately high $T_c$) are also shown.

Modern de Haas - van Alphen investigations yielded the possibility to
observe directly some bands with large effective masses [288,549,550]. Thus
the substances under consideration provide an extremely interesting example
of strong renormalization of electron characteristics owing to interelectron
correlations. Standard band structure calculations of heavy-fermion systems
usually greatly underestmate the values of $N(E_{\mathrm{F}})$. A
satisfactory agreemeent may be achieved in a semiphenomenolgical way by
introducing large phase shifts corresponding to the strong resonance
scattering of electron states at the Fermi level (see [550,551]).

The simplest theoretical model describing the formation of heavy- fermion
state is the s-f exchange model. It should be stressed that, unlike the case
of systems with strong Hubbard correlations (Sect.4.6), the bare interaction
between current carriers and localized moments, which leads to the anomalous
behaviour, is rather weak. However, owing to resonance character of s-f
scattering near the Fermi level, the effective interaction in the
many-electron system tends to infinity. Thus we deal with an essentially
many-particle problem. In the next Section, we start the consideration of
this problem from the case of one magnetic d(f)-impurity.

\section{The one-centre Kondo effect}

It is believed now that the main cause for anomalous properties of
heavy-fermion systems is the Kondo effect. This effect was first discussed
in connection with the problem of resistivity minimum in diluted alloys of
transition metals. Even in ``pure'' samples of copper, gold and zinc, an
increase of resistivity was observed at low temperatures below 10-20 K. It
was established experimentally that this phenomenon is closely related to
the presence of a small amount ($10^{-2}-10^{-3}\%$) of impurities of
transition metals (Cr,Fe,Mn), which retain magnetic moment in the host
metal. Such a strong effect cannot be explained within simple one-electron
approximations for impurity resistivity. Kondo [552] demonstrated that in
the third order of perturbation theory the s-d exchange interaction of
conduction electrons with localized moments results in a singular $\ln T$%
-correction to resistivity owing to many-body effects (Fermi statistics).
When combined with the usual low temperature $T^5$-contribution owing to
electron-phonon scattering, this correction does yield the minimum of
resistivity. Minimizing the expression
$$
\rho =Ac\ln T+BT^5
\eqno{(6.1)}
$$
with $c$ being the impurity concentration we obtain $T_{\min }\sim c^{1/5}$,
i.e a weak $c$-dependence.

Consider the occurence of Kondo anomalies in the s-d exchange model with one
impurity atom
$$
\mathcal{H}=\sum_{\mathbf{k}\sigma }t_{\mathbf{k}}c_{\mathbf{k}\sigma
}^{\dagger }c_{\mathbf{k}\sigma }-I\sum_{\mathbf{kk}^{\prime }\sigma \sigma
^{\prime }}(\mathbf{S\sigma }_{\sigma \sigma ^{\prime }})c_{\mathbf{k}\sigma
}^{\dagger }c_{\mathbf{k}^{\prime }\sigma ^{\prime }}
\eqno{(6.2)}
$$
The lowest-order matrix element of the elastic s-d scattering ($t_{\mathbf{k}%
}=t_{\mathbf{k}^{\prime }}=E_{\mathrm{F}}$) do not differ from those of the
usual potential impurity scattering:
$$
\langle \mathbf{k}^{\prime }\uparrow |T|\mathbf{k}\uparrow \rangle =-IS^z
\eqno{(6.3)}
$$
In the second order of perturbation theory, two types of scattering
processes contribute to the matrix element (6.3): \\1) An electron passes
from the state $|\mathbf{k}\uparrow \rangle $ into the state $|%
\mathbf{k}^{\prime }\uparrow \rangle $. The
intermediate state $|\mathbf{k}^{\prime \prime }\sigma \rangle $ should be
empty. \\2) An electron from the occupied state $|\mathbf{k}^{\prime \prime
}\sigma \rangle $ passes into the state $|\mathbf{k}^{\prime }
\uparrow \rangle $, and then an electron from the state $|%
\mathbf{k}\uparrow \rangle $ passes into the state $|\mathbf{k}%
^{\prime \prime }\sigma \rangle $. The sign of this contribution is opposite
to that of the first contribution because of the antisymmetry of the
many-electron wave function.

The whole expression for the second order contribution reads
$$
\langle \mathbf{k}^{\prime } \uparrow |T|\mathbf{k}\uparrow \rangle
^{(2)}=I^2(S^zS^z+S^{-}S^{+})\sum_{\mathbf{k}^{\prime \prime }}\frac{1-n_{%
\mathbf{k}^{\prime \prime }}}{t_{\mathbf{k}}-t_{\mathbf{k}^{\prime \prime }}}
\eqno{(6.4)}
$$
$$
-I^2(S^zS^z+S^{-}S^{+})\sum_{\mathbf{k}^{\prime \prime }}\frac{1-n_{%
\mathbf{k}^{\prime \prime }}}{t_{\mathbf{k}}-t_{\mathbf{k}^{\prime \prime }}}
$$
$$
=I^2S(S+1)\sum_{\mathbf{k}^{\prime \prime }}\frac 1{t_{\mathbf{k}}-t_{%
\mathbf{k}^{\prime \prime }}}+I^2S^z\sum_{\mathbf{k}^{\prime \prime }}\frac{%
1-2n_{\mathbf{k}^{\prime \prime }}}{t_{\mathbf{k}}-E}
$$
where we have used the commutation relation $[S^{+},S^{-}]=2S^z$; similar
expressions may be obtained for other matrix elements $\langle \mathbf{k}%
\sigma |T|\mathbf{k}^{\prime }\sigma ^{\prime }\rangle $. The first term in
the right-hand side of (6.4) yields only a small correction to the potential
scattering. At the same time, the second term, which occurs because of
non-commutativity of spin operators and contains Fermi distribution
functions, contains a large logarithmic factor which diverges as $E$
approaches the Fermi energy:
$$
\sum_{\mathbf{k}^{\prime \prime }}\frac{1-2n_{\mathbf{k}^{\prime \prime }}}{%
t_{\mathbf{k}}-t_{\mathbf{k}^{\prime \prime }}}=\int dE^{\prime }\rho
(E^{\prime })\frac{1-2f(E^{\prime })}{E^{\prime }-E}\approx 2\rho \ln \frac
W{\max \{|E|,T\}}
\eqno{(6.5)}
$$
where $W$ is of order of conduction band width, $E$ is referred to the Fermi
level, $\rho (E)$ is the bare density of states of conduction electrons with
one spin projection, $\rho =\rho (E_{\mathrm{F}})$. The total contribution
of magnetic scattering to resistivity is obtained after averaging of the
squared matrix elements over the localized spin projections:
$$
\rho _{\mathrm{sd}}=\rho _{\mathrm{sd}}^{(0)}\left( 1-4I\rho \ln \frac
WT\right) ,\qquad \rho _{\mathrm{sd}}^{(0)}\sim I^2S(S+1)
\eqno{(6.6)}
$$
(at calculating resistivity, $|E|\sim T$). Thus the singular Kondo
contribution occurs in the third order in $I$.

Similar perturbation calculations may be performed for other physical
properties [552]. The magnetic susceptibility is diminished by logarithmic
corrections of the second order:
$$
\chi =\frac{S(S+1)}{3T}\left( 1-4I^2\rho ^2\ln \frac WT\right)
\eqno{(6.7)}
$$
The s-d contribution to specific heat occurs in the fourth order [560]
$$
C_{\mathrm{sd}}(T)=16\pi ^2S(S+1)I^4\rho ^4\left( 1-8I\rho \ln \frac
WT\right)
\eqno{(6.8)}
$$
The logarithmic term in (6.6) results in an resistivity increase with
lowering $T$ for $I<0$. This sign of the s-d exchange parameter takes place
for magnetic impurities in noble metals where the effective s-d exchange is
in fact due to combined action of the s-d hybridization and Coulomb
interaction (see Appendix N):
$$
I=V^2\left( \frac 1\Delta -\frac 1{\Delta +U}\right)
\eqno{(6.9)}
$$
where $V$ is hybridization matrix element, $\Delta $ is the position of the
d-level calculated from $E_{\mathrm{F}}$, $U$ is the one-site Coulomb
interaction.

The resistivity increase is suppressed at very low temperatures by magnetic
ordering of impurities owing to long-range RKKY-interaction among them (in
the ordered phase, the orientation of spins becomes fixed and the scattering
becomes ineffective). For d-impurities the third-order correction (6.6) is
in most cases sufficient to describe experimental data since at not too
small c the higher order contributions of perturbation theory are small up
to the magnetic ordering temperature. On the other hand, rare-earth
impurities (e.g., Ce, Yb, Sm, Tm in Y or La-based hosts) may be considered
as isolated ones up to $c\sim 1\%$; even at larger concentrations the
interaction among them does not necessarily result in usual magnetic
ordering, but leads to the formation of ``dense'' Kondo systems [545].
Therefore the problem of accurate treatment of many-electron effects owing
to s-d(f) exchange interaction at low temperatures (the Kondo problem)
occurs.

Summing up the leading logarithmic terms yields 
$$
\rho _{\mathrm{sd}}=\rho _{\mathrm{sd}}^{(0)}\left( 1+2I\rho \ln \frac
WT\right) ^{-2}  
\eqno{(6.10)}
$$
In the case of ``ferromagnetic'' s-d exchange $I>0$ this ``parquet''
approximation [14] solves the Kondo problem. However, in more important case 
$I<0$ this approximation yields a divergence of resistivity at the
temperature 
$$
T_{\mathrm{K}}=W\exp \frac 1{2I\rho }  
\eqno{(6.11)}
$$
which is called the Kondo temperature.

Unlike the critical temperature of a ferromagnet or superconductor, the
Kondo temperature does not correspond to a phase transition, but is just a
characteristic energy scale for a crossover between high-and low-temperature
regions. The consideration of the region $T<T_{\mathrm{K}}$ is a very
difficult and beautiful mathematical problem. The case $T\ll T_{\mathrm{K}}$
was investigated within phenomenological Fermi-liquid theory [553,10] and
analytical renormalization group methods [554,555]. The numerical solution
was obtained by Wilson with the use of renormalization group approach [556].
Finally, under some simplifying approximations (which reduce the problem to
one dimension) the exact solution of the one-impurity s-d model was obtained
by Andrei and Wiegmann with the use of the Bethe ansatz [557,558].

It turns out that at $T\rightarrow 0$ the effective (renormalized) s-d
interaction becomes infinitely strong, so that the impurity magnetic moment
is totally compensated (screened) by conduction electrons. Strictly
speaking, in the usual s-d model with zero orbital momentum (6.2) such a
compensation occurs only for $S=1/2$, and for a general $S$ the Kondo effect
results in the contraction of the impurity spin, $S\rightarrow S-1/2$.
However, in a realistic situation of degenerate electron bands, the number
of ``scattering channels'' for conduction electrons is sufficient to provide
the screening.

The resistivity tends at $T\rightarrow 0$ to a finite unitarity limit (which
corresponds to the maximum possible phase shift of $\pi /2$), the
corrections at low $T$ being proportional to $(T/T_{\mathrm{K}})^2$
[14,552,558]: 
$$
\rho _{\mathrm{sd}}=\frac 3\pi \left( \rho v_{\mathrm{F}}e\right)
^{-2}\left( 1-\frac{\pi ^2T^2}{T_{\mathrm{K}}^2}+O\left( \frac T{T_{\mathrm{K%
}}}\right) ^4\right)  
\eqno{(6.12)}
$$
The specific heat of the system has a maximum at $T\sim T_{\mathrm{K}}$ and
behaves linearly at $T\rightarrow 0$: 
$$
C_{\mathrm{sd}}(T)=\frac \pi 3\frac T{T_{\mathrm{K}}}\left( 1+O\left( \frac
T{T_{\mathrm{K}}}\right) ^2\right)  
\eqno{(6.13)}
$$
which is reminiscent of electronic specific heat with $E_{\mathrm{F}%
}\rightarrow \pi T_{\mathrm{K}}.$

The magnetic entropy at $T=0$, 
\[
S(0)=R\ln (2S+1) 
\]
is removed owing to the magnetic moment screening rather than to magnetic
ordering. The magnetic susceptibility 
$$
\chi =\frac{(g\mu _B)^2}{2\pi T_{\mathrm{K}}}\left( 1-O\left( \frac{T^2}{T_{%
\mathrm{K}}^2}\right) \right)  
\eqno{(6.14)}
$$
demonstrates the Pauli behaviour (in contrast with the Curie law (6.7) at $%
T>T_{\mathrm{K}}$) and is greatly enhanced, as well as specific heat. These
results may be described in terms of a narrow many-particle Abrikosov-Suhl
resonance at the Fermi level with a width of order $T_{\mathrm{K}}$ and a
height of order $1/T_{\mathrm{K}}$, so that $T_{\mathrm{K}}$ plays the role
of the effective degeneracy temperature. Thus a new Fermi-liquid state is
formed which is characterized by large many-electron renormalizations.

The interpolation formula for $\chi (T)$ may be presented in a form of the
Curie-Weiss law with a negative paramagnetic Curie temperature, $|\theta
|\sim T_{\mathrm{K}}$ . It should be noted in this connection that the
difference between transition metal impurities, which retain magnetic moment
in a given host, from ``non-magnetic'' ones has quantitative rather than
qualitative nature. The second case may be viewed as that of high $T_{%
\mathrm{K}}$ of order 10$^2$-10$^4$ K, sometimes higher than the melting
point (in the case of usual Pauli susceptibility $T\rightarrow E_{\mathrm{F}%
} $). Similar considerations may be applied to pure substances where local
magnetic moments do not exist at low temperatures (although concrete
theoretical models may be quite different). For enhanced Pauli paramagnets
like Pd, Pt, UAl$_2$ , where the Curie-Weiss holds at high temperatures, one
introduces, instead of the Kondo temperature, the so-called spin-fluctuation
temperature.

\section{The Kondo temperature for d-impurities}

Due to the strong exponential dependence on the model parameters, the Kondo
temperature varies in a large interval from 10$^{-2}$ to 10$^4$ K. Values of 
$T_{\mathrm{K}}$ for transition metal impurities in copper and gold
[559,560], which are determined from anomalies of various physical
properties (resistivity, thermopower, specific heat, magnetic susceptibilty)
are presented in Fig.6.2. One can see a characteristic $V$-shape dependence
in the 3d-series with a sharp minimum in the middle of the series ($n=5$).
In the papers [559] these data were interpreted within the Schrieffer model 
$$
\mathcal{H}_{\mathrm{sd}}=-\frac In\sum_{\mathbf{kk}^{\prime }\sigma }%
\mathbf{S\sigma }_{\sigma \sigma ^{\prime }}\sum_mc_{\mathbf{k}lm\sigma
}^{\dagger }c_{\mathbf{k}^{\prime }lm\sigma ^{\prime }}  
\eqno{(6.15)}
$$
[561] which is rather artificial since it does not take into account
scattering by orbital degrees of freedom, despite they should play an
important role [562]. The model (6.15) yields the following $n$-dependence
of $T_{\mathrm{K}}$ 
$$
T_{\mathrm{K}}=W\exp \left( -\frac n{2|I|\rho }\right)  
\eqno{(6.16)}
$$
Consider the calculation of the Kondo temperature with account of the
orbital scattering for the localized degenerate d-level (the configuration d$%
^n$, $n<5$) at neglecting intraconfiguration splitting. Then the dependence
on quantum numbers of ME terms vanishes, so that the Hamiltonian of the
one-impurity model takes the form 
$$
\mathcal{H} =\sum_{\mathbf{k}m}t_{\mathbf{k}}c_{\mathbf{k}m}^{\dagger }c_{%
\mathbf{k}m}-I\sum_{\mathbf{kk}^{\prime }mm^{\prime }m_i}[X(\{m_1\ldots
m_{n-1},m^{\prime }\},\{m_1\ldots m_{n-1},m\})  
$$
$$
-\frac 1{2[l]}\delta _{mm^{\prime }}]c_{\mathbf{k}m}^{\dagger }c_{%
\mathbf{k}^{\prime }m^{\prime }}  
\eqno{(6.17)}
$$
where all the indices in the sets $\{{m}_i{\}}$ ($m_i=1,2...2[l]$
include both spin and orbital indices, $[l]=2l+1$, $l=2$ for d-electrons)
are different, X-operators mark all the possible transitions, the second
term in the brackets is subtracted to exclude the potential scattering, $I<0$. 
For $l=0$ the model (6.17) is reduced to the usual s-d exchange model
(6.2) with $S=1/2$.

The Kondo temperature is determined from the pole of the $T$-matrix defined
by 
$$
\langle \!\langle c_{\mathbf{k}^{\prime }m}|c_{\mathbf{k}m}^{\dagger
}\rangle \!\rangle _E=\frac{\delta _{\mathbf{kk}^{\prime }}}{E-t_{\mathbf{k}}%
}+\frac{T_{\mathbf{kk}^{\prime }}(E)}{(E-t_{\mathbf{k}})(E-t_{\mathbf{k}%
^{\prime }})}  
\eqno{(6.18)}
$$
Write down the equation of motion 
$$
(E-t_{\mathbf{k}^{\prime }})\langle \!\langle c_{\mathbf{k}^{\prime }m}|c_{%
\mathbf{k}m}^{\dagger }\rangle \!\rangle _E=\delta _{\mathbf{kk}^{\prime
}}-I\sum_{\mathbf{p}}\Gamma _{\mathbf{kp}m}(E)  
\eqno{(6.19)}
$$
$$
\Gamma _{\mathbf{kq}m}(E) =\sum_{m_1\ldots m_{n-1}m^{\prime }}\langle
\langle [X(\{m_1\ldots m_{n-1},m^{\prime }\},\{m_1\ldots m_{n-1},m\}) 
\eqno{(6.20)} 
$$
$$
-\frac 1{2[l]}\delta _{mm^{\prime }}]c_{\mathbf{q}m}|c_{\mathbf{k}%
m}^{\dagger }\rangle \!\rangle _E 
$$
Performing the simplest decoupling in the equation of motion for $\Gamma $
(which is an analogue of the Nagaoka decoupling, see [552]) we derive 
$$
(E-t_{\mathbf{q}})\Gamma _{\mathbf{kq}m}(E) =
-I\{1-[l]^{-1}(2[l]-n+1)^{-1}+\frac 14[l]^{-2}  
$$
$$
+(n-2)\sum_{m^{\prime }m^{\prime \prime }\mathbf{p}}\langle c_{\mathbf{%
p}m^{\prime \prime }}^{\dagger }c_{\mathbf{q}m^{\prime }}[X(\{m_1\ldots
m_{n-1},m^{\prime }\},\{m_1\ldots m_{n-1},m\})  
$$
$$
-\frac 1{2[l]}\delta _{m^{\prime }m^{\prime \prime }}]\rangle \}\sum_{%
\mathbf{k}}\langle \!\langle c_{\mathbf{k}^{\prime }m}|c_{\mathbf{k}%
m}^{\dagger }\rangle \!\rangle _E-I\{(2[l]-2n+2)n_{\mathbf{q}}  
$$
$$
-n-1-\frac 1{2[l]}\}\sum_{\mathbf{p}}\Gamma _{\mathbf{kp}m}(E)  
\eqno{(6.21)}
$$
Solving the equation (6.21) we obtain the estimate for $T_{\mathrm{K}}$ from
the divergence in the ``parquet'' approximation, which corresponds to the
second Born approximation for resistivity, 
$$
T_{\mathrm{K}}=W\exp \left( -\frac 1{2[l]-2n+2}\frac 1{|I|\rho }\right) 
\eqno{(6.22)}
$$
For the second half of the d-series $n>5$ the degeneracy of the
configuration d$^n$ is smaller than that of d$^{n-1}$. Thus we have to
consider the situation where the level d$^n$ lies above the Fermi level,
i.e. $\Delta >0$ (otherwise, the Kondo effect is absent). Then we may pass
to the hole representation and reproduce the result (6.22) with the
replacement $n\rightarrow 2[l]-n$. The formula (6.22) may be fitted to
experimental data of Fig.6.2 with $|I|\rho =1/16$, whereas using (6.16)
yields the unreasonably large value $|I|\rho =1/4$.

The dependence on ME term quantum numbers, which is neglected in above
consideration, seems to be important since the distance between different
terms in free atoms is of order of several eV. An account of ME term
splitting is performed in Appendix N within the degenerate Anderson model.
The result has the form 
$$
T_{\mathrm{K}}=W\exp \left( -\left( \frac{[S][L]}{[S^{\prime }][L^{\prime }]}%
-1\right) ^{-1}\left( n^{1/2}G_{S^{\prime }L^{\prime }}^{SL}\right)
^{-2}\frac 1{|I|\rho }\right) ,\qquad I=\frac{v_l^2(\mathbf{k}_{\mathrm{F}})}%
\Delta  
\eqno{(6.23)}
$$
Although the general picture of ion levels (especially in a crystal field)
is very complicated, the occurence of the squared fractional parentage
coefficients $\left( n^{1/2}G_{S^{\prime }L^{\prime }}^{SL}\right) ^{-2}$ in
(6.23) is expected to lead to a further sharpening of the dependence $T_{%
\mathrm{K}}(n)$. Indeed, the $n$-dependence of fractional parentage
coefficients on the average has a minimum in a middle of the d-series. Such
a dependence is due to that the total number of ME terms is maximum near $%
n=5 $ by combinatorial reasons, and for a given $n$ the values of $G^2$
satisfy the sum rules (A.8). However, direct use of the formula (6.23),
which yields a strongly oscillating dependence $T_K(n)$ in contradiction
with experimental data, probably overestimates the role of ME term effects.

At derivation of (6.23) we have restricted ourselves to the case of $LS$%
-splitting and neglected crystal-field (CF) effects which may be very
important. In particular, CF results in that $L=0$ for some d-impurities
(e.g., for V, Ni [555]). The CF splitting may be taken into consideration in
a similar way, but this requires more cumbersome calculations with the use
of Clebsh-Gordan and fractional parentage coefficients for a point group
(these may be found, e.g., in [565]). At accurate estimations of $T_{\mathrm{%
K}}$ , the influence of several groups of degenerate levels (e.g.,
corresponding to different atomic or crystal-field split terms) should be
also taken into account. The expression for two level groups ($\Delta _1$
with degeneracy $N_1$ and $\Delta _2$ with degeneracy $N_2$) has the form
[565] 
$$
T_{\mathrm{K}}=T_{\mathrm{K}}^{(0)}\left( \frac{T_{\mathrm{K}}^{(0)}}{\Delta
_2-\Delta _1}\right) ^{N_2/N_1}  
\eqno{(6.24)}
$$
where 
\[
T_{\mathrm{K}}^{(0)}=W\exp \left( \frac{\Delta _1}{\rho V^2(N_1+N_2)}\right) 
\]
However, such calculations require a detailed information on the electronic
structure of Kondo impurities.

The role of variation of the interconfiguration splitting should be also
discussed. It was supposed above that the effective s-d parameters, i.e. the
values of $v$, $\Delta $ and $\rho $ do not depend on the configuration d$^n$%
. However, it is well known that in the many-electron picture these
configurations possess different stability (e.g., the value of $\Delta $
should be related to atomic ionization potentials). In particular, the
spherically symmetric configuration d$^5$ is rather stable which may result
in a large value of $|\Delta |$ thereby lowering $T_{\mathrm{K}}$ for
manganese.

\section{Spin dynamics and electronic properties of Kondo lattices}

The anomalous rare-earth and actinide compounds are classified as
concentrated Kondo systems or Kondo lattices since the formation of
low-temperature Kondo state gives most natural explanation of their unusual
properties (large values of $\gamma $ and $\chi (0)$). Most of these
compounds exhibit $\ln T$-contribution in resitivity at high temperatures,
but have a metallic ground state with $\rho (T\rightarrow 0)\sim T^2$ .
However, examples of insulating Kondo lattices are known too. The system
CeNiSn possesses at low temperatures extremely small energy gap of order of
a few K [566,567]. Partial replacing of Ni by Cu results in heavy-fermion
metallic behaviour [568]. A similar situation takes place for the narrow-gap
compound CeRhSb [569]: the system Ce$_{1-x}$Th$_x$RhSb demonstrates at $%
x\sim 0.4$ the huge value $\gamma >1$ J/mol K$^2$ [570]. Narrow energy gaps
are found also for NdBiPt [571], Ce$_3$Bi$_3$Pt$_4$ [572].

The picture of the insulating Kondo lattice is also used sometimes for the
narrow-gap intermediate valence semiconductors SmB$_6$, SmS (golden phase)
[545], which are discussed in Sect. 6.5. The formation of the insulating
Kondo state may be described in terms of the coherent Kondo scattering, the
Abrikosov-Suhl resonance being transformed into the narrow many-electron gap.

Consider the Kondo effect manifestations for the periodic lattice of
localized f-moments within the framework of the s-f exchange model (G.2).
This case differs from that of a single Kondo impurity by the presence of
intersite exchange interactions and, consequently, of spin dynamics which
tends to weaken usual Kondo divergences and also leads to some new effects.

We apply perturbation theory in the s-f exchange parameter, the Heisenberg
f-f interaction being taken into account exactly. The calculation of the
second-order contribution to the electron self-energy yields [367] 
$$
\Sigma _{\mathbf{k}}^{(2)}(E)=I^2\sum_{\mathbf{q}}\int d\omega K_{\mathbf{q}%
}(\omega )\left( \frac{1-n_{\mathbf{k}+\mathbf{q}}}{E-t_{\mathbf{k}+\mathbf{q%
}}+\omega }+\frac{n_{\mathbf{k}+\mathbf{q}}}{E-t_{\mathbf{k}+\mathbf{q}%
}-\omega }\right)  
\eqno{(6.25)}
$$
where $K_{\mathbf{q}}(\omega )$ is the spectral density of the localized
spin subsystem, defined by 
$$
\langle \mathbf{S}_{-\mathbf{q}}(t)\mathbf{S}_{\mathbf{q}}\rangle
=\int\limits_{-\infty }^{+\infty }d\omega e^{i\omega t}K_{\mathbf{q}}(\omega
)  
\eqno{(6.26)}
$$
One can see that in the presence of spin dynamics Kondo-like divergences in
the self-energy arise already in the second order. They are formally due to
the Fermi functions: 
$$
\sum_{\mathbf{q}}\frac{n_{\mathbf{k}+\mathbf{q}}}{E-t_{\mathbf{k}+\mathbf{q}%
}\pm \bar \omega }\simeq \rho \ln \frac W{\max \{|E|,T,\bar \omega \}} 
\eqno{(6.27)}
$$
where $\bar \omega $ is the characteristic spin-fluctuation frequency. As
follows from the spectral representation for the commutator Green's
functions (E.18), 
\[
K_{\mathbf{q}}(\omega )=-\frac 1\pi N_{\mathrm{B}}(\omega )\Im \chi _{%
\mathbf{q}\omega } 
\]
$$
\chi _{\mathbf{q}\omega }=\langle \!\langle \mathbf{S}_{\mathbf{q}}|\mathbf{S%
}_{-\mathbf{q}}\rangle \!\rangle _\omega ,\qquad 
\Im \chi _{\mathbf{q}\omega }=-\Im \chi _{\mathbf{q}-\omega }  
\eqno{(6.28)}
$$
Then in the classical limit $\omega \ll T$ one has 
\[
K_{\mathbf{q}}(\omega )=K_{\mathbf{q}}(-\omega )=-\frac 1\pi \frac \omega T%
\Im \chi _{\mathbf{q}\omega } 
\]
so that the terms with the Fermi functions cancel out mutually. However, in
the quantum case $\Sigma (E)$ varies sharply in an energy region of order $%
\bar \omega $ near $E_{\mathrm{F}}$. This leads to an appreciable
renormalization of the Green's function residue $Z$ (see (G.53)) and,
consequently, of the electronic effective mass $m^{*}$ and specific heat.
These renormalizations vanish at $T\gg \bar \omega $. In particular, the
result (6.25) with $I\rightarrow U$ yields the spin-fluctuation (paramagnon)
renormalization in the Hubbard model [573].

Expressions, similar to (6.25) may be also obtained in other situations. For 
$$
K_{\mathbf{q}}(\omega )\sim [1-f(\Delta _{\mathrm{cf}})]\delta (\omega
+\Delta _{\mathrm{cf}})+f(\Delta _{\mathrm{cf}})\delta (\omega -\Delta _{%
\mathrm{cf}})  
\eqno{(6.29)}
$$
the formula (6.25) describes effects of the interaction with
crystalline-electric-field excitations [263], $\Delta _{\mathrm{cf}}$ being
the CF level splitting. In the case of the electron-phonon interaction $I$
is replaced by the corresponding matrix element, and 
$$
K_{\mathbf{q}}(\omega )=[1+N_{\mathrm{B}}(\omega _{\mathbf{q}})]\delta
(\omega +\omega _{\mathbf{q}})+N_{\mathrm{B}}(\omega _{\mathbf{q}})\delta
(\omega -\omega _{\mathbf{q}})  
\eqno{(6.30)}
$$
A spectral density of the form (6.29) with $\Delta $, which depends weakly
on $\mathbf{q}$, corresponds to localized spin fluctuations. The
renormalization of $m^{*}$ owing to such fluctuations is much stronger than
that owing to ``soft'' paramagnons because of smallness of fluctuation phase
volume in the latter case.

Thus the definition of the Kondo effect in systems exhibiting dynamics is
non-trivial. The condition $Z\ll 1$, characteristic for the Kondo lattices,
may be satisfied not only due to the usual Kondo effect (the formation of
the Abrikosov-Suhl resonance at $T<T_{\mathrm{K}}$), but also due to the
interaction with low-energy spin- or charge-density fluctuations.

The result $m^{*}\sim 1/\bar \omega $, which follows from (6.25), (6.27),
does not change in form when account is taken of higher orders terms even
for arbitrarily small $\Delta $. This problem is investigated in [574] for a
simple model which describes the interaction with local excitations of a
two-level system ($\bar \omega =\Delta $). After collecting all the singular
terms of the type under consideration, we obtain in the $(n+1)$-th order the
singular factors $\ln ^n|(E\pm \bar \omega )/\bar \omega |$, i.e. the
divergences are shifted from E and the cutoff parameter is $\bar \omega $
rather than the bandwidth $W$. Besides that, all the singularities are
cancelled at $\bar \omega \rightarrow 0$ due to the factors of the type 
$\tanh (\Delta /2T)$.

Now we consider the true Kondo divergences, corresponding to another
sequence of singular terms which describes spin-flip processes and starts
from the third order in the s-f parameter. These divergences do not vanish
in the absence of dynamics and do yield at $E\rightarrow 0$ the factors $\ln
(W/\max \{\bar \omega ,T\})$. The corresponding contribution to the
imaginary part of the self-energy with account of spin dynamics reads 
$$
\Im \Sigma _{\mathbf{k}}^{(3)}(E)=2\pi I^3\rho (E)\int d\omega \sum_{%
\mathbf{q}}K_{\mathbf{q}}(\omega )\frac{n_{\mathbf{k}+\mathbf{q}}}{E-t_{%
\mathbf{k}+\mathbf{q}}-\omega }  
\eqno{(6.31)}
$$
(the real part of the singular contribution is absent [367]). The quantity
(6.31) determines the damping of one-particle state and, consequently, the
relaxation rate $\tau ^{-1}(E)$. Comparing (6.31) with (6.4) we see that
spin dynamics results in a smearing of logarithmic term in resistivity.
Using, e.g., the simplest diffusion approximation for the spectral density, 
$$
K_{\mathbf{q}}(\omega )=\frac{S(S+1)}\pi \frac{D_s\mathbf{q}^2}{\omega +(D_s%
\mathbf{q}^2)^2}  
\eqno{(6.32)}
$$
with $D_s$ being the spin diffusion constant, we obtain 
$$
\delta \tau ^{-1}(E)=4\pi I^3\rho ^2S(S+1)\ln \frac{E^2+\bar \omega ^2}{W^2}
\eqno{(6.33)}
$$
where $\bar \omega =4D_sk_{\mathrm{F}}^2$ . Thus in resistivity 
$$
\ln T\rightarrow \frac 12\ln (T^2+\bar \omega )\approx \ln (T+a\bar \omega
),\qquad a\sim 1  
\eqno{(6.34)}
$$
A similar replacement occurs in other physical properties (e.g., in
paramagnetic susceptibility and specific heat). In this connection, shifts
in $\rho (T)$ and $c(T)$ curves with changing content of components in
heavy-fermion systems were discussed in [367].

Now we discuss the thermoelectric power $\alpha (T)$ in Kondo lattices. At
moderately high (as compared to $T_{\mathrm{K}}$ ) temperatures $\alpha (T)$
is usually large and has an extremum (a maximum at $\alpha <0$, a minimum at 
$\alpha <0$). Large Kondo contributions to $\alpha (T)$ correspond to the
anomalous odd contribution to$\tau ^{-1}(E)$ [552], which should arise, by
analytical properties of $\Sigma (E)$, from the logarithmic singularity in Re%
$\Sigma (E)$ [367]. Although such singularity is absent in the usual Kondo
problem, it occurs in the presence of the potential scattering $V$ which
leads to emergence of complex factors 
$$
1+V\sum_{\mathbf{k}}(E-t_{\mathbf{k}}+i0)^{-1}  
\eqno{(6.35)}
$$
which ``mix'' Im$\Sigma $ and Re$\Sigma $ in the incoherent regime. Then
spin dynamics leads to the replacements (6.36) 
$$
\ln \frac{|E|}W\rightarrow \frac 12\ln \frac{E^2+\bar \omega ^2}{W^2},\qquad 
\mathrm{sign}E\rightarrow \frac 2\pi \tan ^{-1}\frac E{\bar \omega } 
\eqno{(6.36)}
$$
in Im$\Sigma $ and Re$\Sigma $ respectively, and the anomalous contribution
to $\alpha (T)$ reads 
$$
\alpha (T)\sim \frac{I^3V}{e\rho (T)}\int dE\frac ET\frac{\partial f(E)}{%
\partial E}\tan ^{-1}\frac E{\bar \omega }\sim \frac{I^3V}{e\rho (T)}\frac
T{\max \{T,\bar \omega \}}  
\eqno{(6.37)}
$$
Thus the quantity $\bar \omega $ plays the role of a characteristic
fluctuating magnetic field which is introduced in [552] to describe
thermoelectric power of diluted Kondo systems.

Near the magnetic ordering point $T_{\mathrm{M}}$, the quantity $\bar \omega 
$ contains non-analytic contributions, proportional to $(T-T_{\mathrm{M}%
})^{1-\alpha }$ ($\alpha $ is the critical index for specific heat).
Therefore the Fisher-Langer transport properties anomalies near magnetic
phase transitions [575] should be enhanced in Kondo magnets. This is
confirmed by experimental data on Ce$_3$Al [576], UCu$_5$ [577], Ce$_{1-x}$La%
$_x$In$_3$ [578] where appreciable breaks of thermopower at the Neel point
were observed (Fig.6.3).

\section{Ground state of the Kondo lattices}

The calculations of the previous Section correspond to the temperature
region $T\gg T_{\mathrm{K}}$ . As well as in the case of one Kondo impurity,
below $T_{\mathrm{K}}$ the perturbation theory regime passes into the strong
coupling regime where magnetic momnets are suppressed and a new Fermi-liquid
state with greatly enhanced electronic effective mass occurs.

Besides the Kondo temperature, one introduces sometimes the second energy
scale - the coherence temperature $T_{\mathrm{coh}}$, which corresponds to
onset of coherent Kondo scattering by different lattice sites. This is
usually small in comparison with $T_{\mathrm{K}}$. The picture of the
coherent state formation enables one to treat experimental data on
low-temperature anomalies of thermoelectric power in heavy-fermion systems
[545,579,580]. With decreasing $T$ below the above-discussed
high-temperature extremum, $\alpha (T)$ often changes its sign, has an
extremum again and vanishes linearly at $T\rightarrow 0$ (see Fig.6.4). Such
a behaviour may be attributed to occurence of a pseudogap with reversing the
sign of the quantity $dN(E)/dE$ at the Fermi level, which determines the $%
\alpha (T)$ sign (Sect.5.4). Besides that, formation of the coherent state
is indicated by the positive magnetoresistivity and a sharp negative peak in
the Hall coefficient.

To describe the formation of the singlet Kondo state in the strong coupling
region we may use the simplest Hamiltonian of the $SU(N)$ Anderson-lattice
model 
$$
\mathcal{H} =\sum_{\mathbf{k}m}t_{\mathbf{k}}c_{\mathbf{k}m}^{\dagger }c_{%
\mathbf{k}m}+\Delta \sum_{im}X_i(mm)  
$$
$$
+V\sum_{\mathbf{k}m}\left[ c_{\mathbf{k}m}^{\dagger }X_{\mathbf{k}%
}(0m)+X_{-\mathbf{k}}(m0)c_{\mathbf{k}m}\right]  
\eqno{(6.38)}
$$
($m=1...N$). This model is convenient at describing the interconfiguration
transitions f$^0$-f$^1$ (cerium, $J=5/2$) or f$^{14}$-f$^{13}$ (ytterbium, $%
J=7/2$) and is treated often within the $1/N$-expansion. A more realistic
model of s-f hybridization with inclusion of two (generally speaking,
magnetic) configurations is investigated in Appendix N. The model (6.38) may
be mapped by a canonical transformation, which excludes the hybridization,
onto the Coqblin- Schrieffer model 
$$
\mathcal{H}_{\mathrm{CS}}=\sum_{\mathbf{k}m}t_{\mathbf{k}}c_{\mathbf{k}%
m}^{\dagger }c_{\mathbf{k}m}-I\sum_{imm^{\prime }}X_i(mm^{\prime })c_{%
\mathbf{k}m^{\prime }}^{\dagger }c_{\mathbf{k}m}  
\eqno{(6.39)}
$$
\[
I=\frac{V^2}\Delta 
\]
However, it is easier to treat the Kondo-lattice state in the
strong-coupling regime directly in the model (6.38). To avoid difficulties
owing to complicated commutation relations for the X-operators, the
representation may be used [581,551] 
$$
X_i(m0)=f_{im}^{\dagger }b_i^{\dagger },\qquad
X_i(m^{\prime }m)=f_{im^{\prime }}^{\dagger }f_{im},\qquad
X_i(00)=b_i^{\dagger }b_i  
\eqno{(6.40)}
$$
where $f^{\dagger }$ are the Fermi operators, $b^{\dagger }$ are the
auxiliary (slave) Bose operators. This representation enables one to satisfy
the needed commutation relations (A.35). At the same time, according to
(A.28), one has to require fulfilment of the subsidiary condition 
$$
\sum_mX_i(mm)+X_i(00)=\sum_mf_{im}^{\dagger }f_{im}+b_i^{\dagger }b_i=1 
\eqno{(6.41)}
$$
Then the parameter $\langle b_i\rangle $ renormalizes the hybridization
matrix elements.

A description of the crossover to the coherent regime was constructed within
a modified SU(N) Anderson model [582]. The temperature dependence of
effective hybridization parameter was obtained in the form 
\[
V_{\mathrm{ef}}^2\sim \langle b_i^{\dagger }b_i\rangle \sim \varphi (T)
\]
$$
\varphi (T)=(N+e^{-T_{\mathrm{K}}/T}+1)^{-1}=\left\{ 
\begin{array}{ll}
1 & ,\quad T\ll T_{\mathrm{coh}} \\ 
O(1/N) & ,\quad T_{\mathrm{coh}}\ll T\ll T_{\mathrm{K}}
\end{array}
\right.   
\eqno{(6.42)}
$$
with the coherence temperature $T_{\mathrm{coh}}=T_{\mathrm{K}}/\ln N$.

Below we describe a simple approach to the problem of the ground state of
Kondo lattices, which uses directly the X-operator representation [367]. (An
alternative treatment of the Kondo-lattice state, which starts directly from
the s-f exchange model with $S=1/2$ and uses the pseudofermion
representation for spin operators, is performed in Appendix O and Sect.6.6.)
Consider the retarded Green's function for the localized level states 
$$
G_{\mathbf{k}m}^f(E)=\langle \!\langle X_{\mathbf{k}}(0m)|X_{-\mathbf{k}%
}(m0)\rangle \!\rangle _E  
\eqno{(6.43)}
$$
Write down the Dyson equation in the form 
$$
G_{\mathbf{k}m}^f(E)=R\left[ E-\Delta -\Sigma _{\mathbf{k}}^f(E)\right] ^{-1}
\eqno{(6.44)}
$$
where 
$$
R=N_0+N_m=1-\frac{N-1}Nn_f  
\eqno{(6.45)}
$$
The equation of motion for $G_{\mathbf{k}m}^f(E)$ has the form 
$$
(E-\Delta )G_{\mathbf{k}m}^f(E)=R\left( 1+V\langle \!\langle c_{\mathbf{k}%
m}|X_{-\mathbf{k}}(m0)\rangle \!\rangle _E+V\sum_{m^{\prime }\neq m}\langle
\!\langle X_{\mathbf{q}}(m^{\prime }m)c_{\mathbf{k-q}m^{\prime }}|c_{\mathbf{%
k}m}^{\dagger }\rangle \!\rangle _E\right)  
\eqno{(6.46)}
$$
Here we have carried out the decoupling of the term which corresponds to
processes without change of m and yields the formation of hybridization gap,
the corresponding spectrum being given by 
$$
(E-t_{\mathbf{k}})(E-\Delta )=V^2R  
\eqno{(6.47)}
$$
Further we neglect for simplicity the influence of this gap, which is
possible provided that the latter lies well below the Fermi level (besides
that, these contributions are formally small in the inverse degeneracy of
f-level, $1/N$).

The terms with $m\neq m^{\prime }$ contribute to Kondo divergences. Carrying
out decouplings in the equations for the Green's function in the right- hand
side of (6.46) we obtain 
$$
\Sigma _{\mathbf{k}}^f(E)=(N-1)V^2\sum_{\mathbf{q}}\frac{n_{\mathbf{q}}}{%
E-t_{\mathbf{q}}}\approx (N-1)\rho V^2\ln \frac W{|E|}  
\eqno{(6.48)}
$$
Here we have applied again the approximation (6.5).

Then the Green's function (6.44) has the pole, which is exponentially close
to the Fermi level, 
$$
|E|=|\Delta ^{*}|=T_{\mathrm{K}}\approx W\exp \left( \frac{|\Delta |}{%
(N-1)\rho V^2}\right)  
\eqno{(6.49)}
$$
Near this pole we may expand 
$$
G_{\mathbf{k}m}^f(E)\approx \frac Z{E-\Delta ^{*}}  
\eqno{(6.50)}
$$
where the residue of the Green's function, determining the inverse effective
mass, is 
$$
Z=R\left( 1-\frac{\partial \Sigma _{\mathbf{k}}^f(E)}{\partial E}\right)
_{E=\Delta ^{*}}^{-1}\approx \frac R{N-1}\frac{T_{\mathrm{K}}}{\rho V^2}\ll 1
\eqno{(6.51)}
$$
Thus the pole (6.49) determines a characteristic low-energy scale --- the
Kondo temperature. Note that one has to neglect, in spirit of the 
$1/N$-expansion, unity in comparison with $N$ to obtain the result which is
correct in the case $N=2$, which corresponds to the s-f model with $S=1/2$.
Similar calculations may be performed within the models with realistic
atomic configurations (N.9), (N.5) to obtain the results (N.23), (N.24).
These formulas have a simple physical meaning: we have in the exponent the
ratio of degeneracies of the multiplets $\Gamma _n$ and $\Gamma _{n-1}$.

The results for the Kondo temperature in the present approximation is valid
also for the case of one magnetic impurity. Then the value of $T_{\mathrm{K}%
} $ obtained determines the position of the Abrikosov-Suhl resonance.
Exponential dependence on external parameters (in particular, on the f-level
position $\Delta $) makes difficult to establish experimentally reliable
correlations of $T_{\mathrm{K}}$ with ME quantum numbers. However, the
expression (N.23) enables one to explain very low values of $T_{\mathrm{K}}$
, which are observed for Tm impurities (these are not obtained within the
large-$N$ approach [565]). Indeed, in this case both the configurations $%
\Gamma _n$ and $\Gamma _{n-1}$ are magnetic: $J=7/2$ for Tm$^{2+}$ and $J=6$
for Tm$^{3+}$ , so that the ratio $(J-J^{\prime })/(2J^{\prime }+1)$ in the
exponent of (N.23) is small.

In a periodic lattice of f-moments (anomalous rare earth and actinide
compounds), the Green's function of conduction electrons has the form 
$$
\langle \!\langle c_{\mathbf{k}m}|c_{\mathbf{k}m}^{\dagger }\rangle
\!\rangle _E=\left[ E-t_{\mathbf{k}}-V^2G_{\mathbf{k}m}^f(E)\right] ^{-1} 
\eqno{(6.52)}
$$
As follows from comparison of (6.52) with (6.47), (6.50), the effective
hybridization parameter near $E_{\mathrm{F}}$ is estimated as $v^{*}\sim
VZ^{1/2}\sim (T_{\mathrm{K}}/\rho )^{1/2}$. Thus, instead of the
Abrikosov-Suhl resonance, a Kondo gap (or pseudogap), which has a width of
order of $T_{\mathrm{K}}$, and corresponding density-of-states peaks occur
near the Fermi level, the latter being, generally speaking, shifted from the
gap. Such a picture of the energy spectrum is confirmed for metalic
heavy-fermion compounds, by direct far-infrared and point-contact
spectroscopy data [583].

\section{Intermediate valence systems}

A number of rare-earth elements (Ce, Sm, Eu, Tm, Yb and possibly Pr) do not
possess a stable valence, but vary it in different compounds. In some
systems these elements may produce so-called mixed (or intermediate) valent
state which is characterized by non- integer number of f-electrons per atom.
Such a situation may occur provided that the configurations 4f$^n$(5d6s)$^m$
and 4f$^{n-1}$(5d6s)$^{m+1}$ are nearly degenerate, so that
interconfiguration fluctuations are strong. In metallic systems, this
corresponds to the f-level located near the Fermi energy, f-states being
hybridized with conduction band states.

The intermediate valence (IV) state is characterized by the single line in
Moessbauer experiments (the time scale of these measurements is about 10$%
^{-11}$s), which has an intermediate position. On the other hand, in X-ray
experiments (the time is about 10$^{-16}$s) two lines are seen, which
correspond to configurations f$^n$ and f$^{n-1}$. A peculiar feature of the
transition into IV state is also the change of the lattice parameter to a
value which is intermediate between those for corresponding integer-valent
states. Such transformations (e.g., under pressure) are as a rule sharp
(first-order) transitions. Besides that, IV compounds possess at low
temperatures substantially enhanced electronic specific heat and magnetic
susceptibitlity. At high $T$, $\chi (T)$ obeys the Curie-Weiss law with the
effective moment which is intermediate beween the the values for the
corresponding atomic configurations.

Examples of IV compounds are metallic compounds YbAl$_2$, YbZn, YbCu$_2$Si$%
_2 $, YbAgCu$_4$, CeN, CeBe$_{13}$, CeSn$_3$, EuCu$_2$Si$_2$ (see reviews
[512,520,584]) and narrow-gap semiconductors TmSe, SmB$_6$, the ``golden''
phase of SmS (under pressure above 6 kbar), YbB$_{12}$.

The metallic cerium in the $\alpha $-phase was earlier considered as an IV
system [584] which seemed to be confirmed by the volume change at the $%
\gamma -\alpha $ transition under pressure. However, it is believed now that
this transition is in fact the Mott-Hubbard transition in the f-band
(delocalization of 4f-electrons due to overlap of f-states at different
sites) without a considerable valence change [585]. This is confirmed by
both spectroscopic data [586] and band calculations [587]. Besides that, the
energy difference of the f$^0$ and f$^1$ configurations in metallic cerium
is too high for explanation of small heats for formation of metallic Ce
alloys. According to paper [588], the ``tetravalent'' state of cerium
(without localized f-electrons) should be attributed to f$^2$ rather than to
f$^0$ configuration. As follows from atomic calculations (see [81]), a
collapse of f-electrons in cerium takes place at small variations of the
atomic potential. This phenomenon is accompainied by a strong decrease of
f-bandwidth.

For d-metals and their compounds, one can hardly use the term ``intermediate
valence'' since the hybridization and, consequently, the widths of d-peaks
are as a rule large, and valence fluctuations are too rapid. Thus we shall
observe in any experiment the ion states with non-integer number of
d-electrons.

The 5f-ions in actinide compounds demonstrate a great variety of valence
states -- from 1$+$ to 7$+$ [371], hybridization being also rather strong.
Only the compounds with ion binding have relatively stable valence
configurations. Thus the situation for actinides is more close to
3d-elements than to rare earths. Delocalization of 5f-electrons and analogy
between intermediate valence in actinides and cerium is discussed in
[81,589].

The ``homogeneous'' intermediate valence should be distinguished from the
inhomogeneous case where lattice sites corresponding to different valences
are inequivalent and the interconfigurational fluctuations are very slow.
Such a situation takes place, e.g., in the compound Sm$_3$S$_4$ which is
characterized by static charge ordering (similar to the well-known case of
magnetite Fe$_3$O$_4$). The inhomogeneous mixed-valent state demonstrates
two distinct lines not only in X-ray, but also in in Moessbauer spectra.
This case is opposite to rapid valence fluctuations in d-metals. In a number
of systems, e.g., EuM$_2$Si$_2$ (M = Fe, Co, Ni, Pd, Cu), EuPd$_2$P$_2$
[584], a transition from inhomogeneous mixed-valent state to the homogeneous
one is observed with increasing temperature.

Consider simplest theoretical models for description of the IV state. The
Hamiltonian of spinless Falicov-Kimball model reads 
$$
\mathcal{H}=\sum_{\mathbf{k}}\left[ t_{\mathbf{k}}c_{\mathbf{k}}^{\dagger
}c_{\mathbf{k}}+\Delta f_{\mathbf{k}}^{\dagger }f_{\mathbf{k}}+V\left( c_{%
\mathbf{k}}^{\dagger }f_{\mathbf{k}}+f_{\mathbf{k}}^{\dagger }c_{\mathbf{k}%
}\right) \right] +G\sum_if_i^{\dagger }f_ic_i^{\dagger }c_i  
\eqno{(6.53)}
$$
where $G$ is the parameter of on-site d-f Coulomb repulsion, we neglect for
simplicity the $\mathbf{k}$-dependence of hybridization. This Hamiltonian
enables one to take simply into account strong on-site f-f repulsion (in the
spinless model, doubly-occupied states are forbidden by the Pauli principle)
and is convenient at description of valence phase transitions, the
interaction $G$ being important for many-electron ``exciton'' effects. The
Falicov-Kimball model may be generalized by inclusion of Coulomb interaction
at different sites, which permits to describe charge ordering.

Unlike the model (6.53), the periodic Anderson model takes into account spin
degrees of freedom and permits to describe magnetic properties, but includes
explictitly the Hubbard repulsion. At neglecting orbital degeneracy the
Anderson Hamiltonian has the form 
$$
\mathcal{H}=\sum_{\mathbf{k}\sigma }\left[ t_{\mathbf{k}}c_{\mathbf{k}\sigma
}^{\dagger }c_{\mathbf{k}\sigma }+\Delta f_{\mathbf{k}\sigma }^{\dagger }f_{%
\mathbf{k}\sigma }+V\left( c_{\mathbf{k}\sigma }^{\dagger }f_{\mathbf{k}%
\sigma }+f_{\mathbf{k}\sigma }^{\dagger }c_{\mathbf{k}\sigma }\right)
\right] +U\sum_if_{i\uparrow }^{\dagger }f_{i\uparrow }f_{i\downarrow
}^{\dagger }f_{i\downarrow }  
\eqno{(6.54)}
$$
The intermediate valent situation corresponds to the situation where the
width of f-peak owing to hybridization, $\Gamma =\pi V^2\rho $, is small in
comparison with the distance $|\Delta |=|\varepsilon _{\mathrm{F}}-E_{%
\mathrm{F}}|$. On the contrary, the Kondo-lattice (heavy-fermion) state may
be considered as the nearly integral limit of the IV state (the valence
change does not exceed of a few percents). In a sense, IV systems may be
treated as Kondo lattices with high values of $T_{\mathrm{K}}$ [545]. Unlike
the Kondo lattice state, not only spin, but also charge fluctuations play an
important role in the IV state.

The excitation spectrum in IV systems may be obtained via diagonalizing the
Hamiltonian (6.53) in the simplest Hartree-Fock approximation which
corresponds to the picture of ``exciton condensate'' (formation of
electron-hole pairs) [590]. This is achieved, similar to the
superconductivity theory, by the Bogoliubov transformation 
$$
c_{\mathbf{k}}^{\dagger } =\cos \frac{\theta _{\mathbf{k}}}2\alpha _{%
\mathbf{k}}^{\dagger }+\sin \frac{\theta _{\mathbf{k}}}2\beta _{-\mathbf{k}}
$$
$$
f_{\mathbf{k}}^{\dagger } =\cos \frac{\theta _{\mathbf{k}}}2\beta _{-%
\mathbf{k}}-\sin \frac{\theta _{\mathbf{k}}}2\alpha _{\mathbf{k}}^{\dagger }
\eqno{(6.55)}
$$
Substituting (6.55) into (6.53) we derive 
$$
\langle \mathcal{H}\rangle  =\sum_{\mathbf{k}}\left[ t_{\mathbf{k}}n_{%
\mathbf{k}}^\alpha +\Delta (1-n_{\mathbf{k}}^\beta )\right] +\frac 12\sum_{%
\mathbf{k}}{}^{\prime }(t_{\mathbf{k}}-\Delta )(1-\cos \theta _{\mathbf{k}})
$$
$$
+V\sum_{\mathbf{k}}{}^{\prime }\sin \theta _{\mathbf{k}}+\frac 14G\left[
1-\left( \sum_{\mathbf{k}}{}^{\prime }\cos \theta _{\mathbf{k}}\right)
^2\right] -\frac 14G\left( \sum_{\mathbf{k}}{}^{\prime }\sin \theta _{%
\mathbf{k}}\right) ^2  
\eqno{(6.56)}
$$
where the fourth and fifth terms correspond to the Hartree and Fock
decoupling, $n_{\mathbf{k}}^{\alpha ,\beta }=f(E_{\mathbf{k}}^{\alpha ,\beta
}\mp \zeta )$ are the Fermi distribution functions for ``electron'' and
``hole'' excitations, 
\[
\sum_{\mathbf{k}}{}^{\prime }\ldots \equiv \sum_{\mathbf{k}}\left( 1-n_{%
\mathbf{k}}^\alpha -n_{\mathbf{k}}^\beta \right) \ldots 
\]
The one-particle energies are given by 
$$
E_{\mathbf{k}}^{\alpha ,\beta }=\frac{\delta \langle \mathcal{H}\rangle }{%
\delta n_{\mathbf{k}}^{\alpha ,\beta }}=\frac 12\left[ E_{\mathbf{k}%
}^{\alpha ,\beta }\pm (t_{\mathbf{k}}+\Delta )\right]   
\eqno{(6.57)}
$$
where 
\[
E_{\mathbf{k}}=\left[ X^2+(t_{\mathbf{k}}+Y)^2\right] ^{1/2},\qquad
X=2V-G\sum_{\mathbf{k}}{}^{\prime }\sin \theta _{\mathbf{k}},\qquad
Y=\Delta +G\sum_{\mathbf{k}}{}^{\prime }\cos \theta _{\mathbf{k}}
\]
The hybridization picture of the electron spectrum (Fig.6.5) is confirmed by
optical spectra and other data for IV semiconductors SmB$_6$, SmS and YbB$%
_{12}$ [591] and by investigation of transport properties for TmSe [592].

In the case of the semiconductor, the chemical potential is determined from
the condition 
$$
\sum_{\mathbf{k}}\left( n_{\mathbf{k}}^\alpha -n_{\mathbf{k}}^\beta \right)
=0  
\eqno{(6.58)}
$$
The renormalized hybridization parameter X determines the ``direct'' energy
gap 
$$
|X|=\min_{\mathbf{k}}\left( E_{\mathbf{k}}^\alpha -E_{\mathbf{k}}^\beta
\right)  
\eqno{(6.59)}
$$
which is observed in optical transitions. Unlike the case of usual
semiconductors, extrema of valence and conduction band do not coincide. The
width of the hybridization gap is estimated in terms of $X$ and bandwidth $W$
as $\delta \sim X^2/W$. The quantity $Y$ is the position of f-level in the
Hartree-Fock approximation. Varying (6.56) with respect to $\theta _{\mathbf{%
k}}$ we obtain 
\[
\sin \theta _{\mathbf{k}}=-\frac X{E_{\mathbf{k}}},\qquad \cos \theta _{%
\mathbf{k}}=\frac{t_{\mathbf{k}}-Y}{E_{\mathbf{k}}} 
\]
and the equations for $X$ and $Y$ take the form 
$$
X=2V\left( 1-G\sum_{\mathbf{k}}{}^{\prime }\frac 1{E_{\mathbf{k}}}\right)
^{-1}  
\eqno{(6.60)}
$$
$$
Y=\Delta +G\left( 1-\sum_{\mathbf{k}}{}^{\prime }\frac{t_{\mathbf{k}}-Y}{E_{%
\mathbf{k}}}\right)  
\eqno{(6.61)}
$$
These equations have, generally speaking, several IV solutions with various
values of the energy gap. Consider the case $T=0$. For a narrow-gap IV state
with $|X|\ll W$ the quantity $Y$ equals approximately to the Fermi energy
and the equation (6.60) takes the form 
$$
X=2V\left( 1-2\lambda \ln \frac{aW}{|X|}\right) ^{-1},\qquad \lambda =G\rho
,\qquad a\sim 1  
\eqno{(6.62)}
$$
The solution to 
$$
X_1=2V\left( 1-2\lambda \ln \frac{aW}{|V|}\right) ^{-1}  
\eqno{(6.63)}
$$
describes the state with a hybridization gap, which is renormalized by
correlation effects. It should be noted that the logarithmic divergences in
the narrow-gap state are cut at the width of the energy gap, so that, unlike
metallic Kondo lattices, the strong coupling regime is not achieved.

At $\lambda \geq 1$ the expression (6.63) yields the unique narrow-gap
solution. In this case, there exist also two broad-band solution, which are
approximately (at neglecting $V$) determined from the equation 
$$
G\left. \sum_{\mathbf{k}}\right. ^{\prime }\frac 1{E_{\mathbf{k}}}=1 
\eqno{(6.64)}
$$
describing ``excitonic'' gap states owing to Coulomb interaction.

In the case $\lambda \ll 1$ equation (6.62) has additional narrow-gap
excitonic solutions provided that 
$$
|V|<V_{\mathrm{c}}=aW\lambda \exp \left( -\frac 1{2\lambda }-1\right) 
\eqno{(6.65)}
$$
For $V\ll V_{\mathrm{c}}$ they are given by 
$$
|X_{2,3}|=aW\exp \left( -\frac 1{2\lambda }\right) \pm \frac{|V|}\lambda 
\eqno{(6.66)}
$$
The corresponding total energy is lower than for the state with the
hybridization gap (6.63). Thus, for a given valence there are metastable
insulator states. One can assume that in real IV semiconductors $V\sim V_{%
\mathrm{c}}$ so that the value of the gap is determined by a combined effect
of hybridization and many-electron (exciton) effects. Phase transitions
between states with different gaps were observed in the system TmSe$_{1-x}$Te%
$_x$ [593].

It should be noted that the hybridization gap can in principle be obtained
in band structure calculations. Such calculations with account of
relativistic effects were performed for SmS [594].

The expression for the density of states for the spectum (6.57) in terms of
the bare DOS $\rho (t)$ for the s-band has the form 
$$
N^{\alpha ,\beta }(E)=\left( 1+\frac{X^2}{4(E\mp Y)^2}\right) \rho \left( 
\frac{X^2}{4(Y\mp E)}\pm E\right)  
\eqno{(6.67)}
$$
Near the tops of the bare band and of the lower ``hole'' hybridization band
we have respectively 
$$
\rho (T)=A(t_{\mathrm{max}}-t)^{1/2}  
\eqno{(6.68)}
$$
$$
N^\beta =\widetilde{A}(E_{\mathrm{max}}-E)^{1/2},\qquad \widetilde{A}%
=A\left( \frac{2t_{\mathrm{max}}}X\right) ^2\approx A\frac W\delta 
\eqno{(6.69)}
$$
Taking into account the renormalization of the chemical potential
(calculated from the band top) we obtain for the electronic specific heat
enhancement in a highly doped semiconductor 
$$
\frac{\widetilde{\gamma }}\gamma =\frac{N(\widetilde{\xi })}{\rho (\xi )}=%
\frac{\widetilde{A}}A\left( \frac{\widetilde{\xi }}\xi \right) ^{1/2}=\left( 
\frac{\widetilde{A}}A\right) ^{2/3}=\left( \frac W\delta \right) ^{2/3} 
\eqno{(6.70)}
$$
Thus the states near the gap possess large effective mass. This explains the
large value of the linear specific heat ($\gamma =$ 145 mJ/mol K$^2$) [595]
in the ``golden phase'' of SmS where the current carriers (heavy holes) lie
near the lower edge of the gap.

To consider finite temperature behaviour we treat the simplest case where
the conduction band is symmetric and f-level lies exactly at the conduction
band centre, so that the mean number of f-electrons equals to $1/2$ and the
chemical potential does not depend on temperature. Then the equation (6.61)
takes the form 
$$
X=\frac{2V}{1-GL},\qquad L=\sum_{\mathbf{k}}\frac{1-2n_{\mathbf{k}}}{E_{%
\mathbf{k}}}  
\eqno{(6.71)}
$$
In the narrow-gap state we have 
$$
L=\int\limits_{-W}^W\frac{dE\rho (E)}{(E^2+X^2)^{1/2}}\tanh \frac{%
(E^2+X^2)^{1/2}+E}{4T}\approx \rho \int\limits_{-W/2T}^{W/2T}\frac{dx}x\tanh
x  
\eqno{(6.72)}
$$

At high temperatures, the energy gap logarithmically decreases with $T$: 
$$
X(T)\approx \frac{2V}{\lambda \ln (T/T^{*})},\qquad
T\gg \delta (T=0),T^{*} 
\eqno{(6.73)}
$$
with 
$$
T^{*}=\frac \pi 4\frac W{1.14}\exp \left( -\frac 1\lambda \right)  
\eqno{(6.74)}
$$
being of order of the direct excitonic gap. Such a behaviour explains
qualitatively data on the IV semiconductor SmB$_6$ where an appreciable
temperature dependence of the gap is observed in ESR experiments [596]. At
low temperatures, the corrections are exponentially small, as well as in the
case of superconducting gap.

To calculate the electronic specific heat in the narrow-gap insulating state
we write down the expression for the entropy of quasiparticles $\alpha $ and 
$\beta $%
\[
S=-\sum_{\mathbf{k,}j=\alpha ,\beta }\left[ n_{\mathbf{k}}^j\ln n_{\mathbf{k}%
}^j+(1-n_{\mathbf{k}}^j)\ln (1-n_{\mathbf{k}}^j)\right] 
\]
Then we get 
$$
C(T)=T\frac{\partial S}{\partial T}=-\frac 1T\sum_{\mathbf{k}j}\frac{%
\partial n_{\mathbf{k}}^j}{\partial E_{\mathbf{k}}^j}\left( 1-\frac T2\frac
\partial {\partial T}\right) \left( E_{\mathbf{k}}^j\right) ^2  
\eqno{(6.75)}
$$
At low temperatures $T\ll \delta (0)$ specific heat is exponentially small, 
$$
C(T)\approx \frac{2\rho }TX^2(0)\exp \left( \frac \delta T\right)  
\eqno{(6.76)}
$$
At high temperatures $\delta \ll T<|X|$ we have 
$$
C(T)=\frac \rho TX^2(T)\left[ 1-\frac 2{\ln (T^{*}/T)}\right] +\frac{\pi ^2}%
3\rho T  
\eqno{(6.77)}
$$
The first term in (6.77) dominates over the usual linear term. Thus $C(T)$
should have a maximum at $T\sim \delta $. Such a behaviour of $C(T)$ is
reminiscent of the Schottky anomaly for localized states, but because of
hybridization the dependence $1/T^2$ transforms into $1/T$.

The magnetic properties of the IV state may be investigated after including
spin variables within the model (6.54) [597]. The corresponding energy
spectrum in magnetic field is given by 
$$
E_{\mathbf{k}\sigma }^{\alpha ,\beta }(H)=E_{\mathbf{k}\sigma }^{\alpha
,\beta }(0)-\frac \sigma 2H\left[ \mu _{\mathrm{s}}+\mu _{\mathrm{f}}\pm
(\mu _{\mathrm{s}}-\mu _{\mathrm{f}})\frac{t_{\mathbf{k}}-E_{\mathrm{F}}}{E_{%
\mathbf{k}}}\right]  
\eqno{(6.78)}
$$
where $\mu _{\mathrm{s}}$ and $\mu _{\mathrm{f}}$ are magnetic moments of s-
and f-electrons. The field dependence of the energy gap turns out to be
simple 
$$
\delta (H)=\frac 12\left[ \min_{\mathbf{k}}E_{\mathbf{k}\uparrow }^\alpha
(H)+\min_{\mathbf{k}}E_{\mathbf{k}\uparrow }^\beta (H)\right] \approx \delta
(0)-\mu _{\mathrm{f}}H  
\eqno{(6.79)}
$$
One can see that at $\delta =\mu _{\mathrm{f}}H$ the gap vanishes and a
transition into metallic state takes place. Such a transition was observed
in YbB$_{12}$ at $H\sim $200 kOe [598].

The magnetic susceptibility is given by 
$$
\chi (T)=\lim_{H\rightarrow 0}\sum_\sigma \frac{\sigma \left( \mu _{\mathrm{f%
}}\langle f_\sigma ^{\dagger }f_\sigma \rangle +\mu _{\mathrm{s}}\langle
c_\sigma ^{\dagger }c_\sigma \rangle \right) }H  
\eqno{(6.80)}
$$
At high temperatures $T\gg \delta $, $\chi $ is determined by the
contribution of transitions between hybridization subbands 
$$
\Delta \chi (T) =\frac 12\sum_{\mathbf{k}}\left\{ \left( -\frac{\partial
n_{\mathbf{k}}^\alpha }{\partial E_{\mathbf{k}}}\right) \left[ \mu _{\mathrm{%
f}}\left( 1-\frac{t_{\mathbf{k}}-E_{\mathrm{F}}}{E_{\mathbf{k}}}\right) +\mu
_{\mathrm{s}}\left( 1+\frac{t_{\mathbf{k}}-E_{\mathrm{F}}}{E_{\mathbf{k}}}%
\right) \right] ^2\right.  
$$
$$
\left. +\left( -\frac{\partial n_{\mathbf{k}}^\alpha }{\partial E_{\mathbf{%
k}}}\right) \left[ \mu _{\mathrm{f}}\left( 1+\frac{t_{\mathbf{k}}-E_{\mathrm{%
F}}}{E_{\mathbf{k}}}\right) +\mu _{\mathrm{s}}\left( 1-\frac{t_{\mathbf{k}%
}-E_{\mathrm{F}}}{E_{\mathbf{k}}}\right) \right] ^2\right\}  
$$
$$
\approx \frac{\mu _{\mathrm{f}}^2}T\rho W  
\eqno{(6.81)}
$$
Thus we obtain, owing to strong energy dependence of density of states
(narrow peaks), a Curie-like behaviour with a non-integer magnetic moment.
This mechanism of temperature dependence was also discussed for transition
d-metals [599]. At $T\ll \delta $ the contribution (6.81) is exponentially
small, and $\chi (T)$ is determined by intrasubband transition contribution
which is given by 
$$
\chi (0)=2\left( \mu _{\mathrm{s}}-\mu _{\mathrm{f}}\right) ^2\rho 
\eqno{(6.82)}
$$
Note that at $\mu _{\mathrm{s}}=\mu _{\mathrm{f}}$ expression (6.82)
vanishes since the ground state is singlet.

We see that $\chi (T)$ should also have a maximum at $T\sim \delta $,
characteristic energy scale $\delta $ playing thereby the role of the
``Kondo temperature''. Such a maximum is weakly pronounced in SmB$_6$ and
SmS [512] and clearly observed in YbB$_{12}$ [600], the susceptibility at
low $T$ being masked by Van Vleck contribution of the Sm$^{2+}$ ion or by
paramagnetic impurities.

\section{Magnetic ordering in Kondo lattices and heavy-fermion compounds}

It was traditionally believed for many years that the competition of the
intersite RKKY exchange interaction and the Kondo effect should result in
the formation of either the usual magnetic ordering with large atomic
magnetic moments (as in elemental rare-earth metals) or the non-magnetic
Kondo state with suppressed magnetic moments.

However, experimental investigations of last years have convincingly
demonstrated that magnetic ordering and pronounced spin fluctuations are
widely spread among heavy-fermion systems and other anomalous 4f- and
5f-compounds, which are treated usually as concentrated Kondo systems. Data
on magnetic properties of such systems are presented in the Table 6.1.

The class of ``Kondo'' magnets is characterized by the following features [601]

\noindent (i) Logarithmic temperature dependence of resistivity at $T > T_K$
characteristic of Kondo systems (Fig.6.6).

\noindent (ii) Small value of the magnetic entropy at the ordering point, in
comparison with the value $R\ln (2S+1)$, which corresponds to the usual
magnets with localized moments (Fig.6.7). This phenomenon is connected with
the suppression of magnetic specific heat owing to the Kondo effect (see
Sect.6.1).

\noindent (iii)The ordered magnetic moment $M_s$ is small in comparison with
the ``high-temperature'' moment $\mu $m determined from the Curie constant.
The latter has as a rule ''normal'' value, which is close to that for the
corresponding rare-earth ion (e.g., $\mu _{eff}\simeq 2.5\mu _B$ for Ce$%
^{3+} $ ion). Such a behaviour is reminiscent of weak itinerant magnets (see
Sect.4.4).

\noindent (iv) Paramagnetic Curie temperature $\theta $ is as rule negative (even
for ferromagnets) and exceeds appreciably in absolute value the magnetic
ordering temperature. This behaviour is due to the large single-site Kondo
contribution to the paramagnetic susceptibility ($\chi (T=0)\sim 1/T_K$).
The most bright example is the Kondo ferromagnet CeRh$_3$B$_2$ with $T_C=115$%
K, $\theta =-370$K [284] with moderate $\gamma =$ 16 mJ/mol K$^2.$ (Large
value of $T_C$ in this compound, which exceeds even that for GdRh$_3$B$_2$, 
$T_C=105$K, is not typical and probably connected with strong d-f
hybridization.)

There exist numerous examples of systems (ferromagnets CePdSb, CeSi$_x$, Sm$%
_3$Sb$_4$, Ce$_4$Bi$_3$, NpAl$_2$, antiferromagnets CeAl$_2$, TmS, CeB$_6$,
UAgCu$_4$, some experimental data and bibliography are presented in Table
6.1) where ``Kondo'' anomalies in thermodynamic and transport properties
coexist with magnetic ordering, the saturation moment $M_s$ being of order
of 1$\mu _B$ .

As for heavy-fermion systems themselves, the situation is more complicated.
There exist unambigous evidences for the antiferromagnetism in UCd$_{11}$
and U$_2$Zn$_{17}$ with the same order of magnitude of $M_s$ [507]. For UPt$%
_3$ and URu$_2$Si$_2$, $M_s\simeq $ 2$\div $3 10$^{-2}$ $\mu _B$ [524-526].
Antiferromagnetic ordering with very small $M_s$ was also reported for CeAl$%
_3$ [509,510], UBe$_{13}$ [527], CeCu$_2$Si$_2$ [513,514]. Indications of
possible magnetic transition at 2mK were obtained for CeCu$_6$ [515].
However, the data for CeAl$_3$ and UBe$_{13}$ were not confirmed in papers
[511] and [528] respectively. Generally, a characteristic feature of heavy
fermion magnets is high sensitivity of $M_s$ to external factors such as
pressure and doping by a small amount of impurities. For example, UBe$_{13}$
becomes an antiferromagnet with an appreciable M under the pressure $P>23$
kBar; on the contrary, CeAl$_3$ becomes paramagnetic above $P=3$ kBar
[510]. The moment in UPt$_3$ increases up to values of order 1$\mu _B$ upon
adding 5\% of Pd instead of Pt or 5\% of Th instead of U [602]. A number of
heavy-fermion systems demonstrates metamagnetic transitions in weak magnetic
fields with a sharp increase of magnetic moment [603]. The ``marginal''
situation in the magnetic state of heavy-fermion systems is discussed in
[604] by using experimental data on their critical behaviuor.

The problem of magnetic ordering in the Kondo lattices was investigated in a
number of theoretical works [605-612]. The roles of the Kondo effect and and
the intersite RKKY interaction are determined by the relation of the two
energy scales: the Kondo temperature $T_K=W\exp (1/2I\rho )$ which
determines the crossover from the free-moment regime to the strong coupling
region, and $T_{RKKY}\sim I^2\rho $. The latter quantity is of the order of
magnetic ordering temperature $T_M$ in the absence of the Kondo effect. The
ratio $T_K/T_M$ may vary depending on external parameters and alloy
composition. As an example, Fig.6.8 shows concentration dependences of the
saturation magnetization $M_0$, $T_K$ and $T_C$ in the alloy 
CeNi$_{1-x}$Pd$_x$.

In the non-magnetic case, $T_{RKKY}\sim \overline{\omega }$ with $\overline{%
\omega }$ being a characteristic spin-fluctuation frequency. For most
compounds under consideration, $T_K>T_{RKKY}$ . However, there exist also
anomalous cerium and uranium-based magnets with $T_K\ll T_N$, e.g., CeAl$_2$%
Ga$_2$ [521], UAgCu$_4$ ($T_N$ = 18K, $T_K$ = 3K [530]). This case is close
to that of usual rare-earth magnets with the Kondo effect almost suppressed
by magnetic ordering.

To describe the formation of magnetic Kondo state, we consider Kondo
perturbation corrections to magnetic characteristics with account of spin
dynamics. Calculation of magnetic susceptibility [367] yields (cf. (6.7)) 
$$
\chi =\frac{S(S+1)}{3T}(1-4I^2L)  
\eqno{(6.83)}
$$
where 
$$
L=\frac 1{S(S+1)}\sum_{\mathbf{pq}}\int d\omega K_{\mathbf{p-q}}\frac{n_{%
\mathbf{p}}(1-n_{\mathbf{q}})}{(t_{\mathbf{q}}-t_{\mathbf{p}}-\omega )^2} 
\eqno{(6.84)}
$$
and the spin spectral density is defined by (6.26). A simple estimation of
the integral in (6.84) yields 
$$
\chi =\frac{S(S+1)}{3T}(1-2I^2\ln \frac{W^2}{T^2+\overline{\omega }^2}) 
\eqno{(6.85)}
$$
where the quantity in square brackets describes the suppression of the
effective moment. The Kondo corrections to the magnetic moment in the ferro-
and antiferromagnetic state are obtained by using the standard spin-wave
result 
$$
\delta \overline{S}=-\sum_{\mathbf{q}}\langle b_{\mathbf{q}}^{\dagger }b_{%
\mathbf{q}}\rangle  
\eqno{(6.86)}
$$
and substituting zero-point corrections to magnon occupation numbers 
$$
\delta \langle b_{\mathbf{q}}^{\dagger }b_{\mathbf{q}}\rangle =2I^2S\sum_{%
\mathbf{k}}\frac{n_{\mathbf{k\downarrow }}(1-n_{\mathbf{k-q\uparrow }})}{(t_{%
\mathbf{k\downarrow }}-t_{\mathbf{k-q\uparrow }}-\omega _{\mathbf{q}})^2} 
\eqno{(6.87)}
$$
$$
\delta \left\{ 
\begin{array}{c}
\langle b_{\mathbf{q}}^{\dagger }b_{\mathbf{q}}\rangle  \\ 
\langle b_{\mathbf{q}}^{\dagger }b_{-\mathbf{q}}^{\dagger }\rangle 
\end{array}
\right\} =2I^2S\sum_{\mathbf{k}}\left( \frac{n_{\mathbf{k}}(1-n_{\mathbf{k-q}%
})}{(t_{\mathbf{k}}-t_{\mathbf{k-q}}-\omega _{\mathbf{q}}^2}\pm \frac{n_{%
\mathbf{k}}(1-n_{\mathbf{k+Q-q}})}{(t_{\mathbf{k}}-t_{\mathbf{k+Q-q}%
})^2-\omega _{\mathbf{q}}^2}\right)   
\eqno{(6.88)}
$$
respectively (see Appendix G). The integration in both the cases gives 
$$
\delta \overline{S}/S=-2I^2\rho ^2\ln \frac W{\overline{\omega }}  
\eqno{(6.89)}
$$
The obtained corrections to the ground state moment occur in any conducting
magnets including pure 4f-metals. However, in the latter case they should be
small (of order of 10$^{-2}$). On the other hand, it would be interesting to
search them in rare-earth compounds with high values of $\rho $.

To obtain a self-consistent picture for a magnet with appreciable Kondo
renormalizations, we have to calculate the corrections to characteristic
spin-fluctuation frequencies $\overline{\omega }$. In the paramagnetic
phase, we can use the estimation from the second-order correction to the
dynamic susceptibility 
$$
\omega _{\mathbf{q}}^2=(\stackrel{.}{S}_{-\mathbf{q}}^z,\stackrel{.}{S}_{%
\mathbf{q}}^z)/(S_{-\mathbf{q}}^z,S_{\mathbf{q}}^z)  
\eqno{(6.90)}
$$
with 
$$
(A,B)\equiv \int_0^{1/T}d\lambda \langle \exp (\lambda H)B\exp (-\lambda
H)B\rangle  
\eqno{(6.91)}
$$
The calculation yields [608] 
$$
\omega _{\mathbf{q}}^2=\frac 43S(S+1)\sum_{\mathbf{p}}(J_{\mathbf{q-p}}-J_{%
\mathbf{p}})^2[1-4I^2L(1-\alpha _{\mathbf{q}})]  
\eqno{(6.92)}
$$
Here $L$ is defined by (6.85), 
$$
\alpha _{\mathbf{q}}=\sum_{\mathbf{R}}J_{\mathbf{R}}^2\left( \frac{\sin k_FR%
}{k_FR}\right) ^2[1-\cos \mathbf{qR]/}\sum_{\mathbf{R}}J_{\mathbf{R}%
}^2[1-\cos \mathbf{qR]}  
\eqno{(6.93)}
$$
Since $0<\alpha _{\mathbf{q}}<1$, the Kondo effect results in decreasing $%
\overline{\omega }$ as $T$ is lowered. In the approximation of nearest
neighbours (with the distance $d$) for $J(\mathbf{R})$, the value of $\alpha 
$ does not depend on $\mathbf{q}$: 
$$
\alpha _{\mathbf{q}}=\alpha =\left( \frac{\sin k_Fd}{k_Fd}\right) ^2 
\eqno{(6.94)}
$$
The corrections to the spin-wave frequency in ferromagnetic and
antiferromagnetic phase owing to magnon-magnon interactions are obtained by
using the results of Appendix G, (E.5), (E.13), (6.87), (6.88). Then we
obtain 
$$
\delta \omega _{\mathbf{q}}/\omega _{\mathbf{q}}=-4I^2\rho ^2a\ln \frac W{%
\overline{\omega }}  
\eqno{(6.95)}
$$
where the factor $a$ depends on the type of magnetic ordering.

The above perturbation theory results permit a qualitative description of
the magnetic Kondo-lattice state with a small magnetic moment. Suppose we
lower the temperature starting from the paramagnetic state. As we do it, the
magnetic moment is ``compensated'', but, in contrast with the one-impurity
situation, the degree of compensation is determined by $(T^2+\overline{%
\omega }^2)^{1/2}$ rather than $T$. At the same time, w itself decreases
according to (6.92). This process cannot be described analytically in terms
of perturbation theory. However, one needs to take $\overline{\omega }\sim
T_K$ at $T<T_K$ if he has in view the establishment of an universal energy
scale of the order of $T_K$ . The latter fact is confirmed by a large body
of experimental data on quasielastic electron scattering in Kondo systems,
which demonstrate that at low $T$ the typical ``central-peak'' width $\Gamma 
$ $\sim \overline{\omega }$ is of the same order of magnitude as the Fermi
degeneracy temperature determined from thermodynamic and transport
properties, i.e. $T_K$. Thus the process of the magnetic moment compensation
terminates somewhere at the boundary of the strong-coupling region and
results in the state with a finite (although possibly small) saturation
moment $M_s$.

A quantitative consideration of the Kondo lattice magnetism problem may be
performed within renormalization group approach in the simplest form of
Anderson's ``poor man scaling'' [554]. The above expressions, obtained from
perturbation theory, enable one to write down the renormalization group
equations for the effective s-f parameter and $\overline{\omega }$ [612].
This is achieved by considering the integrals over $\mathbf{k}$ with the
Fermi functions in the Kondo corrections to electron self-energes (G.33),
(G.73) and spin-fluctuation frequencies. To construct a scaling procedure,
one picks out the contributions from the energy layer $C<E<C+\delta C$ near
the Fermi level $E_F=0$. For example, in the case of ferromagnet we have for
the effective splitting in the electron spectrum 
$$
2I_{ef}S=2IS-[\Sigma _{\mathbf{k\uparrow }}^{FM}(E_F)-\Sigma _{\mathbf{%
k\downarrow }}^{FM}(E_F)]_{k=k_F}  
\eqno{(6.96)}
$$
Using (G.34) we derive 
$$
\delta I_{ef} =I^2\sum_{C<t_{\mathbf{k+q}}<C+\delta C}\left( \frac 1{t_{
\mathbf{k+q}}+\omega _{\mathbf{q}}}+\frac 1{t_{\mathbf{k+q}}-\omega _{
\mathbf{q}}}\right)  
$$
$$
=\frac{\rho I^2}{\overline{\omega }}\delta C\ln \left| \frac{C-\overline{%
\omega }}{C+\overline{\omega }}\right|   
\eqno{(6.97)}
$$
where 
\[
\overline{\omega }=4\mathcal{D}k_F^2
\]
$\mathcal{D}$ is the spin-wave stiffness. Introducing the dimensionless
coupling parameters 
$$
g=-2I\rho ,\qquad
g_{ef}(C)=-2I_{ef}(C)\rho   
\eqno{(6.98)}
$$
we obtain the system of renormalization group equations. In the
nearest-neighbour approximation for the intersite Heisenberg exchange
interaction it has a simple form 
$$
\partial g_{ef}(C)/\partial C =-\Phi  
$$
$$
\partial \ln \overline{\omega }_{ef}(C)/\partial C =a\Phi /2  
\eqno{(6.99)}
$$
$$
\partial \ln \overline{S}_{ef}(C)/\partial C =\Phi /2  
$$
with 
$$
\Phi =\Phi (C,\overline{\omega }_{ef}(C))=[g_{ef}^2(C)/C]\phi (\overline{%
\omega }_{ef}(C)/C)  
\eqno{(6.100)}
$$
The scaling function for para-, ferro- and antiferromagnetic phases has the
form 
$$
\phi (x)=\left\{ 
\begin{tabular}{ll}
$x^{-1}\arctan x$ & $\qquad \mathrm{PM}$ \\ 
$\frac 1{2x}\ln \left| \frac{1+x}{1-x}\right| $ & $\qquad \mathrm{FM}$ \\ 
$-x^{-2}\ln |1-x^2|$ & $\qquad \mathrm{AFM}$%
\end{tabular}
\right.   
\eqno{(6.101)}
$$
In all the cases $\phi (0)=1$, which guarantees the correct limit of one
Kondo impurity (Sect.6.1).

The results of investigation of the equations (6.99)-(6.101) [612] are as
follows. Depending on the relation between the one-impurity Kondo
temperature and the bare spin-fluctuation frequency, three regimes are
possible at $I < 0$:

\noindent 
(i) the strong coupling regime where $g_{ef}$ diverges at some $C$ is
roughly determined by the condition
$$
\overline{\omega }<T_K=W\exp (-1/g)  
\eqno{(6.102)}
$$
Then $I_{ef}(C\rightarrow 0)=\infty $, so that all the conduction electrons
are bound into singlet states and spin dynamics is suppressed.

\noindent 
(ii) the regime of a ``Kondo'' magnet with an appreciable, but not total
compensation of magnetic moments. This is realized in the interval $T_K<%
\overline{\omega }<AT_K$ ($A$ is a numerical factor of order unity), which
corresponds to a small interval $\delta g\sim g^2$. In this interval, the
renormalized values of magnetic moment and spin-fluctuation frequency, $%
S_{ef}(0)$ and $\overline{\omega }_{ef}(0)$, increase from zero to
approximately the bare values.

\noindent 
(iii) the regime of ''usual'' magnets with small logarithmic corrections to
the ground state moment (see (6.89)), which occurs at $\overline{\omega }%
>AT_K$. Note that the same situation takes place in the case of
``ferromagnetic'' s-f exchange interaction $I>0$.

High sensitivity of the magnetic state to external factors, which was
discussed above, is explained by that in the case (ii) the magnetic moment
changes strongly at small variations of the bare coupling constant. Of
course, the quantitative description should be different for the realistic
long-range behaviour of the RKKY-interaction. The renormalization of the
latter may be not described by the single constant $\alpha $.

Outwardly, the described mechanism of the formation of magnetic state with
small $M_s$ differs radically from the ordinary mechanism for weak itinerant
ferromagnets which are assumed to be located in the immediate vicinity of
the Stoner instability. Recall, however, that both the energy spectrum of
new Fermi quasiparticles and the effective interaction among them suffer
strong renormalizations. Therefore the inapplicability of the Stoner
criterion with bare parametes for Kondo magnets is practically evident.

Since a continuous transition exists between the highly-correlated Kondo
lattices and the ``usual'' itinerant-electron systems (in particular, Pauli
paramagnets may be viewed as systems with high $T_K$ of order of the Fermi
energy), the question arises about the role which many-electron effects play
in the ``classical'' weak itinerant magnets like ZrZn$_2$ . It may turn out
that the closeness of the ground state to the Stoner instability point, i.e.
the smallness of $M_s$, in the latter systems is not due to accidental bare
values of $N(E_F)$ and Stoner parameter, but determined by their
renormalization. In this context, it would be of interest to describe weak
itinerant magnets not from the ``band'' side, but from the side of local
magnetic moments which are nearly compensated. As it is customary now to
treat UPt Pd , CeSi and CeRh B as weak itinerant magnets (see, e.g., [613]),
the second approach appears already by far less natural than the first. From
the formal point of view, the calculations in the Hubbard model (Appendix
G), which describes itinerant electron systems, are similar to those in the
s-f exchange model if one postulates the existence of local moments.

The simple analytical description of crossover between the high-temperature
region $T>T_K$, where perturbation theory holds, and the strong-coupling
regime is hardly possible. Therefore it is important to discuss
approximations which work at $T<T_K$. A special mean-field approximation for
the ground state of magnetic Kondo lattices is considered in Appendix O.
This approach exploits the pseudofermion representation for spin operators
and reduces the s-f exchange model to an effective hybridization model.

The corresponding energy spectrum contains narrow DOS peaks owing to the
pseudofermion contribution (Fig.6.9). It should be noted that
f-pseudofermions become itinerant in the situation under consideration.
Delocalization of f-electrons in heavy-fermion systems, which is confirmed
by observation of large electron mass in de Haas - van Alphen experiments,
is not simply understandable in the s-f exchange model (unlike the
stituation in the Anderson model with f-states near the Fermi level with
bare s-f hybridization included). This delocalization is analogous to the
occurence of the Fermi excitation branch in the resonating valence bond
(RVB) theory of high-$T_c$ superconductors (see Sect.6.8).

As discussed above in Sect.6.4, the hybridization form of electron spectrum
with the presence of DOS peaks is confirmed by numerous experimental
investigations of Kondo lattices, including heavy-fermion systems. As for
ferromagnetic Kondo systems, of interest are the results on the temperature
dependences of magnetization in Sm$_4$Sb$_3$and Sm$_4$As$_3$ [542], which
turn out to be non-monotonous. Such a behaviour (temperature-induced
ferromagnetism) may be explained by the sharp energy dependence of DOS in
the hybridization model [614].

Because of the dependence of the effective hybridization on spin projection,
there exist, generally speaking, several ferromagnetic solutions. For the
constant bare DOS, only the saturated ferromagnetic state turns out to be
stable (remember that the same situation takes place in the Wohlfarth model,
i.e. the Stoner model with the rectangle DOS). One may assume that in more
general models (e.g., for large degeneracy of electron bands) the role of
this dependence is not so important, so that it may be neglected. Then the
criterion of ferromagnetism $J\sim T_K$ reduces, roughly speaking, to the
Stoner criterion with the replacement of the intrasite interaction parameter
by the intersite exchange parameter $J$, the effective density of states at
the Fermi level $N(E_F)\sim 1/T_K$ being large because of hybridization DOS
peaks. Thus magnetism of Kondo lattices has features of both localized spin
magnets (essential role of intersite exchange interaction) and of itinerant
ones (non-integer value of the magnetic moment and its connection with the
DOS structure).

The influence of spin-fluctuation corrections to the mean-field
approximation was investigated in [608]. Calculating the contribution to the
saturation magnetization from the fluctuations of the Heisenberg interaction
by analogy with (6.88) we obtain
$$
\delta \overline{S}\sim -\frac J{T_K}\ln \frac{T_K}J  
\eqno{(6.103)}
$$
($T_K$ plays the role of a characterisitc energy scale in the electron
spectrum). Thus at $J\sim T_K$ we have $\delta \overline{S}\sim S,$ and
formation of a state with a small moment is possible.

To consider antiferromagnetic ordering of the Kondo lattices one has to pass
to the local coordinate system (see (E.8)). Then the mean-field Hamiltonian
of the f-subsystem in the pseudofermion representation takes the form
$$
H_f=-J_{\mathbf{Q}}\overline{S}\sum_{\mathbf{k}}(f_{\mathbf{k+Q\uparrow }%
}^{\dagger }f_{\mathbf{k\downarrow }}+f_{\mathbf{k\downarrow }}^{\dagger }f_{%
\mathbf{k+Q\uparrow }})  
\eqno{(6.104)}
$$
Unlike the case of a ferromagnet, one may neglect the $\sigma $-dependence
of the hybridization since corrections owing to spin polarization have the
structure $(J_{\mathbf{Q}}\overline{S})^2/(t_{\mathbf{k+Q}}-t_{\mathbf{k}})$%
and are proportional to $(J_{\mathbf{Q}}\overline{S})^2/W$. Thus the
criterion of antiferromagnetism has the usual form
$$
J_{\mathbf{Q}}\chi _{\mathbf{Q}}>1  
\eqno{(6.105)}
$$
with $\chi _{\mathbf{Q}}$ being the non-enhanced ($H_f\rightarrow 0$)
staggered susceptibility of the f-subsystem in the effective hybridization
model (O.2). Using the Bogoliubov transformation (O.8) yields
$$
\chi _{\mathbf{q}} =\sum_{\mathbf{k}}\left[ v_{\mathbf{k}}^2v_{\mathbf{k+q}%
}^2\frac{n_{\mathbf{k}}^\alpha -n_{\mathbf{k+q}}^\alpha }{E_{\mathbf{k}%
}^\alpha -E_{\mathbf{k+q}}^\alpha }+u_{\mathbf{k}}^2u_{\mathbf{k+q}}^2\frac{%
n_{\mathbf{k}}^\beta -n_{\mathbf{k+q}}^\beta }{E_{\mathbf{k}}^\beta -E_{%
\mathbf{k+q}}^\beta }\right.  
$$
$$
-\left. 2u_{\mathbf{k}}^2v_{\mathbf{k+q}}^2\frac{n_{\mathbf{k}}^\beta -n_{%
\mathbf{k+q}}^\alpha }{E_{\mathbf{k}}^\beta -E_{\mathbf{k+q}}^\alpha }\right]
\eqno{(6.106)}
$$
Owing to the hybridization peaks, the contribution from the intersubband
transitions (the last term in brackets of (6.106)) turns out to be large: $%
\chi _{\mathbf{Q}}\sim 1/T_K$. Thus antiferromagnetism occurs at
$$
J_{\mathbf{Q}}>\nu T_K  
\eqno{(6.107)}
$$
where the constant $\nu $ of order of unity is determined by the band
structure.

The electron spectrum of antiferromagnetic Kondo lattices is disturbed by
both hybridization (Kondo) gap and an antiferromagnetic gap. Consider some
experimental examples. For the intermediate-valent semiconductor TmSe the
gap at $E_F$ seems to be of a hybridization nature since there exist
indication of its retaining in the paramagnetic state [592]. As discussed in
Sect.6.4, the insulator Kondo state with very small (of order of 3K) gap
occurs in the compound CeNiSn which is non-magnetic (however, recently
static spin correlations with extremely small local moment of order 10$%
^{-3}\mu _B$ were found in this substance at $T<0.13$ K [615]). The system
YbNiSn has a small canted ferromagnetic moment (see Table 6.1), the absence
of gap being probably connected with weakness of d-f hybridization in
ytterbium compounds.

The compound UNiSn turns out to be an antiferromagnet [496,497], in
contradiction with the band structure calculations [315] which yield a
half-metalic ferromagnetic structure. A transition from metallic
antiferromagnetic state to semiconductor paramagnetic one takes place at 47K
with increasing $T$ (in contrast with the usual picture of the
temperature-induced metal-insulator transition [25]), the gap in the
semiconducting state being rather small (of order of 10K). Most simple
explanation of this phenomenon is that the emergence of the sublattice
magnetization results in a shift of the Fermi level outside the energy gap.
The gap may have either Kondo nature (as in CeNiSn) or usual band origin.
The latter situation takes place in the systems TiNiSn, ZrNiSn where the gap
in the $C_{1b}$ lattice has the ``vacansion'' nature [566]. (This situation
is reminiscent of the Heusler aloys RMnSb, see Sect.4.5; recently, a
hypothesis about the formation of the half-metallic ferromagnetic state in
TiCoSn with $T_C=143$\thinspace K was put forward [616].) However, the gap
in these compounds is large (of order of 10$^3$K). Therefore the Kondo
origin of the gap in UNiSn seems to be more probable. Then the possible
explanation of the metal-insulator transition observed is that the
antiferromagnetic exchange interaction suppresses the Kondo order parameter $%
V\sim \langle c^{\dagger }f\rangle $ (see Appendix O) and, consequently, the
gap [508].

\section{Current carriers in a two-dimensional antiferromagnet}

The interest in many-electron models with strong correlations has been
recently greatly revived in connection with the discovery of high-$T_c$
superconductivity in copper-oxide ceramics La$_{2-x}$Sr$_x$CuO$_4$ [617] and
YBa$_2$Cu$_3$O$_{7-y}$ [618]. The current carriers in these systems move in
weakly coupled CuO$_2$-layers and form rather narrow energy bands. A
characteristic feature of these systems are presence of pronounced spin
fluctuations and, for some compositions, the antiferromagnetic ordering. The
important role of correlations is confirmed by that La$_2$CuO$_4$ is a
typical Mott-Hubbard insulator. This compound yields also the best known
example of a quasi-two-dimensional Heisenberg antiferromagnet with small
magnetic anisotropy.

In the present Section we do not discuss the problem of high-temperature
superconductivity itself, but demonstrate the application of simple ME
models to the description of current carrier states in highly correlated
two-dimensional systems.

The electron states in CuO$_2$-planes of copper-oxide perovskites may be
described by the so called Emery model
$$
\mathcal{H} =\sum_{\mathbf{k}\sigma }[\varepsilon p_{\mathbf{k}\sigma
}^{\dagger }p_{\mathbf{k}\sigma }+\Delta d_{\mathbf{k}\sigma }^{\dagger }d_{%
\mathbf{k}\sigma }+V_{\mathbf{k}}(p_{\mathbf{k}\sigma }^{\dagger }d_{\mathbf{%
k}\sigma }+d_{\mathbf{k}\sigma }^{\dagger }p_{\mathbf{k}\sigma })]
$$
$$
+U\sum_id_{i\uparrow }^{\dagger }d_{i\uparrow }d_{i\downarrow }^{\dagger
}d_{i\downarrow }
\eqno{(6.108)}
$$
where $\varepsilon $ and $\Delta $ are positions of p- and d-levels for Cu
and O ions respectively. The $\mathbf{k}$-dependence of matrix elements of
p-d hybridization for the square lattice is given by
$$
V_{\mathbf{k}}=2V_{pd}(\sin ^2k_x+\sin ^2k_y)^{1/2}
\eqno{(6.109)}
$$
At $|V_{pd}|\ll \varepsilon -\Delta $ the Hamiltonian (6.108) is reduced by
a canonical transformation [619] to the Hubbard model with strong Coulomb
repulsion and the effective Cu-Cu transfer integrals
$$
t_{eff}=\frac{V_{pd}^2}{\varepsilon -\Delta }
\eqno{(6.110)}
$$
At present, a large number of models for high-$T_c$ superconductors are
developed which take into account formation of several hybridized narrow and
wide bands with orbital degeneracy. Here we confine ourselves to a simple
consideration of the current carriers within the s-d exchange model [620].

An important property of two-dimensional (2D) Heisenberg magnets is the
absence of long-range order at finite temperatures since it is destroyed by
long-wavelength fluctuations. At the same time, the strong short-range order
with large correlation length $\xi $ persists up to the temperatures of
order of the intersite exchange parameter $J$. Unlike purely 2D Heisenberg
magnets, quasi-2D compounds possess finite values of magnetic ordering
temperature because of a weak interlayer coupling $J^{\prime }$ and/or
easy-axis magnetic anisotropy which are estimated as
$$
T_M=4\pi |J|S^2\left/ \ln \frac{|J|}{\max \{|J^{\prime }|,|J^z-J^{\perp }|\}}%
\right.
\eqno{(6.111)}
$$
Thus the ordering temperature is small, which is reminiscent to weak
itinerant magnets. In such a situation, a consistent perturbation theory
can be developed. Formulas, which are more exact than (6.111) and yield
a quantitative agreement with experimental data, were obtained in [728].

Experimental data on layered perovskites (including La-Cu-O systems)
demonstrate a pronounced short-range order above the Neel point. A similar
situation takes place in frustrated three-dimensional (3D) antiferromagnets
where long-range magnetic ordering is also partially suppressed (see
Sect.6.8).

Recent developments in the theory of two-dimensional Heisenberg
antiferromagnets have provided a simple and successful description of their
thermodynamic properties. Unlike the usual mean-field approximation, the
self-consistent spin-wave theories (SSWT), based on the non-linear
representations of Schwinger bosons [622,623] or the Dyson-Maleev ideal
magnons [624], yield a smooth transition from the ordered state at $T = 0$
to the finite-temperature state with a strong short-range order, the
correlation length $\xi$ being exponentially large at low $T$. The
short-range order parameter is described as an anomalous average of Bose
operators, and the long-range ordering as the Bose condensation.

It is clear from the physical point of view that the electron spectrum in 2D
systems at low $T$ does not change its form in comparison with the ordered
state and should be determined by short-range order (the situation is
reminiscent of spontaneous spin splitting above the Curie point in strong
itinerant ferromagnets). To obtain a quantitative description we calculate
the electron Green's function. First we treat the broad band case with the
use of perturbation expansion in the s-d exchange parameter $I$.
Substituting the result for the spin spectral density $K_{\mathbf{q}}(\omega
)$ in SSWT (see (P.18), (P.22)) in the expression for electron self-energy
(6.25) we obtain
$$
\Sigma _{\mathbf{k}}(E) =\frac{I^2\overline{S}_{ef}^2}{E-t_{\mathbf{k+Q}}}%
+I^2\overline{S}_{ef}\sum_{q,|\mathbf{q-Q}|>\xi ^{-1}}\left( \frac{1-\phi _{%
\mathbf{q}}}{1+\phi _{\mathbf{q}}}\right) ^{1/2}
$$
$$
\times \left( \frac{1-n_{\mathbf{k+q}}+N_B(\omega _{\mathbf{q}})}{E-t_{%
\mathbf{k+q}}-\omega _{\mathbf{q}}}+\frac{n_{\mathbf{k+q}}+N_B(\omega _{%
\mathbf{q}})}{E-t_{\mathbf{k+q}}+\omega _{\mathbf{q}}}\right)
\eqno{(6.112)}
$$
with
\[
\phi _{\mathbf{q}}=\frac 12(\cos q_x+\cos q_y)
\]
which has the same form as the self-energy of the usual 3D antiferromagnet
at $T\ll T_N$ (G.69). The first term in (6.112) describes the formation of
antiferromagnetic gap (the effective ``sublattice magnetization'' $\overline{%
S}_{ef}(T)$ is determined by the singular contribution to the spin
correlation function and has the linear $T$-dependence (P.21)), and the
second terms corresponds to the interaction with magnons.

Consider the peculiarities of the electron spectrum near the band bottom in
the case of a single current carrier. Summing up higher orders corrections
of perturbation series (i.e. replacing energy denominators by the exact
Green's functions) [520] we derive the self-consistent equation
$$
\Sigma _{\mathbf{k}}(E)=\Phi _{\mathbf{k}}(E)-I^2\overline{S}_{ef}^2/\Phi _{%
\mathbf{k+Q}}(E)
\eqno{(6.113)}
$$
where
$$
\Phi _{\mathbf{k}}(E)=I^2\sum_{|\mathbf{q-Q}|>\xi ^{-1}}\int d\omega K_{%
\mathbf{q}}(\omega )G_{\mathbf{k+q}}(E+\omega )
\eqno{(6.114)}
$$
For $T=0$ we obtain
$$
\Phi _{\mathbf{k}}(E)=I^2\overline{S}_{ef}\sum_{\mathbf{q}}\left( \frac{%
1-\phi _{\mathbf{q}}}{1+\phi _{\mathbf{q}}}\right) ^{1/2}G_{\mathbf{k+q}%
}(E-\omega _{\mathbf{q}})
\eqno{(6.115)}
$$
To solve the equation (6.113) one can use the ``dominant pole''
approximation [625]
$$
G_{\mathbf{k}}(E)=\frac{Z_{\mathbf{k}}}{E-\widetilde{E}_{\mathbf{k}}}%
+G_{inc}(\mathbf{k,}E)
\eqno{(6.116)}
$$
where $G_{inc}$ is the incoherent contribution to the Green's function,
$$
Z_{\mathbf{k}}=\left( 1-\frac \partial {\partial E}\Re \Sigma _{\mathbf{%
k}}(E)\right) _{E=\widetilde{E}_{\mathbf{k}}}^{-1}
\eqno{(6.117)}
$$
is the residue at the pole near the band bottom, corresponding to the
spectrum of new quasiparticles,
$$
\widetilde{E}_{\mathbf{k}}\simeq t_{\min }+Z_{\mathbf{k}}(t_{\mathbf{k}%
}-t_{\min })\simeq t_{\min }+Z|t|k^2
\eqno{(6.118)}
$$
Substituting (6.116) into (6.115) and performing integration over $q$ we
obtain the estimation in the 2D case
$$
Z^{-1}-1\sim I^2/|Jt|
\eqno{(6.119)}
$$
Thus, as $|I|$ increases, the spectral weight passes into the incoherent
contribution and undamped qusiparticles become heavy, so that at $I^2\gg
J|t| $ we have the ``heavy-fermion'' situation with $m^{*}/m=Z^{-1}\gg 1$.
In the 3D case the divergence is weaker, and corrections to effective mass
contain a logarithmic factor:
$$
Z^{-1}-1\sim \frac{I^2S}{t^2}\ln \left| \frac t{JS}\right|
\eqno{(6.120)}
$$
The terms with the Bose functions in (6.112) yield corrections to (6.119)
which are proportinal to $T/\xi J$, i.e. exponentially small. Therefore the
picture of the electron spectrum holds at finite $T\ll J$.

It should be noted that similar results are obtained in the case of the
interaction with acoustic phonons if we replace
\[
I^2\overline{S}_{ef}\left( \frac{1-\phi _{\mathbf{q}}}{1+\phi _{\mathbf{q}}}%
\right) ^{1/2}\rightarrow \Lambda ^2q
\]
with $\Lambda $ being the electron-phonon interaction constant. Then the
estimation for the residue renormalization reads
$$
Z^{-1}-1\sim \Lambda ^2/\overline{\omega }|t|\sim 1
\eqno{(6.121)}
$$
It is instructive to perform a comparison with the case of an usual
paramagnet without strong antiferromagnetic correlations, so that the
electron-spin interaction matrix element is constant at $q\rightarrow 0$
rather than proportional to $q^{1/2}$. In such a situation we may use the
spin diffusion approximation (6.32). Then quasiparticles turn out to possess
a strong damping
$$
\Gamma _{\mathbf{k}}=-I^2\sum_{\mathbf{q}}\Im \int d\omega K_{\mathbf{q}%
}(\omega )\frac{Z_{\mathbf{k+q}}}{\widetilde{E}_{\mathbf{k}}-\widetilde{E}_{%
\mathbf{k+q}}+\omega }
\eqno{(6.122)}
$$
At small $k$ we have
$$
\Gamma _{\mathbf{k}}\sim \frac{I^2JS^3}{Zt^2}\times \left\{
\begin{array}{ll}
|\ln k| & ,\qquad D=2 \\
1 & ,\qquad D=3
\end{array}
\right.
\eqno{(6.123)}
$$
Using the dominant pole approximation (6.116) with $\widetilde{E}_{\mathbf{k}%
}\rightarrow \widetilde{E}_{\mathbf{k}}-i\Gamma _{\mathbf{k}}$ we obtain
$$
Z^{-1}-1\sim \left\{
\begin{array}{ll}
|t/JS|^{1/2} & ,\qquad D=2 \\
|I^2S/tJ|^{1/2} & ,\qquad D=3
\end{array}
\right.
\eqno{(6.124)}
$$
Although the residue of the damped quasiparticles may still be small, it is
difficult to separate them from the background of the incoherent
contribution.

Now we treat the s-d model with strong correlations $|I|\rightarrow \infty $,
which includes as a particular case the Hubbard model with $U\rightarrow
\infty $ (Appendix I). Consider the Green's function
\[
G_{\mathbf{k}\alpha \sigma }(E)=\langle \!\langle g_{\mathbf{k}\alpha \sigma }|
g_{\mathbf{k}\alpha \sigma }^{\dagger }\rangle \!\rangle _E,\qquad
\alpha =\mathrm{sign}I
\]
where the operators $g$ are defined in (I.4). The result of calculation with
account of spin fluctuations has the form [520]
$$
G_{\mathbf{k}\alpha }(E)=\left[ E\left( \Phi _{\mathbf{k}\alpha }(E)-\frac{%
\overline{S}_{ef}^2t_{\mathbf{k+Q}}/(2S+1)^2}{E-\Psi _{\mathbf{k+Q}}(E)t_{%
\mathbf{k+Q}}}\right) ^{-1}-t_{\mathbf{k}}\right] ^{-1}
\eqno{(6.125)}
$$
with
$$
\Psi _{\mathbf{k}\alpha }(E)=P_\alpha +\sum_{|\mathbf{q-Q}|>\xi ^{-1}}\frac{%
t_{\mathbf{k+q}}}{(2S+1)^2}\int d\omega K_{\mathbf{q}}(\omega )\Psi _{%
\mathbf{k+q,}\alpha }^{-1}(E)G_{\mathbf{k+q,}\alpha }(E+\omega )
\eqno{(6.126)}
$$
\[
P_{+}=\frac{S+1}{2S+1},\qquad
P_{-}=\frac S{2S+1}
\]
When neglecting spin fluctuations, $\Psi _\alpha =P_\alpha $ and the
electron spectrum contains two quasiparticle subbands, as well as in the
three-dimensional antiferromagnet (cf. (I.15)):
$$
E_{\mathbf{k}\alpha }^{1,2}=\frac{P_\alpha }2(t_{\mathbf{k}}+t_{\mathbf{k+Q}%
})\pm \left( \frac{P_\alpha ^2}4(t_{\mathbf{k}}-t_{\mathbf{k+Q}})^2+\frac{%
\overline{S}_{ef}^2}{(2S+1)^2}t_{\mathbf{k}}t_{\mathbf{k+Q}}\right) ^{1/2}
\eqno{(6.127)}
$$
In the nearest-neighbour approximation ($t_{\mathbf{k+Q}}=-t_{\mathbf{k}}$)
we have
$$
E_{\mathbf{k}}^{1,2}=\pm \frac{t_{\mathbf{k}}}{2S+1}\left\{
\begin{array}{ll}
\lbrack S^2-\overline{S}_{ef}^2(T)]^{1/2} & ,\qquad \alpha =- \\
\lbrack (S+1)^2-\overline{S}_{ef}^2(T)]^{1/2} & ,\qquad \alpha =+
\end{array}
\right.
\eqno{(6.128)}
$$
The second term in (6.126) (spin-fluctuation corrections) leads to
qualitative changes in the spectrum near the band bottom. To solve the
system (6.125), (6.126) at $T=0$ we employ again the dominant pole
approximation
$$
G_{\mathbf{k}\alpha }(E)=\Psi _{\mathbf{k}\alpha }\left[ \frac{Z_{\mathbf{k}}%
}{E-\widetilde{E}_{\mathbf{k}}}+G_{inc}(\mathbf{k,}E)\right]
\eqno{(6.129)}
$$
The estimate for the residue is analogous to (6.119), $I^2$ being replaced
by $(t/2S)^2$ which is typical for the strong-coupling limit. Then we obtain
$$
Z^{-1}-1\sim \left\{
\begin{array}{ll}
|t/JS|^{1/2} & ,\qquad D=2 \\
S^{-1}\ln |t/JS|^{1/2} & ,\qquad D=3
\end{array}
\right.
\eqno{(6.130)}
$$
Thus a narrow quasiparticle band with the width of order of $|J|$ is formed
near the bare band bottom in the 2D case. This result was firstly obtained
analytically by Kane et al [625] and confirmed by numerical calculations in
the $t-J$ model [626].

We see that strong electron-spin interaction in two-dimensional systems may
result in heavy electron mass formation near the band bottom even in the
case of one current carrier. Simultaneous consideration of this effect and
many-electron Kondo divergences is an interesting physical problem. In this
connection we mention the system Y$_{1-x}$Pr$_x$Ba$_2$Cu$_3$O$_7$ where
increasing $x$ results in suppression of superconductivity and heavy-fermion
behavior [627] and the high-$T_c$ superconductor Nd$_{2-x}$Ce$_x$CuO$_4$
which posseses a logarithmic term in the temperature dependence of
resistivity [628]. Recently [629], heavy-fermion behaviour with very large
$\gamma $ was found in the latter system at $T<0.3K$.

Investigation of electron spectrum in a 2D antiferromagnetic metal was
performed in Ref.[433]. It was demonstrated that a behavior, which is similar
to the marginal Fermi liquid [645] (the linear in $|E-E_F|$ dependence of the
electron damping) takes place in some interval near the Fermi energy due to
intersubband transitions. This results in anomalous temperature dependences
of thermodynamic and transport properties at not too low temperatures.
A similar picture takes place in the 3D case for the ``nesting'' situation
where $t_{\mathbf{k+Q}}-E_F=E_F-t_{%
\mathbf{k}}$

Interaction with lattice degrees of freedom (strong polaron effects) should
also play an important role in the formation of electron spectrum of high-$%
T_c$ superconductors. The investigation of the interaction among electrons
and ions in a double-well potential was performed by Yu and Anderson [530]
with application to A15-superconductors with moderate $T_c$. A pseudo-Kondo
lattice model, which treats the interaction of current carriers with
strongly anharmonic displacements of oxygen atoms (which may be described as
two-level systems), is developed in papers [405,442].

\section{Spin-liquid state in systems with spin and charge degrees of
freedom}

The consideration of the ground state of Kondo lattices (Sect.6.6, Appendix
O) uses essentially the idea of physical reality of auxiliary pseudofermions
that arise when localized-spin operators are ``dismantled". The
pseudofermion representation was applied by Coleman and Andrei [711] to
describe the spin-liquid state in the two-dimensional periodic s-f model.
Similar concepts were extensively used in Anderson's theory of resonating
valence bonds (RVB) for copper-oxide high-$T_c$ superconductors [631]. This
state is characterized by absence of long-range magnetic ordering and
formation of singlet bonds between neighbour spins on the square lattice. As
well as in quantum chemistry (e.g., for the benzol molecule), the bonds can
move (resonate) in the crystal, so that the ground state is a superposition
of the corresponding wavefunctions.

Anderson put forward the idea of separating spin and and charge degrees of
freedom by using the representation of slave Bose and Fermi operators
$$
c_{i\sigma }^{\dagger }=X_i(\sigma ,0)+\sigma X_i(2,-\sigma )=s_{i\sigma
}^{\dagger }e_i+\sigma d_i^{\dagger }s_{i-\sigma }
\eqno{(6.131)}
$$
where $s_{i\sigma }^{\dagger }$ are creation operators for neutral fermions
(spinons) and $e_i^{\dagger }$, $d_i^{\dagger }$ for charged spinless bosons
(holons). The physical sense of such excitations may be explained as follows
[632]. Consider the lattice with one electron per site with strong Hubbard
repulsion, so that each site is neutral. In the ground RVB state each site
takes part in one bond. When a bond becomes broken, two uncoupled sites
occur which possess spins of 1/2. The corresponding excitations (spinons)
are uncharged. On the other hand, the empty site (hole) in the system
carries the charge, but not spin.

The requirement of the Fermi commutation relations for electron operators
yields
$$
e_i^{\dagger }e_i+d_i^{\dagger }d_i+\sum_is_{i\sigma }^{\dagger }s_{i\sigma
}=1
\eqno{(6.132)}
$$
In the half-filled case only spinon excitations with the kinetic energy of
order of $|J|$ are present. At doping the system by holes, there occur the
current carriers which are described by holon operators $e_i^{\dagger }$. In
the simplest gapless version, the Hamiltonian of the system for a square
lattice may be presented as
$$
\mathcal{H}=\sum_{\mathbf{k}}(4t\phi _{\mathbf{k}}-\zeta )e_{\mathbf{k}%
}^{\dagger }e_{\mathbf{k}}+4\sum_{\mathbf{k}}(\Delta +t\delta )\phi _{%
\mathbf{k}}(s_{\mathbf{k}\sigma }^{\dagger }s_{-\mathbf{k-}\sigma }^{\dagger
}+s_{\mathbf{k}\sigma }s_{-\mathbf{k-}\sigma })+...
\eqno{(6.133)}
$$
with $\Delta $ being the RVB order parameter, which is determined by
anomalous averages of the spinon operators, $\delta =\langle e^{\dagger
}e\rangle $ the hole concentration. Thus a spin-liquid state with long-range
magnetic order suppressed, a small energy scale $J$, and a large linear term
in specific heat ($\gamma \sim 1/|J|$), which is owing to existence of the
spinon Fermi surface, can arise in a purely spin systems without conduction
electrons. Experimentally, some data indicate presence of a $T$-linear term
in the insulating phase of copper-oxide systems.

Later, more complicated versions of the RVB theory, which use topological
consideration and analogies with the fractional quantum Hall effect, were
developed (see, e.g., [633]). These ideas led to rather unusual and
beautiful results. For example, it was shown that spinons may obey
fractional statistics, i.e. the wavefunction of the system acquires a
complex factor at permutation of two quasiparticles.

Here we demonstrate the suppression of long-range magnetic order at $T = 0$
and occurence of the spin-liquid state within a simple spin-wave treatment
of a two-dimensional Heisenberg antiferromagnet [601,609]. To this end we
write down the correction to sublattice magnetization due to zero-point
vibrations (see (E.14))
$$
\delta \overline{S}=-\frac 12\sum_{\mathbf{q}}[S(4J_{\mathbf{Q}}-J_{\mathbf{%
Q+q}}-J_{\mathbf{Q-q}}-2J_{\mathbf{q}})/\omega _{\mathbf{q}}-1]
\eqno{(6.134)}
$$
At $q\rightarrow 0$ we have
$$
\omega _{\mathbf{q}}^2\simeq S^2(J_{\mathbf{Q}}-J_0)[\alpha q^2+\beta f(\phi
)q^4]
\eqno{(6.135)}
$$
where $\beta >0$ and $f(\phi )$ is a positive polar-angle function. For $%
\alpha \rightarrow 0$ (frustration situation) we find
$$
\overline{S} =S-a\ln \frac \beta \alpha ,
\eqno{(6.136)}
$$
$$
a =\frac 1{16\pi ^2}(J_{\mathbf{Q}}-J_0)^{1/2}\int_0^{2\pi }\frac{d\phi }{%
[\beta f(\phi )]^{1/2}}
$$
so that $\overline{S}=0$ in some regions of parameters
$$
\alpha <\beta \exp (-S/a)
\eqno{(6.137)}
$$
Thus, unlike the Kondo lattices (cf.(6.89), (6.103)), destruction of
magnetic ordering is due to frustrations of the f-f exchange interaction
itself rather than to Kondo screening of magnetic moments. One may assume
that similar frustration effects (e.g., large next- neighbour exchange
interactions or presence equilateral nearest- neighbour triangles) may
result in total or partial destruction of magnetic moment in some
three-dimensional systems. Such Heisenberg systems, which possess developed
spin fluctuations, reduced magnetic moments and strong short-range order
above the Neel point [634] are reminiscent in a sense of itinerant magnets.

There exist a number of experimental data which indicate realization of a
spin-liquid-type state in three-dimensional d- and f-systems [635,609].
Consider some examples.

The compound YMn$_2$ has a frustrated AFM structure and pronounced
short-range order above $T_N$ [636]. In the system Y$_{1-x}$Sc$_x$Mn$_2$
with $x=0.03$ the long-range order is destroyed, while linear term in
specific heat reaches very large value, $\gamma =$ 140 mJ/mol K$^2$ (a
similar situation occurs under pressure) [637]. This value exceeds by ten
times the result of band calculations and is a record for d-systems.

The Mott insulator NiS$_{2-y}$ [638] exhibits magnetic ordering, but one may
talk of a suppression of the magnetism in the sense that, similar to
magnetic Kondo lattices, $T_N=45K$ is small in comparison with $|\theta
|\simeq 1500$K. The slope of the phase diagram line indicates a high entropy
value for the insulator phase [44], which is characteristic of a
spin-liquid-type state. In metallic NiS$_{2-x}$Se$_x$ near the transition ($%
x\simeq 0.5$), the value of $\gamma =30$ mJ/mol$\cdot $K$^2$ is rather large.

The most striking of the spin-liquid state realization is probably the
intermediate valence semiconductor Sm$_3$Se$_4$ [639], a system with not
only spin, but also charge degrees of freedom, where the ions Sm$^{2+}$ and
Sm$^{3+}$ are distributed over crystallographically equivalent sites in the
Th$_3$P$_4$-lattice. This situation may be described by an effective
(generally speaking, anisotropic) pseudospin Hamiltonian for the charge
degrees of freedom [635]. By contrast with isostructural compounds like Eu$%
_3 $S$_4$, there is no indication of charge ordering up to $T=0$ in Sm$_3$Se$%
_4$, and $\gamma $ has at $T<1$K the giant value 4.5 J/mol$\cdot $K$^2$.

In the system Yb$_4$As$_{3-x}$P$_4$ with the anti-Th$_3$P$_4$ structure,
where charge ordering occurs near 300K, $\gamma $ increases from 200 to 400
mJ/mol$\cdot $K$^2$ as $x$ is varied from 0 to 0.3, the current carrier
concentration $n$ (of order of 10$^{-3}$ per atom) remaining practically
unchanged [640]. Thus we deal with a new class of heavy-fermion systems with
extremely small concentration of current carriers. Here belong also the
compound YbAs with $n\sim 10^{-2}$, $\gamma =$ 270 mJ/mol$\cdot $K$^2$
[538], and possibly YbNiSb [541], YbBiPt [571,641]. In the latter system, $%
\gamma $ reaches the value of 8J/mol$\cdot $K$^2$.

Although in most cases there are no serious grounds in doubting that the
Kondo effect is the major cause of anomalous behaviour of f-systems, which
were discussed above, the role of f-f interaction in the formation of
low-energy spectrum should be also investigated. Apparently, an intermediate
state between the usual magnetic state and the spin liquid arises in some
compounds. In Kondo lattices with a small number of carriers, the exchange
interaction frustrations can be even more important than the Kondo effect.
Thus, the semimetal CeSb (a classical example of a system with competing
exchange interactions) displays a complicated magnetic phase diagram --- the
``devil's staircase'' [642]. According to [643], the system Ce$_{0.8}$La$%
_{0.2}$Sb possesses large electronic specific heat. The ``frustrated''
picture of magnetism, that is reminiscent of spin-glass state (absense of a
marked phase transition) is discussed for CeAl$_3$ [508]. The problem of
role of frustrations and of the Kondo effect in the formation of complicated
magnetic structures and reduced magnetic moments is also of interest for
rare-earth metals (Sect.4.7).

To conclude this Chapter, we discuss briefly some modern concepts in the
theory of systems with strong correlations. The idea of unusual excitation
spectrum in correlated metallic systems was extensively developed by
Anderson [644]. He put forward the concept about the inadequacy of the
Fermi-liquid picture and formation of the Luttinger liquid state. The latter
is characterized by the absence of simple poles for the one-electron Green's
function, i.e. of usual single-particle Fermi excitations). The transition
into the Luttinger liquid is connected with the phase shift at the Fermi
level owing to strong electron scattering.

A phenomenological ``marginal Fermi-liquid'' theory of high-$T_c$
superconductors, which yields close results, was proposed by Varma et al
[645]. The non-Fermi liquid behaviour (unusual power-law temperature
dependence of electronic specific heat, resistivity etc.) is found now for a
number of f-electron systems [646]. Besides that, the power-law divergence
of magnetic susceptibility, $\chi (T)\sim T^{-\alpha }$, is observed in
low-dimensional organic conductors; the analogy with the behaviour of
yttrium-based ceramics was discussed in [647].

In the Luttinger model [648], which was developed for the one-dimensional
electron systems, the bare spectrum contains two linear branches
$$
E_{\mathbf{k}}^{1,2}=\pm v_Fk \qquad (k>0)
\eqno{(6.138)}
$$
which is reminiscent of relativistic models in quantum electrodynamics (the
vacuum of the states with $k<0$ is excluded). The Luttinger Hamiltonian may
be ``bosonized'' by introducing the collective excitation operators ($n=1,2;$
$q>0$)
$$
\rho _n(q)=\sum_kc_{k+q,n}^{\dagger }c_{kn},\qquad
rho _n(-q)=\rho _n^{\dagger }(q)
\eqno{(6.139)}
$$
so that
$$
b_q^{\dagger }=\left( \frac{2\pi }{N|q|}\right) ^{1/2}\times \left\{
\begin{array}{ll}
\rho _1(q) & ,\qquad q>0 \\
\rho _2(q) & ,\qquad q<0
\end{array}
\right.
\eqno{(6.140)}
$$
(the summation in (6.139) goes over both positive and negative $k$, $N$ is
the number of lattice sites). The operators (6.140) turn out to satisfy the
Bose commutation relations, which is connected with account of the
``vacuum'' states in (6.139) [649]. Charge and spin operators may be
introduced by
$$
a_k^{\dagger }=\frac 1{\sqrt{2}}(b_{k\uparrow }^{\dagger }
+b_{k\downarrow }^{\dagger }),\qquad
s_k^{\dagger }=\frac 1{\sqrt{2}}(b_{k\uparrow }^{\dagger }
-b_{k\downarrow }^{\dagger })
\eqno{(6.141)}
$$
In the representation (6.140), both kinetic energy (6.138) and interaction
Hamiltonians contain quadratic terms only. Then the one-electron Green's
function is calculated exactly and turns out to have the singularities of
the form
$$
G_{\mathbf{k}}^{1,2}(E)\sim (E\mp v_Fk)^{\alpha -1}
\eqno{(6.142)}
$$
where the parameter $\alpha >0$ is determined by the electron interaction.
Then the electron distribution function at $T=0$ has, instead of a jump, a
power-law behaviour at the ``Fermi surface'',
$$
\langle c_{kn}^{\dagger }c_{kn}\rangle -\frac 12\sim |k\mp k_F|^\alpha
\mathrm{sign}(k\mp k_F)
\eqno{(6.143)}
$$
Such a behaviour may be derived rigorously for the one-dimensional Hubbard
model which permits rigorous consideration (in this case, in contradiction
with the Landau theory, an arbitarily small interaction leads to a
reconstruction of the ground state, e.g. to the metal-insulator transition
for a half-filled band). However, generalization of these results to
two-dimensional systems of interacting electrons is a very difficult
problem. As discussed by Anderson [644], the violation of the Fermi-liquid
picture may be described in terms of the Hubbard splitting: the states in
the upper Hubbard subband correspond to the anomalous forward scattering.
Modern treatment of the Fermi-liquid state and of its instabilities with the
use of the ``Bose'' representation is given, e.g., in [650].

\chapter*{CONCLUSIONS}
\addcontentsline{toc}{chapter}{CONCLUSIONS}

In the present book we have attempted to consider the whole variety of
physical properties of transition metals. The characteristics of TM are much
more complicated and interesting than those of simple metals. Simultaneous
consideration of all the properties permits to establish some regularities
with filling of d- and f-shells and their connection with the electronic
structure. One might expect similarity of the properties within a given
transition series, since, with some exceptions, the total number of external
sp-electrons does not change.

In fact, chemical and most physical properties (crystal structure etc.) of
rare earths are very close. At the same time, magnetic characteristics of
4f-metals change appreciably with the number of f-electrons. Moreover, as
demonstrated in Sect.4.8.2, there exists some periodicity within the
4f-series, so that the latter forms a miniature ``periodic table". This
regularity is connected with the many-electron term structure of the
4f-shell.

On the other hand, non only magnetic, but also other properties of d-metals
depend strongly on the number of d-electrons $n_d$, and they play an
important role in the formation of a crystal. An appreciable delocalization
of d-states is directly observed in the experimental investigations of the
Fermi surface.

In some anomalous rare earth and actinide systems, f-electrons also take
part in formation of the Fermi surface, due to both one-electron mechanisms
(hybridization) and many-electron effects (as in Kondo lattices). This may
lead to occurence of states with rather exotic properties, e.g. to greatly
enhanced electronic effective masses (Chapter 6). In metallic cerium, direct
f-f overlap plays an important role.

The ``dual" (localized vs. itinerant) nature of d-electrons requires using
various approaches for analyzing physical properties of TM and their
compounds. One can pick out two main approaches. The first one starts from
first-principle one-electron band calculations. According to the
Hohenberg-Kohn theorem, these calculations can provide an accurate
description of some ground state characteristics. At the same time, standard
band calculations, which use the local density aproximation, are often
insufficient for narrow d-, 5f- and especially 4f-bands. Besides that, the
density functional approach is, generally speaking, unable to describe the
whole excitation spectrum and thermodynamic properties.

The band approach can explain anomalies in TM properties, which are
connected with singularities of electron density of states. Presence of such
singularities leads sometimes to considerable modifications of standard
formulas of solid state theory, e.g., for electronic specific heat and
paramagnetic susceptibility. It should be also noted that for ``flat"
regions of spectrum with small values of grad$E(\mathbf{k})$, i.e. for
density of states peaks, many-electron perturbation corrections become large
and the role of correlation effects increases.

The second approach takes into account electron correlations in a
microscopic way starting from the atomic picture. The adequacy of this
picture is evident for strongly localized 4f-states. However, atomic
features are retained to some extent for d-states too. In particular, this
is confirmed by a characteristic ``toothed" $n_d$-dependence of electronic
properties in transition metals and by many-electron term effects in
spectral measurements.

In this connection, the problem of strong (ten-fold) degeneracy of d-states
seems to be important. Altough the degeneracy is lifted at broadening of the
atomic levels into energy bands by the periodic lattice potential, it
surives in some points of the Brillouin zone. This degeneracy is important,
e.g., for the orbital momenta which determine anisotropy of a number of
properties. It should be stressed that such a scheme of degeneracy lift
takes place in the one-electron picture only, and the classification of
electron states changes in the many-electron representation where additional
quantum numbers occur. The corresponding new quasiparticles may possess
different degree of localization and mobility. This may change essentially
the results of the standard band theory.

From the point of view of the qualitative microscopic description,
investigation of simple theoretical models, which include effects of strong
intraatomic electron correlations, turns out to be very useful. Such effects
turn out to be especially bright for some d- and f-compounds. In the case of
narrow bands (large Coulomb interaction) the correlations result in a
radical rearrangement of electron spectrum --- formation of the Hubbard
subbands. A convenient tool for describing atomic statististics of
excitations in such a situation is the formalism of many-electron Hubbard
operators. On the other hand, even small interaction among localized and
itinerant electrons may result in a reconstruction of electron spectrum at
low temperatures owing to peculiarities of resonance scattering in
many-particle systems (the Kondo effect).

Methods which combine band structure calculations and model considerations
seem to be promising. As discussed in the book, such approaches were
developed, e.g, for transition metal oxides and heavy-fermion systems.

In a number of cases, correlation effects are reduced to a modification of
electron spectrum and density of states (e.g., formation of hybridization
gap or Abrikosov-Suhl resonance in intermediate valence and Kondo systems),
so that electron properties may be further calculated in a phenomenological
way with the use of one-electron theory results. However, the modifications
of electron spectrum parameters themselves may not be obtained in the
standard band theory. In particular, the parameters are often strongly
temperature dependent owing to many-electron renormalizatons.

On the other hand, sometimes the excitation spectrum is not described within
the usual quasiparticle picture and has an essentially incoherent nature.
Simple examples are provided by the electron spectrum of the Hubbard
ferromagnet (Appendix J) and two-dimensional conducting antiferromagnet
(Sect.6.7).

The spectrum of highly-correlated systems is often described in terms of
auxiliary (slave) Fermi and Bose operators, which correspond to
quasiparticles with exotic properties (neutral fermions, charged bosons
etc.). Last time such ideas are extensively applied in connection with the
unusual spectra of high-$T_c$ superconductors and heavy-fermion systems.
Investigation of these problems leads to complicated mathematics, which uses
the whole variety of modern quantum field theory methods, and very beautiful
physics. For example, description of the Fermi liquid state in terms of Bose
excitations becomes possible. These concepts change essentially classical
notions of the solid state theory.

\appendix

\chapter{Many-electron creation operators for atomic configurations and
Hubbard's operators}

In this Appendix we consider an operator description of many- electron
systems with large intrasite Coulomb correlations.

At passing to the standard second quantization representation, the
many-electron (ME) wave function of a crystal $\Psi (x_1...x_N)$ ($x=\{%
\mathbf{r}_is_i\},s_i$ is the spin coordinate) is chosen in the form of a
linear combination of the Slater determinants. These are constructed from
the one-electron wave functions $\psi _\lambda (x)$ ($\lambda =\{{\nu \gamma
\}}$, $\nu $ are the indices of lattice sites and $\gamma $ are the
one-electron quantum number sets):
$$
\Psi \left( x_1...x_N\right) =\sum_{\lambda _1...\lambda _N}c\left( \lambda
_1...\lambda _N\right) \Psi _{\lambda _1...\lambda _N}\left(
x_1...x_N\right)
\eqno{(\rm A.1)}
$$
where
$$
\Psi _{\lambda _1...\lambda _N}\left( x_1...x_N\right) =\left( N!\right)
^{-1/2}\sum_P\left( -1\right) ^PP\prod_i\psi _{\lambda _i}\left( x_i\right)
\eqno{(\rm A.2)}
$$
with $P$ being all the possible permutations of $x_i$. The expansion (A.1)
is exact provided that the system of functions $\psi _\lambda $ is complete
[651]. The second quantization representation is introduced by using the
one-electron occupation numbers $n_\lambda $ as new variables:
$$
\Psi \left( x_1...x_N\right) =\sum_{\left\{ n_\lambda \right\} }c\left(
...n_\lambda ...\right) \Psi _{\{n_\lambda \}}\left( x_1...x_N\right)
\eqno{(\rm A.3)}
$$

Then $c\left( ...n_\lambda ...\right) $ plays the role of a new wave
function. The Fermi one-electron creation and annihilation operators are
defined by
$$
a_\lambda c\left( ...n_\lambda ...\right)  =\left( -1\right) ^{\eta
_\lambda }n_\lambda c\left( ...n_\lambda -1...\right)
\eqno{(\rm A.4)}
$$
$$
a_\lambda ^{+}c\left( ...n_\lambda ...\right)  =\left( -1\right) ^{\eta
_\lambda }\left( 1-n_\lambda \right) c\left( ...n_\lambda +1...\right)
$$
with
\[
\eta _\lambda =\sum_{\lambda ^{\prime }>\lambda }n_{\lambda ^{\prime }},\qquad
a_\lambda ^{+}a_\lambda =\hat n_\lambda
\]

Now we try to generalize this approach by introducing quantum numbers of
some electron groups. In particular, we may combine electrons at a given
lattice site ($\Lambda =\left\{ {\nu \Gamma }\right\} $, $\Gamma _i$ are ME
terms) to obtain
$$
\Psi \left( x_1...x_N\right) =\sum_{\{N_\lambda \}}c\left( ...N_\lambda
...\right) \Psi _{\{N_\lambda \}}\left( x_1...x_N\right)
\eqno{(\rm A.5)}
$$
In the case of equivalent electron configuration $l^n$, the ME wave function
of the electron group is constructed as (see [20])
$$
\Psi _{\Gamma _n}\left( x_1...x_N\right) =\sum_{\Gamma _{n-1},\gamma
}G_{\Gamma _{n-1}}^{\Gamma _n}C_{\Gamma _{n-1},\gamma }^{\Gamma _n}\Psi
_{\Gamma _{n-1}}\left( x_1...x_{n-1}\right) \psi _\gamma \left( x_n\right)
\eqno{(\rm A.6)}
$$
where ${C}$ are the Clebsh-Gordan coefficients. In the case of $LS$-coupling
we use the notation
$$
C_{\Gamma _{n-1},\gamma }^{\Gamma _n}\equiv
C_{L_{n-1}M_{n-1},lm}^{L_nM_n}C_{S_{n-1}\mu _{n-1},\frac 12\sigma }^{S_n\mu
_n}
\eqno{(\rm A.7)}
$$
and the summation over $\gamma =\left\{ lm{\sigma }\right\} $ (we omit for
brevity the principal quantum number) stands for the summation over
one-electron orbital projection m and spin projection $\sigma $, but not
over $l$. The case of $jj$-coupling, where strong spin-orbit coupling should
be taken into account in the first place, is applicable for $5f$ actinide
compounds. Here we have $\gamma =j{\mu }$, $\Gamma =JM$ with $j=l+1/2$, so
that the consideration is formally more simple. The quantities
\[
G_{\Gamma _{n-1}}^{\Gamma _n}\equiv G_{S_{n-1}L_{n-1}\alpha
_{n-1}}^{S_nL_n\alpha _n}
\]
are the fractional parentage coefficients ($\alpha $\ are additional quantum
numbers which distinguish different terms with the same $S$, $L$, e.g., the
Racah's seniority; we will omit $\alpha $\ for brevity where it is
possible). They do not depend on momentum projections, and the quantities $%
\left( G_{\Gamma _{n-1}}^{\Gamma _n}\right) ^2$ have the meaning of fracture
of the term $\Gamma _{n-1}$ at formation of the term $\Gamma _n$ (at $n<2$, $%
G\equiv 1$). The fractional parentage coefficients satisfy the orthogonality
relations [20,32]
$$
\sum_{\left\{ S^{\prime }L^{\prime }\alpha ^{\prime }\right\} }G_{S^{\prime
}L^{\prime }\alpha ^{\prime }}^{SL\alpha }G_{S^{\prime }L^{\prime }\alpha
^{\prime }}^{SL\alpha ^{\prime \prime }}=\delta _{\alpha \alpha ^{\prime
\prime }}
\eqno{(\rm A.8a)}
$$
$$
n\sum_{\left\{ SL\alpha \right\} }\left[ S\right] \left[ L\right]
G_{S^{\prime }L^{\prime }\alpha ^{\prime }}^{SL\alpha }G_{S^{\prime
}L^{\prime }\alpha ^{^{\prime \prime }}}^{SL\alpha }=\left( 2\left[ l\right]
+1-n\right) \left[ S^{\prime }\right] \left[ L^{^{\prime }}\right] \delta
_{\alpha ^{\prime }\alpha ^{\prime \prime }}
\eqno{(\rm A.8b)}
$$
where
\[
\left[ A\right] \equiv 2A+1
\]

The relation (A.8b) is obtained from (A.8a) after passing to the hole
representation in the atomic shell. In the case of $jj$-coupling
$$
\left[ S\right] \left[ L\right] \rightarrow \left[ J\right] ,\qquad
2\left[ l\right] \rightarrow [j]
\eqno{(\rm A.9)}
$$

The recurrence relation (A.6) enables one to obtain the ME functions with an
arbitrary electron number $n$. The fractional parentage and Clebsh-Gordan
coefficients ensure the antisymmetry of the function (A.6) with respect to
permutation of electron coordinates.

If the added electron belongs to another shell, we can write down
$$
\Psi _{\Gamma _n}\left( x_1...x_n\right) =h^{-1/2}\sum_{i,\Gamma
_{n-1},\gamma }\left( -1\right) ^{n-i}C_{\Gamma _{n-1},\gamma }^{\Gamma
_n}\Psi _{\Gamma _{n-1}}\left( x_1...x_{i-1},x_{i-1}...x_{n-1}\right) \psi
_\gamma \left( x_i\right)
\eqno{(\rm A.10)}
$$
(unlike the case of equivalent electrons, antisymmetrization is required).
Note, however, that introducing ME functions and operators, which describe
several electron shells, is advisable in the solid state theory only
provided that the interaction between the shells is large in comparison with
the band energies (e.g., in the narrow-band s-d exchange model, see Appendix
I).

The wave function of the the whole crystal (A.5) may be now obtained as an
antisymmetrized product of ME functions for electron groups.

By analogy with (A.6), (A.10) we can introduce ME creation operators for
electron groups [652]. For equivalent electrons
$$
A_{\Gamma _n}^{+}=h^{-1/2}\sum_{\Gamma _{n-1},\gamma }G_{\Gamma
_{n-1}}^{\Gamma _n}C_{\Gamma _{n-1},\gamma }^{\Gamma _n}a_\gamma
^{+}A_{\Gamma _{n-1}}^{+}
\eqno{(\rm A.11)}
$$
At adding an electron from another shell
$$
A_{\Gamma _n}^{+}=\sum_{\Gamma _{n-1},\gamma }C_{\Gamma _{n-1},\gamma
}^{\Gamma _n}a_\gamma ^{+}A_{\Gamma _{n-1}}^{+}
\eqno{(\rm A.12)}
$$
Antisymmetry of the functions
$$
|\Gamma _n\rangle =A_{\Gamma _n}^{+}|0\rangle
\eqno{(\rm A.13)}
$$
is provided by anticommutation of the Fermi operators. Extra factors of $%
(1/n)^{1/2}$ in (A.11), (A.12) in comparison with (A.6), (A.10) respectively
arise at passing from the ${x}$-representation to second quantization one.

In the particular case of two equivalent electrons
$$
A_\Gamma ^{+}=\frac 1{\sqrt{2}}\mathop{\sum_{m_1m_2}}_{\sigma _1\sigma _2}
C_{lm_1,lm_2}^{LM}C_{\frac 12\sigma _1,\frac 12\sigma _2}^{S\mu
}a_{lm_2\sigma _2}^{+}a_{lm_1\sigma _1}^{+}
\eqno{(\rm A.14)}
$$
One can see from (A.14) that only the terms with even $S+L$ are possible
[20]: as follows from the properties of the Clebsh-Gordan coefficients, ${A}%
_\Gamma $ is identically zero if ${S}+L$ is odd.

Unlike the Fermi operators, the commutation relation for the operators $A$
are complicated. For example, we obtain from (A.14)
\[
\lbrack A_\Gamma ,A_\Gamma ^{+}]=\delta _{\Gamma \Gamma ^{\prime
}}+2\sum_{\gamma _1\gamma _2\gamma _3}C_{\gamma _1\gamma _3}^\Gamma
C_{\gamma _2\gamma _3}^{\Gamma ^{\prime }}a_{\gamma _2}^{+}a_{\gamma _1}
\]
Representing the operator products in terms of one-electron Fermi operators,
performing the pairings of the latter and using the orthogonality relations
for the Clebsh-Gordan and fractional parentage coefficients, we obtain
$$
\langle 0|A_{\Gamma ^{\prime }}A_\Gamma ^{+}|0\rangle  =\delta _{\Gamma
\Gamma ^{\prime }}
\eqno{(\rm A.15)}
$$
$$
A_{\Gamma _n^{\prime }}A_{\Gamma _n}^{+}|0\rangle  =\delta _{\Gamma
_n\Gamma _n^{\prime }}|0\rangle   
$$
However, for $m<n$%
\[
A_{\Gamma _m^{\prime }}A_{\Gamma _n}^{+}|0\rangle \neq 0
\]

Therefore the operators (A.11), (A.12) are convenient only for the treatment
of configurations with a fixed number of electrons (e.g., in the
Heitler-London model). At the same time, they are insufficient at
considering problems with electron transfer between different shells or
sites because of the ``non-orthogonality'' for different $n$. In such
situations, it is suitable to define new ME creation operators, which
contain projection factors [653, 654]
$$
\tilde A_\Gamma ^{+}=A_\Gamma ^{+}\prod_\gamma \left( 1-\hat n_\gamma
\right) A_\Gamma   
\eqno{(\rm A.16)}
$$
Formally, the product in (A.16) goes over all the relevant one-electron
states $\gamma $. Howewer, because of the identity $a_\gamma ^{+}\hat
n_\gamma =0$, it is sufficient to retain only those $\gamma $\ which do not
enter the corresponding operator products in $A_\Gamma $. It should be noted
that introducing the ME operators, which depend on all the one-electron
quantum numbers (both for occupied and empty states), is in a sense the next
step in the quantum-field description after the usual second quantization.

We obtain, instead of (A.15), the operator identities 
$$
\tilde A_\Gamma \tilde A_\Gamma ^{+} =\delta _{\Gamma \Gamma ^{\prime
}}\prod_\gamma \left( 1-\hat n_\gamma \right)   
\eqno{(\rm A.17)} 
$$
$$
\tilde A_\Gamma \tilde A_{\Gamma ^{\prime }} =\tilde A_{\Gamma ^{\prime
}}^{+}\tilde A_\Gamma ^{+}=0 \qquad
\left( |\Gamma \rangle \neq 0\right)   
\eqno{(\rm A.18)}
$$
(after reducing to the normal form, the terms with Fermi operators in the
left-hand side of (A.17) are cancelled by the factors $(1-\hat n_\gamma )$).
Thus we may pass to the representation of ME occupation numbers $N_\Gamma $
at a given site:
$$
\tilde A_\Gamma |\Gamma ^{\prime }\rangle  =\delta _{\Gamma \Gamma
^{\prime }}|0\rangle ,\tilde A_\Gamma ^{+}|\Gamma ^{\prime }\rangle =\delta
_{\Gamma ^{\prime }0}|\Gamma \rangle   
\eqno{(\rm A.19)} 
$$
$$
\tilde A_\Gamma ^{+}\tilde A_\Gamma  =\hat N_\Gamma ,\hat N_\Gamma |\Gamma
\rangle =\delta _{\Gamma \Gamma ^{\prime }}|\Gamma \rangle   
$$
$$
\sum_\Gamma \hat N_\Gamma  =1  
$$
Note that, unlike (A.4), only one of the numbers ${N}_\Gamma $ is non-zero,
and the commutation relation for the operators (A.16) differ considerably
from those for the Fermi operators:
$$
\lbrack \tilde A_\Gamma ,\tilde A_\Gamma ^{+}]_{\pm }=\hat N_0\delta
_{\Gamma \Gamma ^{\prime }}\pm \tilde A_{\Gamma ^{\prime }}^{+}\tilde
A_\Gamma ^{+}  
\eqno{(\rm A.20)}
$$
with
\[
\hat N_0=\prod_\gamma \left( 1-\hat n_\gamma \right) =\prod_\Gamma \left(
1-\hat N_\Gamma \right) 
\]
Therefore, at practical calculations, it is convenient to pass from the ME
creation and annihilation operators to X-operators, which possess more
simple properties.

For different $\Gamma $\ and $\Gamma ^{\prime }$, the product
$$
X\left( \Gamma ,\Gamma ^{\prime }\right) =\tilde A_\Gamma ^{+}\tilde
A_{\Gamma ^{\prime }}  
\eqno{(\rm A.21)}
$$
transforms the state $\Gamma ^{\prime }$ into the state $\Gamma $. Such
operators were firstly introduced by Hubbard [31] in the axiomatic way as
generalized projection operators:
$$
X\left( \Gamma ,\Gamma ^{\prime }\right) =|\Gamma \rangle \langle \Gamma
^{\prime }|  
\eqno{(\rm A.22)}
$$
where $|\Gamma \rangle $ are the exact eigenstates of the intraatomic
Hamiltonian.

Using the introduced operators of electron configurations enables one to
obtain explicit expressions for X-operators in terms of one-electron
operators. In particular,
$$
X\left( \Gamma ,0\right) =\tilde A_\Gamma ^{+},\qquad
X\left( \Gamma ,\Gamma \right) =\hat N_\Gamma   
\eqno{(\rm A.23)}
$$
For example, we consider the simplest case of $s$-electrons where $\gamma
=\sigma =\pm (\uparrow ,\downarrow )$, $\Gamma =0,\sigma ,2$ with $|0\rangle 
$ being the empty state (hole) and $|2\rangle $ the doubly-occupied singlet
state (double) on a site. Then we have
$$
X\left( 0,0\right)  =\left( 1-\hat n_{\uparrow }\right) \left( 1-\hat
n_{\downarrow }\right) ,\qquad 
X\left( 2,2\right) =\hat n_{\uparrow }\hat n_{\downarrow } 
$$
$$
X\left( \sigma ,\sigma \right)  =\hat n_\sigma \left( 1-\hat n_{-\sigma
}\right) ,\qquad 
X\left( \sigma ,-\sigma \right) =a_\sigma ^{+}a_{-\sigma }
\eqno{(\rm A.24)} 
$$
$$
X\left( \sigma ,0\right)  =a_\sigma ^{+}\left( 1-\hat n_{-\sigma }\right),
\qquad 
X\left( 2,\sigma \right) =\sigma a_{-\sigma }^{+}\hat n_\sigma  
$$

As follows from (A.17), the simple multiplication rules, postulated by
Hubbard [31], hold
$$
X\left( \Gamma ,\Gamma ^{\prime }\right) X\left( \Gamma ^{\prime \prime
},\Gamma ^{\prime \prime \prime }\right) =\delta _{\Gamma ^{\prime }\Gamma
^{\prime \prime }}X\left( \Gamma ,\Gamma ^{\prime \prime \prime }\right)  
\eqno{(\rm A.25)}
$$
For each number of electrons $n$, the sum rule is satisfied
$$
\sum_{\Gamma _n}X\left( \Gamma ,\Gamma ^{\prime }\right) =\frac
1{n!}\sum_{\gamma _i\neq \gamma _j}\hat n_{\gamma _n}\hat n_{\gamma
_{n-1}}...\hat n_{\gamma _1}\prod_{\gamma \neq \gamma _i}\left( 1-\hat
n_\gamma \right) \equiv X_n  
\eqno{(\rm A.26)}
$$
which may be verified by direct muliplication
$$
X\left( \Gamma ^{\prime },\Gamma _m\right) X_n=\delta _{nm}X\left( \Gamma
^{\prime },\Gamma _m\right)   
\eqno{(\rm A.27)}
$$
Finally, one can prove the completeness relation
$$
\sum_\Gamma X\left( \Gamma ,\Gamma \right) =\sum_nX_n=1  
\eqno{(\rm A.28)}
$$

An arbitrary operator $\hat O$ acting on the electrons at a given site i, is
expressed in terms of X-operators as
$$
\hat O=\sum_{\Gamma \Gamma ^{\prime }}\int dx_1...dx_n\Psi _\Gamma
^{*}\left( x_1...x_n\right) \hat O\Psi _{\Gamma ^{\prime }}\left(
x_1...x_n\right) \tilde A_\Gamma ^{+}\tilde A_{\Gamma ^{\prime
}}=\sum_{\Gamma \Gamma ^{\prime }}\langle \Gamma |\hat O|\Gamma ^{\prime
}\rangle X(\Gamma ,\Gamma ^{\prime })  
\eqno{(\rm A.29)}
$$
It follows from (A.11) that the matrix elements of the one-electron Fermi
operators read
$$
\langle \Gamma _n|a_\gamma ^{+}|\Gamma _{n-1}\rangle =n^{1/2}G_{\Gamma
_{n-1}}^{\Gamma _n}C_{\Gamma _{n-1},\gamma }^{\Gamma _n}  
\eqno{(\rm A.30)}
$$
Then we obtain the representation [32]
$$
a_\gamma ^{+}=\sum_nn^{1/2}\sum_{\Gamma _n\Gamma _{n-1}}G_{\Gamma
_{n-1}}^{\Gamma _n}C_{\Gamma _{n-1},\gamma }^{\Gamma _n}X\left( \Gamma
_n,\Gamma _{n-1}\right)   
\eqno{(\rm A.31)}
$$
In particular, for $s$-electrons one gets
$$
a_\sigma ^{+}=X\left( \sigma ,0\right) +\sigma X\left( 2,-\sigma \right)  
\eqno{(\rm A.32)}
$$

Using (A.31) and the orthogonality relations for Clebsh-Gordan and
fractional parentage coefficients (A.8) we obtain
$$
\sum_\gamma a_\gamma ^{+}a_\gamma  =\sum_{n\Gamma _n}nX\left( \Gamma
_n,\Gamma _n\right)   
\eqno{(\rm A.33)} 
$$
$$
\sum_\gamma a_\gamma a_\gamma ^{+} =\sum_{n\Gamma _n}\left( 2\left[
l\right] -n\right) X\left( \Gamma _n,\Gamma _n\right) 
$$
The fullfilment of Fermi commutation relations is ensured by the identity
$$
n\sum_{\Gamma _{n-1}}G_{\Gamma _{n-1}}^{\Gamma _n}G_{\Gamma _{n-1}}^{\Gamma
_n^{\prime }}C_{\Gamma _{n-1},\gamma ^{\prime }}^{\Gamma _n}C_{\Gamma
_{n-1},\gamma }^{\Gamma _n}+\left( n+1\right) \sum_{\Gamma _{n-1}}G_{\Gamma
_n}^{\Gamma _{n+1}}G_{\Gamma _n^{\prime }}^{\Gamma _{n+1}}C_{\Gamma
_n,\gamma }^{\Gamma _{n+1}}C_{\Gamma _n^{\prime },\gamma ^{\prime }}^{\Gamma
_{n+1}}=\delta _{\gamma \gamma ^{\prime }}\delta _{\Gamma _n\Gamma
_n^{\prime }}  
\eqno{(\rm A.34)}
$$
After multiplying by ${C}_{\gamma \gamma }^{\gamma ^{\prime \prime }}$, and
summing over momentum projections, this identity may be expressed in terms
of $6j$-symbols and used to obtain recurrence relations for fractional
parentage coefficients [32].

Up to now we have considered the algebra of ME operators at one lattice
site. General commutation relations for X-operators at sites $\nu $ and $\nu
^{\prime }$ read
$$
\left[ X_\nu \left( \Gamma ,\Gamma ^{\prime }\right) ,X_\nu \left( \Gamma
^{\prime \prime },\Gamma ^{\prime \prime \prime }\right) \right] _{\pm
}=\delta _{\nu \nu ^{\prime }}\{X\left( \Gamma ,\Gamma ^{\prime \prime
\prime }\right) \delta _{\Gamma ^{\prime }\Gamma ^{\prime \prime }}\pm
X\left( \Gamma ^{\prime \prime },\Gamma ^{\prime }\right) \delta _{\Gamma
\Gamma ^{\prime \prime \prime }}\}  
\eqno{(\rm A.35)}
$$
where the plus sign corresponds to the case where both X-operators have the
Fermi type, i.e. change the number of electrons by an odd number, and minus
sign to all the other cases. For the Fourier-transforms of X-operators,
equation (A.35) takes the form
$$
\left[ X_{\mathbf{k}}\left( \Gamma ,\Gamma ^{\prime }\right) ,X_{-\mathbf{k}%
^{\prime }}\left( \Gamma ^{\prime \prime },\Gamma ^{\prime \prime \prime
}\right) \right] _{\pm }=X_{\mathbf{k-k}^{\prime }}\left( \Gamma ,\Gamma
^{\prime \prime \prime }\right) \delta _{\Gamma ^{\prime }\Gamma ^{\prime
\prime }}\pm X_{\mathbf{k-k}^{\prime }}\left( \Gamma ^{\prime \prime
},\Gamma ^{\prime }\right) \delta _{\Gamma \Gamma ^{\prime \prime \prime }} 
\eqno{(\rm A.36)}
$$

The Hubbard operators may be introduced not only for the eigenstates of the
Hamiltonian with spherical symmetry, but also in the presence of a strong
crystal field. In such a situation, we have to use the irreducible
representations of the point group and corresponding Clebsh-Gordan
coefficients [564].

\chapter{Angular momentum operators and double irreducible tensor operators}

In a number of problems of solid state physics, it is convenient to pass
into the angular-momentum operator representation. The first example of such
approach was introducing spin operators for ${S}=1/2$ by Dirac
$$
a_{\uparrow }^{+}a_{\downarrow } =S^{+},\qquad
a_{\downarrow }^{+}a_{\uparrow }=S^{-}  
\eqno{(\rm B.1)} 
$$
$$
\frac 12\left( a_{\uparrow }^{+}a_{\uparrow }-a_{\downarrow
}^{+}a_{\downarrow }\right)  =S^z 
$$
to represent the electron exchange Hamiltonian for singly-occupied s-states
in the form of a scalar product:
$$
-\sum_{\sigma \sigma ^{\prime }}a_{1\sigma }^{+}a_{1\sigma ^{\prime
}}a_{2\sigma }^{+}a_{2\sigma ^{\prime }}=-\left( \frac 12+2\mathbf{S}_1%
\mathbf{S}_2\right)   
\eqno{(\rm B.2)}
$$

This yielded a basis for the Heisenberg model of magnetism (Sect.4.1), which
was also applied for arbitrary $S$. However, for ${S}>\frac 12$ one has to
consider electrons in orbit-degenerate states and, consequently, to treat
orbital momenta.

Consider the connection of spin and orbital momentum operators for an atomic
shell (1.7) with Hubbard's operators. The momentum operators are diagonal in 
${LS}$-terms:
$$
\langle SL\mu M\alpha |L_q|S^{\prime }L^{\prime }\mu ^{\prime }M^{\prime
}\alpha ^{\prime }\rangle =\delta _{SL\alpha ,S^{\prime }L^{\prime }\alpha
^{\prime }}\delta _{\mu \mu ^{\prime }}\sqrt{L\left( L+1\right) }%
C_{LM^{\prime },1q}^{LM}  
\eqno{(\rm B.3)}
$$
$$
\langle SL\mu M\alpha |S_q|S^{\prime }L^{\prime }\mu ^{\prime }M^{\prime
}\alpha ^{\prime }\rangle =\delta _{SL\alpha ,S^{\prime }L^{\prime }\alpha
^{\prime }}\delta _{MM^{\prime }}\sqrt{S\left( S+1\right) }C_{S\mu ^{\prime
},1q}^{S\mu }  
\eqno{(\rm B.4)}
$$
where we have introduced the cyclic components of a vector
\begin{eqnarray*}
A_{\pm 1} &=&\mp \frac 1{\sqrt{2}}A^{\pm }=\mp \frac 1{\sqrt{2}}\left(
A^x\pm iA^y\right)  \\
A_0 &=&A^z
\end{eqnarray*}

In the presence of spin-orbit coupling, we have to construct the functions
with definite total angular momentum ${J}=L+S$. In the Russel-Saunders
approximation we have
$$
|SLJM_J\rangle =\Psi _{SLJM_J}=\sum_{\mu M}C_{S\mu ,LM}^{JM_J}|SL\mu
M\rangle   
\eqno{(\rm B.5)}
$$
and the corresponding ME creation operator reads
$$
A_{SLJM_J}^{+}=\sum_{\mu M}C_{S\mu ,LM}^{JM_J}A_{SL\mu M}^{+}  
\eqno{(\rm B.6)}
$$
The matrix elements of the total momentum operator within the multiplet with
a given $J$ are
$$
\langle JM|J_q|JM^{\prime }\rangle =\sqrt{J\left( J+1\right) }C_{JM^{\prime
},1q}^{JM}  
\eqno{(\rm B.7)}
$$
Standard components of the vector $I=S,L,J$ are expressed as
$$
I^{+} =\sum_M\left[ \left( I-M\right) \left( I+M+1\right) \right]
^{1/2}X\left( M+1,M\right)   
\eqno{(\rm B.8)}
$$
$$
I^{-} =\sum_M\left[ \left( I-M+1\right) \left( I+M\right) \right]
^{1/2}X\left( M-1,M\right)  
$$
$$
I^z =\sum_MMX\left( M,M\right) 
$$
At passing from exchange Hamiltonians, which are expressed in terms of ME
operators, into momentum representation, we need the inverse transformations
from X-operators to momentum operators. To obtain them, we can write down
expressions for all the powers $(I^\alpha )^k$ with $k=0...2I$ (higher
powers are not linearly independent) and solve this system of $\left(
2I+1\right) ^2$ equations, which is very cumbersome. It is more convenient
to use the irreducible tensor operators
$$
I_q^{\left( k\right) }=\sum_{MM^{\prime }}C_{IM^{\prime },kq}^{IM}X\left(
IM,IM^{\prime }\right) \qquad (q=-k...k)  
\eqno{(\rm B.9)}
$$
so that
$$
S_q =\sqrt{S\left( S+1\right) }S_q^{(1)}  
\eqno{(\rm B.10)} 
$$
$$
L_q^{\left( 1\right) } =\sqrt{L\left( L+1\right) }L_q^{(1)}  
$$
The operators (B.9) are connected with the Stevens equivalent operators in
the crystal field theory (see Sect.1.3). Using the orthogonality relations
for the Clebsh-Gordan coefficients, it is easy to obtain the inverse
relations required
$$
X\left( IM,IM^{\prime }\right) =\sum_{kq}C_{IM^{\prime },kq}^{IM}I_q^{\left(
k\right) }  
\eqno{(\rm B.11)}
$$
The commutation rules for the operators (B.9) have the form
\[
\lbrack I_q^{\left( k\right) },I_{q^{\prime }}^{\left( k\right)
}]_{-}=\sum_{k^{\prime \prime }q^{\prime \prime }}\left( \left( -1\right)
^{k^{\prime \prime }}-\left( -1\right) ^{k+k^{\prime }}\right) \left\{ 
\begin{array}{ccc}
k & k^{\prime } & k^{\prime \prime } \\ 
I & I & I
\end{array}
\right\} \left( \left[ K^{\prime \prime }\right] \left[ I\right] \right)
^{1/2}C_{kq,k^{\prime }q^{\prime }}^{k^{\prime \prime }q^{\prime \prime
}}I_{q^{\prime \prime }}^{\left( k^{\prime \prime }\right) }
\]
with $\left\{ 
\begin{array}{ccc}
. & . & . \\ 
. & . & .
\end{array}
\right\} $ being the $6j$-symbols. Explicit expressions for the irreducible
tensor operators in terms of usual momentum operators are tabulated [41,43].
Besides that, one can apply at analytical calculations the recurrence
relation
$$
\left( -1\right) ^k\left( [k-p][p][I]\right) ^{1/2}\left\{ 
\begin{array}{ccc}
k-p & p & k \\ 
I & I & I
\end{array}
\right\} I^{\left( k\right) }=[I^{\left( k-p\right) }\times I^{\left(
p\right) }]^{\left( k\right) }  
\eqno{(\rm B.12)}
$$
where the tensor product of rank $c$ is defined by
$$
\lbrack A^{\left( a\right) }\times B^{\left( b\right) }]_\gamma ^{\left(
c\right) }=\sum_{\alpha \beta }C_{a\alpha ,b\beta }^{c\gamma }A_\alpha
^{\left( a\right) }B_\beta ^{\left( b\right) }  
\eqno{(\rm B.13)}
$$
To express X-operators, connecting terms with different $L,S,J$ one can use
``hyperbolic'' operators which change the value of the momentum itself
[656]. They satisfy the commutation relations
$$
\lbrack K^{\pm },K^z]=\mp K^{\pm },\qquad
[K^{+},K^{-}]=2K^z  
\eqno{(\rm B.14)}
$$
and have the following non-zero matrix elements
$$
\langle JM|K^{+}|J-1M\rangle  =\left[ \left( J+M\right) \left( J-M\right)
\right] ^{1/2}  
\eqno{(\rm B.15)} 
$$
$$
\langle JM|K^z|JM\rangle  =J+\frac 12
$$
This problem is discussed also by Popov and Loginov [657].

At passing from the $LS$-representation to the $J$-representation for the
multiplet with a given ${J}$, we have to express the products ${L}^{\left(
k\right) }S^{\left( \kappa \right) }$ in terms of operators ${J}^{\left(
p\right) }$. Using (A.11), (B.6) and summing the product of four
Clebsh-Gordan coefficients with the use of a $9j$-symbol we get
$$
S_\xi ^{\left( \kappa \right) }L_q^{\left( k\right) }=\sum_{P\rho }\left\{ 
\begin{array}{ccc}
J & S & L \\ 
J & S & L \\ 
p & \kappa  & k
\end{array}
\right\} \left( [S][L][J][p]\right) ^{1/2}C_{\kappa \xi ,kq}^{P\rho }J_\rho
^{\left( P\right) }  
\eqno{(\rm B.16)}
$$
Note that ${p}+\kappa +k$ is even (in the opposite case, the $9j$-symbol in
(B.15) is zero due to its symmetry properties). For $k=0$ or $\kappa =0$ the 
$9j$-symbol is simplified, so that we obtain
$$
S^{\left( \kappa \right) }=\left( -1\right) ^{J+S+L+\kappa }\left( \left[
S\right] \left[ J\right] \right) ^{1/2}\left\{ 
\begin{array}{ccc}
J & S & L \\ 
S & J & \kappa 
\end{array}
\right\} J^{\left( \kappa \right) }  
\eqno{(\rm B.17)}
$$
$$
L^{\left( k\right) }=\left( -1\right) ^{J+S+L+k}\left( \left[ L\right]
\left[ J\right] \right) ^{1/2}\left\{ 
\begin{array}{ccc}
J & L & S \\ 
L & J & k
\end{array}
\right\} J^{\left( k\right) }  
\eqno{(\rm B.18)}
$$
In particular, at $\kappa =1$ or ${k}=1$, substituting explicit values of $6j
$-coefficients yields
$$
\mathbf{S}=(g-1)\mathbf{J}  
\eqno{(\rm B.19)}
$$
$$
\mathbf{L}=(2-g)\mathbf{J}  
\eqno{(\rm B.20)}
$$
where
\[
g=1+\frac{\left( \mathbf{LS}\right) }{J^2}=1+\frac{J\left( J+1\right)
-S\left( S+1\right) -L\left( L+1\right) }{2J\left( J+1\right) }
\]
is the Lande factor. Thus the well-known de Gennes formula (B.19) (see
Sect.4.7) is obtained from the general relation (B.16).

Now we discuss another approach to description of the Fermi operator
products. Since matrix elements of the Fermi creation operators ${a}%
_{lm\sigma }^{+}$ (A.10) are proportional to the Clebsh-Gordan coefficients,
these operators may be considered as $2(2l+1)$ components of a double tensor
operator acting in the spin and orbital spaces. At calculating matrix
elements of products of Fermi operators at one site, it is convenient to
introduce for a given shell the double irreducible tensor operators with the
components
$$
W_{\xi q}^{\left( \kappa k\right) }=\sum_{mm^{\prime }\sigma \sigma ^{\prime
}}C_{lm^{\prime },kq}^{lm}C_{1/2\sigma ^{\prime },\kappa \xi }^{1/2\sigma
}a_{lm\sigma }^{+}a_{l^{\prime }m^{\prime }\sigma ^{\prime }}  
\eqno{(\rm B.21)}
$$
(we use the definition, slightly different from that in [32]). Any operator
of the form
\[
\widehat{F}=\sum_i\widehat{f}_i
\]
where $\widehat{{f}}$ is an arbitrary one-electron operator, may be
represented in terms of the operators (B.21). In particular,
$$
W^{\left( 00\right) }=\sum_{m\sigma }a_{lm\sigma }^{+}a_{lm\sigma }=\hat n 
\eqno{(\rm B.22)}
$$
is the number-of-particle operator for the shell. The cyclic components of
the total spin and orbital momentum operators (1.7) are given by (cf.(B.10))
$$
S_\xi =\frac{\sqrt{3}}2W_\xi ^{\left( 10\right) }  
\eqno{(\rm B.23)}
$$
$$
L_q=\sqrt{l\left( l+1\right) }W_q^{\left( 01\right) }  
\eqno{(\rm B.24)}
$$
and the inner scalar product
$$
\frac 12\sqrt{3l\left( l+1\right) }\sum_q\left( -1\right) ^qW_{q,-q}^{\left(
11\right) }=\sum_is_il_i  
\eqno{(\rm B.25)}
$$
is proportional to the operator of spin-orbit coupling.

The operators (B.21) for ${k}+\kappa >1$, unlike the operators (B.23),
(B.24), have non-zero matrix elements between terms with different $S$ and $L
$ and are, generally speaking, not reduced to products of the momentum
operators (B.9). However, the connection may be established within a
concrete term with given $L$, $S$ (or $J$). Calculating the matrix element
of the product of Fermi operators with account of (3.10) we obtain in
agreement with the Wigner-Eckart theorem
$$
\langle SL\mu M\alpha |W_{\xi q}^{\left( \kappa k\right) }|S^{\prime
}L^{\prime }\mu ^{\prime }M^{\prime }\alpha ^{\prime }\rangle   
\eqno{(\rm B.26)} 
$$
$$
=\left( \left[ S\right] \left[ L\right] \right) ^{-1/2}C_{L^{\prime
}M^{\prime },kq}^{LM}\langle SL\alpha |W^{\left( \kappa k\right) }|S^{\prime
}L^{\prime }\alpha ^{\prime }\rangle   
$$
where the reduced matrix elements have the form
$$
\langle SL\alpha ||W^{\left( \kappa k\right) }||S^{\prime }L^{\prime
}\alpha ^{\prime }\rangle =
n\left( 2\left[ l\right] \left[ S\right] \left[ S^{\prime }\right] \left[
L\right] \left[ L^{\prime }\right] \right) ^{1/2}  
\eqno{(\rm B.27)} 
$$
$$
\times \sum_{\left\{ \bar S\bar L\bar \alpha \right\} }G_{\bar S\bar L\bar
\alpha }^{SL\alpha }G_{\bar S\bar L\bar \alpha }^{S^{\prime }L^{\prime
}\alpha ^{\prime }}\left\{ 
\begin{array}{ccc}
\frac 12 & \kappa  & \frac 12 \\ 
S & \bar S & S^{\prime }
\end{array}
\right\} \left\{ 
\begin{array}{ccc}
l & k & l \\ 
L & \bar L & L^{\prime }
\end{array}
\right\} \left( -1\right) ^{1/2+l+S+L+\bar S+\bar L+\kappa +k}  
$$
$$
\langle SL\alpha ||W^{\left( 0k\right) }||S^{\prime }L^{\prime }\alpha
^{\prime }\rangle  =n\delta _{SS^{\prime }}\left( \left[ l\right] \left[
S\right] \left[ L\right] \left[ L^{\prime }\right] \right) ^{1/2}  
\eqno{(\rm B.28)}
$$
$$
\times \sum_{\left\{ \bar S\bar L\bar \alpha \right\} }G_{\bar S\bar L\bar
\alpha }^{SL\alpha }G_{\bar S\bar L\bar \alpha }^{SL^{\prime }\alpha
^{\prime }}\left\{ 
\begin{array}{ccc}
l & k & l \\ 
L & \bar L & L^{\prime }
\end{array}
\right\} \left( -1\right) ^{l+L+\bar L+k+1} 
$$
They can be also taken from the tables available (see, e.g., [10, 659]).
Thus we have for a given term
$$
W^{\left( \kappa k\right) }=\left( \left[ S\right] \left[ L\right] \right)
^{1/2}\langle SL\alpha ||W^{\left( \kappa k\right) }||SL\alpha \rangle S_\xi
^{\left( \kappa \right) }L_q^{\left( k\right) }  
\eqno{(\rm B.29)}
$$

Comparing (B.27), (B.29) with (B.3), (B.4), (B.23), (B.24) one can see that
the double irreducible tensor operator approach yields a summation of
fractional parentage coefficients at $k=0$, $\kappa =1$ or $k=1$, $\kappa =0$
[660]. We shall see in Appendix D that this leads to a simplification of
some terms in the exchange Hamiltonian. In particular, substituting the
values of $6j$-symbols at $SL\alpha =S^{\prime }L^{\prime }\alpha ^{\prime }$, 
we derive
$$
\sum_{\left\{ \bar S\bar L\bar \alpha \right\} }\left( G_{\bar S\bar L\bar
\alpha }^{SL\alpha }\right) ^2\left( -1\right) ^{S-\bar S+1/2}\left[
S\right] ^{-1}=\frac 1n \qquad \left( S\neq 0\right)   
\eqno{(\rm B.30)}
$$
$$
\sum_{\left\{ \bar S\bar L\bar \alpha \right\} }\left( G_{\bar S\bar L\bar
\alpha }^{SL\alpha }\right) ^2\bar L\left( \bar L+1\right) =l\left(
l+1\right) +\left( 1-\frac 2n\right) L\left( L+1\right) \qquad
\left( L\neq 0\right)  
\eqno{(\rm B.31)}
$$

The double irreducible tensor operator formalism may be generalized by
considering triple tensor operators $a^{(qls)}$ ($q=s=1/2$) with the
components
$$
a_{1/2m\sigma }^{(qls)} =a_{lm\sigma }^{+}  
\eqno{(\rm B.32)} 
$$
$$
a_{-1/2m\sigma }^{(qls)} =\left( -1\right) ^{l+s-m-\sigma }a_{lm\sigma } 
$$
and constructing from them the operators
$$
T_{\lambda \xi r}^{\left( \Lambda \kappa k\right) }=
\mathop{\sum _{\varphi \varphi ^{\prime }\sigma \sigma ^{\prime }}}_ 
{mm^{\prime }} 
C_{q\varphi
^{\prime },\Lambda \lambda }^{q\varphi }C_{S\sigma ^{\prime },\kappa \xi
}^{S\sigma }C_{lm^{\prime },kr}^{lm}a_{\varphi m\sigma }^{(qls)}a_{\varphi
^{\prime }m^{\prime }\sigma ^{\prime }}^{(qls)}  
\eqno{(\rm B.33)}
$$
In particular, for $\kappa =0$, ${k}=0$, $\Lambda =1$ we obtain the
quasispin operator with the components
$$
Q^{+} =\frac 12\sum_{m\sigma }\left( -1\right) ^{l+1/2-m-\sigma
}a_{lm\sigma }^{+}a_{l-m-\sigma }^{+}  
\eqno{(\rm B.34)}
$$
$$
Q^z =-\frac 12\left( 2l+1-\sum_{m\sigma }a_{lm\sigma }^{+}a_{lm\sigma
}\right) 
$$
which satisfy the usual commutation relations for spin operators. Note that
the operator ${Q}^{+}$ coincides to a numerical factor with the creation
operator (2.14) for the two-electron singlet zero-$L$ state $^1S$ and adds
such a pair to configuration $l^{n-2}$ . The value of quasispin for a given
term sequence equals to maximum value of $Q^z$:
$$
Q=\frac 12\left( 2l+1-v\right)   
\eqno{(\rm B.35)}
$$
where $v$ is the Racah's seniority quantum number, $(n-v)/2$ being the
number of closed electron pairs with zero spin and orbital momenta in the
given ME term. Using the Wigner-Eckart theorem in the quasispin space
enables one to establish additional symmetry of the ME problem (in
particular, to present the dependences of reduced matrix elements and
fractional parentage coefficients on the electron number in terms of $v$)
[32]. This formalism may be useful in the solid state theory at considering
charge fluctuations. Further applications of the group theory to the
classification of many-electron states are discussed in the book [65].

\chapter{Hamiltonian of a crystal with many-electron atoms}

To present derivation of some many-electron (ME) models we treat the general
Hamiltonian of the ME system in a crystal 
$$
\mathcal{H}=\sum_i\left( -\frac{\hbar ^2}{2m}\Delta _{\mathbf{r}_i}+V(%
\mathbf{r}_i)\right) +\frac 12\sum_{i\neq j}\frac{e^2}{|\mathbf{r}_i-\mathbf{%
r}_j|}  
\eqno{(\rm C.1)}
$$
where $V(\mathbf{r})$ is the periodic crystal potential. Further we pass to
the second quantization representation. To this end one has to use
orthogonal wave functions. However, the atomic wave functions 
$$
\varphi _{lm\sigma }(x)=\varphi _{lm}(\mathbf{r})\chi _\sigma
(s)=R_l(r)Y_{lm}(\mathbf{\hat r})\chi _\sigma (s)  
\eqno{(\rm C.2)}
$$
($s$ is the spin coordinate, $R_l$ is the radial wavefunction, $Y$ is the
spherical function, $\mathbf{\hat r}=(\theta ,\phi )$) do not satisfy this
condition at different sites $\nu $. The non-orthogonality problem is one of
most difficult in the magnetism theory [656,661]. Here we use the
orhtogonalization procedure developed by Bogolyubov [651]. To lowest order
in the overlap of atomic wavefunctions the orthogonalized functions read 
$$
\psi _{\nu lm}(\mathbf{r})=\varphi _{\nu lm}(\mathbf{r})-\frac 12\sum_{\nu
^{\prime }\neq \nu }\sum_{l^{\prime }m^{\prime }}\varphi _{\nu ^{\prime
}l^{\prime }m^{\prime }}(\mathbf{r})\int d\mathbf{r}^{\prime }\varphi _{\nu
^{\prime }l^{\prime }m^{\prime }}^{*}(\mathbf{r}^{\prime })\varphi _{\nu lm}(%
\mathbf{r})  
\eqno{(\rm C.3)}
$$
The non-orthogonality corrections may be neglected at considering intrasite
interactions and two-site Coulomb (but not exchange) matrix elements.

Then we obtain the many-electron Hamiltonian of the polar model [651,662,
663] in the general case of degenerate bands 
$$
\mathcal{H} =\sum_{\nu lm\sigma }\varepsilon _la_{\nu lm\sigma }^{\dagger
}a_{\nu lm\sigma }+\sum_{\nu ^{\prime }\neq \nu }\sum_{l_im_i\sigma }\beta
_{\nu _1\nu _2}(l_1m_1,l_2m_2)a_{\nu _1l_1m_1\sigma }^{\dagger }a_{\nu
_2l_2m_2\sigma }  
\eqno{(\rm C.4)}
$$
$$
+\frac 12\sum_{\nu _il_im_i\sigma _1\sigma _2}I_{\nu _1\nu _2\nu _3\nu
_4}(l_1m_1,l_2m_2,l_3m_3,l_4m_4)a_{\nu _1l_1m_1\sigma _1}^{\dagger }a_{\nu
_2l_2m_2\sigma _2}^{\dagger }a_{\nu _4l_4m_4\sigma _2}a_{\nu _3l_3m_3\sigma
_1}
$$
Here we have used the orthogonality of spin wave functions $\chi _\sigma (s)$, 
$$
\varepsilon _l=\int\limits_0^\infty r^2drR_l(r)\left[ -\frac 1{2mr^2}\frac
d{dr}\left( r^2\frac d{dr}\right) +\frac{l(l+1)}{2mr^2}+v(r)\right] R_l(r) 
\eqno{(\rm C.5)}
$$
are the one-electron levels in the central potential of a given site $v(%
\mathbf{r})$ (we neglect the influence of potentials of other atoms, i.e.
crystal-field effects), 
$$
\beta _{\nu _1\nu _2}(l_1m_1,l_2m_2)=\int d\mathbf{r}\psi _{\nu
_1l_1m_1}^{*}(\mathbf{r})\left( -\frac{\hbar ^2}{2m}\Delta +V(\mathbf{r}%
)\right) \psi _{\nu _2l_2m_2}(\mathbf{r})  
\eqno{(\rm C.6)}
$$
are the transfer matix elements between the sites $\nu _1$ and $\nu _2$, 
$$
I_{\nu _1\nu _2\nu _3\nu _4}(l_1m_1,l_2m_2,l_3m_3,l_4m_4)  
\eqno{(\rm C.7)} 
$$
$$
=\int d\mathbf{r}d\mathbf{r}^{\prime }\psi _{\nu _1l_1m_1}^{*}(\mathbf{r}%
)\psi _{\nu _2l_2m_2}^{*}(\mathbf{r}^{\prime })\frac{e^2}{|\mathbf{r}-%
\mathbf{r}^{\prime }|}\psi _{\nu _3l_3m_3}(\mathbf{r})\psi _{\nu _4l_4m_4}(%
\mathbf{r}^{\prime })  
$$
are the matrix elements of electrostatic interaction. Consider the
electrostatic interactions between two atomic shells on the same lattice
site $\nu _1=\nu _2=\nu _3=\nu _4$. We use the standard expansion 
$$
\frac 1{|\mathbf{r}-\mathbf{r}^{\prime }|}=\sum_{p=0}^\infty \frac{4\pi }{[p]%
}\frac{r_{<}^p}{r_{>}^{p+1}}\sum_{q=-p}^pY_{pq}(\mathbf{\hat r)}Y_{pq}%
\mathbf{(\hat r}^{\prime })  
\eqno{(\rm C.8)}
$$
and the expression for the matrix element of a spherical function 
$$
\int\limits_0^\pi \sin \theta d\theta \int\limits_0^{2\pi }d\varphi
Y_{l_1m_1}^{*}(\mathbf{\hat r})Y_{l_2m_2}(\mathbf{\hat r})Y_{l_3m_3}(\mathbf{%
\hat r})  
\eqno{(\rm C.9)} 
$$
$$
=\left( \frac{[l_2][l_3]}{4\pi [l_1]}\right)
^{1/2}C_{l_20,l_30}^{l_10}C_{l_2m_2,l_3m_3}^{l_1m_1}\equiv \tilde
C_{l_2m_2,l_3m_3}^{l_1m_1}  
$$
The quantity $\tilde C$ vanishes if $l_1+l_2+l_3$ is odd, and for 
$l_1+l_2+l_3=2g$ one has 
\[
C_{l_20,l_30}^{l_10}=(-1)^{l_1+g}[l_1]^{1/2}g!\{(2g+1)\}^{-1/2}\prod
\limits_{i=1}^3\frac{\{(2g-2l_i)!\}^{1/2}}{(g-l_i)!} 
\]
Carrying out the integration, we obtain for the Coulomb term ($l_1=l_3$, $%
l_2=l_4$ ) 
$$
\mathcal{H}_{\mathrm{coul}}(l_1l_2)=%
\sum_pC_{l_10,p0}^{l_10}C_{l_20,p0}^{l_20}F^{(p)}(l_1l_2)(W_1^{(0p)}W_2^{(0p)})
\eqno{(\rm C.10)}
$$
where 
$$
F^{(p)}(l_1l_2)=e^2\int r_1^2dr_1r_2^2dr_2\frac{r_{<}^p}{r_{>}^{p+1}}%
R_{l_1}^2(r_1)R_{l_2}^2(r_2)  
\eqno{(\rm C.11)}
$$
are the Slater parameters, the irreducible tensor operators $W$ are given by
(B.21) and the scalar product is defined by 
$$
\left( A^{(a)}B^{(a)}\right) =\sum_\alpha (-1)^\alpha A_\alpha
^{(a)}B_{-\alpha }^{(a)}\equiv (-1)^a[a]^{1/2}[A^{(a)}\times B^{(a)}]^{(0)} 
\eqno{(\rm C.12)}
$$
The Hamiltonian (C.10) may be expressed in terms of many-electron
X-operators with the use of (B.29) (see Appendix D).

For the exchange integral we obtain 
\[
I_\nu (l_1m_1,l_2m_2,l_3m_3,l_4m_4)=\sum_{pq}\frac{4\pi }{[p]}\tilde
C_{l_2m_3,pq}^{l_1m_1}\tilde C_{l_1m_4,pq}^{l_2m_2}G^{(p)}(l_1l_2) 
\]
where 
$$
G^{(p)}(l_1l_2)=e^2\int \mathbf{r}_1^2d\mathbf{r}_1\mathbf{r}_2^2d\mathbf{r}%
_2R_{l_1}(\mathbf{r}_1)R_{l_2}(\mathbf{r}_2)\frac{\mathbf{r}_{<}^p}{\mathbf{r%
}_{>}^{p+1}}R_{l_2}(\mathbf{r}_1)R_{l_1}(\mathbf{r}_2)  
\eqno{(\rm C.13)}
$$
Transforming the product of Clebsh-Gordan coefficients 
$$
\sum_qC_{l_2m_3,pq}^{l_1m_1}C_{l_1m_4,pq}^{l_2m_2}=\sum_{k\mu }\left\{ 
\begin{array}{ccc}
l_1 & l_2 & p \\ 
l_2 & l_1 & k
\end{array}
\right\} C_{l_1m_4,pq}^{l_1m_1}C_{l_2m_3,pq}^{l_2m_2}(-1)^{k-\mu } 
\eqno{(\rm C.14)}
$$
we derive 
$$
H_{\mathrm{exch}}(l_1l_2) =-\frac 12\sum_{kp\kappa }\left\{ 
\begin{array}{ccc}
l_1 & l_2 & p \\ 
l_2 & l_1 & k
\end{array}
\right\} (-1)^k[k][\kappa ]\left (\frac{[l_2]}{[l_1]}\right )^{1/2}  
\eqno{(\rm C.15)} 
$$
$$
\times \left( C_{l_20,p0}^{l_10}\right) ^2G^{(p)}(l_1l_2)\left(
W_1^{(\kappa k)}W_2^{(\kappa k)}\right)  
$$
For $k=0$ we obtain from (B.22), (B.23) the intraatomic exchange Hamiltonian 
$$
\sum_{\mbox{\boldmath$\kappa $}}\mathcal{H}_{\mathrm{exch}}^
{(\mbox{\boldmath$\kappa $}
0)}(l_1l_2)=-\frac 12\sum_p\left (\frac{[l_2]}{[l_1]}\right )^{1/2}\left(
C_{l_20,p0}^{l_10}\right) ^2G^{(p)}(l_1l_2)[n_1n_2+4(\mathbf{S}_1\mathbf{S}%
_2)]  
\eqno{(\rm C.16)}
$$
and for $\mbox{\boldmath$\kappa $}=0$, $k=1$ from (B.24) the orbital exchange
Hamiltonian 
$$
H_{\mathrm{exch}}^{(01)}(l_1l_2)=-\frac 34\sum_p\left(
C_{l_20,p0}^{l_10}\right) ^2\frac{l_1(l_1+1)+l_2(l_2+1)-p(p+1)}{%
l_1(l_1+1)l_2(l_2+1)[l_1]}G^{(p)}(l_1l_2)(\mathbf{L}_1\mathbf{L}_2) 
\eqno{(\rm C.17)}
$$

In the case of electrostatic interaction between electrons of the same
shell, $F=G$, so that we may represent the Hamiltonian in both forms (C.10)
and (C.15). Thus 
$$
\mathcal{H}(l)=\frac 12\sum_p\left( C_{l_20,p0}^{l_10}\right)
^2F^{(p)}(ll)\left[ (W^{(0p)}W^{(0p)})-n\right]  
\eqno{(\rm C.18)}
$$
where the second term in square brackets arises because of the condition $%
i\neq j$ in (C.1). In the ME representation, the Hamiltonian (C.18) takes
the quasidiagonal form. Calculating the matrix elements of the scalar
product we obtain 
\[
\mathcal{H}(l)=\sum_{SL\mu M}\sum_{\alpha \alpha ^{\prime }}E_{SL}^{\alpha
\alpha ^{\prime }}X(SL\mu M\alpha ,SL\mu M\alpha ^{\prime }) 
\]
with 
$$
E_{SL}^{\alpha \alpha ^{\prime }} =\frac 12\sum_{SLM\mu }\sum_p\left(
C_{l_20,p0}^{l_10}\right) ^2F^{(p)}(ll)\{[S]^{-1}[L]^{-1}\sum_{\bar L\bar
\alpha }\langle SL\alpha \Vert W^{(0p)}\Vert S\bar L\bar \alpha \rangle 
\eqno{(\rm C.19)} 
$$
$$
\times \langle SL\alpha ^{\prime }\Vert W^{(0p)}\Vert S\bar L\bar \alpha
-\delta _{\alpha \alpha ^{\prime }}n\rangle \}  
$$
and the reduced matrix elements are given by (B.27). If several ME terms
with the same values of $L$,$S$ are present (which is typical for $d$-and $f$%
-electrons), additional diagonalization is required. Retaining in (C.19) the
contribution with $p=0$ only we obtain 
$$
E_\Gamma =\frac 12n(n-1)F^{(0)}(ll)  
\eqno{(\rm C.20)}
$$
The contributions with $p=2,4...$ yield the dependence of term energy on the
ME quantum numbers $S$, $L$ according to the Hund rules.

Expressions (C.10), (C.16), (C.18) reduce the problem of electrostatic
interaction between electrons to calculation of the Slater integrals. These
may be considered as parameters, which should be taken from experimental
data (such a procedure is used often in the atomic spectroscopy). In this
case, the wave functions, which enter the Slater integrals, are to be
calculated in a self-consistent way from the corresponding
integro-differential equations [20]. Consider the one-band Hamiltonian of a
crystal in the many-electron representation. Besides the on-site Coulomb
repulsion and the intersite electron transfer (which are taken into account
in the Hubbard model) we take into account the electron transfer owing to
matrix elements of electrostatic interaction (C.7) with $\nu _1\neq \nu _3$, 
$\nu _2=\nu _4$. We confine ourselves to the terms with $\nu _1=\nu _2$ or $%
\nu _2=\nu _3$ (three-site terms are smaller due to decrease of the Coulomb
interaction with distance). Using (B.21), (C.19) and taking into account
corrections owing to the non-orthogonality we obtain 
$$
\mathcal{H} =\sum_{\nu \Gamma }E_\Gamma X_\nu (\Gamma ,\Gamma )+\sum_{\nu
_1\neq \nu _2}\sum_{\Gamma _n\Gamma _{n-1}\Gamma _{n^{\prime }}\Gamma
_{n^{\prime }-1}}B_{\nu _1\nu _2}(\Gamma _n\Gamma _{n-1},\Gamma _{n^{\prime
}}\Gamma _{n^{\prime }-1})  
\eqno{(\rm C.21)} 
$$
$$
\times X_{\nu _1}(\Gamma _n\Gamma _{n-1})X_{\nu _2}(\Gamma _{n^{\prime
}}\Gamma _{n^{\prime }-1})  
$$
where 
$$
B_{\nu _1\nu _2}(\Gamma _n\Gamma _{n-1},\Gamma _{n^{\prime }}\Gamma
_{n^{\prime }-1}) =(nn^{\prime })^{1/2}G_{\Gamma _{n-1}}^{\Gamma
_n}G_{\Gamma _{n^{\prime }-1}}^{\Gamma _{n^{\prime }}}  
\eqno{(\rm C.22)} 
$$
$$
\times \sum_{\gamma _1\gamma _2}C_{\Gamma _{n-1},\gamma _1}^{\Gamma
_n}C_{\Gamma _{n^{\prime }-1},\gamma _2}^{\Gamma _{n^{\prime }}}\delta
_{\sigma _1\sigma _2}\beta _{\nu _1\nu _2}(lm_1,lm_2)  
$$
$$
+[n(n-1)n^{\prime }(n^{\prime }-1)]^{1/2}  
$$
$$
\times [\sum_{\bar \Gamma _{n-1}\bar \Gamma _{n-2}}G_{\bar \Gamma
_{n-1}}^{\Gamma _n}G_{\bar \Gamma _{n-2}}^{\bar \Gamma _{n-1}}G_{\bar \Gamma
_{n-2}}^{\Gamma _{n-1}}G_{\bar \Gamma _{n^{\prime }-1}}^{\Gamma _{n^{\prime
}}} 
$$
$$
\times \sum_{\gamma _1\ldots \gamma _4}G_{\bar \Gamma _{n-1},\gamma
_1}^{\Gamma _n}G_{\bar \Gamma _{n-2},\gamma _2}^{\bar \Gamma _{n-1}}G_{\bar
\Gamma _{n-2},\gamma _3}^{\Gamma _{n-1}}G_{\bar \Gamma _{n^{\prime
}-1},\gamma _4}^{\Gamma _{n^{\prime }}}\delta _{\sigma _1\sigma _3}\delta
_{\sigma _2\sigma _4}  
$$
$$
\times I_{\nu _1\nu _2\nu _3\nu _4}(lm_1,lm_2,lm_3,lm_4)  
$$
$$
+\{\Gamma _n,\Gamma _{n-1}\} \leftrightarrow \{\Gamma _{n^{\prime
}},\Gamma _{n^{\prime }-1}\}  
$$
are the many-electron transfer integrals. E.g., for the $s$-band we have 
$$
\mathcal{H} =U\sum_\nu X_\nu (2,2)+\sum_{\nu _1\nu _2\sigma }\{\beta _{\nu
_1\nu _2}^{(00)}X_{\nu _1}(\sigma ,0)X_{\nu _2}(0,\sigma )+\beta _{\nu _1\nu
_2}^{(22)}X_{\nu _1}(2,\sigma )X_{\nu _2}(\sigma ,2)  
\eqno{(\rm C.23)} 
$$
$$
+\sigma \beta _{\nu _1\nu _2}^{(02)}X_{\nu _1}(\sigma ,0)X_{\nu
_2}(-\sigma ,2)+X_{\nu _1}(2,-\sigma )X_{\nu _2}(0,\sigma )  
$$
where $U=I_{\nu \nu \nu \nu }=F^{(0)}(00)$ is the Hubbard parameter, 
$$
\beta _{\nu _1\nu _2}^{(00)} =\beta _{\nu _1\nu _2}  
\eqno{(\rm C.24)} 
$$
$$
\beta _{\nu _1\nu _2}^{(22)} =\beta _{\nu _1\nu _2}+2I_{\nu _1\nu _1\nu
_2\nu _1} 
$$
$$
\beta _{\nu _1\nu _2}^{(02)} =\beta _{\nu _1\nu _2}^{(20)}=\beta _{\nu
_1\nu _2}+I_{\nu _1\nu _1\nu _2\nu _1}  
$$
are the transfer integral for holes and doubles, and the integral of the
double-hole pair creation; according to (C.3), 
$$
I_{\nu _1\nu _1\nu _2\nu _1}=I_{\nu _1\nu _1\nu _2\nu _1}^{(0)}-\frac U2\int
d\mathbf{r}\varphi _{\nu _2}(\mathbf{r})\varphi
_{\nu 1}(\mathbf{r})  
\eqno{(\rm C.25)}
$$
with $I^{(0)}$ being calculated for the atomic functions $\varphi $. It
should be noted that the dependence of the transfer integrals on the atomic
ME terms may be more complicated if we use at solving the atomic problem the
approaches which are more complicated than in Appendix A. For example, the
general Hartree-Fock approximation (see [20]) yields the radial one-electron
wave functions which depend explicitly on atomic term. In some variational
approaches of the many-electron atom theory (see [664] the ME wavefunction
is not factorized into one-electron ones. Then the transfer integrals are to
be calculated with the use of ME wave functions: 
$$
B_{v_1v_2}(\Gamma _n\Gamma _{n-1},\Gamma _{n^{\prime }}\Gamma _{n^{\prime
}-1}) =\int \prod dx_i\Psi _{v_1\Gamma _{n^{\prime }}}^{*}\Psi _{v_2\Gamma
_{n^{\prime }}}^{*}  
\eqno{(\rm C.26)} 
$$
$$
\times \sum_i\left( -\frac{\hbar ^2}{2m}\Delta _{\mathbf{r}_i}+V(\mathbf{r}%
_i)\right) \Psi _{v_1\Gamma _{n^{\prime }}}\Psi _{v_2\Gamma _{n^{\prime }}} 
$$
In particular, for the $s$-band the integrals (C.24) may be different even
at neglecting interatomic Coulomb interactions and non-orthogonality.
Besides that, the many-configuration approach, which takes into account the
interaction of different electron shells, is sometimes required.

To analyze the $m$-dependence of two-site matrix elements, it is convenient
to transform them into one-site ones by expanding the wave functions $%
\varphi _{\nu _2}(\mathbf{r})$ about the first atom [665]. Passing to the
Fourier transforms 
$$
\varphi _{lm}(\mathbf{r})=\int d\mathbf{k}e^{i\mathbf{kr}}\tilde
R_l(k)Y_{lm}(\mathbf{\hat k})  
\eqno{(\rm C.27)}
$$
(with $\tilde R_l(k)$ being the Fourier transforms of radial functions) and
expanding plane waves in spherical harmonics 
$$
e^{i\mathbf{kr}}=4\pi \sum_{\lambda \mu }i^\lambda j_\lambda (\mathbf{kr}%
)Y_{\lambda \mu }^{*}(\mathbf{\hat k})Y_{lm}(\mathbf{\hat r})  
\eqno{(\rm C.28)}
$$
where $j_\lambda (x)$ are the spherical Bessel function, we obtain 
$$
\varphi _{\nu _2lm}(\mathbf{r})\equiv \varphi _{\nu _1lm}(\mathbf{r+\rho }%
)=4\pi \sum_{\lambda \mu \eta \xi }\tilde C_{lm,\lambda \mu }^{\eta \xi
}R_{l\eta \lambda }(\mathbf{r},\mbox{\boldmath$\rho $})Y_{\lambda \mu }^{*}(\hat \rho
)Y_{\eta \xi }(\mathbf{\hat r})  
\eqno{(\rm C.29)}
$$
with 
$$
R_{l\eta \lambda }(\mathbf{r},\mbox{\boldmath$\rho $})=\int\limits_0^\infty
k^2dkR_l(k)i^{\lambda +\eta }j_\eta (\mathbf{kr})j_\lambda (\mathbf{k}\rho )
\eqno{(\rm C.30)}
$$
As well as in (C.3), we retain only the crystal potential at the site $\nu
_1 $ is retained; in this approximation, non-orthogonality corrections to
one-electron transfer integrals are absent [651]. Then, substituting (C.29)
into (C.6) we pick out the dependence of transfer matrix elements on
magnetic quantum numbers [660] 
$$
\beta _{\nu _1\nu _2}(l_1m_1,l_2m_2)=4\pi \sum_{\lambda \mu }\tilde
C_{l_2m_2,\lambda \mu }^{l_1m_1}\bar \beta _{\nu _1\nu _2}(l_1l_2\lambda
)Y_{\lambda \mu }(\hat \rho _{\nu _1\nu _2})  
\eqno{(\rm C.31)}
$$
where 
$$
\bar \beta _{\nu _1\nu _2}(l_1l_2\lambda )=\int\limits_0^\infty
r^2drR_{l_1}(r)v(r)R_{l_2l_1\lambda }(r\mathbf{,\rho }_{\nu _1\nu _2}) 
\eqno{(\rm C.32)}
$$
One can see that for even $l_1-l_2$ (in particular, for intraband transfer
integrals with $l_1=l_2$ ) $\lambda $ is even, and for odd $l_1-l_2$ (in
particular, for the matrix elements of $s-p$, $p-d$ and $d-f$ hybridization)
the angle dependence with $\lambda =1$ emerges. The Coulomb contributions to
(C.22) turn out to have qualitatively the same anisotropy as (C.31) yields.

In the case of strong intraatomic Coulomb interaction we may retain in the
Hamiltonian (C.21) only two lower ME terms, $\Gamma =\{{SL\}}$ and $\Gamma
=\{{S}^{\prime }{L}^{\prime }\}$. Then the intraatomic Hamiltonian yields a
constant energy shift only and may be omitted. The transfer Hamiltonian may
be represented through the spin and orbital momentum operators corresponding
to the term $\Gamma _n$. To this end, we pick out from X-operators one Fermi
operator with the use of (A.21), (A.17), (A.11). Substituting (C.31) and
transforming products of Clebsh-Cordan coefficients we get 
$$
\mathcal{H} =(4\pi )^{1/2}\left( G_{\bar \Gamma _n}^{\Gamma _{n+1}}\right)
^4\sum_{\nu _im_i\sigma _i\sigma }\sum_{k_1k_2\lambda pqk}\bar \beta _{\nu
_1\nu _2}(ll\lambda )[k_1][k_2]([k][p])^{1/2}  
\eqno{(\rm C.34)}
$$
$$
\times [L^{\prime }]^2[L]^{-1}\left\{ 
\begin{array}{ccc}
L & L & k_1 \\ 
l & l & k_2
\end{array}
\right\} \left\{ 
\begin{array}{ccc}
k_1 & l & l \\ 
k_2 & l & l \\ 
k & p & \lambda 
\end{array}
\right\} (-1)^qC_{l0,\lambda 0}^{l0}C_{lm_2,pq}^{lm_1}  
$$
$$
\times a_{\nu _1lm_1\sigma _1}^{+}P_{\nu _1\sigma _1\sigma }\left[
Y^{(\lambda )}({\hat \rho })\times \left[ L_{\lambda _1}^{(k_1)}\times
L_{\lambda _2}^{(k_2)}\right] ^{(\lambda )}\right] _{-q}^{(p)}P_{\nu
_2\sigma \sigma _2}a_{\nu _2lm_2\sigma _2}  
$$
where 
\[
\hat P_\nu =\frac 1{2[S]}([S^{\prime }]+(-1)^{S-S^{\prime }+1/2}2(\mathbf{S}%
_\nu \mbox{\boldmath$\sigma $}))
\]
Here $\sigma $ are the Pauli matrices, $Y^{(\lambda )}$ is the irreducible
tensor with the components $Y_{\lambda \mu }$ , the vector product of
irreducible tensor operators is defined by (B.14).

\chapter{Interatomic electrostatic interaction and derivation of the
Heisenberg Hamiltonian}

In this Appendix we treat, starting from the general many- electron of a
crystal (C.1), the ``direct'' Coulomb and exchange interaction between two
degenerate atomic shells at different sites. The consideration of this
mechanism enables one to establish general features of Hamiltonian of
magnetic ion systems with unquenched orbital momenta.

Putting in (C.7) $\nu _1l_1=\nu _3l_3$, $\nu _2l_2=\nu _4l_4$, making the
expansion (C.29), transforming the product of Clebsh-Gordan coefficients and
using the multiplication formula for spherical harmonics 
$$
Y_{\lambda _1\mu _1}(\hat \rho )Y_{\lambda _2\mu _2}(\hat \rho
)=\sum_{\lambda \mu }\tilde C_{\lambda _1\mu _1,\lambda _2\mu _2}^{\lambda
\mu }Y_{\lambda \mu }(\hat \rho )  
\eqno{(\rm D.1)}
$$
we obtain, similar to (C.10), the Hamiltonian of the intersite Coulomb
interaction [660] 
$$
\mathcal{H}_{\mathrm{coul}}(\nu _1\nu _2) =(4\pi )^{1/2}\sum_{pb\lambda
}\sum_{\lambda _1\lambda _2\eta _1\eta _2}F^{(p)}(l_1l_2\eta _1\eta
_2\lambda _1\lambda _2)[b][\lambda _1][\lambda _2][l_2]^{1/2}  
\eqno{(\rm D.2)} 
$$
$$
\times (-1)^{\lambda _1}C(l_1pl_1,l_2\lambda _1\eta _1,\eta _2p\eta
_1,l_2\lambda _2\eta _2,\lambda _1\lambda _2\lambda )\left\{ 
\begin{array}{lll}
l_1 & \lambda _1 & \eta _1 \\ 
l_2 & \lambda _2 & \eta _2 \\ 
l & \lambda  & p
\end{array}
\right\} 
$$
$$
\times \left( [W_1^{(0p)}\times W_2^{(ab)}]^{(\lambda )}Y^{(\lambda
)}(\hat \rho )\right)   
$$
where 
$$
F^{(p)}(l_1l_2\eta _1\eta _2\lambda _1\lambda _2)=e^2\int\limits_0^\infty
r_1^2dr_1r_2^2dr_2R_{l_1\eta _1\lambda _1}^{*}(r_2,\rho )\frac{r_{<}^p}{%
r_{>}^{p+1}}R_{l_1\eta _1\lambda _1}(r_2,\rho )R_{l_1}^2(r_1)  
\eqno{(\rm D.3)}
$$
is a generalized Slater integral for two-site interaction, and the notation
is used for brevity 
$$
C(a_1,b_1,c_1,a_2,b_2,c_2,\ldots )=\prod\limits_iC_{a_i0,b_i0}^{c_i} 
\eqno{(\rm D.4)}
$$
One can see from (C.9) that $p$, $\lambda _1+\lambda _2$ and $\lambda $ are
even ($b$ is also even because of time-reversal invariance), $p\leq 2l_1$, $%
b\leq 2l_2$, $\lambda \leq 2(l_1+l_2)$. Further we pass to the many-electron
representation. According to (B.29), for a given $LS$-term we have to
substitute 
$$
W_i^{(ok)}=\left( [S_i][L_i]\right) ^{-1/2}\langle S_iL_i\Vert
W_i^{(ok)}\Vert S_iL_i\rangle L_i^{(k)}  
\eqno{(\rm D.5)}
$$
with the reduced matrix elements being given by (B.28). The transition to
usual vectors is performed with the use of equation (B.13) and factorization
of spherical harmonics according to (D.1). For example, 
$$
\left( L_1^{(2)}L_2^{(2)}\right)  =20\sqrt{\frac 53}%
\prod_{i=1,2}[L_i]^{1/2}(2L_i-1)^{-1}(2L_i+3)^{-1}  
\eqno{(\rm D.6)}
$$
$$
\times \left\{ \frac 13L_1(L_1+1)L_2(L_2+1)-(\mathbf{L}_1\mathbf{L}_2%
\mathbf{)}^2\mathbf{-}\frac 12\mathbf{(L}_1\mathbf{L}_2)\right\}   
$$
$$
\left( L_1^{(2)}Y^{(2)}(\hat \rho )\right) =5(2\pi
)^{-1/2}2L_i-1)^{-1}(2L_i+3)^{-1}\left\{ \frac 13L_1(L_1+1)-\frac{(L_1\rho
)^2}{\rho ^2}\right\}   
\eqno{(\rm D.7)}
$$
In the classical limit ($L_1$, $L_2\gg 1$) we obtain the expansion 
$$
\mathcal{H}_{\mathrm{coul}}(\nu _1\nu _2)=\sum_{\alpha \beta \gamma
}Q_{\alpha \beta \gamma }\frac{(L_1\rho )^\alpha (L_2\rho )^\beta
(L_1L_2)^\gamma }{\rho ^{\alpha +\beta }}  
\eqno{(\rm D.8)}
$$
where the coefficients $Q$ are linear combinations of the Slater integrals
(D.3), the powers of $L_1$, $L_2$ and $\hat \rho $ in each summand are even
and $\alpha +\gamma \leq 2l_1$, $\beta +\gamma \leq 2l_2$ . Anisotropic
terms, which depend on orientation of only one orbital momentum in the
lattice, are present. Vector products do not appear since they are
transformed into scalar ones:
$$
\left( \lbrack \mathbf{L}_1\times \mathbf{L}_2\mathbf{]\rho }\right) ^2%
\mathbf{=}\det \left( 
\begin{array}{lll}
L_1(L_1+1) & (L_1L_2) & (L_1\rho ) \\ 
(L_1L_2) & L_2(L_2+1) & (L_2\rho ) \\ 
(L_1\rho ) & (L_2\rho ) & \rho ^2
\end{array}
\right)  
\eqno{(\rm D.9)}
$$
For concrete electron configurations, the total number of terms in the
series (D.7) is small. So, maximum powers of $L_1$ and $L_2$ do not exceed 2
for p-electrons and 4 for $d$-electrons.

At considering the intersite direct exchange interaction [660, 666, 667] ($%
\nu _1l_1=\nu _4l_4$ , $\nu _2l_2=\nu _3l_3$), which is small in the overlap
of atomic wave functions at different sites, one has to take into account
corrections owing to non-orthogonality since they have the same (second)
order of smallness. Calculating the integral (C.6) for the functions (C.3),
and expanding the wavefunctions, that arise, about the site $\nu $ with the
use of (C.29), we obtain the intersite exchange Hamiltonian [660]
$$
\mathcal{H}_{\mathrm{exch}}(\nu _1\nu _2) =-\frac 12\sum_{k_1k_2\lambda
\kappa }(4\pi )^{1/2}[\kappa ]^{3/2}(-1)^\kappa I(l_1l_2k_1k_2\lambda ) 
\eqno{(\rm D.10)} 
$$
$$
\times \left( [W_1^{(\kappa k_1)}\times W_2^{(\kappa k_2)}]^{(0\lambda
)}Y^{(\lambda )}(\hat \rho _{\nu _1\nu _2})\right)  
$$
where effective exchange parameters are given by
$$
I(l_1l_2k_1k_2\lambda ) =\frac 12[\sum_{p\eta _1\eta _2\lambda _1\lambda
_2}\{G^{(p)}(l_1l_2\eta _1\eta _2\lambda _1\lambda _2)-\Re
[G^{(p)}(l_1l_2\eta _1\eta _2\lambda _2)Z(l_2\eta _2\lambda _2)]  
\eqno{(\rm D.11)}
$$
$$
+\frac 12G^{(p)}(l_1\eta _1\eta _2)Z(l_1\eta _1\lambda _1)Z^{*}(l_2\eta
_2\lambda _2)\}  
$$
$$
\times (-1)^{l_1-l_2}[k_1][k_2][\lambda _1][\lambda _2]\left( \frac{%
[l_2][\eta _1]}{[\lambda ]}\right) ^{1/2}  
$$
$$
\times C(\eta _1pl_1,l_1p\eta _2,l_2\lambda _2\eta _2,l_2\lambda _1\eta
_1,\lambda _1\lambda _2\lambda )  
$$
$$
\times \left\{ 
\begin{array}{lll}
\eta _1 & \eta _2 & k_1 \\ 
l_1 & l_1 & p
\end{array}
\right\} \left\{ 
\begin{array}{lll}
l_2 & \lambda _1 & \eta _1 \\ 
l_2 & \lambda _2 & \eta _2 \\ 
k_2 & \lambda  & k_1
\end{array}
\right\}   
$$
$$
+\frac 14\sum_{p\lambda _i\Lambda _i\eta _ia}F^{(p)}(l_1l_2\eta _1\eta
_2\lambda _2\lambda _3\lambda _4)Z(l_1\eta _1\lambda _1)Z(l_2\eta _3\lambda
_3)  
$$
$$
\times (-1)^{\eta _3}\prod_{i=1}^3[\lambda _i][k_1][k_2][a]\left( \frac{%
[\Lambda _1][\Lambda _2][\eta _1][\eta _3]}{[\lambda ]}\right) ^{1/2} 
$$
$$
\times C(l_1\lambda _1\eta _1,\eta _3pl_1,\eta _1\lambda _4\eta
_4,l_2\lambda _3\eta _3,\eta _2p\eta _4,l_2\lambda _2\eta _2,\lambda
_1\lambda _4\Lambda _1,\Lambda _1\lambda _3\Lambda _2,\Lambda _2\lambda
_2\lambda )  
$$
$$
\times \left\{ 
\begin{array}{lll}
l_1 & \eta _4 & \Lambda _1 \\ 
\lambda _4 & \lambda _1 & \eta _1
\end{array}
\right\} \left\{ 
\begin{array}{lll}
l_2 & a & \Lambda _2 \\ 
\Lambda _1 & \gamma _3 & \eta _3
\end{array}
\right\} \left\{ 
\begin{array}{lll}
l_1 & \eta _3 & \Lambda _1 \\ 
l_1 & \Lambda _1 & \eta _4 \\ 
k & a & \eta _2
\end{array}
\right\} \left\{ 
\begin{array}{lll}
l_2 & \lambda _2 & \eta _2 \\ 
l_2 & \Lambda _2 & a \\ 
k & \lambda  & k_1
\end{array}
\right\} ] 
$$
$$
+\frac 12[\{\nu _1l_1k_1\} \leftrightarrow \{\nu _2l_2k_2\}]^{*}
$$
with $k_1+k_2$ and $\lambda $ being even, $k_1\leq 2l_1$, $k_2\leq 2l_2$ , $%
\lambda \leq 2(l_1+l_2)$, $\kappa \leq 1$. Here we have introduced the
two-site Slater integrals 
$$
G^{(p)}(l_1l_2\eta _1\eta _2\lambda _1\lambda _2)=e^2\int
dr_1r_1^2dr_2r_2^2R_{l_1}(r_1)R_{l_2\eta _2\lambda _2}^{*}(r_2,\rho )\frac{%
r_{<}^p}{r_{>}^{p+1}}R_{l_2\eta _1\lambda _2}(r_1,\rho )R_{l_1}(r_2) 
\eqno{(\rm D.12)}
$$
$$
G^{(p)}(l_1l_2\eta _1\eta _2\lambda _1)=e^2\int
dr_1r_1^2dr_2r_2^2R_{l_1}(r_1)R_{\eta _2}(r_2)\frac{r_{<}^p}{r_{>}^{p+1}}%
R_{l_2\eta _1\lambda _1}(r_1,\rho )R_{l_1}(r_2)  
\eqno{(\rm D.13)}
$$
$$
G^{(p)}(l\eta _1\eta _2)=e^2\int dr_1r_1^2dr_2r_2^2R_{l_1}(r_1)R_{\eta
_2}(r_2)\frac{r_{<}^p}{r_{>}^{p+1}}R_{\eta _1}(r_1)R_{l_1}(r_2)  
\eqno{(\rm D.14)}
$$
$$
F^{(p)}(l_1l_2\eta _1\eta _2\eta _3\eta _4\lambda _2\lambda _3\lambda
_4)=e^2\int dr_1r_1^2dr_2r_2^2R_{l_1}(r_1)R_{l_2\eta _2\lambda
_2}^{*}(r_2,\rho )\frac{r_{<}^p}{r_{>}^{p+1}}R_{l_3\eta _3\lambda
_3}(r_1,\rho )R_{\eta _1\eta _4\lambda _4}(r_2,\rho )  
\eqno{(\rm D.15)}
$$
and the non-orthogonality integrals 
$$
Z(l\eta \lambda )=\int\limits_0^\infty r^2drR_l(r)R_{l\eta \lambda }(r,\rho
)  
\eqno{(\rm D.16)}
$$
Putting $k_1=k_2=\lambda =0$ we obtain, similar to (C.16), the spin
Hamiltonian 
$$
\sum_\kappa \mathcal{H}_{\mathrm{exch}}^{(\kappa 0)}(\nu _1\nu _2)=-\frac
12I(l_1l_2000)[n_1n_2+4(\mathbf{S}_1\mathbf{S}_2)]  
\eqno{(\rm D.17)}
$$
Putting $k_1=k_2=1$, $\kappa =0$ we obtain the orbital exchange Hamiltonian
$$
\mathcal{H}_{\mathrm{exch}}^{(01)}(\nu _1\nu _2) =-\frac
12(l_1(l_1+1)l_2(l_2+1))^{-1/2}  
\eqno{(\rm D.18)} 
$$
$$
\times \{I(l_1l_2110)(\mathbf{L}_1\mathbf{L}_2)+(8\pi
)^{-1/2}I(l_1l_2112)[\frac 13(\mathbf{L}_1\mathbf{L}_2)-\frac{(L_1\rho
)(L_2\rho )}{\rho ^2}]\} 
$$
For all the other contributions, we have to pass to the operators $%
S^{(\kappa )}$ and $L^{(k)}$ for a concrete term with the use of (B.29).
Then we obtain the expansion
$$
\mathcal{H}_{\mathrm{exch}}(\nu _1\nu _2)=-\sum_{\alpha \beta \gamma \kappa
}I_{\alpha \beta \gamma \kappa }\frac{\mathbf{(L}_1\mathbf{\rho )}^\alpha 
(\mathbf{L}_2\mbox{\boldmath$\rho $})^\beta (\mathbf{L}_1\mathbf{L}_2)
^\gamma (\mathbf{S}_1\mathbf{S}_2)^\kappa }{\rho ^{\alpha +\beta }}
\eqno{(\rm D.19)}
$$
with even $\alpha +\beta $, $\alpha +\gamma \leq 2l_1$, $\beta +\gamma \leq
2l_2$, $\kappa =0,1$. It should be stressed that the powers of orbital
momentum operators in (D.7), (D.19) are restricted in the microscopic
treatment not only by $2L$ (as follows from kinematic relations), but also
by $2l$. Higher-order terms in spin operators, e.g., biquadratic exchange
ones, do not occur in (D.19) since the electron spin s equals to 1/2. Such
terms may be obtained from the higher-order corrections owing to non-
orthogonality [656,668]. However, they are considerably smaller in the
overlap of wave functions at different sites.

The above treatment of the two-site problem is, generally speaking,
insufficient for obtaining the exchange Hamiltonian of a crystal. However,
neglect of multi-site terms is justified in the nearest-neighbor
approximation. Corrections to the two-centre approximation are especially
important if some nearest neighbour form equilateral triangles (e.g., for
the fcc and hcp lattices).

It should be noted that the electrostatic interaction may be investigated by
direct using the representation of ME wave functions [655]. We obtain 
$$
\mathcal{H}(\nu _1\nu _2) =\int \prod_{ij}dx_idx_j^{\prime }\Psi _{\nu
_1\Gamma _1}^{*}(\{x_i\})\Psi _{\nu _2\Gamma _2}^{*}(\{x_i^{\prime }\}) 
\eqno{(\rm D.20)} 
$$
$$
\times \sum_{ij}\frac{e^2}{|\mathbf{r}_i-\mathbf{r}_j|}
(1-P_{ij})\Psi _{\nu _1\Gamma _3}(\{x_i\})\Psi _{\nu
_2\Gamma _4}(\{x_i^{\prime }\})X_{\nu _1}(\Gamma _1,\Gamma _3)X_{\nu
_2}(\Gamma _2,\Gamma _4)  
$$
where $P_{ij}$ is the permutation operator. Such an approach does not yield
immediately the results (D.17), (D.18) and requires a summation of
fractional parentage coefficients. However, this approach permits to take
into account the dependence of radial wavefunctions on ME quantum numbers
which occurs, e.g., in the Hartree-Fock approximation and is therefore more
general.

In the case of strong (in comparison with crystal field) spin-orbital
interaction the orbital and spin momenta are coupled into the total momentum 
$J$ (the Russel-Saunders scheme, appropriate for rare-earth ions). In the
Coulomb term, the spin-orbital coupling results in the replacement $%
L_i\rightarrow J_i$ according to (B.18). For the exchange term, one has to
make some additional manipulations. Using (B.16) and performing summation of
Clebsh-Gordan coefficients, we derive the result (D.10) with the replacement 
$$
\lbrack W_1^{(\kappa k_1)}W_2^{(\kappa k_2)}]^{(0\lambda )} \rightarrow
[\kappa ]^{-1/2}\sum_{P_1P_2}[J_1^{(P_1)}\times J_2^{(P_2)}]^{(\lambda
)}\left\{ 
\begin{array}{lll}
P_1 & P_2 & \lambda  \\ 
k_1 & k_2 & \kappa 
\end{array}
\right\}   
\eqno{(\rm D.21)} 
$$
$$
\times \prod_{i=1,2}[p_i]\left( [S_i][L_i][J_i]\right) ^{1/2}\left\{ 
\begin{array}{ccc}
J_i & S_i & L_i \\ 
J_i & S_i & L_i \\ 
P_i & \kappa  & k_i
\end{array}
\right\} \langle S_iL_i\Vert W_i^{(\kappa k_i)}\Vert S_iL_i\rangle  
$$
Then the expansion of the exchange Hamiltonian reads 
$$
\mathcal{H}_{\mathrm{exch}}(\nu _1\nu _2)=-\sum_{\alpha \beta \gamma
}I_{\alpha \beta \gamma }\frac{(J_1\rho )^\alpha (J_2\rho )^\beta
(J_1J_2)^\gamma }{\rho ^{\alpha +\beta }}  
\eqno{(\rm D.22)}
$$
with $\alpha +\gamma \leq 2l_1+1$, $\beta +\gamma \leq 2l_2+1$, $\alpha
+\beta $ being odd.

In the limit of strong spin-orbital coupling under consideration,
antisymmetric Dzyaloshinsky-Moriya-type exchange interaction of the form $%
K_{12}[\mathbf{J}_1\times \mathbf{J}_2]$ may be obtained with account of
crystal-field effects only. Unlike the case of weak spin-orbital coupling
considered by Moriya [669], where the antisymmetric exchange is determined
by matrix elements of orbital momenta, the components of the pseudovector $%
K_{ij}$ are given by matrix elements of electrostatic interaction in the
local coordinate system [666].

Besides the ``potential'' exchange (D.10), we have to consider the
``kinetic'' exchange interaction. We consider the degenerate Hubbard model
(C.21) with large Coulomb interaction. Then in the ground state the electron
states at all sites correspond to the same $SL$-term, so that perturbation
theory is applicable. The kinetic exchange occurs in the second order in the
electron transfer 
$$
\widetilde{\mathcal{H}} =2\sum_{\nu _im_i\sigma _i}\beta _{\nu _1\nu
_2}(lm_1,lm_2)\beta _{\nu _2\nu _1}(lm_3,lm_4)\sum_{\Gamma _n^{(i)}\Gamma
_{n-1}\Gamma _{n+1}}(2E_{\Gamma _n}-E_{\Gamma _{n+1}}-E_{\Gamma _{n-1}})^{-1}
$$
$$
\times \langle \Gamma _{n+1}|a_{\nu _1lm_1\sigma _1}^{+}|\Gamma
_{n-1}\rangle \langle \Gamma _{n-1}|a_{\nu _1lm_4\sigma _2}|\Gamma
_n^{\prime }\rangle \langle \Gamma _n^{\prime \prime }|a_{\nu _2lm_2\sigma
_1}|\Gamma _{n+1}\rangle   
\eqno{(\rm D.26)} 
$$
$$
\times \langle \Gamma _{n+1}|a_{\nu _2lm_3\sigma _2}^{+}|\Gamma
_n^{\prime \prime \prime }\rangle X_{\nu _1}(\Gamma _n,\Gamma _n^{\prime
})X_{\nu _2}(\Gamma _n^{\prime \prime },\Gamma _n^{\prime \prime \prime }) 
$$
where sum over $\Gamma ^{(i)}$ stands for the sum over momentum projections
of the term $\Gamma _n=\{SL{\mu }M\}$. In particular, for the s-band we
obtain the standard result for the Anderson's kinetic exchange (4.8).

Generally speaking, in the problem of kinetic exchange for degenerate bands
we cannot use the double irreducible tensor operators (B.21) since the
denominators in (D.26) depend on ME quantum numbers $S$, $L$ of virtual
states. This may be, however, made if we take into account the dependence of 
$E_\Gamma $ on the number of electrons only (C.20) to obtain 
$$
\widetilde{\mathcal{H}}=\sum_{\nu _1\nu _2}\left\{ \frac{\bar \beta _{\nu
_1\nu _2}^2(ll0)}{F^{(0)}(ll)}\right\} \{n_1n_2+4(\mathbf{S}_1\mathbf{S}%
_2)-[l](n_1+n_2)\}  
\eqno{(\rm D.27)}
$$
The interaction (D.27) is antiferromagnetic and dominates as a rule over the
potential exchange interaction (D.10). It should be noted that the kinetic
exchange interaction survives even in the limit $F^{(2)}=U\rightarrow \infty 
$ owing to non-orthogonality corrections (the second term in (C.25)).

In the general case we have to apply directly the ME operator approach.
Substituting the expressions for matrix elements of Fermi operators (A.30),
taking into account the m-dependence of transfer integrals (C.31) and
performing summation of the Clebsh-Gordan coefficients we obtain [660] 
$$
\widetilde{\mathcal{H}} =(4\pi )^{1/2}n(n+1)
\mathop{\sum_{\nu _ik_i\lambda
_i\lambda }}_{\{S_{n\pm 1},L_{n\pm 1}\}}
\left( G_{\Gamma
_{n-1}}^{\Gamma _n}G_{\Gamma _n}^{\Gamma _{n+1}}\right) ^2\frac{\bar \beta
_{\nu _1\nu _2}(ll\lambda _1)\bar \beta _{\nu _1\nu _2}(ll\lambda _2)}{%
E_{\Gamma _{n+1}}+E_{\Gamma _{n-1}}-2E_{\Gamma _n}}  
\eqno{(\rm D.28)} 
$$
$$
\times C(l\lambda _1l,l\lambda _2l,\lambda _1\lambda _2\lambda
)[l][\lambda _1][\lambda _2][k_1][k_2][L_{n+1}][\lambda
]^{-1/2}(-1)^{L_{n+1}-L_{n-1}}  
$$
$$
\times \left\{ 
\begin{array}{ccc}
L & L & k_1 \\ 
l & l & L_{n-1}
\end{array}
\right\} \left\{ 
\begin{array}{ccc}
L & L & k_2 \\ 
l & l & L_{n+1}
\end{array}
\right\} \left\{ 
\begin{array}{ccc}
k_1 & k_2 & \lambda  \\ 
l & l & \lambda _1 \\ 
l & l & l_2
\end{array}
\right\} ([L_1^{(k_1)}\times L_2^{(k_2)}]^{(\lambda )}Y^{(\lambda )}(\hat
\rho _{1,2}))  
$$
$$
\times (-1+4(-1)^{S_{n+1}-S_{n-1}}([S_{n+1}][S_{n-1}])^{-1}\mathbf{(S}_1%
\mathbf{S}_2\mathbf{)})  
$$
with $\lambda _1$, $\lambda _2$, $\lambda $ and $k_1+k_2$ being even, $%
\lambda _1$, $\lambda _2\leq 2l$, $k_1$, $k_2$, $k\leq 2l$, $2L$. Passing to
the usual vectors, we obtain the multipole expansion in the same form as
(D.19). The Hamiltonian (D.28) contains only bilinear terms in spin
operators. The biquadratic exchange may be obtained in the fourth order of
perturbation theory [661,668-670] which also corressponds to higher order
corrections in the overlap parameter.

The sign of contributions of virtual configurations $\Gamma _{n-1}$ and $%
\Gamma _{n+1}$ to the effective exchange parameter ($k_1=k_2=0$) is
determined by their spins. The coupling is antiferromagnetic if $%
S_{n+1}=S_{n-1}=S_n\pm \frac 12$ and ferromagnetic if $S_{n+1}-S_{n-1}=\pm 1$.

Similar rules for the coupling between orbital momenta ($k_1=k_2=1$, $%
\lambda _1=\lambda _2=0$) are obtained after substituting explicit values of 
$6j$-coefficients in (D.28). The orbital exchange interaction is
``antiferromagnetic'' if both the differences 
$$
\Delta _{\pm }=L(L+1)+l(l+1)-L_{n\pm 1}(L_{n\pm 1}+1)  
\eqno{(\rm D.29)}
$$
have the same sign and ``ferromagnetic'' in the opposite case.

In real situations, the form of exchange Hamiltonians is strongly modified
by crystal field (CF) which quenches, at leat partially, orbital momenta.
Even in the case of intermediate CF, one has to consider, instead of the
many-electron SL-terms, the corresponding irreducible representations of
point groups. Besides that, the overlap between partially occupied 
$d(f)$-shells and, consequently, the direct exchange is as a rule small, so one has
to take into account more complicated ``superexchange'' mechanisms via
non-magnetic atoms [661]. The kinetic exchange may be treated as a
particular case of the superexchange interaction (the indirect interaction
via valence band). The case of a narrow-band metal or semiconductor with
non-integer band filling where exchange interaction is mediated by current
carriers (the ``double exchange'' situation) [668] may be described within
the Hubbard and $s-d$ exchange models with strong correlations. The
corresponding Hamiltonians (D.28), (I.10) describe the interaction of
electrons with spin and orbital degrees of freedom. In this case the
exchange interaction is not reduced to a Heisenbergian form.


\chapter{Spin waves in Heisenberg magnets and the Green's function method}

The exponential behaviour of magnetization at low temperatures in the
mean-field approximation (4.20) contradicts experimental data on both
insulator and metallic ferromagnets. At $T\ll T_C$ the excitation spectrum
and thermodynamics of the Heisenberg model may be investigated in detail.
This is achieved by passing from spin operators $S^z,S^{\pm }=S^x\pm iS^y$
to Bose operators which correspond to weakly interacting excitations in the
ferromagnet -- spin waves (magnons). The simplest way to do this is using
the Holstein-Primakoff representation
$$
S^z =S-b^{\dagger }b,\qquad S^{\dagger }=(2S)^{1/2}(1-\frac
1{2S}b^{\dagger }b)^{1/2}b,
\eqno{(\rm E.1)}
$$
\[
S^{-}=(2S)^{1/2}b^{\dagger }(1-\frac 1{2S}b^{\dagger }b)
\]
In fact, this representation is well defined only on the basis of physical
states where spin deviations (magnon occupation numbers) on a site do not
exceed $2S.$ Unlike the representation (E.1), the Dyson- Maleev
representation of ideal bosons
$$
S^z =S-b^{\dagger }b,
\eqno{(\rm E.2)}
$$
\[
S^{\dagger }=(2S)^{1/2}(1-\frac 1{2S}b^{\dagger }b)^{1/2}b,\qquad
S^{-}=(2S)^{1/2}b^{\dagger }
\]
does not contain irrational functions of Bose operators, but is
not-Hermitian. However, this lead to errors which are exponentially small at
low temperatures.

Performing expansion of (E.1) in the quasiclassical parameter $1/2S$ we
reduce the Heisenberg Hamiltonian (4.9) to the form
\[
\mathcal{H}=\sum_{\mathbf{q}}\omega _{\mathbf{q}}b_{\mathbf{q}}^{\dagger }b_{%
\mathbf{q}}+\frac 12\sum_{\mathbf{pqr}}(J_{\mathbf{q}}+J_{\mathbf{p}}-2J_{%
\mathbf{q-p}})b_{\mathbf{q}}^{\dagger }b_{\mathbf{r}}^{\dagger }b_{\mathbf{p}%
}b_{\mathbf{q+r-p}}+...
\]
$$
\omega _{\mathbf{q}}=2S(J_0-J_{\mathbf{q}})
\eqno{(\rm E.3)}
$$
The first term of the Hamiltonian (E.3) describes the system of magnons with
the frequencies $\omega _{\mathbf{q}}$, and the second term the dynamical
interaction (magnon-magnon scattering). At small $q$ for a cubic lattice we
have
$$
\omega _{\mathbf{q}}=Dq^2,\qquad D=\frac 13J_0S\approx T_c/2S(S+1)
\eqno{(\rm E.4)}
$$
The quantity $D$ is called the spin stiffness constant. The temperature
dependent correction to the spin-wave spectrum owing to magnon-magnon
interaction is obtained by decoupling the second term of (E.3) with the use
of the Wick theore
$$
\delta \omega _{\mathbf{q}}=2\sum_{\mathbf{p}}(J_{\mathbf{q}}+J_{\mathbf{p}%
}-J_{\mathbf{p-q}}-J_0)N_{\mathbf{q}},N_{\mathbf{q}}=\langle b_{%
\mathbf{q}}^{\dagger }b_{\mathbf{q}}\rangle
\eqno{(\rm E.5)}
$$
and is proportional to $T^{5/2}$. The temperature dependence of
magnetization is given by the Bloch law
$$
\delta \langle S^z\rangle =-\sum_{\mathbf{p}}N_{\mathbf{p}}=-\frac{v_0}{2\pi
^2}\int_0^\infty \frac{q^2dq}{\exp (Dq^2/T)-1}=-\zeta \left( \frac 32\right)
\frac{v_0}{8\pi ^{3/2}}\left( \frac TD\right) ^{3/2}
\eqno{(\rm E.6)}
$$
where $v_0$ is the lattice cell volume, $\zeta (x)$ is the Riemann
zeta-function.

Now we consider the case of an antiferromagnet. We treat a general case of
the spiral structure, characterized by the wavevector $\mathbf{Q}$. The
corresponding classical spin configuration is written as
$$
\langle S_i^x\rangle =S\cos \mathbf{Q}R_{i},\qquad \langle S_i^y\rangle
=S\sin \mathbf{Q}R_{i},\qquad \langle S_i^z\rangle =0
\eqno{(\rm E.7)}
$$
In particular, for the usual ``chess'' ordering in the simple cubic lattice
one has $\mathbf{Q}=(\pi \pi \pi )$, and in the case of ferromagnetic planes
with alternating magnetization direction $\mathbf{Q}=(00\pi )$. It is
convenient to introduce the local coordinate system where spins at each site
are directed along the new local $z$-axis:
$$
S_i^x \rightarrow S_i^z\cos \mathbf{QR}_i-S_i^y\sin \mathbf{QR}_i,
\eqno{(\rm E.8)}
$$
\[
S_i^y\rightarrow S_i^z\sin \mathbf{QR}_i+S_i^x\cos \mathbf{QR}_i,\qquad
S_i^z\rightarrow -S_i^x
\]
Passing from the spin operators in the local coordinate system to the
operators of spin deviations with the use of (E.1) we represent the
Heisenberg Hamiltonian as
$$
\mathcal{H} =\sum_{\mathbf{q}}[C_{\mathbf{q}}b_{\mathbf{q}}^{\dagger }b_{%
\mathbf{q}}+\frac 12D_{\mathbf{q}}(b_{\mathbf{q}}b_{\mathbf{-q}}+b_{\mathbf{q%
}}^{\dagger }b_{\mathbf{-q}}^{\dagger })]+...
\eqno{(\rm E.9)}
$$
\[
C_{\mathbf{q}}=S[2J_{\mathbf{Q}}-\frac 12(J_{\mathbf{Q+q}}
+J_{\mathbf{Q-q}})-J_{\mathbf{q}}],
\]
\[
D_{\mathbf{q}}=S[\frac 12(J_{\mathbf{Q+q}}+J_{\mathbf{Q-q}})-J_{\mathbf{q}}]
\]
The Hamiltonian (E.9) can be diagonalized by the Bogoliubov-Tyablikov
transformation
$$
b_{\mathbf{q}} =u_{\mathbf{q}}\beta _{\mathbf{q}}-v_{\mathbf{q}}\beta _{%
\mathbf{-q}}^{\dagger },
\eqno{(\rm E.10)}
$$
\[
u_{\mathbf{q}}^2=\frac 12\left( \frac{C_{\mathbf{q}}}{\omega _{\mathbf{q}}}%
+1\right) ,v_{\mathbf{q}}^2=\frac 12\left( \frac{C_{\mathbf{q}}}{\omega _{%
\mathbf{q}}}-1\right)
\]
to obtain
$$
\mathcal{H}=\mathrm{const}+\sum_{\mathbf{q}}\omega _{\mathbf{q}}\beta _{%
\mathbf{q}}^{\dagger }\beta _{\mathbf{q}}+...
\eqno{(\rm E.11)}
$$
where the spin-wave spectrum reads
$$
\omega _{\mathbf{q}}=(C_{\mathbf{q}}^2-D_{\mathbf{q}}^2)^{1/2}=2S\{[J_{%
\mathbf{Q}}-\frac 12(J_{\mathbf{Q}+\mathbf{q}}+J_{\mathbf{Q}-\mathbf{q}%
})][J_{\mathbf{Q}}-J_{\mathbf{q}}]\}^{1/2}
\eqno{(\rm E.12)}
$$
The spin-wave frequency tends to zero at both $q\rightarrow 0$ and $\mathbf{%
q\rightarrow Q},$ the $q$-dependence being linear, unlike the case of a
ferromagnet (we do not discuss here magnetic anisotropy which results in a
disturbation of spin-wave spectrum at small $q$, $|\mathbf{q-Q}|$
[15,16]). Using the Wick theorem for the quartic terms, which are neglected
in the Hamiltonian (E.9), we may obtain corrections to $C_{\mathbf{q}}$
and $D_{\mathbf{q}}$ owing to magnon-magnon interaction
$$
\delta C_{\mathbf{q}} =-\frac 12\sum_{\mathbf{p}}(4J_{\mathbf{Q}}+2J_{%
\mathbf{Q}+\mathbf{q-p}}+2J_{\mathbf{Q-q-p}}-2J_{\mathbf{p}}-2J_{\mathbf{Q}}
\eqno{(\rm E.13)}
$$
\[
-J_{\mathbf{Q+q}}-J_{\mathbf{Q-q}}--2J_{\mathbf{Q+p}})\langle b_{\mathbf{p}%
}^{\dagger }b_{\mathbf{p}}\rangle -\sum_{\mathbf{p}}(J_{\mathbf{Q+p}}-J_{%
\mathbf{p}})\langle b_{\mathbf{p}}^{\dagger }b_{-\mathbf{p}}\rangle
\]
\[
\delta D_{\mathbf{q}}=-\frac 12\sum_{\mathbf{p}}(J_{\mathbf{Q+q}}+2J_{%
\mathbf{Q+p}}+J_{\mathbf{Q-q}}-2J_{\mathbf{q}}-2J_{\mathbf{p}})\langle b_{%
\mathbf{p}}^{\dagger }b_{\mathbf{p}}\rangle
\]
\[
+\sum_{\mathbf{p}}(J_{\mathbf{Q+p}}+J_{\mathbf{p}})\langle b_{\mathbf{p}%
}^{\dagger }b_{-\mathbf{p}}^{\dagger }\rangle
\]
The low-temperature behaviour of the sublattice magnetization is given by
$$
\overline{S}=\langle S^z\rangle =S-\langle b^{\dagger }b\rangle =S-\sum_{%
\mathbf{q}}\left[ u_{\mathbf{q}}^2N_{\mathbf{q}}+v_{\mathbf{q}}^2(1+N_{%
\mathbf{q}})\right]
\eqno{(\rm E.14)}
$$
so that
\[
\delta \overline{S}(T)\sim -\left( T/T_N\right) ^2
\]
The spin-wave approach enables one to obtain an interpolation description of
the Heisenberg ferromagnet in the whole temperature region. This was made by
Tyablikov [357] within the Green's function method. Define the
anticommutator (commutator) double-time retarded Green's function for
operators $A$ and $B$ by [671]
$$
\langle \langle A|B\rangle \rangle _E^{\pm }=\int_{-\infty
}^0dt e^{iEt}\langle [e^{i\mathcal{H}t}A e^{-i\mathcal{H}t},B]_{\pm
}\rangle ,\qquad \Im E>0
\eqno{(\rm E.15)}
$$
The Green's function (E.15) satisfies the equation of motion
$$
E\langle \langle A|B\rangle \rangle _E^{\pm }=\langle [A,B]_{\pm }\rangle
+\langle \langle [A,\mathcal{H}]|B\rangle \rangle _E^{\pm }
\eqno{(\rm E.16a)}
$$
or
$$
E\langle \langle A|B\rangle \rangle _E^{\pm }=\langle [A,B]_{\pm }\rangle
+\langle \langle A|[\mathcal{H},B\rangle \rangle _E^{\pm }
\eqno{(\rm E.16b)}
$$
and is thereby expressed in terms of more complicated Green's functions. As
one can see from (E.16), it is convenient to use the commutator Green's
function in the case of Bose-type operators $A,B$ (where the commutator $%
[A,B]$, which stands in the average, is a ``simpler'' operator), and the
anticommutator ones in the case of Fermi-type operators. In the cases of
free bosons and fermions, where the Hamiltonian has a diagonal form, the
equations (E.16) are closed and we find
$$
\langle \langle b_{\mathbf{q}}|b_{\mathbf{q}}^{\dagger }\rangle \rangle
_\omega ^{-}=\frac 1{\omega -\omega _{\mathbf{q}}},\qquad \langle \langle c_{%
\mathbf{k}}|c_{\mathbf{k}}^{\dagger }\rangle \rangle _E^{+}=\frac 1{E-E_{%
\mathbf{k}}}
\eqno{(\rm E.17)}
$$
Calculation of the Green's function permits to restore the corresponding
thermodynamical averages by using the spectral representation
$$
\langle BA\rangle =-\frac 1\pi \int_\infty ^\infty dE\frac 1{e^{E/T}\pm
1}\Im \langle \langle A|B\rangle \rangle _{E+i\delta }^{\pm }
\eqno{(\rm E.18)}
$$
In particular, we obtain from (E.17)
$$
\langle b_{\mathbf{q}}^{\dagger }b_{\mathbf{q}}\rangle =N_B(\omega _{\mathbf{%
q}}),\qquad \langle c_{\mathbf{k}}^{\dagger }c_{\mathbf{k}}\rangle =f(E_{%
\mathbf{k}})
\eqno{(\rm E.19)}
$$

In the general case, the infinite sequence of equations (E.16) may be
''decoupled'' by reducing higher-order Green's functions to simpler ones. In
the case of interacting quasiparticles the Green's function are expressed by
the Dyson equations
$$
\langle \langle b_{\mathbf{q}}|b_{\mathbf{q}}^{\dagger }\rangle \rangle
_\omega =[\omega -\omega _{\mathbf{q}}-\Pi _{\mathbf{q}}(\omega )]^{-1}
\eqno{(\rm E.20)}
$$
\[
\langle \langle c_{\mathbf{k}}|c_{\mathbf{k}}^{\dagger }\rangle \rangle
_E=[E-E_{\mathbf{k}}-\Sigma _{\mathbf{k}}(E)]^{-1}
\]
The real part of the self-energy $\Sigma $ (or of the polarization operator
$\Pi $) yields the energy shift, and the imaginary part determines the
quasiparticle damping. Provided that the interaction Hamiltonian contains a
small parameter, the method of equations of motion permits to construct the
perturbation expansion in a convenient form. In particular, applying both
equations (E.16) we obtain the expression for the self-energy
\[
\Sigma _{\mathbf{k}}(E)=\Lambda _{\mathbf{k}}+\langle \langle [c_{\mathbf{k}%
},\mathcal{H}_{int}]-\Lambda _{\mathbf{k}}c_{\mathbf{k}}|[\mathcal{H}%
_{int},c_{\mathbf{k}}^{\dagger }]-\Lambda _{\mathbf{k}}c_{\mathbf{k}%
}^{\dagger }\rangle \rangle _E^{irr},
\]
$$
\Lambda _{\mathbf{k}}=\langle \left\{ [c_{\mathbf{k}},\mathcal{H}%
_{int}],c_{\mathbf{k}}^{\dagger }\right\} \rangle
\eqno{(\rm E.21)}
$$
where the symbol `irr' means that the divergent contributions, containing
the denominators $(E-\varepsilon _{\mathbf{k}})^n$ , should be omitted at
further calculations of the irreducible Green's function (E.21). In this
book we apply the method of double-time retarded Green's functions to
various many-electron models describing highly-correlated d- and f-systems.
In transition metal theory, this technique turns out to be very useful since
we meet with operators which do not possess simple commutation relations, so
that standard diagram expansions [27] are inapplicable. Here belong
many-electron X-operators of the Fermi and Bose type (Appendix A) and spin
operators (Appendix B). Below we describe briefly the derivation of the
Tyablikov equation for magnetization in the Heisenberg model for spin $%
S=1/2$(the general case is discussed in Appendix F within framework of
the Hubbard X-operator approach). Write down the equation of motion for the
commutator transverse spin Green's function
\[
(\omega -H)\langle \langle S_{\mathbf{q}}^{+}|S_{\mathbf{-q}}^{-}\rangle
\rangle _\omega =2\langle S^z\rangle +2\sum_{\mathbf{p}}(J_{\mathbf{p-q}}-J_{%
\mathbf{p}})\langle \langle S_{\mathbf{p}}^zS_{\mathbf{q-p}}^{+}|S_{\mathbf{%
-q}}^{-}\rangle \rangle _\omega
\]
Performing the simplest decoupling at different lattice sites we derive
$$
\langle \langle S_{\mathbf{q}}^{+}|S_{\mathbf{-q}}^{-}\rangle \rangle
_\omega =\frac{2\langle S^z\rangle }{\omega -\omega _{\mathbf{q}}},\omega
_{\mathbf{q}}=2\langle S^z\rangle (J_0-J_{\mathbf{q}})+H
\eqno{(\rm E.22)}
$$
Presence of the factor $2\langle S^z\rangle $ in comparison with (E.17) is
connected with non-Bose commutation relation for spin operators. The result
(E.22) is valid for arbitrary $S.$ For $S=1/2$, using (E.18) and the
identity
$$
S_i^z=\frac 12-S_i^{-}S_i^{+}
\eqno{(\rm E.23)}
$$
we obtain the self-consistent equation for $\langle S^z\rangle $%
\[
\langle S^z\rangle =\frac 12[1+2\sum_{\mathbf{q}}N_B(\omega _{\mathbf{q}%
})]^{-1}
\]
which may be transformed to the form
$$
\langle S^z\rangle =\frac 12\left[ \sum_{\mathbf{q}}\coth \frac{2\langle
S^z\rangle (J_0-J_{\mathbf{q}})+H}T\right] ^{-1}
\eqno{(\rm E.24)}
$$

The equation (E.24) differs from the mean-field equation (4.14) by the
presence of dispersion ($q$-dependence) of excitation spectrum. It describes
satisfactorily thermodynamics of the Heisenberg model at both high and low
temperatures, although higher-order terms in the low-$T$ expansion do not
quite agree with the rigorous Dyson's results (in particular, the spurious $%
T $ -term in magnetization occurs). Numerous attempts to improve the
spin-wave region description by using more complicated decoupling procedures
(see [357]) resulted in fact in deterioration of the interpolation. The
values of ferromagnetic and paramagnetic Curie temperature in the Tyablikov
approximation for arbitrary $S$ value are given by
$$
T_c=\frac{S(S+1)}3\left( \sum_{\mathbf{q}}\frac 1{J_0-J_{\mathbf{q}}}\right)
^{-1}
\eqno{(\rm E.25)}
$$
$$
\theta =\frac{S(S+1)}3J_0
\eqno{(\rm E.26)}
$$
and are somewhat different (the mean-field result (4.18) is obtained from
(E.25) by averaging the denominator over $\mathbf{q}$, i.e. in the lowest
order in $1/z$). The corresponding expressions for an antiferromagnet are
obtained from (E.25), (E.26) by the replacement $J_0\rightarrow J_{\mathbf{q}%
}$ . To end this Appendix, we discuss one more method in the theory of
Heisenberg magnets [672]. To take into account kinematical interactions,
this approach introduces auxiliary pseudofermion operators c into the
Dyson-Maleev representation (E.2):
$$
S^z =S-b^{\dagger }b-(2S+1)c^{\dagger }c,\qquad S^{-}=(2S)^{1/2}b^{\dagger
}
\eqno{(\rm E.27)}
$$
\[
S^{\dagger }=(2S)^{1/2}\left( 1-\frac 1{2S}b^{\dagger }b\right)
b-2(2S+1)(2S)^{1/2}bc^{\dagger }c
\]
The distribution function of the dispersionless ``fermions'' turns out to be
$$
\langle c_i^{\dagger }c_i\rangle =-N_B((2S+1)\widetilde{H}),\qquad
\widetilde{H}=H=2J_0\langle S^z\rangle
\eqno{(\rm E.28)}
$$
We see that the pseudofermion operators exclude non-physical states, so that
an extrapolation to high temperatures becomes possible. In particular,
neglecting dispersion of spin-wave excitations we obtain from (E.27), (E.28)
$$
\langle S^z\rangle =S-N_B(\widetilde{H})+(2S+1)N_B((2S+1)\widetilde{H}%
)=S-B_S(S\widetilde{H}/T)
\eqno{(\rm E.29)}
$$
Thus we have rederived the mean-field equation for the magnetization (4.15).

\chapter{Hubbard operator approach in the Heisenberg model}

Using the Hubbard operators yields a possibility to obtain in a simple way
main results of the theory of Heisenberg magnets. In particular, this
formalism enables one to take into account the strong single-ion magnetic
anisotropy in the zeroth order approximation [673-677]. The Hamiltonian of
the Heisenberg model with an arbitrary single-site anisotropy has the form
$$
\mathcal{H}=-\sum_{ij}J_{ij}S_iS_j+\mathcal{H}_a=-\sum_{\mathbf{q}}J_{%
\mathbf{q}}S_{\mathbf{q}}S_{-\mathbf{q}}+\mathcal{H}_a
\eqno{(\rm F.1)}
$$
$$
\mathcal{H}_a=\sum_i\varphi (S_i^x,S_i^y,S_i^z,H)
\eqno{(\rm F.2)}
$$
where $J_{\mathbf{q}}$ are the Fourier transforms of exchange parameters, $H$
is the external magnetic field, $\varphi $ is an arbitary function. Note
that, as follows from the results of Appendix C, the Hamiltonan (F.1) with $%
S\rightarrow L$ describes the system of interacting orbital momenta in the
intermediate crystal field, $\mathcal{H}_a$ having the sense of the
crystalline electric field Hamiltonian. Thus the anisotropic Heisenberg
model may be applied to the problem of quenching of orbital momenta. First,
we consider the easy-axis ferromagnet where
$$
\mathcal{H}_a=-\sum_i\varphi (S_i^z)-H\sum_iS_i^z
\eqno{(\rm F.3)}
$$
We use the representation of momentum operators in terms of Hubbard
operators (Appendix B)
\[
S_i^{\pm }=\sum_{M=-S}^S\gamma _S(\pm M)X_i(M\pm 1,M),\qquad
S_i^z=\sum_{M=-S}^SMX(M,M)
\]
$$
\gamma _S(M)\equiv [(S-M)(S+M+1)]^{1/2}
\eqno{(\rm F.4)}
$$
Then the anisotropy Hamiltonian takes the diagonal form
$$
\mathcal{H}_a=-\sum_{iM}[\varphi (M)+HM]X_i(M,M)
\eqno{(\rm F.5)}
$$
It is convenient to introduce the commutator Green's functions
$$
G_{\mathbf{q}}(\omega ) =\langle \langle S_{\mathbf{q}}^{+}|S_{-\mathbf{q}%
}^{-}\rangle \rangle _\omega ,
\eqno{(\rm F.6)}
$$
\[
G_{\mathbf{q}M}(\omega )=\langle \langle X_{\mathbf{q}}(M+1,M)|S_{\mathbf{-q}%
}^{-}\rangle \rangle _\omega
\]
Write down the equation of motion
\[
(\omega -H-\varphi (M+1)+\varphi (M)-2J_0\langle S^z\rangle )G_{\mathbf{q}%
M}(\omega )
\]
$$
=\gamma _S(M)(N_{M+!}-N_M)[1-J_{\mathbf{q}}G_{\mathbf{q}}(\omega )]
\eqno{(\rm F.7)}
$$
where we have carried out a simplest decoupling, which corresponds to the
Tyablikov's decoupling at different sites (E.22),
\[
N_M=\langle X(M,M)\rangle
\]
After summing over $M$ one obtains
$$
G_{\mathbf{q}}(\omega )=\frac{\Phi _S(\omega )}{1+J_{\mathbf{q}}\Phi
_S(\omega )}
\eqno{(\rm F.8)}
$$
where
$$
\Phi _S(\omega )=\sum_M\frac{\gamma _S^2(M)(N_{M+1}-N_M)}{\omega -H-\varphi
(M+1)+\varphi (M)-2J_0\langle S^z\rangle }
\eqno{(\rm F.9)}
$$
The structure of expession (F.8) is reminiscent of the Green's functions in
the itinerant electron magnetism theory (see Appendices G,H). The excitation
spectrum is determined by the equation
$$
1+J_{\mathbf{q}}\Phi _S(\omega )=0
\eqno{(\rm F.10)}
$$
and contains $2S$ branches. The expressions (F.7), (F.8) enable one to
calculate the occupation numbers $N_{M}$ and obtain a self-consistent
equation for the magnetization
$$
\langle S^z\rangle =\sum_MMN_M
\eqno{(\rm F.11)}
$$
For simplicity we restrict ourselves to analyzing this equation in the
isotropic case where reduces to (E.22). Using the multiplication rule for
the Hubbard operators and the spectral representation (E.18) we obtain
$$
\langle S_i^{-}X_i(M+1,M)\rangle =\gamma _S(M)N_M
\eqno{(\rm F.12)}
$$
\[
=\gamma _S(M)(N_{M+1}-N_M)\sum_{\mathbf{q}}N_B(\omega _{\mathbf{q}})
\]
Then the system for $N_M$ reads
$$
N_M=(N_{M+1}-N_M)P_S,\qquad \sum_MN_M=1,\qquad P_S\equiv \sum_{\mathbf{q}%
}N_B(\omega _{\mathbf{q}})
\eqno{(\rm F.13)}
$$
Solving this system we derive the equation for $\langle S^z\rangle $ [675]
$$
\langle S^z\rangle =SB_S(-S\ln (P_S/[1+P_S]))
\eqno{(\rm F.14)}
$$
The equation (F.14) has the form, which is somewhat different from the
standard one [357], and is more convenient. Now we discuss the general case
[675-677]. The Hamiltonian (F.2) may be diagonalized to obtain
$$
\mathcal{H}_a|\psi _\mu \rangle =E_\mu |\psi _\mu \rangle
\eqno{(\rm F.15)}
$$
The eigenfunctions $\psi _\mu $ are be expanded in the eigenfunctions $%
|M\rangle $ of the operator $S^z$:
$$
|\psi _\mu \rangle =\sum_Mc_{\mu M}|M\rangle
\eqno{(\rm F.16)}
$$
Then the spin operators are presented in terms of the Hubbard operators as
$$
S_i^{+} =\sum_{\mu \mu ^{/}M}c_{\mu ^{/}M}c_{\mu ,M+1}^{*}\gamma
_S(M)X_i(\mu ,\mu ^{/})
\eqno{(\rm F.17)}
$$
$$
S_i^z =\sum_{\mu \mu ^{/}M}c_{\mu ^{/}M}c_{\mu ,M}^{*}MX_i(\mu ,\mu ^{/})
\eqno{(\rm F.18)}
$$
For $H||\langle S^z\rangle $, in the simplest mean-field approximation we
may take into account the intersite exchange interaction by replacing $H$ to
the effective field
$$
\widetilde{H}=H+2J_0\langle S^z\rangle
\eqno{(\rm F.19)}
$$
Then the occupation numbers $N_{M}$ are given by
$$
N_M=\sum_\mu |c_{\mu M}|^2N_\mu =\frac{\sum_\mu |c_{\mu M}|^2\exp (-E_\mu /T)%
}{\sum_\mu \exp (-E_\mu /T)}
\eqno{(\rm F.20)}
$$

Consider the easy-plane ferromagnet where
$$
\mathcal{H}_a=2D\sum_i(S_i^x)^2-H\sum_iS_i^z,\qquad D>0
\eqno{(\rm F.21)}
$$
The results turn out to be essentially different for integer and
half-integer spins. (Such a difference is typical for quantum spin systems,
see, e.g., [678].) This is due to the fact that the ground state of an ion
with an integer $S$ is singlet, and excited magnetic states are separated
by an energy gap. In particular, for $S=1$ we have [673]
$$
|\psi _{1,3}\rangle  =\cos \theta |\pm 1\rangle \pm \sin \theta |\mp
1\rangle
\eqno{(\rm F.22)}
$$
\[
E_{1,3}=D\mp (H^2+D^2)^{1/2},\qquad E_2=2D
\]
where
$$
\cos 2\theta =\frac H{(H^2+D^2)^{1/2}},\qquad \sin 2\theta =-\frac
D{(H^2+D^2)^{1/2}}
\eqno{(\rm F.23)}
$$
According to (F.10),(F.20) the equation for the magnetization has the form
$$
\langle S^z\rangle =\frac{\cos 2\widetilde{\theta }\sinh [(\widetilde{H}%
^2+D^2)^{1/2}/T]}{\cosh [(\widetilde{H}^2+D^2)^{1/2}/T]+\frac 12\exp (-D/T)}
\eqno{(\rm F.24)}
$$
At zero temperature and magnetic field we obtain
$$
\langle S^z\rangle =\left\{
\begin{array}{ll}
\lbrack 1-(D/2J_0)^2]^{1/2} & ,\qquad D<2J_0 \\
0 & ,\qquad D>2J_0
\end{array}
\right.
\eqno{(\rm F.25)}
$$
Thus at $D>2J_0$ the ground state $|\psi _1\rangle $ is a singlet
superposition of the states $|1\rangle $ and $|-1\rangle $ $(\sin \theta
=\cos \theta )$, and ferromagnetic ordering is suppressed. Easy-plane
higly-anisotropic systems exhibit also a non-trivial behaviour of
magnetization at changing magnetic field [679,680], which is reminiscent of
the Hall conductivity behaviour in the situation of the quantum Hall effect
[681]. For half-integer $S$, the ground state is ordered at arbitrary $D$.
However, the value of magnetization is diminished due to mixing of states
with different $M$ by the anisotropy field. So, for $S=3/2$ we have
$$
|\psi _{1,4}\rangle  =\cos \theta _{\pm }|\pm 3/2\rangle \mp \sin \theta
_{\pm }|\mp 1/2\rangle
\eqno{(\rm F.26)}
$$
\[
|\psi _{2,3}\rangle =\cos \theta _{\mp }|\pm 1/2\rangle \mp \sin \theta
_{\mp }|\mp 3/2\rangle
\]
\[
E_{1,4}=\frac 12(5D\mp H)\mp E_{\pm },\qquad E_{2,3}=\frac 12(5D\pm H)\mp
E_{\mp }
\]
where
\[
\cos \theta _{\pm }=[H\pm D+E_{\pm }]/[3D^2+(H\pm D+E_{\pm })^2]^{1/2}
\]
$$
E_{\pm }=[(H\pm D)^2+3D^2]^{1/2}
\eqno{(\rm F.27)}
$$
One can see that the lowest level $|\psi _1\rangle $ is a mixture of the
states with $M=3/2$ and $M=-1/2$. The magnetization at $T=0$ is given by
$$
\langle S^z\rangle =\frac 32-\frac{6D^2}{3D^2+(\widetilde{H}+D+\widetilde{E}%
_{+})^2}
\eqno{(\rm F.28)}
$$
At $D\gg J$ one obtains from (F.28) $\langle S^z\rangle =1$, and the ground
state is a doublet.

For cubic ansotropic ferromagnets we have
$$
\mathcal{H}_a=-\frac 12K\sum_i\left[ (S_i^x)^4+(S_i^y)^4+(S_i^z)^4\right]
-H\sum_iS_i^z
\eqno{(\rm F.29)}
$$
At $S\leq 3/2$ the single-ion cubic anisotropy does not occur by
kinematical reasons. For $S=2$ one obtains
\[
|\psi _{1,2}\rangle =\cos \theta |\pm 2\rangle \pm \sin \theta |\mp
2\rangle ,\qquad |\psi _{3,4}\rangle =|\pm 1\rangle ,\qquad |\psi _S\rangle
=|0\rangle
\]
$$
E_{1,2} =C\mp E,\qquad E_{3,4}=C+K\mp H,\qquad E_S+C-K
\eqno{(\rm F.30)}
$$
\[
C=-\frac{21}2K,\qquad E=(H^2+K^2)^{1/2}
\]
where
$$
\cos \theta =\frac{2H+E}{[K^2+(2H+E)^2]^{1/2}},\qquad \sin \theta =\frac
D{[K^2+(2H+E)^2]^{1/2}}
\eqno{(\rm F.31)}
$$
We treat the case where $K>0$, so that the easy magnetization direction
corresponds to the $z$ axis. Then the ground state is $|\psi _1\rangle $.
The equation for the magnetization has the form
$$
\langle S^z\rangle =\frac{2\cos 2\widetilde{\theta }\sinh (\widetilde{%
\varepsilon }/T)+\exp (-K/T)\sin (\widetilde{H}/T)}{\cosh (\widetilde{%
\varepsilon }/T)+\frac 12\exp (K/T)+\exp (-K/T)\cos (\widetilde{H}/T)}
\eqno{(\rm F.32)}
$$
The ground state magnetization reads
$$
\langle S^z\rangle =2[1-(K/8J_0)^2]^{1/2}
\eqno{(\rm F.33)}
$$
and vanishes at $K>8J.$ The ferromagnetic ordering at large $K$ does not
also occur for $S=4$. Thus the orbital momenta $L=2$ and $L=4$ are quenched
by the cubic crystal field. The intraatomic orbital exchange interaction,
which is required for unquenching, is determined by (F.19), (F.32). For
$S=5/2,3$ and $7/2$ the ground state turns out to be always ordered, the
values of $\langle S^z\rangle $ at $T=0$ in the strong anisotropy limit
being 11/6, 3/2 and 7/6 respectively [676].


\chapter{Electron-magnon interaction in magnetic metals}

In this Appendix we calculate the spectrum of single-particle and spin-wave
excitations in metallic magnets. To this end we use many-electron models
which permit to describe effects of interelectron correlations. The simplest
model of such a type is the Hubbard model. In the case of a non-degenerate
band its Hamiltonian reads
$$
\mathcal{H}=\sum_{\mathbf{k}\sigma }t_{\mathbf{k}}c_{\mathbf{k}\sigma
}^{\dagger }c_{\mathbf{k}\sigma }+\mathcal{H}_{\mathrm{int}},
\eqno{(\rm G.1)}
$$
\[
\mathcal{H}_{\mathrm{int}}=U\sum_ic_{i\uparrow }^{\dagger }c_{i\uparrow
}c_{i\downarrow }^{\dagger }c_{i\downarrow }
\]
with $U$ being the on-site repulsion parameter. The Hubbard model was widely
used to consider itinerant electron ferromagnetism since this takes into
account the largest term of the Coulomb interaction -- the intraatomic one.
Despite apparent simplicity, this model contains a very reach physics, and
its rigorous investigation is a very difficult problem.

Besides the Hubbard model, it is sometimes convenient to use for theoretical
description of magnetic metals the s-d(f) exchange model. The s-d exchange
model was first proposed for transition d-metals to consider peculiarities
of their electrical resistivity [265]. This model postulates existence of
two electron subsystems: itinerant ``s-electrons'' which play the role of
current carriers, and localized ``d-electrons'' which give the main
contribution to magnetic moment. Such an assumption may be hardly justified
quantitatively for d-metals, but it may be useful at qualitative
consideration of some physical properties (especially of transport
phenomena). At the same time, the s-f model provides a good description of
magnetism in rare-earth metals and their compounds with well-localized
4f-states.

The Hamiltonian of the s-d(f) model in the simpest version has the form
\[
\mathcal{H}=\sum_{\mathbf{k}\sigma }t_{\mathbf{k}}c_{\mathbf{k}\sigma
}^{\dagger }c_{\mathbf{k}\sigma }-\sum_{\mathbf{q}}J_{\mathbf{q}}\mathbf{S}%
_{-\mathbf{q}}\mathbf{S}_{\mathbf{q}}+\mathcal{H}_{\mathrm{int}}
\]
$$
\equiv \mathcal{H}_s+\mathcal{H}_{d(f)}+\mathcal{H}_{\mathrm{int}}
\eqno{(\rm G.2)}
$$
\[
\mathcal{H}_{\mathrm{int}}=-I\sum_{i\sigma \sigma ^{\prime }}(\mathbf{S}_i%
\mathbf{\sigma }_{\sigma \sigma ^{\prime }})c_{i\sigma }^{\dagger
}c_{i\sigma ^{\prime }}
\]
where $\mathbf{S}_{\mathbf{q}}$ are operators for localized spins, $\mathbf{%
\sigma }$ are the Pauli matrices, $I$ is the parameter of the s-d(f)
exchange interaction which is assumed to be contact (derivation of the
s-d(f) model in a more general situation is considered in Appendix K), $J_{%
\mathbf{q}}$ are the Fourier transforms of the exchange parameters between
localized spins. In rare earth metals the latter interaction is usually the
indirect RKKY exchange via conduction electrons which is due to the same s-f
interaction. However, at constructing perturbation theory, it is convenient
to include this interaction in the zero-order Hamiltonian.

Altough more complicated in its form, the s-d model turns out to be in some
respect simpler than the Hubbard model since it permits to construct the
quasiclassical expansion in the small parameter $1/2S$. Within simple
approximations, the results in the s-d(f) and Hubbard models differ as a
rule by the replacement $I\rightarrow U$ only.

Below we perform a systematic investigation of spin-wave and electron
spectra of conducting ferro- and antiferromagnets within the above models.
We demonstrate similarities and differences in comparison with
localized-moment isulator magnets which are described by the Heisenerg model.

\section{Ferromagnets}

Here we consider spin-wave theory of the Hubbard ferromagnet using the
Stoner spin-split state as the zero-order approximation following mainly to
paper [338]. (The limit of strong correlations is discussed in Sect.4.6 and
Appendices H,J.) The simplest Hartree-Fock (Stoner) approximation in the
Hubbard model, which corresponds formally to first-order perturbation theory
in $U$, yields the electron spectrum of the form
$$
E_{\mathbf{k}\sigma }=t_{\mathbf{k}}+Un_{-\sigma }=t_{\mathbf{k}}+U(\frac
n2-\sigma \langle S^z\rangle )\equiv t_{\mathbf{k}\sigma }
\eqno{(\rm G.3)}
$$
so that we have for the spin splitting
$$
\Delta =U(n_{\uparrow }-n_{\downarrow })=2U\langle S^z\rangle
\eqno{(\rm G.4)}
$$
and $U$ plays the role of the Stoner parameter. The Hartree-Fock decoupling
does not take into account correctly the formation of doubles, i.e. doubly
occupied states at a site. (This can be made within the Hubbard
many-electron approach [28-31], see Appendix H). Note that this error does
not play a role for the saturated ferromagnetic state.

Unlike the Stoner theory, the Hubbard model enables one to describe
spin-wave excitations in an itinerant ferromagnet. To this end we present
the interaction Hamiltonian in the form
$$
\mathcal{H}_{\mathrm{int}}=\frac U2\sum_{\mathbf{k}\sigma }c_{\mathbf{k}%
\sigma }^{\dagger }c_{\mathbf{k}\sigma }-\frac U2\sum_{\mathbf{q}}(S_{-%
\mathbf{q}}^{-}S_{\mathbf{q}}^{+}+S_{\mathbf{q}}^{+}S_{-\mathbf{q}}^{-})
\eqno{(\rm G.5)}
$$
where we have introduced the Fourier components of spin density operators
$$
S_{\mathbf{q}}^{+} =\sum_{\mathbf{k}}c_{\mathbf{k}\uparrow }^{\dagger }
c_{\mathbf{k}+\mathbf{q}\downarrow },\qquad
S_{\mathbf{q}}^{-}=\sum_{\mathbf{k}}c_{\mathbf{k}\downarrow }^{\dagger }
c_{\mathbf{k+q}\uparrow }
\eqno{(\rm G.6)}
$$
\[
S_{\mathbf{q}}^z=\frac 12\sum_{\mathbf{k}}(c_{\mathbf{k}\uparrow }^{\dagger
}c_{\mathbf{k}+\mathbf{q}\uparrow }-c_{\mathbf{k}\downarrow }^{\dagger }c_{%
\mathbf{k}+\mathbf{q}\downarrow })
\]
The first term in (G.5) yields a renormalization of the chemical potential
and may be omitted. Consider the spin Green's function
\[
G_{\mathbf{q}}(\omega )=\langle \langle S_{\mathbf{q}}^{+}|S_{-\mathbf{q}%
}^{-}\rangle \rangle _\omega
\]
Writing down the sequence of equations of motion for this we derive
$$
G_{\mathbf{q}}(\omega )=2\langle S^z\rangle +\sum_{\mathbf{k}}(t_{\mathbf{k+q%
}}-t_{\mathbf{k}})\langle \langle c_{\mathbf{k}\uparrow }^{\dagger }c_{%
\mathbf{k}+\mathbf{q}\downarrow }|S_{-\mathbf{q}}^{-}\rangle \rangle _\omega
\eqno{(\rm G.7)}
$$
\[
(\omega -t_{\mathbf{k+q}}+t_{\mathbf{k}}-\Delta )\langle \langle c_{\mathbf{k%
}\uparrow }^{\dagger }c_{\mathbf{k}+\mathbf{q}\downarrow }|S_{-\mathbf{q}%
}^{-}\rangle \rangle _\omega
\]
$$
=(n_{\mathbf{k\uparrow }}-n_{\mathbf{k+q\downarrow }})[1-UG_{\mathbf{q}%
}(\omega )]
\eqno{(\rm G.8)}
$$
where we have introduced the irreducible Green's function
$$
L_{\mathbf{kqp}} =\delta \langle \langle c_{\mathbf{k}\uparrow }^{\dagger
}S_{\mathbf{p}}^{+}c_{\mathbf{k}+\mathbf{q}-\mathbf{p\uparrow }}-c_{\mathbf{%
k+p}\downarrow }^{\dagger }S_{\mathbf{p}}^{+}c_{\mathbf{k}+\mathbf{q}%
\downarrow }
$$
$$
-\delta _{\mathbf{pq}}(n_{\mathbf{k\uparrow }}-n_{\mathbf{k+q\downarrow }%
})S_{\mathbf{q}}^{+}|S_{-\mathbf{q}}^{-}\rangle \rangle _\omega
\eqno{(\rm G.9)}
$$
(the symbol $\delta $ means that the Hartree-Fock decouplings have to be
excluded). Substituting (G.8) into (G.7) we get
$$
G_{\mathbf{q}}(\omega )=\frac{\langle S^z\rangle -\Omega _{\mathbf{q}%
}(\omega )/U}{\omega -\Omega _{\mathbf{q}}(\omega )-\pi _{\mathbf{q}}(\omega
)}
\eqno{(\rm G.10)}
$$
$$
\Omega _{\mathbf{q}}(\omega )=U\sum_{\mathbf{k}}\frac{t_{\mathbf{k+q}}-t_{%
\mathbf{k}}}{t_{\mathbf{k+q}}-t_{\mathbf{k}}+\Delta -\omega }(n_{\mathbf{%
k\uparrow }}-n_{\mathbf{k+q\downarrow }})
\eqno{(\rm G.11)}
$$
where the self-energy $\pi $ is determined by the Green's function (G.9).
When neglecting $\pi $ we come to the random phase approximation (RPA).
Unlike the standard form
$$
G_{\mathbf{q}}(\omega )=\frac{\Pi _{\mathbf{q}}(\omega )}{1-U\Pi _{\mathbf{q}%
}(\omega )}
\eqno{(\rm G.12)}
$$
with
\[
\Pi _{\mathbf{q}}(\omega )=\sum_{\mathbf{k}}\frac{n_{\mathbf{k\uparrow }}-n_{%
\mathbf{k+q\downarrow }}}{\omega +t_{\mathbf{k\uparrow }}-t_{\mathbf{%
k+q\downarrow }}}
\]
the representation (G.10) yields explicitly the magnon (spin-wave) pole
$$
\omega _{\mathbf{q}}\simeq \Omega _{\mathbf{q}}(0)=\sum_{\mathbf{k}\sigma }%
\mathcal{A}_{\mathbf{kq}}^\sigma n_{\mathbf{k}\sigma }
\eqno{(\rm G.13)}
$$
where
$$
\mathcal{A}_{\mathbf{kq}}^\sigma =\sigma U\frac{t_{\mathbf{k+q}}-t_{\mathbf{k%
}}}{t_{\mathbf{k+q}}-t_{\mathbf{k}}+\sigma \Delta }
\eqno{(\rm G.14)}
$$
has the meaning of the electron-magnon interaction amplitude. Expanding in $%
q $ we get
\[
\omega _{\mathbf{q}}=D_{\alpha \beta }q_\alpha q_\beta
\]
where
$$
D_{\alpha \beta }=\frac U\Delta \sum_{\mathbf{k}}\left[ \frac{\partial ^2t_{%
\mathbf{k}}}{\partial k_\alpha \partial k_\beta }(n_{\mathbf{k}\uparrow }+n_{%
\mathbf{k}\downarrow })-\frac 1\Delta \frac{\partial t_{\mathbf{k}}}{%
\partial k_\alpha }\frac{\partial t_{\mathbf{k}}}{\partial k_\beta }(n_{%
\mathbf{k}\uparrow }-n_{\mathbf{k}\downarrow })\right]
\eqno{(\rm G.15)}
$$
are spin-wave stiffness tensor components. For a weak ferromagnet ($\Delta
\ll E_F,\ U$) we derive
$$
D_{\alpha \beta }=\frac{U\Delta }4\sum_{\mathbf{k}}\left[ \frac{\partial
^2t_{\mathbf{k}}}{\partial k_\alpha \partial k_\beta }\frac{\partial ^2n_{%
\mathbf{k}}}{\partial t_{\mathbf{k}}^2}+\frac 16\frac{\partial t_{\mathbf{k}}%
}{\partial k_\alpha }\frac{\partial t_{\mathbf{k}}}{\partial k_\beta }\frac{%
\partial ^3n_{\mathbf{k}}}{\partial t_{\mathbf{k}}^3}\right]
\eqno{(\rm G.16)}
$$
so that $D\sim \Delta $. The magnon damping in the RPA is given by
$$
\gamma _{\mathbf{q}}^{(1)}(\omega )=-\Im \Omega _{\mathbf{q}}(\omega )=\pi
U\Delta \omega \sum_{\mathbf{k}}\left( -\frac{\partial n_{\mathbf{k}\uparrow
}}{\partial t_{\mathbf{k}\uparrow }}\right) \delta (\omega -t_{\mathbf{%
k+q\downarrow }}+t_{\mathbf{k\uparrow }})
\eqno{(\rm G.17)}
$$
$$
\gamma _{\mathbf{q}}^{(1)}\equiv \gamma _{\mathbf{q}}^{(1)}(\omega _{\mathbf{%
q}})\simeq \pi U\Delta \omega _{\mathbf{q}}N_{\uparrow }(E_F)N_{\downarrow
}(E_F)\theta (\omega _{\mathbf{q}}-\omega _{-})
\eqno{(\rm G.18)}
$$
with $\theta (x)$ being the step function. Here $\omega _{-}=\omega (q_0)$
is the threshold energy which is determined by the condition of entering
into the Stoner continuum (decay into the Stoner excitations, i.e.
electron-hole pairs), $q_0$ being the minimal (in $\mathbf{k}$) solution to
the equation
$$
t_{\mathbf{k+q}_0\downarrow }=t_{\mathbf{k\uparrow }}=E_F
\eqno{(\rm G.19)}
$$
The quantity $\omega _{-}$ determines a characteristic energy scale
separating two temperature regions: the contributions of spin waves (poles
of the Green's function (G.10)) dominate at $T<\omega _{-}$, and those of
Stoner excitations (its branch cut) at $T>\omega _{-}$.

In the case of weak ferromagnets, the contribution of the branch cut of the
spin Green's function may be approximately treated as that of a paramagnon
pole at imaginary $\omega $, and we obtain
$$
q_0=k_{F\uparrow }-k_{F\downarrow },\qquad \omega _{-}=D(k_{F\uparrow
}-k_{F\downarrow })^2\sim \Delta ^3\sim T_c^2/E_F
\eqno{(\rm G.20)}
$$
Since $q_0$ is small, we have at smal $q>q_0$ , instead of (G.18),
$$
\gamma _{\mathbf{q}}^{(1)}(\omega _{\mathbf{q}})\simeq \frac{U\Delta \omega }%
q\frac{v_0}{4\pi }(m^{*})^2\equiv A/q
\eqno{(\rm G.21)}
$$
The estimation (G.20) holds also for the s-d(f) exchange model with the
indirect RKKY-interaction where
$$
D\sim T_c/S\sim I^2S/E_F
\eqno{(\rm G.22)}
$$
The damping at very small $q<q_0$ (where (G.17) vanishes) is due to the
two-magnon scattering processes. To consider these we have to calculate the
function $\pi $ to leading order in the fluctuating part of the Coulomb
interaction. Writing down the equation of motion for the Green's function
(G.9) we obtain
\[
\Pi _{\mathbf{q}}(\omega )=\sum_{\mathbf{pk}}(A_{\mathbf{kq}}^{\uparrow
})^2[B(\mathbf{k\uparrow ,k+q-p\uparrow ,}\omega _{\mathbf{p}}-\omega )+B(%
\mathbf{k+p\downarrow ,k+q\downarrow ,}\omega _{\mathbf{p}}-\omega )
\]
$$
-B(\mathbf{k+p}\mathbf{\downarrow ,}\mathbf{k}\uparrow \mathbf{,}\omega _{%
\mathbf{p}})-B(\mathbf{k+q\downarrow ,k+q-p\downarrow ,}\omega _{\mathbf{p}%
})]
\eqno{(\rm G.23)}
$$
where
\[
B(\mathbf{k}^{\prime }\sigma ^{\prime },\mathbf{k}\sigma ,\omega )=\frac{N_{%
\mathbf{p}}(n_{\mathbf{k}^{\prime }\sigma ^{\prime }}-n_{\mathbf{k}\sigma
})+n_{\mathbf{k}^{\prime }\sigma ^{\prime }}(1-n_{\mathbf{k}\sigma })}{%
\omega -t_{\mathbf{k}^{\prime }\sigma ^{\prime }}+t_{\mathbf{k}\sigma }}
\]
The magnon damping needede is given by the imaginary part of (G.23). After
some transformations we derive
\[
\gamma _{\mathbf{q}}^{(2)}(\omega )=\pi \sum_{\mathbf{kp}\sigma }(A_{\mathbf{%
kq}}^{\uparrow })^2(n_{\mathbf{k}\sigma }-n_{\mathbf{k+q-p}\sigma })\left[
N_{\mathbf{p}}-N_B(\omega _{\mathbf{p}}-\omega )\right]
\]
$$
\times \delta (\omega +t_{\mathbf{k}}-t_{\mathbf{k+q-p}}-\omega _{%
\mathbf{p}})
\eqno{(\rm G.24)}
$$
Integration for the isotropic electron spectrum gives [682,683]
$$
\gamma _{\mathbf{q}}^{(2)}(\omega )=\frac{v_0^2}{12\pi ^3}\frac{q^4}{%
4\langle S^z\rangle ^2}\sum_\sigma k_{F\sigma }^2\times \left\{
\begin{array}{ll}
\omega _{\mathbf{q}}/35 & ,\qquad T\ll \omega _{\mathbf{q}} \\
(T/4)\left( \ln (T/\omega _{\mathbf{q}})+\frac 53\right) & ,\qquad T\gg
\omega _{\mathbf{q}}
\end{array}
\right.
\eqno{(\rm G.25)}
$$
Real part of (G.23) describes the temperature dependence of the spin
stiffness owing to two-magnon processes (besides the simplest $T^2$%
-contribution which occurs from the temperature dependence of the Fermi
distribution functions in (G.10)). The spin-wave conribution connected with
the magnon distribution functions is proportional to $T$. More interesting
is the non-analytical many-electron contribution owing to the Fermi
functions:
$$
\delta D_{\alpha \beta } =\frac 1{4\langle S^z\rangle ^2}\sum_{\mathbf{pk}}%
\frac{\partial t_{\mathbf{k}}}{\partial k_\alpha }\frac{\partial t_{\mathbf{k%
}}}{\partial k_\beta }\left[ \frac{n_{\mathbf{k}\downarrow }(1-n_{\mathbf{k-p%
}\uparrow })}{t_{\mathbf{k}}-t_{\mathbf{k-p}}-\omega _{\mathbf{p}}}+\right.
\eqno{(\rm G.26)}
$$
\[
\left. +\frac{n_{\mathbf{k+p}\downarrow }(1-n_{\mathbf{k}\downarrow })}{t_{%
\mathbf{k+p}}-t_{\mathbf{k}}-\omega _{\mathbf{p}}}-\frac{n_{\mathbf{k+p}%
\downarrow }(1-n_{\mathbf{k}\uparrow })}{t_{\mathbf{k+p\downarrow }}-t_{%
\mathbf{k\uparrow }}-\omega _{\mathbf{p}}}-\frac{n_{\mathbf{k}\downarrow
}(1-n_{\mathbf{k-p}\downarrow })}{t_{\mathbf{k\downarrow }}-t_{\mathbf{%
k-p\uparrow }}-\omega _{\mathbf{p}}}\right]
\]
Performing integration for parabolic spectra of electrons and magnons ($t_{%
\mathbf{k}}=k^2/2m^{*}$, $\omega _{\mathbf{q}}=Dq^2$) yields
\[
\delta D(T)=\left( \frac{\pi v_0T}{12\langle S^z\rangle m^{*}}\right)
^2\frac 1D\left[ \sum_\sigma N_\sigma ^2(E_F)\ln \frac T{\omega _{+}}\right.
\]
$$
-\left. 2N_{\uparrow }(E_F)N_{\downarrow }(E_F)\ln \frac{\max (\omega _{-},T)%
}{\omega _{+}}\right]
\eqno{(\rm G.27)}
$$
with
$$
\omega _{\pm }=D(k_{F\uparrow }\pm k_{F\downarrow })^2,\qquad
N_\sigma (E_F)=m^{*}v_0k_F/2\pi ^2
\eqno{(\rm G.28)}
$$
At $\omega _{-}\ll T\ll \omega _{+}$ we have
$$
\delta D(T)=\left( \frac{\pi v_0T}{12\langle S^z\rangle m^{*}}\right)
^2\frac 1D\left[ N_{\uparrow }(E_F)-N_{\downarrow }(E_F)\right] ^2\ln \frac
T{\omega _{+}}
\eqno{(\rm G.29)}
$$
It should be noted that the correction (G.27) dominates at low temperatures
over the above-mentioned $T^2$-correction, which demonstrates an important
role of corrections to the RPA approximation. Unfortunately, the $T^2\ln T$%
-term has not yet to be considered at analyzing magnon spectra of
ferromagnetic metals. We see that temperature dependences of spin-wave
characterisitics in conducting magnets differ considerably from those in the
Heisenberg model.

To obtain corrections to the Stoner approximation for the electron spectrum
(G.3) we have to calculate the one-electron Green's function
$$
G_{\mathbf{k\sigma }}(E)=\langle \langle c_{\mathbf{k}\sigma }|c_{\mathbf{k}%
\sigma }^{\dagger }\rangle \rangle _E=[E-t_{\mathbf{k}\sigma }-\Sigma _{%
\mathbf{k}\sigma }(E)]^{-1}
\eqno{(\rm G.30)}
$$
One obtains for the self-energy [573]
$$
\Sigma _{\mathbf{k}\sigma }(E)=U^2\sum_{\mathbf{q}}\int_{-\infty }^\infty
\frac{\mathrm{d}\omega }\pi \Im \langle \langle S_{\mathbf{q}}^\sigma |S_{-%
\mathbf{q}}^{-\sigma }\rangle \rangle _\omega \frac{N_B(\omega )+n_{\mathbf{%
k+q,}-\sigma }}{E-t_{\mathbf{k+q,}-\sigma }+\omega }
\eqno{(\rm G.31)}
$$
Retaining only the magnon pole contribution to the spectral density (i.e.
neglecting the spin-wave damping) we may put
$$
-\frac 1\pi \Im \langle \langle S_{\mathbf{q}}^\sigma |S_{-\mathbf{q}%
}^{-\sigma }\rangle \rangle _\omega =2\sigma \langle S^z\rangle \delta
(\omega -\sigma \omega _{\mathbf{q}})
\eqno{(\rm G.32)}
$$
so that
$$
\langle S_{-\mathbf{q}}^{-}S_{\mathbf{q}}^{+}\rangle =2\langle S^z\rangle N_{%
\mathbf{p}}
\eqno{(\rm G.33)}
$$
Then we get
\[
\Sigma _{\mathbf{k\uparrow }}(E)=U\Delta \sum_{\mathbf{q}}\frac{N_{\mathbf{q}%
}+n_{\mathbf{k+q\downarrow }}}{E-t_{\mathbf{k+q\downarrow }}+\omega _{%
\mathbf{q}}}
\]
$$
\Sigma _{\mathbf{k\downarrow }}(E)=U\Delta \sum_{\mathbf{q}}\frac{1+N_{%
\mathbf{q}}-n_{\mathbf{k-q\uparrow }}}{E-t_{\mathbf{k-q\uparrow }}-\omega _{%
\mathbf{q}}}
\eqno{(\rm G.34)}
$$
The results (G.34) are valid in the s-d model ($U\rightarrow I$) to first
order in the small parameter $1/2S$ [323].Taking into account the relation
$$
\langle S^z\rangle =S_0-\sum_{\mathbf{p}}N_{\mathbf{p}}
\eqno{(\rm G.35)}
$$
where $S$ is the saturation magnetization one obtains for the spin-wave
correction to the electron energy
$$
\delta E_{\mathbf{k}\sigma }(T) =\sum_{\mathbf{q}}A_{\mathbf{kq}}^\sigma
N_{\mathbf{q}}
\eqno{(\rm G.36)}
$$
\[
=\frac{v_0}{2\langle S^z\rangle }\frac{\xi (5/2)}{32\pi ^{3/2}}\left( \frac
TD\right) ^{5/2}\left[ \frac{\partial ^2t_{\mathbf{k}}}{\partial k_x^2}%
-\frac \sigma {U\langle S^z\rangle }\left( \frac{\partial t_{\mathbf{k}}}{%
\partial k}\right) ^2\right]
\]
The $T^{5/2}$-dependence of the electron spectrum owing to magnons is weaker
than the $T^{3/2}$-dependence of the magnetization. This fact is due to
vanishing of electron-magnon interaction amplitude $A$ at zero magnon
wavevector, which is connected with the symmetry of exchange interaction.
Such a weakening of temperature dependence of the spin splitting was
observed in iron [145]. It should be noted that the same $T^{5/2}$%
-dependence takes place also in a ferromagnet with the Hubbard subbands
[337,338].

The one-electron occupation numbers are obtained via the spectral
representation for the anticommutator Green's function (G.30):
$$
\langle c_{\mathbf{k\uparrow }}^{\dagger }c_{\mathbf{k\uparrow }}\rangle
=f(t_{\mathbf{k}}+\Re \Sigma _{\mathbf{k\uparrow }}(t_{\mathbf{k\uparrow }%
}))
\eqno{(\rm G.37)}
$$
\[
+U\Delta \sum_{\mathbf{p}}\frac{N_{\mathbf{p}}(n_{\mathbf{k+p\downarrow }%
}-n_{\mathbf{k\uparrow }})+n_{\mathbf{k+p\downarrow }}(1-n_{\mathbf{%
k\uparrow }})}{(t_{\mathbf{k+p\downarrow }}-t_{\mathbf{k\uparrow }}-\omega _{%
\mathbf{p}})^2}
\]
$$
\langle c_{\mathbf{k\downarrow }}^{\dagger }c_{\mathbf{k\downarrow }}\rangle
=f(t_{\mathbf{k}}+\Re \Sigma _{\mathbf{k\downarrow }}(t_{\mathbf{%
k\downarrow }}))
\eqno{(\rm G.38)}
$$
\[
+U\Delta \sum_{\mathbf{p}}\frac{N_{\mathbf{p}}(n_{\mathbf{k\downarrow }}-n_{%
\mathbf{k-p\uparrow }})+n_{\mathbf{k\downarrow }}(1-n_{\mathbf{k-p\uparrow }%
})}{(t_{\mathbf{k-p\uparrow }}-t_{\mathbf{k\downarrow }}+\omega _{\mathbf{p}%
})^2}
\]
where the second term comes from the imaginary part of the self-energy.
Retaining only the magnon contributions up to $T^{3/2}$ we get
$$
\langle c_{\mathbf{k}\sigma }^{\dagger }c_{\mathbf{k}\sigma }\rangle
\simeq n_{\mathbf{k}\sigma }\frac{S_0+\langle S^z\rangle }{2S_0}+n_{%
\mathbf{k-}\sigma }\frac{S_0-\langle S^z\rangle }{2S_0}
\eqno{(\rm G.39)}
$$
with $\langle S^z\rangle $ defined by (G.35). Thus, despite the presence of
the spin splitting, electron occupation numbers have a strong $T^{3/2}$%
-dependence rather than an exponential one (as in the Stoner theory). This
dependence arises because of thermal magnon emission and absorption
processes. Thus the conduction electron spin polarization
$$
P=\frac{n_{\uparrow }-n_{\downarrow }}{n_{\uparrow }+n_{\downarrow }}
\eqno{(\rm G.40)}
$$
is equal to the relative magnetization, which is obvious for an itinerant
ferromagnet. However, such a behaviour takes place also in arbitrary
conducting ferromagnets, e.g., for ferromagnetic semiconductors which are
describe by the s-d(f) exchange model [329]. Formally, the $T^{3/2}$%
-dependence of $P(T)$ is due to the strong temerature dependence of the
residues of the electron Green's functions and to the occurence of the
non-quasiparticle states in ``alien'' spin subband owing to electron-magnon
scattering. The picture of the density of states is discussed in more
details in Sect.4.5.

Consider corrections to the magnetization $\langle S^z\rangle $. We have
$$
\langle S^z\rangle =\frac n2-\sum_{\mathbf{q}}\langle S_{\mathbf{q}}^{-}S_{%
\mathbf{q}}^{+}\rangle -\langle \widehat{n}_{i\uparrow }\widehat{n}%
_{i\downarrow }\rangle
\eqno{(\rm G.41)}
$$
The first average involved in (G.41) is calculated from the spectral
representation of the RPA Green's function (G.10):
$$
\langle S_{-\mathbf{q}}^{-}S_{\mathbf{q}}^{+}\rangle =2S_0N_{\mathbf{q}}\qquad
(q<q_0)
\eqno{(\rm G.42)}
$$
$$
\langle S_{-\mathbf{q}}^{-}S_{\mathbf{q}}^{+}\rangle =\frac 1\pi
\int_{-\infty }^\infty \mathrm{d}\omega \frac{N_B(\omega )\gamma _{\mathbf{q}%
}^{(1)}(\omega )(\Delta -\omega )/U}{[\omega -\Re \Omega _{\mathbf{q}%
}(\omega )]^2+[\gamma _{\mathbf{q}}^{(1)}(\omega )]^2}\qquad (q>q_0)
\eqno{(\rm G.43)}
$$
Using the identity
$$
N_B(t_{\mathbf{k+q\downarrow }}-t_{\mathbf{k\uparrow }})(n_{\mathbf{%
k\uparrow }}-n_{\mathbf{k+q\downarrow }})=n_{\mathbf{k+q\downarrow }}(1-n_{%
\mathbf{k\uparrow }})
\eqno{(\rm G.44)}
$$
we derive from (G.43)
$$
\langle S_{-\mathbf{q}}^{-}S_{\mathbf{q}}^{+}\rangle =\sum_{\mathbf{k}}%
\frac{(t_{\mathbf{k+q\downarrow }}-t_{\mathbf{k\uparrow }})^2n_{\mathbf{%
k+q\downarrow }}(1-n_{\mathbf{k\uparrow }})}{(t_{\mathbf{k+q\downarrow }}-t_{%
\mathbf{k\uparrow }}-\omega _{\mathbf{q}})^2+\left[ \gamma _{\mathbf{q}%
}^{(1)}(t_{\mathbf{k+q\downarrow }}-t_{\mathbf{k\uparrow }})\right] ^2}
\eqno{(\rm G.45)}
$$
In contradiction with (G.41), (G.42), the true Bloch spin-wave contribution
to magnetization should be given by (G.35) since every magnon decreases $%
\langle S^z\rangle $ by unity. The agreement may be restored by allowing not
only the magnon pole, but also branch cut contributions. Replacing in (G.45)
$n_{\mathbf{k}\sigma }\rightarrow \langle c_{\mathbf{k}\sigma }^{\dagger }c_{%
\mathbf{k}\sigma }\rangle $ and using (G.39) we obtain
$$
\delta \langle S^z\rangle _{SW} =-\sum_{\mathbf{q}}N_{\mathbf{q}}\left[
2S_0+\right.
\eqno{(\rm G.46)}
$$
\[
\left. +\frac 1{2S_0}\sum_{\mathbf{kk}^{\prime }}\frac{(t_{\mathbf{k}}-t_{%
\mathbf{k}^{\prime }})^2(n_{\mathbf{k\uparrow }}-n_{\mathbf{k\downarrow }})}{%
(t_{\mathbf{k}^{\prime }\downarrow }-t_{\mathbf{k\uparrow }}-\omega _{%
\mathbf{k-k}^{\prime }})^2}(1-n_{\mathbf{k}^{\prime }\uparrow }+n_{\mathbf{k}%
^{\prime }\downarrow })\right] \simeq -\sum_{\mathbf{q}}N_{\mathbf{q}}
\]
where we have neglected the spin splitting in the denominator.

In the semi-phenomenological manner, it is convenient to introduce
``magnon'' operators which satisfy on the average the Bose commutation
relations:
$$
b_{\mathbf{q}}=(2S_0)^{-1/2}S_{\mathbf{q}}^{+},\qquad
b_{\mathbf{q}}^{\dagger }=(2S_0)^{-1/2}S_{-\mathbf{q}}^{-}
\eqno{(\rm G.47)}
$$
Then we have
$$
\delta \langle S^z\rangle =-\sum_{\mathbf{q}}\langle b_{\mathbf{q}%
}^{\dagger }b_{\mathbf{q}}\rangle =\frac 1{(2S_0)}\sum_{\mathbf{q}}\langle
S_{-\mathbf{q}}^{-}S_{\mathbf{q}}^{+}\rangle
\eqno{(\rm G.48)}
$$
Performing integration over $\omega $ in (G.43) at $T=0$ we obtain
$$
\delta \langle S^z\rangle =-\frac 1\pi \sum_{\mathbf{q}}\frac{\gamma _{%
\mathbf{q}}^{(1)}}{\omega _{\mathbf{q}}}\ln \frac W{\omega _{\mathbf{q}}}
\eqno{(\rm G.49)}
$$
with $W$ being the bandwidth. This contribution describes zero-point
decrease of the magnetization due to the ground-state magnon damping which
is owing to the Stoner excitations. For parabolic electron and magnon
spectra, neglecting the damping in the denominator of (G.43) we obtain at
low temperatures $T<\omega _{-}$%
\[
\delta \langle S^z\rangle _{cl}\simeq -U\Delta \sum_{\mathbf{kk}^{\prime }}%
\frac{n_{\mathbf{k}^{\prime }\downarrow }(1-n_{\mathbf{k\uparrow }})}{(t_{%
\mathbf{k}^{\prime }\downarrow }-t_{\mathbf{k\uparrow }}-\omega _{\mathbf{k-k%
}^{\prime }})^2}
\]
\[
=-\left( \frac{m^{*}v_0}{2\pi ^2}\right) ^2\frac{U\Delta }{4D}\left[ \omega
_{+}\ln \frac W{\omega _{\mathbf{+}}}-\omega _{-}\ln \frac W{\omega _{%
\mathbf{-}}}\right.
\]
$$
+\left. \frac{2\pi ^2}3T^2\left( \frac 1{\omega _{\mathbf{-}}}-\frac
1{\omega _{\mathbf{+}}}\right) \right]
\eqno{(\rm G.50)}
$$
(G.50) with $\omega _{\pm }$ defined in (G.28). For a weak ferromagnet, the
temperature correction to $\langle S^z\rangle $ in (G.50) is proportional to
$(T/T_C)^2$, in agreement with the self-consistent renormalization theory
[296,26]. It should be stressed that the $T^2$-correction obtained is much
larger than the Stoner contribution of the order of $(T/E_F)^2$.

An account of the damping at low $T$ influences numerical factors in (G.50)
only. At the same time, at high $T>\omega _{-}$ the damping in the
denominator dominates at small $q$ in the case of a weak ferromagnet. Taking
into account (G.21) we obtain from (G.43)
$$
\delta \langle S_{-\mathbf{q}}^{-}S_{\mathbf{q}}^{+}\rangle =\frac \Delta
{\pi U}\int_{-\infty }^\infty \mathrm{d}\omega N_B(\omega )\int_0^\infty
\frac{\omega Aqdq}{(Dq^2)^2+A^2\omega ^2/q^2}\sim \left( \frac T{E_F}\right)
^{4/3}
\eqno{(\rm G.51)}
$$
Thus we get from (G.41) the $T^{4/3}$-contribution to the magnetization,
which agrees with the result of the phase transition scaling theory near $%
T=T_C$. For a ferromagnet with well-localized magnetic moments the damping
may be neglected and we derive [716]
$$
\delta \langle S^z\rangle _{el}\sim -I^2\omega _{-}\ln (T/\omega _{-})
\eqno{(\rm G.52)}
$$

Consider the renormalization of electronic specific heat in an itinerant
ferromagnet due to interaction with spin fluctuations. Integration in (G.34)
at $T=0$ gives
$$
\Re \Sigma _\sigma (k_{F\sigma },E)=-\frac{U\Delta }{\omega _{+}-\omega _{-}}%
N_{-\sigma }(E_F)\sum_{\alpha =\pm }\alpha (E-\omega _\alpha )\ln \frac{%
|E-\omega _\alpha |}W
\eqno{(\rm G.53)}
$$
Then the inverse residue of the electron Green's function
\[
Z_{\mathbf{k}\sigma }^{-1}(E)=1-\frac \partial {\partial E}\Re
\Sigma _{\mathbf{k}\sigma }(E)
\]
takes the form
$$
Z_\sigma ^{-1}(k_{F\sigma },E_F)=1+\frac{U\Delta }{\omega _{+}-\omega _{-}}%
N_{-\sigma }(E_F)\ln \left| \frac{E-\omega _{+}}{E-\omega _{-}}\right|
\eqno{(\rm G.54)}
$$
The quantity (G.54) determines the renormalization of the electron effective
mass owing to the electron-magnon interaction. Thus we obtain for the
coefficient at the linear term in the electronic specific heat at $T\ll
\omega _{-}$
\[
\gamma _\sigma =\gamma _\sigma ^{(0)}/Z_\sigma (k_{F\sigma },E_F)=\frac{\pi
^2}3N_\sigma (E_F)\left[ 1+\right.
\]
$$
\left. +\frac{U\Delta }{\omega _{+}-\omega _{-}}N_{-\sigma }(E_F)\ln \frac{%
\omega _{+}}{\omega _{-}}\right]
\eqno{(\rm G.55)}
$$
For weak itinerant ferromagnets we have
$$
\ln \frac{\omega _{+}}{\omega _{-}}\simeq -2\ln (UN(E_F)-1)
\eqno{(\rm G.56)}
$$
so that the expression (G.55) describes the paramagnon enhancement of the
specific heat [297,573] discussed in Sect.4.4. The numerical factor in
(G.55) is inexact in this limit because of neglecting longitudinal spin
fluctuations (see [26]). On the other hand, our consideration is not
restricted to the case of weak ferromagnets. This is important since a
considerable enhancement of specific heat owing to spin fluctuations is
observed in a number of strong ferromagnets. For example, the experimental
value of g in the system CeFe$_{1-x}$Co$_x$S$_2$ (where a ferro-antiferro
transition takes place at increasing $x$) in the ferromagnetic phase, $%
\gamma =48$ mJ/mol K$^2$ at $x=0$, exceeds by two times the theoretical
value obtained from the calculated density of states, and is considerably
larger than the values in para- and antiferromagnetic phases [684].

Other thermodynamic properties may be investigated by calculating the free
energy of the system. At low $T<\omega _{-}$ the many-electron (branch cut)
contribution reads
$$
F_{el} =\frac 1{2S_0}\sum_{\mathbf{q>q}_0}\omega _{\mathbf{q}}\langle S_{-%
\mathbf{q}}^{-}S_{\mathbf{q}}^{+}\rangle
\eqno{(\rm G.57)}
$$
\[
\simeq U\Delta \sum_{\mathbf{kk}^{\prime }}\frac{n_{\mathbf{k}^{\prime
}\downarrow }(1-n_{\mathbf{k\uparrow }})}{t_{\mathbf{k\uparrow }}-t_{\mathbf{%
k}^{\prime }\downarrow }+\omega _{\mathbf{k-k}^{\prime }}}\equiv
F_{el}(0)+\delta F_{el}(T)
\]
where
$$
F_{el}(0)=\frac 1\pi \sum_{\mathbf{q}}\gamma _{\mathbf{q}}^{(1)}\ln \frac
W{\omega _{\mathbf{q}}}\approx \frac 1{8D}\left( \frac{m^{*}v_0}{2\pi ^2}%
\right) ^2\left[ \omega _{+}^2\ln \frac W{\omega _{+}}-\omega _{-}^2\ln
\frac W{\omega -}\right]
\eqno{(\rm G.58)}
$$
$$
\delta F_{el}(T)=-\frac{U\Delta }{\omega _{+}-\omega _{-}}N_{\uparrow }(E_F)%
\frac{\pi ^2}3T^2\ln \frac{\omega _{+}}{\max (\omega _{-},T)}
\eqno{(\rm G.59)}
$$
The spin-wave contribution to the free energy has the form, usual for the
Bose excitations with the square dispersion law
$$
\delta F_{SW} =-\frac 23\delta \langle \mathcal{H}\rangle _{SW},
\eqno{(\rm G.60)}
$$
\[
\delta \langle \mathcal{H}\rangle _{SW}=\sum_{q<q_0}\omega _{\mathbf{q}}N_{%
\mathbf{q}}=\frac{3v_0}{16\pi ^{3/2}}\xi \left( \frac 52\right) \frac{T^{5/2}%
}{D^{3/2}}
\]
Temperature-dependent corrections to physical properties are obtained from
(G.59), (G.60). Differentiating (G.59) over $T$ we obtain
$$
\delta C_{el} =-\frac \partial {\partial T}\delta F_{el}(T)
\eqno{(\rm G.61)}
$$
\[
=U^2\frac{2\langle S^z\rangle }{\omega _{+}-\omega _{-}}N_{\uparrow
}(E_F)N_{\downarrow }(E_F)\frac{2\pi ^2}3T\ln \frac{\omega _{+}}{\max
(\omega _{-},T)}
\]
Thus at $T\gg \omega _{-}$ we obtain instead of (G.55) the $T\ln T$%
-dependence of specific heat.

Consider the local magnetic moment at a site
$$
\langle S^z\rangle =\frac 34(n-2N_2),\qquad N_2=\langle n_{i\uparrow
}n_{i\downarrow }\rangle
\eqno{(\rm G.62)}
$$
The number of doubles $N_2$ may be determined by using the Hellman-Feynman
theorem $N_2=\partial F/\partial U$. We obtain at $T<\omega _{-}$%
$$
\delta \langle S^z\rangle _{el}(T)\sim -(T/\Delta )^2\delta \langle
S^z\rangle _{el}(T)\sim -(T/T_c)^{5/2}
\eqno{(\rm G.63)}
$$
Thus the temperature dependence of the spin-wave contribution to $\langle
\mathbf{S}^2\rangle $ is weaker than that to $\langle S^z\rangle $, which
justifies neglecting the former in the above discussion of the magnetization
(G.41). At high $T$, the local moment has the $T$-dependence:
$$
\delta \langle S_i^2\rangle =\delta \sum_{\mathbf{q}}\langle S_{-\mathbf{q}%
}^{-}S_{\mathbf{q}}^{+}\rangle \sim (T/E_F)^{4/3}
\eqno{(\rm G.64)}
$$
As one can see from (G.55), the enhancement of effective mass and electronic
specific heat owing to spin fluctuations is absent in the half-metallic
state (Sect.4.5). We shall demonstrate that the specific heat of a
conducting ferromagnet may contain spin-fluctuation contributions of another
nature. Write down a general expression for the specific heat in the s-d
exchange model in terms of the total energy
$$
C(T) =\frac{\partial \langle \mathcal{H}\rangle }{\partial T}=\frac
\partial {\partial T}\int \mathrm{d}EEf(E)N_t(E)
\eqno{(\rm G.65)}
$$
\[
=\frac{\pi ^2}3N_t(E)T+\int \mathrm{d}EEf(E)\frac \partial {\partial
T}N_t(E,T)
\]
where
\[
N_t(E)=-\frac 1\pi \sum_{\mathbf{k}\sigma }\Im G_{\mathbf{k}\sigma }(E)
\]
is the total sensitiy of states. The first term in the right-hand side of
(G.65) yields the standard result of the Fermi-liquid theory. The second
term is due to the energy dependence of the density of states. Such a
dependence occurs in the conducting ferromagnet owing to non-quasiparticle
(incoherent) states (Sect.4.5). Substituting (4.87) into this term we derive
[338]
$$
\delta C_\sigma (T)=2\sigma I^2\langle S^z\rangle \sum_{\mathbf{kq}}\frac{%
f(t_{\mathbf{k+q},-\sigma }-\sigma \omega _{\mathbf{q}})}{(t_{\mathbf{k+q}%
,-\sigma }-t_{\mathbf{k},\sigma })^2}\frac \partial {\partial T}n_{\mathbf{%
k+q},-\sigma }
\eqno{(\rm G.66)}
$$
At low temperatures
\[
f(t_{\mathbf{k+q},\downarrow }-\omega _{\mathbf{q}})=1,\qquad
f(t_{\mathbf{k+q},\uparrow }-\omega _{\mathbf{q}})=0
\]
(G.67) Thus the non-quasiparticle states with $\sigma =\downarrow $ do not
contribute to linear specific heat since they are empty at $T=0$. In the
half-metallic state the non-quasiparticle contributions (G.66) with $\sigma
=\uparrow $ are present for $I<0$ only, and we obtain
$$
\delta C_{\uparrow }(T)=\frac{2\pi ^2}3I^2\langle S^z\rangle N_{\downarrow
}(E_F)T\sum_{\mathbf{k}}\frac 1{(t_{\mathbf{k\uparrow }}-E_F)^2}
\eqno{(\rm G.67)}
$$
To avoid misunderstanding, it should be stressed that presence of such
contributions to specific heat means inapplicability of the Fermi-liquid
description in terms of dynamical quasiparticles only, which are determined
by poles of Green functions. It may be shown rigorously that the entropy of
interacting Fermi systems at low $T$ is expressed in terms of Landau
quasiparticles with the energies, determined as variational derivatives of
the total energy with respect to occupation numbers [685]. Thus, even in the
presence of non-pole contributions to the Green functions, the description
of thermodynamics in terms of statistical quasiparticles [685] holds.
(However, the quasiparticle description is insufficient for spectral
characteristics, e.g., optical and emission data.) The anomalous $\gamma T$%
-term is determined by the difference of the spectra of statistical and
dynamical quasiparticles.

Similar contributions to specific heat in the Hubbard model with strong
correlations are discussed in the paper [338] too. They dominate in the
enhancement of specific heat for half-metallic ferromagnets and may be
important, besides the effective mass enhancement (G.55), for ``usual''
magnets with well-defined local moments.

\section{Antiferromagnets}

To consider the electron and magnon spectrum of a metallic antiferromagnet
in the s-d(f) exchange model, we have to pass to the local coordinate system
according to (E.8). Then the Hamiltonian of the s-d(f) exchange interaction
takes the form
$$
\mathcal{H}_{sd} =-I\sum_{\mathbf{kq}}[S_{\mathbf{q}}^x(c_{\mathbf{%
k+q\downarrow }}^{\dagger }c_{\mathbf{k\downarrow }}-c_{\mathbf{k+q}\uparrow
}^{\dagger }c_{\mathbf{k\uparrow }})
\eqno{(\rm G.68)}
$$
\[
+iS_{\mathbf{q}}^y(c_{\mathbf{k-Q\downarrow }}^{\dagger }c_{\mathbf{%
k-Q\uparrow }}-c_{\mathbf{k+q}\uparrow }^{\dagger }c_{\mathbf{k-Q\downarrow }%
})
\]
\[
+S_{\mathbf{q}}^z(c_{\mathbf{k+q\uparrow }}^{\dagger }c_{\mathbf{%
k-Q\downarrow }}+c_{\mathbf{k-Q\downarrow }}^{\dagger }c_{\mathbf{%
k-q\uparrow }})]
\]
Passing to the magnon representation with the use of (E.1), (E.10) and
calculating the electron self-energy to second order in $I$ we obtain
$$
\Sigma _{\mathbf{k}}(E) =\frac{I^2\overline{S}^2}{E-t_{\mathbf{k-Q}}}
\eqno{(\rm G.69)}
$$
\[
+\frac 12I^2S\sum_{\mathbf{q}}\left\{ (u_{\mathbf{q}}-v_{\mathbf{q}%
})^2\left[ \frac{1-n_{\mathbf{k-q}}+N_{\mathbf{q}}}{E-t_{\mathbf{k-q}%
}-\omega _{\mathbf{q}}}+\frac{n_{\mathbf{k-q}}+N_{\mathbf{q}}}{E-t_{\mathbf{%
k-q}}+\omega _{\mathbf{q}}}\right] \right.
\]
\[
+(u_{\mathbf{q}}+v_{\mathbf{q}})^2\left. \left[ \frac{1-n_{\mathbf{k+q-Q}%
}+N_{\mathbf{q}}}{E-t_{\mathbf{k+q-Q}}-\omega _{\mathbf{q}}}+\frac{n_{%
\mathbf{k+q-Q}}+N_{\mathbf{q}}}{E-t_{\mathbf{k+q-Q}}+\omega _{\mathbf{q}}}%
\right] \right\}
\]
with $\overline{S}$ being the sublattice magnetization. In the mean-field
approximation the electron spectrum contains two split antiferromagnetic
subbands:
$$
E_{\mathbf{k}}^{1,2}=\frac 12(t_{\mathbf{k}}+t_{\mathbf{k-Q}})\pm \frac
12[(t_{\mathbf{k}}-t_{\mathbf{k-Q}})^2+4I^2\overline{S}^2]^{1/2}
\eqno{(\rm G.70)}
$$
The fluctuation correction (second term in (G.68)) are important, besides
the dependence $\overline{S}(T)$, at calculating the temperature dependence
of electron spectrum. Consider the magnon corrections owing to the Bose
functions
$$
\delta E_{\mathbf{k}}(T)=I^2S\sum_{\mathbf{q}}\left[ -2\frac{u_{\mathbf{q}%
}^2+v_{\mathbf{q}}^2}{t_{\mathbf{k}}-t_{\mathbf{k-Q}}}+\frac{(u_{\mathbf{q}%
}-v_{\mathbf{q}})^2}{t_{\mathbf{k}}-t_{\mathbf{k-q}}}+\frac{(u_{\mathbf{q}%
}+v_{\mathbf{q}})^2}{t_{\mathbf{k}}-t_{\mathbf{k+q-Q}}}\right] N_{\mathbf{q}}
\eqno{(\rm G.71)}
$$
The corrections to the band bottom ($t_{\mathbf{k}}=t_{\min }$ ) owing to
sublattice magnetization (the first term in square brackets) and transverse
fluctuations have opposite signs. The contribution of fluctuations prevails
which results in a ``blue'' shift of conduction band bottom with decreasing
temperature, as observed in antiferromagnetic semiconductors [352,686], in
contrast to the ``red'' shift in ferromagnetic semiconductors (cf.(G.36)).
(The same situation takes place at high temperatures [686,687].) In
particular, in the nearest-neighbour approximation for simple lattices where
$t_{\mathbf{k+Q}}=-t_{\mathbf{k}}$ the fluctuation contribution is larger in
absolute value by two times. The integration yields ($z$ is the
nearest-neighbour number)
$$
\delta E_{\min }(T)=-\frac 3{16}\frac{I^2S}W\left( \frac z2\right)
^{3/2}(S+1)^{3/2}\left( \frac T{T_N}\right) ^2
\eqno{(\rm G.72)}
$$
The $T^2$-dependence of electron spectrum (the same as of the sublattice
magnetization) is due to linear dispersion of spin-wave spectrum and $q^{-1}$%
-dependence of electron-magnon interaction amplitude, which are specific for
antiferromagnets. Therefore, the behavior of electron spectrum in a
ferrimagnet is qualitatively similar to that in ferromagnets (the $T^{5/2}$%
-dependence) [687].

We write down also the third-order many-electron contribution to the
self-energy, which describes renormalization of the antiferromagnetic gap
due to Kondo-like divergences [367] (see Chapter 6)
$$
\delta \Sigma _{\mathbf{k}}^{(3)}(E) =2I^3S^2\sum_{\mathbf{q}}n_{\mathbf{%
k+q}}(E-t_{\mathbf{k+q}})\frac 1{(E-t_{\mathbf{k+q}})^2-\omega _{\mathbf{q}%
}^2}
\eqno{(\rm G.73)}
$$
\[
\times \left( \frac 1{t_{\mathbf{k+q}}-t_{\mathbf{k-Q+q}}}-\frac 1{E-t_{%
\mathbf{k+Q}}}\right)
\]

To investigate magnon spectrum we calculate the retarded commutator Green's
functions
\[
\Gamma _{\mathbf{q}}(\omega )=\langle \langle b_{\mathbf{q}}|b_{\mathbf{q}%
}^{\dagger }\rangle \rangle _\omega ,\bar \Gamma _{\mathbf{q}}(\omega
)=\langle \langle b_{-\mathbf{q}}^{\dagger }|b_{\mathbf{q}}^{\dagger
}\rangle \rangle _\omega
\]
Writing down the sequence of equations of motion to second order in $I$ and
performing the simplest possible decouplings we derive [716] (cf.(E.9))
$$
\Gamma _{\mathbf{q}}(\omega )=\frac{\omega +C_{\mathbf{q-}\omega }}{(\omega
-C_{\mathbf{q}\omega })(\omega +C_{\mathbf{q-}\omega })+D_{\mathbf{q}\omega
}^2}
\eqno{(\rm G.74)}
$$
$$
\bar \Gamma _{\mathbf{q}}(\omega )=\frac{D_{\mathbf{q}\omega }}{(\omega -C_{%
\mathbf{q}\omega })(\omega +C_{\mathbf{q-}\omega })+D_{\mathbf{q}\omega }^2}
\eqno{(\rm G.75)}
$$
\[
C_{\mathbf{q}\omega }=S(J_{\mathbf{Q+q},\omega }^{tot}+J_{\mathbf{q}\omega
}^{tot}-2J_{\mathbf{Q}0}^{tot})+\sum_{\mathbf{p}}[C_{\mathbf{p}}\Phi _{%
\mathbf{pq}\omega }
\]
$$
-(C_{\mathbf{p}}-D_{\mathbf{p}})\Phi _{\mathbf{p}00}+\phi _{\mathbf{pq}%
\omega }^{+}+\phi _{\mathbf{pq}\omega }^{-}]+g_{\mathbf{q}}
\eqno{(\rm G.76)}
$$
\[
D_{\mathbf{q}\omega }=D_{\mathbf{q-}\omega }=S(J_{\mathbf{q}\omega
}^{tot}-J_{\mathbf{Q+q},\omega }^{tot})+\sum_{\mathbf{p}}D_{\mathbf{p}}\Phi
_{\mathbf{pq}\omega }+h_{\mathbf{q}}
\]
where the s-d exchange contributions of the first order in $1/2S$ correspond
to the RKKY approximation
$$
J_{\mathbf{q}\omega }^{tot}=J_{\mathbf{q}}+J_{\mathbf{q}}^{\mathrm{RKKY}%
}(\omega )=J_{\mathbf{q}}+I^2\sum_{\mathbf{k}}\frac{n_{\mathbf{k}}-n_{%
\mathbf{k-q}}}{\omega +t_{\mathbf{k}}-t_{\mathbf{k-q}}}
\eqno{(\rm G.77)}
$$
$J_{\mathbf{q}}^{\mathrm{RKKY}}(\omega )$ being the $\omega $-dependent Fourier
transform of the integral of indirect exchange interaction via conduction
electrons (cf.(K.1)). The function $\Phi $, which determines the
second-order corrections, is given by
$$
\Phi _{\mathbf{pq}\omega }=(\phi _{\mathbf{pq}\omega }^{+}-\phi _{\mathbf{pq}%
\omega }^{-})/\omega _{\mathbf{p}}
\eqno{(\rm G.78)}
$$
\[
\phi _{\mathbf{pq}\omega }^{\pm }=I^2\sum_{\mathbf{k}}\frac{n_{\mathbf{k}%
}(1-n_{\mathbf{k+p-q}})+N_B(\pm \omega _{\mathbf{p}})(n_{\mathbf{k}}-n_{%
\mathbf{k+p-q}})}{\omega +t_{\mathbf{k}}-t_{\mathbf{k+p-q}}\mp \omega _{%
\mathbf{p}}}
\]
(note that $\phi _{\mathbf{pq}\omega }^{+}=-\phi _{\mathbf{pq}-\omega }^{-},$
so that the corrections of order of $(1/2S)^2$ violate the equality $C_{%
\mathbf{q}\omega }=C_{\mathbf{q-}\omega }$) , $\omega _{\mathbf{p}}$ is the
magnon frequency to zeroth order in $I$ and $1/2S$. We have taken into
account in (G.76) the equations of motion of the type
$$
(\omega -t_{\mathbf{k}}+t_{\mathbf{k+q}})\langle \langle c_{\mathbf{%
k+q\uparrow }}^{\dagger }c_{\mathbf{k\uparrow }}|b_{\mathbf{q}}^{\dagger
}\rangle \rangle _\omega =I\left( \frac S2\right) ^{1/2}(n_{\mathbf{k+q}}-n_{%
\mathbf{k}})\langle \langle b_{\mathbf{q}}|b_{\mathbf{q}}^{\dagger }\rangle
\rangle _\omega
\eqno{(\rm G.79)}
$$
and the expressions for the static correlation functions that occurred in
the equations of motion
$$
I\sum_{\mathbf{k}}\langle c_{\mathbf{k-Q}\downarrow }^{\dagger }c_{\mathbf{%
k\uparrow }}\rangle =-S(J_{\mathbf{Q}0}^{tot}-J_{\mathbf{Q}})
\eqno{(\rm G.80)}
$$
$$
I\sum_{\mathbf{k}}\langle b_{\mathbf{p}}^{\dagger }(c_{\mathbf{k-p}\uparrow
}^{\dagger }c_{\mathbf{k\uparrow }}-c_{\mathbf{k-p}\downarrow }^{\dagger }c_{%
\mathbf{k\downarrow }})\rangle =-(2S)^{1/2}(C_{\mathbf{p}}-D_{\mathbf{p}%
})\Phi _{\mathbf{p}00}
\eqno{(\rm G.81)}
$$
These are obtained by calculating the corresponding retarded Green's
functions and using the spectral representation (E.8). The functions
$$
g_{\mathbf{q}} =\sum_{\mathbf{p}}[(2J_{\mathbf{Q}}+2J_{\mathbf{q-p}}-2J_{%
\mathbf{p}}-J_{\mathbf{Q+q}}-J_{\mathbf{q}})\langle b_{\mathbf{p}}^{\dagger
}b_{\mathbf{p}}\rangle -2J_{\mathbf{p}}\langle b_{-\mathbf{p}}b_{\mathbf{p}%
}\rangle ]
\eqno{(\rm G.82)}
$$
\[
h_{\mathbf{q}}=\sum_{\mathbf{p}}[(J_{\mathbf{Q+q}}-J_{\mathbf{q}})\langle b_{%
\mathbf{p}}^{\dagger }b_{\mathbf{p}}\rangle -2J_{\mathbf{q-p}}\langle b_{-%
\mathbf{p}}b_{\mathbf{p}}\rangle ]
\]
describe the ``direct'' magnon-magnon interaction,The s-d exchange
contributions to the averages in (G.82) are obtained by using (G.74), (G.75)
to the first order in $1/2S$ and the spectral representation
$$
\left\{
\begin{tabular}{l}
$\langle b_{\mathbf{q}}^{\dagger }b_{\mathbf{q}}\rangle $ \\
$\langle b_{\mathbf{q}}^{\dagger }b_{-\mathbf{q}}^{\dagger }\rangle $%
\end{tabular}
\right\} =-\frac 1\pi \int_{-\infty }^\infty d\omega N_B(\omega )\Im \left\{
\begin{array}{c}
\Gamma _{\mathbf{q}}(\omega ) \\
\bar \Gamma _{\mathbf{q}}(\omega )
\end{array}
\right\}
\eqno{(\rm G.83)}
$$

The energy denominators in (G.77), (G.78) do not take into account the band
splitting which comes from AFM ordering. At the same time, it is important
to separate the contributions of the transitions within and between AFM
subbands. Such a separation may be performed by taking into account the AFM
splitting $\Delta =2|I|\overline{S}$ ($\overline{S}$ is the sublattice
magnetization) in the zero-order approximation, e.g. within perturbation
theory in $1/2S.$ However, the corresponding expressions are very cumbersome
(see the end of this Appendix). Therefore we use the simple perturbation
theory in $I$ bearing in mind that the transitions between AFM subbands
correspond to the electron quasimonentum transfer $\mathbf{q}\sim
\mathbf{Q}$. Generally speaking, the intersubband contributions to the spectrum
characteristics and thermodynamic properties are more singular, but in fact
they should be cut off because of the AFM splitting. The corresponding
threshold value of the magnon quasimomentum transfer is estimated as
\[
\min |\mathbf{q-Q}|=q_0=\Delta /v_F
\]
($v_F$ is the electron velocity at the Fermi level). This quantity
determines a characteristic temperature and energy scale
$$
T^{*}=\omega (q_0)=cq_0\sim (\Delta /v_F)T_N
\eqno{(\rm G.84)}
$$
with $c$ being the magnon velocity and the magnon spectrum is given by the
pole of (G.74),
\[
\omega _{\mathbf{q}}^2=\Omega _{\mathbf{q}}^2(\omega _{\mathbf{q}})=C_{%
\mathbf{q}}^2(\omega _{\mathbf{q}})-D_{\mathbf{q}}^2(\omega _{\mathbf{q}})
\]

The dependence $J_{\mathbf{q}}^{\mathrm{RKKY}}(\omega )$, which is lost in
the standard method of canonical transformation [265], is important at
calculating the magnon damping. The spin-wave damping owing to one-magnon
decay processes, which is determined by imaginary part of (G.77), reads at
small $q$
$$
\gamma _{\mathbf{q}}^{(1)}=\pi S[\frac AL\omega _{\mathbf{q}}+B\psi (q)]
\eqno{(\rm G.85)}
$$
where $L=2S(J_0-J_{\mathbf{Q}}),$ the function $\psi $ describes entering
the ``Stoner continuum'', $\psi (q<q_0)=0,$ $\psi (q\gg q_0)=1$,
$$
A=cI^2\lim_{\mathbf{q}\rightarrow 0}q\sum_{\mathbf{k}}\delta (t_{\mathbf{k}%
})\delta (t_{\mathbf{k-q}})
\eqno{(\rm G.86)}
$$
$$
B=LI^2\sum_{\mathbf{k}}\delta (t_{\mathbf{k}})\delta (t_{\mathbf{k-Q}}),
\eqno{(\rm G.87)}
$$
$t_{\mathbf{k}}$ being referred to the Fermi level. Generally speaking, $A$
depends on the direction of the vector $\mathbf{q}$. For an isotropic
electron spectrum one has
$$
A=cI^2v_0\{4\pi ^2|k^{-1}\partial t_{\mathbf{k}}/\partial k|_{k=k_F}^2\}^{-1}
\eqno{(\rm G.88)}
$$
where $v_0$ is the lattice cell volume.

One can see that the one-magnon damping (G.85) is finite at arbitrarily
small $q$ (in contrast with the FM case), but becomes considerably larger
when intersubband transitions begin to work ($q>q_0$). Thus the ratio $%
\gamma /\omega |_{q\rightarrow 0}$ makes up about $I^2S/W^2$ and does not
depend on wavevector and electron concentration [689]. The same situation,
which is similar to the case of electron-phonon interaction, takes place for
an itinerant antiferromagnet [690]. The linear dependence of damping on
wavevector was observed, e.g., in the antiferromagnet Mn$_{0.9}$Cu$_{0.1}$
[691]. An account of relaxation of conduction electrons, which occurs at
disordering, results in a change of the $q$-dependence. The calculation in
such a situation [689] yields $\gamma \sim \omega ^2$ at small $q$, which is
in agreement with hydrodynamics.

Similar to (G.20), we obtain from (G.74), (G.77) and the spectral
representation the logarithmic correction to magnon occupation numbers and
sublattice magnetization (6.88), (6.89).

The damping owing to two-magnon scattering processes is determined by the
imaginary part of the function (G.78). The intersubband transitions turn out
to contain smaller powers of $\omega $ and $T$, but contribute at $\max
(T,\omega _{\mathbf{q}})>T^{*}$ only. Using the identity
\[
n(\epsilon )[1-n(\epsilon ^{\prime })]=N(\epsilon -\epsilon ^{\prime
})[n(\epsilon ^{\prime })-n(\epsilon )]\label{Nn}
\]
and expanding in $\omega =\omega _{\mathbf{q}}$ and $\omega _{\mathbf{p}}$
we obtain from the pole of (G.77)
$$
\gamma _{\mathbf{q}}^{(2)} =\frac \pi 2I^2\sum_{\mathbf{k,p}}\sum_{\alpha
,\beta =\pm }\left( \frac{C_{\mathbf{q}}-\alpha D_{\mathbf{q}}}\omega \frac{%
C_{\mathbf{p}}+\alpha D_{\mathbf{p}}}{\omega _{\mathbf{p}}}-\beta \right)
\eqno{(\rm G.89)}  \label{Gam2gen}
$$
\[
\times (\omega -\beta \omega _{\mathbf{p}})[N_B(\omega _{\mathbf{p}%
})-N_B(\omega _{\mathbf{p}}-\beta \omega )]\delta (t_{\mathbf{k}})\delta (t_{%
\mathbf{k+p-q}})
\]
Integration at $T\ll \omega $ with account of leading temperature
corrections gives
$$
\gamma _{\mathbf{q}}^{(2)}=\frac{v_0}{24\pi c^3}\left[ (\frac 95A\omega
+\tilde B)\omega ^2+4\pi ^2(2A\omega +\tilde B)T^2\right]
\eqno{(\rm G.90)}
$$
with $\tilde B(\omega \gg T^{*})=B,$ $\tilde B(\omega \ll T^{*})=0.$ At $%
\omega \ll T$ we find
$$
\gamma _{\mathbf{q}}^{(2)}=\frac{v_0}{2\pi c^3}\left[ 6\zeta (3)AT+\frac{\pi
^2}3\tilde B\right] T^2
\eqno{(\rm G.91)}
$$
with $\tilde B(T\gg T^{*})=B,$ $\tilde B(T\ll T^{*})=0,\zeta (3)\simeq 1.2$.

The non-analytic corrections to the sublattice magnetization and magnon
frequency come from both inter- and intrasubband transitions. The
intrasubband contribution to the magnon velocity reads [716]
$$
(\delta c/c)_1=-\frac{v_0}{3\pi c^3}AT^2\ln \frac{\overline{\omega }}T
\eqno{(\rm G.92)}
$$
For $T>T^{*}$ the intersubband contribution to the sublattice magnetization
has the form
$$
(\delta \overline{S}_{el})_2=-\frac{v_0}{\pi ^2c^3}SLBT^{*}\ln \frac T{T^{*}}
\eqno{(\rm G.93)}
$$

The above results are valid also in the Hubbard model. In the case of small $%
\overline{S}$ the magnon damping plays an important role at calculating
temperature dependences of magnetic and thermodynamic properties at not too
low temperatures are determined by contribution of spin fluctuations with
small $\mathbf{|q-Q|}$ and, as well as for a ferromagnet, the magnon damping
plays an important role. However, unlike (G.21), the damping at $\mathbf{%
q\rightarrow Q}$ does not contain the factor of $|\mathbf{q-Q|}^{-1}$
(However, such a dependence occurs in some $q$-region provided that the
electron spectrum approximately satisfies the ``nesting'' condition $t_{%
\mathbf{k+Q}}=-t_{\mathbf{k}}$ in a large part of the Fermi surface.) Then
we have to replace in the denominator of the Green's functions (G.74)
$$
\frac 1{\omega ^2-\Omega _{\mathbf{q}}^2(\omega )}\rightarrow \frac{A\omega
}{[\omega ^2-\Omega _{\mathbf{q}}^2(\omega )]^2+A^2\omega ^2}
\eqno{(\rm G.94)}
$$
which yields for a weak itinerant antiferromagnet
$$
\delta \overline{S}\sim -\int \mathrm{d}\omega N_B(\omega )\sum_{\mathbf{q}}%
\frac{A\omega }{\omega _{\mathbf{q}}^4+A^2\omega ^2}\sim -\left( \frac
T{E_F}\right) ^{3/2}
\eqno{(\rm G.95)}
$$

Now we consider the electron and magnon spectrum of the Hubbard
antiferromagnet with strong correlations. Occurence of AFM ordering results
in splitting the bare electron band into two Slater subbands [44] which are
described by new electron operators [692,693]
$$
\alpha _{\mathbf{k}}^{\dagger } =A_{\mathbf{k}}c_{\mathbf{k+Q/}2\mathbf{%
\uparrow }}^{\dagger }+B_{\mathbf{k}}c_{\mathbf{k-Q/}2\mathbf{\downarrow }%
}^{\dagger },
\eqno{(\rm G.96)}
$$
\[
\beta _{\mathbf{k}}^{\dagger }=A_{\mathbf{k}}c_{\mathbf{k-Q/}2\mathbf{%
\downarrow }}^{\dagger }-B_{\mathbf{k}}c_{\mathbf{k+Q/}2\mathbf{\uparrow }%
}^{\dagger },
\]
\[
A_{\mathbf{k}},B_{\mathbf{k}}=\frac 12\left( 1\mp \frac{\tau _{\mathbf{k}}}{%
E_{\mathbf{k}}}\right) ,\qquad E_{\mathbf{k}}=(\tau _{\mathbf{k}}^2+U^2%
\overline{S}^2)^{1/2}
\]
\[
\theta _{\mathbf{k}},\tau _{\mathbf{k}}=\frac 12(t_{\mathbf{k+Q/}2}\pm t_{%
\mathbf{k-Q/}2})
\]
In the Hartree-Fock approximation, the transformation (G.96) reduces the
Hubbard Hamiltonian to the diagonal form
$$
\mathcal{H}=\sum_{\mathbf{k}}(E_{\mathbf{k}}^\alpha \alpha _{\mathbf{k}%
}^{\dagger }\alpha _{\mathbf{k}}+E_{\mathbf{k}}^\beta \beta _{\mathbf{k}%
}^{\dagger }\beta _{\mathbf{k}})
\eqno{(\rm G.97)}
$$
where the one-particle energies are given by
$$
E_{\mathbf{k}}^{\alpha ,\beta }=\theta _{\mathbf{k}}\mp E_{\mathbf{k}}
\eqno{(\rm G.98)}
$$
The quantity
\[
\overline{S}=\sum_{\mathbf{k}}\langle c_{\mathbf{k\uparrow }}^{\dagger }c_{%
\mathbf{k+Q\downarrow }}\rangle
\]
which determines the AFM splitting, satisfies the self-consistency equation
$$
1=\frac U2\sum_{\mathbf{k}}\frac{n_{\mathbf{k}\alpha }-n_{\mathbf{k}\beta }}{%
(\tau _{\mathbf{k}}^2+U^2\overline{S}^2)^{1/2}}
\eqno{(\rm G.99)}
$$
If the Coulomb interaction is strong enough, the whole energy band is split,
so that the gap occurs in the all directions. In particular, for one
electron per atom, a metal-insulator transition takes place. Provided that
the ``nesting'' condition $t_{\mathbf{k}}-E_F=E_F-t_{\mathbf{k+Q}}$ holds
for a given vector $\mathbf{Q}$, the insulator state is favourable for
arbitrarily small $U$, the gap being exponentially small [692].

To obtain the spin-wave corrections to the Hartree-Fock approximation we
pass to local coordinate system for electron operators
$$
d_{\mathbf{k}\sigma }^{\dagger }=\frac 1{\sqrt{2}}(c_{\mathbf{k+Q/}2\mathbf{%
\uparrow }}^{\dagger }+\sigma c_{\mathbf{k-Q/}2\mathbf{\downarrow }%
}^{\dagger })
\eqno{(\rm G.100)}
$$
and represent the Hubbard Hamiltonian in the form
$$
\mathcal{H}=\sum_{\mathbf{k}\sigma }[(\theta _{\mathbf{k}}+\frac U2)d_{%
\mathbf{k}\sigma }^{\dagger }d_{\mathbf{k}\sigma }+\tau _{\mathbf{k}}d_{%
\mathbf{k}\sigma }^{\dagger }d_{\mathbf{k-}\sigma }]-\frac U2\sum_{\mathbf{q}%
}(S_{-\mathbf{q}}^{-}S_{\mathbf{q}}^{+}+S_{\mathbf{q}}^{+}S_{-\mathbf{q}%
}^{-})
\eqno{(\rm G.101)}
$$
where
$$
S_{\mathbf{q}}^\sigma =\sum_{\mathbf{k}}d_{\mathbf{k}\sigma }^{\dagger }d_{%
\mathbf{k+q,-}\sigma },\qquad S_{\mathbf{q}}^z=\frac 12\sum_{\mathbf{k}%
\sigma }\sigma d_{\mathbf{k}\sigma }^{\dagger }d_{\mathbf{k+q,}\sigma }
\eqno{(\rm G.102)}
$$
Calculating the transverse spin Green's function with antiferromagnetic gap
being taken into account in the zero-order order approximation we obtain for
the magnon spectrum [693]
$$
\omega _{\mathbf{q}}^2 =\frac{U^4\overline{S}^2}4\sum_{\mathbf{k}}\left[
\frac 2{E_{\mathbf{k}}}(n_{\mathbf{k}\alpha }-n_{\mathbf{k}\beta })+\left( 1-%
\frac{U^2\overline{S}^2}{E_{\mathbf{k}}E_{\mathbf{k+q}}}\right) \right.
\eqno{(\rm G.103)}
$$
\[
\times \left. \left( \frac{n_{\mathbf{k}\alpha }-n_{\mathbf{k+q}\alpha }}{E_{%
\mathbf{k}}^\alpha -E_{\mathbf{k+q}}^\alpha }+\frac{n_{\mathbf{k}\beta }-n_{%
\mathbf{k+q}\beta }}{E_{\mathbf{k}}^\beta -E_{\mathbf{k+q}}^\beta }\right)
+2\left( 1+\frac{U^2\overline{S}^2}{E_{\mathbf{k}}E_{\mathbf{k+q}}}\right)
\left( \frac{n_{\mathbf{k}\alpha }-n_{\mathbf{k+q}\beta }}{E_{\mathbf{k}%
}^\alpha -E_{\mathbf{k+q}}^\beta }\right) \right] ^2
\]
\[
-\frac{U^4\overline{S}^2}4\left [\sum_{\mathbf{k}}\frac{\tau _{\mathbf{k}}\tau _{%
\mathbf{k+q}}}{E_{\mathbf{k}}E_{\mathbf{k+q}}}\left( \frac{n_{\mathbf{k}%
\alpha }-n_{\mathbf{k+q}\alpha }}{E_{\mathbf{k}}^\alpha -E_{\mathbf{k+q}%
}^\alpha }+\frac{n_{\mathbf{k}\beta }-n_{\mathbf{k+q}\beta }}{E_{\mathbf{k}%
}^\beta -E_{\mathbf{k+q}}^\beta }-2\frac{n_{\mathbf{k}\alpha }-n_{\mathbf{k+q%
}\beta }}{E_{\mathbf{k}}^\alpha -E_{\mathbf{k+q}}^\beta }\right) \right ]^2
\]
For small $U$ we may expand (G.103) to obtain the result (G.74)-(G.77) with $%
I\rightarrow U.$ In the case of large $U$ and half-filled conduction band we
obtain the result (E.12) with
$$
J_{\mathbf{q}}=-\frac 2U\sum_{\mathbf{k}}t_{\mathbf{k}}t_{\mathbf{k+q}}
\eqno{(\rm G.104)}
$$
(G.94) being the kinetic exchange integral (see Sect.5.1). Finally, in the
case of $U\rightarrow \infty $ and non-integer band filling, where the
current carriers result in the non-Heisenbergian double exchange
interaction, we obtain
\[
\omega _{\mathbf{q}}^2=\left\{ J_{\mathbf{Q}}^H-J_{\mathbf{q}}^H+\sum_{%
\mathbf{k}}\left[ (\Theta _{\mathbf{k+q}}-\Theta _{\mathbf{k}})n_{\mathbf{k}%
}+\tau _{\mathbf{k}}(\tau _{\mathbf{k+q}}+\tau _{\mathbf{k}})\frac{n_{%
\mathbf{k+q}}-n_{\mathbf{k}}}{\Theta _{\mathbf{k+q}}-\Theta _{\mathbf{k}}}%
\right] \right\}
\]
$$
\times \left\{ \frac 12(J_{\mathbf{Q}}^H-J_{\mathbf{Q+q}}^H-J_{\mathbf{Q-q}%
}^H)\right.
\eqno{(\rm G.105)}
$$
\[
\left. +\sum_{\mathbf{k}}\left[ (\Theta _{\mathbf{k+q}}-\Theta _{\mathbf{k}%
})n_{\mathbf{k}}-\tau _{\mathbf{k}}(\tau _{\mathbf{k+q}}-\tau _{\mathbf{k}})%
\frac{n_{\mathbf{k+q}}-n_{\mathbf{k}}}{\Theta _{\mathbf{k+q}}-\Theta _{%
\mathbf{k}}}\right] \right\}
\]
where
\[
n_{\mathbf{k}}=n_{\mathbf{k}\beta }=f(\Theta _{\mathbf{k}}),\qquad
n_{\mathbf{k}\alpha }\equiv 1
\]
and the Heisenberg interaction $J^H$ is introduced to stabilize the
antiferromagnetic state.

The spin-wave corrections to electron spectrum are determined, as well as
for the ferromagnet (cf.(G.13), (G.36), (G.60)) by the electron-magnon
interaction amplitude
$$
\frac{\delta E_{\mathbf{k}}^i}{\delta N_{\mathbf{q}}}=\frac{\delta \omega _{%
\mathbf{k}}}{\delta n_{\mathbf{k}i}}=\frac{\delta ^2\langle \mathcal{H}%
_{sw}\rangle }{\delta N_{\mathbf{q}}\delta n_{\mathbf{k}i}}
\eqno{(\rm G.106)}
$$
The corresponding correction to the free energy reads
$$
\delta \mathcal{F}_{sw}=-\frac 13\delta \langle \mathcal{H}_{sw}\rangle
=-\frac 13\sum_{\mathbf{q}}\omega _{\mathbf{q}}N_{\mathbf{q}}\sim -\left(
\frac T{T_N}\right) ^4
\eqno{(\rm G.107)}
$$
(the difference with (G.60) is due to the linear dispersion law of spin
waves). The temperature dependence of the local moment is given by the
Hellmann-Feynman theorem:
$$
\delta \langle \mathbf{S}_i^2\rangle =-\frac 32\delta N_2=-\frac 32\frac
\partial {\partial U}\delta \mathcal{F}_{sw}\sim T^4
\eqno{(\rm G.108)}
$$

The considered Hartree-Fock-type approximation with fluctuation corrections
yields the same results for electron and magnon spectra in the case of s-d
model (with the replacement $U\rightarrow I$). In particular, for $%
|I|\rightarrow \infty $ the electron spectrum with account of spin-wave and
many-electron corrections reads [693]
\[
E_{\mathbf{k}}^\alpha =-I(S+n_\beta ),\qquad
E_{\mathbf{k}}^\beta =-I(S+1-n_\alpha )
\]
Thus the generalized Hatree-Fock approximation yields correct ``atomic''
values $E=\pm IS$, $\pm I(S+1)$ (Appendix I) for integer values $n_\alpha $,
$n_\beta $.


\chapter{The Hubbard model with strong correlations}

The Hubbard model with a degenerate band is applicable in some cases for
d-electrons in transition metals and their compounds. The corresponding
Hamiltonian has the form
$$
\mathcal{H}=\sum_{\mathbf{k}m\sigma }t_{\mathbf{k}}a_{\mathbf{k}lm\sigma
}^{\dagger }a_{\mathbf{k}lm\sigma }+\mathcal{H}_{\mathrm{int}}
\eqno{(\rm H.1)}
$$
where $t_{\mathbf{k}}$ is the band energy. For simplicity, we do not take
into account $m$-dependence of transfer integrals, i.e. we neglect
crystal-field effects and retain in (C.31) the contribution with $\lambda =0$
only. Such an approximation enables one to treat in the simplest situation
effects of many-electron term structure in the electron spectrum. In the
many-electron representation of Hubbard's operators (A.22) the interaction
Hamiltonian takes the diagonal form
$$
\mathcal{H}_{\mathrm{int}}=\sum_{i\Gamma }E_\Gamma X_i(\Gamma ,\Gamma )
\eqno{(\rm H.2)}
$$
where the energies $E_\Gamma $ are given by (C.19) and do not depend on
momentum projections. However, the treatment of the kinetic energy term
becomes more difficult because of complicated commutation relations for $X$%
-operators (A.36).

Consider the one-electron Green's function. According to (A.31),
$$
G_{\mathbf{k}\gamma }(E)=\langle \langle a_{\mathbf{k}\gamma }|a_{\mathbf{k}%
\gamma }^{\dagger }\rangle \rangle _E=\sum_{n\Gamma _n\Gamma
_{n-1}}n^{1/2}G_{\Gamma _{n-1}}^{\Gamma _n}C_{\Gamma _{n-1},\gamma }^{\Gamma
_n}\langle \langle X_{\mathbf{k}}(\Gamma _{n-1},\Gamma _n)|a_{\mathbf{k}%
\gamma }^{\dagger }\rangle \rangle _E
\eqno{(\rm H.3)}
$$
In the equation of motion for the Green's function in the right-hand side of
(H.3) we perform the simplest decoupling which corresponds to the
``Hubbard-I'' decoupling [28,29,31] at different lattice sites:
\[
(E-E_{\Gamma _n}+E_{\Gamma _{n-1}})\langle \langle X_{\mathbf{k}}(\Gamma
_{n-1},\Gamma _n)|a_{\mathbf{k}\gamma }^{\dagger }\rangle \rangle _E
\]
\[
=n^{1/2}G_{\Gamma _{n-1}}^{\Gamma _n}C_{\Gamma _{n-1},\gamma }^{\Gamma
_n}(N_{\Gamma _n}+N_{\Gamma _{n-1}})\left[ 1+t_{\mathbf{k}}G_{k\gamma
}(E)\right]
\]
Then we obtain
$$
G_{\mathbf{k}\gamma }(E)=\frac{\Phi _\gamma (E)}{1-t_{\mathbf{k}}\Phi
_\gamma (E)}
\eqno{(\rm H.4)}
$$
where
$$
\Phi _\gamma (E)=\sum_{n\Gamma _n\Gamma _{n-1}}n\left( G_{\Gamma
_{n-1}}^{\Gamma _n}C_{\Gamma _{n-1},\gamma }^{\Gamma _n}\right) ^2\frac{%
N_{\Gamma _n}+N_{\Gamma _{n-1}}}{E-E_{\Gamma _n}+E_{\Gamma _{n-1}}}
\eqno{(\rm H.5)}
$$
Note that the expressions (H.4), (H.5) have the structure which is
reminiscent of (F.8),(F.9). As well as in the Appendix F, our consideration
may be easily generalized to include effects of single-site crystal field
(see also [29]). If we use for $E_\Gamma $ the approximation (C.20), the
fractional parentage coefficients in (H.5) may be summed up and the
dependence on ME quantum numbers $L,S$ vanishes, which corresponds to the
approximation [29].

In the absence of magnetic and orbital ordering the occupation numbers $%
N_\Gamma $ in (H.5) do not depend on spin projections and we have
$$
\Phi _\Gamma (E)=\sum_{n\Gamma _n\Gamma _{n-1}}\frac
n{2[l]}([S_{n-1}][L_{n-1}])^{-1}(G_{\Gamma _{n-1}}^{\Gamma _n})^2\frac{%
N_{\Gamma _n}+N_{\Gamma _{n-1}}}{E-E_{\Gamma _n}+E_{\Gamma _{n-1}}}
\eqno{(\rm H.6)}
$$

The excitation spectrum is given by
$$
1-t_{\mathbf{k}}\Phi _\gamma (E)=0
\eqno{(\rm H.7)}
$$
Thus the intersite electron transfer results in a smearing of each
transition between atomic levels into an energy Hubbard subband. These
subbands are separated by correlation gaps. In particular, for $s$-band we
obtain the spectrum which contains in ferromagnetic region four subbands
$$
E_{\mathbf{k}\sigma }^{1,2}=\frac 12\left\{ t_{\mathbf{k}}+U\mp \left[ (t_{%
\mathbf{k}}-U)^2+4t_{\mathbf{k}}U(N_{-\sigma }+N_2)\right] ^{1/2}\right\}
\eqno{(\rm H.8)}
$$
For a more general model (C.23),
$$
E_{\mathbf{k}\sigma }^{1,2}=\frac 12\left[ \beta _{\mathbf{k}%
}^{(00)}(N_0+N_\sigma )+\beta _{\mathbf{k}}^{(22)}(N_2+N_{-\sigma
})+U\right] \mp \frac 12\left\{ \left[ \beta _{\mathbf{k}}^{(00)}(N_0+N_%
\sigma )\right. \right.
\eqno{(\rm H.9)}
$$
\[
-\left. \left. \beta _{\mathbf{k}}^{(22)}(N_2+N_{-\sigma })-U\right]
^2+4\left| \beta _{\mathbf{k}}^{(02)}\right| ^2(N_0+N_\sigma
)(N_2+N_{-\sigma })\right\} ^{1/2}
\]
The expression (H.8), (H.9) may be also rewritten in terms of one-electron
occupation numbers since
\[
N_\sigma +N_2=n_\sigma ,N_\sigma +N_0=1-n_{-\sigma }
\]
One can see that the dependence of the spectrum on the occupation numbers
does not reduce to the constant shift of subbands, as it takes place in the
Stoner approximation (G.3). The Hubbard-I spectrum has the most simple form
in the case of large $U$ where
$$
E_{\mathbf{k}\sigma }^1=(1-n_{-\sigma })t_{\mathbf{k}},\qquad E_{\mathbf{k}%
\sigma }^2=t_{\mathbf{k}}n_{-\sigma }+U
\eqno{(\rm H.10)}
$$

One can assume that some subbands are in fact ill-defined because of large
damping. Indeed, it is demonstrated in Appendix J for the saturated
ferromagnetic state that in higher-order approximations some energy
denominators are replaced by resolvents, and the corresponding states have a
non-quasiparticle nature. A similar situation takes place in the Hubbard-III
approximation [30,694,695] where damping is finite and large at Fermi level
(see (H.17), (J.24)).

Consider the Hubbard model with $U\rightarrow \infty $, $n=1-c<1$ ($%
c=N_0=n_0 $ is the hole concentration, $N_2=0$) with inclusion of the
external magnetic field (4.96). In the ``Hubbard-I'' approximation'' we
obtain the Green's function
$$
\langle \langle X_{\mathbf{k}}(\sigma 0)|X_{-\mathbf{k}}(0\sigma )\rangle
\rangle _E =\frac{c+n_\sigma }{E-\tau _{\mathbf{k}\sigma }-\sigma H/2}
\eqno{(\rm H.11)}
$$
\[
\tau _{\mathbf{k}\sigma }=(c+n_\sigma )\varepsilon _{\mathbf{k}%
}=[(1+c)/2+\sigma \langle S^z\rangle ]\varepsilon _{\mathbf{k}}
\]
and, with the use of the spectral representation (E.18), the corresponding
expressions for the occupation numbers
$$
n_{\mathbf{k}\sigma }\equiv \langle X_{-\mathbf{k}}(0\sigma )X_{\mathbf{k}%
}(\sigma 0)\rangle =\left( \frac{1+c}2+\sigma \langle S^z\rangle \right)
f(\tau _{\mathbf{k}\sigma }+\frac 12h\sigma )
\eqno{(\rm H.12)}
$$

In principle, the equations (H.4), (H.5) may be used to investigate the
magnetic ordering in the Hubbard model. However, the Hubbard-I approximation
is hardly satisfactory in this problem since it is difficult to formulate a
reasonable criterion of ferromagnetism by direct using the expressions for
one-electron Green's functions like (H.4), (H.11). The first attempt of this
kind was made by Hubbard [28], who found no magnetic solutions for simple
bare densities of states (however, the situation may change considerably in
the case of degenerate d-bands [696]). The $X$-operator approach clarifies
the causes of this failure. In particular, one can see that the
approximation (H.11) violates the kinematical requirements (A.25) since it
is impossible to satisfy at $\langle S^z\rangle \neq 0$ the relation
$$
\sum_{\mathbf{k}}n_{\mathbf{k}\sigma }=\langle X(00)\rangle =c
\eqno{(\rm H.13)}
$$
for both spin projections $\sigma $. Besides that, as discussed above, such
Stoner-like approaches do not describe the formation of LMM and are
physically unsatisfactory. Therefore we use in Sect 4.5 the approach which
is based on the spin Green's function
$$
G_{\mathbf{q}}(\omega )=\langle \langle X_{\mathbf{q}}(+-)|X_{-\mathbf{q}%
}(-+)\rangle \rangle _\omega =\langle \langle S_{\mathbf{q}}^{+}|S_{-\mathbf{%
q}}^{-}\rangle \rangle _\omega
\eqno{(\rm H.14)}
$$
To calculate this we write down the sequence of equations of motion
$$
(\omega -H)G_{\mathbf{q}}(\omega ) =n_{\uparrow }-n_{\downarrow }+\sum_{%
\mathbf{k}}(\varepsilon _{\mathbf{k-q}}-\varepsilon _{\mathbf{k}})
\eqno{(\rm H.15)}
$$
\[
\times \langle \langle X_{\mathbf{q-k}}(0-)X_{\mathbf{k}}(+0)|X_{-%
\mathbf{q}}(-+)\rangle \rangle _\omega
\]
\[
(\omega -\tau _{\mathbf{k}\uparrow }+\tau _{\mathbf{k-q}\uparrow
}-H)\langle \langle X_{\mathbf{q-k}}(0-)X_{\mathbf{k}}(+0)|X_{-\mathbf{q}%
}(-+)\rangle \rangle _\omega
\]
\[
=n_{\mathbf{k}\uparrow }-n_{\mathbf{k-q}\downarrow }+(\varepsilon _{%
\mathbf{k-q}}n_{\mathbf{k-q}\downarrow }-\varepsilon _{\mathbf{k}}n_{\mathbf{%
k}\uparrow })G_{\mathbf{q}}(\omega )
\]
where we have performed the simplest decoupling, which corresponds to
neglecting fluctuations of hole occupation numbers. Substituting (H.15) into
(H.14) we obtain the expression (4.97).

Fluctuation correction to the electron spectrum (H.4) for a ferromagnet with
$s$-band were investigated in [337,338] within the $1/z$-expansion, $z$
being the nearest-neighbour number. They are expressed in terms of
one-particle occupation numbers and spin and charge correlation functions.
For the degenerate Hubbard model, correlation functions of orbital operators
$L^{(k)}$ will also occur.

For paramagnetic phase, the gap in the spectrum (H.8) persists for
arbitrarily small $U$. To describe the metal-insulator transition which
takes place at $U\sim W$ (with $W$ being the bandwidth) more complicated
self-consistent approximations for the one-electron Green's functions were
used. First description of this type was proposed by Hubbard [30], and a
more simple approximation was used by Zaitsev [697].

The Hubbard-III expression for the one-electron Green's function in the case
of half-filled band may be represented in the form [695]
$$
G_{\mathbf{k}}(E)=\left[ E-t_{\mathbf{k}}-\Sigma (E)\right] ^{-1}
\eqno{(\rm H.16)}
$$
the electron self-energy being determined in a self-consistent way in terms
of the exact resolvent
$$
\Sigma (E) =\frac{U^2}{16\Psi }R(E)/\left[ 1+\Sigma (E)R(E)+ER(E)(\frac
1{4\Psi }-1)\right]
\eqno{(\rm H.17)}
$$
\[
R(E)=\sum_{\mathbf{k}}G_{\mathbf{k}}(E)
\]
where $\Psi =3/4$. The expression (H.17) holds also for the classical ($S$ $%
\rightarrow \infty $) s-d exchange model (see Appendix I) if we put $\Psi
=1/4$, $U\rightarrow |IS|$. Then (H.17) is simplified and coincides with
the coherent potential approximation (CPA) result in the disordered alloy
theory [435]. Evolution of electron spectrum vs. interaction parameter is
shown in Fig.H1.

Some shortcomings of the approximations [30,697] (violence of analytical
properties of the Green's functions and inconsistent description of
thermodynamic quantities) are discussed in [694,695] from the point of view
of the 1/z-expansion. Construction of a correct physical pictrure of the
Mott-Hubbard transition is up to now a serious physical problem. Last time,
the large-$d$ ($d$ is the space dimensionality) approximation is widely used
in this problem [705]. Such approaches yield sometimes two phase
transitions: at $U>U_{c1}$ the Fermi-liquid picture breaks, and at $%
U>U_{c2}>U_{c1}$ the system passes into the insulator state. We have seen
that the one-electron (Hartree-Fock) and Hubbard approaches yield
essentially different results for electron spectrum.

The failure of one-electron approach in the case of large $U$ may be
demonstrated by treatment of the case of small electron concentrations $n$
[353]. Consider the expansion of the one-electron Green's function in the
electron occupation numbers. In the one-electron representation ($f_{\mathbf{%
k}}$ = $\langle a_{\mathbf{k}\sigma }^{\dagger }a_{\mathbf{k}\sigma }\rangle
$) we obtain the equation of motion
$$
(E-t_{\mathbf{k}})\langle \langle a_{\mathbf{k}\uparrow }|a_{\mathbf{k}%
\uparrow }^{\dagger }\rangle \rangle _E =1+U\sum_{\mathbf{k}_1\mathbf{k}%
_2}F_{\mathbf{k}}(\mathbf{k}_1\mathbf{k}_2E)
\eqno{(\rm H.18)}
$$
\[
F_{\mathbf{k}}(\mathbf{k}_1\mathbf{k}_2E)=\langle \langle a_{\mathbf{k}%
_1\downarrow }^{\dagger }a_{\mathbf{k}_2\downarrow }a_{\mathbf{k+k}_1\mathbf{%
-k}_2\uparrow }|a_{\mathbf{k}\uparrow }^{\dagger }\rangle \rangle _E
\]
To lowest-order in $f_{\mathbf{k}}$ we obtain the closed integral equation
$$
(E-t_{\mathbf{k+k}_1\mathbf{-k}_2}+t_{\mathbf{k}_1}-t_{\mathbf{k}_2})F_{%
\mathbf{k}}(\mathbf{k}_1,\mathbf{k}_2,E)
\eqno{(\rm H.19)}
$$
\[
=\delta _{\mathbf{k}_1\mathbf{k}_2}f_{\mathbf{k}}+U(1-f_{\mathbf{k}_2})\sum_{%
\mathbf{p}}F_{\mathbf{k}}(\mathbf{k,p,}E)
\]
Solving this we derive the expression for the electron self-energy
$$
\Sigma _{\mathbf{k}}(E)=U\sum_{\mathbf{k}_1}f_{\mathbf{k}_1}\left[ 1-U\sum_{%
\mathbf{k}_2}\frac{(1-f_{\mathbf{k}_2})(1-f_{\mathbf{k+k}_1\mathbf{-k}_2})}{%
E-t_{\mathbf{k+k}_1\mathbf{-k}_2}+t_{\mathbf{k}_1}-t_{\mathbf{k}_2}}\right]
^{-1}
\eqno{(\rm H.20)}
$$
which is finite at $U\rightarrow \infty $ [348]. At the same time, according
to (E.18), (H.18) the number of doubles
$$
N_2 =\langle a_{i\uparrow }^{\dagger }a_{i\uparrow }a_{i\downarrow
}^{\dagger }a_{i\downarrow }\rangle =-\frac 1\pi \int dEf(E)\Im \sum_{%
\mathbf{kk}_1\mathbf{k}_2}F_k(\mathbf{k}_1\mathbf{k}_2E)
\eqno{(\rm H.21)}
$$
\[
=-\frac 1{\pi U}\int dEf(E)\Im \sum_{\mathbf{k}}\frac{\Sigma _{\mathbf{k}}(E)%
}{E-t_{\mathbf{k}}}
\]
behaves as $1/U$ in this limit. Then the Hellmann-Feynman theorem $%
N_2=\partial \mathcal{E}/\partial U$ yields a divergence of the ground-state
energy $\mathcal{E}$%
$$
\mathcal{E}(U)-\mathcal{E}(0)=\int_0^\infty dUN_2(U)\sim \ln U
\eqno{(\rm H.22)}
$$
This divergence indicates formation of the Hubbard subbands and inadequacy
of the one-electron picture at large $U$. On the other hand, calculation in
the ME representation yields [353]
$$
\langle \langle X_{\mathbf{k}}(\sigma 2)|X_{-\mathbf{k}}(2\sigma )\rangle
\rangle _E\simeq \sum_{\mathbf{k}_1}f_{\mathbf{k}_1}\left[ E+t_{\mathbf{k}%
_1}-U-\sum_{\mathbf{k}_2}\frac{(t_{\mathbf{k+k}_1\mathbf{-k}_2}+t_{\mathbf{k}%
_2})^2}{E-t_{\mathbf{k+k}_1\mathbf{-k}_2}+t_{\mathbf{k}_1}-t_{\mathbf{k}_2}}%
\right] ^{-1}
\eqno{(\rm H.23)}
$$
so that
$$
N_2 =-\frac 1\pi \Im \sum_{\mathbf{k}}\int dEf(E)\Im \langle \langle X_{%
\mathbf{k}}(\sigma 2)|X_{-\mathbf{k}}(2\sigma )\rangle \rangle _E
\eqno{(\rm H.24)}
$$
\begin{eqnarray*}
&\simeq &\frac 1{U^2}\sum_{\mathbf{kk}_1\mathbf{k}_2}(t_{\mathbf{k+k}_1%
\mathbf{-k}_2}+t_{\mathbf{k}_2})^2f_{\mathbf{k}_1}f(t_{\mathbf{k+k}_1\mathbf{%
-k}_2}-t_{\mathbf{k}_1}+t_{\mathbf{k}_2}) \\
&\simeq &\frac{n^2}{U^2}t_{\min }
\end{eqnarray*}
Thus we obtain the correct asymptotics $N_2\sim 1/U^2$.

\chapter{Narrow-band s-d exchange model and t-J model}

At considering the electron transfer in narrow degenerate bands, one can
use, besides the Hubbard model, the s-d exchange model with strong
correlations. This model corresponds to the case where the current carriers
do not belong to the same energy band where magnetic moments are formed.
Such a situation takes place in some magnetic semiconductors and insulators
[668].

Unlike the degenerate Hubbard model, the standard Hamiltonian of the s-d
exchange model (G.2) does not include orbital degrees of freedom. In the
case of large s-d exchange parameter $|I|$ it is convenient to pass to the
atomic representation [698-700]. Substituting the values of the
Clebsh-Gordan coefficients, corresponding to the coupling of momenta $S$ and
1/2, we obtain the eigenfunctions of $\mathcal{H}_{sd}$
$$
|M\rangle \equiv |SM\rangle |0\rangle ,\qquad |M2\rangle \equiv |SM\rangle
|2\rangle
\eqno{(\rm I.1)}
$$
$$
|\mu \pm \rangle =\left( \frac{S\pm \mu +1/2}{2S+1}\right) ^{1/2}|S,\mu
-\frac 12\rangle |\uparrow \rangle \pm \left( \frac{S\mp \mu +1/2}{2S+1}%
\right) ^{1/2}|S,\mu +\frac 12\rangle |\downarrow \rangle
\eqno{(\rm I.2)}
$$
where $|m\alpha \rangle $ are the singly-occupied states with the total
on-site spin $S+\alpha /2$ and its projection m. Then $\mathcal{H}_{sd}$
takes the diagonal form
$$
\mathcal{H}_{sd}=-IS\sum_{\mu =-S-1/2}^{S+1/2}\sum_iX_i(\mu +,\mu
+)+I(S+1)\sum_{\mu =-S+1/2}^{S-1/2}\sum_iX_i(\mu -,\mu -)
\eqno{(\rm I.3)}
$$
The one-electron operators are expressed in terms of $X$-operators as
$$
c_{i\sigma }^{\dagger }=\sum_\alpha (g_{i\sigma \alpha }^{\dagger
}+h_{i\sigma \alpha }^{\dagger })
\eqno{(\rm I.4)}
$$
\[
g_{i\sigma +}^{\dagger }=\sum_M\left\{ (S+\sigma M+1)/(2S+1)\right\}
^{1/2}X_i(M+\frac \sigma 2,+;M)
\]
\[
g_{i\sigma -}^{+}=\sum_M\sigma \left\{ (S-\sigma M)/(2S+1)\right\}
^{1/2}X_i(M+\frac \sigma 2,-;M)
\]
\[
h_{i\sigma +}^{\dagger }=\sum_M\left\{ (S+\sigma M)/(2S+1)\right\}
^{1/2}X_i(M2;M-\frac \sigma 2,-)
\]
\[
h_{i\sigma +}^{\dagger }=\sum_M\sigma \left\{ (S-\sigma M+1)/(2S+1)\right\}
^{1/2}X_i(M2;M-\frac \sigma 2,+)
\]
In the limit $I$ $\rightarrow \alpha \infty $ and for conduction electron
concentration $n<1$ one has to retain in (I.4) only the terms with $%
g_{i\alpha }$ and omit the Hamiltonian $\mathcal{H}_{sd}$, which yields a
constant energy shift, to obtain
$$
\mathcal{H}=\sum_{\mathbf{k}\sigma }t_{\mathbf{k}}g_{\mathbf{k}\sigma \alpha
}^{\dagger }g_{\mathbf{k}\sigma \alpha }+\mathcal{H}_d,\qquad \alpha =%
\mathrm{sign}I
\eqno{(\rm I.5)}
$$
For $n>1$, we have to retain only terms with $h_{i\alpha }$ and pass to the
``hole'' representation by introducing new localized spins $\widetilde{S}%
=S\pm 1/2$. Then the Hamiltonian takes the same form (I.6) with the
replacement [700]
$$
t_{\mathbf{k}}\rightarrow -t_{\mathbf{k}}([\widetilde{S}]/[S])
\eqno{(\rm I.6)}
$$
At theoretical consideration of higly-correlated compounds, e.g.
copper-oxide high-$T_c$ superconductors the $t-J$ model (the $s$-band
Hubbard model with $U\rightarrow \infty $ and Heisenberg exchange included)
is now widely used. Its Hamiltonian in ME representation reads
$$
\mathcal{H}=-\sum_{ij\sigma }t_{ij}X_i(0\sigma )X_j(\sigma
0)-\sum_{ij}J_{ij}\left\{ X_i(+-)X_j(-+)\right.
\eqno{(\rm I.7)}
$$
\[
+\left. \frac 14\left[ X_i(++)-X_i(--)\right] \left[ X_j(++)-X_j(--)\right]
\right\}
\]
At derivation of the $t-J$ model from the large-$U$ Hubbard model, $%
J=-4t^2/U $ is the antiferromagnetic kinetic exchange integral. However, it
is convenient sometimes J as an independent variable. In particular, one
treats sometimes the ``supersymmetrical'' case with $t=J$ [701] which
permits to use non-trivial mathematical methods.

One can see that the model (I.7) is a particular case of the s-d exchange
model, corresponding to $I\rightarrow -\infty $, $S=1/2$, $t_k$ in (I.5)
being replaced by $2t$ (the factor of $2$ occurs because of equivalence of
both electrons in the Hubbard model). The s-d model with arbitrary $S$ is
sometimes more convenient since this enables one to use at calculations,
besides the small parameter $1/z$ ($z$ is the nearest- neighbour number),
the quasiclassical parameter $1/2S$. The quasiclassical s-d model with $S\gg
1$ in the atomic representation was used to investigate the metal-insulator
transition [695] (Appendix H).

The Hamiltonian (I.5), similar to (C.33), may be expressed in terms of the
Fermi and localized spin operators. To this end, we pick out a conduction
electron operator from X-operators with the use of (A.21), (A.11):
$$
g_{i\sigma \alpha }^{\dagger }=\sum_{\sigma ^{\prime }}c_{i\sigma ^{\prime
}}^{\dagger }(1-n_{i,-\sigma ^{\prime }})\left[ P_\alpha \delta _{\sigma
\sigma ^{\prime }}+\frac \alpha {2S+1}(\mathbf{S}_i\mathbf{\sigma }_{\sigma
\sigma ^{\prime }})\right]
\eqno{(\rm I.8)}
$$
where
$$
P_{+}=\frac{S+1}{2S+1},\qquad P_{-}=\frac S{2S+1}
\eqno{(\rm I.9)}
$$
This result was obtained by Kubo and Ohata [702] by a canonical
transformation. Using properties of the Pauli matrices we get
$$
\mathcal{H} =\sum_{ij\sigma \sigma ^{\prime }}t_{ij}\left\{ [\frac
14P_\alpha ^2+\frac{\mathbf{S}_i\mathbf{S}_j}{(2S+1)^2}]\delta _{\sigma
\sigma ^{\prime }}+\frac 12\frac \alpha {(2S+1)^2}P_\alpha (\mathbf{S}_i%
\mathbf{+S}_j\mathbf{)\sigma }_{\sigma \sigma ^{\prime }}\right.
\eqno{(\rm I.10)}
$$
\[
+\left. \frac{2i}{(2S+1)^2}\mathbf{\sigma }_{\sigma \sigma ^{\prime }}[%
\mathbf{S}_i\mathbf{\times S}_j]\right\} c_{i\sigma }^{\dagger
}(1-n_{i,-\sigma })(1-n_{i,-\sigma ^{\prime }})c_{j\sigma ^{\prime }}+%
\mathcal{H}_d
\]
The terms with vector products (cf. (K.8)) describe anisotropic electron
scattering and may be important at considering transport phenomena in narrow
bands, e.g., the extraordinary Hall effect. The Hamiltonian in the form
(I.10) may be also useful at treating the states with the anomalous
``chiral'' order parameters, which are now extensively investigated within
the two-dimensional Heisenberg and $t-J$ models (see, e.g., [703]).

The Hamiltonian (I.5) is more convenient at considering simple
approximations within $1/z$-expansion [694]. Performing the simplest
decoupling we obtain the electron spectrum in the ferromagnetic phase
$$
E_{\mathbf{k}\sigma \alpha }=\left( P_\alpha -\frac{\alpha \sigma }{2S+1}%
\langle S^z\rangle \right) t_{\mathbf{k}}
\eqno{(\rm I.11)}
$$
This expression demonstrates a strong dependence of electron spectrum on
magnetic ordering. Rigorous expression for spin-down Green's function at $%
\alpha =+$, $T=0$ [699] has the non-quasiparticle form
$$
\langle \langle g_{\mathbf{k}\downarrow +}|g_{\mathbf{k}\downarrow
+}^{\dagger }\rangle \rangle _E=\left[ E-t_{\mathbf{k}}-2S(\sum_{\mathbf{q}%
}\frac 1{E-t_{\mathbf{q}}})^{-1}\right] ^{-1}
\eqno{(\rm I.12)}
$$

The magnon spectrum in the model (I.5) was calculated in [79,80,83]. The
result to leading order in $1/z$ reads
$$
\omega _{\mathbf{q}} =\frac 1{2S}\sum_{\mathbf{k}}(t_{\mathbf{k}-\mathbf{q}%
}-t_{\mathbf{k}})f(t_{\mathbf{k}}),\qquad \alpha =+
\eqno{(\rm I.13)}
$$
\[
\omega _{\mathbf{q}}=\frac 1{2S+1}\sum_{\mathbf{k}}(t_{\mathbf{k}-\mathbf{q}%
}-t_{\mathbf{k}})f(\frac{2S}{2S+1}t_{\mathbf{k}}),\qquad \alpha =-
\]

In an antiferromagnet with the spiral magnetic structure, corresponding to
the wavevector $\mathbf{Q}$, we have to pass in the $s-d$ Hamiltonian to the
local coordinate system by using (E.8), (G.90). Then, passing from the
operators $d_{i\sigma }^{\dagger }$ to the ME operators, we obtain instead
of (I.5)
$$
\mathcal{H} =\sum_{\mathbf{k}\sigma }\left( \theta _{\mathbf{k}}g_{\mathbf{k}%
\sigma \alpha }g_{\mathbf{k}\sigma \alpha }+\tau _{\mathbf{k}}g_{\mathbf{k}%
\sigma \alpha }g_{\mathbf{k},-\sigma ,\alpha }\right) +\mathcal{H}_d
\eqno{(\rm I.14)}
$$
\[
\tau _{\mathbf{k}}=\frac 12(t_{\mathbf{k}+\mathbf{Q}/2}-t_{\mathbf{k}-%
\mathbf{Q}/2}),\qquad
\theta _{\mathbf{k}}=\frac 12(t_{\mathbf{k}+\mathbf{Q}/2}
+t_{\mathbf{k}-\mathbf{Q}/2})
\]
Performing the ``Hubbard-I'' decoupling, we obtain for the electron spectrum
[687]
$$
E_{\mathbf{k}}^{1,2}=P_\alpha \theta _{\mathbf{k}}\pm \left\{ (\frac{%
\overline{S}}{2S+1}\theta _{\mathbf{k}})^2+[P_\alpha ^2-(\frac{\overline{S}}{%
2S+1})^2]\tau _{\mathbf{k}}\right\} ^{1/2}
\eqno{(\rm I.15)}
$$
with $\overline{S}$ being the sublattice magnetization. In the
nearest-neighbor approximation ($\theta _{\mathbf{q}}$ $=0$) for $I>0$ the
band at $T=0$ is narrowed by $(2S+1)^{1/2}$ times. At the same same time,
for $I<0$ (and also in the $t-J$ model) the electron may not pass to
neighbour sites in the approximation under consideration, and its motion is
possible due to quantum effects only. This problem is discussed in Sect.6.7.

Fluctuation corrections to the spectrum (I.15) are discussed in [687,620].
The result for the magnon spectrum obtained in the ME representation turn
out to coincide with that of the generalized Hartree-Fock approximation
(G.95).

\chapter{APPENDIX J Electron states and spin waves in the narrow-band
Hubbard ferromagnet}

The electron and magnon spectra of a Hubbard ferromagnet with strong
correlations ($U\rightarrow \infty $), which is described by the Hamiltonian
$$
\mathcal{H}=\sum_{\mathbf{k}\sigma }\varepsilon _{\mathbf{k}}X_{-\mathbf{k}%
}(0\sigma )X_{\mathbf{k}}(\sigma 0)
\eqno{(\rm J.1)}
$$
may be investigated rigorously in the case of small concentrations of holes $%
c=1-N_e/N$ (almost half-filled band) and low temperatures. This is formally
achieved by expansion in hole and magnon occupation numbers [333,699,700].
Consider the one-particle Green's functions
$$
G_{\mathbf{k}\sigma }=\langle \langle X_{\mathbf{k}}(\sigma 0)|X_{-\mathbf{k}%
}(0\sigma )\rangle \rangle _E
\eqno{(\rm J.2)}
$$
Using the commutation relation
\[
\left[ X_{\mathbf{k}}(+0),\mathcal{H}\right] =\sum_{\mathbf{p}}\varepsilon _{%
\mathbf{k-p}}\left\{ [X_{\mathbf{p}}(00)+X_{\mathbf{p}}(++)]X_{\mathbf{k-p}%
}(+0)+X_{\mathbf{p}}(+-)X_{\mathbf{k-p}}(-0)\right\}
\]
we write down for $\sigma =\uparrow $ the equation of motion (E.16a)
$$
(E-\varepsilon _{\mathbf{k}})G_{\mathbf{k}\uparrow }(E) =1-n_{\downarrow
}+\sum_{\mathbf{pq}}(\varepsilon _{\mathbf{k-p}}-\varepsilon _{\mathbf{k+q-p}%
})
\eqno{(\rm J.3)}
$$
\[
\times \langle \langle X_{-\mathbf{q}}(-+)X_{\mathbf{p}}(+-)X_{\mathbf{k+q-p}%
}(+0)|X_{-\mathbf{k}}(0+)\rangle \rangle _E
\]
Here we have taken into account the kinematical relations
$$
X_{\mathbf{p}}(++)+X_{\mathbf{p}}(--) =\delta _{\mathbf{p}0}-X_{\mathbf{p}%
}(--),\qquad
X_{\mathbf{p}}(--)=\sum_{\mathbf{q}}X_{-\mathbf{q}}(-+)X_{\mathbf{p+q}}(+-),
\eqno{(\rm J.5)}
$$
\[
X_{\mathbf{k-p}}(-0)=\sum_{\mathbf{q}}X_{-\mathbf{q}}(-+)X_{\mathbf{k+q-p}%
}(+0)
\]
which follow from (A.28), (A.25), reduced the operator products to the
``normal form'' where all the opeators $X(-+)$ stand at the left, and
retained only the terms, which are linear in spin deviations and neglected
the terms which are proportional to the hole concentrations. Introducing the
function
$$
\varphi _{\mathbf{kqp}}(E)=\langle \langle X_{-\mathbf{q}}(-+)X_{\mathbf{p}%
}(+-)X_{\mathbf{k+q-p}}(+0)|X_{-\mathbf{k}}(0+)\rangle \rangle _E/\left[
(E-\varepsilon _{\mathbf{k}})N_{\mathbf{q}}\right]
\eqno{(\rm J.6)}
$$
with
\[
N_{\mathbf{q}}=\langle X_{-\mathbf{q}}(-+)X_{\mathbf{q}}(+-)\rangle
\]
we derive in the same way the closed integral equation
$$
(E-\varepsilon _{\mathbf{k+q-p}})\varphi _{\mathbf{kqp}}(E) =(E-\varepsilon
_{\mathbf{k}})(\delta _{\mathbf{pq}}-1)+
\eqno{(\rm J.7)}
$$
\[
\sum_r(\varepsilon _{\mathbf{k+q-p-r}}-\varepsilon _{\mathbf{%
k+q-r}})\varphi _{\mathbf{kqr}}(E)
\]
which describes the hole-magnon scattering. Writing down the Dyson equation
$$
G_{\mathbf{k}\uparrow }(E) =\frac{1-n_{\downarrow }}{E-\varepsilon _{\mathbf{%
k}}}\left[ 1-\frac 1{E-\varepsilon _{\mathbf{k}}}\Sigma _{\mathbf{k}%
}(E)\right] \simeq (1-n_{\downarrow })\left[ E-\varepsilon _{\mathbf{k}%
}-\Sigma _{\mathbf{k}}(E)\right] ^{-1}
\eqno{(\rm J.8)}
$$
we obtain for the self-energy
$$
\Sigma _{\mathbf{k}}(E)=\sum_{\mathbf{pq}}(\varepsilon _{\mathbf{k-p}%
}-\varepsilon _{\mathbf{k+q-p}})\varphi _{\mathbf{kqp}}(E)
\eqno{(\rm J.9)}
$$

The magnon spectrum is obtained from the spin Green's function. The equation
of motion for this has the form (cf.(H.15))
$$
\omega G_{\mathbf{q}}(\omega ) =1-c+\sum_{\mathbf{kp}}(\varepsilon _{\mathbf{%
k-p}}-\varepsilon _{\mathbf{k}})
\eqno{(\rm J.10)}
$$
\[
\times \langle \langle X_{-\mathbf{k}}(0+)X_{\mathbf{q-k+p}}(+-)X_{\mathbf{k}%
}(+0)|X_{-\mathbf{q}}(-+)\rangle \rangle _\omega
\]
Performing calcualations at $T=0$ to lowest orderin magnon occupation
numbers
\[
n_{\mathbf{k}}=\langle X_{-\mathbf{k}}(0+)X_{\mathbf{k}}(+0)\rangle
=f(\varepsilon _{\mathbf{k}})
\]
we obtain the magnon self-energy. This turns out to be determined by the
same function $\varphi $
$$
G_{\mathbf{q}}(\omega )=(1-c)\left[ \omega -\sum_{\mathbf{kp}}(\varepsilon _{%
\mathbf{k-p}}-\varepsilon _{\mathbf{k+q-p}})\varphi _{\mathbf{kqp}}(\omega
+\varepsilon _{\mathbf{k}})\right] ^{-1}
\eqno{(\rm J.11)}
$$
To lowest order in the small parameter $1/z$ (each order in $1/z$
corresponds to a summation over a wavevector) we obtain for the temperature
correction to the the electron spectrum and for the magnon frequency
$$
\delta \varepsilon _{\mathbf{k}}(T) =\sum_{\mathbf{q}}(\varepsilon _{\mathbf{%
k-q}}-\varepsilon _{\mathbf{k}})N_{\mathbf{q}}
\eqno{(\rm J.12)}
$$
\[
\omega _{\mathbf{q}}=\sum_{\mathbf{k}}(\varepsilon _{\mathbf{k-q}%
}-\varepsilon _{\mathbf{k}})n_{\mathbf{k}}
\]
The equation (J.7) may be solved exactly for concrete lattices. For the
simple cubic lattice in the nearest-neighbour approximation we have for the
band bottom shift
$$
\delta \varepsilon _0=\kappa \frac{3\zeta (5/2)v_0}{32\pi ^{3/2}m^{*}}\left(
\frac TD\right) ^{5/2}
\eqno{(\rm J.13)}
$$
and for the spin-wave frequency at small $q$
$$
\omega _{\mathbf{q}}=Dq^2,\qquad D=\kappa c|t|
\eqno{(\rm J.14)}
$$
where $\kappa $ is expressed in terms of the lattice Green's function,
\[
\kappa =\frac{1-A}{1+A}\approx 0.656
\]
\[
A=\sum_{\mathbf{q}}\frac{\cos q_x}{3-\cos q_x-\cos q_y-\cos q_z}\approx
0.208
\]

It is convenient to calculate more complicated characteristics at finite
temperatures with the use of the expansion in $1/z$ [700]. The magnon
damping is given by (cf.(G.23)-(G.25))
$$
\gamma _{\mathbf{q}} =\pi \sum_{\mathbf{kp}}(\varepsilon _{\mathbf{k-q}%
}-\varepsilon _{\mathbf{k}})^2\left[ n_{\mathbf{k+q-p}}(1-n_{\mathbf{k}%
})+(n_{\mathbf{k+q-p}}-n_{\mathbf{k}})N_{\mathbf{p}}\right]
\eqno{(\rm J.15)}
$$
\[
\times \delta (\omega +\varepsilon _{\mathbf{k}}-\varepsilon _{\mathbf{k+q-p}%
}-\omega _{\mathbf{p}})
\]
and turn out to be finite at $T=0$, unlike the case of Heisenberg
ferromagnet. The temperature dependence of spin-wave stiffness reads
(cf.(G.27))
$$
\delta D(T) =\frac{\pi ^2}{12}T^2N(E_F)\frac d{dE_F}\left( \frac{\partial
^2\varepsilon _{\mathbf{k}}}{\partial k^2}\right) _{k=k_F}
\eqno{(\rm J.16)}
$$
\[
-\frac{5\pi ^{1/2}}{12}\zeta (5/2)\left( \frac{v_0}{4\pi ^2}\right) ^2%
\frac{k_F}{m^{*}}\left( \frac TD\right) ^{5/2}-\frac{v_0^2k_F^4}{144\pi
Dk_F^2}T^2\ln \frac{4Dk_F^2}T 
\]

The magnon Green's function permits to calculate the spin-wave correction to
magnetization 
$$
\langle S^z\rangle =\frac 12(n_{\uparrow }-n_{\downarrow })=\frac{1-c}%
2-n_{\downarrow }  
\eqno{(\rm J.17)}
$$
\[
=\frac{1-c}2+\frac 1\pi \sum_{\mathbf{q}}\int d\omega N_B(\omega )\Im G_{%
\mathbf{q}}(\omega ) 
\]
The corrections owing to the damping turn out to be cancel the factor $1-c$
in the denominator of (J.11) and we obtain to terms of order of $1/z^2$ 
$$
\langle S^z\rangle =\frac{1-c}2-\sum_{\mathbf{p}}N_{\mathbf{p}}  
\eqno{(\rm J.18)}
$$
Thus the ground state is really saturated ferromagnetic, and magnetization
at low temperatures obeys the usual Bloch $T^{3/2}$-law.

As follows from (J.9), spin-up states propagate freely on the background of
the saturated ferromagnetic ordering, and corrections to the spectrum at low
temperatures are proportional to $T^{5/2}$. The situation is more
interesting for spin-down states. Using again the kinematical relation
(A.25) we obtain 
$$
G_{\mathbf{k}\downarrow }(E)=\sum_{\mathbf{q}}\langle \langle X_{-\mathbf{q}%
}(-+)X_{\mathbf{k}+\mathbf{q}}(+0)|X_{-\mathbf{k}}(0+)\rangle \rangle _E 
\eqno{(\rm J.19)}
$$
The simplest decoupling in the equation of motion for the Green's function
in the right-hand side of (J.19) yields 
$$
G_{\mathbf{k}\downarrow }^0(E)=\sum_{\mathbf{q}}\frac{N_{\mathbf{q}}+n_{%
\mathbf{k}+\mathbf{q}}}{E-\varepsilon _{\mathbf{k}+\mathbf{q}}+\omega _{%
\mathbf{q}}}  
\eqno{(\rm J.20)}
$$
The Green's function (J.20) has a purely non-quasiparticle nature. Because
of the weak $k$-dependence of the corresponding distribution function at $%
T\rightarrow 0$, 
$$
\langle X_{-\mathbf{k}}(0-)X_{\mathbf{k}}(-0)\rangle =-\frac 1\pi \int
dEf(E)\Im G_{\mathbf{k}\downarrow }^0(E)\simeq c  
\eqno{(\rm J.21)}
$$
the non-quasiparticle states possess a small mobility and do not carry
current [333,338]. Thus the ``spin-down'' excitations are reminiscent of
Anderson's spinons (Sect.6.8) which are also described by the Green's
function with zero residue [631]. The non-quasiparticle contribution to the
density of states turns out to be appreciable: 
$$
N_{\downarrow }(E,T =0)=\sum_{\mathbf{k}\sigma }n_{\mathbf{k+q}}\delta
(E-\varepsilon _{\mathbf{k+q}}+\omega _{\mathbf{q}})  
\eqno{(\rm J.22)}
$$
\[
=\left\{ 
\begin{array}{ll}
N_{\uparrow }(E) & ,\qquad E_F-E\gg \omega _{\max } \\ 
0 & ,\qquad E>E_F
\end{array}
\right. 
\]
The sense of this result is as follows. The states well below the Fermi
level (of holes) do not possess spin polarization, since elect- rons with
any spin projection may be put into them (the holes are spinless). However,
from states above the Fermi level, only spin-up electrons may be extracted
in the saturated ferromagnetic state under consideration. The large
non-quasiparticle $T$-linear contribution to specific heat may be also
picked up [338]. However, it has more complicated origin in comparison with
the spinon contribution [631] since the non-quasiparticle states are absent
at the Fermi level.

A more advanced decoupling yields the result [338] 
$$
G_{\mathbf{k}\downarrow }(E)=\left\{ E-\varepsilon _{\mathbf{k}}+\left[ G_{%
\mathbf{k}\downarrow }^0(E)\right] ^{-1}\right\} ^{-1}  
\eqno{(\rm J.23)}
$$
At small $c$, the Green's function (J.23) has no poles below the Fermi
level, so that the above conclusions are not changed qualitatively. However,
with increasing $c$, it acquires a spin-polaron pole below $E_F$ , and the
saturated ferromagnetism is destroyed [332].

The expression (J.23) should be compared with the corresponding result for
the paramagnetic phase in the ``Hubbard-III'' approximation (cf.(H.16)) 
$$
G_{\mathbf{k}\downarrow }(E)=\left\{ E-\varepsilon _{\mathbf{k}}+\frac{1-c}%
2\left[ G_{\mathbf{k}\downarrow }^0(E)\right] ^{-1}\right\} ^{-1}  
\eqno{(\rm J.24)}
$$
Unlike (J.23), equation (J.24) does not contain Fermi functions, so that the
incoherent (non-quasiparticle) states do not vanish at $E_F$ .

One may expect that in the non-saturated ferromagnetic state (or at high
temperatures) the expression (J.20) should be replaced by the Hubbard-I
result 
$$
G_{\mathbf{k}\downarrow }(E)=\frac{c+n_{\downarrow }}{E-\varepsilon _{%
\mathbf{k}}(c+n_{\downarrow })}  
\eqno{(\rm J.25)}
$$
which describes usual quasiparticle states with a narrowed band. Note that
the expression (J.25) may be obtained by averaging nominator and denominator
in (J.20) over $\mathbf{q}$, which corresponds to the large-$z$ approximation.
As follows from (J.22), the expressions (J.8), (J.21), unlike (J.25), enable
one to satisfy the sume rule (H.13) since 
$$
\sum_{\mathbf{k}}\langle X_{-\mathbf{k}}(0-)X_{\mathbf{k}}(-0)\rangle
=\int_{-\infty }^{E_F}dEN_\sigma (E)=c=n_0  
\eqno{(\rm J.26)}
$$
Thus the non-quasiparticle nature of spin-down current carriers is
intimately related to the description of the saturated ferromagnetic state.


\chapter{$s-f$ exchange model and indirect exchange interaction in rare
earths}

For rare-earth metals, where 4f-electrons are well localized, the $s-f$
exchange model may provide a basis of a quantitative theory of magnetic
properties. In particular, the main mechanism of exchange between 4f-shells
in rare earths and their conducting compounds is the
Ruderman-Kittel-Kasuya-Yosida (RKKY) indirect interaction via conduction
electrons, which occurs in the second order of perturbation theory in the $%
s-f$ exchange parameter. Excluding from the simplest Hamiltonian of the $s-f$
model (G.2) the $s-f$ exchange interaction by a canonical transformation
[265], we obtain the effective Heisenberg Hamiltonian
$$
\mathcal{H}_f=-\sum_{\mathbf{q}}J_{\mathbf{q}}^{\mathrm{RKKY}}\mathbf{S}_{-%
\mathbf{q}}\mathbf{S}_{\mathbf{q}},\qquad J_{\mathbf{q}}^{\mathrm{RKKY}%
}=I^2\sum_{\mathbf{k}}\frac{n_{\mathbf{k}}-n_{\mathbf{k+q}}}{t_{\mathbf{k+q}%
}-t_{\mathbf{k}}}.
\eqno{(\rm K.1)}
$$
In real space, the RKKY exchange integrals have an oscillating and slowly
decreasing dependence on distance. Performing integration for free electrons
we obtain
$$
J_{ij}^{\mathrm{RKKY}}=\frac{9\pi n^2I^2}{2v_0^2E_{\mathrm{F}}}F(2k_{\mathrm{%
F}}|\mathbf{R}_i-\mathbf{R}_j|),\qquad F(x)=\frac{x\cos x-\sin x}{x^4}.
\eqno{(\rm K.2)}
$$

Now we consider a more realistic model of 4f-metals. For most rare-earths
(except for Eu and Sm), the matrix elements of intersite exchange
interactions are small in comparison with the distances between
LSJ-multiplets, and the Russell-Saunders coupling scheme is a good
approximation. Using for simplicity the representation of s-type plane waves
for conduction electrons, we derive for the $s-f$ Hamiltonian (cf.(D.20))
$$
\mathcal{H}_{sf}=\sum_{\mathbf{kk}^{\prime }\sigma \sigma ^{\prime
}}\sum_{\nu \Gamma _1\Gamma _2\gamma _1\gamma _2}e^{ i(\mathbf{k}-
\mathbf{k}^{\prime })\mathbf{R}_\nu } \langle \gamma _1,\mathbf{k}%
\sigma |\sum_{ic}\frac{e^2}{|\mathbf{r}_i-\mathbf{r}_c|}|\gamma _2,\mathbf{k}%
^{\prime }\sigma ^{\prime }\rangle
$$
$$
\times \langle \Gamma _1|a_{\nu \gamma _1}^{\dagger }a_{\nu \gamma
_2}|\Gamma _2\rangle X_\nu (\Gamma _1,\Gamma _2)c_{\mathbf{k}\sigma
}^{\dagger }c_{\mathbf{k}^{\prime }\sigma ^{\prime }},
\eqno{(\rm K.3)}
$$
where $c_{\mathbf{k}\sigma }^{\dagger }$ are creation operators for
conduction electrons, $\gamma _i=\{lm_i\}$, $\Gamma _i=\{SLJM_i\}$, $p_{ic}$
are operators of permutation of conduction and localized electrons.
Expanding the plane waves according to (C.28) and using (C.7) we obtain the
series in $\lambda $, $\lambda ^{\prime }$ with the ``Slater integrals''
$$
F_{\lambda \lambda ^{\prime }}^{(p)}(\mathbf{kk}^{\prime })=e^2\int
r_1^2dr_1r_2^2dr_2R_l^2(r_1)\frac{r_{<}^p}{r_{>}^{p+1}}R_l(r_2)j_\lambda
(kr_2)j_{\lambda ^{\prime }}(k^{\prime }r_1),
\eqno{(\rm K.4)}
$$
$$
G_{\lambda \lambda ^{\prime }}^{(p)}(\mathbf{kk}^{\prime })=e^2\int
r_1^2dr_1r_2^2dr_2R_l^2(r_1)j_\lambda (kr_2)\frac{r_{<}^p}{r_{>}^{p+1}}%
R_l(r_2)j_{\lambda ^{\prime }}(k^{\prime }r_1),
\eqno{(\rm K.5)}
$$
where $l=3$ for f-electrons. The small parameter of the expansion is $k_{%
\mathrm{F}}r_f\sim 0.2$ where $r_f$ is the radius of the 4f-shell. The
matrix elements that arise may be calculated by the method of double
irre-ducible tensor operators and expressed in terms of matrix elements of
the total angular momentum $J$, as demonstrated in Appendix D. In
particular, for the zeroth-order term we obtain from (B.19)
$$
\mathcal{H}_{sf}(00)=-\frac{4\pi }{[l]}\sum_{\nu \sigma \sigma ^{\prime
}}G_{00}^{(0)}\left[ \frac n2\delta _{\sigma \sigma ^{\prime }}+(g-1)(%
\mbox{\boldmath$\sigma $}_{\sigma \sigma ^{\prime }}\mathbf{J}_\nu )\right]
c_{\nu \sigma }^{\dagger }c_{\nu \sigma ^{\prime }}.
\eqno{(\rm K.6)}
$$
Higher-order terms of the expansion are anisotropic and have the structure
$$
\mathcal{H}_{sf}^{\mathrm{coul}}=\sum_{\nu \mathbf{kk}^{\prime }\sigma
\sigma ^{\prime }}e^{i(\mathbf{k}-\mathbf{k}^{\prime })\mathbf{R}_\nu }
c_{\mathbf{k}\sigma }^{\dagger }c_{\mathbf{k}^{\prime }\sigma
^{\prime }}
$$
$$
\times \left( B_0+B_1\left[ 3\left\{ (\mathbf{k}J_\nu ),(\mathbf{k}^{\prime
}J_\nu )\right\} -2(\mathbf{kk}^{\prime })J(J+1)\right] +\ldots \right) ,%
\eqno{(\rm K.7)}
$$
$$
\mathcal{H}_{sf}^{\mathrm{exch}}=\sum_{\nu \mathbf{kk}^{\prime }\sigma
\sigma ^{\prime }}\left( A_0\delta _{\sigma \sigma ^{\prime }}+A_1(\mathbf{%
\sigma }_{\sigma \sigma ^{\prime }}\mathbf{J}_\nu )+iA_2([\mathbf{kk}%
^{\prime }]\mathbf{J}_\nu )\delta _{\sigma \sigma ^{\prime }}\right.
$$
$$
+A_3\left\{ (\mathbf{kJ}_\nu \mathbf{),(k}^{\prime }\mathbf{J}_\nu )\right\}
+A_4\left[ (\mathbf{k\sigma }_{\sigma \sigma ^{\prime }}\mathbf{)(k}^{\prime
}\mathbf{J}_\nu )+(\mathbf{k}^{\prime }\mbox{\boldmath$\sigma $}_{\sigma
\sigma ^{\prime }}\mathbf{)(kJ}_\nu )\right]
$$
$$
+A_5\left[ (\mathbf{k\sigma }_{\sigma \sigma ^{\prime }}\mathbf{)(kJ}_\nu )
+(\mathbf{k}^{\prime }\mbox{\boldmath$\sigma $}_{\sigma \sigma ^{\prime
}})(\mathbf{k} ^{\prime }\mathbf{J}_\nu )\right]
$$
$$
+A_6\left[ (\mathbf{kJ}_\nu )^2+(\mathbf{k}^{\prime }\mathbf{J}_\nu
)^2\right] +iA_7\left\{ (\mbox{\boldmath$\sigma $}_{\sigma \sigma ^{\prime
}}\mathbf{J}_\nu ),([\mathbf{kk}^{\prime }]\mathbf{J}_\nu )\right\} +\ldots ,
\eqno{(\rm K.8)}
$$
where $\{\ ,\ \}$ is anticommmutator. The maximum power $q$ of momentum
operator $\mathbf{J}$ is determined by the maximum value of $\lambda $,
retained in the expansion (C.28), $q=\min \{2J,2\lambda +1\}$. The terms
with the vector products $[\mathbf{kk}^{\prime }]$ describe the anisotropic
electron scattering and are important in the theory of the anomalous Hall
effect. The coefficients of the expansion (K.8) are given by [704]
$$
A_1=\frac 27(g-1)\left[ G_{00}^{(3)}+(\mathbf{kk}^{\prime })\eta _1\right]
-D_1\left[ \frac 9{35}(\mathbf{kk}^{\prime })\eta _2+\frac 27\sqrt{6}
G_{02}^{(3)}\right] ,
$$
$$
A_2=\frac 1{28}(g-2)\eta _3-\frac{9\sqrt{5}}{70}\eta _2D_2,\qquad
A_3=\frac{9 \sqrt{5}}{35}D_2\eta _2,
$$
$$
A_4=\frac{27}{70}\eta _2D_1,\qquad
A_5\frac{3\sqrt{6}}7G_{02}^{(3)}D_1,\qquad
A_6=-\frac{3\sqrt{5}}7G_{02}^{(3)}D_2,
$$
$$
A_7=A_8=\frac{9\sqrt{15}}{14}\eta _3(2J+1)\left[ \frac{(2J-2)!}{(2J+3)!}
\right] ^{1/2}\left\{
\begin{array}{ccc}
L & J & S \\
L & J & S \\
1 & 2 & 1
\end{array}
\right\} \langle SL||W^{(11)}||SL\rangle ,
\eqno{(\rm K.9)}
$$
where the irreducible matrix elements are determined by (B.27),
$$
\eta _1=\frac 95G_{11}^{(2)}+\frac 43G_{11}^{(4)},\qquad \eta _2=\frac
95G_{11}^{(2)}+\frac 59G_{11}^{(4)},\qquad \eta _3=\frac
95G_{11}^{(2)}-G_{11}^{(4)},
$$
$$
D_1=\left[ \frac{2J+1}{J(J+1)}\right] ^{1/2}\left\{
\begin{array}{ccc}
L & J & S \\
L & J & S \\
2 & 1 & 1
\end{array}
\right\} \langle SL||W^{(12)}||SL\rangle ,
$$
$$
D_2=(-1)^{S+L+J}\frac 2{\sqrt{3}}\frac{2J+1}{(2S+1)^{1/2}}\langle
SL||W^{(02)}||SL\rangle \left\{
\begin{array}{ccc}
L & J & S \\
J & L & 2
\end{array}
\right\} .
\eqno{(\rm K.10)}
$$

The Hamiltonian of indirect f-f interaction is obtained in the second order
in $\mathcal{H}_{sf}$ and has the same structure as (D.22). The main
conributions may be written down in the form [389]
$$
\mathcal{H}_{ff}(\nu _1\nu _2)=-I_1(g-1)^2\mathbf{(J}_1\mathbf{J}_2)
-I_2D_1(g-1)\left[ (\mathbf{J}_1\mathbf{J}_2)-3(\mbox{\boldmath$\rho
$}_{12}\mathbf{J}
_{12}^2\right]
$$
$$
-I_3nD_3\left[ (\mathbf{J}_1\mathbf{J}_2)-3(\mbox{\boldmath$\rho
$}_{12}\mathbf{J} _2)^2/\rho _{12}^2\right] ,
\eqno{(\rm K.11)}
$$
where $I_i$ are linear combinations of the integrals of the type
$$
B_{\mu ^{\prime }\lambda ^{\prime \prime }\lambda ^{\prime \prime \prime
}p^{\prime }}^{\mu \lambda \lambda ^{\prime }p}=\int\limits_0^\infty
k^2dkn_k\int\limits_{-\infty }^\infty \frac{k^{\prime 2}dk^{\prime }}{%
k^2-k^{\prime 2}}j_\mu (k\rho _{12})j_{\mu ^{\prime }}(k^{\prime }\rho
_{12})G_{\lambda \lambda ^{\prime }}^{(p)}(kk^{\prime })G_{\lambda ^{\prime
\prime }\lambda ^{\prime \prime \prime }}^{(p^{\prime })}(kk^{\prime }).%
\eqno{(\rm K.12)}
$$
The largest term of this expansion, which is proportional to $(g-1)^2$,
corresponds to the usual exchange between spins according to the de Gennes
formula (B.19). The dependence of the f-f exchange parameter $J_{\mathrm{eff}%
}\sim (g-1)^2$ is in a good agreement with experimental data on the
paramagnetic Curie temperatures in the series of rare-earth metals. The
orbital contributions to the f-f interaction, which are proportional to $D_1$
and $D_2$ and vanish at $L=0$, are considerably smaller. A still smaller
term of the purely orbital exchange is obtained in the second order in $A_2$%
:
$$
\mathcal{H}_{ff}^{\prime }(\nu _1\nu _2)=-I_4(g-2)^2(\mathbf{J}_1\mathbf{J}%
_2)=-I_4(\mathbf{L}_1\mathbf{L}_2).
\eqno{(\rm K.13)}
$$
The orbital terms may give an appreciable contribution to the exchange
anisotropy of crystal magnetization (Sect.4.8).


\chapter{Spin-orbital interaction}

Besides exchange interactions, an important role in magnetic crystals
belongs to the relativistic spin-orbital interaction (SOI). The latter,
although being weak, results in a partial unquenching of orbital momenta and
is responsible for anisotropy of magnetic and other properties. SOI is
especially important for transport phenomena, e.g., anomalous
halvanomagnetic and thermomagnetic properties.

The operator of SOI for an electron with the quasimomentum $\mathbf{p}$ and
spin $\mathbf{s}$ in the potential $V(\mathbf{r})$ has the form
$$
\mathcal{H}_{\mathrm{so}}=\frac \hbar {2m^2c^2}[\nabla V\mathbf{,p}]\mathbf{s%
}.
\eqno{(\rm L.1)}
$$
For the Coulomb interaction
$$
V(\mathbf{r})=-Ze^2/r
$$
we obtain
$$
\mathcal{H}_{\mathrm{so}}=\lambda (r)(\mathbf{ls)},
\eqno{(\rm L.2)}
$$
where
$$
l=\frac 1\hbar [\mathbf{r},\mathbf{p}],\qquad \lambda (r)=\frac{Ze^2\hbar ^2%
}{2m^2c^2r^3}
$$
For estimating the value of $\lambda (r)$ one can use hydrogen-like
wavefunctions to obtain
$$
\lambda _{nl}=\frac{Z^4\hbar c\alpha ^2}{n^3l(l+1/2)(l+1)}\mathrm{Ry}%
\eqno{(\rm L.3)}
$$
with $\mathrm{Ry}=13.6$ eV being the Rydberg constant, $\alpha =1/137$. SOI
increases rapidly with increasing atomic number. We have $\lambda \sim
10^{-14}$ erg for 3d-electrons and $\lambda \sim 10^{-13}$ erg for
4f-electrons in rare earths. The quantity $\lambda $ may be also estimated
from the data of atomic spectroscopy, which turn out to agree qualitatively
with the theory, provided that effective values of $Z$ are used.

At considering SOI, the question about degeneracy of electron states is
essential. Evidently, for non-degenerate wavefunctions (e.g., for Bloch
electrons in a crystal) diagonal matrix elements of $\mathcal{H}_{\mathrm{so}%
}$ (L.2) vanish, so that corrections to the energy, which are linear in $%
\mathcal{H}_{\mathrm{so}}$, are absent. On the other hand, in the degenerate
case (e.g, for a free atom) a splitting of levels in the first order in $%
\mathcal{H}_{\mathrm{so}}$ takes place.

Besides the proper SOI (orbital electron current in the magnetic field of
its own spin momentum), there exists also the interaction of orbital current
with spins of other electrons
$$
\mathcal{H}_{\mathrm{so}}^{\prime }=\frac \hbar {m^2c^2}\sum_{i\neq
j}[\nabla V_{ij},\mathbf{p}_i\mathbf{]s}_j,
\eqno{(\rm L.4)}
$$
where $i$, $j$ are electron numbers,
$$
V_{ij}=\frac{e^2}{|\mathbf{r}_{ij}|},\qquad \nabla V_{ij}=\frac{e^2\mathbf{r}%
_{ij}}{|\mathbf{r}_{ij}|^3},\qquad \mathbf{r}_{ij}=\mathbf{r}_i-\mathbf{r}_j.%
\eqno{(\rm L.5)}
$$
In the case of two electrons, one of which has zero orbital momentum and
moves closely to the nucleus, we can put $\mathbf{r}_1=0$, $\mathbf{r}_{12}=%
\mathbf{r}_2$. Then (L.5) takes the form [20]
$$
\mathcal{H}_{\mathrm{so}}^{\prime }=-\lambda ^{\prime }(\mathbf{l}_2\mathbf{s%
}_1),
\eqno{(\rm L.6)}
$$
with $\lambda ^{\prime }>0$ being proportional to $Z^3$ rather than $Z^4$
(as in (L.3)). The latter fact results in that the interaction between spin
and foreign orbit is more important for light elements. As a rule, for
3d-electrons $\lambda ^{\prime }$ is smaller than $\lambda $ by two-three
orders of magnitude, but, as we shall demonstrate below, the corresponding
interaction may play an important role because of its singular $\mathbf{k}$%
-dependence.

For a single electron, the ``proper'' SOI makes favourable the antiparallel
interactions of its spin and orbital momenta, but the interaction (L.6)
orientates its orbital momentum parallel to spin of other electrons.
However, for more than half-filled shell the sign of $\lambda $ changes and
the state with the total momentum $\mathbf{J=L+S}$ has the lowest energy.

Now we discuss SOI in periodic crystals. In the localized Heisenberg model
the treatment is close to that for isolated atoms. On the other hand, for
crystals containing 3d-elements the situation changes drastically. As
discussed in Sect.1.3, the local crystal potential may quench orbital
momenta in the case of crystals with low symmetry only. At the same time, in
real d-systems with cubic or hexagonal local symmetry, which possess
degenerate irreducible representations of the point group, the degeneracy is
lifted by the periodic potential in the band picture (Sect.4.8). Therefore
the case of quenched orbital momenta is of interest. Then only off-diagonal
matrix elements of SOI are non-zero and the perturbed wavefunctions read
$$
\Psi _\gamma =\Psi _\gamma ^{(0)}+\sum_{\gamma ^{\prime }\neq \gamma }\frac{%
\langle \gamma ^{\prime }|\mathcal{H}_{\mathrm{so}}|\gamma \rangle }{%
E_\gamma -E_{\gamma ^{\prime }}}\Psi _{\gamma ^{\prime }}^{(0)}
\eqno{(\rm L.7)}
$$
with $\gamma =\{\mathbf{k}m\sigma \}$ are the states of magnetic d-electrons
in the degenerate d-band.

The wavefunctions (L.7) may be used to calculate corrections to various
physical quantities, owing to SOI. In particular, such corrections to matrix
elements of electrostatic interaction between conduction and localized
electrons will be anisotropic. It is these corrections which cause anomalous
transport phenomena in magnetic crystals. The role of the proper SOI and the
interaction (L.6) is different for different situations and concrete
effects. For halvanomagnetic effects in d-magnets one can consider two cases

(a) The mobility of d-electrons is large, so that they determine
halvanovamagnetic effects directly. Then the proper SOI for itinerant
d-electrons plays the dominant role.

(b) There exist two electron groups - conduction s-electrons with small
magnetization and ``magnetic'' d-electons with small mobility. Then, among
four possible types of SOI (s-s, d-d, s-d and d-s), the proper d-d
interaction and the s-d interaction of s-electron orbit with d-electron spin
are most important. The s-s and d-s interactions yield small contributions
because of smallness of magnetization for s-electrons.

Finally, we derive the Hamiltonian of s-d model with account of SOI of d-d
and s-d type in the second quantization representation. We consider the case
of the strong crystal field which destroys total orbital magnetic momenta of
d-electrons. We have
$$
\mathcal{H}=\sum_{\mathbf{k}\sigma }t_{\mathbf{k}}c_{\mathbf{k}\sigma
}^{\dagger }c_{\mathbf{k}\sigma }+\sum_\gamma E_\gamma a_\gamma ^{\dagger
}a_\gamma
$$
$$
+\mathop{\sum_{\mathbf{k}\sigma \mathbf{k}^{\prime }\sigma ^{\prime }}}_{\gamma
_1\gamma _2 }
\left[ I(\mathbf{k}\sigma ,\gamma _1,\mathbf{k}^{\prime
}\sigma ^{\prime },\gamma _2)-I(\mathbf{k}\sigma ,\gamma _1,\gamma _2,%
\mathbf{k}^{\prime }\sigma ^{\prime })\right.
$$
$$
+\left. L(\mathbf{k}\sigma ,\gamma _1,\mathbf{k}^{\prime }\sigma ^{\prime
},\gamma _2)-L(\mathbf{k}\sigma ,\gamma _1,\gamma _2,\mathbf{k}^{\prime
}\sigma ^{\prime })\right] a_{\gamma _1}^{\dagger }a_{\gamma _2}c_{\mathbf{k}%
\sigma }^{\dagger }c_{\mathbf{k}^{\prime }\sigma ^{\prime }},
\eqno{(\rm L.8)}
$$
where the Coulomb and exchange matrix elements
$$
I(1,2,3,4)=\int dxdx^{\prime }\psi _1^{*}(x)\psi _2^{*}(x^{\prime })\frac{e^2%
}{|\mathbf{r}-\mathbf{r}^{\prime }|}\psi _3(x)\psi _4(x^{\prime })
\eqno{(\rm L.9)}
$$
($x=\{r,s\}$) are to be calculated for the wavefunctions with account of
SOI, and the spin-orbital matrix elements are
$$
L(1,2,3,4)=-\frac{e^2\hbar }{m^2c^2}\int dxdx^{\prime }\psi _1^{*}(x)\psi
_2^{*}(x^{\prime })\frac{[\mathbf{r}-\mathbf{r}^{\prime },\mathbf{p]s}%
^{\prime }}{|\mathbf{r}-\mathbf{r}^{\prime }|^3}\psi _3(x)\psi _4(x^{\prime
}).
\eqno{(\rm L.10)}
$$
In the representation $\gamma =\{\mathbf{k}m\sigma \}$ the diagonal matrix
elements of $\mathcal{H}_{\mathrm{so}}(dd)$ vanish and the off-diagonal ones
are obtained within perturbation theory with the use of (L.7). Substituting
(L.7) into the exchange part of (L.9) we derive to linear approximation in $%
\mathcal{H}_{\mathrm{so}}$ the correction
$$
\mathcal{H}_{sd}^{(1)}=-\lambda \sum \frac{\langle \mathbf{k}_1m_1\sigma _1|%
\mathbf{ls}|\mathbf{k}_1m^{\prime }\sigma _2\rangle }{E_{\mathbf{k}_1m_1}-E_{%
\mathbf{k}^{\prime }m_1^{\prime }}}
$$
$$
\times I^{(0)}(\mathbf{kk}_1m_1^{\prime },\mathbf{k}_2m_2\mathbf{k}^{\prime
})a_{\mathbf{k}_1m_1\sigma _1}^{\dagger }a_{\mathbf{k}_2m_2\sigma }c_{%
\mathbf{k}\sigma }^{\dagger }c_{\mathbf{k}^{\prime }\sigma _2}.
\eqno{(\rm L.11)}
$$
The matrix elements in (L.11) are calculated, similar to Appendix K, with
the use of the plane-wave representation for conduction electrons. It is
convenient to use the $m$-representation for d-electron functions (the
functions, corresponding to irrdeducible representations of a point group
are expressed as their linear combinations). This representation enables one
to introduce in a simple way spin operators for localized d-electrons:
$$
a_{m\sigma }^{\dagger }a_{m\sigma }=(1+\sigma 2s^z)\varphi _m(l^z),
$$
$$
a_{m\pm }^{\dagger }a_{m\mp }=s^{\pm }\varphi _m(l^z),
\eqno{(\rm L.12)}
$$
where
$$
\varphi _0=1+\frac 14(l^z)^2[(l^z)^2-5],
$$
$$
\varphi _{\pm 1}=\frac 16l^z(l^z\pm 1)[4-(l^z)^2],
$$
$$
\varphi _{\pm 2}=\frac 1{24}l^z(l^z\pm 1)[(l^z)^2-1],
\eqno{(\rm L.13)}
$$
and the off-diagonal products $a_{m\sigma }^{\dagger }a_{m^{\prime }\sigma }$
are expressed in terms of operators $l^{\pm }$. Then we obtain for a cubic
crystal
$$
\mathcal{H}_{sd}^{(1)}=\frac \lambda {\Delta E}\sum_{ll^{\prime }}\sum_{\nu
\mathbf{kk}^{\prime }\sigma }F_{ll^{\prime }}(\mathbf{l}_\nu ,\mathbf{s}_\nu
)e^{i(\mathbf{k-k}^{\prime })\mathbf{R}_\nu } c_{\mathbf{k}%
\sigma }^{\dagger }c_{\mathbf{k}^{\prime }\sigma },
\eqno{(\rm L.14)}
$$
where we have put for simplicity $\Delta E_{\gamma \gamma ^{\prime }}(%
\mathbf{k})=\Delta E=\mathrm{const}$. The expansion term with $l=l^{\prime
}=1$ reads
$$
\mathcal{H}_{sd}^{(1)}(11)=-i\frac \lambda {\Delta E}\frac 65\pi
e^2\sum_{\nu \mathbf{kk}^{\prime }\sigma }G_{11}^{(1)}(\mathbf{k},\mathbf{k}%
^{\prime })[\mathbf{kk}^{\prime }]\mathbf{s}_\nu \langle \varphi (l_\nu
)\rangle
$$
$$
\times e^{i(\mathbf{k-k}^{\prime })\mathbf{R}_\nu }
c_{\mathbf{k}^{\prime }\sigma }^{\dagger }c_{\mathbf{k}\sigma },
\eqno{(\rm L.15)}
$$
(where the radial integral $G$ is given by (K.5)) and describes anisotropic
scattering.

Now we calculate matrix elements of the ``improper'' SOI (L.10). Replacing
for simplicity squares of wavefunctions of d-electrons by $\delta $%
-functions (which is possible since orbital momenta of d-electrons do not
enter) we derive
$$
L(\mathbf{kk}^{\prime })=i\lambda \sum_\nu e^{i(\mathbf{k-k}%
^{\prime })\mathbf{R}_\nu } \frac{[\mathbf{kk}^{\prime }]}{(\mathbf{k-k%
}^{\prime })^2}\mathbf{s}_\nu
\eqno{(\rm L.16)}
$$
with $\lambda ^{\prime }\sim 10^{-16}$ erg. The expression (L.16) does not
yield a dependence on orientation of localized spin in the crystal. However,
such a dependence will occur for more complicated d-wavefunctions.

Combining (L.15) with (L.16) we write down the correction to the Hamiltonian
of the s-d model owing to SOI as
$$
\mathcal{H}_{sd}^{\prime }=\sum_{\nu \mathbf{kk}^{\prime }\sigma }
e^{i(\mathbf{k-k}^{\prime })\mathbf{R}_\nu } (\mathbf{s}_\nu
\Lambda _{\mathbf{kk}^{\prime }})c_{\mathbf{k}\sigma }^{\dagger }c_{\mathbf{k%
}^{\prime }\sigma },
\eqno{(\rm L.17)}
$$
where
$$
\Lambda _{\mathbf{kk}^{\prime }}^z=i\lambda \bar l\frac{I_{\mathbf{kk}%
^{\prime }}^{(1)}}{\Delta E}\frac{[\mathbf{kk}^{\prime }]_z}
{kk^{\prime }}
+i\lambda ^{\prime }\frac{[\mathbf{kk}^{\prime }]_z}{(\mathbf{k-k}%
^{\prime })^2},
\eqno{(\rm L.18)}
$$
$$
\bar l=\frac 29\langle (l^z)^2[4-(l^z)^2]\left\{ \frac
14(l^z)^2[(l^z)^2-1]+2(l^z)^2[4-(l^z)^2]\right\} \rangle ,
\eqno{(\rm L.19)}
$$
$I^{(1)}$ is defined by (L.15). Although $|\lambda ^{\prime }|\ll |\lambda |$,
the role of the second term in (L.17) may be important provided that small
values of $|\mathbf{k-k}^{\prime }|$ yield the dominant contribution, as it
takes place for extraordinary transport phenomena at low temperatures.


\chapter{The density matrix technique for derivation of transport equations
and the theory of the anomalous Hall effect}

Mathematical description of transport phenomena deals with an balance
equation for the distribution function of current carriers in external
electric, magnetic and thermal fields (5.14). In simple cases such equations
may be obtained from a simple physical consideration of the electron motion
in the $k$-space due to external fields and collisions. However, in more
complicated situations, where higher orders in the scattering amplitude
become important, such simple arguments are not sufficient. Since often we
do meet with this situation for transport phenomena in magnetic crystals,
constructing a general method for derivation of transport equations is
needed. Most convenient is the approach which uses the density matrix. The
equations of motion for the latter quantity have usual quantum mechanics
form and may be simply reduced to transport equations.

Define the density matrix operator
$$
\widehat{\rho }=e^{-\beta \mathcal{H}}/\mathrm{Sp}e^{-\beta \mathcal{H}%
},\qquad \beta \equiv 1/T
\eqno{(\rm M.1)}
$$
where $\mathcal{H}$ is the total Hamiltonian of the system including current
carriers, scattering system and external fields. The coefficient in (M.1) is
determined by the normalization condition
\[
\mathrm{Sp}\widehat{\rho }=1
\]
The operator \thinspace $\widehat{\rho }$, as well as $\mathcal{H}$, may be
written down in any quantum representation. In the solid state theory, it is
convenient to use the second quantization representation. Then the symbol Sp
means the summation over all the possible occupation numbers of
quasiparticles in the system. For example, in the case of electron-phonon
system
$$
\mathrm{Sp}...=\sum_{\{n_{\mathbf{k}}\}\{N_{\mathbf{q}}\}}...
\eqno{(\rm M.2)}
$$
where $n_{\mathbf{k}}=0,1$ are the electron occupation numbers and $N_{%
\mathbf{q}}$ are the phonon ones.

Average value of a physical quantity $A$ is obtained as
$$
\langle A\rangle =\mathrm{Sp}(\widehat{\rho }A)
\eqno{(\rm M.3)}
$$
Calculate for example the average occupation numbers for non-interacting
conduction electrons. We find
\[
\langle n_{\mathbf{k}}\rangle =\langle c_{\mathbf{k}}^{\dagger }c_{\mathbf{%
k}}\rangle =\sum_{\{n_{\mathbf{k}^{\prime }}\}}\langle ...n_{\mathbf{k}%
}...\mid \prod_{\mathbf{k}^{\prime }}\exp (-\beta \varepsilon _{\mathbf{k}%
^{\prime }}c_{\mathbf{k}^{\prime }}^{\dagger }c_{\mathbf{k}^{\prime }})
\]
\[
\times c_{\mathbf{k}}^{\dagger }c_{\mathbf{k}}\mid ...n_{\mathbf{k}%
}...\rangle /\sum_{\{n_{\mathbf{k}^{\prime }}\}}\langle ...n_{\mathbf{k}%
}...\mid \prod_{\mathbf{k}^{\prime }}\exp (-\beta \varepsilon _{\mathbf{k}%
^{\prime }}c_{\mathbf{k}^{\prime }}^{\dagger }c_{\mathbf{k}^{\prime }})\mid
...n_{\mathbf{k}}...\rangle
\]
\[
=\sum_{n_{\mathbf{k}}}\exp (-\beta \varepsilon _{\mathbf{k}})n_{\mathbf{k}%
}/\sum_{n_{\mathbf{k}}}\exp (-\beta \varepsilon _{\mathbf{k}})=(\exp (\beta
\varepsilon _{\mathbf{k}})+1)^{-1}
\]
i.e. we derive the Fermi distribution function.

Provided that the Hamiltonian of the system is written in the form
\[
\mathcal{H=H}_0+\mathcal{H}^{\prime }
\]
with $\mathcal{H}$' being a perturbation, the equilibrium density matrix $%
\widehat{\rho }(\mathcal{H})$ may be expanded in the powers of H', which is
required at calculating the field term in the transport equation. To this
end one may use the theorem about expansion of exponential operators [706]:
\[
\langle n|\exp (-\beta (\mathcal{H}_0+\mathcal{H}^{\prime }))|n^{\prime
}\rangle =\delta _{nn^{\prime }}\exp (-\beta E_n^0)
\]
\[
-\frac{\exp (-\beta E_n^0)-\exp (-\beta E_n^0)}{E_n^0-E_{n^{\prime }}^0}%
\langle n|\mathcal{H}^{\prime }|n^{\prime }\rangle +\sum_{n^{\prime \prime }}%
\frac{\langle n|\mathcal{H}^{\prime }|n^{\prime \prime }\rangle \langle
n^{\prime \prime }|\mathcal{H}^{\prime }|n^{\prime }\rangle }{E_{n^{\prime
\prime }}^0-E_{n^{\prime }}^0}
\]
$$
\times \left[ \frac{\exp (-\beta E_{n^{\prime }}^0)-\exp (-\beta E_n^0)}{%
E_n^0-E_{n^{\prime }}^0}-\frac{\exp (-\beta E_{n^{\prime \prime }}^0)-\exp
(-\beta E_n^0)}{E_n^0-E_{n^{\prime \prime }}^0}\right]
\eqno{(\rm M.4)}
$$
where $E_n^0$ are the eigenvalues of $\mathcal{H}_0$. In the second
quantization representation, the numbers n mean the occupation number sets,
and summing over them yields an expression of $\widehat{\rho }(\mathcal{H})$
in terms of powers of $\mathcal{H}$' and $\widehat{\rho }(\mathcal{H})=%
\widehat{\rho }_0$. In an representation of the wavefunctions $|l\rangle $
we have
\[
\widehat{\rho }=\widehat{\rho }^{(0)}+\widehat{\rho }^{(1)}+\widehat{\rho }%
^{(2)}+...
\]
$$
\widehat{\rho }_{ll^{\prime }}^{(0)}=\rho _l\delta _{ll^{\prime }},\qquad
\widehat{\rho }_{ll^{\prime }}^{(1)}=\frac{\rho _l-\rho _{l^{\prime }}}{%
\varepsilon _l-\varepsilon _{l^{\prime }}}\mathcal{H}_{ll^{\prime }}^{\prime
}
\eqno{(\rm M.6a)}
$$
where
$$
\widehat{\rho }_{ll^{\prime }}^{(2)} =\gamma \rho _l\delta _{ll^{\prime
}}+\sum_{l^{\prime \prime }}\frac{\mathcal{H}_{ll^{\prime \prime }}^{\prime }%
\mathcal{H}_{l^{\prime \prime }l^{\prime }}^{\prime }}{\varepsilon
_l-\varepsilon _{l^{\prime }}}\left( \frac{\rho _l-\rho _{l^{\prime }}}{%
\varepsilon _l-\varepsilon _{l^{\prime }}}-\frac{\rho _l-\rho _{l^{\prime
\prime }}}{\varepsilon _l-\varepsilon _{l^{\prime \prime }}}\right)
\eqno{(\rm M.6b)}
$$
$$
\gamma =\sum_{ll^{\prime }}\frac{|\mathcal{H}_{ll^{\prime }}^{\prime }|^2}{%
\varepsilon _l-\varepsilon _{l^{\prime }}}\left[ \frac{\rho _l-\rho
_{l^{\prime }}}{\varepsilon _l-\varepsilon _{l^{\prime }}}-\frac{\partial
\rho _l}{\partial \varepsilon _l}\right]
\eqno{(\rm M.6c)}
$$

The equation of motion for the density matrix operator has the usual form
$$
\frac{\partial \widehat{\rho }}{\partial t}=[\mathcal{H},\widehat{\rho }]
\eqno{(\rm M.7)}
$$
In any representation $|l\rangle $, we may write down the system of
equations for diagonal and off-diagonal matrix elements. Under certain
conditions this system may be reduced to transport equations of the lowest
Born and next-order approximations. In this way, transport equations were
obtained for elastic scattering by impurities [458], and further for
scattering by phonons [460] and spin inhomogeneities [466]. The theory in
the case of arbitrarily strong scattering amplitude, but small impurity
concentration was developed by Luttinger [707]. However, for a complicated $%
\mathcal{H}_0$, calculations in the matrix form are rather cumbersome even
for impurity scatterng in the second Born approximation. Therefore it is
very convenient for practical calculations to derive transport equations in
an operator form without concretizing the Hamiltonian. Now we consider this
technique constructed in [471,472].

The only requirement to the Hamiltonian is the possibility of the
representation
$$
\mathcal{H}_t=\mathcal{H}_0+\mathcal{H}_Ee^{st}+\mathcal{H}^{\prime }
\eqno{(\rm M.8)}
$$
where $\mathcal{H}_0$ has a diagonal form in the $n$-representation, $%
\mathcal{H}_E$ is the energy of the system in the adiabatically included ($%
s\rightarrow 0$) electric field $\mathbf{E}$, $\mathcal{H}^{\prime }$ is an
off-diagonal part. The total density matrix operator is represented as
$$
\widehat{\rho }_t=\widehat{\rho }+e^{st}(\widehat{f_a}-\widehat{f_b})
\eqno{(\rm M.9)}
$$
where $\widehat{\rho }$ is the equilibrium density matrix in the absence of
electric field, $\widehat{f_a}$ and $\widehat{f_b}$ are the diagonal and
off-diagonal components of the correction. Taking into account the relations
\[
\lbrack \mathcal{H}_0,\widehat{f_a}]=[\mathcal{H}^{\prime },\widehat{f_a}%
]_a=[\mathcal{H}_0,\widehat{f_b}]_a
\]
we obtain to linear approximation in $E$%
$$
s\widehat{f_a}=[\mathcal{H}_E,\widehat{\rho }]_a+[\mathcal{H}^{\prime },
\widehat{f_b}]_a
\eqno{(\rm M.10)}
$$
$$
\widehat{f_b}=\widehat{L}^{-1}\left\{ [\mathcal{H}_E,\widehat{\rho }]_b+[%
\mathcal{H}^{\prime },\widehat{f_a}+\widehat{f_b}]_b\right\}
\eqno{(\rm M.11)}
$$
where
$$
\widehat{L}^{-1}=\frac 1{is-\Delta },\qquad \Delta \widehat{f_b}=[\mathcal{H}%
_0,\widehat{f_b}]
\eqno{(\rm M.12)}
$$
$\Delta $ is the difference of the eigenvalues of $\mathcal{H}$ in the
corresponding states. The system (M.10), (M.11) may be solved by the
iteration method. Substituting (M.11) into (M.10) we derive
$$
is\widehat{f_b}=[\mathcal{H}_E,\widehat{\rho }]_a+[\mathcal{H}^{\prime },%
\widehat{L}^{-1}\left\{ [\mathcal{H}_E,\widehat{\rho }]_b+[\mathcal{H}%
^{\prime },\widehat{f_a}+\widehat{f_b}]_b\right\} ]_a
\eqno{(\rm M.13)}
$$
Repeating the procedure we have after the $n$-th iteration
\[
is\widehat{f_a}=[\mathcal{H}_E,\widehat{\rho }]_a+[\mathcal{H}^{\prime },%
\widehat{L}^{-1}\left\{ [\mathcal{H}_E,\widehat{\rho }]_b+[\mathcal{H}%
^{\prime },\widehat{f_a}]_b\right.
\]
$$
+\left. [\mathcal{H}^{\prime },\widehat{L}^{-1}\left\{ [\mathcal{H}_E,%
\widehat{\rho }]_b+[\mathcal{H}^{\prime },\widehat{f_a}]_b+[\mathcal{H}%
^{\prime },\widehat{L}^{-1}\left\{ \left\{ [\mathcal{H}_E,\widehat{\rho }%
]_b+[\mathcal{H}^{\prime },\widehat{f_a}]_b+...\right. \right. \right.
\right\}
\eqno{(\rm M.14)}
$$
\[
+\left. \left. [\mathcal{H}^{\prime },\widehat{L}^{-1}\left\{ ...\left\{ [%
\mathcal{H}_E,\widehat{\rho }]_b+[\mathcal{H}^{\prime },\widehat{f_a}+
\widehat{f_b}]_b\right\} ...\right\} \right\} \right\} ]_a
\]
where $\{{...\{}$ stands for ${n}$ curly brackets. The iteration
procedure allows one to obtain the solution to any desired order in
$\mathcal{H}^{\prime }$ provided that the expansion of $\widehat{f_a}$ in
$\mathcal{H}^{\prime }$ starts from a lower degree of $\mathcal{H}^{\prime }$
than that of $\widehat{f_b}$. Then, to accuracy of
$(\mathcal{H}^{\prime })^{n+1}$, we may neglect the term
$[\mathcal{H}^{\prime },\widehat{f_b}]$ in (M.14) and solve for
$\widehat{f_a}$, and further find $\widehat{f_b}$ from (M.10).

Consider the first iteration. To lowest order in $\mathcal{H}^{\prime }$ we
have
$$
\lbrack \mathcal{H}_E,\widehat{\rho }^{(0)}]_b+[\mathcal{H}^{\prime },%
\widehat{L}^{-1}[\mathcal{H}^{\prime },\widehat{f_a}]_b]_a=0
\eqno{(\rm M.15)}
$$
Taking into account the identity
$$
\widehat{L}_{nn^{\prime }}^{-1}+\widehat{L}_{n^{\prime }n}^{-1}=-\left(
\frac 1{\Delta _{nn^{\prime }}}+\frac 1{\Delta _{n^{\prime }n}}\right) -i\pi
[\delta (\Delta _{nn^{\prime }})+\delta (\Delta _{n^{\prime }n})]
\eqno{(\rm M.16)}
$$
multiplying by $n_\lambda =a_\lambda ^{\dagger }a_\lambda $ and summing
over the occupation numbers we obtain from (M.15)
$$
C_\lambda ^{(0)}+T_\lambda ^{(0)}=0
\eqno{(\rm M.17)}
$$
where
$$
C_\lambda ^{(0)}=\mathrm{Sp}\{[\mathcal{H}_E,\widehat{\rho }^{(0)}]n_\lambda
\}
\eqno{(\rm M.18)}
$$
is the field term,
$$
T_\lambda ^{(0)} =\mathrm{Sp}\left\{ [\mathcal{H}^{\prime },\widehat{L}%
^{-1}[\mathcal{H}^{\prime },\widehat{f_a}]_b]_an_\lambda \right\}
\eqno{(\rm M.19)}
$$
\[
=2\pi i\mathrm{Sp}\{\mathcal{H}^{\prime }\widehat{f}_a\underline{\mathcal{H%
}}^{\prime }-\mathcal{H}^{\prime }\underline{\mathcal{H}^{\prime }}\widehat{f%
}_a\}\underline{\delta (\Delta )}n_\lambda
\]
is the collision term (the $\delta $-functions correspond to
underlined operator $\mathcal{H}^{\prime }$). One can see that $\widehat{f_a}%
=\widehat{f}^{(-2)}\sim $ $(\mathcal{H}^{\prime })^{-2}$. Thus we
obtain an usual Boltzmann-type transport equation.

In the next order approximation with respect to $\mathcal{H}^{\prime }$ we
have to take into account terms with $\widehat{f}^{(-1)}\sim (\mathcal{H}%
^{\prime })^{-1}$. One obtains
$$
C_\lambda ^{(1)}+C_\lambda ^{\prime (1)}=T_\lambda
^{(0)}(f^{(-1)})+T_\lambda ^{(0)}(f^{(-2)})=0
\eqno{(\rm M.20)}
$$
where
$$
C_\lambda ^{(1)}=\mathrm{Sp}\{[\mathcal{H}_E,\widehat{\rho }^{(1)}]n_\lambda
\}
\eqno{(\rm M.21)}
$$
$$
C_\lambda ^{\prime (1)}=2\pi i\mathrm{Sp}\{[\mathcal{H}^{\prime },\widehat{%
\rho }^{(0)}\underline{\mathcal{H}_E}-\mathcal{H}^{\prime }\underline{%
\mathcal{H}_E}\widehat{\rho }^{(0)}]n_\lambda \delta (\Delta )\}
\eqno{(\rm M.22)}
$$
and, in the case of purely imaginary product $\mathcal{H}^{\prime }\mathcal{H%
}^{\prime }\mathcal{H}^{\prime }$ (e.g., for the spin-orbital interaction),
$$
T_\lambda ^{(1)}=6\pi ^2\mathrm{Sp}\{\widehat{f_a}\underline{\delta
(\Delta )\mathcal{H}^{\prime }}n_\lambda \mathcal{H}^{\prime }\underline{%
\mathcal{H}^{\prime }\delta (\Delta )}\}
\eqno{(\rm M.23)}
$$
We see that one of important advantages of the method is a simple formation
of $\delta $-function terms.

To second order in $\mathcal{H}^{\prime }$%
$$
C_\lambda ^{(2)}+C_\lambda ^{\prime (2)}+C_\lambda ^{\prime \prime
(2)}+T_\lambda ^{(0)}(f^{(0)})+T_\lambda ^{(1)}(f^{(-)})+T_\lambda
^{(2)}(f^{(-2)})=0
\eqno{(\rm M.24)}
$$
where
$$
C_\lambda ^{(2)}=\mathrm{Sp}\{[\mathcal{H}_E,\widehat{\rho }^{(2)}]n_\lambda
\}
\eqno{(\rm M.25)}
$$
$$
C_\lambda ^{\prime (2)}=\mathrm{Sp}\{[\mathcal{H}^{\prime },\widehat{L}^{-1}[%
\mathcal{H}_E,\widehat{\rho }^{(1)}]]n_\lambda \}
\eqno{(\rm M.26)}
$$
$$
C_\lambda ^{\prime \prime (2)}=\mathrm{Sp}\{[\mathcal{H}^{\prime },\widehat{L%
}^{-1}[\mathcal{H}^{\prime },\widehat{L}^{-1}[\mathcal{H}_E,\widehat{\rho }%
^{(0)}]]]n_\lambda \}
\eqno{(\rm M.27)}
$$
$$
T_\lambda ^{(2)}=\mathrm{Sp}\{[\mathcal{H}^{\prime },\underline{\widehat{L}%
^{-1}[\mathcal{H}^{\prime },\widehat{f_a}]}_b[\underline{\underline{[%
\mathcal{H}^{\prime },\widehat{n}_\lambda ]\widehat{L}^{-1}},\mathcal{H}%
^{\prime }]_b\widehat{L}^{-1}}\}
\eqno{(\rm M.28)}
$$
In the matrix form this result was derived by Kohn and Luttinger [478].
Further we consider the transport equations for concrete scattering
mechanisms.

\section{Impurity scattering}

The Hamiltonian of electron-impurity system is given by
$$
\mathcal{H}_0=\sum_l\varepsilon _lc_l^{\dagger }c_{l},\qquad
\mathcal{H}_E=e\mathbf{E}\sum_{ll^{\prime }}\mathbf{r}_{ll^{\prime }}
c_l^{\dagger }c_{l^{\prime }},\qquad
\mathcal{H}^{\prime }=\sum_{ll^{\prime }}V_{ll^{\prime }}c_l^{\dagger }
c_{l^{\prime }}
\eqno{(\rm M.29)}
$$
where $l=\{{n\mathbf{k}\}}$, $n$ is the band index
$$
V_{ll^{\prime }}=\sum_{i=1}^{n_i}e^{i(\mathbf{k-k}^{\prime })\mathbf{R}%
_i}\varphi _{ll^{\prime }}
\eqno{(\rm M.30)}
$$
$$
\varphi _{ll^{\prime }}=\int \mathrm{d}\mathbf{r}e^{i(\mathbf{k-k}^{\prime
})\mathbf{r}}u_l^{*}(\mathbf{r})u_{l^{\prime }}(\mathbf{r})\varphi (\mathbf{r%
})
\eqno{(\rm M.31)}
$$
$\varphi (\mathbf{r})$ is the one-impurity potential, $n_i$ is the number
of impurities, $u_l$ are the Bloch factors in (2.1). Averaging over random
distribution of the impurities yields
$$
\langle |V_{ll^{\prime }}|^2\rangle =n_i|\varphi _{ll^{\prime }}|^2\qquad
(l\neq l^{\prime })
\eqno{(\rm M.32)}
$$
The diagonal matrix elements of the impurity potential may be included into
and are not important at derivation of the transport equation. Consider the
lowest order equation (M.17). Substituting $\mathcal{H}_E$ and $\mathcal{H}%
^{\prime }$ into (M.18), (M.19) we obtain
$$
C_l^{(0)}=eE^\alpha \sum_{l^{\prime }}(\rho _{ll^{\prime
}}^{(0)}r_{l^{\prime }l}^\alpha -r_{ll^{\prime }}^\alpha \rho _{l^{\prime
}l}^{(0)})=e\mathbf{E}[\widehat{\rho }^{(0)},\mathbf{r}]_l=ie\mathbf{E}\frac{%
\partial n_l}{\partial \mathbf{k}}
\eqno{(\rm M.33)}
$$
(hereafter we use the notation $n_l=\rho _l^{(0)}=f(\varepsilon _l)$ for the
equilibrium Fermi distribution function),
$$
T_l^{(0)}=2\pi i\sum_{l^{\prime }}|\varphi _{ll^{\prime }}|^2(f_{l^{\prime
}}-f_l)\delta (\varepsilon _{l^{\prime }}-\varepsilon _l)
\eqno{(\rm M.34)}
$$
with
$$
f_l=\mathrm{Sp}(\widehat{f}_ac_l^{\dagger }c_l)
\eqno{(\rm M.35)}
$$
The expressions (M.33), (M.34) yield the field and collision terms of the
transport equation for the elastic scattering by impurities in the Born
approximation. The results in the two next approximation were obtained by
Luttinger [459]. The second Born approximation transport equation for
impurity scattering reads (first-order corrections to the field term vanish)
$$
T_l^{(1)}(f^{(-2)})+T_l^{(0)}(f^{(-1)})=0
\eqno{(\rm M.36)}
$$
where
\[
T_l^{(1)}=-2\pi \sum_{l^{\prime }l^{\prime \prime }}\delta (\varepsilon
_l-\varepsilon _{l^{\prime }})[(L^{(-1)})_{ll^{\prime \prime }}\langle
V_{ll^{\prime \prime }}V_{l^{\prime \prime }l^{\prime }}V_{l^{\prime
}l}\rangle
\]
$$
+(L^{(-1)})_{ll^{\prime \prime }}^{*}\langle V_{ll^{\prime }}V_{l^{\prime
}l^{\prime \prime }}V_{l^{\prime \prime }l}\rangle (f_l-f_{l^{\prime }})]
\eqno{(\rm M.37)}
$$
and $T_l^{(0)}$ is given by (M.34).

Consider the case where the spin-orbital interaction is present. Picking out
in (M.37) the imaginary part, which is linear in SOI, and averaging over
impurities we obtain
$$
T_l^{(2)}(f^{(-2)})=-(2\pi )^2n_i\sum_{l^{\prime }l^{\prime \prime }}\delta
(\varepsilon _l-\varepsilon _{l^{\prime }})\delta (\varepsilon
_l-\varepsilon _{l^{\prime \prime }})(f_l^{(-2)}-f_{l^{\prime }}^{(-2)})\Im
(\varphi _{ll^{\prime }}\varphi _{l^{\prime }l^{\prime \prime }}\varphi
_{l^{\prime \prime }l})
\eqno{(\rm M.38)}
$$
Assuming that the effective radius $r_0$ of the potential $\varphi (\mathbf{%
r})$ is small $(k_Fr_0\ll 1)$ we may put
$$
\varphi _{\mathbf{kk}^{\prime }}\approx \int d\mathbf{r}\varphi (\mathbf{r}%
)=\overline{\varphi }
\eqno{(\rm M.39)}
$$
Then, using the effective mass approximation we obtain the solution to the
first Born approximation (only intraband transitions should be taken into
account due to the delta functions)
$$
f_l^{(-2)}=-\tau _0eE\frac{\partial n_l}{\partial \varepsilon _l}v_l
\eqno{(\rm M.40)}
$$
$$
\tau _0^{-1}=\frac{n_i}{2\pi }\overline{\varphi }^2\frac{(2m*)^{3/2}}{\hbar
^4}E_F^{1/2}
\eqno{(\rm M.41)}
$$
To calculate corrections owing to SOI we have to expand the matrix elements
(M.31) in small $|\mathbf{k-k}^{\prime }|$:
\[
\mathrm{Im}(\varphi _{ll^{\prime }}\varphi _{l^{\prime }l^{\prime \prime
}}\varphi _{l^{\prime \prime }l})=\frac i2\varphi _{kk^{\prime }}\varphi
_{k^{\prime }k^{\prime \prime }}\varphi _{k^{\prime \prime }k}
\]
$$
\times (k_\alpha ^{\prime }-k_\alpha ^{\prime \prime })(k_\beta ^{\prime
}-k_\beta )\left( \frac{\partial J_\alpha ^l}{\partial k_\beta }-\frac{%
\partial J_\beta ^l}{\partial k_\alpha }\right)
\eqno{(\rm M.42)}
$$
where
$$
J_\alpha ^l=J_\alpha ^{nn}(\mathbf{k}),\qquad
J_\alpha ^{nn^{\prime }}=\int d\mathbf{r}u_{n\mathbf{k}}^{*}(\mathbf{r})
\frac{\partial u_{n^{\prime }\mathbf{k}}(\mathbf{r})}{\partial k_\alpha }
\eqno{(\rm M.43)}
$$
is proportional to SOI and purely imaginary.

Substituting (M.40), (M.42) into (M.38) we derive
\[
\frac{in_i\overline{\varphi }^3E_F^2\tau _0(2m^{*})^3}{12\pi \hbar ^7}%
\frac{\partial n_l}{\partial \varepsilon _l}eE^\alpha k_\beta \left( \frac{%
\partial J_\alpha ^l}{\partial k_\beta }-\frac{\partial J_\beta ^l}{\partial
k_\alpha }\right)
\]
$$
+2\pi n_i\overline{\varphi }^2\sum_{\mathbf{k}}\delta (\varepsilon _{n%
\mathbf{k}}-\varepsilon _{n\mathbf{k}^{\prime }})(f_{n\mathbf{k}}^{(-1)}-f_{n%
\mathbf{k}^{\prime }}^{(-1)})=0
\eqno{(\rm M.44)}
$$
This equation has the structure of an usual Born equation with modified
field term. Its solution has the form
$$
f_l^{(-1)}=-\frac{iE_F}{3n_i\overline{\varphi }}\frac{\partial n_l}{\partial
\varepsilon _l}eE^\alpha k_\beta \left( \frac{\partial J_\alpha ^l}{\partial
k_\beta }-\frac{\partial J_\beta ^l}{\partial k_\alpha }\right)
\eqno{(\rm M.45)}
$$
Now we estimate the quantities $J$. To first order of perturbation theory in
SOI we obtain
$$
u_{n\mathbf{k}}=u_{n\mathbf{k}}^{(0)}+\sum_{n^{\prime }\neq n}\frac{\langle
n^{\prime }|\mathcal{H}_{S0}|n\rangle }{\varepsilon _{n\mathbf{k}%
}-\varepsilon _{n\mathbf{k}^{\prime }}}u_{n^{\prime }\mathbf{k}}
\eqno{(\rm M.46)}
$$
$$
J_\alpha ^l=-2i\sum_{n^{\prime }\neq n}\frac{r_{nn^{\prime }}^\alpha \langle
n^{\prime }|\mathcal{H}_{S0}|n\rangle }{\varepsilon _{n\mathbf{k}%
}-\varepsilon _{n\mathbf{k}^{\prime }}}
\eqno{(\rm M.47)}
$$
where matrix elements of coordinate are connected with those of
quasimomentum
$$
r_{nn^{\prime }}^\alpha (\mathbf{k})=\frac{-iP_{nn^{\prime }}^\alpha (%
\mathbf{k})}{m^{*}(\varepsilon _{n\mathbf{k}}-\varepsilon _{n\mathbf{k}%
^{\prime }})}
\eqno{(\rm M.48)}
$$
Putting for simplicity $\varepsilon _{n\mathbf{k}}-\varepsilon _{n^{\prime }%
\mathbf{k}}=\Delta =\mathrm{const}$ and substituting the expression for $%
\mathcal{H}_{so}$ we derive
\[
J_\alpha ^l=\frac 1{m^{*}\Delta ^2}\sum_{n^{\prime }\neq n}\left( \left(
\mathcal{H}_{S0}\right) _{nn^{\prime }}P_{n^{\prime }n}^\alpha
-P_{nn^{\prime }}^\alpha \left( \mathcal{H}_{S0}\right) _{n^{\prime
}n}\right)
\]
$$
=\frac \hbar {m^{*}\Delta ^2}[\mathcal{H}_{so,}p^\alpha ]_{n^{\prime }n}
\eqno{(\rm M.49)}
$$
Using the Poisson equation
$$
\Delta V(\mathbf{r})=-4\pi e^2\rho (\mathbf{r})
\eqno{(\rm M.50)}
$$
($\rho (\mathbf{r})$ is the charge density) and (L.1) we obtain
$$
J_\alpha ^l=-i\frac{2\pi \hbar ^4e^2\rho _{eff}}{3m^2c^2m^{*}\Delta ^2M(0)}[%
\mathbf{kM}]^\alpha
\eqno{(\rm M.51)}
$$
where $M$ is the magnetization,
$$
\rho _{eff}=\int d\mathbf{r}|u_{n\mathbf{k}}(\mathbf{r})|^2\rho (\mathbf{r}%
)
\eqno{(\rm M.52)}
$$
Substituting (M.51) into (M.45) we find the Hall conductivity
$$
\sigma _{yx} =\frac{e\overline{v}_y}{a_0^3E_x}=\frac
e{a_0^3E_x}\sum_lv_y^lf_l^{(-1)}
\eqno{(\rm M.53)}
$$
\[
=t\frac{e^2n}{m^{*}}\frac{2\pi }{3\Delta ^2}\mu _B^2\hbar \rho _{eff}\frac
M{M(0)}\frac{E_F}{3n_i\overline{\varphi }}
\]
where $t$ is the number of bands, $n=k_F^3/6\pi ^2$ is the electron
concentration. Taking into account the expression for the diagonal component
of the conductivity tensor
\[
\sigma _{xx}=\frac{e\overline{v_x}}{a_0^3E_x}=\frac
e{a_0^3E_x}\sum_lv_x^lf_l^{(-2)}
\]
$$
=-\tau _0e^2\sum_l(v_l)^2\frac{\partial n_l}{\partial \varepsilon _l}=t\frac{%
e^2n\tau _0}{m^{*}}
\eqno{(\rm M.54)}
$$
we obtain for the Hall coefficient the result (5.124).

\section{Scattering by phonons}

In the case of electron-phonon interaction we have
$$
\mathcal{H}_0=\sum_l\varepsilon _lc_l^{\dagger }c_l+\sum_{\mathbf{q}}\omega
_{\mathbf{q}}b_{\mathbf{q}}^{\dagger }b_{\mathbf{q}}
\eqno{(\rm M.55)}
$$
$$
\mathcal{H}^{\prime }=\sum_{ll^{\prime }\mathbf{q}}(Q_{ll^{\prime }\mathbf{q}%
}c_l^{\dagger }c_{l^{\prime }}b_{\mathbf{q}}+Q_{ll^{\prime }\mathbf{q}%
}^{*}c_{l^{\prime }}^{\dagger }c_lb_{\mathbf{q}}^{\dagger })
\eqno{(\rm M.56)}
$$
where
\[
Q_{ll^{\prime }\mathbf{q}}=\frac 23i(2M\omega _{\mathbf{q}%
})^{-1/2}qC_{nn^{\prime }}\delta _{\mathbf{k-k}^{\prime },\mathbf{q}}
\]
$\omega _{\mathbf{q}}$ is the phonon frequency, $b_{\mathbf{q}}^{\dagger }$
and $b_{\mathbf{q}}$ are the Bose creation and annihilation operators for
phonons, $C$ is the Bloch constant in the conductivity theory [1]. Inclusion
of the second term in $\mathcal{H}_0$ takes into account the inelastic
character of electron-phonon scattering.

Consider the lowest-order approximation (M.17). Then the field term is
determined by (M.33), and the collision term is calculated from (M.19):
\[
T_l^{(0)}=2\pi i\sum_{l^{\prime }\mathbf{q}}|Q_{ll^{\prime }\mathbf{q}%
}|^2\left\{ [(f_{l^{\prime }}-f_l)N_{\mathbf{q}}-f_l(1-n_{l^{\prime
}})+f_{l^{\prime }}n_{l^{\prime }}]\delta (\varepsilon _{l^{\prime
}}-\varepsilon _l+\omega _{\mathbf{q}})\right.
\]
$$
+\left. [(f_{l^{\prime }}-f_l)(1+N_{\mathbf{q}})+f_l(1-n_{l^{\prime
}})+f_{l^{\prime }}n_{l^{\prime }}]\delta (\varepsilon _{l^{\prime
}}-\varepsilon _l-\omega _{\mathbf{q}}\right\}
\eqno{(\rm M.57)}
$$
Here we have carried out decouplings of many-particle density matrices, so
that $N_{\mathbf{q}}$ and $n_{n\mathbf{k}}$ are the equilibrium Bose and
Fermi functions.

The lowest-order contribution to AHE owing to the phonon scattering is
described by the equation of the second order in $\mathcal{H}$'. The
corresponding corrections to the distribution function, which are linear in
SOI, occur both from field and collision terms. To simplify the
consideration, we follow to paper [460] and do not take into account all
such corrections, but restrict ourselves to the equation
$$
C_l^{\prime (2)}+T_l^{(0)}(f^{(0)})=0
\eqno{(\rm M.58)}
$$
(see (M.24)) and neglect $C^{(2)}$, $C^{\prime \prime (2)}$ and $T^{(2)}$.
The contributions neglected apparently do not influence qualitative
results.

To calculate $C_l^{\prime (2)}$ we have to expand equilibrium
distribution functions up to the first order in the scattering amplitude.
Performing the decoupling of many-particle averages with the use of the
theorem about expansion of exponential operators (M.6) we obtain
\[
\sum_{l^{\prime }\mathbf{q}}|Q_{ll^{\prime }\mathbf{q}}|^2[\delta
(\varepsilon _{l^{\prime }}-\varepsilon _l+\omega _{\mathbf{q}})\varphi
_{ll^{\prime }\mathbf{q}}^{(0)}-\delta (\varepsilon _{l^{\prime
}}-\varepsilon _l-\omega _{\mathbf{q}})\varphi _{ll^{\prime }\mathbf{q}%
}^{(0)}]
\]
\[
+e\mathbf{E}\sum_{l^{\prime }\mathbf{q}}|Q_{ll^{\prime }\mathbf{q}%
}|^2\beta n_{l^{\prime }}(1-n_l)(\mathbf{r}_{l^{\prime }}-\mathbf{r}%
_l)[\delta (\varepsilon _{l^{\prime }}-\varepsilon _l+\omega _{\mathbf{q}%
})N_{\mathbf{q}}
\]
$$
+\delta (\varepsilon _{l^{\prime }}-\varepsilon _l-\omega _{\mathbf{q}%
})(1+N_{\mathbf{q}})]=0
\eqno{(\rm M.59)}
$$
where
$$
\varphi _{ll^{\prime }\mathbf{q}}^{(i)}=N_{\mathbf{q}}(n_{l^{\prime
}}-n_l)+f_l^{(i)}(1-n_{l^{\prime }})+f_{l^{\prime }}^{(i)}
\eqno{(\rm M.60)}
$$
For high temperatures $T\gg \omega _{\mathbf{q}}$ we have
\[
\beta n_{l^{\prime }}(1-n_{l^{\prime }})=-\frac{\partial n_l}{\partial
\varepsilon _l}
\]
so that the solution, which is linear in SOI, has the form
$$
f_l^{(0)}=ieE^\alpha J_\alpha ^l\frac{\partial n_l}{\partial \varepsilon _l}
\eqno{(\rm M.61)}
$$
where the quantity $J$ (see (M.43)) is determined by the diagonal part of
the coordinate, which is due to SOI [459]:
$$
r_{ll^{\prime }}^\alpha =i\delta _{ll^{\prime }}\frac \partial {\partial
k_\alpha }+iJ_\alpha ^{nn^{\prime }}(\mathbf{k})\delta _{\mathbf{kk}^{\prime
}}
\eqno{(\rm M.62)}
$$
The off-diagonal part of $f^{(0)}$ is obtained from the equation (M.11):
\[
f_{l^{\prime }l}^{(0)}=\frac 1{\Delta _{l^{\prime }l}}\left\{ eE^\alpha [%
\widehat{\rho }^{(0)},\mathbf{r}^\alpha ]_{l^{\prime }l}-\sum_{l^{\prime
\prime }\mathbf{q}}\left[ Q_{l^{\prime }l^{\prime \prime }\mathbf{q}%
}Q_{ll^{\prime \prime }\mathbf{q}}\frac{\varphi _{l^{\prime \prime }l\mathbf{%
q}}^{(-2)}}{\Delta _{l^{\prime \prime }l\mathbf{q}}^{+}}-Q_{l^{\prime \prime
}l^{\prime }\mathbf{q}}^{*}Q_{l^{\prime \prime }l\mathbf{q}}\frac{\varphi
_{ll^{\prime \prime }\mathbf{q}}^{(-2)}}{\Delta _{l^{\prime \prime }l\mathbf{%
q}}^{-}}\right. \right.
\]
$$
-\left. \left. Q_{ll^{\prime \prime }\mathbf{q}}Q_{l^{\prime \prime
}l^{\prime }\mathbf{q}}^{*}\frac{\varphi _{l^{\prime }l^{\prime \prime }%
\mathbf{q}}^{(-2)}}{\Delta _{l^{\prime }l^{\prime \prime }\mathbf{q}}^{+}}%
+Q_{ll^{\prime \prime }\mathbf{q}}Q_{l^{\prime }l^{\prime \prime }\mathbf{q}}%
\frac{\varphi _{ll^{\prime \prime }\mathbf{q}}^{(-2)}}{\Delta _{l^{\prime
}l^{\prime \prime }\mathbf{q}}^{-}}\right] \right\}
\eqno{(\rm M.63)}
$$
where
\[
\Delta _{l^{\prime }l}=\varepsilon _l-\varepsilon _{l^{\prime }}-is,\qquad
\Delta _{l^{\prime }l\mathbf{q}}^{\pm }=\varepsilon _l-\varepsilon
_{l^{\prime }}\pm \omega _{\mathbf{q}}-is,\qquad s\rightarrow 0
\]
Using (M.61) and (M.63) we can calculate average velocity of current
carriers to zeroth order in $Q$:
$$
\overline{\mathbf{v}}=\sum_lf_l^{(0)}\mathbf{v}^l+\sum_{l^{\prime }\neq
l}f_{l^{\prime }l}^{(0)}\mathbf{v}^{ll^{\prime }}\equiv \overline{\mathbf{v}}%
_a\mathbf{+}\overline{\mathbf{v}}_b
\eqno{(\rm M.64)}
$$
$$
\mathbf{v}^l=\frac{\partial \varepsilon _l}{\partial \mathbf{k}},v_\alpha
^{ll^{\prime }}=-(\varepsilon _l-\varepsilon _{l^{\prime }})J_\alpha
^{nn^{\prime }}\delta _{\mathbf{kk}^{\prime }}
\eqno{(\rm M.65)}
$$
Taking into account (M.62) and the relation
$$
\sum_{n^{\prime }}\left( J_\beta ^{nn^{\prime }}J_\alpha ^{n^{\prime
}n}-J_\alpha ^{nn^{\prime }}J_\beta ^{n^{\prime }n}\right) =\frac{\partial
J_\beta ^l}{\partial k_\alpha }-\frac{\partial J_\alpha ^l}{\partial k_\beta
}
\eqno{(\rm M.66)}
$$
we obtain
$$
\overline{\mathbf{v}}_a=ieE^\alpha \sum_l\frac{\partial n_l}{\partial
\varepsilon _l}\mathbf{v}^lJ_\alpha ^l
\eqno{(\rm M.67)}
$$
$$
\overline{v}_{b\beta }=ieE^\alpha \sum_ln_l\left( \frac{\partial J_\beta ^l}{%
\partial k_\alpha }-\frac{\partial J_\alpha ^l}{\partial k_\beta }\right)
\eqno{(\rm M.68)}
$$
where we have neglected interband transitions. Integrating (M.68) by parts
and combining with (M.67) we get the final expression for the average
velocity
$$
\overline{v}_\alpha =ie\mathbf{E}\sum_l\frac{\partial n_l}{\partial
\varepsilon _l}\mathbf{v}^lJ_\alpha ^l
\eqno{(\rm M.69)}
$$
Substituting the expression (M.51) into (M.69) we obtain
$$
\sigma _{yx}^{ph}=-\frac{2\pi }3\frac{e^4\hbar ^4}{m^{*}(mc\Delta )^2}\frac{%
M^z}{M(0)}\rho _{eff}\sum_l\frac{\partial n_l}{\partial \varepsilon _l}%
v_x^lk_x
\eqno{(\rm M.70)}
$$
Integrating over $\mathbf{k}$ and averaging over subbands yields, similar to
(M.53)
$$
\sigma _{yx}^{ph}=\frac{2\pi }3\frac{e^4n\hbar ^4}{(mc\Delta )^2}t\rho
_{eff}\langle \frac 1{m^{*}}\rangle \frac{M^z}{M(0)}
\eqno{(\rm M.71)}
$$
Then the spontaneous Hall coefficient reads
$$
R_S^{ph}=-\frac{\sigma _{yx}}{4\pi M^z}\rho ^2=-\frac 23\frac{\mu
_B^2e^2n\hbar }{\Delta ^2}\rho _{eff}t\langle \frac 1{m^{*}}\rangle \frac{%
\rho ^2}{M(0)}
\eqno{(\rm M.72)}
$$

\section{Scattering by spin inhomogeneities}

We describe the interaction of conduction electrons with magnetic moments
within the s-d exchange model with inclusion of spin-orbital interaction
(L.17). In the case of high temperatures we include into $\mathcal{H}$ the
interaction of electrons with the mean field:
$$
\mathcal{H}_d=-W_M\sum_iS_i^z,\qquad W_M=2J_0\langle S^z\rangle
\eqno{(\rm M.73)}
$$
In the lowest Born approximation we derive
\[
-eE^\alpha \frac{\partial n_{l\pm }}{\partial k_\alpha }=2\pi
\sum_{l^{\prime }}\left\{ |I_{ll^{\prime }}\pm \Lambda _{ll^{\prime
}}^z|^2(f_{l^{\prime }\pm }^{(-2)}-f_{l\pm }^{(-2)})K^{zz}\delta
(\varepsilon _{l\pm }-\varepsilon _{l^{\prime }\pm })+\right.
\]
\[
+|I_{ll^{\prime }}|^2(\Phi _{l\pm ,l^{\prime }\pm }^{(-2)}K^{\pm \mp
}-\Phi _{l^{\prime }\pm l\pm }^{(-2)}K^{\mp \pm })\delta (\varepsilon _{l\pm
}-\varepsilon _{l^{\prime }\pm }\pm W_M)+
\]
\[
\Lambda _{l^{\prime }l}^{+}\Lambda _{l^{\prime }l}^{-}\left[ (\Phi _{l\pm
,l^{\prime }\pm }^{(-2)}K^{+-}-\Phi _{l^{\prime }\pm l\pm
}^{(-2)}K^{-+})\delta (\varepsilon _{l\pm }-\varepsilon _{l^{\prime }\pm
}+W_M)\right.
\]
$$
+\left. \left. (\Phi _{l\pm ,l^{\prime }\pm }^{(-2)}K^{-+}-\Phi
_{l^{\prime }\pm l\pm }^{(-2)}K^{+-})\delta (\varepsilon _{l\pm
}-\varepsilon _{l^{\prime }\pm }-W_M)\right] \right\}
\eqno{(\rm M.74)}
$$
where
$$
\Phi _{\lambda ^{\prime }\lambda }^{(i)} =f_\lambda ^{(i)}(1-n_{\lambda
^{\prime }})-f_{\lambda ^{\prime }}^{(i)}n_\lambda
\eqno{(\rm M.75)}
$$
\[
\varepsilon _{l\pm }=\varepsilon _l\pm I_{ll}\langle S^z\rangle
\]
and we have introduced the on-site averages
$$
K^{zz}=\langle (S^z)^2\rangle -\langle S^z\rangle ^2,\qquad K^{\pm \mp
}=\langle S^{\pm }S^{\mp }\rangle =\langle (S\pm S^z)(S\mp S^z+1)\rangle
\eqno{(\rm M.76)}
$$
neglecting intersite spin correlators in spirit of mean-field
approximation. The equation (M.74) was first derived in [466] by the matrix
method. Omitting spin-orbital terms we obtain in the one-band approximation $%
(l=\mathbf{k})$ for $I_{\mathbf{kk}^{\prime }}=\mathrm{const}$ the result by
Kasuya [422]
$$
-eE^\alpha \frac{\partial n_{k\pm }}{\partial k_\alpha } =2\pi I^2\sum_{%
\mathbf{k}^{\prime }}[(f_{\mathbf{k}^{\prime }\pm }-f_{\mathbf{k}\pm
})K^{zz}\delta (\varepsilon _{\mathbf{k\pm }}-\varepsilon _{\mathbf{k}%
^{\prime }\pm })
\eqno{(\rm M.77)}
$$
\[
+(\Phi _{\mathbf{k\pm ,k}^{\prime }\mp }^{(-2)}K^{\pm \mp }-\Phi _{%
\mathbf{k}^{\prime }\mathbf{\pm ,k\pm }}^{(-2)}K^{\mp \pm })\delta
(\varepsilon _{\mathbf{k\pm }}-\varepsilon _{\mathbf{k}^{\prime }\pm }\pm
W_M)]
\]
The trial solution to the equation (M.77) has the standard form
$$
f_{\mathbf{k\pm }}^{(-2)}=-eE^\alpha \tau _0^{\pm }(\varepsilon _{\mathbf{%
k\pm }})\frac{\partial n_{\mathbf{k\pm }}}{\partial \varepsilon _{\mathbf{%
k\pm }}}v_x
\eqno{(\rm M.78)}
$$
On substituting (M.78) into (M.77) and performing integration over angles we
derive
$$
\tau _0^{\pm }(\varepsilon )=\frac{2\pi }{v_0}\frac{\hbar ^4}{I^2\varepsilon
}(2m^{*})^{-3/2}\varphi ^{\pm }\left( \frac \varepsilon T,\frac{W_M}T\right)
\eqno{(\rm M.79)}
$$
where
$$
\varphi ^{\pm }(\varepsilon ,\eta )=\left[ K^{zz}+K^{\mp \pm }\frac{%
1+e^{-\varepsilon }}{1+e^{-\varepsilon \pm \eta }}\right]
\eqno{(\rm M.80)}
$$
In particular,
$$
\varphi ^{+}(0,\eta )=\varphi ^{-}(0,\eta )\equiv \varphi (\eta
)=[K^{zz}+2K^{-+}(1+e^\eta )]^{-1}
\eqno{(\rm M.81)}
$$
since
\[
K^{+-}=e^{W_M/T}K^{-+}
\]
For $S=1/2$ one has
$$
\lbrack \varphi (\eta )]^{-1}=3\left( \frac 14-\langle S^z\rangle ^2\right)
\eqno{(\rm M.82)}
$$
Now we may introduce the transport relaxation time
$$
\tau _0=-\frac 1{3\pi }\frac \hbar {I^2}\int d\varepsilon \frac{\partial
f(\varepsilon )}{\partial \varepsilon }(\varphi ^{-}+\varphi ^{+})\simeq
\frac 2{3\pi }\frac \hbar {I^2}E_F\varphi \left( \frac{W_M}T\right)
\eqno{(\rm M.83)}
$$
Then the magnetic resistivity reads
$$
\rho _{mag}=\left( \frac{e^2n\tau _0}{m^{*}}\right) ^{-1}=\frac{3\pi }2\frac{%
m^{*}}{e^2n}\frac{I^2}{\hbar E_F}\left[ \varphi \left( \frac{W_M}T\right)
\right] ^{-1}
\eqno{(\rm M.84)}
$$

Now we consider the case of low temperatures $T\ll T_c$ where one can use
the spin-wave approximation. Using the Holstein-Primakoff representation for
spin operators we derive
$$
\mathcal{H}^{\prime } =-(2S)^{1/2}I\sum_{\mathbf{kq}}(c_{\mathbf{%
k+q\uparrow }}^{\dagger }c_{\mathbf{k\downarrow }}b_{\mathbf{q}}+c_{\mathbf{%
k\downarrow }}^{\dagger }c_{\mathbf{k+q\uparrow }}b_{\mathbf{q}}^{\dagger })
\eqno{(\rm M.85)}
$$
\[
-I\sum_{\mathbf{kpq}\sigma }\sigma (I+\sigma \Lambda _{\mathbf{k,k+q-p}%
}^z)c_{\mathbf{k}\sigma }^{\dagger }c_{\mathbf{k+q-p}\sigma }b_{\mathbf{q}%
}^{\dagger }b_{\mathbf{p}}
\]

The lowest-order collision term is calculated similar to case of
electron-phonon interaction. We have
\[
T_{\mathbf{k\pm }}^{(0)}=2\pi I^2S\sum_{\mathbf{q}}N_{\mathbf{q}}\left[
\frac{f(\pm \varepsilon _{\mathbf{k\pm }})\exp (\beta \omega _{\mathbf{q}})}{%
f(\pm \varepsilon _{\mathbf{k\pm }}-\omega _{\mathbf{q}})}f_{\mathbf{k+q\pm }%
}-\frac{f(\pm \varepsilon _{\mathbf{k\pm }}-\omega _{\mathbf{q}})}{f(\pm
\varepsilon _{\mathbf{k\pm }})}f_{\mathbf{k+q\pm }}\right]
\]
$$
\times \delta (\varepsilon _{\mathbf{k+q\mp }}-\varepsilon _{\mathbf{k\pm }%
}\pm \omega _{\mathbf{q}})
\eqno{(\rm M.86)}
$$
The solution to the corresponding transport equation (M.17) which describes
scattering by spin waves has the form
$$
f_{\mathbf{k}\sigma } =\frac 1T\frac{\partial n_{\mathbf{k}\sigma }}{%
\partial \varepsilon _{\mathbf{k}\sigma }}\chi _{\alpha \sigma }(\varepsilon
_{\mathbf{k}\sigma })k_\alpha
\eqno{(\rm M.87)}
$$
\[
\chi _{\alpha \sigma }(\varepsilon )=C_{\alpha \sigma }+\psi _{\alpha \sigma
}(\varepsilon )
\]
The energy dependence of $\chi (\varepsilon )$ is needeed to satisfy the
integral equation. However, it turns out that, as well as in the case of
phonon scattering [1], this dependence results in temperature corrections to
resistivity of higher orders only; however, it is important at considering
the anomalous Hall effect. Substituting (M.87) into the transport equation
yields for the isotropic electron spectrum
$$
-eE^x\frac{\partial n_{\mathbf{k}}}{\partial \varepsilon _{\mathbf{k}}}
=2I^2S\frac{ka_0}{4\pi }\left( \frac{\partial \varepsilon _{\mathbf{k}}}{%
\partial \mathbf{k}}\right) ^{-1}\frac T{T_C}\int_\delta ^\infty
dxN_B(T_x)f(\pm \varepsilon _{\mathbf{k}})f(\pm \varepsilon _{\mathbf{k}%
}-Tx)
\eqno{(\rm M.88)}
$$
\[
\times e^{\pm \beta \varepsilon _{\mathbf{k}}}\{\chi _{x\pm
}(\varepsilon _{\mathbf{k}}\mp Tx)(1-\alpha x)-\chi _{x\pm }(\varepsilon _{%
\mathbf{k}})\}
\]
where we have neglected the small spin splitting,
\[
\delta =\frac T{T_0},\qquad T_0\sim \left( \frac I{E_F}\right)
^2T_c(k_Fa_0)\rightarrow 0,\qquad \alpha =\frac T{2T_c}(ka_0)^{-2}
\]
Integrating (M.88) over $\varepsilon $ we evaluate the constant $C$%
$$
C_x=-eE^x\frac{4\pi ka_0}{I^2S\gamma }\left( \frac{\partial \varepsilon _{%
\mathbf{k}}}{\partial \mathbf{k}}\right) ^2\left( \frac{T_c}T\right) ^2
\eqno{(\rm M.89)}
$$
with
\[
\gamma =\int_{-\infty }^\infty dy\overline{\gamma }(y),\qquad \overline{%
\gamma }(y)=\int_\delta ^\infty dxN_B(Tx)f(Ty)f(T(x+y))e^y
\]
Then the equation for $\psi $ takes the form
\[
\int_0^\infty dxN_B(Tx)f(\varepsilon )f(\varepsilon -Tx)e^{\beta
\varepsilon }[\psi _x(\varepsilon -Tx)-\psi _x(\varepsilon )]
\]
$$
=-\frac{2\pi eE^x}{I^2Ska_0}\left( \frac{\partial \varepsilon _{\mathbf{k}}%
}{\partial \mathbf{k}}\right) ^2\frac{T_c}T\frac{\partial f(\varepsilon )}{%
\partial \varepsilon }
\eqno{(\rm M.90)}
$$
One can see that $\psi _x\sim T_c/T$, so that the contribution of the
function $\psi $ to resistivity is proportional to $(T/T_c)^3$ and may be
neglected. Integrating over $\mathbf{k}$ we derive for the conductivity
$$
\sigma _{xx}=\frac 4{3\pi \gamma }\frac{e^2n}{I^2Sm^{*}}\frac{a_0}{v_F}%
(k_Fa_0)^3\left( \mathbf{k}\frac{\partial \varepsilon _{\mathbf{k}}}{%
\partial \mathbf{k}}\right) _{k_F}\left( \frac{T_c}T\right) ^2
\eqno{(\rm M.91)}
$$
which agrees with the result (5.62).

The next-order approximation transport equation is important for
consideration of the contribution to the anomalous Hall effect in
ferromagnets owing to magnetic scattering. At high temperatures we obtain
[466]
$$
\sum_{l^{\prime }}|I_{ll^{\prime }}|^2K^{zz}(f_{l\pm }^{(-1)}-f_{l^{\prime
}\pm })\delta (\varepsilon _{l\pm }-\varepsilon _{l^{\prime }\pm })
\eqno{(\rm M.92)}
$$
\[
+(K^{\pm \mp }\Phi _{_{l\pm },_{l^{\prime }\pm }}^{(-1)}-K^{\mp \pm
}\Phi _{_{l^{\prime }\pm },l\pm }^{(-1)})\delta (\varepsilon _{l\pm
}-\varepsilon _{l^{\prime }\pm }\pm W_M)
\]
\[
2\pi i\sum_{l^{\prime }}W_{_{l\pm }}^{(1)}\Lambda _{ll^{\prime }}^z(f_{l\pm
}^{(-2)}-f_{l^{\prime }\pm }^{(-2)})\delta (\varepsilon _{l\pm }-\varepsilon
_{l^{\prime }\pm })=0
\]
where $f^{(-2)}$ is the solution to the equation (M.74),
$$
W_{_l\pm }^{(1)} =\sum_{l^{\prime \prime }}|I_{ll^{\prime \prime
}}|^2\left\{ \langle (S^z-\langle S^z\rangle )^3\rangle \delta (\varepsilon
_{l\pm }-\varepsilon _{l^{\prime \prime }\pm })\right.
\eqno{(\rm M.93)}
$$
\[
+\left[ \langle (S^z-\langle S^z\rangle )S^{\pm }S^{\mp }\rangle \pm
(2K^{zz}-K^{\pm \mp })n_{l^{\prime \prime }\mp }\pm K^{\mp \pm }n_{l\pm
}\right.
\]
\[
+\left. \left. 2\langle S^z\rangle n_{l^{\prime \prime }\mp }n_{l\pm
}\right] \delta (\varepsilon _{l\pm }-\varepsilon _{l^{\prime \prime }\pm
}\pm W_M)\right\}
\]
In the one-band approximation, on substituting into (M.92) the solution to
the lowest-order transport equation (M.78) and the matrix elements of SOI
(L.18) we obtain
\[
-eE^x\frac{\partial n_{\mathbf{k\pm }}}{\partial \varepsilon _{\mathbf{%
k\pm }}}v_yW_{\pm }^{(1)}\tau _0^{\pm }(\varepsilon _{\mathbf{k\pm }})\frac{%
a_0^3km^{*}}{4\pi \hbar ^3}\left( \frac{64}{35}\overline{l}\frac{I^{(1)}}{%
\Delta E}\lambda +\frac 12\lambda ^{\prime }\right)
\]
\[
+I^2\sum_{\mathbf{k}}\left[ K^{zz}(f_{\mathbf{k\pm }}^{(-1)}-f_{\mathbf{k}%
^{\prime }\mathbf{\pm }}^{(-1)})\delta (\varepsilon _{\mathbf{k\pm }%
}-\varepsilon _{\mathbf{k}^{\prime }\mathbf{\pm }})\right.
\]
$$
+\left. (K^{\pm \mp }\Phi _{\mathbf{k\pm },\mathbf{k}^{\prime }\mathbf{\pm }%
}^{(-1)}-K^{\mp \pm }\Phi _{\mathbf{k}^{\prime }\mathbf{\pm },\mathbf{k\pm }%
}^{(-1)})\delta (\varepsilon _{\mathbf{k\pm }}-\varepsilon _{\mathbf{k}%
^{\prime }\mathbf{\pm }})\right] =0
\eqno{(\rm M.94)}
$$
The solution to (M.94) is searched as
$$
f_{\mathbf{k}\sigma }^{(-1)}=-eE^x\tau ^\sigma (\varepsilon _{\mathbf{k}%
\sigma })\frac{\partial n_{\mathbf{k\sigma }}}{\partial \varepsilon _{%
\mathbf{k\sigma }}}v_y
\eqno{(\rm M.95)}
$$
Then we derive
$$
\tau ^\sigma (\varepsilon _{\mathbf{k}\sigma })=\frac{m^{*\mathbf{k}}}{%
2\hbar }a_0^3\lambda _{eff}W_\sigma ^{(1)}[\tau _0^\sigma (\varepsilon _{%
\mathbf{k}})]^2
\eqno{(\rm M.96)}
$$
with
$$
\lambda _{eff}=\frac{32}{35}\frac{I^{(1)}}{\Delta E}\overline{l}\lambda
+\lambda ^{\prime }
\eqno{(\rm M.97)}
$$
Now we can calculate the Hall current
\[
j_y=2e\sum_{\mathbf{k}\sigma }f_{\mathbf{k}\sigma }^{(-1)}v_y
\]
The result for the Hall coefficient reads
$$
R_S^{mag} =\frac{9\pi }{32}\left( \frac I{E_F}\right) ^2\frac{m^{*}\lambda
_{eff}}{e^2n\hbar }\frac S{M(0)}\{K^{zz}+(4\langle S^z\rangle )^{-1}[2K^{zz}
\eqno{(\rm M.98)}
$$
\[
-\frac 12(K^{+-}+K^{-+})+\langle S^z\rangle \coth \frac{\beta W_M}2]\frac{%
\sinh \beta W_M-\beta W_M}{\cosh \beta W_M-1}\}
\]
In particular, using the equation for magnetization (4.14) we obtain for $%
S=1/2$%
\[
R_S^{mag}=\frac{9\pi }{64}\left( \frac I{E_F}\right) ^2\frac{m^{*}\lambda
_{eff}}{e^2n\hbar }\frac{1/4-\langle S^z\rangle ^2}{M(0)}
\]
$$
\times \left( 1+\coth \frac{\beta W_M}2\frac{\sinh \beta W_M-\beta W_M}{%
\cosh \beta W_M-1}\right)
\eqno{(\rm M.99)}
$$
At neglecting the weak temperature dependence of the function in the square
brackets the result (M.99) may be represented in the form (5.130).

The transport equation of the next Born approximation with account of linear
corrections in the spin-orbital interaction, which describes electron-magnon
scattering at low temperatures, has the form
$$
T_\lambda ^{(0)}(f^{(-1)})+T_\lambda ^{(1)}(f^{(-2)})=0
\eqno{(\rm M.100)}
$$
where the field term $C^{(1)}$ is zero (as well as for impurity
scattering), $T^{(0)}$ is given by (M.86), the solution to the lowest Born
equation is given by (M.87). To calculate $T$ we write down for (M.85)
\[
\mathcal{H}^{\prime }=\mathcal{H}^{\prime \prime }+\mathcal{H}^{\prime
\prime \prime }
\]
where $\mathcal{H}^{\prime \prime }$ contains the exchange terms and $%
\mathcal{H}^{\prime \prime \prime }$the spin-orbital ones. Using (M.23) and
the properties
$$
\mathcal{H}_{n_1n_2}^{\prime \prime }=\mathcal{H},\qquad
\mathcal{H}_{n_1n_2}^{\prime \prime \prime }
=-\mathcal{H}_{n_2n_1}^{\prime \prime \prime }
\eqno{(\rm M.101)}
$$
calculating the commutators and performing the decouplings, which are
valid to the order under consideration, we obtain
\[
T_{\mathbf{k\pm }}^{(1)}=3\pi I^2S\sum_{\mathbf{pq}}N_{\mathbf{q}}N_{\mathbf{%
p}}\left\{ \Lambda _{\mathbf{k,k+q-p}}^z\delta (\varepsilon _{\mathbf{%
k-p\mp }}-\varepsilon _{\mathbf{k\pm }}\pm \omega _{\mathbf{q}})\delta
(\varepsilon _{\mathbf{k-p\mp }}-\varepsilon _{\mathbf{k\pm }}\mp \omega _{%
\mathbf{q}}\pm \omega _{\mathbf{p}})\right.
\]
\[
\left[ \frac{f(\pm \varepsilon _{\mathbf{k\pm }})f(\pm \varepsilon _{\mathbf{%
k\pm }}+\omega _{\mathbf{q}}-\omega _{\mathbf{p}})}{[f(\pm \varepsilon _{%
\mathbf{k\pm }}-\omega _{\mathbf{p}})]^2}e^{\beta (\omega _{\mathbf{q}%
}+\omega _{\mathbf{p}})}f_{\mathbf{k-p\mp }}-\frac{f(\pm \varepsilon _{%
\mathbf{k\pm }})e^{\beta \omega _{\mathbf{p}}}}{f(\pm \varepsilon _{\mathbf{%
k\pm }}+\omega _{\mathbf{q}}-\omega _{\mathbf{p}})}f_{\mathbf{k+q-p\mp }%
}\right]
\]
$$
\left. +\Lambda _{\mathbf{k-q,k-p}}^z\delta (\varepsilon _{\mathbf{k-p\mp }%
}-\varepsilon _{\mathbf{k\pm }}\pm \omega _{\mathbf{p}})\delta (\varepsilon
_{\mathbf{k-q\mp }}-\varepsilon _{\mathbf{k\pm }}\pm \omega _{\mathbf{p}})%
\frac{f(\pm \varepsilon _{\mathbf{k\pm }}-\omega _{\mathbf{q}})}{f(\pm
\varepsilon _{\mathbf{k\pm }}-\omega _{\mathbf{p}})}e^{\beta \omega _{%
\mathbf{p}}}f_{\mathbf{k-p\mp }}\right\}
\eqno{(\rm M.102)}
$$
An interesting feature of the equation (M.100) should be noted. Substituting
$f^{(2)}\sim C_x = \mathrm{const}$ into (M.102) yields zero after integration
over $\varepsilon $ in the lowest order in $q/k$, and higher orders terms in
$q$ result in higher powers of $T/T_c$. Therefore one has to take into account
the energy dependence of the function (M.87). Then the solution of (M.100)
yields [472]
$$
f_{\mathbf{k}\sigma }^{(-1)}=\frac{3a_0}{16\pi }\left( \frac{\partial
\varepsilon _{\mathbf{k}\sigma }}{\partial \mathbf{k}}\right) ^{-1}\frac
T{T_c}\left[ 4.4\lambda \overline{l}\frac{I^{(1)}}{\Delta E}\frac
T{T_C}-0.24\lambda ^{\prime }-v(\mp \varepsilon )\right] \frac{\partial n_{%
\mathbf{k}\sigma }}{\partial \varepsilon _{\mathbf{k}\sigma }}k_y
\eqno{(\rm M.103)}
$$
with
\[
v(\varepsilon )=\lambda \overline{l}\frac{I^{(1)}}{\Delta E}\int_\delta
^\infty d\omega N_B(\omega )\omega e^{\beta \omega }\frac{f(\varepsilon
+\omega )}{f(\varepsilon )}-\lambda ^{\prime }\ln \left( 1+e^{\beta \omega
}\right)
\]
Calculating $\sigma _{yx}$ with the use of (M.103) and using the expression
for the diagonal conductivity (M.91) we find the expression for the Hall
coefficient
$$
R_S^{mag} =-\frac{3\pi }{512}\frac{2\hbar I^2S}{e^2k_FM(0)}\left( \mathbf{k%
}\frac{d\varepsilon _{\mathbf{k}}}{d\mathbf{k}}\right) _{k_F}^{-3}
\eqno{(\rm M.104)}
$$
\[
\times \left[ 1.1\lambda \overline{l}\frac{I^{(1)}}{\Delta E}\left( \frac
T{T_c}\right) ^4+0.8\lambda ^{\prime }\left( \frac T{T_c}\right) ^3\right]
\]
where the largest $T^3$-term occurs due to the energy dependence of $\psi
(\varepsilon )$.


\chapter{Degenerate Anderson model}

The periodic Anderson model describes the situation where highly correlated
d(f)-electrons do not participate directly in the band motion, but are
hybridized with the conduction band states. Such a situation takes place in
a number of rare-earth and actinide compounds (Chapter 6). The hybridization
(many-configuration) picture is sometimes useful also for discussing some
properties of transition d-metals and other d-electron systems. For example,
strong p-d hybridization takes place in copper-oxide high-$T_c$
superconductors (Sect.6.7). At neglecting spin-orbit coupling, which is
appropriate for transition metals and their compounds, we write down the
Anderson-lattice Hamiltonian in the form
$$
H=H_0+\sum_{\mathbf{k}\sigma }t_{\mathbf{k}}c_{\mathbf{k}\sigma }^{\dagger
}c_{\mathbf{k}\sigma }+\sum_{\mathbf{k}lm\sigma }\left( V_{\mathbf{k}lm}c_{%
\mathbf{k}\sigma }^{\dagger }a_{\mathbf{k}lm\sigma }+\mathrm{h.c.}\right)
\eqno{(\rm N.1)}
$$
where $H_0$ is the Hamiltonian of intrasite interaction between d-electrons.
A symmetry analysis of hybridization mixing in various situations is
performed in the reviews [708,565]. To simplify the model consideration, we
describe the states of conduction electrons by plane waves. Using the
expansion in spherical harmonics (C.28) we obtain for the matrix element of
hybridization
$$
V_{\mathbf{k}lm}=i^lY_{lm}^{*}(\widehat{\mathbf{k}})v_l(k)
\eqno{(\rm N.2)}
$$
where
$$
v_l(k)=4\pi \int r^2drR_l(r)v(r)j_l(kr)
\eqno{(\rm N.3)}
$$
and $v(r)$ is a spherically symmetric potential of a given site. In the
limit of jj-coupling (actinide compounds) one has to replace in (N.1) $%
lm\sigma \rightarrow j\mu $ with $j=l\pm 1/2$ being the total electron
momentum and $m$ its projection.

In the case of strong correlations for d-electrons it is convenient to pass
to the representation of the Hubbard operators which reduces $H_0$ to the
diagonal form (H.2). Retaining two lowest terms $\Gamma _n=\{{SL\}}$, $%
\Gamma _{n-1}=\{{S^{\prime }L^{\prime }\}}$ for the configurations $d^n$ and
$d^{n-1}$ and defining new conduction electron operators
$$
d_{\mathbf{k}lm\sigma }^{\dagger } =\sum_{\mu \mu ^{\prime }MM^{\prime
}}C_{s\mu ^{\prime },\sigma /2}^{s\mu }C_{L^{\prime }M^{\prime },lm}^{LM}X_{%
\mathbf{k}}(SL\mu M,S^{\prime }L^{\prime }\mu ^{\prime }M^{\prime })
\eqno{(\rm N.4)}
$$
\[
c_{\mathbf{k}lm\sigma }^{\dagger }=i^lY_{lm}^{*}(\widehat{\mathbf{k}})c_{%
\mathbf{k}\sigma }^{\dagger }
\]
we present the Hamiltonian (N.1) in the form
$$
H=H_0+\sum_{\mathbf{k}\sigma }\left\{ t_{\mathbf{k}}c_{\mathbf{k}\sigma
}^{\dagger }c_{\mathbf{k}\sigma }+\widetilde{v}_l(k)\left( c_{\mathbf{k}%
lm\sigma }^{\dagger }d_{\mathbf{k}lm\sigma }+\mathrm{h.c.}\right) \right\}
\eqno{(\rm N.5)}
$$
where
$$
H_0 =\Delta \sum_{\mathbf{k}lm\sigma }d_{\mathbf{k}lm\sigma }^{\dagger }d_{%
\mathbf{k}lm\sigma }+\mathrm{const}
\eqno{(\rm N.6)}
$$
\[
\Delta =E_{SL}-E_{S^{\prime }L^{\prime }}-\varsigma
\]
(we have passed to the grand canonical ensemble by introducing the chemical
potential $\varsigma $), effective hybridization parameters are given by
$$
\widetilde{v}_l(k)=n^{1/2}G_{S_{n-1}L_{n-1}}^{S_nL_n}v_l(k)
\eqno{(\rm N.7)}
$$

Now we discuss rare-earth systems. Because of strong Coulomb interaction
between 4f-electrons, formation of the f-bands, containing 14 electron
states, is non-realistic. Thus one has to use the model with two
configurations $f^n$ and $s(d)f^{n-1}$, which corresponds to delocalization
of only one f-electron per atom. In the Russel-Saunders scheme we may
confine ourselves to two lower multiplets of the 4f-ion, $\Gamma _n=SLJ$ and
$\Gamma _{n-1}=S^{\prime }L^{\prime }J^{\prime }$. Passing in (N.1), (A.31)
to the J-representation with the use of (B.5) and summing the product of
Clebsh-Gordan coefficients
$$
\sum_{\mu \mu ^{\prime }MM^{\prime }}C_{S\mu ,LM}^{JM_J}C_{S^{\prime }\mu
^{\prime },L^{\prime }M^{\prime }}^{J^{\prime }M_J^{^{\prime }}}C_{L^{\prime
}M^{\prime },lm}^{LM}C_{S^{\prime }\mu ^{\prime },\sigma /2}^{S\mu }
\eqno{(\rm N.8)}
$$
\[
=\sum_{j\mu }\left\{
\begin{array}{ccc}
S & L & J \\
S^{\prime } & L^{\prime } & J^{\prime } \\
1/2 & l & j
\end{array}
\right\} \left( [j][j^{\prime }][L]\right) ^{1/2}C_{\sigma /2,lm}^{j\mu
}C_{J^{\prime }M_J^{^{\prime }},j\mu }^{JM_J}
\]
we derive the Hamiltonian, which is formally similar to that in the case of
jj-coupling,
$$
H=\sum_{\mathbf{k}j\mu }\left[ \Delta f_{\mathbf{k}j\mu }^{\dagger }f_{kj\mu
}+t_{\mathbf{k}}c_{\mathbf{k}j\mu }^{\dagger }c_{\mathbf{k}j\mu }+\widetilde{%
v}_j(k)\left( c_{\mathbf{k}j\mu }^{\dagger }f_{\mathbf{k}j\mu }+\mathrm{h.c.}%
\right) \right]
\eqno{(\rm N.9)}
$$
where we have introduced new electron operators
$$
f_{\mathbf{k}j\mu }^{\dagger } =\sum_{M_JM_{J^{\prime }}}C_{J^{\prime
}M_J^{^{\prime }},j\mu }^{JM_J}X_{\mathbf{k}}(SLJM_J,S^{\prime }L^{\prime
}J^{\prime }M_J^{\prime }),
\eqno{(\rm N.10)}
$$
\[
c_{\mathbf{k}j\mu }^{\dagger }=i^l\sum_{m\sigma }C_{\sigma
/2,lm}^{j\mu }Y_{lm}^{*}(\widehat{\mathbf{k}})c_{\mathbf{k}\sigma }^{\dagger
}
\]
and the effective hybridization parameters are expressed in terms of a
9j-symbol
$$
\widetilde{v}_j(k)=\left\{
\begin{array}{ccc}
S & L & J \\
S^{\prime } & L^{\prime } & J^{\prime } \\
1/2 & l & j
\end{array}
\right\} \left( [j][j^{\prime }][L]\right) ^{1/2}G_{S^{\prime }L^{\prime
}}^{SL}v_l(k)
\eqno{(\rm N.11)}
$$
Thus hybridization effects in ME systems depend strongly on ME quantum
numbers $S,L,J$ and, consequently, on the atomic number [709]. Such a
dependence in the rare-earth element series is similar to de Gennes
correlation for the $s-f$ exchange parameter and paramagnetic Curie
temperature. Its experimental investigation, e.g., in spectroscopic data, is
of great interest.

In the case where $|\Delta |$ is large in comparison with the width of the $%
d(f)$ level we may exclude the hybridization term from the Hamiltonians
(N.5), (N.9) by a canonical transformation to obtain, respectively,
$$
H =-\frac 1\Delta \sum v_l(k)v_l(k^{\prime })C_{L^{\prime }M^{\prime
},lm}^{LM}C_{S^{\prime }\mu ^{\prime },\sigma /2}^{S\mu }C_{L^{\prime
}M^{\prime \prime \prime },lm^{\prime }}^{LM^{\prime \prime }}C_{S^{\prime
}\mu ^{\prime \prime \prime },\sigma ^{\prime }/2}^{S\mu ^{\prime \prime }}
\eqno{(\rm N.12a)}
$$
\[
\left[ X_{\mathbf{k-k}^{\prime }}\left( SL\mu ^{\prime \prime }M^{\prime
\prime },SL\mu M\right) \delta _{\mu ^{\prime }\mu ^{\prime \prime \prime
}}\delta _{M^{\prime }M}\right.
\]
\[
\left. +X_{\mathbf{k-k}^{\prime }}\left( S^{\prime }L^{\prime }\mu ^{\prime
}M^{\prime },S^{\prime }L^{\prime }\mu ^{\prime \prime \prime }M^{\prime
\prime \prime }\right) \delta _{\mu \mu ^{\prime \prime }}\delta
_{MM^{\prime \prime }}\right] c_{\mathbf{k}lm\sigma }^{\dagger }c_{\mathbf{k}%
^{\prime }lm^{\prime }\sigma ^{\prime }}
\]
$$
H =-\frac 1\Delta \sum v_j(k)v_{j^{\prime }}(k^{\prime })C_{J^{\prime
}M^{\prime },j\mu }^{JM}C_{J^{\prime }M^{\prime \prime \prime },j^{\prime
}\mu ^{\prime }}^{JM^{\prime \prime }}
\eqno{(\rm N.12b)}
$$
\[
\times \left[ X_{\mathbf{k-k}^{\prime }}\left( JM^{\prime \prime },JM\right)
\delta _{M^{\prime }M^{\prime \prime }}+X_{\mathbf{k-k}^{\prime }}\left(
J^{\prime }M^{\prime },J^{\prime }M^{\prime \prime \prime }\right) \delta
_{MM^{\prime \prime }}\right] c_{\mathbf{k}j\mu }^{\dagger }c_{\mathbf{k}%
^{\prime }j^{\prime }\mu ^{\prime }}
\]
For $\Delta <0$ ($\Delta >0$) the filling of the level is $n$ ($n-1$) and
only the first (second) term in the brackets of (N.12) should be retained.

The expressions (N.12) describe the exchange interaction of conduction
electrons with $d(f)$-electrons. It should be noted that in the case under
consideration the interaction is higly anisotropic because of spherical
harmonics which enter (N.4), (N.10). This should result in a strong
anisotropy of indirect RKKY-type $f-f$ interaction which is obtained in the
second order in the s-f exchange. Such an anisotropy is observed in a number
of rare-earth and actinide compounds. Using the identity
$$
\sum_{M^{\prime }}C_{L^{\prime }M^{\prime },lm}^{LM}C_{L^{\prime }M^{\prime
},lm^{\prime }}^{LM^{\prime \prime }}=\sum_{pq}(-1)^{p-q}[p]\left( \frac{[L]%
}{[l]}\right) ^{1/2}\left\{
\begin{array}{ccc}
l & l & p \\
L & L^{\prime } & L^{\prime }
\end{array}
\right\} C_{LM,pq}^{LM^{\prime \prime }}C_{lm^{\prime },p-q}^{lm}
\eqno{(\rm N.13)}
$$
the Hamiltonians (N.12) may be decomposed into sum of terms which correspond
to interaction of conduction electrons with different multipole components
of orbital and spin (or total momentum) degrees of freedom.

As well as in the usual $s-f$ exchange model [552], perturbation theory
expansion in the models (N.12) yields logarithmic corrections to various
physical quantities which indicates reconstruction of the system state at
low temperatures. In particular, such a correction to electron self-energy
and resistivity occurs in the third order in $I$. Unfortunately, the
complicated tensor structure of the Hamiltonians (N.12) prevents the
calculation of the unique energy scale for the infrared divergences (the
Kondo temperature). However, such a calculation may be performed in the case
where the energy of the d(f) level $\Delta $ does not depend on the
many-electron term and is determined by the number of electrons only (see
Sect.6.2).

Consider the anticommutator retarded Green's function of localized
d-electrons (H.3) in the non-magnetic phase of the model (N.1). The simplest
decoupling yields
$$
G_{\mathbf{k}lm}(E)=\left[ \Phi (E)-\frac{|V_{\mathbf{k}lm}|^2}{E-t_{\mathbf{%
k}}}\right] ^{-1}
\eqno{(\rm N.14)}
$$
where the function $\Phi $ is defined by (H.5). The corresponding energy
spectrum contains a system of subbands, separated by hybridization gaps (or
pseudogaps, provided that $V(\mathbf{k})$ vanishes for some $\mathbf{k}$)
which are surrounded by density-of-states peaks. In the model with strong
correlations (N.7), we have
$$
E_{\mathbf{k}}^{1,2}=\frac 12(t_{\mathbf{k}}+\Delta )\pm \left[ \frac 14(t_{%
\mathbf{k}}-\Delta )^2+|\widetilde{V}_{\mathbf{k}lm}|^2\right]
\eqno{(\rm N.15)}
$$
with
$$
\widetilde{V}_{\mathbf{k}lm}=i^lY_{lm}^{*}(\widehat{\mathbf{k}})\widetilde{v}%
_l(k)\left\{ \frac{[S][L]}{2[l]}\left( N_{SL}+N_{S^{\prime }L^{\prime
}}\right) \right\} ^{1/2}
\eqno{(\rm N.16)}
$$
One can see that the width of hybridization gap depends appreciably on the
many-electron occupation numbers (in particular, on the position of
d-level). The approximation (N.14) does not take into account spin-flip
processes, which result in the Kondo effect and can change substantially the
structure of electron spectrum near the Fermi level. To take into account
the Kondo anomalies, we perform a more accurate calculation of the Green's
function. For brevity we consider the model (N.9); in the model (N.5)
\[
\lbrack J]\rightarrow [S][L],\qquad
\widetilde{v}_j\rightarrow \widetilde{v}_l
\]

It is convenient to use the operators, which are averaged over angles of the
vector $\mathbf{k}$:
$$
f_{kjm}^{\dagger }=\int d\widehat{\mathbf{k}}f_{\mathbf{k}jm}^{\dagger },
\qquad
c_{kjm}^{\dagger }=\int d\widehat{k}c_{\mathbf{k}jm}^{\dagger }
\eqno{(\rm N.17)}
$$
Following to the consideration of the $SU(N)$ Anderson model in Sect.6.4, we
write down the equation of motion
$$
(E-\Delta )\langle \langle f_{\mathbf{k}jm}|f_{kjm}^{\dagger }\rangle
\rangle _E =R_j\left( 1+\widetilde{v}_j(k)\langle \langle c_{\mathbf{k}%
jm}|f_{kjm}^{\dagger }\rangle \rangle _E\right)
\eqno{(\rm N.18)}
$$
\[
+\sum_{j^{\prime }m^{\prime }\mu M\mathbf{q}}\widetilde{v}_{j^{\prime
}}(q)C_{J^{\prime }M^{\prime },,jm}^{J\mu }\langle \langle \left[
\sum_{M^{\prime }}C_{J^{\prime }M^{\prime },jm^{\prime }}^{J\mu }X_{\mathbf{%
k-q}}\left( J^{\prime }M,J^{\prime }M^{\prime }\right) \right.
\]
\[
+\left. \sum_{\mu ^{\prime }}C_{J^{\prime }M,j^{\prime }m^{\prime }}^{J\mu
^{\prime }}X_{\mathbf{k-q}}(J\mu ^{\prime },J\mu )\right] c_{\mathbf{q}%
j^{\prime }m^{\prime }}|f_{kjm}^{\dagger }\rangle \rangle
\]
where we have carried out a decoupling for the term, which describes the
processes without changing $m$,
$$
R_j=\frac 1{[j]}\left\{ \frac{[J]}{[J^{\prime }]}-\left( \frac{[J]}{%
[J^{\prime }]}-1\right) \sum_M\langle X\left( JM,JM\right) \rangle \right\}
\eqno{(\rm N.19)}
$$
Further we neglect for simplicity the above-discussed influence of
hybridization gap, which is possible provided that the latter lies far below
the Fermi level (note that the corresponding contributions are formally
small in the inverse degeneracy of f-level, $1/N$). Carrying out decouplings
in the equations for the Green's function in the right-hand side of (N.18)
we obtain
$$
(E-t_{\mathbf{q}})\langle \langle X_{\mathbf{k-q}}\left( J\mu ^{\prime
},J\mu \right) c_{\mathbf{q}j^{\prime }m^{\prime }}|f_{kjm}^{\dagger
}\rangle \rangle _E
\eqno{(\rm N.20)}
$$
\[
=\widetilde{v}_{j^{\prime }}(q)n_q\sum_{M^{\prime }}C_{J^{\prime
}M^{\prime \prime },,j^{\prime }m^{\prime }}^{J\mu ^{\prime }}\langle
\langle X_{\mathbf{k-q}}\left( J^{\prime }M,J\mu ^{\prime \prime }\right)
|f_{kjm}^{\dagger }\rangle \rangle _E,
\]
\[
(E-t_q)\langle \langle X_{k-q}\left( J^{\prime }M,J^{\prime }M^{\prime
}\right) c_{\mathbf{q}j^{\prime }m^{\prime }}|f_{kjm}^{\dagger }\rangle
\rangle _E
\]
\[
=-\widetilde{v}_{j^{\prime }}(q)n_q\sum_{\mu ^{\prime \prime }}C_{J^{\prime
}M^{\prime },,j^{\prime }m^{\prime }}^{J\mu ^{\prime \prime }}\langle
\langle X_{\mathbf{k-q}}\left( J^{\prime }M,J\mu ^{\prime \prime }\right)
|f_{kjm}^{\dagger }\rangle \rangle _E
\]
with $n_k=\langle c_{kjm}^{\dagger }c_{kjm}\rangle =f(t_k)$ being the Fermi
distribution functions. Substituting (N.20) into (N.19), averaging over
angles and using the orthogonality relations for the Clebsh-Gordan
coefficients, we find
$$
\langle \langle f_{kjm}|f_{kjm}^{\dagger }\rangle \rangle _E=R_J\left[
E-\Delta -\Sigma _f(E)\right] ^{-1}
\eqno{(\rm N.21)}
$$
$$
\Sigma _f(E)=2\rho \sum_j\widetilde{v}_j^2(k_F)\frac{J-J^{\prime }}{%
[J^{\prime }]}\ln \left| \frac WE\right|
\eqno{(\rm N.22)}
$$
where we have used the approximation (6.5). At $J>J^{\prime }$ the Green's
function (N.21) has the pole
$$
|\Delta ^{*}|=T_K\approx W\exp \left\{ -\left( \frac{[J]}{[J^{\prime }]}%
-1\right) ^{-1}|\Delta |\left[ \rho \sum_j\widetilde{v}_j^2(k_F)\right]
^{-1}\right\}
\eqno{(\rm N.23)}
$$
The usual Kondo effect corresponds to the total compensation of magnetic
moment ($J^{\prime }=0$). At $J^{\prime }>J$ the pole (N.23) is absent (the
strong coupling regime does not occur) since the model under consideration
is mapped by a canonical transformation into a Coqblin-Schrieffer model with
a positive exchange parameter. An analogue of the result (N.23) for
d-impurities has the form
$$
T_K=W\exp \left\{ -\left( \frac{[S][L]}{[S^{\prime }][L^{\prime }]}-1\right)
^{-1}\frac{|\Delta |}{\rho \widetilde{v}_j^2(k_F)}\right\}
\eqno{(\rm N.24)}
$$
Note that the formula (N.24) satifies the condition of the particle-hole
symmetry ($n\rightarrow n^{\prime }=2[l]+1-n,\Delta \rightarrow -\Delta $)
owing to the relation for the fractional parentage coefficients
$$
\left\{ G_{S^{\prime }L^{\prime }}^{SL}(n^{\prime },n^{\prime }-1)\right\}
^2=\frac{n[S^{\prime }][L^{\prime }]}{n^{\prime }[S][L]}\left\{
G_{SL}^{S^{\prime }L^{\prime }}(n,n-1)\right\} ^2
\eqno{(\rm N.25)}
$$
At neglecting the dependence $\Delta (LS),$ all the ME terms of
configurations $d^n$ and $d^{n-1}$ contribute equally to spin-flip
processes, and, consequently, to the Kondo temperature. Then the
coefficients $G$ which enter (N.7) may be summed up in the equations of
motion with the use of orthogonality relations (A.8), (A.9) to obtain for
the Kondo temperature
$$
T_K\approx W\exp \left[ -\left( 2[l]+1-2n\right) ^{-1}\left( |I|\rho \right)
^{-1}\right] ,\qquad I=\frac{\widetilde{v}_l^2(k_F)}\Delta
\eqno{(\rm N.26)}
$$
Note that the expression (N.26) may be represented in the the form, which is
similar to (N.24),
$$
T_K\approx W\exp \left[ -\left( \frac{2[l]+1-2n}n-1\right) ^{-1}\frac
1{n|I|\rho }\right]
\eqno{(\rm N.27)}
$$
with the factor $(2[l]+1-n)/n$ being again the ratio of statistical weights
for the configurations $d^n$ and $d^{n-1}$:
\[
\frac{2[l]+1-n}n=\frac{(2[l])!}{n!(2[l]-n)!}/\frac{(2[l])!}{(n-1)!(2[l]-n+1)!%
}
\]

The result (N.26) differs from the result of high-temperature perturbation
theory (6.22) by an unity in the denominator of the exponential only. Such a
difference is typical for the calculation of the Kondo temperature in the
degenerate Anderson model [565,574] and is explained by that this approach
is justified, strictly speaking, only in the large-$N$ limit.

\chapter{Mean-field approximation for the ground state of magnetic Kondo
lattices}

To construct the mean-field approximation describing the ground state of
Kondo lattices we use the Abrikosov pseudofermion representation for
localized spins $S=1/2$ (which coincides with the Dirac representation
(B.1))
$$
\mathbf{S}_i=\frac 12\sum_{\sigma \sigma ^{\prime }}f_{i\sigma }^{\dagger }%
\mathbf{\sigma }_{\sigma \sigma ^{\prime }}f_{i\sigma ^{\prime }}
\eqno{(\rm O.1)}
$$
with the subsidiary condition
\[
f_{i\uparrow }^{\dagger }f_{i\uparrow }+f_{i\downarrow }^{\dagger
}f_{i\downarrow }=1
\]

Making the saddle-point approximation for the path integral describing the
spin-fermion interacting system [711] one can reduce the Hamiltonian of the $%
s-f$ exchange interaction to the effective hybridization model:
$$
-I\sum_{\sigma \sigma ^{\prime }}c_{i\sigma }^{\dagger }c_{i\sigma }\left(
\mathbf{\sigma }_{\sigma \sigma ^{\prime }}\mathbf{S}_i-\frac 12\delta
_{\sigma \sigma ^{\prime }}\right) \rightarrow f_i^{\dagger
}V_ic_i+c_i^{\dagger }V_i^{\dagger }f_i-\frac 1{2I}\mathrm{Sp}%
(V_iV_i^{\dagger })
\eqno{(\rm O.2)}
$$
where the vector notations are used
\[
f_i^{\dagger }=(f_{i\uparrow }^{\dagger },f_{i\downarrow }^{\dagger
}),\qquad c_i^{\dagger }=(c_{i\uparrow }^{\dagger },c_{i\downarrow
}^{\dagger })
\]
$V$ is the effective hybridization matrix which is determined from a minimum
of the free energy. Coleman and Andrei [711] considered the formation of a
spin-liquid state in the two-dimensional situation. Here we treat the more
simple case of ferromagnetic ordering following to [608]. The Heisenberg
Hamiltonian of the $f$-subsystem $H_f$ is considered in the mean-field
approximation. For a ferromagnet we have ($\overline{S}=\langle S^z\rangle $)
$$
H_f=-J_0\overline{S}\sum_{i\sigma }\sigma f_{i\sigma }^{\dagger }f_{i\sigma
},V_{\sigma \sigma ^{\prime }}=V_\sigma \delta _{\sigma \sigma ^{\prime }}
\eqno{(\rm O.3)}
$$
and the Hamiltonian of the $s-f$ model takes the form
$$
H-\zeta n =\sum_{\mathbf{k}\sigma }[(t_{\mathbf{k}}-\zeta )c_{\mathbf{k}%
\sigma }^{\dagger }c_{\mathbf{k}\sigma }+w_\sigma f_{\mathbf{k}\sigma
}^{\dagger }f_{\mathbf{k}\sigma }
\eqno{(\rm O.4)}
$$
\[
+V_\sigma (c_{\mathbf{k}\sigma }^{\dagger }f_{\mathbf{k}\sigma }+f_{%
\mathbf{k}\sigma }^{\dagger }c_{\mathbf{k}\sigma })+\mathrm{const}
\]
where
\[
w_\sigma =w-\sigma J_0\overline{S},
\]
$w$ being the energy of ``$f$-level''. The equations for $w$, the chemical
potential $\zeta $ and $\overline{S}$ read
$$
n_\sigma ^{\dagger }\equiv \sum_{\mathbf{k}}\langle f_{\mathbf{k}\sigma
}^{\dagger }f_{\mathbf{k}\sigma }\rangle =\frac 12+\sigma \overline{S}
\eqno{(\rm O.5)}
$$
$$
n=\sum_{\mathbf{k}}\langle c_{\mathbf{k}\sigma }^{\dagger }c_{\mathbf{k}%
\sigma }\rangle
\eqno{(\rm O.6)}
$$
One can see that $-w$ plays the role of the chemical potential for
pseudofermions (note that the numbers of electrons and pseudofermions are
conserved separately). After the minimization one obtains the equation for $%
V_\sigma $%
$$
V_\sigma =2I\sum_{\mathbf{k}}\langle f_{\mathbf{k}\sigma }^{\dagger }c_{%
\mathbf{k}\sigma }\rangle
\eqno{(\rm O.7)}
$$
Diagonalizing the Hamiltonian (O.4) by a canonical transformation
$$
c_{\mathbf{k}\sigma }^{\dagger } =u_{\mathbf{k}\sigma }\alpha _{\mathbf{k}%
\sigma }^{\dagger }-v_{\mathbf{k}\sigma }\beta _{\mathbf{k}\sigma }^{\dagger
},\qquad f_{\mathbf{k}\sigma }^{\dagger }=u_{\mathbf{k}\sigma }\beta _{%
\mathbf{k}\sigma }^{\dagger }-v_{\mathbf{k}\sigma }\alpha _{\mathbf{k}\sigma
}^{\dagger }
\eqno{(\rm O.8)}
$$
\[
u_{\mathbf{k}\sigma }=\cos (\theta _{\mathbf{k}\sigma }/2),\qquad v_{\mathbf{%
k}\sigma }=\sin (\theta _{\mathbf{k}\sigma }/2)
\]
with
$$
\sin \theta _{\mathbf{k}\sigma }=\frac{2V_\sigma }{E_{\mathbf{k}\sigma }}%
,\qquad \cos \theta _{\mathbf{k}\sigma }=\frac{t_{\mathbf{k}}-\zeta
-w_\sigma }{E_{\mathbf{k}\sigma }}
\eqno{(\rm O.9)}
$$
we obtain the energy spectrum of a hybridization form
$$
E_{\mathbf{k}\sigma }^{\alpha ,\beta }=\frac 12\left( t_{\mathbf{k}}-\zeta
+w_\sigma \pm E_{\mathbf{k}\sigma }\right)
\eqno{(\rm O.10)}
$$
Then the equations (O.5)-(O.7) take the form
$$
n_\sigma ^f=\frac 12\sum_{\mathbf{k}}\left[ \left( 1-\cos \Theta _{\mathbf{k}%
\sigma }\right) n_{\mathbf{k}\sigma }^\alpha +\left( 1+\cos \Theta _{\mathbf{%
k}\sigma }\right) n_{\mathbf{k}\sigma }^\beta \right]
\eqno{(\rm O.11)}
$$
$$
n=\frac 12\sum_{\mathbf{k}\sigma }\left[ \left( 1+\cos \Theta _{\mathbf{k}%
\sigma }\right) n_{\mathbf{k}\sigma }^\alpha +\left( 1-\cos \Theta _{\mathbf{%
k}\sigma }\right) n_{\mathbf{k}\sigma }^\beta \right]
\eqno{(\rm O.12)}
$$
$$
1=-2I\sum_{\mathbf{k}}\left. \left( n_{\mathbf{k}\sigma }^\beta -n_{\mathbf{k%
}\sigma }^\alpha \right) \right/ E_{\mathbf{k}\sigma }
\eqno{(\rm O.13)}
$$
At small $|V_\sigma |,|w_\sigma |$ and $T=0$ we have
$$
\cos \Theta _{\mathbf{k}\sigma }\approx \mathrm{sign}\left( t_{\mathbf{k}%
}-\zeta -\omega _\sigma \right)
\eqno{(\rm O.14)}
$$
so that the equations (O.11), (O.12) are simplified. The edges of the
hybridization gaps in spin subbands are given by
$$
E_{\mathbf{k}\sigma }^\alpha >w_\sigma +V_\sigma ^2/\zeta ,\qquad E_{\mathbf{%
k}\sigma }^\beta <w_\sigma -V_\sigma ^2/(W-\zeta )
\eqno{(\rm O.15)}
$$
Further we consider different types of ferromagnetic solutions. We confine
ourselves to the case where the conduction electron concentration $n<1$ (the
results for $n>1$ are obtained after the particle-hole transformation). As
follows from (O.15), at not too large $\bar S$ for both $\sigma $ the
condition
$$
w_\sigma >V_\sigma ^2/(W-\zeta )
\eqno{(\rm O.16)}
$$
holds, i.e the chemical potential lies below the energy gap, as well as in
the non-magnetic case. Define the function $\zeta (c)$ by
$$
c=2\int\limits_0^{\zeta (c)}dE\rho (E)
\eqno{(\rm O.17)}
$$
with $\rho (E)$ ($0<E<W$) being the bare density of states. Then the
equation (O.12) takes the form $\zeta (n)=\zeta $, and (O.11) and (O.13)
yield
$$
\lambda _\sigma \equiv V_\sigma ^2/\omega _\sigma =\zeta (n+2n_\sigma
^f)-\zeta (n)
\eqno{(\rm O.18)}
$$
$$
1=-2I\int\limits_0^{\zeta +\lambda _\sigma }\frac{dE\rho (E)}{\left[
(E-\zeta -w_\sigma )^2+4V_\sigma ^2\right] ^{1/2}}
\eqno{(\rm O.19)}
$$
Calculating the integral in (O.19) to leading and next-order terms in $1/\ln
|W/V_\sigma |$ we obtain
$$
1=-4I\rho \ln \left| \frac C{V_\sigma }\right| -2I\left[ \int\limits_C^\zeta
dE\frac{\rho (E+\zeta )}E+\int\limits_{C+\zeta }^{\zeta (n+2n_\sigma ^f)}dE%
\frac{\rho (E)}{E-\zeta }\right]
\eqno{(\rm O.20)}
$$
where $C$ is a cutoff parameter which does not enter final results. In the
leading approximation $V_\sigma $ does not depend on $\sigma $ and we have
$$
|V_\sigma |\sim (WT_K)^{1/2},\qquad T_K\equiv W\exp \frac 1{2I\rho }
\eqno{(\rm O.21)}
$$
Retaining next order terms we calculate the ratio and obtain the
self-consistent equation for magnetization
$$
\tanh \left( \frac 1{4\rho }\int\limits_{\zeta (n+1-2\bar S)}^{\zeta
(n+1+2\bar S)}dE\frac{\rho (E)-\rho }{E-\zeta }\right) =\frac{J_0\bar S}w
\eqno{(\rm O.22)}
$$
where
$$
\omega \approx V^2/[\zeta (n+1)-\zeta ]\sim T_K
\eqno{(\rm O.23)}
$$
The equation (O.22) has no non-trivial solutions for the bare DOS $\rho (E)=%
\mathrm{const}$. However, solutions with $\bar S\neq 0$ may occur for some $%
\rho (E)$ if both the left and right-hand side of (O.22) are of order unity,
i.e. $J_0\sim T_K$ (see Fig.O.1).

Provided that
$$
w_{\downarrow }>V_{\downarrow }^2/(W-\zeta ),\qquad w_{\uparrow
}<-V_{\uparrow }^2/\zeta
\eqno{(\rm O.24)}
$$
i.e. $\zeta $ lies in the energy gap for $\sigma =\uparrow $, we obtain the
``half-metallic'' ferromagnetic solution with
$$
n_{\uparrow }^f=1-n/2,\qquad n_{\downarrow }^f=n/2,\qquad \bar S=(1-n)/2
\eqno{(\rm O.25)}
$$
(see Fig.O.2). Such a solution exists under the condition
$$
-\varphi /\zeta <\left[ \zeta (2n)-\zeta \right]
^{-1}-J_0(1-n)/V_{\downarrow }^2<\varphi /(W-\zeta )
\eqno{(\rm O.26)}
$$
$$
\varphi \equiv \left| \frac{V_{\uparrow }}{V_{\downarrow }}\right| ^2=\exp
\left( \frac 1\rho \int\limits_{\zeta (2n)}^W\frac{dE\rho (E)}{E-\zeta }%
\right)
\eqno{(\rm O.27)}
$$
For $\rho (E)=\mathrm{const}$ (O.26) takes the form
$$
J_0(1-n)<T_K
\eqno{(\rm O.28)}
$$
and the corresponding total energy
$$
\mathcal{E}=\frac{n^2}{4\rho }-\frac n2T_K-J_0\bar S^2=\mathcal{E}_{\mathrm{%
non-mag}}-J_0\bar S^2
\eqno{(\rm O.29)}
$$
is always lower than that of non-magnetic Kondo state. Thus the formation of
the state of half-metallic Kondo ferromagnet is energetically favourable. In
this state each conduction electron compensates one localized spin, and the
magnetic ordering is due to exchange interaction between non-compensated
moments. Such a picture is reminiscent of the stuation in the narrow-band
ferromagnet in the Hubbard or $s-d$ exchange model with large intrasite
interaction (Sect.4.6). In our case the bare interaction is small, but
effective interaction is large in the strong coupling regime.

The expression (O.29) should be compared with the energy of the usual
ferromagnetic state with the Kondo effect suppressed:
$$
\mathcal{E}=\frac{n^2}{4\rho }-J_0/4\qquad (V_\sigma =0,\bar S=1/2)
\eqno{(\rm O.30)}
$$
We see that the latter state becomes energetically favourable at
$$
J_0(1-\frac n2)>T_K
\eqno{(\rm O.31)}
$$
At the critical point, a first-order transition takes place.

Third type of possible ferromagnetic solutions correspond to the situation
of large energy splitting where $\zeta $ lies in the lower hybridization
subband for $\sigma =\downarrow $ and in the upper one for $\sigma
=\downarrow $ (Fig.O.3). However, such solutions (at least for $\rho (E)=%
\mathrm{const}$) are energetically unfavourable [608].

\chapter{Schwinger and Dyson-Maleev representations in the theory of
two-dimensional Heisenberg antiferromagnets}

The Schwinger boson representaton for spin operators has the form, which is
similar to the Abrikosov representation (O.1), the Fermi operators being
replaced by Bose operators $b_{i\sigma }$:

\[
\mathbf{S}_i=\frac 12\sum_{\sigma \sigma ^{\prime }}b_{i\sigma }^{\dagger }%
\mathbf{\sigma }_{\sigma \sigma ^{\prime }}b_{i\sigma ^{\prime }}
\]
$$
S_i^{+}=b_{i\uparrow }^{\dagger }b_{i\downarrow },\qquad
S_i^{-}=b_{i\downarrow }^{\dagger }b_{i\uparrow },\qquad S_i^z=\frac
12\left( b_{i\uparrow }^{\dagger }b_{i\uparrow }-b_{i\downarrow }^{\dagger
}b_{i\downarrow }\right)
\eqno{(\rm P.1)}
$$

For the given localized spin $S$, these operators should satisfy at each
site $i$ the subsidiary condition (constraint)
$$
\sum_\sigma b_{i\sigma }^{\dagger }b_{i\sigma }=2S
\eqno{(\rm P.2)}
$$
It should be noted that the ''hyperbolic'' operators (B.14) which change the
value of momentum may be easily represented in terms of the Schwinger bosons
[656]:
$$
K^{+}=b_{\uparrow }^{\dagger }b_{\downarrow }^{\dagger },\qquad
K^{-}=b_{\downarrow }b_{\uparrow },\qquad K^z=\frac 12\left( b_{\uparrow
}^{\dagger }b_{\uparrow }+b_{\downarrow }^{\dagger }b_{i\downarrow }+1\right)
\eqno{(\rm P.3)}
$$
The representation (P.1) turns out to be convenient at considering
low-dimensional systems which do not possess long-range ordering at finite
temperatures, but demonstrate developed spin fluctuations (short-range
order).

Consider the ``mean field'' approximation for the two-dimensional Heisenberg
model within the self-consistent spin-wave theory [623]. (This differs
drastically from the usual mean field approximation of Sect 4.1 since
permits to describe the strong short-range order above the ordering point.)
The constraint (P.2) is taken into account in the average by introducing the
Lagrange multiplier $\lambda $ which does not depend on $i$. Anomalous
averages $\langle b_{i\uparrow }b_{j\downarrow }\rangle $ describe singlet
pairing of bosons, i.e. the short-range order parameters. Occurence of
long-range ordering with the wavevector $\mathbf{Q}$ corresponds to the
Bose-Einstein condensation with the quasimomenta $\mathbf{k}=\pm \mathbf{Q}%
/2 $. It is convenient to introduce the interaction with small external
magnetic field $H$. Then the Bogoliubov transformation 
\[
b_{\mathbf{Q}/2+\mathbf{k}\uparrow }=\cosh \frac{\Theta _{\mathbf{k}}}%
2\alpha _{\mathbf{k}}-\sinh \frac{\Theta _{\mathbf{k}}}2\beta _{-\mathbf{k}%
}^{\dagger } 
\]
$$
b_{\mathbf{Q}/2-\mathbf{k}\downarrow }=\cosh \frac{\Theta _{\mathbf{k}}}%
2\beta _{-\mathbf{k}}-\sinh \frac{\Theta _{\mathbf{k}}}2\alpha _{\mathbf{k}%
}^{\dagger }  
\eqno{(\rm P.4)}
$$
reduces the Heisenberg Hamiltonian to the diagonal form 
$$
H_d=\sum_{\mathbf{k}}\left( E_{\mathbf{k}}^\alpha \alpha _{\mathbf{k}%
}^{\dagger }\alpha _{\mathbf{k}}+E_{\mathbf{k}}^\beta \beta _{\mathbf{k}%
}^{\dagger }\beta _{\mathbf{k}}\right)  
\eqno{(\rm P.5)}
$$
where, for the square lattice with the parameter $a_0=1$, 
$$
E_{\mathbf{k}}^{\alpha ,\beta }=E_{\mathbf{k}}\mp \left( \frac
12H-2|J|\langle S^z\rangle \right) ,\qquad E_{\mathbf{k}}=(\lambda ^2-\gamma
_{\mathbf{k}}^2)^{1/2}  
\eqno{(\rm P.6)}
$$
$$
\sinh \Theta _{\mathbf{k}}=\gamma _{\mathbf{k}}/E_{\mathbf{k}},\qquad \cosh
\Theta _{\mathbf{k}}=\lambda /E_{\mathbf{k}}  
\eqno{(\rm P.7)}
$$
with 
\[
\gamma _{\mathbf{k}}=\frac 12\gamma (\sin k_x+\sin k_y) 
\]
Equations for $\lambda $ and $\gamma $, which are obtained from (P.7), have
the form 
$$
2S+1=\frac 1N\sum_{\mathbf{k}}\frac \lambda {E_{\mathbf{k}}}(1+N_{\mathbf{k}%
\alpha }+N_{\mathbf{k}\beta })  
\eqno{(\rm P.8)}
$$
$$
1=\frac 1{2N}\sum_{\mathbf{k}}\frac{|J|}{E_{\mathbf{k}}}(\sin k_x+\sin
k_y)^2(1+N_{\mathbf{k}\alpha }+N_{\mathbf{k}\beta })  
\eqno{(\rm P.9)}
$$
where $N_{\mathbf{k}\alpha ,\beta }=N_B(E_{\mathbf{k}}^{\alpha ,\beta })$, $%
N $ is the number of lattice sites which should be retained explicitly in
the problem under consideration.

At $T=0$ we have $\lambda =\gamma $ and $N_{\mathbf{k}\alpha }$ (but not $N_{%
\mathbf{k}\beta }$) contains the condensate term at $E_{\mathbf{k}}^\alpha
=0 $, i.e. 
$$
\mathbf{k}=\pm \mathbf{Q}/2,\qquad E_{\mathbf{k}}=\frac 12H-2J\langle
S^z\rangle \sim H  
\eqno{(\rm P.10)}
$$
Then we obtain 
$$
N_{\pm \mathbf{Q}/2}=N\langle S^z\rangle =NE_{\mathbf{Q}/2}n_B/\lambda 
\eqno{(\rm P.11)}
$$
with $2n_B$ being the density of condensed bosons. Equation (P.8) yields 
$$
n_B=S+\frac 12-\frac 1{2N}\sum_{\mathbf{k}}\left[ 1-\frac 14(\sin k_x+\sin
k_y)^2\right] ^{-1/2}=S-0.197  
\eqno{(\rm P.12)}
$$
so that $n_B$ equals the ground state sublattice magnetization of the Neel
antiferromagnet $\bar S(0)$ with account of zero-point spin-wave corrections.

At finite $T$ the boson spectrum (P.6) contains the gap and the condensate
is absent. Then we may put $H=0$, $N_{\mathbf{k}\alpha ,\beta }=N_{\mathbf{k}%
}$ from the beginning.

Consider the spin spectral density 
$$
K_{\mathbf{q}}(\omega )=-\frac 1\pi N_B(\omega )\mathrm{Im}\langle \langle
S_{\mathbf{q}}|S_{-\mathbf{q}}\rangle \rangle _\omega  
\eqno{(\rm P.13)}
$$
in the Schwinger boson representation. The spin Green's function are
expressed in terms of polarization operators of non-interacting bosons $%
\alpha $ and $\beta $, and we derive 
\[
K_{\mathbf{q}}(\omega )=\frac 14\sum_{\mathbf{k}}\sum_{\nu ,\mu =\alpha
,\beta }\left\{ (2-\delta _{\mu \nu })\cosh ^2\frac{\Theta _{\mathbf{k}%
}-\Theta _{\mathbf{k+q}}}2N_{\mathbf{k}\mu }(1+N_{\mathbf{k+q}\nu })\right. 
\]
\[
\times \delta (\omega +E_{\mathbf{k+q}}^\nu -E_{\mathbf{k}}^\mu )+(1+\delta
_{\mu \nu })\sinh ^2\frac{\Theta _{\mathbf{k}}-\Theta _{\mathbf{k+q}}}2[N_{%
\mathbf{k}\mu }N_{\mathbf{k+q}\nu } 
\]
$$
\times \left. \delta (\omega -E_{\mathbf{k+q}}^\nu -E_{\mathbf{k}}^\mu
)+(1+N_{\mathbf{k}\mu })(1+N_{\mathbf{k+q}\nu })\delta (\omega +E_{\mathbf{%
k+q}}^\nu +E_{\mathbf{k}}^\mu )]\right\}  
\eqno{(\rm P.14)}
$$
As follows from (P.11), (P.14), the spectral density contains at $T=0$, $%
H\rightarrow 0$ the delta-function contribution 
$$
\delta K_{\mathbf{q}}(\omega )=\frac 32n_B^2N\delta _{\mathbf{qQ}}\delta
(\omega )  
\eqno{(\rm P.15)}
$$
The factor of $3/2$ in (P.15) should in fact be omitted since it is an
artifact of the mean field approximation which yields, because of inaccurate
account of the condition (P.2), 
\[
\langle S_i^2\rangle =\frac 32S(S+1) 
\]
violating thereby the corresponding sum rule at the site.

At finite temperatures we have 
$$
E_{\mathbf{k}}^2=\frac 12\lambda ^2\left[ (\mathbf{k}\mp \mathbf{Q}/2)^2+\xi
^{-2}\right] ,\qquad \mathbf{k}\rightarrow \pm \mathbf{Q}/2  
\eqno{(\rm P.16)}
$$
where the correlation length is given by 
$$
\xi \sim \exp (\pi \lambda n_B/2T)  
\eqno{(\rm P.17)}
$$
At $\mathbf{q}=\mathbf{Q}$ the integral in (P.14) is ``almost'' divergent at
the points $\mathbf{k}=\pm \mathbf{Q}/2$ with the cutoff scale $|\mathbf{k}%
\mp \mathbf{Q}/2|\sim \xi ^{-1}$. Using the expansion $N_{\mathbf{k}}\cong
T/E_{\mathbf{k}}$ we present the corresponding singular contribution in the
form 
$$
\delta K_{\mathbf{q}}(\omega )=\frac 32\left[ \frac{2T}{\pi \lambda }\ln \xi
\right] ^2\Delta _{\mathbf{q}}\Delta _\omega \approx \frac 32n_B^2\Delta _{%
\mathbf{q}}\Delta _\omega  
\eqno{(\rm P.18)}
$$
where $\Delta _{\mathbf{q}}$ and $\Delta _\omega $ are $\delta (\mathbf{q}-%
\mathbf{Q})$ and $\delta (\omega )$ like functions smoothed on the scales $%
\xi ^{-1}$ and $\omega _\xi \sim J/\xi $ respectively. At $T\ll J$ we have $%
\omega _\xi \ll T$ and we may neglect the smoothing which yields a formal
description of strong short-range order. To obtain the temperature
dependence of the coefficient in (P.18) (``sublattice magnetization'') we
estimate the intensity of the Ornstein-Cernike peak at $\mathbf{q}=\mathbf{Q}
$ in the static spin correlator 
$$
S_{\mathrm{eff}}(T)=\left( \sum_{|\mathbf{q}-\mathbf{Q}|<q_0}\langle \mathbf{%
S}_{-\mathbf{q}}\mathbf{S}_{\mathbf{q}}\rangle \right) ^{1/2}  
\eqno{(\rm P.19)}
$$
where $q_0\gg \xi ^{-1}$ is a cutoff wavevector. Using the result of the
scaling consideration [712], which yields the correct preexponential factor 
$$
\xi =C_\xi \exp \left( \pi c\bar S(0)/\sqrt{2}T\right)  
\eqno{(\rm P.20)}
$$
where $c$ is the magnon velocity and $C_\xi \cong 0.01/(2\pi )\ll 1$ is a
numerical factor. Substituting (P.20) into (P.19), (P.18) and neglecting $%
\ln q$ in comparison with $\ln C_\xi $ we obtain the linear dependence [713] 
$$
S_{\mathrm{eff}}(T)=\bar S(0)-\frac{\sqrt{2}T}{\pi c}|\ln C_\xi |  
\eqno{(\rm P.21)}
$$

The contributions to spectral density from antiferromagnetic spin waves are
determined by the terms, which are linear in $n_B$ . Putting $E_{\mathbf{k}%
}\rightarrow 0$, $N_{\mathbf{k}}=T/E_{\mathbf{k}}$ , but retaining $E_{%
\mathbf{k+q}}$ and vice versa, and performing integration over $\mathbf{k}$,
we derive 
\[
\delta _1K_{\mathbf{q}}(\omega )=\frac 32\frac T{E_{\mathbf{q+Q}/2}}\frac
1N\sum_{\mathbf{k}}\frac{\lambda ^2-\gamma _{\mathbf{k}}\gamma _{\mathbf{k+q}%
}}{E_{\mathbf{k}}^2}[(1+N_{\mathbf{q+Q}/2})\delta (\omega +E_{\mathbf{q+Q}%
/2})+N_{\mathbf{q+Q}/2}\delta (\omega -E_{\mathbf{q+Q}/2})] 
\]
$$
\approx \frac 32n_B\left( \frac{1-\varphi _{\mathbf{q}}}{1+\varphi _{\mathbf{%
q}}}\right) ^{1/2}\{[1+N_B(\omega _{\mathbf{q}})]\delta (\omega +\omega _{%
\mathbf{q}})+N_B(\omega _{\mathbf{q}})\delta (\omega -\omega _{\mathbf{q}})\}
\eqno{(\rm P.22)}
$$
where, for $q$, $|\mathbf{q}-\mathbf{Q|}\gg \xi ^{-1}$, 
$$
\omega _{\mathbf{q}}=\lambda (1-\varphi _{\mathbf{q}}^2)^{1/2}\approx E_{%
\mathbf{q+Q}/2},\qquad \varphi _{\mathbf{q}}\equiv \frac 12(\cos q_x+\cos
q_y)  
\eqno{(\rm P.23)}
$$
so that $\omega _{\mathbf{q}}$ is the renormalized magnon frequency.

For comparison, we consider the application of the Dyson-Maleev
representation (E.2) to the same problem. In the case of a two-sublattice
antiferromagnet this has the form 
\[
S_l^{-}=(2S)^{1/2}a_l^{\dagger },\qquad S_l^{+}=(2S)^{1/2}(1-\frac
1{2S}a_l^{\dagger }a_l)a_l 
\]
\[
S_l^z=S-a_l^{\dagger }a_l,\qquad l\in A 
\]
$$
S_m^{-} =(2S)^{1/2}b_m^{\dagger },\qquad S_m^{+}=(2S)^{1/2}b_m^{\dagger
}(1-\frac 1{2S}b_m^{\dagger }b_m)b_m  
\eqno{(\rm P.24)} 
$$
\[
S_m^z=-S+b_m^{\dagger }b_m,\qquad m\in B 
\]
In the self-consistent approach [624] one puts on each site the condition $%
\langle S^z\rangle =0$, i.e. 
$$
\langle a_l^{\dagger }a_l\rangle =\langle b_l^{\dagger }b_l\rangle =S 
\eqno{(\rm P.25)}
$$
Using the Bogoliubov transformation 
\[
a_{\mathbf{k}}=\cosh \frac{\Theta _{\mathbf{k}}}2\alpha _{\mathbf{k}}-\sinh 
\frac{\Theta _{\mathbf{k}}}2\beta _{-\mathbf{k}}^{\dagger } 
\]
$$
b_{-\mathbf{k}}^{\dagger }=\cosh \frac{\Theta _{\mathbf{k}}}2\beta _{-%
\mathbf{k}}^{\dagger }-\sinh \frac{\Theta _{\mathbf{k}}}2\alpha _{\mathbf{k}}
\eqno{(\rm P.26)}
$$
we diagonalize the Hamiltonian to obtain 
$$
H=\sum_{\mathbf{k}}\omega _{\mathbf{k}}(\alpha _{\mathbf{k}}^{\dagger
}\alpha _{\mathbf{k}}+\beta _{\mathbf{k}}^{\dagger }\beta _{\mathbf{k}}) 
\eqno{(\rm P.27)}
$$
with 
\[
\langle \alpha _{\mathbf{k}}^{\dagger }\alpha _{\mathbf{k}}\rangle =\langle
\beta _{\mathbf{k}}^{\dagger }\beta _{\mathbf{k}}\rangle =N_{\mathbf{k}%
}=N_B(\omega _{\mathbf{k}}) 
\]
$$
\omega _{\mathbf{k}}=(\lambda ^2-\gamma ^2\varphi _{\mathbf{k}%
}^2)^{1/2},\qquad \tanh 2\Theta _{\mathbf{k}}=\gamma \varphi _{\mathbf{k}%
}/\lambda  
\eqno{(\rm P.28)}
$$
Here $\Sigma ^{\prime }$ stands for the sum over reduced Brillouin zone. The
equations for $\lambda $ and $\gamma $ read 
$$
S+\frac 12=\sum_{\mathbf{k}}{}^{\prime }\frac \lambda {\omega _{\mathbf{k}%
}}(1+2N_{\mathbf{k}})  
\eqno{(\rm P.29)}
$$
$$
1=\sum_{\mathbf{k}}{}^{\prime }\frac{|J|}{\omega _{\mathbf{k}}}(\cos
k_x+\cos k_y)(1+2N_{\mathbf{k}})  
\eqno{(\rm P.30)}
$$
At $T=0$ (in more general case, in the presence of the long-range ordering), 
$\lambda =\gamma +O(1/N)$, so that the spectrum is gapless. The $1/N$%
-corrections describe the Bose condensation: 
$$
\left( 1-\gamma ^2/\lambda ^2\right) ^2=Nn_B  
\eqno{(\rm P.31)}
$$
As follows from the constraint (P.25) and the structure of the Dyson-Maleev
transformation, the transvese spin correlation function $\langle S_{\mathbf{q%
}}^{+}S_{-\mathbf{q}}^{-}\rangle $ vanishes. Thus the corresponding spectral
density reads 
\[
K_{\mathbf{q}}(\omega )=K_{\mathbf{q}}^{zz}(\omega )=\frac 1N\sum_{\mathbf{k}%
}{}^{\prime }\left\{ \cosh \frac{\Theta _{\mathbf{k}}-\Theta _{\mathbf{k+q}}}%
2N_{\mathbf{k}}(1+N_{\mathbf{k+q}})\right. 
\]
\[
\times \delta (\omega +\omega _{\mathbf{k+q}}-\omega _{\mathbf{k}})+\frac
12\sinh ^2\frac{\Theta _{\mathbf{k}}-\Theta _{\mathbf{k+q}}}2\{N_{\mathbf{k}%
}N_{\mathbf{k+q}}\delta (\omega -\omega _{\mathbf{k+q}}-\omega _{\mathbf{k}%
}) 
\]
$$
+\left. (1+N_{\mathbf{k}})(1+N_{\mathbf{k+q}})\delta (\omega +\omega _{%
\mathbf{k+q}}+\omega _{\mathbf{k}})]\right\}  
\eqno{(\rm P.32)}
$$
The delta-function contribution is 
$$
\delta K_{\mathbf{q}}(\omega )=Nn_B^2\delta _{\mathbf{qQ}}\delta (\omega ) 
\eqno{(\rm P.33)}
$$
Unlike (P.15), the expession (P.33) does not contain the superfluous factor $%
3/2$. On the other hand, in contrast with the ``isotropic'' Schwinger
representation, the Dyson-Maleev representation violates the rotational
invariance even in the paramagnetic phase.

The versions of self-consistent spin-wave theory (SSWT), based on non-linear
boson representations, yield a gap in the spin-wave spectrum in the absence
of long-range order. This contradicts, e.g., to exact results for
one-dimensional lattice with half-integer spin $S$ where the spectrum is
gapless (for integer spin, the spectrum contains so called Haldane gap [714]
which is reproduced qualitatively by SSWT). However, as discussed in [622],
such approaches are satisfactory in the cases where the ground state
possesses long-range order.

It should be noted that both approaches under consideration describe the
long-range order in terms of delta-function singularity of the spin
correlation function, the sublattice magnetization being always zero. The
possibility of such a description in the three-dimensional situation for a
wide range of physical properties, including local characteristics, is
demonstrated in [715]. In particular, in such an antiferromagnetic state
without sublattices the Green's function of a nuclear spin at each site
turns out to have both poles $\pm AS$ ($A$ is the hyperfine interaction
parameter).

Applications of the non-linear boson representations to systems with a weak
interlayer coupling and magnetic anisotropy are discussed in [713].




\chapter*{FIGURE CAPTIONS}
\addcontentsline{toc}{chapter}{FIGURE CAPTIONS}

Fig.1.1. The effective potential for a hydrogen-like atom.

Fig.1.2. The effective potential $V_{eff}$ and radial wavefunctions $%
P_{4f}(r)$ for BaI (broken lines) and LaI (solid lines) [18].

Fig.1.3. (a) energy levels (eV) of external atomic electrons in the Gd atom
[19] (b) radial distribution of the Hartree-Fock charge density for 4f-,
5s-, 5p- and 6s-electrons of the Gd$^{3+}$ ion [16].

Fig.1.4. Experimental $L_{2,3} M _{45}  M_{45} $ Auger spectra from copper
and zinc. Atomic multiplet structures are shown as lines under spectra [36].

Fig.1.5. Effect of multiplets on the $L_3VV$ Auger spectra of nickel. Lower
panel: XPS VB after subtraction of satellite $N(E)$ and its self-fold $D^0(E)
$. Middle and top panels: experimental spectrum and calculated contributions
from individual d$^8$ terms with different values of Slater integrals [37].

Fig.1.6. Interpolation scheme of levels for Ni$^{2+}$(d$^8$) ion in the
crystal field. The numbers in brackets stand for degeneracy of a state.

Fig.1.7. Fragment of level scheme for Fe$^{2+}$(d$^6$) ion. Position of
levels in the ground state corresponds to unquenched orbital momenta for
both strong and intermediate CF. Unlike Fig.1.6, interpolation is not shown,
since not all corresponding levels are included.

Fig.1.8. Resonance photoemission experiment on CuO (a) spectra with the
photon energies sweeping to the point of the 2p reesonance (b) the resonance
intensity (c) a number of spectra on extendend scale [47].

Fig.1.9. XPS-valence band spectrum of MnO and analysis of the spectra by
cluster calculations of two groups [48,49].

Fig.1.10. Comparison of experimental XPS valence band spectra of NiO with
the results of cluster calculations [50,51].

Fig.2.1. Total and partial densities of states for titanium [78].

Fig.2.2. Total and partial densities of states for zirconium [78].

Fig.2.3. Total and partial densities of states for hafnium [78].

Fig.2.4. Total and partial densities of states for vanadium [78].

Fig.2.5. Total and partial densities of states for chromium [78].

Fig.2.6. Total and partial densities of states for rhodium [78].

Fig.2.7. Total and partial densities of states for platinum [78].

Fig.2.8. The unhybridized canonical bands for the fcc lattice [56].

Fig.2.9. The unhybridized canonical bands for the bcc lattice [56].

Fig.2.10. The unhybridized canonical bands for the ideal hcp lattice [56].

Fig.2.11. Densities of states (per spin) for the unhybridized canonical
bands shown in Figs.2.8-2.10.

Fig.2.12. The density of states for s-band in the bcc lattice in the
nearest-neighbour approximation [79].

Fig.2.13. The density of states for s-band in the fcc lattice in the
nearest-neighbour approximation [79].

Fig.2.14. The density of states for s-band in the fcc lattice with the ratio
of next-nearest and nearest-neighbour transfer integrals $\gamma = t_2/t_1 =
0.8$ and 3 [80].

Fig.2.15. s-, p- and d-projected densities of states for fcc palladium and
bcc niobium [56].

Fig.2.16. Density of states of molybdenum for the lowest six bands [82].

Fig.2.17. Energy bands of vanadium, calculated for two possivle choices of
the crystal potential [83].

Fig.2.18. X-ray emission spectra of vanadium [57] (a) $K$-band (b) $L_{III}$%
-band (c) $M_{III}$-band; 1 the experimental data, 2 and 3 results of
calculations with different perturbation operators.

Fig.2.19. X-ray emission spectra of Y and Zr; solid line shows experimental
data, and broken line the results of calculations [57].

Fig.2.20. Experimental X-ray emission bands of 5d-metals [57].

Fig.2.21. $N_{VI,VII}$ emission spectra of Ir (a) and Pt (b). Lower curves
are experimental data [57].

Fig.2.22. Comparison of BIS and density of states curves for 3d- metals and
Cu. The dashed curves correspond to the unbroadened DOS, the solid lines to
the DOS broadened with Gaussian and Lorentz broadening to simulate
instrumental and lifetime broadening respectively. The upper dotted curve is
the measured BIS spectrum [102].

Fig.2.23. The same data for 4d metals and Ag [102].

Fig.2.24. Calculated spin up and down densities of states of ferromagnetic
iron [24].

Fig.2.25. Calculated spin up and down densities of states of ferromagnetic
nickel [24].

Fig.2.26. Spin and angle-resolved distribution curves form Fe (100) at 60eV
photon enegy for two different temperatures ($T/T_C = 0.3$ and 0.85) [105].
The arrows refer to the spontaneous magnetization direction.

Fig.2.27. Angle-resolved energy distribution curve showing
temperature-dependent exchange splitting $\delta E_{ex} = 1.09 \Delta_{ex}$
in Ni [108]. According to the experimental conditions, only the uppermost
d-band is observed.

Fig.2.28. Differences of correlation energies $\xi $ for transitions 4f$%
^{n+1}$5d6s$\rightarrow $4f$^n$5d6s in free rare earth atoms and for
transitions 3d $^{n+1}\rightarrow $ 3d$^n$ between ground state multiplets
of iron group atoms [133].

Fig.2.29. The Fermi surface of Pd (a) electron surface at the point $\Gamma$
(b) multiconnected hole tubes [11].

Fig.2.30. (a) Theoretical model of the Fermi surface of gadolinium by
Freeman, Dimmock and Watson (b) The Fermi surface of Tb in the one-band
scheme: hole surface in the third zone and the electron surface in the
fourth zone [11].

Fig.3.1. Melting points of the IV, V and VI period elements of the periodic
table [235].

Fig.3.2. Heat of fusion of the IV, V and VI period elements of the periodic
table [235]. Open points are estimated values.

Fig.3.3. Boiling points of the IV, V and VI period elements of the periodic
table [235].

Fig.3.4. (a) Boiling points of rare earth metals (b) Cohesive energy of rare
earth metals. Open points are estimated values [235].

Fig.3.5. Linear coefficient of thermal expansion of 3d-metals.

Fig.3.6. (a) Linear coefficient of thermal expansion of rare earth metals
(b) Atomic volume of rare earth metals. Open points are estimated values
[235].

Fig.3.7. Debye temperature at $T = 0$ as determined from specific heat data
for the fourth, fifth and sixth periods of the periodic table [235]. Open
points are estimated values.

Fig.3.8. (a) The measured cohesive energies of the d-metal series. (b) The
corresponding calculated valence bond energies [236].

Fig.3.9. Components of the cohesive energy for the 3d and 4d transition
metals. The experimental value is denoted by the open box, and the
calculated one by the filled box [242].

Fig.3.10. The calculated coefficients of thermal expansion for non- magnetic
cubic metals together with experimental points [243].

Fig.3.11. Crystal structures of elemental metals at low temperatures, $n_d$
being the number of d-electrons in d-series [245].

Fig.3.12. Structural energy differences $\mu S^2\Delta $ ($\Delta
=E_{bcc(hcp)}-E_{fcc}$, $\mu $ is the d-band mass, $S$ is the atomic
Wigner-Seitz radius), obtained from canonical d-bands, as a function of the
canonical d occupation [245].

Fig.3.13. Structural energy differences obtained from canonical d-bands as
functions of the canonical d occupation in the $n_d$-range, corresponding to
the lanthanide crystal-structure sequence [245].

Fig.3.14. Diagram of the most stable close-packed crystal structure as a
function of the d-occupation number. Two estimates of the actual
d-occupation numbers for the d-metals together with the experimental crystal
structures are given below [245].

Fig.3.15. Structural energy differences for 3d, 4d and 5d rows calculated at
the experimentally observed equilibrium volume and plotted as functions of
the d-occupation numbers [245].

Fig.3.16. Calculated bcc-fcc and hcp-fcc structural energy differences
(solid and broken lines) for the 4d-metals [245] compared with the enthalpy
differences derived from phase-diagram studies (open circles).

Fig.3.17. Phonon dispersion curves in Nb and Mo [257].

Fig.3.18. Experimental specific heat of vanadium. The dash-dotted line
correspond to the Dulong-Petit value ($R^{\prime}= R/M$, $M$ is the atomic
weight) [239].

Fig.3.19. Experimental specific heat of zirconium [239].

Fig.3.20. Experimental specific heat of gadolinium. The inset shows the
behaviour near the Curie point. The dependence near the melting point $T_m$
is typical for all the heavy rare-earth metals [239].

Fig.3.21. Temperature dependence of the electron-phonon enhancement
coefficient ($c_e(T) = \gamma_0[1 + \lambda(T)]T$) in the Debye model [262].

Fig.3.22. Experimental temperature dependence $\gamma(T)/\gamma_0$ [262].

Fig.3.23. Temperature dependence $\gamma(T)/\gamma_0$, calculated from the
empirical density of states. The latter is obtained in the rigid-band model
from the experimental $\gamma_0$ for alloys [262].

Fig.3.24. Temperature dependence $\gamma(T)/\gamma_0$, obtained from the
calculated density of states [262].

Fig.3.25. Experimental specific heat of iron [239].

Fig.3.26. Experimental specific heat of nickel according to the data of
various authors [239].

Fig.4.1. Paramagnetic susceptibility at 293K in 3d, 4d and 5d rows as a
function of valence electron number, $z = n_s + n_d$ [269].

Fig.4.2. Temperature dependence of magnetic susceptibility for transition
metals of the third group [270]: (a) scandium (b) yttrium (c) lutetium.

Fig.4.3. Temperature dependence of magnetic susceptibility for transition
metals of the fourth group according to data of various aithors [270]: (a)
titanium (b) hafnium (c) zirconium.

Fig.4.4. Temperature dependence of magnetic susceptibility for transition
metals of the fifth group [270]: (a) vanadium (b) niobium (c) tantalum.

Fig.4.5. Temperature dependence of magnetic susceptibility for transition
metals of the sixth group [270]: (a) molybdenum (b) tungsten.

Fig.4.6. Temperature dependence of magnetic susceptibility for paramagnetic
transition metals of the eighth group [270]: (a) ruthenium and osmium (b)
rhodium and iridium (c) palladium and platinum.

Fig.4.7. The dependence of the ratio $p_C /p_s$ on the Curie temperature
(the Rhodes-Wolfarth curve) [26].

Fig.4.8. Density of states (eV$^{-1}$, right-hand axis) and integrated
density of states (dotted line, left-hand axis) of paramagnetic iron [24].
Energy is referred to the Fermi level.

Fig.4.9. Comparison between equation (4.86) and experimental Curie
temperatures for iron group metals and intermetallic systems. The asymptotes
corresponds to pure fluctuation behaviour and the tangent to pure Stoner
behaviour.

Fig.4.10. Density of states (Ry$^{-1})$ in PtMnSb [307] (a) spin up (b) spin
down.

Fig.4.11. Density of states for the Heusler alloy Co$_2$MnSn [311].

Fig.4.12. Partial densities of states (1/eV atom, energy in eV) for CrO$_2$
[313]. The solid line corresponds to 3d states of Cr, dotted and broken
lines to 2s and 2p states of oxygen (a) spin up (b) spin down.

Fig.4.13. Partial densities of states (1/Ry atom) for Mn$_4$N [324].

Fig.4.14. Density of states in the s-d model in the case of empty conduction
band ($I > 0$). At $T = 0$ (solid line) the spin-polaron tail of spin-down
states reaches the band bottom. The broken line corresponds to finite
temperatures [329].

Fig.4.15. Density of states in a half-metallic ferromagnet with $I>0$.
Non-quasiparticle states with $\sigma =\downarrow $ are absent below the
Fermi level [329].

Fig.4.16. Density of states in a half-metallic ferromagnet with $I<0$.
Non-quasiparticle states with $\sigma =\uparrow $ occur below the Fermi
level [329].

Fig.4.17. Magnetic structures of heavy rare-earth metals and the
corresponding temperature intervals.

Fig.4.18. The Hill diagram for uranium compounds which demonstrates
dependence of type and and temperature of magnetic ordering on the distance
between uranium ions; 1 - antiferromagnetic ordering, 2 - ferromagnetic
ordering, 3 - temperature-independent paramagnetism [371].

Fig.4.19. Positions of vectors representing the orbital $\widetilde{l}$ and
spin $s$ angular momenta in 3d-ferromagnets (weak spin-orbit coupling): (a)
in the absence of an external field (b) in the presence of the field H
directed along the hard axis.

Fig.4.20. Anisotropic distribution of the f-electron density expected for
the f$^1$ and f$^6$ configurations with $L = 3$. The solid curve in the
right part represents an orbital hole which appears when f$^1$ is replaced
by f$^6$. In the presence of magnetic anisotropy, the hole and its vector $%
L^{\prime}$ are rotated by 90$^o$ (dashed curve) relative to the electron
[39].

Fig.5.1. The temperature dependence of electrical resistivity of scandium
according to data of various authors [239].

Fig.5.2. The temperature dependence of electrical resistivity of vanadium
[239].

Fig.5.3. The electrical resistivity of chromium according to data of various
authors (1,3,4); the calculated anomaly near the Neel point (2,5) [239].

Fig.5.4. The electrical resistivity of rhenium in the direction of the
hexagonal axis and perpendicular to it, and for a polycrystal (3) [239].

Fig.5.5. (a) The electrical resistivity of ruthenium in the direction of the
hexagonal axis and perpendicular to it, and for polycrystalline sample; (b)
the anisotropy of resistivity (1) and diffusivity (2), and the ratio $c/a$
(3) [239].

Fig.5.6. The electrical resistivity of iridium (1); 2) the calculation based
on the Mott s-d model [239].

Fig.5.7. The electrical resistivity of platinum (1); 2) the temperature
coefficient of resistivity $\alpha$ [239].

Fig.5.8. The electrical resistivity of nickel (1); 2) the anomaly near the
melting point; 3) the temperature coefficient of resistivity as a function
of $t = (T - T_C)/T_C$ near the Curie point [239].

Fig.5.9. Coefficient at the $T^2$-term $A$ vs $\gamma^2$ for transition
metal elements [410].

Fig.5.10. Anomaly of the resistivity $\rho (T)$ ($10^{-6}\Omega $ cm) at the
Neel point ($T_N$ = 96K) in $\alpha $-Mn [265].

Fig.5.11. Temperature dependences of resistivity in the Heusler alloys TMnSb
(T = Cu (A), Au (B), Co (D), Ni (E), Pt (F)) and PtMnSn (C) [331].

Fig.5.12. The concentration dependence of resistivity ($\mu\Omega$ cm) in
Pd-Au alloys.

Fig.5.13. Residual resistivity of NiCo$_{1-x}$Rh$_x$ alloy. The large
deviation from the ``Matthiessen rule'' (linear dependence) is accounted for
strongly different values of $\alpha _{Co}$ and $\alpha _{Rh}$%
; the solid curve corresponds to their values of 13 and 0.3 respectively
[436].

Fig.5.14. The typical temperature dependences of thermoelectric power ($\mu$%
V/K) in metals (a) noble metals and refractory d-metals in a wide
temperature region (b) noble metals at intermediate temperatures (c)
alcaline metals at low temperatures [8].

Fig.5.15. A summary of thermoelectric power data for transition metals, U
and Th [438]. Roman numerals indicate the periodic table column.

Fig.5.16. The anomaly of thermoelectric power of chromium at the Neel point
[438].

Fig.5.17. The Hall coefficient as a function of the number of valence
electrons for polycrystalline transition metals and alloys according to data
of different authors [443].

Fig.5.18. The temperature dependence of the Hall coefficient in scandium
according to data of different authors [239].

Fig.5.19. The temperature dependence of the Hall coefficient in lanthanum
[239].

Fig.5.20. The temperature dependence of the Hall coefficient in titanium
[239].

Fig.5.21. The temperature dependence of the Hall coefficient in zirconium
[239].

Fig.5.22. The temperature dependence of the Hall coefficient in hafnium
[239].

Fig.5.23. The temperature dependence of the Hall coefficient in molybdenum
[239].

Fig.5.24. The temperature dependence of the Hall coefficient in rhenium
[239].

Fig.5.25. The temperature dependence of the Hall coefficient in ruthenium
according to data of different authors [239].

Fig.5.26. The temperature dependence of the Hall coefficient in rhodium
[239].

Fig.5.27. The temperature dependence of the Hall coefficient in platinum
[239].

Fig.5.28. The temperature dependences of the Hall coefficient in the field $%
H = 13.5$ kOe for Ta (open squares) and V (different measurement series,
other symbols) [446].

Fig.5.29. Temperature dependence of the Hall coefficient of polycrystalline $%
\alpha$-Mn and $\alpha$-Pu [443].

Fig.5.30. The temperature dependence of the Hall coefficient of
polycrystalline Cu [443].

Fig.5.31. Reduced Kohler diagram showing the transverse magnetoresistance of
some metals in the high-field region [448].

Fig.5.32. The schematical behaviour of Hall resistivity as a function of the
field $H$ in ferromagnetic metals. The dependence $\rho_H(B)$, which
demonstrates expicitly the separation of the normal and spontaneous Hall
effect, has the same form.

Fig.5.33. The temperature dependences of the spontaneous Hall coefficient in
iron group metals [451].

Fig.5.34. The extraordinary Hall resistivities of gadolinium, terbium and
dysprosium as a function of $T/T_N$. The dashed curve is the theoretical
plot obtained from the results by Maranzana. Above $T_N$, the values of $R_s$
for heavy rare earths are shown which are temperature independent [15].

Fig.5.35. The extraordinary Hall coefficient of dysprosium with the magnetic
field along the $a$-axis as a function of spin-disorder resistivity [15].

Fig.5.36. The field dependence of $\Delta\rho/\rho$ in ferromagnetic nickel
for parallel and perpendicular orientations of current and magnetic field
[265].

Fig.6.1. Coefficient at the $T^2$-term $A$ vs $\gamma^2$ for anomalous rare
earth and actinide compounds, and some d-systems [410].

Fig.6.2. Experimental data on the Kondo teperature for d-impurities in
copper and gold [559,560].

Fig.6.3. Anomaly of thermoelectric power in CeIn at the Neel point [578].

Fig.6.4. The temperature dependences of thermoelectric power of CeCu$_2$Si$_2
$ and CeAl$_3$; inset shows the dependence for CeAl$_3$ between 0.15 and
350K on a logarithmic scale [579].

Fig.6.5. The energy spectrum and density of states in the effective
hybridization model of an intermediate valent system.

Fig.6.6. Temperature dependences of resisitivity for the system Ce$_x$La$%
_{1-x}$Ge$_2$ (which is ferromagnetic at $0.4 < x < 1$), the logarithmic
scale being used. Arrows show the value of $T_C$ [601].

Fig.6.7. Temperature dependences of magnetic entropy for the system Ce$_x$La$%
_{1-x}$Ge$_2$. Arrows show the value of $T_C$ [601].

Fig.6.8. Concentration dependences of saturation moment $M_0$ (solid line), $%
T_C$ (broken line) and $T_K$ (schematically, dash-dotted line) in the system
CeNi$_x$Pd$_{1-x}$ [601].

Fig.6.9. The density of states picture in the mean-field approximation for
the s-f exchange model (the non-magnetic phase) [608].

Fig.H.1. Evolution of one-electron density of states (the metal-insulator
transition) with increasing the interaction parameter $J = 2IS$ in the
classical s-d model in the Hubbard-III approximation for semielliptic (a)
and rectangular (b) bare DOS, and simple cubic lattice (c) [695].

Fig.O.1. The density of states picture for the ferromagnetic solution with a
small spin splitting.

Fig.O.2. The density of states picture for the saturated (half-metallic)
ferromagnetic solution.

Fig.O.3. The density of states picture for the ferromagnetic solution where
the spin splitting exceeds the energy gap.

\chapter*{TABLES}
\addcontentsline{toc}{chapter}{TABLES}

Table 1.1. Atomic configurations and ground terms for the free atoms and M$%
^{2+}$ ions, and third ionization potentials (eV) in d-series.

\noindent
\begin{tabular}{|l|ccccccccc|}
\hline
3d & Sc & Ti & V & Cr & Mn & Fe & Co & Ni & Cu \\
4d & Y & Zr & Nb & Mo & Tc & Ru & Rh & Pd & Ag \\
5d & La & Nb & Ta & W & Re & Os & Ir & Pt & Au \\ \hline
3d & d$^1$s$^2$ & d$^2$s$^2$ & d$^3$s$^2$ & d$^5$s & d$^5$s$^2$ & d$^6$s$^2$
& d$^7$s$^2$ & d$^8$s$^2$ & d$^{10}$s \\
4d & d$^1$s$^2$ & d$^2$s$^2$ & d$^4$s & d$^5$s & d$^5$s$^2$ & d$^7$s & d$^8$s
& d$^{10}$ & d$^{10}$s \\
5d & d$^1$s$^2$ & d$^2$s$^2$ & d$^3$s$^2$ & d$^4$s & d$^5$s$^2$ & d$^6$s$^2$
& d$^7$s$^2$ & d$^9$s & d$^{10}$s \\ \hline
3d & ${}^2$D$_{3/2}$ & ${}^3$F$_2$ & ${}^4$F$_{3/2}$ & ${}^7$S$_3$ & ${}^6$S$%
_{5/2}$ & ${}^5$D$_4$ & ${}^4$F$_{9/2}$ & ${}^3$F$_4$ & ${}^2$S$_{1/2}$ \\
4d & ${}^2$D$_{3/2}$ & ${}^2$F$_2$ & ${}^6$D$_{1/2}$ & ${}^7$S$_3$ & ${}^6$D$%
_{9/2}$ & ${}^5$F$_5$ & ${}^4$F$_{9/2}$ & ${}^1$S$_o$ & ${}^2$S$_{1/2}$ \\
5d & ${}^2$D$_{3/2}$ & ${}^3$F$_2$ & ${}^4$F$_{3/2}$ & ${}^5$D$_0$ & ${}^6$S$%
_{5/2}$ & ${}^5$D$_4$ & ${}^4$F$_{9/2}$ & ${}^4$D$_3$ & ${}^2$S$_{1/2}$ \\
\hline
M$^{2+}$, & d$^1$ & d$^2$ & d$^3$ & d$^4$ & d$^5$ & d$^6$ & d$^7$ & d$^8$ & d%
$^9$ \\ \hline
3d, 4d, 5d & ${}^2$D$_{3/2}$ & ${}^3$F$_2$ & ${}^4$F$_{3/2}$ & ${}^5$D$_0$ &
${}^6$S$_{5/2}$ & ${}^5$D$_4$ & ${}^4$F$_{9/2}$ & ${}^3$F$_4$ & ${}^2$D$%
_{5/2}$ \\ \hline
3d & 24.75 & 27.47 & 29.31 & 30.95 & 33.69 & 30.64 & 33.49 & 35.16 & 36.83
\\
4d & 20.51 & 22.98 & 25.04 & 27.13 & 31.9 & 28.46 & 31.05 & 32.92 & 34.82 \\
5d & 19.18 & 28.1 & 22.3 & 24.1 & 26 & 25 & 27 & 28.5 & 30.5 \\ \hline
\end{tabular}

Table 1.2. Atomic configurations and ground terms for free atoms and R$^{3+}$
ions of rare earths and actinides. Rare earths and heavy actinides (starting
from Am) are characterized by the most stable valence of $3+$. Beside that,
the $4+$ state is possible for Ce, Tb and Pr, and the $2+$ state for Nd, Sm,
Eu, Tm and Yb. Light actinides exhibit a large variety of valence states.

\noindent
\begin{tabular}{|l|ccccccc|}
\hline
& La & Ce & Pr & Nd & Pm & Sm & Eu \\
R & f$^o$d & f$^1$d & f$^3$ & f$^4$ & f$^5$ & f$^6$ & f$^7$ \\
& D$^2_{3/2}$ & H$^3_4$ & I$^4_{9/2}$ & I$^5_4$ & H$^6_{5/2}$ & F$^7_0$ & S$%
^8_{7/2}$ \\ \hline
R$^{3+}$ & f$^0$ & f$^1$ & f$^2$ & f$^3$ & f$^4$ & f$^5$ & f$^6$ \\
& $^1$S$_0$ & $^2$F$_{5/2}$ & $^3$H$_4$ & $^4$I$_{9/2}$ & $^5$I$_4$ & $^6$H$%
_{5/2}$ & $^7$F$_0$ \\ \hline
& Gd & Tb & Dy & Ho & Er & Tm & Yb \\
R & f$^7$d & f$^8$d & f$^{10}$ & f$^{11}$ & f$^{12}$ & f$^{13}$ & f$^{14}$
\\
& $^9$D$_2$ & $^8$H$_{17/2}$ & $^5$I$_8$ & $^4$I$_{15/2}$ & $^3$H$_6$ & $^2$F%
$_{7/2}$ & $^1$S$_0$ \\ \hline
R$^{3+}$ & f$^7$ & f$^8$ & f$^9$ & f$^{10}$ & f$^{11}$ & f$^{12}$ & f$^{13}$
\\
& S$^8_{7/2}$ & F$^7_6$ & H$^6_{15/2}$ & I$^5_8$ & I$^4_{15/2}$ & H$^3_6$ & F%
$^2_{7/2}$ \\ \hline
\end{tabular}

\noindent
\begin{tabular}{|l|cccccccccc|}
\hline
& Ac & Th & Pa & U & Np & Pu & Am & Cm & Bk & Cf \\
R & f$^0$d$^1$ & f$^0$d$^2$ & f$^2$d$^1$ & f$^3$d$^1$ & f$^5$d$^0$ & f$^6$d$%
^0$ & f$^7$d$^1$ & f$^7$d$^1$ & f$^8$d$^1$ & f$^{10}$d$^0$ \\
& $^2$D$_{3/2}$ & $^3$F$_2$ & $^4$K$_{11/2}$ & $^2$L$_6$ & $^6$H$_{5/2}$ & $%
^7$F$_0$ & $^8$S$_{7/2}$ & $^9$D$_2$ & $^8$H$_{17/2}$ & $^5$J$_8$ \\ \hline
R$^{3+}$ & f$^0$ & f$^1$ & f$^2$ & f$^3$ & f$^4$ & f$^5$ & f$^6$ & f$^7$ & f$%
^8$ & f$^9$ \\
& $^1$S$_0$ & $^2$F$_{5/2}$ & $^3$H$_4$ & $^4$I$_{9/2}$ & $^5$I$_4$ & $^6$H$%
_{5/2}$ & $^7$F$_0$ & $^8$S$_{7/2}$ & $^7$F$_6$ & $^6$H$_{15/2}$ \\ \hline
\end{tabular}

Table 1.3. Probabilities of Auger transitions in Cu (10$^{-4}$a.u.) [38].\\

\noindent
\begin{tabular}{|c|ccccc|}
\hline
Term & $^1S$ & $^1G$ & $^3P$ & $^1D$ & $^3F$ \\ \hline
Multiplicity & 1 & 9 & 9 & 5 & 21 \\
$L_3M_{45}M_{45}$ & 1.15 & 2.88 & 1.39 & 7.40 & 16.18 \\
$L_2M_{45}M_{45}$ & 0.58 & 21.43 & 0.69 & 3.69 & 8.10 \\ \hline
\end{tabular}

Table 2.1. Calculated characteristics of band structure for d-metals: the
number of d-electrons n$_{\mathrm{d}}$, position of the Fermi level E$_{%
\mathrm{d}}$ (eV) and density of states N(E$_{\mathrm{F}}$) (eV$^{-1}$at$%
^{-1}$) [78], the model parameters [13]: average energy E$_{\mathrm{d}}$
(relative to the s-band bottom, eV) and the bandwidth W$_{\mathrm{d}}$ (eV);
experimental data on W$_{\mathrm{d}}$ from PES and IPES [40] are presented
for 3d-metals only.

\noindent
\begin{tabular}{|l|ccccccccc|}
\hline
3d & Sc & Ti & V & Cr & Mn & Fe & Co & Ni & Cu \\
4d & Y & Zr & Nb & Mo & Tc & Ru & Rh & Pd & Ag \\
5d & La & Hb & Ta & W & Re & Os & Ir & Pt & Au \\ \hline
&  &  &  &  &  & 4.6$^{\uparrow}$ & 4.8$^{\uparrow}$ & 4.82$^{\uparrow}$ &
\\
& 1.76 & 2.90 & 3.98 & 4.96 & 5.98 &  &  &  & 9.91 \\
n$_{\mathrm{d}}$ &  &  &  &  &  & 2.34$_{\downarrow}$ & 3.06$_{\downarrow}$
& 4.15$_{\downarrow}$ &  \\
& 1.68 & 2.96 & 4.10 & 5.07 & 6.23 & 7.24 & 7.99 & 8.96 & 10.01 \\
& - & 2.69 & 3.78 & 4.73 & 5.73 & 6.70 & 7.65 & 8.74 & 9.89 \\ \hline
& 5.82 & 8.03 & 9.26 & 10.6 & 10.3 & 10.2 & 9.53 & 8.71 & 7.90 \\
E$_{\mathrm{F}}$ & 5.39 & 7.42 & 9.06 & 10.6 & 10.3 & 10.1 & 8.61 & 7.06 &
6.30 \\
& - & 7.87 & 9.31 & 11.3 & 10.9 & 11.6 & 10.4 & 8.68 & 7.32 \\ \hline
&  &  &  &  &  & 0.85$^{\uparrow}$ & 0.14$^{\uparrow}$ & 0.15$^{\uparrow}$ &
\\
& 1.76 & 0.54 & 1.61 & 0.65 & 2.72 &  &  &  & 0.15 \\
&  &  &  &  &  & 0.24$_{\downarrow}$ & 0.66$_{\downarrow}$ & 1.54$%
_{\downarrow}$ &  \\
N(E$_{\mathrm{F}}$) & 1.69 & 0.53 & 1.20 & 0.52 & 0.91 & 0.81 & 1.32 & 2.31
& 0.06 \\
& - & 0.41 & 1.04 & 0.36 & 0.70 & 0.65 & 0.90 & 1.67 & 0.18 \\ \hline
& 7.05 & 7.76 & 8.13 & 8.01 & 7.91 & 7.64 & 7.36 & 6.91 & 5.90 \\
E$_{\mathrm{d}}$ & 6.75 & 7.17 & 7.29 & 7.12 & 6.67 & 6.02 & 5.08 & 4.52 &
2.49 \\
& - & 9.12 & 9.50 & 9.45 & 8.99 & 8.38 & 7.35 & 6.51 & 5.18 \\ \hline
& 5.13 & 6.08 & 6.77 & 6.56 & 5.60 & 4.82 & 4.35 & 3.78 & 2.80 \\
W$_{\mathrm{d}}$ & 6.59 & 8.37 & 9.72 & 9.98 & 9.42 & 8.44 & 6.89 & 5.40 &
3.63 \\
& - & 9.56 & 11.2 & 11.4 & 11.0 & 10.3 & 8.71 & 7.00 & 5.28 \\ \hline
W$^{\mathrm{exp}}_{\mathrm{d}}$(3d) & 6.2 & 6.6 & 6.8 & 6.5 & 8.5 & 8.5 & 6.9
& 5.4 & 2.6 \\ \hline
\end{tabular}

Table 2.2. Partial s,p,d densities of states at the Fermi level and the
number of d-electrons in 3d-metals [78].

\noindent
\begin{tabular}{|l|ccccccccccc|}
\hline
& Sc & Ti & V & Cr & Mn & Fe$^\uparrow $ & Fe$_{\downarrow}$ & Co$_{\uparrow}
$ & Co$_{\downarrow}$ & Ni$_{\uparrow}$ & Ni$_{\downarrow}$ \\ \hline
s & 0.34 & 0.07 & 0.43 & 0.08 & 0.19 & 0.22 & 0.02 & 0.36 & 0.06 & 0.28 &
0.16 \\
p & 6.06 & 0.80 & 2.46 & 0.72 & 0.31 & 0.14 & 0.04 & 0.23 & 0.44 & 0.24 &
0.09 \\
$\mathrm{t_{2g}}$ & 13.70 & 4.61 & 18.0 & 6.52 & 17.9 & 10.5 & 2.80 & 0.65 &
5.54 & 1.25 & 13.07 \\
$\mathrm{e_g}$ & 10.12 & 2.69 & 3.95 & 2.31 & 19.1 & 1.07 & 0.41 & 1.22 &
3.48 & 0.74 & 7.94 \\
$\mathrm{n_d}$ & 1.76 & 2.9 & 3.98 & 4.96 & 5.98 & 4.6 & 2.34 & 4.8 & 3.06 &
4.82 & 4.15 \\ \hline
\end{tabular}

Table 2.3. Experimental (angle-resolved photoemission) and calculated energy
on the $\Gamma \Delta L$ axis in palladium (in eV relative to the Fermi
level) [90].

\noindent
\begin{tabular}{|l|ccccc|}
\hline
k-point, & Exper. & RAPW & RAPW & HFS & Pseudopot. \\
band number & [90] & [91] & [92] & [93] & [94] \\ \hline
$\Gamma_{2,3,4} $ & $-2.55\pm 0.15$ & $-2.79$ & $-2.49$ & $-2.59$ & $-2.56$
\\
L$_{2,3}$ & $-2.4\pm 0.2$ & $-2.98$ & $-2.62$ & $-2.70$ & $-2.66$ \\
$\Gamma_{5,6}$ & $-1.15\pm 0.1$ & $-1.17$ & $-$ & $-2.70$ & $-2.66$ \\
L$_4$ & $-0.4\pm 0.2$ & $-0.14$ & $-$ & $-0.06$ & $-0.09$ \\
L$_5$ & $-0.1\pm 0.1$ & $+0.05$ & $-$ & $-0.06$ & $-0.09$ \\
L$_7$ & $+7.7\pm 0.3$ & $+7.30$ & $-$ & $-$ & $-$ \\
$\Gamma_7$ & $+18.4\pm 0.5$ & $+17.71$ & $-$ & $-$ & $-$ \\
$\Gamma_8$ & $+21.7\pm 0.5$ & $+21.65$ & $-$ & $-$ & $-$ \\ \hline
\end{tabular}

Table 2.4. Positions of density of states peaks, and widths of empty and
occupied part of the band W$^\pm $ for 3d and 4d metals (in eV) [101].

\noindent
\begin{tabular}{|l|cccccccc|}
\hline
3d & Sc & Ti & V & Cr & Mn & Fe & Co & Ni \\
4d & Y & Zr & Nb & Mo & Tc & Ru & Rh & Pd \\ \hline
& 1.8, 3.8 & 1.0, 2.8 & 2.3 & 1.0 & 1.4 & 1.6 & 0.5 & 0.3 \\
E$^{\mathrm{exp}}_{\mathrm{peak}}$ & 2.0, 5.4 & 0.9, 4.4 & 3.5 & 2.0 & - &
1.2 & 0.4 & 0.2 \\ \hline
& 1.4, 3.5 & 0.8, 3.1 & 2.5 & 1.2 &  & 1.6 & 0.5 & 0.3 \\
E$^{\mathrm{theor}}_{\mathrm{peak}}$ & 1.3, 4.6 & 0.9, 4.6 & 3.4 & 1.8 & - &
1.3 & 0.5 & 0.1 \\ \hline
& 1.7 & 3.1 & 3.2 & 4.5 & - & 5.3 & 5.5 & 5.0 \\
W$^{-}_{\mathrm{theor}}$ & 2.0 & 3.1 & 3.5 & 5.4 & - & 6.2 & 5.8 & 4.8 \\
\hline
& 4.0 & 3.65 & 4.0 & 2.5 & - & 2.6 & 1.15 & 0.4 \\
W$^{+}_{\mathrm{theor}}$ & 5.3 & 5.3 & 5.6 & 3.9 & - & 1.95 & 1.2 & 0.4 \\
\hline
& 4.4-4.7 & 3.2-3.8 & 3.4-3.8 & 1.8-2.3 & 3.5 & 2.9-3.5 & 1.4 & 0.4 \\
W$^{+}_{\mathrm{exp}}$ & 6.1 & 5.2-5.7 & 5.2-5.9 & 3.9-4.5 & - & 1.95 & 1.1
& 0.4 \\ \hline
W$^{+}_{\mathrm{theor}}\ -$ & -0.5 & 0.0 & +0.4 & +0.5 & - & -0.6 & -0.25 &
0.0 \\
$-$ W$^{+}_{\mathrm{exp}}$ & -0.8 & -0.1 & 0 & -0.3 & - & 0.0 & 0.1 & 0.0 \\
\hline
\end{tabular}

Table 2.5. Values of $\Delta _{-}=E_{\mathrm{F}}-\varepsilon (4f^n)$ and $%
\Delta _{+}=\varepsilon (4f^{n+1})-E_{\mathrm{F}}$ for rare-earth metals.

\noindent 
\begin{tabular}{|l|ccccccccccccccc|}
\hline
$\Delta_{\pm }$(eV) & $\gamma $-Ce & Pr & Nd & Pm & Sm & Eu & Gd & Tb & Dy & 
Ho & Er & Tm & Yb & Lu & La \\ \hline
$\Delta_{-}$ & 2.1 & 4.1 & 5.1 & - & 5.6 & 1.5 & 8.7 & 4.6 & 5.6 & 6.0 & 6.0
& 6.2 & 0.7 & - & - \\ 
$\Delta_{-}$ & 1.9 & 3.8 & 5.1 & - & 5.5 & 1.9 & 8.3 & 3.3 & 4.7 & 5.5 & 5.4
& 5.4 & 1.1 & 7.5 & - \\ 
$\Delta_{-}$ & 1.9 & 3.3 & 4.5 & 4.5 & 4.5 & - & 7.0 & 1.9 & 3.4 & 4.8 & 4.8
& 5.2 & - & 8.5 & - \\ 
$\Delta_{+}$ & 3.3 & 2.6 & 2.1 & - & 0.5 & - & 3.3 & 2.1 & 1.8 & 1.5 & 1.5 & 
1.3 & - & - & - \\ 
$\Delta_{+}$ & 3.1 & 2.0 & 1.3 & - & 0.2 & - & 3.2 & 2.1 & 1.4 & 1.6 & 1.6 & 
1.0 & - & - & 4.7 \\ \hline
\end{tabular}

Table 2.6. Characteristics of Fermi surfaces of some simple and transition
metals. $e_{\mathrm{N}}(i)$ and $h_{\mathrm{N}}(i)$ stand for electron and
hole orbits at the point N in the Brillouin zone with the number i, jg -
jungle gym, mc - multiconnected. S are areas of cross sections; in some
cases the corresponding $k_{\mathrm{F}}$ values are given in brackets. $A_0$%
, $k_0$ are the cross section and Fermi quasimomentum for free electrons.

\noindent 
\begin{tabular}{|l|cccccc|}
\hline
Metal & Orbits & S, $\mbox{\AA }^{-2}$ & ($k_{\mathrm{F}}$,$\mbox{\AA }%
^{-1}$) & m$_c$/m & m/m & Ref. \\ 
& theor. & exp. & theor. & exp. &  &  \\ \hline
Li & e$_\Gamma $(1) (110) & 0.976$A_0$ & (1.06$k_0$) &  & 1.48 & calc.147-149
\\ 
& spher. (100) & 0.993$A_0$ & (0.98$k_0$) &  & 1.65 & dHvA 150 \\ 
& (111) & 1.011$A_0$ & (0.99$k_0$) &  & 1.82 &  \\ \hline
Na & e$_\Gamma $(1) & $A_0$ & $A_0$ & 1 & 1 & calc.147-149 \\ 
& spher. &  &  &  &  & dHvA 11 \\ \hline
Ca & h(1) &  &  &  &  & calc.151,152 \\ 
& e$_L$(2) &  &  &  &  & dHvA 153 \\ \hline
Sc & h$_\Gamma $(3,4) &  &  &  &  & calc.154 \\ 
& e$_{\Gamma K}$(4), h$_{MK}$(4) &  &  &  &  &  \\ \hline
Y &  & 0.03-2.18 &  &  &  & calc.155,156 \\ 
&  &  &  &  &  & dHvA 156,157 \\ \hline
La & h$_{\Gamma A}$(5), h$_\Gamma A$(6) &  &  &  &  & calc.158 \\ 
& $e_{mc}$(7) &  &  &  &  &  \\ 
& $e_{AH}$(7), $e_{AH}$(8) &  &  &  &  &  \\ \hline
Ti & h$_L$(3,4),e$_\Gamma $(5,6) &  &  &  &  & calc.159-161 \\ 
& or h$_\Gamma $(3,4), & 0.10-0.72 & 0.18-0.64 & 0.65-1.95 & 1.6-2.8 & dHvA
162 \\ 
& h$_A$(3), e$_H$(5,6) &  &  &  &  &  \\ \hline
Zr & h$_\Gamma $(3,4) & 0.22-1.34 & 2.0-2.5 &  & 0.8, 1.2 & calc.163 \\ 
& e$_H$(5,6) & 2.29-3.36 &  &  &  & dHvA 164,165 \\ \hline
V & h$_\Gamma $(2), h$_{jg}$(3), &  &  &  &  & calc.165-167 \\ 
&  & 0.24-0.72 & 0.32-0.64 &  &  & dHvA 168-171 \\ 
& h$_N$(3), closed &  &  &  &  &  \\ \hline
Nb & h$_\Gamma $(2), & 0.117- & 0.138- & 0.57- & 1.12- & calc.172,173 \\ 
& h$_{jg,N}$(3) & 2.68 & 1.862 & 1.92 & 1.60 & dHvA 171,174 \\ 
& h$_{jg,\Delta }$ &  &  &  &  &  \\ \hline
Ta & h$_\Gamma $(2), & 0.292- & 0.279- & 0.84- & 1.09, & calc.172,173 \\ 
& h$_{jg,N}$(3) & 2.00 & 2.00 & 1.66 & 1.35 & dHvA 175 \\ 
& h$_{jg,\Delta }$ &  &  &  &  &  \\ \hline
Cr & h$_H$(3), &  &  &  &  & calc.176-181 \\ 
& e$_\Gamma $(4), & (0.109- & (0.17- &  &  & dHvA 182,183 \\ 
& e$_{\Gamma H}$(5) & 0.226) & 0.27) &  &  &  \\ \hline
Mo & h$_H$(3) & 0.06- & 0.05- &  &  & calc.82,184 \\ 
& h$_N$(3),e$_\Gamma $(4), & 2.25 & 2.48 & 0.30 & 0.4-1.5 & dHvA 145, \\ 
& e$_{\Gamma H}$(5) &  &  &  &  & 185,186 \\ \hline
W & h$_H$(3) & 0.014- & 1-1.45 & 0.9 &  & calc.177 \\ 
& h$_N$(3) & 0.43(2$\pi$/a)$^2$ & 0.06- &  & 0.25-1 & dHvA 185,187 \\ 
& e$_\Gamma $(4) &  & 0.23 &  &  &  \\ \hline
$\gamma -$ &  &  &  &  &  & calc 188,189 \\ \hline
Re & h(5,6,7) & 0.48-1.18 & 0.42-1.6 &  &  & calc.190 \\ 
& h$_L$(7) & 4.57-6.11 &  &  &  & dHvA 191 \\ 
& e$_{\Gamma A}$(8) & 6.6 & 4-6.7 &  &  &  \\ 
& e(9) &  &  &  &  &  \\ \hline
Fe & h$_{H\uparrow }$ & 0.041- & 0.037 &  &  & calc.192-194 \\ 
& h$_{N\uparrow }$ & 4.16 & 4.16 & 0.36 & 0.71 & 145 \\ 
& e$_{\Gamma \downarrow }$ &  &  &  &  & dHvA 195,196 \\ \hline
Co & e$_{\Gamma \uparrow }$, e$_{\Gamma \downarrow }$ & 0.007- & 0.01-0.16 & 
0.1 & 0.08 & calc.197-200 \\ 
& e$_{ML\downarrow }$ & 1.42 &  &  & 0.2 & dHvA 201-203 \\ 
& e$_{L\downarrow }$ & 0.025- &  &  &  &  \\ 
&  & 4.13 &  &  &  &  \\ \hline
Ni & e(5),e(6) & 0.04-0.75 & 0.03-8.03 & 2.9,3.7 & 1-1.9 & calc.204-206, \\ 
& h(3),h(4) &  &  &  &  & dHvA 207-209 \\ 
&  &  &  &  &  & 145 \\ \hline
Cu & neck, & 0.036$A_0$ & 0.035$A_0$ & 0.41 & 0.46 & calc.210 \\ 
& belly & 0.98$A_0$ & 0.98$A_0$ &  &  & dHvA 211 \\ \hline
Ru & e$_\Gamma $, e$_{mc}$, &  &  &  &  & calc.11 \\ 
& h$_\Gamma $, h$_{LM}$ &  &  &  &  & dHvA 212,213 \\ \hline
Os & h$_{LM}$(7) & (0.057- & 0.015-0.030 &  & 0.12-0.2 & dHvA 214 \\ 
& e$_\Gamma $(9) & 0.119) & 1.71-2.19 &  & - &  \\ 
& e$_\Gamma $(10) &  & 1.33-1.64 &  & 1.2 &  \\ 
& h$_{LM,mc}$ &  & 0.73-1.18 &  & 1.1-1.5 &  \\ \hline
Rh & e$_{\Gamma 1}$ & 1.92-2.5 & 2.03-2.46 & 1.43-2.27 &  & calc.215 \\ 
& e$_{\Gamma 2}$ & 4.04-5.47 & 4.30 & 3.04-4.35 &  & dHvA 216-219 \\ 
& h$_{X1}$ & 0.16-0.27 & 0.15-0.25 & 0.35-0.54 &  &  \\ 
& h$_{X2}$ & 0.32-0.46 & 0.46 & 0.91-1.33 & 1.2 &  \\ 
& h$_L$ & 0.025 & 0.022 & 0.11 & 0.11 &  \\ \hline
Ir & e$_{\Gamma 1}$,e$_{\Gamma 2}$, & similar & similar & 1.34-3.02 &  & 
calc.215,222 \\ 
& h$_{X1}$, h$_{X2}$ & to Rh & to Rh & 0.21-1.05 &  & dHvA 216-221 \\ 
& h$_L$ &  &  &  &  &  \\ \hline
Pd & e$_\Gamma $(s) & 2.25-2.58 & 2.30-2.59 & 1.34-1.47 & 2.2 & calc.215 \\ 
& h$_X$(d) & 0.10-0.72 & 0.10-0.60 & 0.41-0.7 & 0.63-1.91 & dHvA 223-225 \\ 
&  &  & 0.05-0.08 &  &  &  \\ 
& h$_{mc}$(d) & 0.25, 1.32 & 0.28 & 1.37, 2.18 & 2.40-8.1 &  \\ 
&  &  & 0.24-0.98 & 9.1, 10.53 &  &  \\ 
& h$_L$ & 0.02-0.03 &  &  & 0.84, 1.21 &  \\ \hline
Pt & e$_\Gamma $(s) & 2.41-3.00 & 2.48-3.09 & 1.42-2.05 & 2.06-3.16 & 
calc.215 \\ 
& h$_{\mathrm{open}}$(d) & 0.25-1.22 & 0.25-0.78 & 0.91-6.23 & 1.53-3.62 & 
dHvA 226,227 \\ 
& h$_{\mathrm{closed}}$(d) & 0.004 & 0.15 & 0.30 & 0.36 &  \\ \hline
UPt3 & a-axis & 0.057-0.63 & 0.058-0.63 & 1.6-5.3 & 25-90 & dHvA 228 \\ 
& b-axis & 0.041-0.21 & 0.044-0.27 & 1.3-4.2 & 13-50 &  \\ \hline
\end{tabular}

Table 3.1. Atomic volume V$_a$, Debye temperature $\Theta _D$, melting point
T$_m$, boiling point T$_B$, linear coefficient of thermal expansion $\alpha $%
, self-diffusion activation energy Q, heat of fusion $\Delta H_f$, cohesive
energy $\Delta H^0$, Young's modulus Y, shear modulus $\mu $, Poisson's
ratio $\sigma $, bulk modulus B, Leibfried, modified Leibfried and Bragg
numbers, L, L$^{\prime }$ and $\mathcal{B}$, for 3d, 4d and 5d transition
metals and neighbour elements. For some metals, estimated values are
presented [235,238].

\noindent 
\begin{tabular}{|l|ccccccccccc|}
\hline
3d & Ca & Sc & Ti & V & Cr & Mn & Fe & Co & Ni & Cu & Zn \\ 
4d & Sr & Y & Zr & Nb & Mo & Tc & Ru & Rh & Pd & Ag & Cd \\ 
5d & Ba & La & Hf & Ta & W & Re & Os & Ir & Pt & Au & Hg \\ \hline
& d$^0$ & d$^1$s$^2$ & d$^2$s$^2$ & d$^3$s$^2$ & d$^5$s & d$^5$s$^2$ & d$^6$s%
$^2$ & d$^7$s$^2$ & d$^8$s$^2$ & d$^{10}$s & d$^{10}$ \\ 
& d$^0$ & d$^1$s$^2$ & d$^2$s$^2$ & d$^4$s & d$^5$s & d$^5$s$^2$ & d$^7$s & d%
$^8$s & d$^{10}$ & d$^{10}$s & d$^{10}$ \\ 
& d$^0$ & d$^1$s$^2$ & d$^2$s$^2$ & d$^3$s$^2$ & d$^4$s$^2$ & d$^5$s$^2$ & d$%
^6$s$^2$ & d$^7$s$^2$ & d$^9$s & d$^{10}$s & d$^{10}$ \\ \hline
& 25.9 & 15.0 & 10.6 & 8.5 & 7.23 & 7.39 & 7.1 & 6.6 & 6.59 & 7.09 & 9.17 \\ 
V$_a$, & 33.9 & 19.9 & 14.0 & 10.8 & 9.39 & 8.63 & 8.18 & 8.29 & 8.88 & 10.3
& 13.0 \\ 
cm$^3$/mol & 38.1 & 22.5 & 13.4 & 10.8 & 9.55 & 8.86 & 8.44 & 8.52 & 9.09 & 
10.2 & 14.1 \\ \hline
& 234 & 470 & 426 & 326 & 598 & 418 & 457 & 452 & 427 & 342 & 316 \\ 
$\Theta _D$, & 147 & 268 & 289 & 241 & 459 & 351 & 600 & 480 & 283 & 228 & 
352 \\ 
K & 110 & 142 & 256 & 247 & 388 & 429 & 500 & 425 & 234 & 165 & 75 \\ \hline
& 1112 & 1812 & 1941 & 2178 & 2148 & 1512 & 1808 & 1765 & 1726 & 1356 & 692
\\ 
T$_m$, K & 1045 & 1775 & 2123 & 2741 & 2888 & 2443 & 2553 & 2233 & 1825 & 
1234 & 594 \\ 
& 998 & 1193 & 2495 & 3271 & 3653 & 3433 & 3300 & 2716 & 2042 & 1336 & 234
\\ \hline
& 1765 & 3537 & 3586 & 3582 & 2918 & 2368 & 3160 & 3229 & 3055 & 2811 & 1175
\\ 
T$_B$,K & 1645 & 3670 & 4650 & 4813 & 5785 & 5300 & 4325 & 3960 & 3200 & 2468
& 1038 \\ 
& 1910 & 3713 & 4575 & 5760 & 6000 & 6035 & 5300 & 4820 & 4100 & 3240 & 630
\\ \hline
& 22.4 & 10.0 & 8.35 & 8.3 & 8.4 & 22.6 & 11.7 & 12.4 & 12.7 & 16.7 & 29.7
\\ 
$\alpha \cdot 10^6$, & 20 & 12.0 & 5.78 & 7.07 & 4.98 & 8.06 & 9.36 & 8.40 & 
11.5 & 19.2 & 30.6 \\ 
K$^{-1}$ & 18.8 & 10.4 & 6.01 & 6.55 & 4.59 & 6.63 & 4.7 & 6,63 & 8.95 & 14.1
& 61 \\ \hline
&  &  & 48.0 & 91.5 & 73.2 &  & 64.0 & 61.9 & 67.0 & 48.9 & 23.9 \\ 
Q$\cdot 10^3$, &  & 61.1 & 52.0 & 98.0 & 96.9 &  &  &  & 63.5 & 45.8 &  \\ 
kcal/mol &  & 40.8 & 43.7 & 100 & 120 &  &  &  & 66.8 & 39.4 &  \\ \hline
& 2.07 & 3.70 & 3.42 & 3.83 & 3.47 & 3.50 & 3.67 & 3.70 & 4.21 & 3.12 & 1.76
\\ 
$\Delta H_f$, & 2.19 & 2.73 & 3.74 & 4.82 & 6.66 & 5.42 & 5.67 & 4.96 & 4.10
& 2.78 & 1.48 \\ 
kcal/mol & 1.83 & 1.48 & 4.39 & 5.76 & 8.42 & 7.86 & 7.56 & 6.22 & 4.70 & 
2.95 & 0.55 \\ \hline
& 42.1 & 80.1 & 112 & 122 & 94.5 & 66.9 & 99.4 & 102 & 102 & 80.8 & 31.0 \\ 
$\Delta H^0$ & 39.3 & 97.6 & 146 & 174 & 157 & 152 & 154 & 133 & 89.9 & 68.3
& 26.8 \\ 
kcal/mol & 42.8 & 102 & 145 & 187 & 200 & 186 & 187 & 159 & 135 & 87.6 & 15.4
\\ \hline
& 0.20 & 0.81 & 1.08 & 1.34 & 2.48 & 2.02 & 2.14 & 2.10 & 1.97 & 1.26 & 0.94
\\ 
Y$\cdot 10^{-6}$, & 0.14 & 0.66 & 0.94 & 1.07 & 3.34 & 3.76 & 4.20 & 3.70 & 
1.26 & 0.82 & 0.63 \\ 
kg/cm$^3$ & 0.13 & 0.39 & 1.40 & 1.85 & 4.05 & 4.7 & 5.50 & 5.38 & 1.74 & 
0.74 & 0.28 \\ \hline
& 0.07 & 0.32 & 0.40 & 0.47 & 1.19 & 0.78 & 0.83 & 0.78 & 0.76 & 0.46 & 0.38
\\ 
$\mu \cdot 10^{-6}$, & 0.05 & 0.26 & 0.35 & 0.38 & 1.18 & 1.45 & 1.63 & 1.50
& 0.52 & 0.29 & 0.25 \\ 
kg/cm$^2$ & 0.05 & 0.15 & 0.54 & 0.70 & 1.56 & 1.82 & 2.14 & 2.14 & 0.62 & 
0.28 & 0.10 \\ \hline
& 0.31 & 0.27 & 0.34 & 0.36 & 0.21 & 0.24 & 0.28 & 0.33 & 0.30 & 0.34 & 0.29
\\ 
$\sigma $ & 0.3 & 0.2 & 0.3 & 0.35 & 0.30 & 0.29 & 0.29 & 0.27 & 0.37 & 0.37
& 0.30 \\ 
& 0 28 & 0.29 & 0.30 & 0.35 & 0.28 & 0.29 & 0.28 & 0.26 & 0.38 & 0.42 & 0.36
\\ \hline
& 0.15 & 0.58 & 1.07 & 1.65 & 1.94 & 0.61 & 1.72 & 1.95 & 1.90 & 1.33 & 0.61
\\ 
B$\cdot 10^{-6}$, & 0.12 & 0.37 & 0.85 & 1.74 & 2.78 & 3.03 & 3.27 & 2.76 & 
1.84 & 1.03 & 0.48 \\ 
kg/cm$^2$ & 0.10 & 0.25 & 1.11 & 2.04 & 3.30 & 3.79 & 4.26 & 3.62 & 2.84 & 
1.77 & 0.29 \\ \hline
& 4.80 & 3.20 & 3.42 & 4.66 & 2.12 & 2.24 & 2.60 & 2.87 & 2.90 & 3.51 & 1.69
\\ 
L & 4.90 & 2.88 & 3.69 & 5.62 & 2.21 & 1.65 & 1.62 & 1.52 & 3.34 & 3.49 & -
\\ 
& 4.44 & 2.95 & 2.91 & 3.67 & 2.08 & 1.80 & 1.55 & 1.26 & 3.06 & 3.94 & 3.51
\\ \hline
& 4.28 & 2.85 & 3.04 & 4.15 & 1.88 & 2.00 & 2.32 & 3.35 & 3.39 & 4.10 & 1.97
\\ 
L$^{\prime }$ & 4.36 & 2.56 & 3.29 & 5.00 & 1.97 & 1.93 & 1.90 & 2.78 & 3.90
& 4.07 & - \\ 
& 3.96 & 2.63 & 2.59 & 3.27 & 1.85 & 2.11 & 1.81 & 1.47 & 3.57 & 4.61 & 4.10
\\ \hline
& 4.21 & 3.08 & 2.84 & 3.86 & 1.61 & 2.44 & 2.49 & 2.84 & 3.84 & 3.81 & 2.03
\\ 
$\mathcal{B}$ & 4.84 & 2.09 & 3.06 & 4.66 & 2.40 & 1.73 & 1.70 & 1.59 & 3.54
& 3.70 & - \\ 
& 3.84 & 1.73 & 2.42 & 3.05 & 2.26 & 1.95 & 1.67 & 1.36 & 3.32 & 4.11 & 3.81
\\ \hline
\end{tabular}

Table 3.2. Atomic volume V$_a$ (cm$^3$/mol), Debye temperature $\Theta _D$%
(K), melting point T$_m$(K), boiling point T$_B$(K), linear coefficient of
thermal expansion $\alpha $ (10$^{-6}$K$^{-1}$), heat of fusion $\Delta H_f$
(kcal/mol), cohesive energy $\Delta H^0$ (kcal/mol), Young's modulus Y(10$^6$%
kg/cm$^2$), shear modulus $\mu $ (10$^6$kg/cm$^2$), Poisson's ratio $\sigma $%
, bulk modulus B (10$^6$kg/cm$^2$) for rare earth metals [235].

\noindent 
\begin{tabular}{|l|ccccccc|}
\hline
& La & Ce($\gamma$) & Pr & Nd & Pm & Sm & Eu \\ \hline
& f$^0$d & f$^1$d & f$^3$ & f$^4$ & f$^5$ & f$^6$ & f$^7$ \\ \hline
V$_a$ & 22.54 & 17.03 & 20.82 & 20.59 & 20.33 & 19.95 & 28.98 \\ 
$\Theta _D$ & 142 & 146 & 85 & 159 & 158 & 116 & 127 \\ 
T$_m$ & 1193 & 1070 & 1208 & 1297 & 1308 & 1345 & 1099 \\ 
T$_B$ & 3713 & 3972 & 3616 & 2956 & 2730 & 2140 & 1971 \\ 
$\Delta H_f$ & 1.48 & 1.24 & 1.65 & 1.71 & 1.94 & 2.06 & 2.20 \\ 
$\Delta H^0$ & 101.9 & 97.9 & 85.8 & 75.9 & 64 & 50.2 & 42.9 \\ 
$\alpha $ & 10.4 & 8.5 & 6.79 & 9.98 & 9.0 & 10.8 & 33.1 \\ 
Y & 0.387 & 0.306 & 0.332 & 0.387 & 0.43 & 0.348 & 0.155 \\ 
$\mu $ & 0.152 & 0.122 & 0.138 & 0.148 & 0.17 & 0.129 & 0.060 \\ 
$\sigma $ & 0.288 & 0.248 & 0.305 & 0.306 & 0.278 & 0.352 & 0.286 \\ 
B & 0.248 & 0.244 & 0.312 & 0.333 & 0.360 & 0.300 & 0.150 \\ \hline
& Gd & Tb & Dy & Ho & Er & Tm & Yb \\ \hline
& f$^7$d & f$^8$d & f$^{10}$ & f$^{11}$ & f$^{12}$ & f$^{13}$ & f$^{14}$ \\ 
\hline
V & 19.94 & 19.26 & 18.99 & 18.75 & 18.46 & 18.13 & 24.87 \\ 
$\Theta _D$ & 170 & 150 & 172 & 114 & 134 & 127 & 118 \\ 
T$_m$ & 1585 & 1629 & 1680 & 1734 & 1770 & 1818 & 1097 \\ 
T$_B$ & 3540 & 3810 & 3011 & 3228 & 3000 & 2266 & 1970 \\ 
$\Delta H_f$ & 2.44 & 2.46 & 2.49 & 3.38 & 2.62 & 4.22 & 1.83 \\ 
$\Delta H^0$ & 82.7 & 89.9 & 66.9 & 70.5 & 70.7 & 58.3 & 40.3 \\ 
$\alpha $ & 8.28 & 10.3 & 10.0 & 10.7 & 12.3 & 13.3 & 24.96 \\ 
Y & 0.573 & 0.586 & 0.644 & 0.684 & 0.748 & 0.77 & 0.182 \\ 
$\mu $ & 0.227 & 0.233 & 0.259 & 0.272 & 0.302 & 0.31 & 0.071 \\ 
$\sigma $ & 0.259 & 0.261 & 0.243 & 0.255 & 0.238 & 0.235 & 0.284 \\ 
B & 0.391 & 0.407 & 0.392 & 0.404 & 0.419 & 0.405 & 0.135 \\ \hline
\end{tabular}

Table 3.3. Atomic volume V$_a$ (cm$^3$/mol), Debye temperature $\Theta _D$%
(K), melting point T$_m$(K), boiling point T$_B$(K), linear coefficient of
thermal expansion $\alpha $ (10$^{-6}$K$^{-1}$), heat of fusion $\Delta H_f$
(kcal/mol), cohesive energy $\Delta H^0$ (kcal/mol), Young's modulus Y(10$^6$%
kg/cm$^2$), shear modulus $\mu $ (10$^6$kg/cm$^2$), Poisson's ratio $\sigma $%
, bulk modulus B (10$^6$kg/cm$^2$) for 5f-elements [235].

\noindent 
\begin{tabular}{|l|cccccccccc|}
\hline
& Ac & Th & Pa & U & Np & Pu & Am & Cm & Bk & Cf \\ \hline
& f$^0$d$^1$ & f$^0$d$^2$ & f$^2$d$^1$ & f$^3$d$^1$ & f$^5$d$^0$ & f$^6$d$^0$
& f$^7$d$^0$ & f$^7$d$^1$ & f$^8$d$^1$ & f$^{10}$d$^0$ \\ \hline
V$_a$ & 22.56 & 19.79 & 15.03 & 13.16 & 13.11 & 12.06 & 17.78 & 12.8 &  & 
\\ 
$\Theta _D$ & 124 & 170 & 159 & 200 & 121 & 171 &  &  &  &  \\ 
T$_m$ & 1323 & 2024 & 1698 & 1404 & 910 & 913 & 1473 & 913 &  &  \\ 
T$_B$ & 3200 & 4500 & 4680 & 3950 & 4150 & 3727 &  &  &  &  \\ 
$\Delta H_f$ & 3.03 & 3.56 & 2.99 & 2.47 & 1.60 & 0.68 &  &  &  &  \\ 
$\Delta H^0$ & 104 & 136.7 & 132 & 125 & 113 & 91.8 &  &  &  &  \\ 
$\alpha $ & 14.9 & 11.2 & 7.3 & 12.6 & 27.5 & 55 & 7.5 &  &  &  \\ 
Y & 0.35 & 0.76 & 1.02 & 1.90 & 1.02 & 0.98 &  &  &  &  \\ 
$\mu $ & 0.138 & 0.284 & 0.398 & 0.75 & 0.406 & 0.446 &  &  &  &  \\ 
$\sigma $ & 0.269 & 0.285 & 0.282 & 0.245 & 0.255 & 0.15 &  &  &  &  \\ 
B & 0.25 & 0.553 & 0.078 & 1.007 & 0.694 & 0.546 &  &  &  &  \\ \hline
\end{tabular}

Table 3.4. Polymorphic transformations in Ca, Sr, d-metals, rare earths and
actinides [238,139]. The temperature intervals (K) for stablility of crystal
structures under normal pressure are given in parentheses.

\noindent 
\begin{tabular}{|l|c|}
\hline
Ca & fcc (0-737), hcp (737-1123) \\ \hline
Sc & hcp (0-1223), bcc (1223-1811) \\ \hline
Ti & hcp (0-1158), bcc (1158-1938) \\ \hline
Mn & compl.cubic A12 (0-1000), compl.cubic A13 (1000-1365), fcc with \\ 
& tetragonal distortions (1000-1365), bcc (1405-1517) \\ \hline
Fe & bcc ($\alpha $, 0-1183), fcc ($\gamma $, 1183-1163), bcc ($\delta $,
1663-1912) \\ \hline
Co & hcp (0-700), fcc (700-1763) \\ \hline
Sr & fcc (0-506), hcp (506-813), bcc (813-1163) \\ \hline
Y & hcp (0-1763), bcc (1763-1773) \\ \hline
Zr & hcp (0-1135), bcc (1135-2128) \\ \hline
La & hcp (0-583), fcc (583-1137), bcc (1137-1193) \\ \hline
Hf & hcp (0-2050), bcc (2050-2222) \\ \hline
Ce & fcc ($\alpha $, 0-116), dhcp ($\beta $, 116-263), fcc ($\gamma $,
263-1003), bcc ($\delta $, \\ 
& 1003-1068) \\ \hline
Pr & dhcp (0-1071), bcc (1071-1208) \\ \hline
Nd & dhcp (0-1141), bcc (1141-1297) \\ \hline
Pm & dhcp (0-1163), bcc (1163-1315) \\ \hline
Sm & rhomboedr.Sm-type (0-1190), fcc (1190-1345) \\ \hline
Eu & bcc (0-1099) \\ \hline
Gd & hcp (0-1535), bcc (1535-1585) \\ \hline
Tb & hcp (0-1560), bcc (1560-1633) \\ \hline
Dy & hcp (0-1657), bcc (1657-1682) \\ \hline
Ho & hcp (0-1701), bcc (1701-1743) \\ \hline
Er & hcp (0-1795) \\ \hline
Tm & hcp (0-1818) \\ \hline
Yb & hcp (0-1065), bcc (1065-1097) \\ \hline
Lu & hcp (0-1929) \\ \hline
Th & fcc (0-1400), bcc (1400-1750) \\ \hline
Pa & bct (0-1170), bcc (1170-1575) \\ \hline
U & orthorhomb.(0-662), tetr.(662-672), bcc (772-1132) \\ \hline
Np & orthorhomb.(0-278), tetr.(278-577), bcc (577-637) \\ \hline
Pu & monoclynic ($\alpha $, 0-122), monoclynic ($\beta $, 122-206),
orthorhomb. \\ 
& (206-310), fcc (310-458), bct (458-480), bcc (480-641) \\ \hline
Am & dhcp (0-1079), fcc (1079-1176) \\ \hline
Cm & dhcp, fcc (0-1340) \\ \hline
Bk & dhcp, fcc (0-986) \\ \hline
Cf & dhcp ($\alpha $, 0-600), fcc ($\beta $, 600-725), fcc ($\gamma $, >725)
\\ \hline
\end{tabular}

Table 3.5. Theoretical (obtained from band calculations) and experimental
values of the coefficient of the linear specific heat in d-metals [78].

\noindent 
\begin{tabular}{|l|ccccccccc|}
\hline
3d & Sc & Ti & V & Cr & Mn & Fe & Co & Ni & Cu \\ 
4d & Y & Zr & Nb & Mo & Tc & Ru & Rh & Pd & Ag \\ 
5d & La & Nb & Ta & W & Re & Os & Ir & Pt & Au \\ \hline
& 5.25 & 1.40 & 4.31 & 1.67 & 6.49 & 2.63 & 2.08 & 4.13 & 0.70 \\ 
$\gamma _{theor}$ & 5.46 & 1.37 & 3.44 & 1.40 & - & 2.00 & 3.24 & 5.58 & 0.62
\\ 
mJ/mol K$^2$ & - & 1.15 & 2.96 & 0.99 & 1.91 & 1.65 & 2.20 & 5.18 & 0.69 \\ 
\hline
& 10.7 & 3.35 & 9.26 & 1.40 & 9.20 & 4.98 & 4.73 & 7.02 & 0.69 \\ 
$\gamma _{exp}$ & 10.2 & 2.80 & 7.79 & 2.0 & - & 3.3 & 4.9 & 9.42 & 0.65 \\ 
mJ/mol K$^2$ & 10.1 & 2.16 & 5.90 & 1.3 & 2.3 & 2.4 & 3.1 & 6.8 & 0.73 \\ 
\hline
\end{tabular}

Table 4.1. Magnetic susceptibility ($10^{-6}$ emu/mol) of paramagnetic
d-metals with a cubic lattice. To exclude the influence of magnetic
impurities, the signs of $d\chi /dT$ are given at not too low temperatures.
References may be found in the review [270].

\noindent 
\begin{tabular}{|l|ccccccccccc|}
\hline
& V & Nb & Mo & Tc & Rh & Pd & Lu & Ta & W & Ir & Pt \\ \hline
& d$^3$s$^2$ & d$^3$s$^2$ & d$^5$s$^1$ & d$^5$s$^2$ & d$^8$s$^1$ & d$^{10}$
& d$^1$s$^2$ & d$^3$s$^2$ & d$^4$s$^2$ & d$^7$s$^2$ & d$^9$s$^1$ \\ \hline
$\chi $ & 300 & 212 & 89.2 & 270 & ?07 & 550 & 336 & 162 & 53.3 & 24.1 & 192
\\ 
$d\chi /dT$ & $-$ & $-$ & $+$ & $-$ & $+$ & $-$ & $-$ & $-$ & $+$ & $+$ & $-$
\\ \hline
\end{tabular}

Table 4.2. Longitudinal ($H_{\parallel } c$) and transverse ($H_{\perp } c$)
magnetic susceptibility ($10^{-6}$ emu/mol) of paramagnetic d-metals with a
hcp lattice. The susceptibility of a polycrystal is given by $\chi = (\chi
_{\parallel } + 2\chi _{\perp })/3$. Two sets of data correspond to results
of various authors (see [270]).

\noindent 
\begin{tabular}{|l|cccccccc|}
\hline
& Sc & Y & Ti & Zr & Hf & Re & Ru & Os \\ \hline
& d$^1$s$^2$ & d$^1$s$^2$ & d$^2$s$^2$ & d$^2$s$^2$ & d$^2$s$^2$ & d$^4$s$^2$
& d$^7$s$^1$ & d$^6$s$^2$ \\ \hline
$\chi _{\parallel }$ & 294, 281 & 174, 270 & 169 & 147, 151 & 95 & 68.3 & 
35.2 & 5.4 \\ 
$\chi _{\perp }$ & 232, 298 & 220, 445 & 145 & 86, 100 & 63 & 73.0 & 44.2 & 
12.6 \\ 
$d\chi /dT$ & $-$ & $+$ & $+$ & $+$ & $+$ & $+$ & $+$ & $+$ \\ \hline
\end{tabular}

Table 4.3. Values of the ground state moment $p_s$ determined from the
saturation magnetization ($M_0 = p_s\mu _B$), ferromagnetic and paramagnetic
Curie temperature $T_{\mathrm{C}}$ and $\Theta $, and paramagnetic moment $%
p_C$, determined from the Curie constant ($C = \mu _{\mathrm{eff}}^2/3 =
p_C(p_C + 2) \mu _{\mathrm{B}}^2/3$) for some d- and f-metals, and their
alloys and compounds. Rough estimates for non-ferromagnetic transition
metals Pd and Pt, where the Curie-Weiss law holds approximately, are also
presented.

\noindent 
\begin{tabular}{|l|ccccc|}
\hline
& $p_s$ & $T$ K & $p$ & $\Theta $ K & Ref. \\ \hline
Fe & 2.22 & 1044 & 2.3 & 1101 &  \\ 
Co & 1.71 & 1388 & 2.3 & 1411 &  \\ 
Ni & 0.62 & 627 & 0.9 & 649 &  \\ 
CrBr$_3$ & 3.0 & 37 & 3.0 & 37 & 280 \\ 
CoS$_2$ & 0.84 & 116 & 1.00 &  &  \\ 
CrO$_2$ & 4.0 & 400 &  &  &  \\ 
Pd & - & - & 0.6 & -200 &  \\ 
Pd $+ 1 \%$ Fe & 0.084 & 50 & 0.72 &  & 280 \\ 
Pt & - & - & 0.6 & -1000 &  \\ 
ZrZn$_2$ & 0.12 & 21 & 0.66 & 35 & 280 \\ 
Sc$_3$In & 0.045 & 5.5 & 0.22 & 16 & 280 \\ 
Ni$_3$Al & 0.075 & 41 & 0.64 &  & 280 \\ 
Co$_2$MnSi & 5.10 & 1034 & 2.03 & 1044 & 281 \\ 
Co$_2$MnSn & 5.37 & 826 & 3.35 & 870 & 281 \\ 
NiMnSb & 4.2 & 728 & 2.06 & 910 & 282 \\ 
PtMnSb & 3.96 & 572 & 3.56 & 670 & 282 \\ 
Pd$_2$MnSn & 4.22 & 189 & 4.05 & 201 & 283 \\ 
PtMnSn & 3.5 & 330 & 4.2 & 350 & 282 \\ 
Gd & 7.13 & 289 & 7.05 & 316 &  \\ 
Tb & 9.34 & 221 & 8.2 & 232 &  \\ 
EuO & 6.80 & 69.4 & 6.8 & 76 & 280 \\ 
EuRh$_3$B$_2$ & 0.56 & 46 & 3.5 & -40 & 284 \\ 
CeRh$_3$B$_2$ & 0.38 & 115 & 2.2 & -373 & 284 \\ \hline
\end{tabular}

Table 4.4. Calculated and experimental values of the spin splitting $\Delta $
(eV), the Curie temperature $T_{\mathrm{C}}$ (K) and the ratio $\Delta /T_{%
\mathrm{C}}$ for iron group metals according to data of various authors
[291,304].

\noindent 
\begin{tabular}{|l|ccc|}
\hline
& Fe & Co & Ni \\ \hline
$\Delta ^{(1)}$ [292] & 2 & 1.54 & 0.58 \\ 
$\Delta ^{(2)}$ [293] & 1.41 & 1.11 & 0.38 \\ 
$\Delta ^{exp}$ [104] & 1.5 & 1.1 & 0.3 \\ 
$T_{\mathrm{C}}^{(1)}$ & 5300 & 4000 & 2900 \\ 
$T_{\mathrm{C}}^{(2)}$ & 2560 & 2240 & 1790 \\ 
$T_{\mathrm{C}}^{exp}$ & 1040 & 1390 & 630 \\ 
$\delta ^{(1)}$ & 4.38 & 4.47 & 2.32 \\ 
$\delta ^{(2)}$ & 6.39 & 5.75 & 2.46 \\ 
$\delta ^{(exp)}$ & 16.7 & 9.16 & 5.5 \\ \hline
\end{tabular}

Table 4.5. Values of the Stoner and renormalized Curie points, $T_{\mathrm{S}%
}$ and $T_{\mathrm{C}}$, spin-fluctuation temperature $T_{\mathrm{sf}}$, and
the fluctuating magnetic moment at $T_{\mathrm{C}}^{calc}$ for iron group
metals; $t_{\mathrm{C}} = T_{\mathrm{C}}^{exp}/T_{\mathrm{S}}$ [304].

\noindent 
\begin{tabular}{|l|cccccc|}
\hline
& $T_{\mathrm{S}}$ & $T_{\mathrm{sf}}$ & $T_{\mathrm{C}}^{calc}$ & $T_{%
\mathrm{C}}^{exp}$ & $\langle m^2 \rangle ^{1/2}$ & $t_{\mathrm{C}}$ \\ 
\hline
Fe & 2560 & 1293 & 1068 & 1043 & 1.52 & 0.41 \\ 
Co & 2240 & 2439 & 1436 & 1388 & 0.93 & 0.62 \\ 
Ni & 1790 & 759 & 656 & 631 & 0,42 & 0.35 \\ \hline
\end{tabular}

Table 4.6. Magnetic anisotropy constants ($10^5$ erg/cm$^3$) and anisotropy
field (Oe) for iron group metals and gadolinium. Data of [265] with some
corrections are used.

\noindent 
\begin{tabular}{|l|cccc|}
\hline
& Fe (bcc) & Ni(fcc) & Co (hcp) & Gd (hcp) \\ \hline
K$_1$(293K) & - & - & 43 & - \\ 
K$_1$(4.2K) & - & - & 77 & -8.5 \\ 
K$_2$(293K) & 4.8 & -0.49 & 12 & - \\ 
K$_2$(4.2K) & 6 & -12 & 10 & 25 \\ 
K$_3$(293K) & 2 & 0.4 & - & - \\ 
K$_3$(4.2K) & - & 6 &  &  \\ 
H$_a$(293K) & 560 & 205 & 9500 & - \\ \hline
\end{tabular}

Table 4.7. Magnetic anisotropy constants, 10$^8$ erg/cm$^3$ at low
temperatures [15,381], and anisotropy of paramagnetic Curie temperature $%
\Delta \Theta = \Theta _{\parallel } - \Theta _{\perp }$, K [39] for heavy
rare earth metals. The corresponding theoretical estimates are presented for
the crystal field and exchange mechanism.

\noindent 
\begin{tabular}{|l|ccccc|}
\hline
& Tb & Dy & Ho & Er & Tm \\ \hline
$K_1^{exp}$ & -5.5 & -5 & -2.2 & - & - \\ 
$K_2^{exp}$ & -0.45 & 0.54 & -1.7 & - & - \\ 
$K_1^{cf}$ & -5.5 & -5.05 & -1.98 & 1.97 & 5.5 \\ 
$K_1^{exch}$ & -5.5 & -4.6 & -1.43 & 1.1 & 1.85 \\ 
$\Delta \Theta ^{exp}$ & 44 & 48 & 15 & -29 & -58 \\ 
$\Delta \Theta ^{cf}$ & 44 & 38 & 15 & -16 & -44 \\ 
$\Delta \Theta ^{exch}$ & 48 & 40 & 12 & -25 & -24(-??) \\ \hline
\end{tabular}

Table 4.8. Values of orbital momentum and the type of magnetic anisotropy
for rare earth ions.

\noindent 
\begin{tabular}{|l|c|ccc|ccc|ccc|ccc|c|}
\hline
R$^{3+}$ & Ce & Pr & Nd & Pm & Sm & Eu & Gd & Tb & Dy & Ho & Er & Tm & Yb & 
\\ \hline
& f$^1$ & f$^2$ & f$^3$ & f$^4$ & f$^5$ & f$^6$ & f$^7$ & f$^8$ & f$^9$ & f$%
^{10}$ & f$^{11}$ & f$^{12}$ & f$^{13}$ &  \\ \hline
& F & H & I & I & H & F & S & F & H & I & I & H & F &  \\ 
L & 3 & 5 & 6 & 6 & 5 & 3 & 0 & 3 & 5 & 6 & 6 & 5 & 3 &  \\ \hline
\end{tabular}

Table 5.1. Electrical resistivity $\rho $ ($\mu \Omega $ cm) at room
temperatures, the coefficient at the $T^2$-term $A$ ($10^{-6}$ $\mu \Omega $
cm/K$^2$), the coefficient at linear specific heat $\gamma $ (mJ/mol K$^2$),
and the ratios $A/\gamma ^2$ and $\rho /\gamma $ in 3d, 4d and 5d rows
[406,412,413].

\noindent 
\begin{tabular}{|l|ccccccccc|}
\hline
3d & Sc & Ti & V & Cr & Mn & Fe & Co & Ni & Cu \\ 
4d & Y & Zr & Nb & Mo & Tc & Ru & Rh & Pd & Ag \\ 
5d & La & Nb & Ta & W & Re & Os & Ir & Pt & Au \\ \hline
& 52 & 48 & 20 & 12 & 144 & 10.2 & 6 & 7.4 & 1.7 \\ 
$\rho $ & 67 & 43 & 15 & 5. & 20 & 7.5 & 5 & 10.8 & 1.6 \\ 
& 62 & 34 & 13 & 5. & 20 & 10.6 & 5.3 & 10.8 & 2.3 \\ \hline
& - & - & 3.3$^a$ & - & - & - & - & - & - \\ 
$A$ & 100$^b$ & 80 & 32 & 2 & - & 2.7 &  & 33 & - \\ 
& - & 15 & 70$^c$ & 1$^c$ & 4-5$^c$ & 0.2-0.5 & 0.9 & 12-19 & - \\ \hline
& 9 & 4.5 & 8.5 & 2 & 17 & 5 & 5.1 & 7 & 1 \\ 
$\gamma $ & 8 & 2.8 & 7.8 & 2 & - & 2.95 & 4.65 & 9.57 &  \\ 
& - & 2.16 & 6.3 & 0.84 & 2.3 & 2.35 & 3.14 & 6.41 &  \\ \hline
& - & - & 0.045 & - & - & - & - & - & - \\ 
$A/\gamma ^2$ & 1.6 & 10.2 & 0.53 & 0.5 & - & 0.31 & - & 0.36 & - \\ 
&  & 3.2 & 1.78 & 1.41 & 0.95 & 0.36 & 0.09 & 0.38 & - \\ \hline
& 6 & 11 & 2.2 & 6 & 8.5 & 2 & 1.2 & 1.05 & 1 \\ 
$\rho /\gamma $ & 8 & 15.5 & 1.9 & 2.75 & - & 2.6 & 1.1 & 1.13 & - \\ 
& - & 15.75 & 2.15 & 6.48 & 8.68 & 4.5 & 1.7 & 1.68 & - \\ \hline
\end{tabular}

$^a \rho /\rho$ (4.2K) $= 1400$ $^b \rho /\rho (4.2K) = 700$ $^c$After
excluding dimensional effect, $A < 0.05 10^{-6}$ $\mu \Omega $cm/K$^2$

Table 6.1. Electronic specific heat and magnetic characteristics
(paramagnetic suceptibility, Neel or Curie (in brackets) point, paramagnetic
Curie temperature and the ground state moment) of some anomalous rare-earth
and actinide compounds (heavy-fermion, Kondo lattice and intermediate
valence systems)

\noindent 
\begin{tabular}{|l|cccccc|}
& $\gamma $ & $\chi (0)$ & $T_N(T_C)$ & $M_s$ & $-\theta $ & Ref. \\ \hline
& mJ/mol K$^2$ & 10$^{-3}\frac{\mathrm{emu}}{\mathrm{mol}}$ & K & $\mu _B$ & 
K &  \\ 
CeAl$_3$ & 1620 & 36 & 1.2? & 0.3? & 46 & 507-511 \\ 
CeAl$_2$ & 135 & 44 & 3.8 & 0.89$^a$ & 32 & 512 \\ 
CeCu$_2$Si$_2$ & 1100 & 7 & 0.8 &  & 164 & 507,513,514 \\ 
CeCu$_6$ & 1450 & 27 &  &  & 45 & 507,515 \\ 
CeCu$_2$ & 90 &  & 3.5 &  &  & 516 \\ 
CeCu$_5$ & 100 &  & 3.9 &  &  & 517 \\ 
CeCu$_2$Ge$_2$ & 100 &  & 4.1 & 1 & 18 & 518 \\ 
CeIn$_3$ & 260 & 11 & 10 & 0.6 & 50 & 512 \\ 
CeInCu$_2$ & 1200 & 20 & 1.6 & 0.1 & 20 & 519 \\ 
CePb$_3$ & 1000$^b$ &  & 1.1 & 0.1 &  & 520 \\ 
CeCu$_4$Al & 2000($<$1K) &  &  &  & 25 & 520 \\ 
CeCu$_4$Ga & 1900 & 20 &  &  & 26 & 520 \\ 
CeCu$_3$Al$_2$ & 540(1.6K) & 29 &  &  & 40 & 520 \\ 
CeCu$_3$Ga$_2$ & 730(1.5K) &  &  &  &  & 520 \\ 
CeAl$_2$Ga$_2$ & 80 &  & 8.5 & 1.2 & 18 & 521 \\ 
CeInPt$_4$ & 2500 &  &  &  & 225 & 520 \\ 
CePtSi & 800 &  &  &  & 47 & 520 \\ 
CePtSi$_2$ & 1700 (1.2K) &  &  &  & 17 & 520 \\ 
CePt$_2$Sn$_2$ & 3500$^c$ &  & 0.88 &  & 25 & 520 \\ 
CePtIn & 700 &  &  &  & 73 & 520 \\ 
CePdIn & 330 &  & 1.8 &  &  & 520 \\ 
CePdSn &  &  & 7.5 &  & 68 & 522 \\ 
CePdSb &  &  & (17) & 1.2 & -10 & 520 \\ 
CeRu$_2$Si$_2$ & 350 & 17 &  &  & 54 & 520 \\ 
CeZn$_{11}$ & 2500 &  & 2 &  &  & 523 \\ 
CeRh$_3$B$_2$ & 16 &  & (115) & 0.37 & 370 & 284 \\ 
UPt$_3$ & 450 & 7 & 5.0 & 0.02 & 200 & 524 \\ 
URu$_2$Si$_2$ & 180 &  & 17.5 & 0.03 & 65 & 525,526 \\ 
UBe$_{13}$ & 1100 & 15 &  &  & 53-70 & 507,527,528 \\ 
UZn$_{8.5}$ & 535 & 12.5 & 9.7 & 0.8 & 105 & 507,526,529 \\ 
UCd$_{11}$ & 800 & 45 & 5.0 &  & 23 & 507,526 \\ 
UAgCu$_4$ & 310 &  & 18 &  &  & 520,530 \\ 
UNiAl & 164 &  & 19 & 0.8 &  & 531 \\ 
UNi$_2$Al$_3$ &  &  & 5.2 & 0.24 &  & 532 \\ 
UPd$_2$Al$_3$ & 150 &  & 14 & 0.85 & 47 & 533 \\ 
UPdIn & 280 &  & 20$^d$ &  &  & 534,535 \\ 
U$_2$PtSi$_3$ & 400 &  & (8) &  &  & 536 \\ 
YbCuAl & 260 & 25.5 &  &  & 34 & 512 \\ 
YbAs$^f$ & 270 &  & 0.6 & 0.82 &  & 537,538 \\ 
YbP &  &  & 0.4 & 0.79 &  & 537 \\ 
YbSb &  &  & 0.3 & 0.63 &  & 537 \\ 
YbSi &  &  & 1.5 & 0.2 &  & 539 \\ 
YbPdCu$_4$ & 200 &  & 0.8 &  &  & 520 \\ 
YbNiSn & 300 &  & (5.5) & 0.4$^e$ & 65 & 540 \\ 
YbNiSb$^f$ & 175 &  & 0.8 &  & 13 & 541 \\ 
Sm$_4$As$_3$ &  &  & (160) & 2.5? &  & 542 \\ 
Sm$_4$Sb$_3$ &  &  & (16) & 1.5? &  & 542 \\ 
TmS &  &  & 8.9 & 4.0 &  & 512 \\ 
TmSe &  &  & 3 & 1.7 &  & 512 \\ 
PrCu$_2$Si$_2$ & 225 &  & 21 & 2.5 &  & 543 \\ 
NpBe & 900 & 56 & 3.4 &  & 42 & 507 \\ 
NpAl$_2$ & 193 &  & (57) &  &  & 544 \\ \hline
\end{tabular}

$^a$Maximum value in modulated-moment structure.

$^b\gamma \simeq $ 200 above the Neel temperature

$^c$Above the Neel temperature

$^d$Small canted ferromagnetic moment 0.3$\mu _B$ below 7K.

$^e$Canted ferromagnetic moment.

$^f$Low carrier concentration.

Table 6.1. Electronic specific heat and magnetic characteristics
(paramagnetic suceptibility, Neel or Curie (in brackets) point, paramagnetic
Curie temperature and the ground state moment) of some anomalous rare-earth
and actinide compounds (heavy-fermion, Kondo lattice and intermediate
valence systems)

\noindent 
\begin{tabular}{|l|cccccc|}
& $\gamma$ & $\chi (0)$ & $T_N(T_C)$ & $M_s$ & $-\theta $ & Ref. \\ 
hline & mJ/mol K$^2$ & 10$^{-3}$emu/mol & K & $\mu $ K &  &  \\ 
CeAl$_3$ & 1620 & 36 & 1.2? & 0.3? & 46 & 507-511 \\ 
CeAl$_2$ & 135 & 44 & 3.8 & 0.89$^a$ & 32 & 512 \\ 
CeCu$_2$Si$_2$ & 1100 & 7 & 0.8 &  & 164 & 507,513,514 \\ 
CeCu$_6$ & 1450 & 27 &  &  & 45 & 507,515 \\ 
CeCu$_2$ & 90 &  & 3.5 &  &  & 516 \\ 
CeCu$_5$ & 100 &  & 3.9 &  &  & 517 \\ 
CeCu$_2$Ge$_2$ & 100 &  & 4.1 & 1 & 18 & 518 \\ 
CeIn$_3$ & 260 & 11 & 10 & 0.6 & 50 & 512 \\ 
CeInCu$_2$ & 1200 & 20 & 1.6 & 0.1 & 20 & 519 \\ 
CePb$_3$ & 1000$^b$ &  & 1.1 & 0.1 &  & 520 \\ 
CeCu$_4$Al & 2000($<$1K) &  &  &  & 25 & 520 \\ 
CeCu$_4$Ga & 1900 & 20 &  &  & 26 & 520 \\ 
CeCu$_3$Al$_2$ & 540(1.6K) & 29 &  &  & 40 & 520 \\ 
CeCu$_3$Ga$_2$ & 730(1.5K) &  &  &  &  & 520 \\ 
CeAl$_2$Ga$_2$ & 80 &  & 8.5 & 1.2 & 18 & 521 \\ 
CeInPt$_4$ & 2500 &  &  &  & 225 & 520 \\ 
CePtSi & 800 &  &  &  & 47 & 520 \\ 
CePtSi$_2$ & 1700 (1.2K) &  &  &  & 17 & 520 \\ 
CePt$_2$Sn$_2$ & 3500$^c$ &  & 0.88 &  & 25 & 520 \\ 
CePtIn & 700 &  &  &  & 73 & 520 \\ 
CePdIn & 330 &  & 1.8 &  &  & 520 \\ 
CePdSn &  &  & 7.5 &  & 68 & 522 \\ 
CePdSb &  &  & (17) & 1.2 & -10 & 520 \\ 
CeRu$_2$Si$_2$ & 350 & 17 &  &  & 54 & 520 \\ 
CeZn$_{11}$ & 2500 &  & 2 &  &  & 523 \\ 
CeRh$_3$B$_2$ & 16 &  & (115) & 0.37 & 370 & 284 \\ 
UPt$_3$ & 450 & 7 & 5.0 & 0.02 & 200 & 524 \\ 
URu$_2$Si$_2$ & 180 &  & 17.5 & 0.03 & 65 & 525,526 \\ 
UBe$_{13}$ & 1100 & 15 &  &  & 53-70 & 507,527,528 \\ 
UZn$_{8.5}$ & 535 & 12.5 & 9.7 & 0.8 & 105 & 507,526,529 \\ 
UCd$_{11}$ & 800 & 45 & 5.0 &  & 23 & 507,526 \\ 
UAgCu$_4$ & 310 &  & 18 &  &  & 520,530 \\ 
UNiAl & 164 &  & 19 & 0.8 &  & 531 \\ 
UNi$_2$Al$_3$ &  &  & 5.2 & 0.24 &  & 532 \\ 
UPd$_2$Al$_3$ & 150 &  & 14 & 0.85 & 47 & 533 \\ 
UPdIn & 280 &  & 20$^d$ &  &  & 534,535 \\ 
U$_2$PtSi$_3$ & 400 &  & (8) &  &  & 536 \\ 
YbCuAl & 260 & 25.5 &  &  & 34 & 512 \\ 
YbAs$^f$ & 270 &  & 0.6 & 0.82 &  & 537,538 \\ 
YbP &  &  & 0.4 & 0.79 &  & 537 \\ 
YbSb &  &  & 0.3 & 0.63 &  & 537 \\ 
YbSi &  &  & 1.5 & 0.2 &  & 539 \\ 
YbPdCu$_4$ & 200 &  & 0.8 &  &  & 520 \\ 
YbNiSn & 300 &  & (5.5) & 0.4$^e$ & 65 & 540 \\ 
YbNiSb$^f$ & 175 &  & 0.8 &  & 13 & 541 \\ 
Sm$_4$As$_3$ &  &  & (160) & 2.5emu/mol &  & 542 \\ 
Sm$_4$Sb$_3$ &  &  & (16) & 1.5emu/mol &  & 542 \\ 
TmS &  &  & 8.9 & 4.0 &  & 512 \\ 
TmSe &  &  & 3 & 1.7 &  & 512 \\ 
PrCu$_2$Si$_2$ & 225$^c$ &  & 21 & 2.5 &  & 543 \\ 
NpBe$_{13}$ & 900 & 56 & 3.4 &  & 42 & 507 \\ 
NpAl$_2$ & 193 &  & (57) &  &  & 544 \\ \hline
\end{tabular}

$^a$Maximum value in modulated-moment structure.

$^b$$\gamma $ 200 above the Neel temperature

$^c$Above the Neel temperature

$^d$Small canted ferromagnetic moment $0.3\mu _{\mathrm{B}}$ below 7K.

$^e$Canted ferromagnetic moment.

$^f$Low carrier concentration.


\begin{thebibliography}{800}
\addcontentsline{toc}{chapter}{BIBLIOGRAPHY}

\bibitem{1}
A.Sommerfeld and H.Bethe, Electronentheorie der Metalle, Handb.Phys,
Vol.XXIV, ed. H.Geiger and K.Scheel, Springer, Berlin, 1933

\bibitem{2}
A.H.Wilson, Theory of Metals, Cambridge, 1953

\bibitem{3}
F.Seitz, Modern Theory of Solids, McGraw Hill, New York, 1940

\bibitem{4}
R.E.Peierls, Quantum Theory of Solids, Oxford, Clarendon Press, 1955

\bibitem{5}
C.G.Kittel, Introduction to Solid State Physics, Wiley, New York, 1953

\bibitem{6}
J.M.Ziman, Principles of the Theory of Solids, London, 1964

\bibitem{7}
J.M.Ziman, Electrons and Phonons, Clarendon Press, Oxford, 1960

\bibitem{8}
F.J.Blatt, Physics of Electronic Conduction in Solids, McGraw Hill, New
York, 1968

\bibitem{9}
W.A.Harrison, Solid State Theory, McGraw Hill, New York, 1970

\bibitem{10}
I.M.Lifshitz, M.Ya.Azbel and M.I.Kaganov, Electron Theory of Metals, New
York, Consultants Bureau, 1973

\bibitem{11}
A.P.Cracknell and K.C.Wong, The Fermi Surface, Clarendon Press, Oxford,
1973

\bibitem{12}
N.W.Ashcroft and N.D.Mermin, Solid State Physics, Holt, Rienart and
Winston, 1976

\bibitem{13}
W.A.Harrison, Electronic Structure and the Properties of Solids,
Freeman, San Francisko, 1980

\bibitem{14}
A.A.Abrikosov, Theory of Metals, Moscow, Nauka, 1987

\bibitem{15}
B.Coqblin, The Electronic Structure of Rare Earth Metals and Alloys, Academic
Press, 1977

\bibitem{16}
K.N.R.Taylor and M.I.Darby, Physics of Rare Earth Solids, Chapman and Hall,
London, 1972

\bibitem{17}
S.A.Nikitin, Magnetic Properties of Rare Earth Metals and their Alloys,
Moscow State University, 1989; K.P.Belov, M.A.Belyanchikova, R.Z.Levitin and
S.A.Nikitin, Rare Earth Ferro- and Antiferromagnets, Moscow, Nauka, 1965

\bibitem{18}
D.C.Griffin, K.Andrew and R.D.Cowan, Phys.Rev.177, 62 (1969)

\bibitem{19}
A.J.Freeman, In: Magnetic Properties of Rare Earth Metals, Ed.R.J.Elliot,
Plenum Press, 1972, p.245

\bibitem{20}
I.I.Sobelman, Introduction to the Theory of Atomic Spectra, Pergamon Press,
Oxford, 1973

\bibitem{21}
P.Soederlind, O.Eriksson, J.M.Wills and B.Johansson, Phys.Rev.B48, 9212
(1993)

\bibitem{22}
U.Benedict, W.A.Grosshans and W.B.Holtzapfel, Physica B144, 14 (1986)

\bibitem{23}
U.Benedict, J.R.Peterson, R.G.Haire and C.Dufour, J.Phys.F14, L43 (1984)

\bibitem{24}
V.L.Moruzzi, J.F.Janak and A.R.Williams, Calculated Electronic Properties of
Metals, Pergamon Press, New York, 1978

\bibitem{25}
N.F.Mott, Metal-Insulator Transitions, London, Taylor and Francis, 1974

\bibitem{26}
T.Moriya, Spin Fluctuations in Itinerant Electron Magnetism. Springer,
Berlin, 1985

\bibitem{27}
A.Abrikosov, L.Gorkov and I.Dzyaloshinsky, Methods of Quantum Field Theory in
Statistical Physics, Prentice Hall, New York, 1963

\bibitem{28}
J.Hubbard, Proc.Roy.Soc. A276, 238 (1963)

\bibitem{29}
J.Hubbard, Proc.Roy.Soc. A277, 237 (1963)

\bibitem{30}
J.Hubbard, Proc.Roy.Soc. A281, 401 (1964)

\bibitem{31}
J.Hubbard, Proc.Roy.Soc. A285, 542 (1965)

\bibitem{32}
B.R.Judd, Operator Techniques in Atomic Spectroscopy, Mc Graw Hill, New York,
1963; Second Quantization and Atomic Spectroscopy, John Hopkins University
Press, Baltimore, 1967.

\bibitem{33}
S.Fraga, K.Saxena and J.Karwowski, Handbook of Atomic Data, Elsevier,
Amsterdam, 1976

\bibitem{34}
C.E.Moore, Atomic Energy Levels, US National Bureau of Standards Circular No
1463, vol.1-3, 1958

\bibitem{35}
R.Dagis and D.Segzda, Fiz.Tverd.Tela 36, 705 (1994)

\bibitem{36}
S.P.Kowalczuk, R.A.Pollak, F.R.McFeely, L.Ley and D.A.Shirley, Phys.Rev.B8,
2387 (1973)

\bibitem{37}
P.A.Bennett, J.C.Fuggle and F.U.Hillebrecht, Phys.Rev.B27, 2194 (1983)

\bibitem{38}
E.Antonides et al, Phys.Rev.B15, 1669 (1977)

\bibitem{39}
Yu.P.Irkhin, Usp.Fiz.Nauk 154, 321 (1988) [Sov.Phys.Usp.31, 163 (1988)].

\bibitem{40}
R.J.H.Kappert, H.R.Borsje and J.C.Fuggle, J.Magn.Magn.Mat.100, 363 (1991)

\bibitem{41}
S.A.Altshuler and B.M.Kozyrev, Electron paramagnetic resonance of
transition group compounds (in Russian), Moscow, Nauka, 1972

\bibitem{42}
K.I.Kugel and D.I.Khomsky, Usp.Fiz.Nauk 136, 621 (1982)

\bibitem{43}
R.S.Knox and A.Gold, Symmetry in the Solid State, Benjamin, New York,
1964

\bibitem{44}
B.H.Brandow, Adv.Phys.26, 651 (1977)

\bibitem{45}
S.Huefner, Adv.Phys.43, 183 (1994)

\bibitem{46}
B.Koiller and L.M.Falicov, J.Phys.C7, 299 (1974)

\bibitem{47}
L.H.Tjeng, C.T.Chen, G.Ghijsen, R.Rudolf and F.Sette, Phys.Rev.Lett.67,
501 (1991)

\bibitem{48}
A.Fujimori et al, Phys.Rev.B42, 7580 (1990)

\bibitem{49}
J.van Elp, P.H.Potze, H.Eskes, R.Berger and G.A.Sawatzky, Phys.Rev.B44,
1530 (1991)

\bibitem{50}
J.van Elp, H.Eskes, P.Kuiper and G.A.Sawatzky, Phys.Rev.B44, 5927 (1991)

\bibitem{51}
A.Fujimori, F.Minami and S.Sugano, Phys.Rev.B29, 5225 (1984); A.Fujimori
and F.Minami, Phys.Rev.B30, 957 (1984)

\bibitem{52}
J.M.Ziman, The Calculation of Bloch Functions, In: Solid State Physics,
ed.H.Ehrenreich et al., vol.26, Academic Press, 1971.

\bibitem{53}
J.O.Dimmock, In: Solid State Physics, ed.H.Ehrenreich et al., vol.26,
Academic Press, 1970, p.103

\bibitem{54}
T.L.Loucks, APW-Method, Benjamin, Menlo Park, California, 1967

\bibitem{55}
W.Heine, M.Cohen and D.Weaire, Solid State Physics, ed.H.Ehrenreich et
al., vol.24, Academic Press, 1970.

\bibitem{56}
A.R.Mackintosh and O.K.Andersen, The Electronic Structure of Transition
Metals, in: Electrons at the Fermi Surface, Ed.M.Springford, Cambridge
University Press, 1980, p.149

\bibitem{57}
V.V.Nemoshkalenko and V.N.Antonov, Methods of Calculation Physics in the
Solid State Theory. Band Theory of Metals (in Russian), Kiev, Naukova Dumka,
1985

\bibitem{58}
J.C.Slater and G.F.Koster, Phys.Rev.94, 1498 (1954)

\bibitem{59}
P.Hohenberg and W.Kohn, Phys.Rev.B136, 864 (1964)

\bibitem{60}
J.A.Moriarty, Phys.Rev.B1, 1363 (1970); B5, 2066 (1972)

\bibitem{61}
M.D.Girardeau, J.Math.Phys., 12, 165 (1971)

\bibitem{62}
Z.A.Gursky and B.A.Gursky, Fiz.Metallov Metalloved.50, 928 (1980)

\bibitem{63}
G.B.Bachelet, D.R.Hamann and M.Schluter, Phys.Rev.B26, 4199 (1982)

\bibitem{64}
K.-M.Ho, S.G.Louie, J.R.Chelikowsky and M.L.Cohen, Phys.Rev.B15, 1755
(1977)

\bibitem{65}
B.A.Greenberg, M.I.Katsnelson, V.G.Koreshkov et al., phys.stat.sol.(b)
158, 441 (1990); A.S.Ivanov, M.I.Katsnelson, A.G.Mikhin et al, Phil.Mag.69,
1183 (1994)

\bibitem{66}
T.L.Loucks, Phys.Rev.139, A1181 (1965)

\bibitem{67}
O.K.Andersen, Phys.Rev.B12, 864 (1975)

\bibitem{68}
E.Wimmer, H.Krakauer, M.Weinert and A.J.Freeman, Phys.Rev.B24, 864 (1981)

\bibitem{69}
J.C.Slater, The Self-Consistent Field for Molecules and Solids, Mc Graw
Hill, New York, 1974

\bibitem{70}
W.Kohn and L.J.Sham, Phys.Rev.A140, 1133 (1965); L.J.Sham and W.Kohn,
Phys.Rev. 145, 561 (1966)

\bibitem{71}
O.Gunnarsson and B.I.Lundqvist, Phys.Rev.B13, 4274 (1976)

\bibitem{72}
R.O.Jones and O.Gunnarsson, Rev.Mod.Phys, 61, 689 (1989)

\bibitem{73}
L.Hedin and B.I.Lundqvist, J.Phys.C4, 2064 (1971)

\bibitem{74}
U.von Barth and L.Hedin, J.Phys.C5, 1629 (1972)

\bibitem{75}
T.Ziegler, A.Rauk and E.J.Baerends, Theor.Chim.Acta, 43, 261 (1977)

\bibitem{76}
U.von Barth, Phys.Rev.A20, 1963 (1979)

\bibitem{77}
A.Svane and O.Gunnarsson, Phys.Rev.B37, 9919 (1988); Phys.Rev.Lett.65,
1148 (1990); Europhys.Lett.7, 171 (1988); A.Svane, Phys.Rev.Lett., 68, 1900
(1992)

\bibitem{78}
D.A.Papaconstantopoulos, Handbook of Bandstructure of Elemental Solids,
New York, Plenum Press, 1986

\bibitem{79}
R.J.Jelitto, J.Phys.Chem.Sol., 30, 609 (1969)

\bibitem{80}
R.H.Swendsen and H.Callen, Phys.Rev.B6, 2860 (1972)

\bibitem{81}
S.V.Vonsovsky, M.I.Katsnelson and A.V.Trefilov, Fiz.Metallov
Metalloved.76, N3, 3 (1993)

\bibitem{82}
D.Koelling, F.M.Mueller, A.J.Arko and J.B.Ketterson, Phys.Rev.B10, 4889
(1974)

\bibitem{83}
J.Friedel, Canad.J.Phys.34, 1190 (1956)

\bibitem{84}
W.A.Harrison, Phys.Rev.B28, 550 (1983)

\bibitem{85}
N.F.Mott and K.W.H.Stevens, Phil.Mag.2, 1364 (1957)

\bibitem{86}
W.M.Lomer and W.M.Marshall, Phil.Mag.3, 185 (1958)

\bibitem{87}
E.O.Wollan, Phys.Rev.117, 387 (1960)

\bibitem{88}
J.B.Goodenough, Phys.Rev.120, 67 (1960)

\bibitem{89}
L.F.Matheiss, Phys.Rev.134, A970 (1964)

\bibitem{90}
F.J.Himpsel and D.Eastman, Phys.Rev.B18, 5236 (1978)

\bibitem{91}
O.K.Andersen, Phys.Rev.B2, 883 (1970)

\bibitem{92}
N.E.Christensen, Phys.Rev.B14, 3446 (1976)

\bibitem{93}
F.M.Mueller, A.J.Freeman, J.O.Dimmock and A.M.Furduya, Phys.Rev.B1, 4617
(1970)

\bibitem{94}
S.G.Louie, Phys.Rev.Lett.40, 1525 (1978)

\bibitem{95}
N.I.Mazin, E.G.Maksimov, S.I.Rashkeev and Yu.A.Uspensky, Proc.
P.N.Lebedev Physical Institute of Academy of Sciences, vol.190, Moscow,
Nauka, 1988, p.3

\bibitem{96}
K.D.Sevier, Low Energy Electron Spectrometry, New York, Wiley, 1972

\bibitem{97}
R.F.Davis, R.S.Williams, S.D.Kevan, P.S.Wehner and D.A.Shirley,
Phys.Rev.B31, 1997 (1985)

\bibitem{98}
P.S.Wehner, R.S.Williams, S.D.Kevan, D.Denley and D.A.Shirley,
Phys.Rev.B19, 6164 (1979)

\bibitem{99}
K.A.Mills, R.F.Davis, S.D.Kevan, G.Thornton and D.A.Shirley,
Phys.Rev.B22, 581 (1980)

\bibitem{100}
V.V.Nemoshkalenko and V.N.Antonov, Methods of Calculation Physics in the
Solid State Theory. Band Theory of Metals (in Russian), Kiev, Naukova Dumka,
1985

\bibitem{101}
J.F.van der Veen, F.J.Himpsel and D.E.Eastman, Phys.Rev.B22, 4226 (1980)

\bibitem{102}
W.Speier et al., Phys.Rev.B30, 6921 (1984)

\bibitem{103}
H.Hoechst et al, Z.Phys.B42, 99 (1981)

\bibitem{104}
D.E.Eastman, F.J.Himpsel and J.A.Knapp, Phys.Rev.Lett.44, 95 (1980)

\bibitem{105}
E.Kisker, K.Schroeder, W.Gudat and M.Campagna, Phys.Rev.B31, 329 (1985)

\bibitem{106}
R.Clauberg, E.M.Haines and P.Feder, J.Magn.Magn.Mater. 54-57, 622 (1986)

\bibitem{107}
E.Kisker, J.Magn.Magn.Mater.45, 23 (1984)

\bibitem{108}
D.E.Eastman, F.J.Himpsel and J.A.Knapp, Phys.Rev.Lett.40, 1514 (1978);
F.J.Himpsel, J.A.Knapp and D.E.Eastman, Phys.Rev.B19, 2919 (1979)

\bibitem{109}
K.Kakizaki, J.Fujii, K.Shimada et al, Phys.Rev.Lett.72, 2781 (1994)

\bibitem{110}
M.Pessa, P.Heinmann and H.Neddermeyer, Phys.Rev.B14, 3488 (1976)

\bibitem{111}
G.Bush, M.Campagna and H.C.Siegmann, Phys.Rev.B4, 746 (1971)

\bibitem{112}
P.Genoval, A.A.Manuel, E.Walker and M.Peter, J.Phys.:Cond.Mat. 3, 4201
(1991)

\bibitem{113}
C.M.Schneider, P.Schuster, M.S.Hammond and J.Kirschner,
Europhys.Lett.16, 689 (1991)

\bibitem{114}
P.Guletsky, Yu.A.Knyazev, M.M.Kirillova and L.M.Sandratsky
Fiz.Metallov Metalloved.67, 279 (1989)

\bibitem{115}
T.Jo and G.A.Sawatzky, Phys.Rev.B43, 8771 (1991)

\bibitem{116}
H.Martensson and P.O.Wilsson, Phys.Rev.B30, 3047 (1984)

\bibitem{117}
A.Liebsh, Phys.Rev.B23, 5203 (1981)

\bibitem{118}
D.R.Penn, Phys.Rev.Lett.42, 921 (1979)

\bibitem{119}
R.H.Victora and L.M.Falicov, Phys.Rev.Lett.55, 1140 (1985)

\bibitem{120}
E.Jensen and D.M.Wieliczka, Phys.Rev.B30, 7340 (1984)

\bibitem{121}
O.Gunnarson and K.Schoenhammer, Phys.Rev.B28, 4315 (1983)

\bibitem{122}
Yu.P.Irkhin, In: Electron Structure and Physical Propertties of Rare
Earths and Actinides, Ural Science Centre, Sverdlovsk, 1981, p.50

\bibitem{123}
J.T.Waber and A.C.Switendick, Proc.V-th Rare Earth Research Conf., Iowa
State University, 1965, v.B2, p.75.

\bibitem{124}
G.Murhopadhyay and C.K.Majumdar, J.Phys.C2, 924 (1969)

\bibitem{125}
W.E.Pickett, A.J.Freeman and D.D.Koelling, Phys.Rev.B23, 1266 (1981)

\bibitem{126}
G.Johansen and A.R.Mackintosh, Solid State Comm.8, 121 (1970)

\bibitem{127}
B.N.Harmon, J.de Phys.40, 65 (1979)

\bibitem{128}
M.M.Islam and D.J.Newman, J.Phys.F5, 939 (1975) Tm

\bibitem{129}
T.A.Matveeva and R.F.Egorov, Fiz.Metallov Metalloved.51, 717, 950 (1981)

\bibitem{130}
O.V.Farberovich, G.P.Nizhnikova, S.V.Vlasov and E.P.Domashevskaya,
phys.stat.sol(b) 121, 241 (1984)

\bibitem{131}
H.J.F.Jansen, A.J.Freeman and R.Monnier, Phys.Rev.B31, 4092 (1985)

\bibitem{132}
D.M.Bylander and L.Kleinman, Phys.Rev.B49, 1608 (1994); B50, 1363 (1994)

\bibitem{133}
J.F.Herbst, D.N.Lowy and R.E.Watson, Phys.Rev.B6, 1913 (1972)

\bibitem{134}
J.F.Herbst, R.E.Watson and J.W.Wilson, Phys.Rev.B13, 1439 (1976); B17,
3089 (1978)

\bibitem{135}
E.I.Zabolotsky, Yu.P.Irkhin and L.D.Finkelstein, Fiz.Tverd.Tela, 16,
1142 (1974)

\bibitem{136}
P.O.Heden, H.Logfren and S.B.M.Hagstrom, Phys.Rev.Lett.26, 432 (1971)

\bibitem{137}
C.Bounele, R.C.Karnatek and C.Jorgensen, Chem.Phys.Lett.14, 145 (1972)

\bibitem{138}
A.J.Freeman, D.D.Koelling, In: The Actinides: Electron Structure and
Related Properties, ed. A.J.Freeman and J.B.Darby, New York, Academic Press,
1974, p.51

\bibitem{139}
B.V.Karpenko, In: Electron Structure and Physical Propertties of Rare
Earths and Actinides, Ural Science Centre, Sverdlovsk, 1981, p.86

\bibitem{140}
S.C.Keeton, T.L.Loucks, Phys.Rev.146, 429 (1966)

\bibitem{141}
D.D.Koelling, A.J.Freeman, Solid State Comm.9, 1369 (1971);
Phys.Rev.B12, 5622 (1975)

\bibitem{142}
P.Weinberger, A.M.Boring and J.L.Smith, Phys.Rev.B31, 1964 (1985)

\bibitem{143}
O.Eriksson and J.M.Wills, Phys.Rev.B45, 3198 (1992)

\bibitem{144}
J.Kittel, Quantum Theory of Solids, Wiley, New York, 1963

\bibitem{145}
C.G.Lonzarich, Fermi Surface Studies in Ferromagnets, in: Electrons at
the Fermi Surface, Ed.M.Springford, Cambridge University Press, 1980, p.225

\bibitem{146}
A.P.Cracknell, Band Structures and Fermi Surfaces of Metallic Elements,
Landolt-Boernstein New Series, Band 13c, Springer, 1984

\bibitem{147}
F.S.Ham, Phys.Rev.128, 2524 (1962)

\bibitem{148}
D.R.Jennison, Phys.Rev.B16, 5147 (1977)

\bibitem{149}
C.B.So, K.Takegahara and S.Wang, J.Phys.F7, 105 (1977)

\bibitem{150}
M.J.G.Lee, Phys.Rev.178, 953 (1969)

\bibitem{151}
S.L.Altmann, A.P.Cracknell, Proc.Phys.Soc. 84, 761 (1964)

\bibitem{152}
P.Blaha and J.Callaway, Phys.Rev.B32, 7664 (1985)

\bibitem{153}
J.H.Condon, J.A.Marcus, Phys. Rev. 134, A 446 (1964)

\bibitem{154}
C.S.Fleming, T.L.Loucks, Phys.Rev. 173, 685 (1968)

\bibitem{155}
T.L.Loucks, Phys.Rev. 144, 504 (1966)

\bibitem{156}
T.L.Loucks, Phys.Rev.A139, 1181 (1965)

\bibitem{157}
R.C.Young, P.C.Yordan, D.W.Jones and V.J.Hems, J.Phys.F4, L84 (1984)

\bibitem{158}
G.S.Fleming, S.H.Liu and T.E.Loucks, Phys.Rev.Lett.21, 1524 (1968)

\bibitem{159}
O.Jepsen, Phys.Rev.B12, 2988 (1975)

\bibitem{160}
K.M.Welch and E.H.Hygh, Phys.Rev.B9, 1993 (1974)

\bibitem{161}
S.L.Altmann and C.J.Bradley, Proc. Phys. Soc. 92, 764 (1967)

\bibitem{162}
A.J.Hughes and J.Callaway, Phys. Rev. 136, A 1390 (1964)

\bibitem{163}
T.L.Loucks, Phys. Rev. 159, 544 (1967)

\bibitem{164}
A.C.Thorsen and A.S.Joseph, Phys. Rev.131, 2078 (1963)

\bibitem{165}
P.M.Everett, Phys.Rev.B20, 1419 (1979)

\bibitem{166}
L.L.Boyer, D.A.Papaconstantopoulos and B.M.Klein, Phys.Rev.B15, 3685
(1977)

\bibitem{167}
D.G.Laurent, C.S.Wang and J.Gallaway, Phys.Rev.B17, 455 (1978)

\bibitem{168}
R.D.Parker and M.H.Halloran, Phys.Rev.B9, 4130 (1974)

\bibitem{169}
R.A.Phillips, Phys.Lett.A36, 361 (1971)

\bibitem{170}
A.C.Thorsen, T.G.Berlincourt, Phys. Rev. Lett. 7, 244 (1961)

\bibitem{171}
G.B.Scott, M.Springford, J.R.Stockton, Phys.Lett.A27, 655 (1968);
G.B.Scott and M.Springford, Proc.Roy.Soc.A320, 115 (1970) 172.

\bibitem{172}
L.F.Mattheiss, Phys. Rev. B1, 373 (1970)


\bibitem{173}
L.L.Boyer, D.A.Papaconstantopoulos and B.M.Klein, Phys.Rev.B15, 3685
(1977)

\bibitem{174}
M.Cardona and L.Ley, Photoemission in Solids (Berlin:Springer) 1978

\bibitem{175}
M.H.Halloran et al, Phys.Rev.B1, 366 (1970)

\bibitem{176}
T.L.Loucks, Phys.Rev.139, A1181 (1965)

\bibitem{177}
L.F.Mattheiss, Phys. Rev.139, A 1893 (1965)

\bibitem{178}
D.G.Laurent, J.Callaway, J.L.Fry and N.E.Brener, Phys.Rev.B23, 4877
(1981)

\bibitem{179}
L.F.Mattheiss and D.R.Hamann, Phys.Rev.B33, 823 (1986)

\bibitem{180}
S.Asano and J.Yamashita, J.Phys.Soc.Jpn, 23, 714 (1967)

\bibitem{181}
N.I.Kulikov and E.I.Kulatov, J.Phys.F12, 2292 (1982)

\bibitem{182}
K.Okumara and I.M.Tepleton, Proc.R.Soc.A287, 89 (1965)

\bibitem{183}
M.J.G.Lee, Phys.Rev.178, 953 (1969)

\bibitem{184}
J.B.Ketterson, D.D.Koelling, J.C.Shaw and Windmiller, Phys.Rev.B11,
1447 (1975)

\bibitem{185}
D.M.Sparlin and J.A.Marcus, Phys. Rev. 144, 484 (1966)

\bibitem{186}
J.A.Hoekstra and J.L.Stanford, Phys.Rev.B8, 1416 (1973)

\bibitem{187}
R.F.Girvan, A.V.Gold and R.A.Phillips, J.Phys.Chem.Sol.29, 1485 (1968)

\bibitem{188}
G.S.Fleming, S.H.Liu and T.E.Loucks, Phys.Rev.Lett.21, 1524 (1968)

\bibitem{189}
G.C.Fletcher, J. Phys.C2, 1440 (1969)

\bibitem{190}
L.F.Mattheiss, Phys.Rev.151, 450 (1966)

\bibitem{191}
A.C.Thorsen, A.S.Joseph, L.I.Valby, Phys.Rev.150, 523 (1966)

\bibitem{192}
J.H.Wood, Phys. Rev. 126, 517 (1962)

\bibitem{193}
R.A.Tawil and J.Callaway, Phys.Rev.B7, 4242 (1973)

\bibitem{194}
J.Callaway and C.S.Wang, Phys.Rev.B16, 2095 (1977)

\bibitem{195}
A.V.Gold, L.Hodges, P.T.Panousis and D.R.Stone, Int.J.Magn.2, 357 (1971)

\bibitem{196}
D.R.Baraff, Phys.Rev.B8, 3439 (1973)

\bibitem{197}
S.Wakoh, J.Yamashita, J. Phys Soc. Japan 38, 1151 (1970)

\bibitem{198}
C.M.Singal and T.P.Das, Phys.Rev.B16, 5068 (1977) TB

\bibitem{199}
J.R.Anderson and J.F.Shirber, J.Appl.Phys.52, 1630 (1980)

\bibitem{200}
N.I.Kulikov and E.I.Kulatov, J.Phys.F12, 2267 (1982)

\bibitem{201}
F.Batallan, S.Rosenman and C.Summers, Phys.Rev.B11, 545 (1975)

\bibitem{202}
I.Rosenman and F.Batallan, Phys.Rev.B5, 1340 (1972)

\bibitem{203}
J.R.Anderson, J.J.Hudak and D.R.Stone, AIR.Conf.Proc.5, 477 (1972)

\bibitem{204}
J.Callaway and C.S.Wang, Phys.Rev.B7, 1096 (1973); C.S.Wang and
J.Callaway, Phys.Rev.B9, 4897 (1974)

\bibitem{205}
R.Prasad, S.K.Joshi and S.Auluck, Phys.Rev.B16, 1765 (1977); S.Ahuja,
S.Auluck and B.Johansson, Physica Scripta 50, 573 (1994)

\bibitem{206}
J.R.Anderson, D.A.Papaconstantopoulos, L.L.Boyer and J.E.Schriber,
Phys.Rev.B20, 3172 (1979)

\bibitem{207}
E.I.Zornberg, Phys.Rev.B1, 244 (1970)

\bibitem{208}
D.S.Tsui, Phys.Rev.164, 669 (1967)

\bibitem{209}
G.N.Kamm and J.R.Anderson, Phys.Rev.B2, 2944 (1970)

\bibitem{210}
J.F.Cooke, H.L.Davies and R.F.Wood, Phys.Rev.Lett.25, 28 (1970)

\bibitem{211}
P.T.Coleridge and I.M.Templeton, J.Phys.F2, 643 (1972)

\bibitem{212}
P.T.Coleridge, Phys. Lett. 22, 367 (1966);. J.Low Temp.Phys.1, 577
(1969)

\bibitem{213}
E.S.Alekseyev, V.A.Ventsel, O.A.Voronov, A.I.Likhter and
M.V.Magnitskaya, Zh.Eksp.Teor.Fiz.76, 215 (1979) [Sov.Phys.JETP 49, 110
(1979)]

\bibitem{214}
G.N.Kamm, Phys.Rev.B2, 2944 (1970)

\bibitem{215}
O.K.Andersen, A.R.Mackintosh, Solid State Commun. 6, 285 (1968)

\bibitem{216}
P.T.Coleridge, Phys. Lett. 15, 223 (1965)

\bibitem{217}
P.T.Coleridge, Proc. Roy. Soc. A295, 458 (1966)

\bibitem{218}
J.J.Grodski, A.E.Dixson, Solid State Commun.7, 735 (1969)

\bibitem{219}
S.Hornfeldt, Solid State Commun. 8, 673 (1970)

\bibitem{220}
N.V.Volkenshtein, V.A.Novoselov, V.E.Startsev and Yu.N.Tsiovkin,
Zh.Eksp.Teor.Fiz.60, 1733 (1971); Sov.Phys.JETP 33, 959 (1971)

\bibitem{221}
G.O.Arbman and S.Hornfeldt, J.Phys.F2, 1033 (1972)

\bibitem{222}
G.O.Arbman and S.Hornfeldt, J.Phys.F2, 1033 (1972) M.Cooper and
B.Williams, Philos.Mag.26, 1441 (1972)

\bibitem{223}
J.J.Vuillemin, Phys. Rev. 144, 396 (1966)

\bibitem{224}
J.J.Vuillemin, M.G.Priestley, Phys. Rev. Lett. 15, 307 (1965)

\bibitem{225}
D.H.Dye, S.A.Campbell, G.W.Grabtree, J.B.Ketterson, N.B.Sandesara and
J.J.Vuillemin, Pnys.Rev.B23, 462 (1981)

\bibitem{226}
J.B.Ketterson and L.R.Windmiller, Phys.Rev.B1, 4747 (1970); B2, 4813
(1970)

\bibitem{227}
D.H.Dye, J.B.Ketterson and G.W.Crabtree, J.Low Temp.Phys.30, 813 (1978)

\bibitem{228}
N.F.Christensen et al, J.Magn.Magn.Mat.76-77, 23 (1988)

\bibitem{229}
S.Tanuma, W.R.Datars, H.Doi, A.Dunsworth, Solid State Commun.8, 1107
(1970)

\bibitem{230}
R.C.Young, R.G.Jordan and D.W.Jones, Phys.Rev.Lett.31, 1473 (1973)

\bibitem{231}
S.C.Keeton and T.L.Loucks, Phys. Rev. 168, 672 (1968)

\bibitem{232}
S.C.Keeton and T.L.Loucks, Phys. Rev. 146, 428 (1966)

\bibitem{233}
A.C.Thorsen and A.S.Joseph, L.E.Valby, Phys. Rev. 162, 574 (1967)

\bibitem{234}
R.P.Gupta and T.L.Loucks, Phys. Rev. Lett. 22, 458 (1969)

\bibitem{235}
K.A.Gschneidner, Jr., Physical Properties and Interrelationships of
Metallic and Semimetallic Elements, In: Solid State Physics, ed.H.Ehrenreich
et al., vol.16, Academic Press, 1964, p.275

\bibitem{236}
B.Johansson and M.S.S.Brooks, Theory of Cohesion in Rare Earths and
Actinides, In: Handbook Phys.Chem. Rare Earths, ed.K.A.Gschneidner,Jr. et
al, vol.17, p.149

\bibitem{237}
Properties of Elements, ed.M.E.Dritz, Moscow, Metallurgiya, 1985

\bibitem{238}
Properties of Elements, ed.G.V.Samsonov, Moscow, Metallurgiya, 1976

\bibitem{239}
V.E.Zinov'yev, Thermophysical Properties of Metals at High
Temperatures, Moscow, Metallurgiya, 1989

\bibitem{240}
A.M.Oles, Phys.Rev.B23, 271 (1981)

\bibitem{241}
J.Friedel and C.M.Sayers, J.de Phys.38, 697 (1977)

\bibitem{242}
C.D.Gelatt, H.Ehrenreich and R.E.Watson, Phys.Rev.B15, 1613 (1977)

\bibitem{243}
V.L.Moruzzi, J.F.Janak and K.Schwarz, Phys.Rev.B37, 790 (1988)

\bibitem{244}
H.L.Skriver and M.M.Rosengaard, Phys.Rev.B45, 9410 (1992)

\bibitem{245}
H.L.Skriver, Phys.Rev.B31, 1909 (1985)

\bibitem{246}
V.A.Finkel, Structure of Rare Earth Metals (in Russian), Moscow,
Metallurgiya, 1978

\bibitem{247}
C.M.Hurd and J.E.A.Alderson, Solid State Comm., 12, 375 (1974)

\bibitem{248}
W.Hume-Rothery, The Metallic State, New York, University Press, 1931

\bibitem{249}
H.Jones, Proc.Phys.Soc., 49, 423 (1937)

\bibitem{250}
N.Engel, Trans.Amer.Soc.Metals, 57, 610 (1964)

\bibitem{251}
G.V.Samsonov, I.F.Pryadko and L.F.Pryadko, Electron Localization in
Solids (in Russian), Moscow, Nauka, 1976

\bibitem{252}
W.Hume-Rothery, Progr.Mater.Sci., 13, 229 (1968)

\bibitem{253}
D.G.Pettifor, J.Phys.C3, 367 (1970)

\bibitem{254}
J.W.Davenport, R.E.Watson and M.Weinert, Phys.Rev.B32, 4883 (1985)

\bibitem{255}
M.Sigalas, D.Papaconstantopoulos and N.C.Bacalis, Phys.Rev.B45, 5777
(1992)

\bibitem{256}
J.M.Wills and O.Eriksson, Phys.Rev.B45, 13879 (1992)

\bibitem{257}
G.Gladstone, M.A.Jensen and J.R.Schrieffer, In: Superconductivity,
ed.R.D.Parks, New York, 1969

\bibitem{258}
J.A.Moriarty, Phys.Rev.B49, 12431 (1994)

\bibitem{259}
M.M.Steiner, R.C.Albers and L.J.Sham, Phys.Rev.B45, 13272 (1992)

\bibitem{260}
G.Grimwall, J.Phys.Chem.Sol., 29, 1221 (1968)

\bibitem{261}
J.Osuzu, Thesis, Nagoya University, 1968

\bibitem{262}
M.Shimizu, Rep.Progr.Phys., 44, 329 (1981)

\bibitem{263}
P.Fulde and M.Loewenhaupt, Adv.Phys., 34, 589 (1985)

\bibitem{264}
L.M.Noskova, E.V.Rosenfeld and Yu.P.Irkhin, Fiz.Nizkikh Temperatur, 11,
469 (1985)

\bibitem{265}
S.V.Vonsovsky, Magnetism, New York, Wiley, 1974

\bibitem{266}
M.Dixon, F.E.Hoare, T.M.Holden and D.E.Moody, Proc.Roy.Soc.A285, 561
(1965)

\bibitem{267}
R.E.Pawell and E.E.Stansbury, J.Phys.Chem.Sol.26, 757 (1965)

\bibitem{268}
E.V.Rosenfeld, A.A.Siventsev, Yu.P.Irkhin and L.M.Noskova,
Fiz.Tverd.Tela 33, 202 (1991)

\bibitem{269}
S.A.Nemnonov, Fiz.Metallov Metalloved.19, 550 (1965)

\bibitem{270}
E.V.Galoshina, Usp.Fiz.Nauk 113, 105 (1974)

\bibitem{271}
R.Kubo and Y.Obata, J.Phys.Soc.Jpn 11, 547 (1956)

\bibitem{272}
M.Yasui, Physica B149, 139 (1988)

\bibitem{273}
S.Tikadzumi, Physics of Magnetism, Wiley, New York, 1964

\bibitem{274}
R.Huguenin and D.Baldock, Phys.Rev.Lett.Phys.Rev.Lett.16, 795 (1966);
R.Huguenin, G.P.Pells and D.N.Baldock, J.Phys.F1, 281 (1971)

\bibitem{275}
R.Peierls, Z.Phys.80, 763 (1933)

\bibitem{276}
F.Ducastelle and F.Cyrot-Lackmann, J.de Phys.Colloque 32, C1-534 (1971)

\bibitem{277}
M.Matsumoto, J.Staunton and P.Strange, J.Phys.:Cond.Mat.3, 1453 (1991)

\bibitem{278}
S.G.Das, Phys.Rev.B13,3978 (1976)

\bibitem{279}
A.H.MacDonald, K.L.Liu and S.H.Vosko, Phys.Rev.B16, 777 (1977)

\bibitem{280}
P.Rhodes and E.P.Wolfarth, Proc.Roy.Soc. A273, 247 (1963)

\bibitem{281}
H.Ido, J.Magn.Magn.Mat.54-57, 937 (1986); H.Ido, S.Yasuda, J.de
Phys.Colloque C8, 49, 141 (1988)

\bibitem{282}
M.J.Otto, R.A.M. van Woerden, P.J. van der Valk et al.
J.Phys.:Cond.Mat.1, 2341 (1989)

\bibitem{283}
P.J.Webster, J.Phys.Chem.Sol., 32, 1221 (1971)

\bibitem{284}
R.Vijayaraghavan, J.Magn.Magn.Mater.47-48, 561 (1985); M.Kasaya et al,
J.Magn.Magn.Mater.76-77, 347 (1988)

\bibitem{285}
E.S.Stoner, Proc.Roy.Soc.A165, 372 (1938); A169, 339 (1939)

\bibitem{286}
E.P.Wohlfarth, Phil.Mag.42, 374 (1951)

\bibitem{287}
D.M.Ceperley and B.J.Adler, Phys.Rev.Lett.45, 566 (1980)

\bibitem{288}
V.Yu.Irkhin, M.I.Katsnelson and A.V.Trefilov, J.Magn.Magn.Mat.117, 210
(1992); J.Phys.:Cond.Mat.5, 8763 (1993)

\bibitem{289}
G.H.O.Daalderop, M.H.Boon and F.M.Mueller, Phys.Rev.B41, 9803 (1990)

\bibitem{290}
T.Jarlborg and A.J.Freeman, Phys.Rev.B22, 2332 (1980); T.Jarlborg,
A.J.Freeman and D.D.Koelling, J.Magn.Magn.Mat.23, 291 (1981)

\bibitem{291}
Yu.P.Irkhin, E.V.Rosenfeld and A.A.Siventsev, Fiz.Tverd.Tela 33, 1646
(1991)

\bibitem{292}
O.Gunnarson, J.Phys.F6, 587 (1976)

\bibitem{293}
A.M.Oles and G.Stollhoff, J.Magn.Magn.Mat.54-57, 1045 (1986)

\bibitem{294}
E.V.Rosenfeld and Yu.P.Irkhin, Fiz.Metallov Metalloved. 57, 837 (1984);
Yu.P.Irkhin and E.V.Rosenfeld, Solid State Comm., 44, 1371 (1982)

\bibitem{295}
C.Herring, Magnetism, vol.4, New York, Academic Press, 1966.

\bibitem{296}
T.Moriya and A.Kawabata, J.Phys.Soc.Jpn, 34, 639 (1973); 35, 669 (1973)

\bibitem{297}
I.E.Dzyaloshinskii and P.S.Kondratenko, Sov.Phys.JETP 43, 1036 (1976)

\bibitem{298}
J.A.Hertz and M.A.Klenin, Phys.Rev.B10, 1084 (1974); Physica 91B, 49
(1977)

\bibitem{299}
K.K.Murata and S.Doniach, Phys.Rev.Lett.29, 285 (1982)

\bibitem{300}
J.Hubbard, Phys.Rev.B19, 2626 (1979); B20, 4584 (1979)

\bibitem{301}
H.Hasegawa, J.Phys.Soc.Jpn 49, 178; 963 (1980)

\bibitem{302}
E.A.Turov and V.I.Grebennikov, Physica B149, 150 (1988)

\bibitem{303}
A.A.Siventsev and Yu.P.Irkhin, Fiz.Tverd.Tela 35, 1965 (1993)

\bibitem{304}
P.Mohn and E.P.Wohlfarth, J.Phys.F17, 2421 (1987); E.P.Wohlfarth and
P.Mohn, Physica B149, 145 (1988)

\bibitem{305}
G.G.Lonzarich and L.Taillefer, J.Phys.C18, 4339 (1985)

\bibitem{306}
R.A.de Groot, F.M.Mueller, P.G.Mueller, P.G. van Engen and
K.H.J.Bushow, Phys.Rev.Lett.50, 2024 (1983)

\bibitem{307}
R.A.de Groot, F.M.Mueller, P.G. van Engen and K.H.J.Bushow,
J.Appl.Phys.55(6), 2151 (1984)

\bibitem{308}
R.A.de Groot and K.H.J.Buschow, J.Magn.Magn.Mat.54-57, 1377 (1986)

\bibitem{309}
J.Kuebler, Physica B+C 127, 257 (1984)

\bibitem{310}
R.A.de Groot, A.M.von der Kraan and K.H.J.Buschow, J.Magn.Magn.Mat.61,
330 (1986)

\bibitem{311}
J.Kuebler, J.R.Williams and C.B.Sommers, Phys.Rev.B28, 1745 (1983)

\bibitem{312}
S.Fujii, S.Sugimura, S.Ishida and S.Asano, J.Phys.:Cond.Mat. 2, 8583
(1990)

\bibitem{313}
K.Schwarz, J.Phys.F16, L211 (1986)

\bibitem{314}
E.Kulatov and I.Mazin, J.Phys.:Cond.Mat. 2, 343.

\bibitem{315}
R.C.Albers, A.M.Boring, G.H.O.Daalderop and F.M.Mueller, Phys.Rev.B36,
3661 (1987)

\bibitem{316}
S.V.Halilov and E.T.Kulatov, J.Phys.: Cond.Mat.3, 6363 (1991);
Zh.Eksp.Theor.Fiz.98, 1778 (1989)

\bibitem{317}
A.Yanase, K.Siratori, J.Phys.Soc.Jpn 53, 312 (1984) (1988)

\bibitem{318}
V.Yu.Irkhin and M.I.Katsnelson, Usp.Fiz.Nauk 164, 705 (1994)
[Sov.Phys.Uspekhi 37, 565 (1994)]

\bibitem{319}
K.Schwarz, O.Mohn, P.Blaha, J.Kuebler, J.Phys.F14, 2659 (1984)

\bibitem{320}
S.S.Jaswal, Phys.Rev.B41, 9697 (1990); S.S.Jaswal, W.B.Yelon,
G.C.Hadjipanayis et al., Phys.Rev.Lett. 67, 644 (1991)

\bibitem{321}
B.I.Min, J.-S.Kang, J.H.Hong et al., Phys.Rev.B48, 6317 (1993)

\bibitem{322}
S.K.Malik, P.J.Arlinghaus, W.E.Wallace, Phys.Rev.B25, 6488 (1982)

\bibitem{323}
J.Inoue and M.Shimizu, J.Phys.F15, 1511 (1985).

\bibitem{324}
S.Matar, P.Mohn, G.Demazeau and B.Siberchicot, J.de Phys.49, 1761

\bibitem{325}
S.Fujii, S.Ishida and S.Asano, J.Phys.:Cond.Mat.4, 1575 (1992)

\bibitem{326}
A.Sakuma, J.Phys.Soc.Jpn 60, 2007 (1991)

\bibitem{327}
Y.Noda and Y.Ishikawa. J.Phys.Soc.Jpn 40, 690, 699 (1976)

\bibitem{328}
K.Tajima and Y.Ishikawa, P.J.Webster et al. J.Phys.Soc.Jpn 43, 483
(1977)

\bibitem{329} M.I.Auslender and V.Yu.Irkhin, J.Phys.C18, 3533 (1985)

\bibitem{330} E.Kisker, G.Baum, A.Mahan, W.Raith and B.Reihl, Phys.Rev.B18, 2256
(1978)

\bibitem{331} M.J.Otto, R.A.M. van Woerden, P.J. van der Valk et al. J.Phys.:
Cond.Mat.1, 2351 (1989)

\bibitem{332} J.A.Hertz and D.M.Edwards, J.Phys.F3, 2174 (1973); D.M.Edwards and
J.A.Hertz, J.Phys.F3, 2191 (1973)

\bibitem{333} V.Yu.Irkhin and M.I.Katsnelson, Fiz.Tverd.Tela 25, 3383 (1983)
[Sov.Phys.-Solid State 25, 1947 (1983)]

\bibitem{334} G.L.Bona, F.Meier, M.Taborelli, E.Bucher and P.H.Schmidt, Solid State
Comm.56, 391 (1985)

\bibitem{335} T.Moriya, J.Phys.Soc.Jpn 19, 681 (1964)

\bibitem{336} E.A.Turov and M.P.Petrov, Nuclear Magnetic Resonance in Ferro- and
Antiferromagnets [in Russian], Moscow, Nauka, 1969

\bibitem{337} V.Yu.Irkhin and M.I.Katsnelson, Fiz.Metallov Metalloved. 66, 41 (1988)

\bibitem{338} V.Yu.Irkhin and M.I.Katsnelson, J.Phys.:Cond.Mat.2, 7151 (1990)

\bibitem{339} H.Enokiya, J.Phys.Soc.Jpn 31, 1037 (1971)

\bibitem{340} M.M.Matsuura, J.Phys.Soc.Jpn 21, 886 (1966)

\bibitem{341} K.Terakura, T.Oguchi, A.R.Williams and J.Kuebler, Phys.Rev.B30, 4734
(1984)

\bibitem{342} Z.Szotek, W.M.Temmerman and H.Winter, Phys.Rev.B47, 4029 (1993)

\bibitem{343} V.I.Anisimov, A.Zaanen and O.K.Andersen, Phys.Rev.B44, 943 (1991)

\bibitem{344} L.L.Chase, Phys.Rev.B10, 2226 (1974)

\bibitem{345} S.Lauer, A.X.Trautwein and F.E.Harris, Phys.Rev.B29, 6774 (1984)

\bibitem{346} G.L.Zhao, J.Callaway and M.Hayashibara, Phys.Rev.B48, 15781 (1994)

\bibitem{347} H.S.Jarrett, W.H.Cloud, H.J.Bouchard, S.R.Butler, C.G.Frederick and
J.L.Gillson, Phys.Rev.Lett.21, 617 (1968)

\bibitem{348} J.Kanamori, Progr.Theor.Phys.30, 275 (1963)

\bibitem{349} Y.Nagaoka, Phys.Rev.147, 392 (1966)

\bibitem{350} D.N.Khomsky, Fiz.Metallov Metalloved.29, 31 (1970)

\bibitem{351} P.B.Vissher, Phys.Rev.B10, 932, 943 (1974)

\bibitem{352} E.L.Nagaev, Physics of Magnetic Semiconductors, Moscow, Mir, 1983

\bibitem{353} M.I.Auslender, V.Yu.Irkhin and M.I.Katsnelson, J.Phys.C21, 5521 (1988)

\bibitem{354} J.Hubbard and K.P.Jain, J.Phys.C1, 1650 (1968)

\bibitem{355} A.Sakurai, Progr.Teor.Phys.39, 312 (1968)

\bibitem{356} J.J.Field, J.Phys C5, 664 (1972)

\bibitem{357} S.V.Tyablikov, Methods of Quantum Theory of Magnetism [in Russian],
Moscow, Nauka, 1975

\bibitem{358} G.Chrobok, M.Hofmann, G.Regenfus and P.Sizmann, Phys.Rev.B15, 429 (1975)

\bibitem{359} A.Vaterlaus, F.Millani and F.Meier. Phys.Rev.Lett.65, 3041 (1990)

\bibitem{360} S.A.Nikitin, A.S.Andreenko, A.K.Zvezdin and A.F.Popov,
Zh.Eksp.Teor.Fiz.76, 2159 (1979)

\bibitem{361} K.Yosida and A.Watabe, Progr.Theor.Phys.28, 361 (1962)

\bibitem{362} W.E.Evenson and S.V.Lin, Phys.Rev.178, 783 (1969)

\bibitem{363} R.J.Elliott and F.A.Wedgwood, Proc.Roy.Soc.81, 846 (1963); 84, 63 (1964)

\bibitem{364} P.G.de Gennes, Journ.Phys.Rad.23, 630 (1932)

\bibitem{365} H.Miwa, Proc.Phys.Soc.85, 1197 (1965)

\bibitem{366} A.J.Fedro and T.Arai, Phys.Rev.B6, 911 (1972)

\bibitem{367} V.Yu.Irkhin and M.I.Katsnelson, Z.Phys.B75, 67 (1989)

\bibitem{368} I.E.Dzyaloshinsky, Zh.Eksp.Teor.Fiz.46, 1420 (1964); 47, 336, 992 (1964)

\bibitem{369} J.Bohr and D.Gibbs, Physica A140, 349 (1986)

\bibitem{370} S.K.Godovikov, Thesis, Moscow State University, 1994

\bibitem{371} K.G.Gurtovoy and R.Z.Levitin, Usp.Fiz.Nauk, 153, 193 (1987)

\bibitem{372} P.G.Huray, S.E.Nave and R.G.Haire, J.Less-Common Met, 93, 393 (1983)

\bibitem{373} S.E.Nave et al, Physica B130, 225 (1985)

\bibitem{374} I.V.Solovyev, A.I.Liechtenstein, V.A.Gubanov, V.A.Antropov and
O.K.Andersen, Phys.Rev.B43, 14414 (1991)

\bibitem{375} I.V.Solovyev, Thesis, Sverdlovsk, 1991

\bibitem{376} E.A.Turov and A.I.Mitsek, Zh.Eksp.Teor.Fiz.37, 1127 (1959)

\bibitem{377} Ch.Kittel and J.H.Van Vleck, Phys.Rev.118, 1231 (1960)

\bibitem{378} H.B.Callen and E.R.Callen, J.Phys.Chem.Sol.27, 1271 (1966)

\bibitem{379} H.Brooks, Phys.Rev.58, 909 (1940)

\bibitem{380} E.I.Kondorsky and E.Straube, Pis'ma ZhETF 17, 41 (1973)

\bibitem{381} E.I.Kondorsky, Band Theory of Magnetism (in Russian), Moscow University
Press, 1976 (pt.1), 1977 (pt.2)

\bibitem{382} R.Gersdorf, Phys.Rev.Lett.40, 344 (1978)

\bibitem{383} G.H.O.Daalderop, P.J.Kelly and M.F.H.Schuurmans, Phys.Rev.B41, 11919
(1990)

\bibitem{384} O.Eriksson, G.W.Fernando, R.C.Albers and A.M.Boring, Solid State Comm.
78, 801 (1991); O.Eriksson, A.M.Boring, R.C.Albers, G.W.Fernando and
B.R.Cooper, Phys.Rev.B45, 2868 (1992)

\bibitem{385} H.J.F.Jansen, J.Appl.Phys.67, 4555 (1990)

\bibitem{386} D.C.Johnston and J.H.Cho, Phys.Rev.B42, 8710 (1990) (1990)

\bibitem{387} Yu.P.Irkhin, E.V.Rosenfeld and L.D.Finkelstein, Fiz.Tverd.Tela 35, 2769
(1993)

\bibitem{388} G.Di Castro, L.F.Feiner and M.Grilly, Phys.Rev.Lett.66, 3209 (1991)

\bibitem{389} Yu.P.Irkhin, V.V.Druzhinin and A.A.Kazakov, Zh.Eksper.Teor.Fiz. 54,
1183 (1968) [Sov.Phys.JETP 27, 633 (1968)].

\bibitem{390} S.K.Godovikov, M.G.Kozin, V.V.Turovtsev and V.S.Spinel,
phys.stat.sol.(b) 78, 103 (1976)

\bibitem{391} Yu.P.Irkhin and E.V.Rosenfeld, Fiz.Tverd.Tela 16, 485 (1974)

\bibitem{392} A.S.Ermolenko et al, Zh.Eksp.Teor.Fiz.69, 1743 (1975)

\bibitem{393} Yu.P.Irkhin, E.I.Zabolotskii and E.V.Rosenfeld, Fiz.Metallov
Metalloved.49, 1216 (1980)

\bibitem{394} Yu.P.Irkhin, Thesis, Sverdlovsk, 1968

\bibitem{395} V.N.Novogrudsky and I.G.Fakidov, Zh.Eksp.Teor.Fiz.47, 40 (1964)

\bibitem{396} E.A.Turov, V.G.Shavrov and Yu.P.Irkhin, Zh.Eksp.Teor.Fiz.47, 296 (1964)

\bibitem{397} S.S.Levina, V.N.Novogrudsky and I.G.Fakidov, Zh.Eksp.Teor.Fiz.45, 52
(1963)

\bibitem{398} E.A.Turov, Transport, Optical and Acoustical Properties of
Antiferromagnets (in Russian), Sverdlovsk, 1990

\bibitem{399} K.B.Vlasov et al, Fiz.Metallov Metalloved.42, 513 (1976);
Fiz.Tverd.Tela 22, 1656 (1980)

\bibitem{400} K.B.Vlasov et al, Fiz.Metallov Metalloved.44, 1005 (1977); 53, 493
(1982); Fiz.Tverd.Tela 24, 1338 (1982)

\bibitem{401} L.A.Boyarsky, Fiz.Nizkhikh temperatur, 22, 912 (1996).

\bibitem{402} R.Kubo, J.Phys.Soc.Jpn 12, 570 (1957)

\bibitem{403} H.Nakano, Progr.Theor.Phys.17, 145 (1957)

\bibitem{404} H.Mori, Progr.Theor.Phys., 34, 399 (1965)

\bibitem{405} V.Yu.Irkhin, M.I.Katsnelson and A.V.Trefilov, Physica C160, 397 (1989)

\bibitem{406} N.V.Volkenshtein, V.P.Dyakina and V.E.Startsev, phys.stat.sol.(b) 57, 9
(1973)

\bibitem{407} N.V.Volkenshtein et al, Fiz.Metallov Metalloved.67, 734 (1989)

\bibitem{408} G.Baber, Proc.Roy.Soc.A158, 383 (1937)

\bibitem{409} J.Appel, Phil.Mag.90, 1071 (1963)

\bibitem{410} M.J.Rice, Phys.Rev.Lett.20, 1439 (1968)

\bibitem{411} P.A.Schroeder, Physica B+C 109-110, 1901 (1982)

\bibitem{412} N.V.Volkenshtein, V.A.Novoselov and V.E.Startsev, Zh.Eksp.Teor.Fiz.60,
1078 (1971); V.E.Startsev, Thesis, Sverdlovsk, 1985

\bibitem{413} N.V.Volkenshtein et al, Fiz.Nizkikh Temp.10, 841 (1989)

\bibitem{414} F.J.Blatt and H.G.Satz, Helv.Phys.Acta, 33, 1007 (1960)

\bibitem{415} Yu.Kagan and A.P.Zhernov, Zh.Eksp.Teor.Fiz.50, 1107 (1966)

\bibitem{416} K.Ueda and T.Moriya, J.Phys.Soc.Jpn 39, 605 (1975)

\bibitem{417} A.H.MacDonald, Can.J.Phys.60, 710 (1982)

\bibitem{418} N.F.Mott, Proc.Roy.Soc.A153, 699 (1936); A157, 368 (1936); Adv.Phys.13,
325 (1964)

\bibitem{419} A.H.Wilson, Proc.Roy.Soc.A167, 580 (1938)

\bibitem{420} Yu.P.Irkhin, Fiz.Metallov Metalloved.6, 214, 586 (1958)

\bibitem{421} M.D.Wilding and E.W.Lee, Proc.Phys.Soc.85, 955 (1965)

\bibitem{422} T.Kasuya, Progr.Theor Phys. 16, 58 (1956)

\bibitem{423} Sh.Sh.Abelsky and E.A.Turov, Fiz.Metallov Metalloved.10, 801 (1960)

\bibitem{424} S.V.Vonsovsky and Yu.A.Izyumov, Usp.Fiz.Nauk, 58, 3 (1962)

\bibitem{425} E.A.Turov, Izv. AN SSSR (Ser.Fiz.) 19, 474 (1955)

\bibitem{426} T.Kasuya, Progr.Theor Phys. 22, 227 (1959)

\bibitem{427} M.Roessler, phys.stat.sol. 9, K27 (1965)

\bibitem{428} V.S.Lutovinov, M.Yu.Reizer and M.A.Savchenko, Zh.Eksp.Theor.Fiz.77, 707
(1979)

\bibitem{429} A.R.Mackintosh, Phys.Lett.4, 140 (1963)

\bibitem{430} R.V.Colvin and S.Arais, phys.stat.sol.4, 37 (1964)

\bibitem{431} A.A.Berdyshev and I.N.Vlasov, Fiz.Metallov Metalloved. 10, 628 (1960)

\bibitem{432} M.Roesler, phys.stat.sol. 8, K31 (1965)

\bibitem{433} V.Yu.Irkhin and M.I.Katsnelson, Phys.Rev.B52, 6181 (1995)

\bibitem{434} N.V.Ryzhanova, Sh.Sh.Abelsky and Yu.P.Irkhin, Fiz.Metallov
Metalloved.56, 843 (1983)

\bibitem{435} H.Ehrenreich and L.Schwartz, In: Solid State Physics, vol.31,
ed.F.Seitz and D.Turnbull, New York, Academic Press, 1976

\bibitem{436} I.A.Campbell and A.Fert, In: Ferromagnetic Materials, ed.P.Wohlfarth,
North-Holland, vol.3, 1982, p.747; A.Fert and I.A.Campbell, J.Phys.F6, 849
(1976)

\bibitem{437} Proc.First Int.Symp.on Magnetic Multilayers (Kyoto, 1993),
J.Magn.Magn.Mat.126 (1993)

\bibitem{438} M.V.Vedernikov, Adv.Phys.18, 337 (1969)

\bibitem{439} L.T.Rayevskaya, E.V.Rosenfeld and Yu.P.Irkhin, Fiz.Metallov
Metalloved.66, 73 (1988)

\bibitem{440} K.R.Krylov, A.I.Ponomarev, I.M.Tsidilkovsky et al, Phys.Lett.A131, 203
(1988)

\bibitem{441} R.C.Yu et al, Phys.Rev.B37, 7963 (1988)

\bibitem{442} V.Yu.Irkhin, M.I.Katsnelson and A.V.Trefilov, Europhys.Lett.15, 649
(1991); Zh.Eksp.Theor.Fiz.105, 1733 (1994)

\bibitem{443} C.M.Hurd, The Hall Effect in Metals and Alloys, Plenum, New York, 1972

\bibitem{444} M.J.Ziman, Phys.Rev.121, 1320 (1961)

\bibitem{445} T.Matsuda, J.Phys.Chem.Sol.30, 859 (1969); A.Hasegawa and T.Kasuya,
J.Phys.Soc.Jpn 28, 75 (1970)

\bibitem{446} N.V.Volkenshtein et al, Zh.Eksp.Teor.Fiz.89, 172 (1985)

\bibitem{447} E.I.Kondorsky, Sov.Phys.JETP 28, 1256 (1969)

\bibitem{448} E.Fawcett, Adv.Phys.13, 139 (1964)

\bibitem{449} E.M.Pugh, Phys.Rev.36, 1503 (1930)

\bibitem{450} I.K.Kikoin, Sow.Phys.9, 1 (1936)

\bibitem{451} N.V.Volkenshtein and G.V.Fedorov, Zh.Eksp.Teor.Fiz.38, 64 (1960)

\bibitem{452} T.Penney et al, J.Magn.Magn.Mat.54-57, 370 (1986)

\bibitem{453} F.Lapierre et al, J.Magn.Magn.Mat.63-64, 338 (1987)

\bibitem{454} V.Rudnitskii, Zh.Eksp.Teor.Fiz.9, 262 (1939); 10, 744 (1940)

\bibitem{455} R.Hulme, Proc.Roy.Soc.A135, 237 (1932)

\bibitem{456} A.G.Samoilovich and B.L.Kon'kov, Zh.Eksp.Teor.Fiz. 20, 783 (1950)

\bibitem{457} R.Karplus and J.M.Luttinger, Phys.Rev. 95, 1154 (1954)

\bibitem{458} W.Kohn and J.M.Luttinger, Phys.Rev.108, 590 (1957).

\bibitem{459} J.M.Luttinger, Phys.Rev.112, 739 (1958)

\bibitem{460} Yu.P.Irkhin and V.G.Shavrov, Zh.Eksp.Teor.Fiz.42, 1233 (1962)
[Sov.Phys.-JETP 15, 854 (1962)]

\bibitem{461} H.R.Leribaux, Phys.Rev.150, 384 (1966)

\bibitem{462} A.B.Granovskii and E.I.Kondorskii, Fiz.Metallov Metalloved. 39, 718
(1975)

\bibitem{463} A.N.Voloshinsky and L.F.Savitskaya, Zh.Eksp.Theor.Fiz.61, 2018 (1971)

\bibitem{464} Yu.P.Irkhin and Sh.Sh.Abelskii, Fiz.Metallov Metalloved.14, 641 (1962)

\bibitem{465} J.Kondo, Progr.Theor.Phys.27, 772 (1962); 28,846 (1962)

\bibitem{466} Yu.P.Irkhin and Sh.Sh.Abelskii, Fiz.Tverd.Tela 6, 1635 (1964)
[Sov.Phys. - Solid State 5, 1283 (1964)]

\bibitem{467} L.F.Savitskaya and A.N.Voloshinsky, Fiz.Metallov Metalloved.33, 849
(1972)

\bibitem{468} E.I.Kondorskii, Zh.Eksp.Teor.Fiz.48, 506 (1965)

\bibitem{469} L.Berger, Phys.Rev.B2, 4559 (1970); B5, 1862 (1972); B8, 2351 (1973);
P.Nozieres and C.Lewiner, J.Phys.(Paris) 34, 901 (1973)

\bibitem{470} S.K.Lyo and T.Holstein, Phys.Rev.Lett.29, 423 (1972)

\bibitem{471} Yu.Kagan and L.A.Maksimov, Fiz.Tverd.Tela 7, 530 (1965) [Sov.Phys. -
Solid State 7, 422 (1965)]

\bibitem{472} Yu.P.Irkhin, A.N.Voloshinskii and Sh.Sh.Abelskii, phys.stat.sol. 22,
309 (1967)

\bibitem{473} R.Huguenin and D.Rivier, Helv.Phys.Acta 38, 900 (1965)

\bibitem{474} E.I.Kondorskii, Zh.Eksp.Teor.Fiz.55, 558 (1968) [Sov.Phys.- JETP 28,
291 (1968)]; Zh.Eksp.Teor.Fiz.55, 2367 (1968) [Sov.Phys.- JETP 28, 1256
(1968)]

\bibitem{475} A.A.Abdurakhmanov, Transport effects in ferromagnetic metals (in
Russian), Rostov University Press, 1978

\bibitem{476} F.E.Maranzana, Phys.Rev.160, 421 (1967)

\bibitem{477} B.Giovannini, Proc.Conf. sur l'effet Hall extraordinaire, ed.G.Cohen et
al, University of Geneva, 1973, p.82

\bibitem{478} A.Fert, J.Phys.Lett.35, L107 (1974)

\bibitem{479} A.Fert and P.M.Levy, Phys.Rev.B36, 1907 (1987); P.M.Levy, Phys.Rev.B38,
6779 (1988)

\bibitem{480} N.O.Kourov, Yu.N.Tsiovkin and N.V.Volkenshtein, Fiz.Tverd.Tela 24, 1059
(1981)

\bibitem{481} J.Smit, Physica 17, 612 (1951); 21, 817 (1955); 24, 39 (1958)

\bibitem{482} V.Marsocci, Phys.Rev.137, A1842 (1965)

\bibitem{483} Vu Dinh Ky, phys.stat.sol.15, 739 (1966)

\bibitem{484} A.Kundt, Wied.Ann. 49, 257 (1893)

\bibitem{485} H.R.Hulme, Proc.Roy.Soc.A135, 237 (1952)

\bibitem{486} C.Kittel, Phys.Rev.53, A208 (1951)

\bibitem{487} P.N.Argyres, Phys.Rev.97, 334 (1965)

\bibitem{488} B.R.Cooper, Phys.Rev.139, A1504 (1965)

\bibitem{489} A.N.Voloshinsky and G.A.Bolotin, Fiz.Metallov Metalloved.17, 481 (1964)

\bibitem{490} L.E.Gurevich and I.N.Yassievich, Fiz.Tverd.Tela 6, 3341 (1964)

\bibitem{491} E.I.Kondorsky and A.V.Vedyaev, J.Appl.Phys.39, 559 (1968)

\bibitem{492} D.Weller et al, J.de Phys.Colloque C8, 49, 41 (1988)

\bibitem{493} A.V.Druzhinin, Yu.V.Knyazev and G.A.Bolotin, Fiz.Metallov Metalloved.
52, 1194 (1981)

\bibitem{494} R.A. de Groot, P.G. van Engen, P.P.J. van Engelen, K.H.J.Buschow,
J.Magn.Magn.Mat.86, 326 (1990)

\bibitem{495} J.H.Wijngaard, C.Haas, R.A.de Groot, Phys.Rev.B40, 9318 (1989)

\bibitem{496} H.Fuiji, H.Kawanaka, T.Takabatake et al, J.Phys.Soc.Jpn 58, 3481 (1989)

\bibitem{497} M.Yethirai, R.A.Robinson, J.J.Rhyne, J.A.Gotaas and K.H.J.Bushow,
J.Magn.Magn.Mat.79, 355 (1993)

\bibitem{498} G.S.Krinchik and M.V.Chetkin, Zh.Eksp.Teor.Fiz.36, 1924 (1959)

\bibitem{499} V.N.Mayevsky and G.A.Bolotin, Fiz.Metallov Metalloved. 40, 258 (1975)

\bibitem{500} A.V.Druzhinin, Thesis, Sverdlovsk, 1986

\bibitem{501} G.S.Krinchik and R.D.Nuralieva, Fiz.Metallov Metalloved.7, 697 (1959)

\bibitem{502} E.I.Kondorsky and R.P.Vasilyeva, Zh.Eksp.Teor.Fiz.45, 401 (1963)

\bibitem{503} E.I.Kondorsky, Zh.Eksp.Teor.Fiz.45, 511 (1963); 46, 2080 (1964)

\bibitem{504} A.N.Voloshinsky and N.V.Ryzhanova, Fiz.Metallov Metalloved. 40, 7
(1975); 41, 903 (1976)

\bibitem{505} P.A.Lee, T.M.Rice, J.W.Serene, L.J.Sham and J.W.Wilkins,
Comm.Cond.Mat.Phys.12, 99 (1986)

\bibitem{506} A.de Visser and J.J.M.Franse, J.Magn.Magn.Mat.100, 204 (1991)

\bibitem{507} G.R.Stewart, Rev.Mod.Phys. 56, 755 (1984)

\bibitem{508} S.Barth et al, Phys.Rev.B39, 11695 (1989)

\bibitem{509} H.Nakamura, Y.Kitaoka, K.Asayama and J.Floquet, J.Phys.Soc.Jpn 57, 2644
(1988)

\bibitem{510} D.Jaccard, J.Sierro, J.P.Brison and J.Floquet, J.de Phys. Coll.C8, 49,
741 (1988)

\bibitem{511} W.H.Wong and W.G.Clark, J.Magn.Magn.Mater.108, 175 (1992); O.Avenel,
J.S.Xia, B.Andraka et al., Phys.Rev.B45, 5695 (1992)

\bibitem{512} J.M.Lawrence, P.S.Riseborough and R.D.Parks, Rep.Progr.Phys.44, 1 (1981)

\bibitem{513} H.Nakamura, Y.Kitaoka, H.Yamada and K.Asayama, J.Magn.Magn.Mater.
76-77, 517 (1988); H.Nakamura, Y.Kitaoka, T.Iwai and K.Asayama,
J.Phys.Cond.Mat.4, 473 (1992)

\bibitem{514} Y.J.Uemura et al, Phys.Rev.B39, 4726 (1989); C.Tieng, Phys.Rev.B43, 83
(1991)

\bibitem{515} C.Jin et al, Bull.Am.Phys.Soc.36, 717 (1991)

\bibitem{516} E.Gratz, E.Bauer, B.Barbara et al, J.Phys.F15, 1975 (1985)

\bibitem{517} E.Bauer, E.Gratz and C.Schmitzer, J.Magn.Magn.Mater.63-64, 37 (1987).

\bibitem{518} F.R. de Boer et al, J.Magn.Magn.Mater.63-64, 91 (1987)

\bibitem{519} S.Takagi, T.Kimura, N.Sato, T.Satoh and T.Kasuya, J.Phys.Soc.Jpn, 57,
1562 (1988).

\bibitem{520} D.T.Adroya and S.K.Malik, J.Magn.Magn.Mat.100, 126 (1991) (see also
corresponding references therein)

\bibitem{521} D.Gignoux et al, J.Magn.Magn.Mat.74, 1 (1988)

\bibitem{522} D.T.Adroya, S.K.Malik, B.T.Padalia and R.Vijayaraghavan,
Sol.St.Comm.66, 1201 (1988)

\bibitem{523} Y.Nakazawa et al, Techn.Rep.ISSP A2663 (1993)

\bibitem{524} G.Aeppli, F.Bucher, C.Broholm, J.K.Kjems, J.Baumann and J.Hufnagl,
Phys.Rev.Lett. 60, 615 (1988)

\bibitem{525} T.T.M.Palstra et al., Phys.Rev.Lett.55, 2727 (1985)

\bibitem{526} J.K.Kjems and C.Broholm, J.Magn.Magn.Mat.76-77, 371 (1988)

\bibitem{527} R.N.Kleiman, D.J.Bishop, R.H.Ott, Z.Fisk and J.L.Smith, Phys.Rev.Lett.
64 1975 (1990)

\bibitem{528} A.de Visser et al, Phys.Rev.B45, 2962 (1992)

\bibitem{529} L.Degiorgi, H.R.Ott, M.Dressel, G.Gruener and Z.Fisk, Europhys.Lett.26,
221 (1994)

\bibitem{530} J.D.Thompson, Z.Fisk and H.R.Ott, J.Magn.Magn.Mat.54-57, 393 (1986)

\bibitem{531} E.Brueck et al Phys.Rev.B49, 8562 (1994)

\bibitem{532} C.Geibel et al, Z.Phys.B83, 305 (1991); A.Schroeder et al,
Phys.Rev.Lett.72, 136 (1994)

\bibitem{533} C.Geibel et al, Z.Phys.B84, 1 (1991)

\bibitem{534} E.Brueck, F.R. de Boer, V.Sechovsky and L.Havela, Europhys.Lett. 7, 177
(1988)

\bibitem{535} E.Sugiura et al, J.Magn.Magn.Mater.90-91, 65 (1990); H.Fujii et al,
J.Magn.Magn.Mater.90-91, 507 (1990)

\bibitem{536} N.Sato et al, J.Phys.Soc.Jpn 60, 757 (1991)

\bibitem{537} P.Bonville et al. J.Magn.Magn.Mater.76-77, 473 (1988)

\bibitem{538} T.Sakon, N.Sato, A.Oyamada et al. J.Phys.Soc.Jpn 62, 2209 (1992)

\bibitem{539} P.Bonville et al, J.Phys.:Cond.Mat.1, 8567 (1989)

\bibitem{540} M.Kasaya, T.Tani, K.Kawata et al, J.Phys.Soc.Jpn 60, 3145 (1991)

\bibitem{541} S.K.Dhar et al, Phys.Rev.B49, 641 (1994)

\bibitem{542} A.Ochiai, T.Suzuki and T.Kasuya, J.Magn.Magn.Mater.52, 13 (1985)

\bibitem{543} E.V. Sampathkumaran et al, Solid St.Comm.78, 971 (1991).

\bibitem{544} V.A.Stolyarov, L.S.Danelyan and L.Yu.Prokofyeva, Fiz.Tverdogo Tela 28,
2474 (1986).

\bibitem{545} N.B.Brandt and V.V.Moshchalkov, Adv.Phys.33, 373 (1984);
V.V.Moshchalkov and N.B.Brandt, Usp.Fiz.Nauk, 149, 585 (1986)

\bibitem{546} F.Steglich, J.Magn.Magn.Mat.100, 186 (1991)

\bibitem{547} E.Bauer, Adv.Phys.40, 417 (1991)

\bibitem{548} K.Miyake, T.Matsuura and C.M.Varma, Solid State Comm.71, 1149 (1989)

\bibitem{549} M.Springford and P.H.P.Reinders, J.Magn.Magn.Mat.76-77, 11 (1988)

\bibitem{550} G.Zwicknagl, Adv.Phys.41, 203 (1992)

\bibitem{551} P.Fulde, J.Keller and G.Zwicknagl, Solid State Physics, vol.41,
ed.F.Seitz et al., New York, Academic Press, 1988, p.1.

\bibitem{552} J.Kondo, Solid State Physics, vol.23, ed.F.Seitz and D.Turnbull, New
York, Academic Press, 1969, p.183

\bibitem{553} P.Nozieres, J.Low Temp.Phys.17, 31 (1974)

\bibitem{554} P.W.Anderson, J.Phys.C3, 2346 (1970)

\bibitem{555} P.Nozieres and A.Blandin, J.Phys.(Paris) 41, 193 (1980)

\bibitem{556} K.G.Wilson, Rev.Mod.Phys.47, 773 (1975);

\bibitem{557} N.Andrei, K.Furuya and J.H.Loewenstein, Rev.Mod.Phys.55, 331 (1983)

\bibitem{558} A.M.Tsvelick and P.B.Wiegmann, Adv.Phys.32, 745 (1983)

\bibitem{559} M.D.Daybell and W.A.Steyert, Rev.Mod.Phys.40, 380 (1968);
J.Appl.Phys.40, 1056 (1969)

\bibitem{560} A.J.Heeger, Solid State Physics, vol.23, ed.F.Seitz and D.Turnbull, New
York, Academic Press, 1969, p.283

\bibitem{561} J.R.Schrieffer, J.Appl.Phys. 38, 1143 (1967)

\bibitem{562} L.L.Hirst, Z.Phys.244, 230 (1971); Int.J.Magn. 2, 213 (1972);
Adv.Phys.21, 295 (1972)

\bibitem{563} V.Yu.Irkhin and Yu.P.Irkhin, Zh.Eksp.Theor.Fiz., 107, 616 (1995).

\bibitem{564} J.S.Griffith, The Theory of Transition-Metal Ions, Cambridge,
University Press, 1961; The Irreducible Tensor Method for Molecular Symmetry
Groups, New Jersey, Prentice-Hall, Englewood Cliffs, 1962

\bibitem{565} D.M.Newns and N.Read, Adv.Phys.36, 799 (1987)

\bibitem{566} F.G.Aliev et al, J.Magn.Magn.Mat.76-77, 295 (1988); Phys.Rev.B47, 729
(1993)

\bibitem{567} M.Kyogaku et al, J.Phys.Soc.Jpn 59, 1728 (1990)

\bibitem{568} T.Takabatake et al, J.Magn.Magn.Mat.76-77, 87 (1988)

\bibitem{569} S.K.Malik and D.T.Adroya, Phys.Rev.B43, 6277 (1991)

\bibitem{570} B.Andraka, Phys.Rev.B49, 348 (1994)

\bibitem{571} P.C.Canfield et al, J.Appl.Phys.70, 5800 (1991)

\bibitem{572} M.F.Hundley et al, Phys.Rev.B42, 6842 (1990); A.Severing, J.D.Thompson,
P.C.Canfield and Z.Fisk, Phys.Rev.B44, 6832 (1991)

\bibitem{573} W.F.Brinkman and S.Engelsberg, Phys.Rev.169, 417 (1968)

\bibitem{574} V.Yu.Irkhin and M.I.Katsnelson, Z.Phys.B70, 371 (1988)

\bibitem{575} M.E.Fisher, J.S.Langer, Phys.Rev.Lett.20, 665 (1968)

\bibitem{576} J.Sakurai, J.Matsuura and Y.Komura, J.Phys.Colloque C8, 49, 783 (1988)

\bibitem{577} Y.Onuki et al, J.Phys.Colloque C8, 49, 481 (1988)

\bibitem{578} T.Ohyama, J.Sakurai and Y.Komura, Sol.St.Comm.60, 975 (1986)

\bibitem{579} G.Sparn et al, J.Magn.Magn.Mat.47-48, 521 (1985)

\bibitem{580} U.Gottwick et al, J.Magn.Magn.Mat.63-64, 341 (1987)

\bibitem{581} P.Coleman, Phys.Rev.B29, 3035 (1984)

\bibitem{582} Y.Ono, T.Matsuura and Y.Kuroda, Physica C159, 878 (1989)

\bibitem{583} F.Marabelly and P.Wachter, J.Magn.Magn.Mat.70, 364 (1989)

\bibitem{584} D.I.Khomsky, Usp.Fiz.Nauk, 129, 443 (1979)

\bibitem{585} B.Johansson, Phil.Mag.34, 469 (1974)

\bibitem{586} D.Wielizka, I.H.Wesver, D.W.Lynch and C.G.Olson, Phys.Rev.B26, 7056
(1983)

\bibitem{587} R.Podloucky and D.Gloetzel, Phys.Rev.B27, 3390 (1983)

\bibitem{588} L.D.Finkelstein, Fiz.Metallov Metalloved., 57, 402 (1984)

\bibitem{589} B.Johansson, J.Magn.Magn.Mat.47-48, 231 (1985)

\bibitem{590} V.Yu.Irkhin and M.I.Katsnelson, Zh.Eksp.Teor.Fiz.90, 1080 (1986)
[Sov.Phys.JETP 63, 631 (1986)]; Solid State Comm. 58, 881 (1986)

\bibitem{591} G.Travaglini and P.Wachter, Phys.Rev.B30, 893, 5877 (1984); P.Wachter
and G.Travaglini, J.Magn.Magn.Mat.47-48, 423 (1985)

\bibitem{592} A.Amato and J.Sierro, J.Magn.Magn.Mat.47-48, 475 (1985)

\bibitem{593} P.Strange, J.Phys.C17, 4273 (1984); F.Lopez-Aguilar and
J.Costa-Quintana, J.Phys.C19, 2185 (1986)

\bibitem{594} B.Bucher, P.Steiner and P.Wachter, Phys.Rev.Lett.67, 2717 (1991)

\bibitem{595} S.S.Bader, N.E.Phillips and D.B.McWhan, Phys.Rev.B7, 4686 (1973)

\bibitem{596} T.S.Altshuler, V.N.Mironov, G.G.Khaliullin and D.I.Khomsky, Pis'ma
ZhETF 40, 28 (1984) [JETP Lett. 40, 754 (1984)]

\bibitem{597} V.Yu.Irkhin and M.I.Katsnelson, Fiz.Tverd.Tela 29, 1461 (1987)

\bibitem{598} Yu.P.Irkhin, Pis'ma ZhETF 32, 205 (1980)

\bibitem{599} K.Sugiyama et al, J.Magn.Magn.Mat.52, 283 (1985)

\bibitem{600} M.Kasaya, F.Ida, N.Takigawa and T.Kasuya, J.Magn.Magn.Mat.47-48, 429
(1985)

\bibitem{601} V.Yu.Irkhin and M.I.Katsnelson, Fiz.Metallov Metalloved.N1, 16 (1991)

\bibitem{602} P.Frings, B.Renker and C.Vettier, J.Magn.Magn.Mater. 63-64, 202 (1987)

\bibitem{603} Y.Onuki et al, J.Magn.Magn.Mater.76-77, 119 (1989); B.Luthi,
P.Thalmeier, G.Bruls and D.Weber, J.Magn.Magn.Mater.90-91, 37 (1990)

\bibitem{604} M.A.Continentino, Phys.Rev.B47, 11587 (1993)

\bibitem{605} C.Lacroix, J.Magn.Magn.Mater.100, 90 (1991)

\bibitem{606} S.M.M.Evans, J.Phys.:Cond.Mat.3, 8441 (1991)

\bibitem{607} S.V.Vonsovsky, V.Yu.Irkhin and M.I.Katsnelson, Physica B163, 321 (1990)

\bibitem{608} V.Yu.Irkhin and M.I.Katsnelson, Z.Phys.B82, 77 (1991)

\bibitem{609} S.V.Vonsovsky, V.Yu.Irkhin and M.I.Katsnelson, Physica B171, 135 (1991)

\bibitem{610} P.Coleman, Physica B171, 3 (1991)

\bibitem{611} P.Fazekas and E.Mueller-Hartmann, Z.Phys.B85, 285 (1991)

\bibitem{612} V.Yu.Irkhin and M.I.Katsnelson, J.Phys.:Cond.Mat.4, 9661 (1992);
Phys.Rev.B56, 8109 (1997)

\bibitem{613} H.Yashima, N.Sato, H.Mori and T.Satoh, Solid State Comm.43, 595 (1982);
S.K.Dhar, K.A.Gschneidner (Jr.), W.H.Lee, P.Klavins and R.N.Shelton,
Phys.Rev.B36, 341 (1987)

\bibitem{614} S.V.Vonsovsky, Yu.P.Irkhin, V.Yu.Irkhin and M.I.Katsnelson,
J.Phys.Colloque C8, 49, 253 (1988)

\bibitem{615} M.Kyogaku, Y.Kitaoka, K.Asayama, T.Takabatake and K.Fujii,
J.Phys.Soc.Jpn 61, 43 (1992); A.Kratzner et al, Europhys.Lett.19, 649 (1992)

\bibitem{616} R.Kuentzler, R.Claud, G.Schmerber and Y.Dossmann,
J.Magn.Magn.Mat.104-107, 1976 (1992)

\bibitem{617} J.G.Bednorz and K.A.Mueller, Z.Phys.B64, 189 (1986)

\bibitem{618} M.K.Wu et al, Phys.Rev.Lett.58, 908 (1987)

\bibitem{619} F.C.Zhang, T.M.Rice, Phys.Rev.B37, 3759 (1988)

\bibitem{620} V.Yu.Irkhin and M.I.Katsnelson, J.Phys.:Cond.Mat.3, 6439 (1991)

\bibitem{621} R.J.Birgeneau, H.J.Guggenheim and G.Shirane, Phys.Rev.B1, 2211 (1970);
R.J.Birgeneau, D.R.Gabbe, H.P.Jenssen et al, Phys.Rev.B38, 6614 (1988)

\bibitem{622} P.Arovas and A.Auerbach, Phys. Rev.B38, 316 (1988)

\bibitem{623} D.Yoshioka, J.Phys.Soc.Jpn 58, 3393 (1989)

\bibitem{624} M.Takahashi, Phys.Rev.B40, 2494 (1989)

\bibitem{625} C.L.Kane, P.A.Lee and N.Read, Phys.Rev.B39, 6880 (1989)

\bibitem{626} K.J. von Szczepanski, P.Horsch, W.Stephan and M.Ziegler, Phys.Rev.B41,
2017 (1990)

\bibitem{627} A.Kebede, C.S.See, J.Schwegler et al, Phys.Rev.B40, 4453 (1989)

\bibitem{628} J.L.Peng, R.N.Shelton and H.B. Radonsky, Phys.Rev.B41, 187 (1989)

\bibitem{629} T.Brugger, T.Schreiner, G.Roth, P.Adelman and G.Czjzek, Phys.Rev.Lett.
71, 2481 (1991)

\bibitem{630} C.C.Yu and P.W.Anderson, Phys.Rev.B29, 6165 (1984)

\bibitem{631} P.W.Anderson, in Frontiers and Borderlines in Many Particle Physics,
ed.J.Schrieffer and R.Broglia, North-Holland, 1988, p.1

\bibitem{632} S.Kivelson, Phys.Rev.B36, 7237 (1987)

\bibitem{633} P.W.Anderson, Int.J.Mod.Phys.4, 181 (1990)??; Yu.A.Izymov,
M.I.Katsnelson and Yu.N.Skryabin, Itinerant Electron Magnetism [in Russian],
Moscow, Nauka, 1994

\bibitem{634} V.Yu.Irkhin, A.A.Katanin and M.I.Katsnelson, J.Phys.:Cond.Mat. 4, 5227
(1992)

\bibitem{635} V.Yu.Irkhin and M.I.Katsnelson, Pis'ma ZhETF 49, 500 (1989);
Phys.Lett.A150, 47 (1990)

\bibitem{636} J.Deportes, B.Ouladdiaf and K.R.A.Ziebeck, J.Magn.Magn.Mat.70, 14 (1987)

\bibitem{637} H.Wada et al, J.Magn.Magn.Mat.70, 17 (1987); M.Shiga,
J.Magn.Magn.Mat.129, 17 (1994)

\bibitem{638} G.Krill et al, J.Phys.C9, 761 (1976); S.Sudo, J.Magn.Magn.Mat.114, 57
(1992)

\bibitem{639} T.Furuno, K.Ando, S.Kunii et al, J.Magn.Magn.Mat.76-77, 117 (1988);
K.Fraas, U.Ahlheim, P.H.P.Reinders et al, J.Magn.Magn.Mater.108, 220 (1992)

\bibitem{640} O.Nakamura, A.Oyamada, A.Ochiai, S.Kunii, T.Takeda, T.Suzuki and
T.Kasuya, J.Magn.Magn.Mater.76-77, 293 (1988)

\bibitem{641}
Z.Fisk et al, Phys.Rev.Lett.67, 3310 (1991)

\bibitem{642}
J.Rossat-Mignod et al, Phys.Rev.B16, 440 (1977); P.Fisher et al,
J.Phys.C11, 345 (1978)

\bibitem{643}
M.Sera, T.Fujita, T.Suzuki and T.Kasuya, In: Valence Instabilities,
North-Holland, 1982, p.435

\bibitem{644}
P.W.Anderson, Phys.Rev.Lett., 64, 239 (1990); 65, 2306 (1990); 66, 3226
(1991); 71, 1220 (1993)

\bibitem{645}
C.M.Varma, P.B.Littlewood, A.Schmitt-Rink, E.Abrahams and
A.E.Ruckenstein, Phys.Rev.Lett.63, 1996 (1989); P.B.Littlewood and
C.M.Varma, Phys.Rev.B46, 405 (1992)

\bibitem{646}
M.B.Maple et al, J.Low Temp.Phys.95, 225 (1994)

\bibitem{647}
A.A.Romanyukha, Yu.N.Shvachko, V.Yu.Irkhin, M.I.Katsnelson, A.A.Koshta
and V.V.Ustinov, Physica C171, 276 (1990)

\bibitem{648}
J.M.Luttinger, J.Math.Phys.15, 609 (1963); G.Mahan, Many-Particle
Physics, New York, Plenum Press (1981)

\bibitem{649}
F.D.M.Haldane, J.Phys.C12, 4791 (1979)

\bibitem{650}
A.H.Castro Neto and E.Fradkin, Phys.Rev.B49, 10877 (1994) A.Houghton,
H.-J.Kwon and J.B.Marston, Phys.Rev.B50, 1351 (1994)

\bibitem{651}
N.N.Bogoliubov, Lectures in Quantum Statistics. In: Collected works,
vol.2, Kiev, Naukova Dumka, 1970 (in Russian)

\bibitem{652}
Yu.P.Irkhin, Zh.Eksper.Teor.Fiz.50, 379 (1966) [Sov.Phys.JETP 23, 253
(1966)].

\bibitem{653}
V.Yu.Irkhin and Yu.P.Irkhin, Fiz.Metallov Metalloved. 76, N4, 49 (1993)

\bibitem{654}
V.Yu.Irkhin and Yu.P.Irkhin, phys.stat.sol.(b) 183, 9 (1994)

\bibitem{655}
Yu.P.Irkhin, Zh.Eksper.Teor.Fiz.66, 1005 (1974)

\bibitem{656}
D.C.Mattic, The Theory of Magnetism, New York, Harper and Row, 1965

\bibitem{657}
V.A.Popov and A.A.Loginov, phys.stat.sol.(b) 84, 83 (1977)

\bibitem{658}
C.W.Nielson and G.F.Koster, Spectroscopic coefficients for the p$^n$, d$^n$,
f$^n$-configurations, Cambridge, 1963

\bibitem{659}
B.G.Wybourne, Symmetry Principles and Atomic Spectroscopy, Wiley, New
York, 1970

\bibitem{660}
V.Yu.Irkhin and Yu.P.Irkhin, Zh.Eksper.Teor.Fiz.104, 3868 (1993)
[Sov.Phys.JETP 77, 858 (1993)]

\bibitem{661}
P.W.Anderson, Solid State Physics, vol.14, ed.F.Seitz and D.Turnbull,
New York, Academic Press, 1963, p.99

\bibitem{662}
S.P.Shubin and S.V.Vonsovsky, Proc. Roy.Soc.A145, 159 (1934)

\bibitem{663}
S.V.Vonsovsky and M.I.Katsnelson, J.Phys.C12, 2043 (1979)

\bibitem{664}
H.Bethe, Intermediate Quantum Mechanics, Benjamin, New York, 1964

\bibitem{665}
A.B.Bolotin and V.K.Shugurov, Zh.Vych.Mat.Mat.Fiziki, 3, 560 (1963);
H.J.Silverstone, J.Chem.Phys. 47, 537 (1967)

\bibitem{666}
P.M.Levy, Phys.Rev. 177, 509 (1969)

\bibitem{667}
V.V.Druzhinin and A.S.Moskvin, Fiz.Metallov Metalloved. 26, 415 (1968)

\bibitem{668}
E.L.Nagaev, Usp.Fiz.Nauk, 136, 61 (1989)

\bibitem{669}
T.Moriya, Phys.Rev.120, 91 (1960)

\bibitem{670}
R.J.Elliott and M.F.Torpe, J.Appl.Phys. 39, 802 (1968)

\bibitem{671}
D.N.Zubarev, Usp.Fiz.Nauk 71, 71 (1960)

\bibitem{672}
V.G.Bar'yakhtar, V.N.Krivoruchko and D.A.Yablonsky, Solid St.Comm.46,
613 (1983); Zh.Eksp.Teor.Fiz.85, 602 (1983)

\bibitem{673}
R.O.Zaitsev, Zh.Eksp.Teor.Fiz.68, 207 (1975)

\bibitem{674}
V.V.Valkov and S.G.Ovchinnikov, Teor.Mat.Fiz. 50, 466 (1982)

\bibitem{675}
V.V.Valkov and T.A.Valkova, Teor.Mat.Fiz. 59, 453 (1984)

\bibitem{676}
V.V.Valkov, T.A.Valkova and S.G.Ovchinnikov, Zh.Eksp.Teor.Fiz. 88, 550
(1985); phys.stat.sol.(b) 142, 255 (1987)

\bibitem{677}
F.P.Onufrieva, Zh.Eksp.Teor.Fiz.80, 2372 (1981); 86, 1691 (1984); 89,
2270 (1985); 94, 232 (1988)

\bibitem{678}
I.Affleck, J.Phys.:Cond.Mat. 1, 3047 (1989)

\bibitem{679}
E.V.Rosenfeld, Pis'ma ZhETF, 24, 60 (1976)

\bibitem{680}
M.I.Kaganov and A.V.Chubukov, Usp.Fiz.Nauk, 153, 537 (1987)

\bibitem{681}
K. von Klitzing, G.Dorda and M.Pepper, Phys.Rev.Lett.45, 494 (1980)

\bibitem{682}
V.P.Silin and A.Z.Solontsov, Fiz.Metallov Metalloved.58, 1080 (1984)

\bibitem{683}
M.I.Auslender, V.Yu.Irkhin, Z.Phys.B56, 301 (1984); B61, 129 (1985)

\bibitem{684}
H.Wada, M.Nishigori and M.Shiga, J.Phys.:Cond.Mat.3, 2083 (1991)

\bibitem{685}
G.M.Carneiro and C.J.Pethick, Phys.Rev.B11, 1106 (1975)

\bibitem{686}
F.Rys, J.C.Helman, W.Baltensperger, Phys.Kond.Mat.6, 105 (1967)

\bibitem{687}
V.Yu.Irkhin, Fiz.Tverd.Tela 28, 3066 (1986)

\bibitem{688}  S.Jin et al, Science 264, 413 (1994)

\bibitem{689}
V.Yu.Irkhin and M.I.Katsnelson, Fiz.Metallov Metalloved.65, 446 (1988)

\bibitem{690}
N.A.Cade and W.Young, Adv.Phys.26, 393 (1977)

\bibitem{691}
M.C.K.Wiltshire, M.M.Elcombe and C.J.Howard, J.Phys.F15, 1595 (1985)

\bibitem{692}
M.I.Katsnelson and V.Yu.Irkhin, J.Phys.C17, 4291 (1984)

\bibitem{693}
V.Yu.Irkhin and A.M.Entelis, J.Phys.:Cond.Mat.1, 4111 (1989)

\bibitem{694}
A.O.Anokhin and V.Yu.Irkhin, phys.stat.sol.(b) 165, 129 (1991)

\bibitem{695}
A.O.Anokhin, V.Yu.Irkhin and M.I.Katsnelson, J.Phys.:Cond.Mat.3, 1475
(1991)

\bibitem{696}
L.A.Maksimov and K.A.Kikoin, Fiz.Metallov Metalloved.28, 43 (1969)

\bibitem{697}
R.O.Zaitsev, Zh.Eksp.Teor.Fiz. 75, 2362 (1978)

\bibitem{698}
M.Sh.Erukhimov and S.G.Ovchinnikov, phys.stat.sol.(b) 123, 105 (1984)

\bibitem{699}
V.Yu.Irkhin and M.I.Katsnelson, Zh.Eksp.Teor.Fiz.88, 522 (1985)

\bibitem{700}
V.Yu.Irkhin and M.I.Katsnelson, J.Phys.C18, 4173 (1985)

\bibitem{701}
P.B.Wiegmann, Phys.Rev.Lett.60, 821 (1988); Y.Chen, D.Foerster and
P.Larkin, Phys.Rev.B46, 5370 (1992)

\bibitem{702}
K.Kubo and N.Ohata, J.Phys.Soc.Jpn, 33, 21 (1972)

\bibitem{703}
O.Narikiyo, K.Kuboki and H.Fukuyama, J.Phys.Soc.Jpn 59, 2443 (1990)

\bibitem{704}
V.V.Druzhinin and Yu.P.Irkhin, Zh.Eksp.Teor.Fiz.51, 1856 (1966)
[Sov.Phys.JETP 24, 1250 (1967)]

\bibitem{705}
A.Georges and G.Kotliar, Phys.Rev.B45, 6479 (1993); X.Y.Zhang and
C.M.Zhang, Phys.Rev.B49, 7929 (1994); M.J.Rozenberg, G.Kotliar and
X.Y.Zhang, Phys.Rev.B49, 10181 (1994); F.J.Ohkawa, J.Phys.Soc.Jpn 61, 1615
(1992)

\bibitem{706}
R.Karplus and J.Schwinger, Phys.Rev.73, 1020 (1948)

\bibitem{707}
J.M.Luttinger and W.Kohn, Phys.Rev.109, 1892 (1958)

\bibitem{708}
L.L.Hirst, Adv.Phys. 27, 231 (1978)

\bibitem{709}
Yu.P.Irkhin, Fiz.Tverd.Tela 30, 1202 (1988)

\bibitem{710}
K.A.Kikoin and L.A.Maksimov, Zh.Eksp.Teor.Fiz. 58, 2184 (1970)

\bibitem{711}
P.Coleman and N.Andrei, J.Phys.:Cond.Mat.1, 4057 (1989)

\bibitem{712}
S.Chakravarty, B.I.Halperin and D.R.Nelson, Phys.Rev.B39, 2344 (1989)

\bibitem{713}
V.Yu.Irkhin, A.A.Katanin and M.I.Katsnelson, Fiz.Metallov Metalloved.,
79, N1, 65 (1995); Phys.Lett.A157, 295 (1991);

\bibitem{714}
F.D.M.Haldane, Phys.Rev.Lett.50, 1153 (1983)

\bibitem{715}
V.Yu.Irkhin and M.I.Katsnelson, Z.Phys.B62, 201 (1986)

\bibitem{716}
V.Yu.Irkhin and M.I.Katsnelson, Phys.Rev.B53, 14008 (1996)

\bibitem{717}
V.Yu.Irkhin and Yu.P.Irkhin, J.Magn.Magn.Mater.164, 119 (1996)

\bibitem{718}
M.Akai, H.Akai and J.Kanamori, J.Phys.Soc.Jpn 54, 4246,4257 (1985)

\bibitem{719}
R.Zeller, J.Phys.F17, 2123 (1987).

\bibitem{720}
N.Stefanou, A.Oswald, R.Zeller and P.H.Dederichs, Phys.Rev.B35, 6911 (1987).

\bibitem{721}
B.Drittler, N.Stefanou, S.Bluegel, R.Zeller and P.H.Dederichs, Phys.Rev.B40,
8203 (1989)

\bibitem{722}
P.H.Dederichs, R.Zeller, H.Akai, H.Ebert, J.Magn.Magn.Mater.100, 241 (1991)

\bibitem{723}
I.Mertig, E.Mrosan and R.Schoepke, J.Phys.F12, 1689 (1982); I.Mertig,
E.Mrosan, R.Zeller and P.H.Dederichs, phys.stat.sol.(b) 117, 619 (1983)

\bibitem{724}
I.Mertig, E.Mrosan, R.Zeller and P.H.Dederichs, Proc.Int.Conf. on the Physics
of Transition Metals (Darmstadt, July 1992), World Scientific, 1992, p.778.

\bibitem{725}
M.D.Daybell, In: Magnetism, vol.5, ed.H.Suhl, New York, Academic Press, 1973,

\bibitem{726}
A.R.Miedema, J.F.Dorleijn, J.Phys.F5, 487,1543 (1975); O.Jaoul, I.A.Campbell
and A.Fert, J.Magn.Magn.Mater.5, 23 (1977); T.Farrell and D.Greig, J.Phys.C3,
138 (1970); F.Mauri, A.Lucasson, P.Lucasson, P.Moser, F.Faudot,
J.Phys.:Cond.Mat.2, 3269 (1990)

\bibitem{727}
V.Yu.Irkhin, Phys.Rev.B57, 13375 (1998)

\bibitem{728}
V.Yu.Irkhin, A.A.Katanin, Phys.Rev.B55, 12318 (1997); Phys.Rev.B57,
379 (1998)


\end{thebibliography}
\end{document}